\documentclass[12pt, oneside]{book}
 \usepackage{latexsym}
 \usepackage{amsmath} % it subsumes: amstext, amsbsy, amsopn
\usepackage[dvips]{graphics}
\usepackage{latexsym,psfig}
\usepackage{amscd}     \usepackage{amsxtra}
\usepackage{upref}     \usepackage{amsthm}
\usepackage{amssymb}   %\usepackage{mathrsfs}
\usepackage{amsfonts}
\usepackage{makeidx}
\makeindex

\begin{document}

\title{{\bf GEOMETRIC VIEW\\
\vskip 0.4cm
 ON\\
\vskip 0.4cm
 PHOTON-LIKE\\
\vskip 0.4cm
OBJECTS}}
%\newpage
%\author{{\bf Stoil Donev}\footnote{e-mail:
\author{{\bf Stoil Donev}\footnote{e-mail:
 sdonev@inrne.bas.bg}, {\bf Maria Tashkova}, \\ Institute for Nuclear Research
and Nuclear Energy,\\ Sofia,  Bulgaria\\}

\date{}
\maketitle

%\newpage
%\begin{center}
%{\bf STOIL DONEV \ \ MARIA TASHKOVA} \\
 %Institute for Nuclear Research and Nuclear Energy,\\ BULGARIA
%\end{center}

\newpage
\vskip 4cm
\begin{flushright}
%\begin{center}
%\hfill\fbox{
%    \begin{minipage}{0.97\textwidth}
%\begin{center}
%\vskip 0.3cm
%\sc{\bf MAY EVERY SCIENTIFIC TRUTH BE RESPECTED,\\
%\vskip 0.3cm
%BUT NO ONE BE TURNED INTO DOGMA}.
%\end{center}
%\vskip 0.3cm
%\end{minipage}} \hfill \end{center}

\sc{\bf MAY EVERY SCIENTIFIC TRUTH BE RESPECTED, \\
\vskip 0.3cm
BUT NO ONE BE TURNED INTO DOGMA}.
\end{flushright}
\vskip 4cm
%{\bf OUR SCIENTIFIC CREED} :
\vskip 0.5cm
\noindent

\pagenumbering{roman}
\frontmatter
\section*{Preface}
This book aims to summarize in a consistent way the authors' results in
attempting to build spatially finite and time-stable models of photon-like
objects through extending Maxwell vacuum equations to local energy-momentum
exchange relations and making use of modern differential geometry. In
particular, we interpret dynamically Frobenius integrability theory of
distributions on manifolds through an appropriate $\varphi$-extension along
$p$-vector fields of the classical Lie derivative, and give interaction
interpretation of the nonintegrability of subdistributions of an integrable
distribution recognizing physically these subdistributions as time-stable
subsystems of the field object considered and formally presented by the
integrable distribution. The space-time propagation of our photon-like object
is, of course, along appropriate symmetry of the representing distribution.

Such a purpose presumes some clarifying of the following two initial moments,
one from mathematical viewpoint, and one from physical viewpoint. These two
moments are related to the mathematical concept of {\it equality}.

Let's begin with the mathematical one.

Equality : $"="$, is one of the basic concepts used almost everywhere in
mathematics. Generally speaking, it means that on
the two sides of $"="$ stays the {\it same thing}. Formally this concept
means that [1] {\it the symbols on the two sides of $"="$ represent the same
element}, where the concept of "element" is understood as it is introduced in
set theory [1]: {\it a set consists of elements which are capable of possessing
certain properties and having certain relations between themselves or with
elements of other sets}. This viewpoint allows to define an element in various
ways, since the same element may be considered as element of different sets.
For example, the real number $2$ can be expressed as
$3-1=\frac63=2sin(\frac{\pi}{2})=\dots$. The important moment here is that if
we are preliminary sure about this, we could write down equation in the sense
$3-x=\frac63$, so our initial categorical confidence that on the two sides
stays the same element leads to the conclusion $x=1$. This is appropriately
extended, of course, to equations/relations where we declare equality of
mathematical quantities of more complicated structure like tensor fields,
differential operators, etc.

Let's turn now to physics.

In theoretical physics the basic two concepts are {\it physical object} and
{\it physical interaction}, i.e. interaction, or mutual
influence, between/among physical objects. More or less, "physical objects"
are mathematically interpreted as elements, and "physical interactions" are
mathematically interpreted as interdependences between/among the corresponding
mathematical images of the physical objects considered.

An important  "difference" between mathematical elements and physical objects
is that the mathematical objects are {\it indestructible} in nature, their
existence has nothing to do with time, while physical objects {\it are
destructible} in nature, and may transform to different ones, so, they are
time-existent entities. This motivates to work in theoretical physics not
directly with "physical objects", but with the mathematical image of their {\it
physical appearance}, also, with mathematical images of
corresponding physical characteristics of the objects considered, these
mathematical images are called {\bf physical quantities}, they represent
their {\it dynamical appearance}. Therefore, physical quantities may have
various mathematical structure: scalars, tensor fields, curvature forms, etc.,
but behind their mathematical nature always stays their {\it physical nature}.
So, any relation in theoretical physics must duly respect these both aspects.
And the adequate understanding here does not reduce to the physical dimension
of the quantities we are going to equalize, although in particular cases
physicists make appropriate compromises putting some dimensional constant on
the one side of the relation in question, for example: $\Delta\varphi=k\rho$,
$\mathrm{rot}\mathbf{E}=k\,\mathbf{j}$. If we ask, for example, "which physical
quantity can be expressed, first, as $\Delta\varphi$, and, second, as $k\rho$",
we could hardly find satisfactory answer. This aspect of any relation in
theoretical physics should not be neglected, and this is very important when
the relation has the sense of {\it equation}, i.e. when an unknown quantity is
meant to be determined by this relation.

In conclusion, we must be {\bf preliminary sure}, that writing down equality
having physical sense, on the two sides of $"="$ we are putting the same
quantity from {\it quantitative} and form {\it qualitative} point of view, e.g.
{\it mass, energy, momentum change, energy flow}, etc. Therefore, the more
universal is a physical quantity, the more useful from theoretical point of
view it is, because, we are able to express it in various ways. This suggests
that a reliable principle of theoretical physics should be: {\it find out such
physical quantities, separate the conservative ones, and when dynamical
equations will be of interest, write down corresponding local balance relations
for them}.

The above principle suggests to pay due respect to the general problem of
theoretical physics: which relations should be trusted as
basic/fundamental from theoretical point of view, those that express directly
verifiable local balance relations of conservative quantities, or those from
which we have been clever enough to formally deduce directly verifiable by
experiment corresponding balance relations of conservative quantities. We
illustrate this by the following example. From phenomenological viewpoint, the
force fields of the Coulomb kind $q\mathbf{E}$, where $"q"$ is the charge of a
charged particle and $\mathbf{E}=\frac{Q}{r^2}\frac{\partial}{\partial r}$ is
the {\it external} electric field and this system is isolated, work good. But,
from theoretical viewpoint, we can ask:

	1. Where in the local balance relation
$\frac{d}{dt}(m_{q}\mathbf{v})=q\,\mathbf{E}$ is the necessarily available in
reality field $\mathbf{E}_q$ of the charge $"q"$, and close to the $q$-particle
its own field $\mathbf{E}_q$ can in NO way be neglected?

	2. Why the $q$-particle should detect the external field if its field
$\mathbf{E}_q$ should be neglected as local factor? Why we should accept
such asymmetry, where one of the fields, $\mathbf{E}_Q$, presents locally, and
the other one, $\mathbf{E}_q$, presents through the integral
$\int_{S^2}*\mathbf{E}_q=q$?

	3. Why distinguished monographs and textbooks avoid the
question : do both fields $\mathbf{E}_q$ and $\mathbf{E}_Q$ interact locally, at
every point where they are well defined? Such a question seems quite reasonable
in view of the {\it same} physical nature of the considered fields.

	4. In what sense the relation
$\frac{d}{dt}(m_{q}\mathbf{v})=q\,\mathbf{E}_{Q}$ is considered as local
energy-momentum balance relation in view of the {\it static} nature of the
external field $\mathbf{E}_{Q}$, therefore its energy must also be static, i.e.
not time-dependent. This field satisfies the vacuum Maxwell equations, so, it
conserves its energy, and {\it its momentum is zero} in the $Q$-proper frame,
because the Poynting vector is zero! Hence, where the theoretically presupposed
and experimentally confirmed mechanical energy-momentum change of the
$(m_q)$-particle will come from?

It is intuitively clear that each of the charged particles guarantees its
stability by means of establishing and supporting a permanent interaction with
the environment leading to appropriate dynamical equilibrium with the
environment, and each of the two fields represents availability of such a
dynamical equilibrium. Clearly, the closer are the particles the stronger will
be the mutual influence of the two fields, so each particle will aim at
corresponding compensation of its disturbed equilibrium with the environment,
and the observed and measured change of the mechanical state allows
corresponding experimental study and appropriate theoretical description.

The above remarks show that making use of the term {\it external} for one of
the fields, and neglecting the local nature of the other field, theorists,
trying to find corresponding mechanical trajectories of the particles, have
transformed the local {\it interaction} between the two fields to {\it action}
of one of the fields upon the charge of the other particle, and all this to
result in the observed corresponding change of its mechanical state.

Note that these remarks stay in force in almost all classical mechanics where,
as a rule, the so called force-fields are static, the kinetic energy of the
test particle can not change at the expense of the energy of the external
{\it static} field no matter how it is introduced: directly, or through the
gradient of the so called potential function.

One possible way out of these problems seems to be to abandon the
interpretation of $q\mathbf{E}$ in the equation
$\frac{d}{dt}(m_q\mathbf{v})=q\,\mathbf{E}$ as vector field on
$\mathbb{R}^3$. It should be something else that to be constructed out of
both fields $\mathbf{E}$ and $\mathbf{E}_q$. Moreover, the static nature of
what is denoted by $\mathbf{E}$ suggests that the induced and observed
mechanical changes are rather {\it configurational}, i.e., caused by integral
for the system factors, than local dynamical.

We turn now to the above mentioned and widely used term {\it potential}. In the
Coulomb case it is usually accepted that the field
$\mathbf{E}=\frac{Q}{r^2}\frac{\partial}{\partial r}$ is generated by
differentiating the function $U=-\frac{Q}{r}$, but what in these two
expressions is $"r"$, is it spatial coordinate so that these quantities
$\mathbf{E},U$ to be considered as local fields? If we assume the coordinate
sense of $r$, then there is not correct theoretical interpretation of the
observable interaction since these fields are static. Of course, if we have two
charges $q,Q$, there are well defined such field quantities, but in the balance
relation $\frac{d}{dt}(m_q\mathbf{v})=q\,\mathbf{E}$ it seems to participate
another quantity, which has rather integral sense than local one, so, in this
balance relation local coordinates should not participate. In fact, in the
experimentally established Coulomb law the {\it distance} $R$ between the two
charged particles participates, and this $R$ is not coordinate, and one
strongly supporting consideration in this direction is that each of the two
fields is not defined inside the 3-volumes occupied by the particles,
otherwise, the topological nature of the electric charge as defined by the
Gauss-Stokes theorem is under question.

Although we shall give a detailed consideration of these inconsistencies
further in the book, our main goal is to find theoretical description of {\it
spatially finite} time-stable objects of photon-like nature, i.e. real entities
that can exist only if they propagate in space in a consistent translational
and periodically rotational manner with constant translational speed equal to
the speed of light in vacuum, and constant intrinsically defined action for one
period. Classical Maxwell pure field equations have not such solutions, since
each scalar component of the electric $\mathbf{E}$ and magnetic $\mathbf{B}$
fields {\it necessarily} satisfies the classical D'Alembert wave equation
$\square\varphi=0$, which does not admit such solutions. These Maxwell
equations a priori assume space-time recognizablity of $\mathbf{E},\mathbf{B}$
as formal tmages of spatially propagating physical subsystems of the field, and
internal local mutual influence, i.e. local energy-momentum exchange, between
the electric and magnetic vector constituents of the electromagnetic field, but
the corresponding local energy density $\frac12(\mathbf{E}^2+\mathbf{B}^2)$
says: each of $\mathbf{E}$ and $\mathbf{B}$ may carry energy, the total local
energy density is a sum of the two energy densities, and {\it there is no local
interaction energy} between $\mathbf{E}$ and $\mathbf{B}$. Moreover, if
$\mathbf{E},\mathbf{B}$ are time-recognizable and spatially propagating
subsystems of a propagating field object, then each of the two vector
components $\mathbf{E},\mathbf{B}$ should carry momentum, but the theory {\it
does not allow this}, the local momentum is proportional to
$\mathbf{E}\times\mathbf{B}$. Finally, the very equations can not be directly
verified since they have not direct energy-momentum balance sense, and physical
science does not have appropriate devices to verify them directly. In the
experiment physics always establishes energy-momentum exchange relations, so it
seems more reliable to assume as basic theoretical equations namely
relations having energy-momentum exchange nature.

Having in view these remarks, the viewpoint that we shall follow in this book
assumes that we have to pass to dynamical equations having direct local
energy-momentum exchange sense between two (or more) appropriately defined
subsystems of the field object considered, and that the way of space-time
propagation has to be defined {\it intrinsically}, i.e. along infinitesimal
symmetry.

In brief our view on choosing appropriate mathematics is the following:
\vskip 0.3cm
{\bf The geometric theory concerned with symmetries and
integrability/nonintegrability of distributions/differential systems on
manifolds represents good machinery in this respect. In particular, if a
time-stable and space-time evolving spatially finite physical system $\Sigma$
is represented formally by an integrable distribution $\Delta$ on a manifold,
then the developed in the book $\varphi$-extended Lie derivative is used
to differentially interconnect the interacting subsystems $\Sigma_i\subset\Sigma,
i=1,2,...$, formally represented by the subdistributions
$\Delta_i\subset\Delta, i=1,2,...$. Now the nonintegrability determined
curvature forms $\Omega_i, i=1,2,...$ of some subdistributions
$\Delta_i\subset\Delta, i=1,2,...$ of $\Delta$  appear as appropriate
mathematical tools performing the interconnections between/among the
interacting subsystems $\Sigma_i\subset\Sigma, i=1,2,...$. The interior
products of the values of $\Omega_i$ with the volume forms of the corresponding
Pfaff systems can be interpreted then as local quantities of available
energy-momentum exchange, justifying in this way the intrinsic dynamical nature
of $\Sigma$ and its space-time evolution}.
 \vskip 0.3cm
We shall show that this program of passing to appropriate direct local
energy-momentum balance relations as dynamical field equations is an
adequate framework and can be correctly carried out in nonrelativistic and
relativistic approaches to description of electromagnetic field objects. We give
priority to the relativistic approach since we can not consider the absolute
character of time in the nonrelativistic approach as sufficiently motivated.
The time coordinate $\xi=ct$ has to be frame dependent, because this view
naturally corresponds to the pragmatic use of $\xi$ in the theory, namely, to
compare the speeds of a class of real processes, relating all of these
processes to one of them that has been chosen for referent one. Therefore, we
must know how to transform the theoretical quantities and relations when
passing to another reference process inside the same class. Moreover, the
transversal character of the time coordinate with respect to the spatial ones
allows to introduce directly additional 1-dimensional space-time subspace,
defined by the space-time propagation vector field $\bar{\zeta}$, and to
connect with every stress generating vector field $X$ the 2-dimensional
subdistribution $X\wedge\bar{\zeta}\subset\Delta$. These naturally arising
2-dimensional subdistributions we consider as models of {\it elementary
recognizable, interacting} and {\it time-stable subsystems}, and their
interaction then is naturally to be described in terms of the corresponding
curvature forms. The zero value of the appropriately introduced and
correspondingly constructed $\varphi$-extended Lie derivative separates those
couples among these 2-dimensional subdistributions which intercommunicate by
means of their curvature forms, on one hand, and protects the recognizability
of each subdistribution, on the other hand. \vskip 0.3cm We pass now to the
summary of the contents of this book.

The book consists of four parts, eleven sections, a retrospect and three
appendices.

The first part, "{\bf Algebraic and Geometric Preliminaries}", introduces the
necessary mathematical concepts from modern point of view in a way we consider
as appropriate for mathematically inclined physicists. We followed the idea that
the intrinsic harmony and consistency of (multilinear)algebra and differential
geometry of smooth manifolds appears to be a good mathematical image of the
existing harmony among physical objects and their interactions that we
permanently meet in the physical world. The time-recognizable multiaspect nature
of any physical object we consider as a sufficient motivation for making use
of appropriate parts of multilinear algebra, such as the symmetric and
antisymmetric ones.

Chapter I is devoted to the quite appropriate for use in theoretical physics
algebraic concepts and relations such as morphisms and isomorphisms of
algebraic structure(s), subspace, (anti)derivation(s), duality. We mention here
the Poincare isomorphism (Sec.1.4.2.), the interior product of
multivectors with exterior forms introducing the physically motivated terms {\it
attraction/sensitivity}, the various brackets (Sec.1.4.3), the associated to a
projection map algebraic curvature and cocurvature, and the properties of
isometries (Sec.1.5.2).

Chapter II introduces the basic concepts of manifold and bundle theories. There
follow tangent and cotangent bundles, vector and covector fields, and the
associated and very important for physics concept of {\it flow} of a vector
field. Then we consider the tensor algebra over a manifold, together with the
extremely important concepts of Lie derivative and its generalizations with
respect to multivector fields and additional multilinear maps for
vector valued differential forms (Sec.2.8.4); exterior derivative, coderivative
and their generalizations to vector valued differential forms. Orientation and
integration of differential forms together with the Stokes formula are given in
Sec.2.9. The basics of Lie groups and their actions on manifolds, together with
the corresponding concepts are given in Sec.2.10.

In Chapter III of this part we give the most important math-concepts used
further in the modelling of photon-like objects. These are: distributions and
codistributions on a manifold, integral manifolds for distributions,
integral and local symmetries of distributions and Frobenius curvature of
distributions. In Sec.3.2.3 we introduce the "curvature interaction
operators", the corresponding "local flows of exchange" between two
nonintegrable distributions and the concept of "dynamical equilibrium" between
two such distributions in terms of the introduced $\varphi$-extended Lie
derivative. Further these concepts are worked out in terms of projections in
tangent bundles explicitly in coordinate bases. In the sections 3.4 and 3.5
these concepts and relations are made compatible with the bundle structure in
the smooth and principal bundle cases: connection forms, curvature forms,
covariant and exterior covariant derivatives. The case of vector valued forms
together with given representation of the Lie group are also considered.
Sections 3.6 and 3.7 are devoted to linear connections in vector bundles,
corresponding covariant derivatives and curvature relations, in particular, in
Sec.3.7.1 is given the $\varphi$-extended version of the covariant Lie
derivative. In Sec.3.7.3 we introduce the new concept of generalized
parallelism. Finally, in Sec. 3.7.4 and 3.7.5 riemannian connections in vector
bundles and in tangent bundles are considered.

\vskip 0.3cm
The second part of the book, named "{\bf Basics of classical mechanics and vacuum
electrodynamics}" consists of 3 chapters and 13 sections. In Chapter IV
we introduce and comment our vision about such basic concepts in theoretical
physics as {\it physical object, physical interaction, admissible and
nonadmissible changes of physical objects, symmetries and corresponding
conservation laws.} We come to the understanding that the concept of
energy-momentum is appropriate enough to be used as universal measure of
interaction between/among physical objects, so, every physical object should
carry energy, and every quantity of energy should be carried by some physical
object. We also come to the conclusion that admissible changes for an object are
those, the projections of which on the very object are not essential, so
equalizing these projections to zero we obtain some of the dynamical
equations describing admissible evolution of the object considered.

Chapter V begins with geometrical formulation of classical mechanics
of point-like objects in presence of external fields. We note that the
beautiful geometry there does not fully corresponds to some beautiful physics
since the external field usually has {\it static} nature, so, its possible
influence on the behavior of the point-like object is under question in view
of the required energy-momentum transfer from the field to the particle, but a
{\it static} physical field can NOT do this: {\it all its physical
characteristics do not change with time and it has ZERO intrinsic momentum}.
Then we discuss the concepts of {\it stress} in terms of stress tensor of
Maxwell type, and the concept of {\it strain} in terms of the Lie derivative of
the euclidean metric along a vector field defined by the stress. Further, after
recalling some formal aspects of special Relativity, we discuss the concept of
field, wave, solitary wave and solitons. The subsection 5.4.3. is important
since there can be found some realizations of our view on differential
equations describing evolution of a field object consisting of interacting
time-recognizable subsystems of field nature.

Chapter VI presents classical vacuum electrodynamics in a way appropriate for a
future development as we see it. We analyze the Coulomb law and come to the
conclusion that the so called Coulomb potential has not local nature, that it
is rather an integral quantity, namely, an integral interaction energy $U$ of
the fields, topologically generated by each of the two charged particles,
and the observed mechanical behavior of the two particles is a consequence of
the tendency of the whole (isolated) system toward configurations with less
values of $U$. Then in Sec.6.1.2 we present the vacuum Maxwell equations as we
see them, in Sec.6.2.2 we consider the duality symmetry, in Sec.6.2.3 we give
invariant definitions of the concepts of Amplitude and Phase of a vacuum
solution in terms of the invariants of the introduced electromagnetic frame.
After considering the relativistic (nonvariational) formulation of Maxwell
vacuum equations in Sec.6.2.4 we pass to more general local conservation
relations and laws in relativistic terms in Sec.6.3 in terms of the extended
Lie derivative and curvature forms. We specially note the final
relations/equations in Sec.6.3.3, which we consider as basic in our approach
since, making use of the $\varphi$-extended Lie derivative representation of
the Frobenius criteria for integrability/nonintegrability of distributions, we
can unify in one relation the view about possible mathematical representation
of the physical notion of compatible and consistent coexistence of a number of
interacting physical field systems which we may perceive/apprehend as one
composite physical field system. Further in Sec.6.4 we give some additional
considerations and views on Maxwell equations and give a glimpse on the gauge
idea.

Chapter VII is devoted to the non-relativistic approach to the developed by the
authors Extended Electrodynamics (EED). We formulate and apply the
understanding that any real time dependent and spatially propagating
electromagnetic field object demonstrates physical appearance
consisting of two interacting subsystems
$(\Sigma_1,\Sigma_2)$, and each subsystem is described by two partner-fields
inside the $\alpha(x,y,z;t)$-family $$ (\mathcal{E},\mathcal{B})=
(\mathbf{E}\,\mathrm{cos}\,\alpha- \mathbf{B}\,\mathrm{sin}\,\alpha; \
\mathbf{E}\,\mathrm{sin}\,\alpha+ \mathbf{B}\,\mathrm{cos}\,\alpha),
$$
$\Sigma_1=(\mathcal{E}_{\alpha_{1}},\mathcal{B}_{\alpha_{1}}),
\Sigma_2=(\mathcal{E}_{\alpha_{2}},\mathcal{B}_{\alpha_{2}})$,
giving the same Maxwell stress-energy tensor. Each partner-field has electric
and magnetic constituents, and each partner-field is determined by the other
through $(\pm\frac{\pi}{2})$ - rotation-like transformation.  Both
partner-fields carry  the same energy-momentum and minimize the sum of the two
squared invariants: $I_1^2+I_2^2\geqslant 0$. The dynamical appearance of a free
real time-dependent electromagnetic field could be considered as establishing
and maintaining local energy-momentum exchange partnership between the two
subsystems, and since these partnering subsystems carry always the same
stress-energy-momen\-tum, the allowed exchange is necessarily simultaneous and
in equal quantities, so, each partner "conserves" its energy-momentum. The
final equations in nonrelativistic terms are put in a frame at the end of
Sec.7.4. In Sec.7.5 we consider some basic properties of the nonlinear
solutions of our equations, in particular we show in nonrelativistic terms that
all nonlinear solutions have zero invariants, they may have finite spatial
carrier, and, so, to present spatially finite solutions of photon-like nature.
In Sec.7.6 we consider the nonlinear analogs of the classical electric and
magnetic fields, introduce the very important concept of {\it scale factor} of
a nonlinear solution and show how a physically understandable analog of the
Planck constant as invariant characteristic of a nonlinear solution naturally
arises.

Chapter VIII presents the relativistic approach to EED. First we prove the well
known Rainich identity for the energy-momentum tensor, then recall some
properties of null differential 2-forms on Minkowski space-time, and in
Sec.8.3.2. we deduce the EED-equations in coordinate free way by means of the
$\varphi$-extended Lie derivative in $\mathbf{d}$-form and in $\delta$-form. Up
to the next Sec.8.4 we study the properties of the equations and their
nonlinear solutions. Sec.8.4. is devoted to the homology properties of the
energy-momentum tensor for null fields. Section 8.5 describes all nonlinear
solutions. In Sec.8.6 we consider the duality properties of the equations and
their solutions for constant and point-dependent parameters of the duality
matrix. Sec.8.7 represents two other views on the EED-equations in terms of
$\Lambda^1(M,\mathcal{G})$-valued, and in terms of appropriately
defined $L_{\Lambda^2(M)}$-valued 1-forms, where $\mathcal{G}$ is the Lie
algebra of the duality group. And in Sec.8.8 we give various ways to define the
spin for a nonlinear solution. It is seen that the spin is always generated by
the internal energy-momentum exchange between the two subsystems. Finally, a
picture of a theoretical example with helical-like spatial structure is given.
\vskip 0.5cm

Part III of the book is named "{\bf Photon-like objects}".

Chapter IX, introduces and comments our physical
{\it notion} of photon-like object (PhLO). It reads as follows:
\begin{center}
\hfill\fbox{
    \begin{minipage}{0.97\textwidth}
\begin{center}
\vskip 0.3cm
{\bf PhLO are real massless time-stable physical objects with an
intrinsically compatible and time-recognizable translational-rotational
dynamical structure.}
\end{center}
\vskip 0.3cm
\end{minipage}} \hfill \end{center}

\vskip 0.3cm
Chapter X represents one of the key moments of this book
, namely, {\it recognizing the geometrical concept
of Frobenius curvature as the most appropriate mathematical object for
theoretical adequate of the physical concept for local field interaction}. So,
from physical point of view, Frobenius integrability of a finite
distribution naturally corresponds to a propagating spatially
finite field object along the external (shuffling) null local symmetry of the
distribution, and Frobenius nonintegrability of the subdistributions naturally
corresponds to internal local field interaction between/among the corresponding
subsystems and is fully described in terms of the well defined curvature forms.

Section 10.1 introduces and gives the corresponding exact formulations of
the important concept of local dynamical equilibrium . Section 10.2
presents the theoretical description of photon-like objects directly in terms
of distributions on Minkowski space-time, gives explicit expressions of the
scale factor of a solution, of the internal curvature forms and their values,
and the interior product projections on the corresponding volume forms,
obtaining in this way the local energy-momentum exchanges between the
subsystems. Section 10.3 gives the formulations in terms of nonlinear
connections: projections, curvature forms, scale factor, proves the dual
invariance of the scale factor, and associates these geometric quantities to
the local energy-momentum exchanges. Finally, Section 10.4 shows how to obtain
the corresponding description in terms of strain tensors defined by the
Minkowski metric and the before introduced space-like 2-dimensional integrable
distribution. It deserves noting the positive definitness of these two
covariant symmetric strain tensors.

Chapter XI shows three ways to generate spatially finite photon-like solutions
of linear equations under additional conditions on the fields.

A part of the new nonlinear solutions found and called by us {\it photon-like},
carry helical-like space-time structure, i.e. at every moment they fill in a
smoothed out finite part of a spatial tube around a circular helix of height
$2\pi\mathcal{L}_o$ and pitch $\mathcal{L}_o$, they propagate along the
prolongation of this helical tube with constant translational velocity $c$ and,
of course, with constant period $T=\frac{2\pi\mathcal{L}_o}{c}$, they carry
finite total energy $E$ and, so, a specific action $\mathfrak{h}=ET$.

In the Retrospect the accent is mainly on the
new visions and their realizations as appropriate tools for describing
essential aspects of mathematical and physical objects.

Appendix A considers a possible way to extend the theory to not photon-like
objects. The extension made is illustrated with 3-dimensional extension of some
popular (1+1)-soliton solutions.

Appendix B considers an attempt to generalization of EED to an unified
description of photon-like objects propagating along various null directions as
well as possible interaction of overlapping photon-like solutions propagating
along the same null direction.

Appendix C gives various interesting from physical point of view applications of
our "generalized parallelism" approach given in Sec.3.7.3, in particular, a
natural nonlinearization of Yang-Mills theory.

Finally, we give a list of our earlier papers related to the subject.

 \vskip 2cm
{\bf References}
\vskip 0.5cm
1. {\bf N. Bourbaki}, {\it Theory of Sets: Summary of results, p.348}, Hermann
Publ., France, 1968

2. {\bf C. Godbillon},  {\it Geometrie differentielle et mecanique
analytiqe}, Hermann, Paris (1969)

\tableofcontents
\newpage
\mainmatter
\part{Algebraic and Geometric Preliminaries}
\chapter{Algebraic concepts and relations. Morphisms.}
\section{Basic Concepts and Structures}
We begin with recalling some initial concepts that are needed before
to introduce the basic concept of linear algebra, namely the concept of
{\it linear space}.
\begin{itemize}
\item Mathematics works basically with two kinds of concepts: {\it sets} and
{\it maps}. A set \index{set} consists of elements which are able to carry
properties and to participate in relations within the given set or with
elements of other sets. These properties and relations are expressible through
the {\it maps}. The set of all admissible maps inside a given set $S$ defines
the \index{internal structure} internal structure of $S$. This structure
defines how an element $x\in S$ exists among the rest elements of $S$.
 \item If the set $S$ consists of the elements $(a,b,c,\dots)$, then we
say that a map $\varphi:S\rightarrow S: b=\varphi(a)$ maps/transforms $a$ to
$b$. Such maps are called sometimes {\it functions}. A map
$\varphi:S\times S\rightarrow S: \varphi(a,b)=c$ is called sometimes {\it
binary}. Maps may compose: $\varphi\circ\varphi(a)=\varphi(\varphi(a))$.
\item The set $\varphi(S)\subset S$ is called {\it image}
of $\varphi:S\rightarrow S$. The set of all elements of $S$ that are mapped to
 the same element $b\in S$ are called {\it opposite} of $b$
with respect to $\varphi$ and are denoted by $\varphi^{-1}(b)$.
\item The map
$\varphi:S\rightarrow W$, where $W$ is in general another set, is called {\it
injective} \index{injective} if different elements of $S$ are mapped to
different elements of $W$; and $\varphi:S\rightarrow W$ is called {\it
surjective} \index{surjective} if every element of $W$ is an image of some
element(s) of $S$. Now, $\varphi:S\rightarrow W$ is called {\it bijective}
\index{bijective} if it is injective and surjective. Every bijective map
$\varphi:S\rightarrow W$ has opposite $\varphi^{-1}:W\rightarrow S:
\varphi^{-1}(\varphi(a))=a, a\in S$. Every set has {\it identity} map:
$id_{S}(a)=a$. \item If $\varphi:S\rightarrow S$, then the sets
$\varphi^{-1}(x)\subset S, x\in S$, do not intersect and their union gives the
whole $S$, so, equivalence relation is established: two elements of $S$ are
$\varphi$-equivalent if they live in the same $\varphi^{-1}(x)$ for some $x\in
S.$  Every map $f:S\rightarrow W$ factors with respect to $\varphi$ if there is
a map $g:S\rightarrow W$, such that $f=g\circ\varphi$. In such a case
  $f$ has the same value on the whole equivalence class
$\varphi^{-1}(x)$ of $x\in S$.

 \item If $\varphi:S\rightarrow S$ and
$\psi:W\rightarrow W$ are bijective and $f:S\rightarrow W$ satisfies
$f\circ\varphi=\psi\circ f$ then $f$ is called {\it splitting operator}
\index{splitting operator} for the couple $(\varphi,\psi)$. If $f$ is bijective
we could write equivalently in such case $\psi=f\circ\varphi\circ f^{-1}$, or
$\varphi=(f^{-1})\circ\psi\circ f$ \item The element $x\in S$ is called {\it
invariant} with respect to $\varphi:S\rightarrow S$ if $\varphi(x)=x$.
\end{itemize} In algebra binary maps are usually exploited, and specific signs
for these maps are introduced: $"+"$, $"\times"$, $"."$, $"\wedge"$, etc. As a
rule all such maps are called "multiplication". \begin{itemize} \item If
$\varphi:S\times S\rightarrow S$, and $\varphi$ is defined everywhere in $S$,
let's denote it simply by a point: $\varphi(x,y)=x.y\in S$. If $x.y=y.x$ is
true for every couple$(x,y)\in S\times S$ then $\varphi$ is called {\it
symmetric}, or {\it commutative}. The map $\varphi$ is called {\it associative}
if $(x.y).z=x.(y.z)$, and the set $S$ is called {\it monoid}. \item  The
element $a\in S$ is called {\it central} for $\varphi$ if $a.x=x.a$ for every
$x\in S$. The element $e\in S$ is called {\it neutral} with respect to
$\varphi$ if $e.x=x.e=x$ for every $x\in S$. Clearly, a $\varphi$-neutral
element is unique in $S$. Usually, the neutral element with respect to
additively written law: $\varphi="+"$ is denoted by $0$, and with respect to
multiplicatively written law: $\varphi="."$ it is denoted by $\mathbf{1}$.
\item Two elements $x,y$ are called {\it symmetric/opposite} with
respect to $\varphi$ if $x.y=e$, then the usual notation is $y=x^{-1}$ or
$x=y^{-1}$.
\item
Each element $a\in S$ defines by means of $\varphi: S\times S\rightarrow S$ a
map $\gamma_{a}: S\rightarrow S$, called {\it left/right translation}:
$\gamma_{a}(x)=a.x; \ \delta_{a}(x)=x.a$. The element $a\in S$ is called {\it
$\varphi$-regular} if $\gamma_{a}$ and $\delta_{a}$ are bijective.
\item
If $\varphi: W\times S\rightarrow S$ then $W$ is usually called a set
of operators on $S$ with respect to $\varphi$. The usual notation is
$\varphi(\alpha,x)=\alpha.x$, and we say that the set $W$ acts on $S$. We
have two partial maps: varying $\alpha\in W$ with fixed $x\in S$ we obtain a
subset $\varphi_{x}(\alpha)\subset S$ called {\it orbit} of $x\in S$ with
respect to the action of $W$ on $S$; varying $x\in S$ and keeping $\alpha$
fixed we obtain another subset $\varphi_{\alpha}(x)\subset S$.
\end{itemize}

{\bf Definition}: We say that on the set $S$ is defined {\it algebraic
structure} \index{algebraic structure} $\sigma(S)$ if a set of maps (unary,
binary, \dots) $\varphi_{1}, \varphi_{2}, \dots$ inside $S$ is given, and a set
of operator sets $W_{1}, W_{2}, ...$ for $S$ is given by the maps $\phi_{i}:
W_{j}\times S\rightarrow S$, so that all additional properties of the elements
of these sets and maps are compatible.

Let $S_1$ and $S_2$ be two sets with corresponding algebraic structures
$\sigma(S_1)$ and $\sigma(S_2)$, and let $f:S_1\rightarrow S_2$ be a bijection.
Then an operation $\psi$ in $S_2$ and
operation $\varphi$ in $S_1$ are called $f$-compatible if
$f_o\varphi=\psi_o f $.

{\bf Definition}: The two algebraic structures $\sigma(S_1)$ and $\sigma(S_2)$
are called {\it $f$-isomorphic} \index{isomorphic algebraic structures} if the
bijection $f$ establishes a bijection between $\sigma(S_1)$ and $\sigma(S_2)$
in the above sense.

If $f$ is not a bijection, but for every $\varphi\in\sigma(S_1)$, denoted by "."
there is a  $\psi\in\sigma(S_2)$, denoted also by ".", such that
$f(x.y)=f(x).f(y)$, we say that the structure $(S_1,\sigma(S_1))$ is represented
in $(S_2,\sigma(S_2))$ through $f$. In such a case the map $f$ is called {\it
homomorphism} \index{homomorphism} of $\sigma(S_1)$ into $\sigma(S_2)$.

In order to come to the basic concept of linear algebra, namely, {\it linear
space}, we need some additional concepts: {\it distributivity, group, ring} and
{\it field}.
\begin{itemize}
\item
Let $\varphi: S\times S\rightarrow S$ and $\phi: W\times S\rightarrow S$ be
given. We say that $\phi$ is {\it distributive} with respect to $\varphi$ if
$\phi(\alpha,\varphi(x,y))=\varphi(\phi(\alpha,x),\phi(\alpha,y))$. In
simplified notation: $\alpha.(x.y)=(\alpha.x).(\alpha.y)$. If $\varphi$ is
additively written, we get $\alpha.(x+y)=\alpha.x+\alpha.y$.
\item
The map $\varphi:S\times S\rightarrow S$ defines a {\it group structure}
\index{group structure} in $S$ if (in simplified notation):

{\bf 1.} $\varphi$ is associative: $x.(y.z)=(x.y).z$;

{\bf 2.} there is a neutral element $e$: $e.x=x.e=x$;

{\bf 3.} for every element $x\in S$ there is symmetric element $y\in S:
x.y=y.x=e$.

A set with a group structure is usually denoted by $G$. Every element $a\in G$
defines internal isomorphism (automorphism) $\alpha_a$
according to: $\alpha_a(x)=a.x.a^{-1}, x\in G$. A group $G$ is called
commutative if for every two elements we have $x.y=y.x$, in such a case the
sign "+" is usually used instead of the point sign. The internal automorphisms
in case of commutative groups are trivialized to identity. If $f:G\rightarrow
H$ is a map between two groups satisfying $f(x.y)=f(x).f(y)$ then $f$ is called
homomorphism of groups.

\item A {\it ring structure} \index{ring structure}
 in a set $S$ is defined by two maps $\varphi:S\times S\rightarrow S$
and $\psi: S\times S\rightarrow S$, such that:

{\bf 1.} $S$ is a commutative group with respect to $\varphi$, so we write
$\varphi=+$;

{\bf 2.} $\psi$ is associative: $\psi(x,\psi(y,z))=\psi(\psi(x,y),z))$, i.e.
$x.(y.z)=(x.y).z$;

{\bf 3.} $\psi$ is left-right distributive with respect to "+":
$$
x.(y+z)=x.y+x.z;\ \ \  (x+y).z=x.z+y.z.
$$
\item
If in the above notations the nonzero elements of $S$ define a commutative
group structure with respect to $\psi$ then we say that $S$ is a {\it field.}
So, a field has neutral element with respect to $\psi$, this neutral element is
usually called {\it unity} and denoted by $\mathbf{1}$.   It may happen that
if we sum up the unit element $k$-times:
$\mathbf{1}+\mathbf{1}+\dots+\mathbf{1}=k\mathbf{1}$ the result to be the
neutral element of $\varphi$. If there is not such natural number $k$ it is
said that the field has {\it characteristic zero}. We shall restrict ourselves
further in the book to work with fields of characteristic zero. \end{itemize}

\section{Linear Spaces}
Linear spaces appear mostly as {\it modules} and {\it vector spaces}.

1. {\bf Module structure} \index{module structure}. It requires two sets
$(\Gamma, V)$, carrying the following algebraic structures: $\Gamma$ is a ring,
$V$ is a commutative group, and $\Gamma$ acts left-right distributively on $V$
(the action is denoted by the point sign, the zero-elements of $V$ and $\Gamma$
are denoted by the same sign "0"), satisfying: $0.x=0\in V$,
$(\alpha+\beta).x=\alpha.x+\beta.x$, $(\alpha.\beta).x=\alpha.(\beta.x),
\alpha,\beta\in \Gamma, x\in V$. The opposite elements in the additive group
structures in $V$ and in $\Gamma$ are denoted by "-", so $x+(-x)=0\in V, x\in
V$, and $\lambda+(-\lambda)=0\in \Gamma, \lambda\in \Gamma$. If $\Gamma$ has
unit element $\mathbf{1}$ and $\mathbf{1}.x=x, x\in V$, the module is called
{\it unitary}.

2. {\bf Vector space structure.} \index{vector space structure}
It differs from the module structure just by
requiring additionally that $\Gamma$ is a field. As an illustration we recall
that the set of real numbers $\mathbb{R}$ is a field, so, $\lambda\in
\mathbb{R}$ has opposite/symmetric element with respect to the addition:
($-\lambda$), and multiplication: ($\lambda^{-1}$), while the set of continuous
real valued functions defined on the interval $[0,1]$, is a ring, in general,
since if such a function has zero-values it can not have opposite element with
respect to the usual multiplication of two such functions. Note that these
functions define vector space with respect to the usual multiplication by real
numbers. Further the elements of $V$ will be called {\it vectors}, and the
elements of $\Gamma$ will be called {\it scalars}.

3. {\bf Linear combinations.} Let $(\Gamma,E)$ denote a linear space,
$x_1,x_2,\dots,x_p$ be vectors, and $\lambda^1,\lambda^2,\dots,\lambda^p$
be scalars. The expression
$\Sigma_{i=1}^p\lambda^ix_i=\lambda^1x_1+\lambda^2x_2+\dots+\lambda^px_p$ is
another vector $x\in E$, and is called {\it linear combination} of the $x_i,
i=1,2,\dots,p$.  A subset $S\subset E$ is called a {\it system of generators}
for $E$ if every vector in $E$ can be represented as a linear combination of
vectors in $S$.

4. {\bf Linear dependence.} \index{linear dependence} A set of elements $x_i,
i=1,2,\dots,p$ in $E$ is called linearly dependent if there exist a system of
scalars $\lambda^i$ such that $\Sigma_{i=1}^p\lambda^ix_i=0$. Hence, if we have
a system of linearly dependent vectors then each one could be represented as a
linear combination of the others. A family of vectors is linearly independent
if it is not linearly dependent.

5. {\bf Basis.} A basis of $E$ is a system of linearly independent
generators of $E$. So, if $\{e_i\}, i=1,2,\dots,n$ is a basis of $E$ then every
vector $x\in E$ can be represented as a linear combination of the kind
$x=\Sigma_{i=1}^n\lambda^ie_i$. It follows that if $E$ has finite system of
generators it has finite basis, and that every family of linearly independent
vectors can be extended to basis. If $E$ has a basis that consists of $n$
elements, then $E$ is called $n$-dimensional since every other basis is also
$n$-dimensional.

6. {\bf Linear mappings - basic terminology.} \index{linear mapping} If $E$ and
$F$ are linear spaces with the same set of scalars $\Gamma$, then a set mapping
$\varphi:E\rightarrow F$ is called a {\it linear mapping} if
$\varphi(x+y)=\varphi(x)+\varphi(y)$ and
$\varphi(\lambda.x)=\lambda.\varphi(x)$. The linear mappings $E\rightarrow
\Gamma$ are called {\it linear functions}. Clearly all linear mappings preserve
the linear combinations. If $\varphi:E\rightarrow F$ is linear and bijective it
is called {\it linear isomorphism} \index{linear isomorphism}, the two spaces
then are called {\it isomorphic} and $\varphi^{-1}$ is the inverse linear
isomorphism. The linear isomorphisms $\varphi:E\rightarrow E$ are called {\it
linear automorphisms}. Clearly, a set of consecutive linear mappings among a
set of linear spaces: $\varphi_{12}:E_1\rightarrow E_2$,
$\varphi_{23}:E_2\rightarrow E_3$,..., $\varphi_{n-1,n}E_{(n-1)}\rightarrow
E_n$
 can be composed to give a linear mapping between the first
and the last linear spaces. A linear mapping $\varphi:E\rightarrow E$ is called
{\it involution} \index{involution} if the composition $\varphi\circ\varphi$
gives the identity of $E$: $\varphi\circ\varphi=id_{E}$, and $\varphi$ is
called {\it projection} if $\varphi\circ\varphi=\varphi$. If
there exists a linear mapping in $E$ such that $\varphi\circ\varphi=-id_{E}$,
then $\varphi$ is called {\it complex structure} \index{complex structure} in
$E$, and $dim\,E=2n$. Finally, linear automorphisms transform basis into basis,
i.e. they act inside the set of bases of the linear space considered. Moreover,
the set of linear isomorphisms in an $n$-dimensional linear space is in
bijective relation with the set of bases in this space.

 7. {\bf Subspaces and factor spaces.} \index{subspace} A subset
$S$ in a linear
space $(E,\Gamma)$ is a {\it subspace} if for every two elements $x,y\in S$ and
for every two elements $\lambda,\mu\in \Gamma$  we get $(\lambda.x+\mu.y)$ is
in $S$, hence $(S,\Gamma)$ is a linear space. Clearly, the zero element of $E$
is in $S$ and in fact in every subspace of $E$. We have the canonical injection
$i:S\rightarrow E$. The set of linear combinations of any subset of $E$ define
a linear space, which is a subspace of $E$. If $E_1$ and $E_2$ are subspaces of
$E$ then their intersection $E_1\cap E_2$ is again a subspace of $E$. If $E_1$
and $E_2$ are two subspaces of $E$ then their {\it direct sum} is defined by
all vectors $z$ that have {\bf unique} representation as  $z=x_1+x_2$, where
$x_1\in E_1$ and  $x_2\in E_2$, so the intersection $E_1\cap E_2$ is empty in
this case. If $E_1$ is a subspace of $E$ then there is another subspace $E_2$ of
$E$ such that $E$ is a direct sum of $E_1$ and $E_2$: $E=E_1\oplus E_1$, and
$E_1$ and $E_2$ are called {\it complimentary} in $E$. It is possible the space
$E$ to be represented as a direct sum of a finite family of not-intersecting
subspaces, so every element $x$ of $E$ acquires the representation
$x=x_1+x_2+\dots +x_n$. We have also  corresponding projections $\pi_i(x)=x_i$
and canonical injections $j_i:E_i\rightarrow E$. Obviously $j_i(\pi_i(x))=x$.

If $E_1$ is a subspace of $E$ then two vectors of $E$ are called equivalent
with respect to $E_1$ if their difference is in $E_1$. Every element of $E$
lives in unique equivalence class, so we get the canonical surjective
projection $\pi:E\rightarrow E/E_1$, where $E/E_1$ denotes the set of
equivalence classes, and there is unique linear structure in $E/E_1$ with
respect to which $\pi$ is a linear mapping. $E_1$ plays the role of zero in
$E/E_1$, the sum in $E/E_1$ is given by $\bar{x}+\bar{y}=\pi(x+y)$, where
$\bar{x}=\pi(x)$ and $\bar{y}=\pi(y)$. The space $E/E_1$ is called {\it factor
space} \index{factor space} of $E$ with respect to $E_1$. Finally, if $(e_i)$ is
a basis of $E$ then $\pi(e_i)$ defines a basis of $E/E_1$ and the dimension of
$E/E_1$ is equal to the difference of the dimensions of $E$ and $E_1$:
$dim(E/E_1)=dim(E)-dim(E_1)$.

 8. {\bf Linear mappings - further properties.} If $\varphi:E\rightarrow F$ is
a linear mapping then the elements of $E$ satisfying $\varphi(x)=0\in F$ define
a subspace of $E$ called {\it kernal} of $\varphi$ and denoted by
$Ker(\varphi)$. Then $\varphi$ is injective iff $Ker(\varphi)$ is the zero
subspace of $F$. On the other hand the image $Im(\varphi)$ defines a subspace in
$F$. Hence, the factor space $E/Ker(\varphi)$ is defined, and the linear
mapping $\varphi$ factorizes to linear isomorphism
$\bar{\varphi}:E/Ker(\varphi)\rightarrow Im(\varphi)$.

The set of linear mappings $E\rightarrow F$ acquires the structure of linear
space under the assumptions: $(\varphi+\psi)(x)=\varphi(x)+\psi(x)$ and
$(\lambda.\varphi)(x)=\lambda.\varphi(x)$. This space is denoted by $L(E,F)$,
its zero is the mapping zero: $0(x)=0$. Hence, the set of linear isomorphisms
of E, denoted by $GL(E)$, acquires a group structure under {\it composition,
identity mapping and inverse isomorphism}:
$(\varphi,\psi)\rightarrow\varphi\circ\psi; \ \
id(E)\circ\varphi=\varphi, \ (\varphi^{-1})\circ\varphi=\varphi\circ\varphi^{-1}=id(E)$.

Recall that he projections $P$ \index{projection map} in a linear space $E$ are
linear mappings satisfying $ P\circ P=P. $ They have the following two
remarkable properties: $$E=Ker(P)\oplus Im(P), \ P=id(Im P)\oplus 0(Ker P). $$
Since $id_{E}\circ id_{E}=id_{E}$, we shall further consider only projections
satisfying $P\neq id_{E}$. We note that every involution $\varphi,
\varphi\circ\varphi=id$, in $E$ can be represented by an appropriate projection
$P$ in $E$ as follows: $\varphi=2P-id(E)$.

9. {\bf Duality.} \index{duality} Let  $E$ and $F$ be two linear spaces on the
same set of scalars $\Gamma$. A function $\phi: E\times F\rightarrow \Gamma$
satisfying the conditions $$ \phi(\lambda x_1+\mu
x_2,y)=\lambda\phi(x_1,y)+\mu\phi(x_2,y), \ \ x_1,x_2\in E, y\in F; $$ $$
\phi(x,\lambda y_1+\mu y_2)=\lambda\phi(x,y_1)+\mu\phi(x,y_2),
\ \ y_1,y_2\in F, x\in E
$$
is called {\it bilinear function} \index{bilinear function} in $E\times F$.
Every such function defines two subspaces \newline $N_E\subset E$ and
$N_F\subset F$ as follows: $$ N_E=\{x|\phi(x,y)=0\} \ \text{for every} \ \ y\in
F; $$ $$ N_F=\{y|\phi(x,y)=0\} \ \text{for every} \ \ x\in E. $$ These
subspaces are called {\it nullspaces} \index{nullspaces} for $\phi$. If these
two subspaces are zero spaces: $N_E=0, \ N_F=0$, then the function $\phi$ is
called {\it non-degenerate}, and is usually denoted by $\langle \,,\rangle$. If
$\phi$ is nondegenerate then the two spaces are called {\it dual} \index{dual
spaces} (with respect to $\phi$) and instead of $(E,F)$ we write $(E,E^*)$. It
is also sometimes $\phi$ called {\it scalar product} between $E$ and $E^*$. If
$(E,E^*)$ is a pair of dual spaces then two injecitve mappings are defined:
$\phi^*: E^*\rightarrow L(E)$ and $\phi_*:E\rightarrow L(E^*)$ as follows
($L(E)$ means here the space of linear mappings from $E$ to $\Gamma$):
$\phi^*(a^*)(x)=\langle a^*,x\rangle, \ \ a^*\in E^*, \ \ x\in E; \ \
\phi_*(a)(x^*)=\langle x^*,a\rangle, \ \ x^*\in E^*, \ \ a\in E.$ The
injectivity of these mappings follows from the nondegeneracy of $\langle \,
,\rangle$.

If $\varphi:E\rightarrow E$ is a linear mapping, then a {\it dual} linear
mapping \index{dual mapping} $\varphi^*$ in $E^*$ is induced according to
$\langle \varphi^*(x^*),x\rangle=\langle x^*,\varphi(x)\rangle, \ \ x\in E, \
x^*\in E^*$, and $\varphi^*$ is unique. The dual mappings satisfy:
$$
(\varphi+\psi)^*=\varphi^*+\psi^*,\ \ (\lambda\varphi)^*=\lambda(\varphi)^*; \
\ (\varphi\circ\psi)^*=\psi^*\circ\varphi^*.
$$
If $E^*=L(E)$ then the injectivity of $\varphi:E\rightarrow E$ leads to
surjectivity of $\varphi^*: E^*\rightarrow E^*$. If $E$ is finite dimensional,
then  $dim E=dim E^*$. Also,
$Im\,\varphi^*$ annihilates $Ker\,\varphi$, and $Ker\,\varphi^*$
annihilates $Im\,\varphi$.

A basis $\{e_i\}, i=1,2,\dots,n$ in $E$ and a basis $\{\varepsilon^j\},
j=1,2,\dots,n$ in $E^*$ are called {\it dual} \index{dual bases} if $\langle
\varepsilon^j,e_i\rangle=\delta_i^j$, where $\delta_i^j$ is the Kroneker
symbol, i.e., the components of $id_{E}$. The dual bases are in a one to one
correspondence. If $x\in E$ has the representation $x=x^ie_i$, then
$\langle\varepsilon^i,x\rangle=x^i$ and we get $x=\langle
\varepsilon^i,x\rangle e_i$. To every linear mapping $\varphi$ in $E$ and to a
basis $\{e_i\}$ a $(n\times n)$-matrix $\varphi_i^j$ corresponds as follows:
$\varphi(e_i)=\varphi _i^je_j$ (summation over the repeated indices), i.e., the
vector $\varphi(e_i)$ is written as linear combination of the same basis. If
$\varphi$ is isomorphism then the matrix $\varphi_i^j$ has nonzero determinant.
If $(e_i)$ and $(\varepsilon^j)$ are dual bases then the matrix of $\varphi^*$
with respect to $(\varepsilon^j)$, defined by
$\varphi^*(\varepsilon^j)=(\varphi^*)^j_i\varepsilon^i$, is the transposed to
$\varphi_i^j$ in the following sense: \begin{eqnarray*}
\langle(\varphi^*)\varepsilon^j,e_i\rangle&=&(\varphi^*)^j_k\langle
\varepsilon^k,e_i\rangle=(\varphi^*)^j_k\delta^k_i=(\varphi^*)_i^j\\
&=&\langle\varepsilon^j,\varphi(e_i)\rangle=\langle
\varepsilon^j,\varphi_i^ke_k\rangle=\varphi_i^k\delta^j_k=\varphi_i^j.
\end{eqnarray*}
Note that the upper subscript indicates the rows of the matrix and the
lower subscript indicates the columns of the matrix. So, although
$\varphi_i^j=(\varphi^*)_i^j$, the action on $\{e_i\}$ and on
$\{\varepsilon^i\}$ is different.

If $\varphi$ transforms the basis $\{e_i\}$ to $\{e'_i\}$ and $\varphi^*$
transforms the cobasis $\{\varepsilon^i\}$ to $\{\varepsilon'^i\}$, where
$\{e_i\}$ is dual to $\{\varepsilon^i\}$ and $\{e'_i\}$ is dual to
$\{\varepsilon'^i\}$, then the matrix of $\varphi^*:
\{\varepsilon^i\}\rightarrow \{\varepsilon'^i\}$ is the inverse to
the matrix of $\varphi:\{e_i\}\rightarrow\{e'_i\}$, and these two
transformations are
called usually {\it contragradient} to each other. It follows also that the
components $x^i$ of a vector $x\in E$ with respect to the basis $\{e_i\}$ are
transformed to $x'^i$ under the transformation $\varphi$ in the same way as the
covectors of the dual to $\{e_i\}$ basis $\{\varepsilon^i\}$:
$x'^i=((\varphi^*)^{-1})^i_jx^j$. Clearly, the inverse isomorphism
$\varphi^{-1}$ of a isomorphism $\varphi$ generates the inverse matrix:
$(\varphi^{-1})^i_j\varphi_i^k=\delta_j^k$. Finally, if $\varphi$ transforms
the basis $\{e_i\}$ to $\{k_i\}$: $\varphi\{e_i\}=\varphi_i^jk_j$; and transforms
the basis $\{e'_i\}$ to $\{k'_i\}$: $\varphi\{e'_i\}=\bar{\varphi}_i^jk'_j$ then
$\varphi_i^j$ transforms to $\bar{\varphi}_i^j$ as follows:
$\bar{\varphi}_i^n=A_i^j\varphi_j^m(B^{-1})_m^n$, where the matrix $A_i^j$
transforms $\{e_i\}$ to $\{e'_i\}$ and $B_i^j$ transforms $\{k_i\}$ to
$\{k'_i\}$.

If $\varphi$ is a linear mapping $E\rightarrow F$ then the dimension of
$Im\,\varphi$ is called {\it rank} \index{rank} of $\varphi$. The isomorphism
$E/Ker(\varphi)\leftrightarrow Im(\varphi)$ leads to
$rank(\varphi)+dim(Ker\varphi)=dim(E)$. It is easily shown that
$rank(\varphi^*)=rank(\varphi)$.

If $V$ is a $n$-dimensional vector space and $W_1\subset V$ is a subspace, then
the following three spaces can be associate to $W_1$. These are

	- its dual $W_1^*$,

	- $W_2\subset V$, such that $W_1\oplus W_2=V$, so $W_2$ is
annihilated by $W_1^*$,

	- the dual space $W_2^*$ to $W_2$, which annihilates $W_1$.

So, if $\{e_1,...,e_p\}, p<n,$ is a basis of $W_1$,
$\{\varepsilon^1,...,\varepsilon^p\}$ is its dual basis of $W_1^*$,
$\{e_{p+1},...,e_n\}$ is a basis of $W_2$, and
$\{\varepsilon^{p+1},...,\varepsilon^n\}$ is its dual basis of $W_2^*$, then we
have the relations:
$$
\langle \varepsilon^i,e_j\rangle =\delta^i_j, \ \
\langle \varepsilon^m,e_j\rangle = 0, \ \
\langle \varepsilon^i,e_m\rangle = 0, \ \
\langle \varepsilon^m,e_s\rangle = \delta^m_s,
$$
$$
\ i,j=1,...,p, \ \ \ m,s=p+1,...,n.
$$

%\newpage
\section {Algebras, Gradations and Homology}
1. {\bf Algebras.} Algebras \index{algebra} are linear spaces $E$ endowed with a
bilinear mapping $\mathfrak{E}:E\times E\rightarrow E$, $\mathfrak{E}$ is
called {\it multiplication} and the values of $\frak{E}$ are called {\it
products}. Let $(A,\frak{A})$ and $(B,\frak{B})$ be two algebras, then a linear
mapping $\varphi:A\rightarrow B$ satisfying
$\varphi(\frak{A}(x,y))=\frak{B}(\varphi(x),\varphi(y)), x,y\in A$ is called
{\it homomorphism} of algebras, and if $A=B$ then $\varphi$ is called {\it
endomorphism}.  If $\varphi$ is resp.(injective, surjective,
bijective) homomorphism then $\varphi$ is called resp. {\it monomorphism,
epimorphism, isomorphism} of algebras. Every element $a\in A$ defines {\it
multiplication operator} $\mu(a)$ given by $\mu(a)x=\frak{A}(a,x)$.

{\bf Remark}: Further the sign of the bilinear mapping will be dropped, so
$\mu(x)y$ will be written just like $xy$.

An algebra is called {\it associative} if $x(yz)=(xy)z$, and {\it commutative}
if $xy=yx$. A subset $S$ of an associative algebra $A$ is called a {\it system of
generators} for $A$ if every element of $A$ can be represented as a linear
combination of products of elements of $S$. If an algebra $A$ contains an element
$e$ such that $ex=xe$ for every $x\in A$ then $e$ is called {\it unit element},
and it is unique. The algebras $A$ that have unit elements are called {\it
division algebras} if to each $x\in A$ corresponds unique element $a^{-1}$ such
that $aa^{-1}=e$. It deserves noting that the set of linear isomorphisms of a
linear space is an algebra with respect to the composition of isomorphisms and
with unit element the identity map.

A {\it subalgebra} \index{subalgebra} $A_1$ of an algebra $A$ is a linear
subspace of $A$ which is closed under multiplication. A subspace $I$ of $A$ is
called {\it ideal} \index{ideal} in $A$ if $I.A\subset I$. If
$\varphi:A\rightarrow B$ is a homomorphism, then $Ker(\varphi)$ is ideal in $A$
and $Im(\varphi)$ is a subalgebra in $B$.

A {\it derivation} \index{derivation} in an algebra $A$ is a linear mapping
$\theta:A\rightarrow A$ satisfying $\theta(xy)=\theta(x)y+x\theta(y), x,y\in
A$. Clearly a derivation $\theta$ sends the unit element of $A$ to the zero
element of $A$. A derivation $\theta$ in $A$ is completely determined by its
action on the basis elements of $A$:
$\theta(e_ie_j)=\theta(e_i)e_j+e_i\theta(e_j)$. We obtain that $Ker(\theta)$ is
subalgebra and that a linear combination of derivations is again a derivation.
The composition $\theta_1\circ\,\theta_2$ of two derivations $\theta_1$ and
$\theta_2$ is {\bf not} always a derivation, but the commutator
$[\theta_1,\theta_2]=\theta_1\circ\,\theta_2-\theta_2\circ\,\theta_1$ is {\bf
always} a derivation. If $\varphi:A\rightarrow A$ is a homomorphism and
the linear map $f:A\rightarrow A$ satisfies
$f(xy)=f(x)\varphi(y)+\varphi(x)f(y)$, then $f$ is called $\varphi$-derivation.

If $\omega$ is an involution in the linear space structure of $A:
\omega\circ\omega=id$, and an endomorphism of $A:
\omega(xy)=\omega(x)\omega(y)$, it is called an involution of the algebra $A$.
A linear mapping $\Omega:A\rightarrow A$ is called {\it antiderivation}
\index{antiderivation} with respect to the involution $\omega$ (or just
$\omega$-antiderivation) if $\Omega(xy)=\Omega(x)y+\omega(x)\Omega(y)$.
Clearly, if $e$ is the unit of $A$ then $\Omega(e)=0$, and linear combinations
of $\omega$-antiderivations is again an $\omega$-antiderivation. It has to be
noted that the commutator of two $\omega$-antiderivations is not always an
antiderivation.

An algebra $A$, satisfying
$$
xy=-yx \ \ \text{and} \ \  (xy)z+(yz)x+(zx)y=0
$$
is called Lie algebra \index{Lie algebra} and denoted by $\mathcal{G}$. Every
associative algebra can be made into Lie algebra if the product is defined by
the commutator, which is usually denoted by $[\,,]$: $(x,y)\rightarrow
[x,y]=xy-yx$. The multiplication operator in such a Lie algebra $\mathcal{G}$
is denoted by $Ad(a), Ad(a)(x)=[a,x]=ax-xa$ and is a derivation in
$\mathcal{G}$. The space of derivations $\theta_1,\theta_2, \dots$ in an
algebra becomes Lie algebra with respect to the commutator:
$[\theta_1,\theta_2]=\theta_1\theta_2-\theta_2\theta_1$. If $Der(\mathcal{G})$
is the space of derivations in $\mathcal{G}$ then the mapping $
\mathcal{G}\rightarrow Der(\mathcal{G})$ defined by $a\rightarrow Ad(a), a\in
\mathcal{G}$ is a Lie algebra homomorphism.

2. {\bf Gradation.} \index{gradation} Let $\mathbb{Z}$ denote the natural
numbers $0,1,2,3,\dots$. If a linear space $E$ can be represented as a direct
sum of the kind $E=\sum^{\infty}_{k-1}E_k, k\in \mathbb{Z}$, we say that $E$ is
a graded linear space. The elements in $E_k$ are called {\it homogeneous} of
degree $k: deg(x\in E_k)=k$. If $\varphi:E\rightarrow F$ is a linear mapping
between two graded spaces such that $\varphi(E_k)\subset F_{k+p}$ we say that
$\varphi$ is {\it homogeneous} of degree $p$. If every $E_k$ of a graded space
$E$ has finite dimension the Poincare series \index{Poincare series} $P_t(E)$ of
$E$ is defined by $P_E(t)=\Sigma_k(dimE_k)t^k$.

Let now $A$ be an algebra such that the linear space $E$ is
$\mathbb{Z}$-graded. Then $A$ is called graded algebra if for any two
homogeneous elements $x,y\in A$ their product $xy$ is also homogeneous and
$deg(xy)=deg(x)+deg(y)$. A graded algebra $A$ is called anticommutative if for
any two homogeneous elements $x,y\in A$ we get $xy=(-1)^{deg(x).deg(y)}yx$. In
every graded algebra we have the canonical involution $\omega$ defined by
$\omega(x)=(-1)^kx, x\in A_k$. This canonical involution assigns to every
derivation $\Omega$ an antiderivation $\mathbf{D}_{\Omega}:A\rightarrow A$ by
$\mathbf{D}_{\Omega}(xy)=(\Omega x).y+(-1)^kx.(\Omega y), \ x\in A_k$.

3. {\bf Homology.}
Consider a linear mapping $D$ in a linear space $E$
having the property $D\circ D=0\in L_E$. Then we
have two related subspaces, $Ker(D)=\{x\in E: D(x)=0\}$ and
$Im(D)=D(E)$. Since in this case $Im(D)$ is a subspace of $Ker(D)$, we can
factor, and the corresponding factor space $H(D,E)=Ker(D)/Im(D)$
is called the {\it homology space} \index{homology} for $D$. The dual linear
map $D^*$ in the dual space $E^*$ has also the property $D^*\circ D^*=0$, so we
obtain the corresponding {\it cohomology space} \index{cohomology}
$H^*(D^*,E^*)$. In such a situation the map $D$ (resp. $D^*$) is called {\bf
boundary operator} (resp {\bf coboundary operator}). The elements of $Ker(D)$
(resp. $Ker(D^*)$) are called {\it cycles} (resp. {\it cocycles}), and the
elements of $Im(D)$ (resp. $Im(D^*)$) are called {\it boundaries} (resp. {\it
coboundaries}).

We note the following two important moments connected with a boundary operator.

First, if $\alpha\in E^*$ is a $D^*$-cocycle and $x\in E$ is a $D$-boundary,
i.e., $D^*(\alpha)\in Ker(D^*)$ and $x\in Im(D)$ we obtain
$$
\langle \alpha,D(x)\rangle=\langle D^*(\alpha),x\rangle=\langle
0,x\rangle=0.
$$
Also, if now $\alpha$ is any element of $Ker\,D^*$ and $x$ is any element of
$E$, then
$$
0=\langle D^*(\alpha),x\rangle=\langle\alpha,D(x)\rangle.
$$
These relations show that the duality between $E$ and $E^*$ generates duality
between the homology/cohomology spaces $H(E;D)$ and $H(E^*;D^*)$.

Second, every linear map
$\varphi:E\rightarrow E$ which commutes with $D$:
$D\circ\varphi=\varphi\circ D$, induces a linear map
$\varphi_\#:H(D)\rightarrow H(D)$. Under composition we get $(\psi\circ
\varphi)_\#=\psi_\#\circ\varphi_\#$.

So, a boundary operator realizes the
general idea of distinguishing some properties of a class of objects the
properties which are important from a definite point of view, and to find those
transformations which keep invariant these properties.

   If $A$ is an algebra, $D$ is a boundary operator in the linear space $A$,
$\omega$ is an involution of $A$ such that $D\circ\omega=\omega\circ D$ and
that $D$ is antiderivation in $A$ with respect to $\omega$, then $(A,D)$ is
called {\it differential} algebra. So, in such a case, the elements of $Ker(D)$
form a subalgebra of $A$ and the elements of $Im(D)$ form an ideal in $Ker(D)$.
A {\it homotopy} operator \index{homotopy} $h$ in $E$ with respect to $D$ is a
linear mapping in $E$ such that $D\circ h+h\circ D=id(E)$, and such an operator
exists in $E$ only if the homology space (under $D$) is zero: $H(D,E)=\{0\}$.

\section{Multilinear Constructions}
\subsection{Tensor Product. Tensor Algebra}
{\bf Remark:} Further all linear spaces are assumed to be {\bf unitary}, i.e.,
the set of scalars must have unity.

 1. {\bf Multilinear mappings.} Let $(p+1)$ linear spaces $(E_i,G;\Gamma),
i=1,2,\dots,p$ be given. Then a mapping $\varphi: E_1\times
E_2\times\dots\times E_p\rightarrow G$ is called $p-linear$ if for each
$i=1,2,\dots,p$ the following relation holds:
$$
\varphi(x_1,\dots,x_{i-1},\lambda x_i+\mu y_i,x_{i+1},\dots,x_p)=
$$
$$
\lambda \varphi(x_1,\dots,x_i,\dots,x_p)+
\mu \varphi(x_1,\dots,y_i,\dots,x_p), \ x_i,y_i\in E_i, \ \lambda,\mu\in
\Gamma.
$$
If $G=\Gamma$ then $\varphi$ is called $p-$linear function. The $p-$linear
mappings may be summed up and multiplied by scalars:
$$
(\varphi+\psi)(x_1.\dots,x_p)=\varphi(x_1,\dots,x_p)+\psi(x_1,\dots,x_p); \
$$
$$
(\lambda \varphi)(x_1,\dots,x_p)=\lambda\varphi(x_1,\dots,x_p).
$$
If $p=2$ we have bilinear mappings.

2. {\bf Tensor product.}
The idea of tensor product \index{tensor product} of linear spaces is to
connect bilinear mappings with linear mappings. It is constructed in the
following way. Let $\varphi:E\times F\rightarrow G$ be a bilinear mapping, and
$H$ is any other linear space. \vskip 0.2cm {\bf Definition.} The pair
$(G,\varphi)$ is called a {\it tensor product} for $E$ and $F$ if the following
conditions hold:

$\otimes_1: Im(\varphi)=G$,

$\otimes_2$: For every bilinear mapping $\psi:E\times F\rightarrow H$ there
exists a  linear mapping  $f:G\rightarrow H$ such that $\psi=f\circ
\varphi$.
\vskip 0.2cm
If the conditions in the definition are satisfied then instead of $G$ we write
$E\otimes F$ and $\varphi(x,y)=x\otimes y$, and from the bilinearity it follows
$$
(\lambda x_1+\mu x_2)\otimes y=\lambda x_1\otimes y+\mu x_2\otimes y;
$$
$$
x\otimes(\lambda y_1+\mu y_2)=\lambda x\otimes y_1+\mu x\otimes y_2.
$$
If $(e_i)$ and $(k_j)$ are bases in $E$ and $F$ correspondingly, then
$(e_i\otimes k_j)$ form a basis of $E\otimes F$, therefore, $dim(E\otimes
F)=dim(E).dim(F)$ and each element $\mathfrak{t}\in E\otimes F$ can be
represented by $\Sigma_{i,j}\mathfrak{t}^{ij}e_i\otimes k_j$.

Having two linear mappings $\varphi:E\rightarrow E'$ and $\psi:F\rightarrow
F'$, a bilinear mapping $E\times F\rightarrow E'\otimes F'$ can be defined by
$(x,y)\rightarrow \varphi(x)\otimes\psi(y)$, so we obtain a linear mapping
$\chi:E\otimes F\rightarrow E'\otimes F'$: $\chi(x\otimes
y)=\varphi(x)\otimes\psi(y)$. Thus we obtain tensor product of linear mappings:
$(\varphi\otimes \psi)(x,y)=\varphi(x)\otimes\psi(y), \ x\in E, y\in F$.

3. {\bf Some properties}.

1. Composition property:
$(\varphi'\otimes\psi')\circ(\varphi\otimes\psi)=
(\varphi'\circ \varphi)\otimes (\psi'\circ \psi)$

2. Image property: $Im(\varphi\otimes\psi)=Im\varphi\otimes Im\psi$,

3. Kernel property: $Ker(\varphi\otimes\psi)=Ker(\varphi)\otimes F+E\otimes
Ker(\psi)$.

Tensor product of several linear spaces is constructed analogically,
just instead of bilinear mappings we make use of multilinear mappings. So,
under $p-$linear mappings we obtain:

-
$(\psi_1\otimes\dots\otimes\psi_p)\circ(\varphi_1\otimes\dots\otimes\varphi_p)=
(\psi_1\circ\varphi_1)\otimes\dots\otimes(\psi_p\circ\varphi_p)$

-
$Im(\varphi_1\otimes\dots\otimes\varphi_p)=Im(\varphi_1)\otimes\dots\otimes
Im(\varphi_p)$,

- $Ker(\varphi_1\otimes\dots\otimes\varphi_p)=\Sigma_{i=1}^{p}E_1\otimes\dots
\otimes Ker\varphi_i\otimes\dots\otimes E_p$

If $E,E^*$ and $F,F^*$ are two pairs of dual spaces then duality between
$E^*\otimes F^*$ and $E\otimes F$ is established by $\langle x^*\otimes
y^*,x\otimes y\rangle=\langle x^*,x\rangle\langle y^*,y\rangle $ . Similarly,
in the $p$-linear case we obtain
$$
\langle x^*_1\otimes\dots\otimes x^*_p,x_1\otimes\dots\otimes x_p\rangle=
\langle x^*_1,x_1\rangle\dots\langle x^*_p,x_p\rangle.
$$

In the finite dimensional case we have the isomorphism $E^*\cong L(E)$, this
enables to define isomorphism between $E^*\otimes F$ and $L(E,F)$.
Denoting this isomorphism by $T:E^*\otimes F\cong L(E,F)$
the definition is:
$$ T[(a^*\otimes
b)](x)=\langle a^*,x\rangle b, \ a^*\in E^*, b\in F, x\in E.
$$
We obtain
$$
\psi\circ T(a^*\otimes
b)=T(a^*\otimes \psi(b)), \ \ \psi\in L(F,E); $$ $$
 T(a^*\otimes b)\circ \psi=T(\psi^*a^*\otimes b), \ \ \psi\in L(F,E).
$$
The composition formula for two elements of $L(E,E)$ looks as follows:
$$
T(a^*\otimes a)\circ T(b^*\otimes b)=\langle a^*,b\rangle T(b^*\otimes a), \ \
a,b\in E; \ a^*,b^*\in E^*=L(E).
$$
Clearly, the linear map $a^*\otimes a$ sends the
whole vector space to the 1-dimensional subspace defined by $a\neq 0$. So, if
$\langle a^*,a\rangle=0$ then the corresponding composition
$(a^*\otimes a)\circ(a^*\otimes a)$ is a boundary map, and $dim(Ker(a^*\otimes
a))=dim(E)-1$. Also, if $dim\,=2n$, and $\{\varepsilon^i\}$, $\{e_i\}$ are
dual bases, then the combination
$J=\Sigma_i(-1)^i(\varepsilon^{2n+1-i}\otimes\,e_i)$ defines complex structure
in $E^{2n} : J\circ J=-id_{E}$.

Finally, we come to the trace formulas:
$$
tr[T(a^*\otimes b)]=\langle a^*,b\rangle \ \ a^*\in E^*, \ b\in E.
$$
$$
tr[T(a^*\otimes a)\circ T(b^*\otimes b)]=\langle a^*,b\rangle tr[T(b^*\otimes
a)]=\langle a^*,b\rangle\langle b^*,a\rangle.
$$

Recall that the set of linear mappings $L(E,E)$ has the structure of
associative algebra (with respect to composition) denoted usually by $A$. Now
a bilinear mapping $\Omega: A\times A\rightarrow L(A,A)$ is defined by
 $\Omega(\varphi\otimes\psi)(\chi)=\varphi\circ\chi\circ\psi$, and the pair
$(L(A,A),\Omega)$ is a tensor product for $A$ and $A$.

Another important property of the algebra $A=L(E,E)$ is that every linear
automorphism $\varphi$ of $E$ determines endomorphism $h_{\varphi}$ of $A$
according to $h_{\varphi}(\alpha)=\varphi\circ\alpha\circ\varphi^{-1}, \
\alpha\in A$, moreover, every endomorphism of $A$ is obtained in this way.

4. {\bf Tensors.} The elements of $\otimes^p(E)=E\otimes E\otimes\dots\otimes E$
($p$ copies of $E$) are called $p-tensors$ over $E$. If an element
$\mathfrak{t}\in\otimes^p(E)$ can be represented as $\mathfrak{t}=x_1\otimes
x_2\otimes\dots\otimes x_p$ then $\mathfrak{t}$ is called {\it decomposable}
\index{decomposable tensor}.
If $u\in\otimes^p(E)$ and $v\in\otimes^q(E)$ are decomposable then the element
$u\otimes v\in\otimes^{p+q}$ is the tensor product of $u$ and $v$ and is given
by
$$
u\otimes v=(x_1\otimes\dots\otimes x_p)\otimes(x_{p+1}\otimes\dots\otimes
x_{p+q}).
$$
This product is associative but not commutative except the case $dimE=1$.

If $(e_i)$ is a basis in $E$ then the products $e_{i_1}\otimes\dots\otimes
e_{i_p}$ form a basis of  $\otimes^p(E)$. If $dim(E)=n$ then
$dim(\otimes^p(E))=n^p$, and we obtain the unique representation of
$u\in\otimes^p(E)$ as follows: $$ u=\Sigma_{i}u^{i_1\dots
i_p}e_{i_1}\otimes\dots\otimes_{i_p}.
$$
Clearly, the direct sum $\Sigma_{p=1}^n\otimes^p(E)$ is a graded algebra.

Let $\varphi$ be a linear mapping in $E$. Then $\varphi$ is extended to
$\otimes^p(E)$ in two ways. First, $\varphi\rightarrow\varphi^{\otimes}$:
$$
\varphi^{\otimes}(x_1\otimes\dots\otimes
x_p)=\varphi(x_1)\otimes\dots\otimes\varphi(x_p).
$$
Second, $\varphi$ defines a derivation
$\theta^{\otimes}(\varphi)$ in $\otimes^p(E)$ as follows:
$$
\theta^{\otimes}(\varphi)(x_1\otimes\dots\otimes x_p)=
\sum_{i=1}^px_1\otimes\dots\otimes\varphi(x_i)\otimes\dots\otimes x_p,
$$
and $\theta^{\otimes}(\varphi)$ is extended to nondecomposable elements by
linearity. Of course, if $\psi$ is another linear mapping in $E$ then
$\theta^{\otimes}(\lambda\varphi+\mu\psi)=\lambda\,\theta^{\otimes}(\varphi)+
\mu\,\theta^{\otimes}(\psi)$. Finally we note the relation
$tr(\varphi\otimes\psi)=tr(\varphi).tr(\psi)$.

If $E^*$ is the dual to $E$ then duality between $\otimes^p(E^*)$ and
$\otimes^p(E)$ is given by
$$
\langle x^{*1}\otimes\dots\otimes x^{*p},x_{1}\otimes\dots\otimes x_{p}\rangle=
\langle x^{*1},x_1\rangle\dots\langle x^{*p},x_p\rangle.
$$
Thus $\langle u^*,v\rangle=v^{i_1\dots i_{p}}u^*_{i_1\dots i_{p}}$. Clearly,
the dual to $\varphi$ mapping $\varphi^*$ generates derivation
$\theta_{\otimes}(\varphi^*)$ in $\otimes^p(E^*)$ and
$\theta_{\otimes}(\varphi^*)$ is dual to $\theta^{\otimes}(\varphi)$.

Having $\otimes^p(E)$ and $\otimes_q(E^*)$ we can form
$\otimes^p_q(E,E^*)=(\otimes^pE)\otimes(\otimes_qE^*)$, these are tensors
of $p^{th}$ {\it contravariant} and $q^{th}$ {\it covariant} degree. In the
bases $\{e_i\}$ and $\{\varepsilon^j\}$ an element $T$ of
$\otimes^p_q(E,E^*)=(\otimes^pE)\otimes(\otimes_qE^*)$ looks as follows
$$
T=T^{i_1 i_2\dots i_p}_{j_1 j_2\dots j_q}e_{i_1}\otimes
e_{i_2}\dots\otimes
e_{i_p}\otimes\varepsilon^{j_1}\otimes\varepsilon^{j_2}\otimes\dots\otimes\varepsilon^{j_q}.
$$

The {\it contraction} operator \index{contraction operator}
$C_i^j$ in $\otimes^p_q(E,E^*)$ acts as follows:
$$
C_i^{j}(x_{1}\otimes\dots\otimes x_{p}\otimes x^{*1}\otimes\dots\otimes x^{*q})
$$
$$
=\langle x^{*j},x_i\rangle x_{1}\otimes\dots\otimes\hat{x_i}\otimes\dots\otimes
x_{p}\otimes x^{*1}\otimes\dots\otimes\hat{x^*_j}\otimes\dots\otimes x^{*q},
$$
where $\hat{x}$ means that these "hatted" elements are missed.

If $\{e_i\}$ and $\{\varepsilon^j\}$ are dual bases then the (1,1)-tensor
$\mathfrak{t}=\varepsilon^i\otimes e_i$ (summation over i=1,\dots,n)
is called {\it unit} tensor \index{unit tensor} for $(E^*,E)$. It is
independent of the couple of dual bases, which is due to the fact that if
$\varphi$ acts in $E$ then $\varphi$ acts in $E^*$ by the contragradient
$(\varphi^{-1})^*$ of $\varphi$. In fact we have the extension
$\varphi^{\otimes}\otimes\varphi_{\otimes}^{-1}$, where
$\varphi_{\otimes}^{-1}=(\varphi^{-1})^*\otimes\dots\otimes(\varphi^{-1})^*
$ of $\varphi$ in
$\otimes^p_q(E,E^*)$: if $z\in\otimes^p_q(E,E^*)$ then
$\varphi(z)=(\varphi^{\otimes}\otimes\varphi_{\otimes}^{-1})(z), z\in
\otimes^p_q(E,E^*)$.

A linear
mapping $\Phi:\otimes^p_q(E,E^*)\rightarrow\otimes^r_s(E,E^*)$ is called {\it
tensorial} if for every linear automorphism $\psi$ of $E$ we have
$\Phi(\psi(z))=\psi(\Phi(z))$, where by the same letter $\psi$ is denoted the
extension of $\psi$ in $\otimes^p_q(E,E^*)$. For example, the contraction
operator is tensorial.

\subsection{Exterior and Symmetric Algebras}
Recall the elementary concepts from group of permutations. Let a set of
$p$ elements be enumerated by the natural numbers $1,2,\dots,p:
x_1,x_2,\dots,x_p$. Then an rearrangement
$(\sigma(1),\sigma(2),\dots,\sigma(p))$ of the parametrizing numbers
$(1,2,\dots,p)$ yields permutation in $x_1,x_2,\dots,x_p$, given by
$$
\sigma(x_1,x_2,\dots,x_p)=(y_1,y_2,\dots,y_p)=
(x_{\sigma^{-1}(1)},x_{\sigma^{-1}(2)},\dots,x_{\sigma^{-1}(p)}).
$$
A permutation is called {\it transposition} if it replaces just two elements:
$x_i\leftrightarrows x_j$. If a transposition replaces two neighboring
elements, i.e. $j=i+1$, then it is called {\it n-transposition}. Clearly, the
composition $\sigma\circ\sigma$ of a transposition $\sigma$ gives the identity,
so {\it transpositions are involutions}. It is known that {\it every
permutation can be represented as appropriate composition of n-transpositions},
and there are many such representations. The number of representations of a
given permutation through n-transpositions may be {\it even} or {\it odd}, so,
the concept of {\it signature} $\varepsilon_{\sigma}$ of a given permutation
$\sigma$ is introduced such that in the {\it even} case it is assumed
$\varepsilon_{\sigma}=1$, and in the {\it odd} case it is assumed
$\varepsilon_{\sigma}=-1$.

Consider now an decomposable element $(x_1\otimes x_2\otimes\dots\otimes x_p)\in
\otimes^p(E)$. Under the action of a permutation $\sigma$ we get
$(x_{\sigma^{-1}(1)}\otimes
x_{\sigma^{-1}(2)},\otimes\dots,\otimes x_{\sigma^{-1}(p)})$.
Denote by
$N^p(E)$ the space generated by all products $x_1\otimes\dots\otimes x_p$ such
that $x_i=x_j$ for at least one pair $i\neq j$. Clearly, every permutation
transforms $N^p(E)$ into itself. It can be shown that if $u\in \otimes^p(E)$
then $(u-\varepsilon_{\sigma}\sigma(u))\in N^p(E)$. Now, since every
permutation can be represented as a composition of n-transpositions we obtain
that if $\tau$ is a n-transposition and $\sigma$ is represented by
$m$ n-transpositions, then $(u-\varepsilon_{\tau\sigma}\tau\sigma(u))\in
N^p(E)$. Thus, we have a projection operator $\pi_{A}:
\otimes^p(E)\rightarrow\otimes^p(E)$, called {\it alternator} :
$\pi_A=\frac{1}{p!}\Sigma_{\sigma}\varepsilon_{\sigma}\sigma$. If $x_i,
i=1,\dots,p$ are linearly independent in $E$ then the products
$(x_{\sigma^{-1}(1)}\otimes x_{\sigma^{-1}(2)},\otimes\dots,\otimes
x_{\sigma^{-1}(p)})$ are linearly independent and we obtain
$$
\pi_A(x_1\otimes x_2\otimes\dots\otimes x_p)=
\Sigma_{\sigma}\varepsilon_{\sigma}
(x_{\sigma^{-1}(1)}\otimes
x_{\sigma^{-1}(2)},\otimes\dots,\otimes x_{\sigma^{-1}(p)})\neq 0.
$$
We obtain also that $Ker(\pi_A)=N^p(E)$, so, if $X^p(E)$ is the image space of
$\pi_A$ we have the direct decomposition $\otimes^p(E)=N^p(E)\oplus X^p(E)$.
The elements of $X^p(E)$ are called {\it skew symmetric} tensors of order $p$.

If $E^*$ is the dual space of $E$ we obtain the action of the permutation
$\sigma$ in $E^*$ according to the duality relation $\langle
u^*,\sigma(u)\rangle=\langle\sigma^{-1}u^*,u\rangle, u^*\in \otimes^p(E^*),
u\in\otimes^p(E)$. Denoting by $\pi^A$ the corresponding alternator in
$\otimes^p(E^*)$ we obtain that $\pi_A$ and $\pi^A$ are dual. This duality
leads to the following duality between the corresponding image spaces:
$$
\langle\pi^A(x^{*1}\otimes\dots\otimes x^{*p}),
\pi_A(x_1\otimes\dots\otimes x_p)\rangle=\frac{1}{p!}\det(\langle
x^{*i},x_j\rangle) .
$$

If $\otimes(E)$ is the tensor algebra over $E$ then the direct sum
$N(E)=\sum N^p(E)$ is a (graded) ideal in $\otimes(E)$, and for two arbitrary
tensors $u\in\otimes^p(E)$ and $v\in\otimes^q(E)$ we obtain
$(u\otimes v-(-1)^{pq}v\otimes u)\in N^{p+q}(E)$. Thus, $\pi_A(u\otimes
v)=(-1)^{pq}\pi_A(v\otimes u)$. Forming the factor algebra $\otimes(E)/N(E)$
with canonical projection $\pi$ we obtain multiplication in $\otimes(E)/N(E)$
by $\pi(a).\pi(b)=\pi(a\otimes b), a,b\in \otimes(E)$. So, for every two
homogeneous elements of degree $p$ and $q$ we have the commutation relation
$u.v=(-1)^{pq}v.u$. Also, denoting by $X(E)$ the direct sum $\sum_pX^p(E)$, then
the isomorphism $\rho: X(E)\cong\otimes(E)/N(E)$
induces a scalar product between $\otimes(E)/N(E)$ and $\otimes(E^*)/N(E^*)$
by $\langle \rho u^*,\rho u\rangle =p!\langle u^*,u\rangle, u^*\in X^p(E^*),
u\in X^p(E)$. For two decomposable elements $x_1\otimes\dots\otimes x_p$ and
$x^{*1}\otimes\dots\otimes x^{*p}$ we obtain
$$
\langle \pi(x^{*1}\otimes\dots\otimes x^{*p}),\pi(x_1\otimes\dots\otimes
x_p)\rangle=\det(\langle x^{*i},x_j\rangle).
$$

The composition $\pi\circ\otimes $ is called {\it exterior product}, so we
have the $p$-th exterior product $\Lambda_p(E)=E\wedge E\dots\wedge E$
and the $p$-th exterior product $\Lambda^p(E^*)=E^*\wedge\dots\wedge E^*$
correspondingly. The decomposable elements of $\Lambda_p(E)$ look like
$x_1\wedge\dots\wedge x_p$ and are called {\it p-vectors} \index{p-vector}
, and the
decomposable elements $x^{*1}\wedge\dots\wedge x^{*p}$ of $\Lambda^p(E^*)$ are
called {\it p-forms} \index{p-form}
. The direct sums $\bigwedge(E)=\sum_p\Lambda_p(E)$ and
$\bigwedge(E^*)=\sum_p\Lambda^p(E^*)$ together with the corresponding exterior
products are called {\it exterior algebras} \index{exterior algebra}
over $E$ and $E^*$ respectively.
For example, $x\wedge y=x\otimes y-y\otimes x$, and $(x_1\wedge
x_2)\wedge(x_3\wedge x_4)=x_1\wedge x_2\wedge x_3\wedge x_4$.

It is important to keep in mind that if $x_1\wedge x_2\wedge\dots\wedge x_p\neq
0$ then all vectors $x_1,\dots,x_p$ are {\it linearly independent}, so {\bf they
define a $p-$dimensional subspace} in $E$. Also, if $x^{*1},\dots,x^{*(n-p)}$
are linearly independent, then $x^{*1}\wedge\dots\wedge x^{*(n-p)}\neq 0$.
Usually, if $\langle x^{*i},x_j\rangle=0 $ the subspace in $E^*$ defined by
$x^{*1}\wedge x^{*2}\wedge\dots\wedge x^{*(n-p)}\neq 0$ is
called {\it orthogonal} to that defined by
$x_1\wedge x_2\wedge\dots\wedge x_p\neq 0$.

 Following the same logic but ignoring
$\varepsilon_{\sigma}$ wherever it appears above, we come to the {\it symmetric
tensors}. The corresponding projection $\pi_{s}$ is called {\it symmetrizer}
and looks like $\pi_s=\frac{1}{p!}\sum_{\sigma}\sigma$. The direct sum of the
image space $Y^p$ of $\pi_s$ for $p$-tensors and the kernal space $Ker(\pi_s)$,
denoted by $M^p(E)$, yields $\otimes^p(E)$. Hence, if $u\in\otimes^p(E)$ then
$\pi_s(u)$ is its symmetric part. If $u^*=x^{*1}\otimes\dots\otimes
x^{*p}\in\otimes^p(E^*)$ and $u=x_1\otimes\dots\otimes x_p\in \otimes^p(E)$
then the duality yields $$ \langle\pi^s(x^{*1}\otimes\dots\otimes
x^{*p}),\pi_s(x_1\otimes\dots\otimes
x_p)\rangle=\frac{1}{p!}\mathrm{perm}(\langle x^{*i},x_j\rangle),
$$
where
$\mathrm{perm}(\alpha_i^j)=
\sum_{\sigma}\alpha^1_{\sigma(1)}\dots\alpha^p_{\sigma(p)}$. Also,
$M(E)=\sum_pM^p(E)$ is graded ideal in $\otimes(E)$. Clearly, if $u,v$ are two
arbitrary elements of $\otimes(E)$ then $u\otimes v-v\otimes u\in M(E)$, and in
the factor algebra $\otimes(E)/M(E)$ we get the multiplication
$\pi(a).\pi(b)=\pi(a\otimes b), a,b\in \otimes(E)$. Finally, the duality
between $\otimes(E)/M(E)$ and $\otimes(E^*)/M(E^*)$ yields
$$
\langle \pi(x^{*1}\otimes\dots\otimes x^{*p}),\pi(x_1\otimes\dots\otimes
x_p)\rangle=\mathrm{perm}(\langle x^{*i},x_j\rangle).
$$
The corresponding composition $\pi\circ\otimes$ is denoted by $\vee$, so,
$x_1\vee\dots\vee x_p\in S_p(E)$ and $x^{*1}\vee\dots\vee x^{*p}\in S^p(E^*)$
denote symmetric decomposable p-vectors and symmetric decomposable p-forms
respectively. For example, $x\vee y=x\otimes y+y\otimes x$, and $(x_1\vee
x_2)\vee(x_3\vee x_4)= x_1\vee x_2\vee x_3\vee x_4$. Finally, the direct sums
$\bigvee(E)=\sum_pS_p(E)$ and $\bigvee(E^*)=\sum_pS^p(E)$ are called {\it
symmetric algebras} \index{symmetric algebra}
 over $E$ and $E^*$ respectively.

If $(e_1,\dots,e_n)$ is a basis in $E, dimE=n$, then all
$(e_{i_1}\wedge
e_{i_2}\wedge\dots\wedge e_{i_p}), 1\leq i_1<i_2<\dots <i_p\leq n$
form a basis
in $\Lambda^p(E)$. So, $dim\,\Lambda^p(E)=n!/p!(n-p)!$, $dim\,\Lambda^n(E)=1$,
$dim\,\bigwedge(E)=2^n$. The same rules are used for $\bigwedge(E^*)$.

In the symmetric case the basis of $S^P(E)$ is formed by all
$(e_{i_1}\vee
e_{i_2}\vee\dots\vee e_{i_p}), \linebreak 1\leq i_1\leq i_2<\dots \leq i_p\leq
n$, $dim\,S^p(E)=(p+n-1)!/p!(n-p)!$.

If $\varphi$ is a linear mapping in $E$ then it induces a mapping
$\Lambda^p\varphi$ in $\Lambda^p(E)$ according to:
$$
\Lambda^p(\varphi)(x_1\wedge\dots\wedge
x_p)=\varphi(x_1)\wedge\dots\wedge\varphi(x_p),
$$
and a mapping $\vee^p$ in $S^p(E)$ according to
$$
\vee^p(\varphi)(x_1\vee\dots\vee x_p)=
\varphi(x_1)\vee\dots\vee\varphi(x_p).
$$
The same is true for $\varphi^*$ in the dual space(s).

Every such linear mapping induces also derivations and antiderivations in the
same way as in the tensor algebra $\otimes(E)$, just instead of $\otimes$ we
write $"\wedge"$, or $"\vee"$ correspondingly.

The duality between $E$ and $E^*$ allows to distinguish the following
antiderivation. Let $h\in E$, then we obtain the derivation
$i(h)$, or $i_h$, in $\Lambda(E^*)$ of degree $(-1)$ (sometimes called
substitution/contraction operator, interior product) \index{interior product}
 according to:
$$
i(h)(x^{*1}\wedge\dots\wedge
x^{*p})=\sum_{i=1}^{p}(-1)^{(i-1)}\langle x^{*i},h\rangle
x^{*1}\wedge\dots\wedge\hat{x^{*i}}\wedge
\dots\wedge x^{*p}.
$$
Clearly, if $u^*\in \Lambda^p(E^*)$ and $v^*\in\Lambda(E^*)$ then
$$
i(h)(u^*\wedge v^*)=(i(h)u^*)\wedge v^*+(-1)^pu^*\wedge i(h)v^*.
$$
Also, we get
$$
i(h)u^*(x_1,\dots,x_{p-1})=u^*(h,x_1,\dots,x_{p-1}), \
$$
$$
i(x)\circ i(y)=-i(y)\circ i(x).
$$

This antiderivation is extended to a mapping $i(h_1\wedge\dots\wedge h_p):
\Lambda^m(E^*)\rightarrow\Lambda^{(m-p)}(E^*)$, $m\geqq p$,
according to
$$
i(h_1\wedge h_2\wedge\dots\wedge h_p)u^*=i(h_p)\circ\dots\circ i(h_1)\,u^*.
$$
Note that this extended mapping is not an antiderivation except for $p=1$.

This mapping is extended to multivectors and exterior forms which are linear
combinations: if $\Psi=\Psi_1+\Psi_2+...$ is an arbitrary multivector on $E$ and
$\Phi=\Phi^1+\Phi^2+...$ is an arbitrary exterior form on $E^*$ then
$i_{\Psi}\Phi$ is defined as extention by linearity, e.g.,
$$
i(\Psi_1+\Psi_2)(\Phi^1+\Phi^2)
=i(\Psi_1)\Phi^1+i(\Psi_1)\Phi^2+i(\Psi_2)\Phi^1+i(\Psi_2)\Phi^2.
$$

If the interior product $i(\Psi)\Phi$
between the $p-$vector $\Psi$ and the $q-$exterior form $\Phi$ is not zero:
$i(\Psi)\Phi\neq 0$, then $\Psi$ and $\Phi$ may be called {\it partners}.

The above relations suggest to talk about {\it attraction/sensitivity} between
a couple of partnering $p-$vector $\Psi$ and a $q-$form $\Phi$ in the following
sense. If we consider the $q$-form $\Phi$ as a volume form on the subspace
$E^q\subset E$, and $\Psi$ is a nonzero $p$-vector on $E^q, p<q$, then the
expression $i(\Psi)\Phi$ is surely different from zero, so, we could say that
$\Phi$ and $\Psi$ {\it feel the presence of each other}. Now, if we consider
$\Phi$ as an usual $p$-form on the vector space $E$, i.e., not as a volume form
on a subspace, it is not necessary that, for different $\Psi$ defined on  $E$,
we must always obtain $i(\Psi)\Phi\neq 0$. In view of this we are going to say
that if $i(\Psi)\Phi\neq 0$, then the $p$-form $\Phi$ is $\Psi-attractive$, or
that $\Psi$ is $\Phi-sensitive$, or that the algebraic flow of $\Psi$ across
$\Phi$ is not zero.

This concept of partnering, or of {\it attractiveness/sensitivity}
\index{attractiveness/sensitivity} is easily extended to vector-valued forms,
i.e. to the space $\Lambda^p(E^*)\otimes W$, where $W$ is another vector space.
In fact, if $\varphi:W\times W\rightarrow W$ is a bilinear map, and $e_i,
i=1,2,...,dim(W)$ is a basis of $W$ we consider the objects
$\Gamma=\Psi_i\otimes e_i$, where $\Psi_i$ are $p$-vectors, and
$\Omega=\Phi^j\otimes e_j$, where $\Phi_j$ are $q$-forms, (summing with respect
to $(i,j)$). Now we form the expression
$i(\Psi_i)\Phi^j\otimes\varphi(e_i,e_j)$. This suggests to consider $\Omega$ as
$\Gamma-attractive$  with
respect to $\varphi$, or, $\Gamma$ as $\Phi-sensitive$ with respect to
$\varphi$, if at least one of the summonds is not zero, and if all summonds are
different from zero we can talk about
$(\Omega\leftarrow\Gamma)-${\it special attraction}, or
$(\Gamma\rightarrow\Omega)$-{\it special sensitivity}.

Finally we note that the two spaces $\Lambda^p(E)\otimes\Lambda^n(E^*)$
and $\Lambda^{n-p}(E^*)$
have the same dimension, so, every nonzero $\omega\in\Lambda^n(E^*)$
generates isomorphism $D^{p}$, called Poincare isomorphism, \index{Poincare
isomorphism} between these two spaces according to
$(u,\omega)\rightarrow i(u)\omega$, where $u\in\Lambda^p(E)$ is a $p$-vector
over $E$. In particular, if $\{e_i\}$ and $\{\varepsilon_j\}$ are dual bases,
the corresponding basis elements
$$
e_{\nu_1}\wedge\dots\wedge e_{\nu_p},\ \ \ \nu_1<\nu_2<...<\nu_p ,
$$
and
$$\varepsilon^{\nu_{p+1}}\wedge\dots\wedge\varepsilon^{\nu_n} , \ \ \
\nu_{p+1}<\nu_{p+2}<...<\nu_n,
$$ are connected according to
$$
D^{p}(e_{\nu_1}\wedge\dots\wedge e_{\nu_p})=
(-1)^{\sigma}\varepsilon^{\nu_{p+1}}\wedge\dots\wedge\varepsilon^{\nu_n},
$$
where $\sigma=\sum_{i=1}^p(\nu_i-i)$. Also,
$$
D_{p}(\varepsilon^{\nu_1}\wedge\dots\wedge \varepsilon^{\nu_p})=
(-1)^{\sigma}e_{\nu_{p+1}}\wedge\dots\wedge e_{\nu_n},
$$
$$
D_{p}(\varepsilon^{\nu_{p+1}}\wedge\dots\wedge \varepsilon^{\nu_n})=
(-1)^{p(n-p)+\sigma}e_{\nu_1}\wedge\dots\wedge e_{\nu_p}.
$$
Clearly, we have
$$
i(e_{\nu_1}\wedge\dots\wedge e_{\nu_p})
D^{p}(e_{\nu_1}\wedge\dots\wedge e_{\nu_p})=0.
$$
Also, we note that in this way every subspace $V^p\subset E$ leads to
defining three other spaces:
$$
(V^p)^*\subset E^*; \ \ D^p(V^p)\subset E^*; \ \
 D_p((V^p)^*)\subset E \ ,
$$ where $D^p(V^p)$ is
orthogonal to $V^p$ and $D_p((V^p)^*)$ is orthogonal to $(V^p)^*$, and
$$
E=V^p\oplus(D^p(V^p))^* \ ; \ E^*=(V^p)^*\oplus D^p(V^p) .
$$
We can say that
$V^p$ and $V_{n-p}^*=D^p(V^p)$ are not attractive/sensitive to
each other, and $V^{n-p}$ and $V_{p}^*$ are not attractive/sensitive to each
other.

Finally, we should not forget that these isomorphisms depend on the chosen
element $\omega\in\Lambda^n(E^*)$, but in what follows we shall omit writing
$\omega$ for clarity.

 These last two formulas allow to make use of any
isomorphism between $E$ and $E^*$ for defining isomorphisms
$\Lambda^p(E)\cong\Lambda^{n-p}(E)$, and
$\Lambda^p(E^*)\cong\Lambda^{n-p}(E^*)$, e.g., the Hodge $*$-operator, where
the isomorphism used is defined by a metric tensor.

For these isomorphisms and their duals
$$
D^p:\Lambda^p(E)\rightarrow\Lambda^{n-p}(E^*), \ \
(D^p)^*: \Lambda^{n-p}(E)\rightarrow\Lambda^p(E^*)
$$
$$
D_p:\Lambda^p(E^*)\rightarrow\Lambda^{n-p}(E), \ \
(D_p)^*: \Lambda^{n-p}(E^*)\rightarrow\Lambda^p(E) ,
$$
the following relations also hold:
$$ (D_p)^*=(D^p)^{-1}=(-1)^{p(n-p)}D_{n-p} \ ; \ \
(D^p)^*=(D_p)^{-1}=(-1)^{p(n-p)}D^{n-p} \ ;
$$
$$ D_{n-p}\circ D^p=(-1)^{p(n-p)}id , \ \
D^{n-p}\circ D_p=(-1)^{p(n-p)}id ,
$$
where $id$ denotes the corresponding identity map. So,
up to a sign factor, $D^p$ and $D_{n-p}$ are inverse linear isomorphisms.
It seems important to have always in mind this possibility to connect with every
subspace $W$ three other subspaces: $W^*, D^p(W)$, and $(D^p(W))^*$.

We make some remarks concerning the concept of symmetry.

Let $E$ be a $n$-dimensional linear space over $\mathbb{R}$ and $x\in E, x\neq
0$, so, $x$ generates 1-dimensional subspace of $E$. If $\varphi: E\rightarrow
E$ is a linear map in $E$ such that $\varphi(x)=x$ we say that $\varphi$ is a
symmetry of $x\in E$. If for each $\lambda\in\mathbb{R}$ we have
$\varphi(x)=\lambda\varphi(x)$, we say that $\varphi$ is a symmetry of the
1-dimensional subspace generated by $x$, or that this 1-dimensional subspace is
invariant with respect to $\varphi$, and for some $\lambda\in \mathbb{R}$
 it is an eigen space of $\varphi$.

Let $E_p$ denote the $p$-dimensional space generated by the
linearly independent elements $x_i, i=1,2,...,p : x_1\wedge x_2\wedge ...
\wedge x_p\neq 0$, and consider the corresponding $\frac{n!}{p!(n-p)!}$ -
dimensional space $\Lambda_p(E)$ of $p$-vectors. We say that $\varphi$
is a {\it symmetry} of $\Lambda_p(E)$ if
$$
\varphi(x_i)\wedge x_1\wedge
x_2\wedge ... \wedge x_i\wedge ...\wedge x_p=0,
$$
i.e. if every $\varphi(x_i)$ is linearly reprsentable by (some or all) of the
generators $x_i, i=1,...,p$, of $\Lambda_p(E)$.

Consider now the symmetry $\varphi$ of $\Lambda_p(E)$ and the two subspaces
$E_r\subset E_p$ and $E_s\subset E_p$, where $r,s<p$. If the restriction of
$\varphi$ to $E_r\subset E_p$ takes values in $E_s\subset E_p$ and the
restriction of $\varphi$ to  $E_s\subset E_p$ takes values in $E_r\subset E_p$,
we say that the symmetry $\varphi$ generates {\it intercommunication} between
$E_r\subset E_p$ and $E_s\subset E_p$. This intercommunication defines also
intercommunication between the corresponding $\Lambda_r(E)$ and $\Lambda_s(E)$.
Hence, having sufficient number of such symmetries of $E_p$ that
intercommunicate the various subspaces of $E_p$ we can talk about structure of
$E_p\subset E$.

Let now $\Phi: E\times E\rightarrow E$ be a {\it bilinear} map. If for each
couple $(x_i,x_j), i,j=1,2,...,p$ of the generators of $\Lambda_p(E)$ we have
that $\Phi(x_i,x_j)$ is linearly representable by generators of $\Lambda_p(E)$
we also say that $\Phi$ is a symmetry of $\Lambda_p(E)$. We may also say that a
linear combination $x=a^ix_i, a^i\in\mathbb{R}$, of generators of
$\Lambda_p(E)$ generates symmetry of $\Lambda_p(E)$ with respect to $\Phi$ if
$\Phi(x,x_i)$ is representable by generators $\Lambda_p(E)$.

Finally, if $z\in E$ can NOT be represented linearly by the generators of
$\Lambda_p(E)$, i.e. $z$ lives out of $E_p$, then, if for each $x\in E_p$ the
value $\Phi(z,x)$ is linearly representable by generators of $\Lambda_p(E)$,
we say that $z\in E$ is {\it external} symmetry of $\Lambda_p(E)$ with respect
to $\Phi$. This last precision is useful when Lie algebras are studied and will
be of use for us when integrability of distributions on manifolds will be later
considered.

 \subsection{Brackets}
The algebraic generalization of the elementary concept of Lie-bracket in a Lie
algebra aims to find those maps in an algebraic structure, which are {\it
tensorial} and carry some properties of {\it (anti)symmetry}. We recall the
purely algebraic Schouten (or, Schouten-Nijenhuis) bracket \index{brackets}
(SN bracket) acting
in $\Lambda(\mathcal{G})$, of a Lie algebra ($\mathcal{G,[\,,]}$). Let
$U=(x_1\wedge\dots\wedge x_p)$ and $V=(y_1\wedge\dots\wedge y_q)$ be two
decomposable elements of $\Lambda^p(\mathcal{G})$ and $\Lambda^q(\mathcal{G})$
respectively. Then their SN-bracket $[U,V]\in\Lambda^{p+q-1}(\mathcal{G})$ is
defined by $$
[U,V]=\sum_{i,j}(-1)^{i+j}[x_i,y_j]x_1\wedge\dots\wedge\hat{x_i}\wedge\dots\wedge x_p
\wedge y_1\wedge\dots\wedge\hat{y_j}\wedge\dots\wedge y_q,
$$
where the "hat" means that this element is skipped. In particular,
$$
[x\wedge y, x\wedge y]=2[x,y]\wedge x\wedge y, \ \
[x\wedge y, z\wedge y]=-[x,y]\wedge y\wedge z-[y,z]\wedge x\wedge y.
$$
This bracket defines a grading and satisfies the following relations:
$$
[U,V]=-(-1)^{(p-1)(q-1)}[V,U],
$$
$$
[U,V\wedge W]=[U,V]\wedge W+(-1)^{(degU-1)degV}V\wedge[U\wedge W].
$$

If the SN-bracket $[U,V]$ is nonzero then $(U,V)$ may be called SN-partners
since their mutually induced change is not zero.

This example suggests to consider all graded derivations of degree $k$ in
$\bigwedge(E)$, where $E$ is a Lie algebra. These are linear mappings
$D:\bigwedge(E)\rightarrow\bigwedge(E)$ satisfying  the two
properties:
$$ D(\Lambda^p(E))\subset\Lambda^{p+k}(E), \ \ \ \ D(P\wedge
Q)=D(P)\wedge Q+(-1)^{kp}P\wedge D(Q),
$$
where $P\in\Lambda^p(E)$. If we
consider now the space of all derivations
$Der(\bigwedge(E))=\sum_kDer_k(\bigwedge(E))$ it turns out that this space is a
graded Lie algebra with respect to the following bracket:
$$
[D_1,D_2]=D_1\circ D_2-(-1)^{k_1k_2}D_2\circ D_1, \ \ D_1\in
Der_{k_1}(\bigwedge(E)), \ \ D_2\in Der_{k_2}(\bigwedge(E)).
$$
Moreover,
$Der(\bigwedge(E))$ is (super)anticommutative:
$[D_1,D_2]=-(-1)^{k_1k_2}[D_2,D_1]$, and the graded Jacobi identity holds:
$$
[D_1,[D_2,D_3]]=[[D_1,D_2],D_3]+(-1)^{k_1k_2}[D_2,[D_1,D_3]].
$$

Another example, extending the graded operator $i(x)$ of degree $(-1)$ in
$\bigwedge (E^*)$, is the following. Consider the space of antisymmetric
multilinear mappings $\Lambda^{k+1}(E)\rightarrow E$, i.e. the space
$\Lambda^{k+1}(E^*)\otimes E$. Now if $K\in\Lambda^{k+1}(E^*)\otimes E$ and
$\omega\in\Lambda^l(E^*)$, and if $K=\alpha\otimes y$, where
$\alpha\in\Lambda^{k+1}(E^*)$, then
$i_K\omega=i(\alpha\otimes y)\omega=\alpha\wedge i(y)\omega$ is of degree
$(k+l)$. Hence, $i_K(\omega)$ satisfies the relation $$
i_K\omega(x_1,\dots,x_{k+l})
$$
$$
=\frac{1}{(k+1)!(l-1)!}
\sum_{\sigma\in
S_{k+l}}\varepsilon_(\sigma)\omega(K(x_{\sigma(1)},\dots,x_{\sigma(k+1)}),
x_{\sigma(k+2)},\dots,x_{\sigma(k+l)}).
$$
Clearly, $i_K(\omega)$ is extensible to $i_K(\omega\otimes y)$ according to
$$
i_K(\omega\otimes y)=i_K(\omega)\otimes y .
$$
Thus to every two $E$-valued multilinear forms $K$ and $L$
on $E$ of degree "k+1" and "l+1" respectively, we can associate their bracket
$[i_K,i_L]$, which satisfies
$$
[i_K,i_L]=i_{K}L-(-1)^{kl}i_LK.
$$

Let now $K,L\in L(\mathcal{G},\mathcal{G})\cong\mathcal{G}^*\otimes\mathcal{G}$.
The bracket looks as follows:
$$
[K,L](x,y)=[K(x),L(y)]-[K(y),L(x)]-L\Big([K(x),y]-[K(y),x]\Big)
$$
$$
-K\Big([L(x),y]-[L(y),x]\Big)+(KL+LK)([x,y]).
$$

In case a derivation $\mathbf{d}:\Lambda^p(\mathcal{G}^*)
\rightarrow\Lambda^{p+1}(\mathcal{G}^*)$
 is given, we consider the $\mathcal{G}$-valued forms: $\alpha\otimes x$ and
$\beta\otimes y, \alpha\in\Lambda^p(\mathcal{G}^*),
\beta\in\Lambda^q(\mathcal{G}^*), \ x,y\in \mathcal{G}$.
Then recalling that $\mathcal{G}$ is a Lie algebra with a bracket
$[x,y]$ and that $\bigwedge(\mathcal{G}^*)\otimes\mathcal{G}$ is a
module over $\bigwedge(\mathcal{G}^*))$ we get:
$$
[\alpha\otimes x,\beta\otimes y]
=\alpha\wedge\beta\otimes[x,y]-i(y)\mathbf{d}\alpha\wedge\beta\otimes x
+(-1)^{pq}i(x)\mathbf{d}\beta\wedge\alpha\otimes y
$$
$$
-\mathbf{d}(i(y)\alpha\wedge\beta)\otimes x+
(-1)^{pq}\mathbf{d}(i(x)\beta\wedge\alpha)\otimes y.
$$
It deserves noting that $[\alpha\otimes x,\alpha\otimes x]$ is NOT necessarily
zero.
\vskip 0.3cm

	{\bf Important remark}.
Note that these relations may be correspondingly adapted for differential forms
on a manifold $M$, valued in the corresponding tangent bundle $T(M)$ and
usually denoted by $\Lambda(M,TM)$, since, according to the above, every such
differential form defines a graded algebraic derivation in $\Lambda(M)$ with
respect to the usual exterior derivative $\mathbf{d}$ in $\Lambda(M)$.
The corresponding bracket operation is called {\it Fr$\ddot{o}$licher-Nijenhuis}
bracket. For details see [7, Sec.16].

\vskip 0.3cm
If $\mathfrak{A}$ and $\mathfrak{R}$
are two algebras then in their tensor product there is a natural algebraic
operation defined by
$(\mathfrak{a_1}\otimes\mathfrak{r_1}).(\mathfrak{a_2}\otimes\mathfrak{r_2})=
(\mathfrak{a_1}.\mathfrak{a_2})\otimes(\mathfrak{r_1}.\mathfrak{r_2})$. As an
example, if $\mathfrak{R}$ is an associative algebra and $L(\mathfrak{R})$
denotes the linear mappings in $\mathfrak{R}$ endowed with the commutator, then
we obtain the operation
$[\mathfrak{r_1}\otimes\varphi_1,\mathfrak{r_2}\otimes\varphi_2]=
\mathfrak{r_1}.\mathfrak{r_2}\otimes[\varphi_1,\varphi_2]$. If $\bigwedge(E^*)$
is the exterior algebra over $E^*$, $\alpha\in\Lambda^p(E^*)$,
$\beta\in\Lambda^{q}(E^*)$, we have
$$
[\alpha\otimes\varphi_1,\beta\otimes\varphi_2]=
\alpha\wedge\beta\otimes[\varphi_1,\varphi_2]=
\alpha\wedge\beta\otimes(\varphi_1\circ\varphi_2-\varphi_2\circ\varphi_1)
$$
$$
=\dots= (\alpha\otimes\varphi_1)\wedge(\beta\otimes\varphi_2)-
(-1)^{pq}(\beta\otimes\varphi_2)\wedge(\alpha\otimes\varphi_1),
$$
i.e., we obtain the so called "super commutator" in $\bigwedge(E^*)\otimes
L(E)$.

Note that the Lie algebraic structure always requires $[A,A]=0$. In order to
define a bracket operation of linear maps such that $[\varphi,\varphi]\neq 0$ in general,
let $\phi$ and $\psi$ be two arbitrary linear maps in a module $\mathfrak{M}$,
and $\mathfrak{B}:\mathfrak{M}\times \mathfrak{M}\rightarrow\mathfrak{M}$ be
just a binar map satisfying
$\mathfrak{B}(\mathbf{x+z},\mathbf{y})=\mathfrak{B}(\mathbf{x},\mathbf{y})+
\mathfrak{B}(\mathbf{z},\mathbf{y})$ and
$\mathfrak{B}(\mathbf{x},\mathbf{y+z})=\mathfrak{B}(\mathbf{x},\mathbf{y})+
\mathfrak{B}(\mathbf{x},\mathbf{z})$, where
$(\mathbf{x},\mathbf{y},\mathbf{z})$ are three arbitrary elements of
$\mathfrak{M}$. Aiming to define the desired bracket, called
$\mathfrak{B}-$bracket, for $\varphi$ and $\psi$ , we consider the expression
$$ \mathcal{A}(\mathfrak{B};\phi,\psi)(\mathbf{x},\mathbf{y})\equiv
\frac12\Big[\mathfrak{B}(\phi(\mathbf{x}),\psi(\mathbf{y}))+
\mathfrak{B}(\psi(\mathbf{x}),\phi(\mathbf{y}))+
\phi\circ\psi(\mathfrak{B}(\mathbf{x},\mathbf{y}))
$$
$$
+\psi\circ\phi(\mathfrak{B}(\mathbf{x},\mathbf{y}))
-\phi(\mathfrak{B}(\mathbf{x},\psi(\mathbf{y})))-
\phi(\mathfrak{B}(\psi(\mathbf{x}),\mathbf{y}))-
\psi(\mathfrak{B}(\mathbf{x},\phi(\mathbf{y})))-
\psi(\mathfrak{B}(\phi(\mathbf{x}),\mathbf{y}))\Big]\ \ .
$$

If now $\varphi$ is a linear map in $\mathfrak{M}$, this bracket allows to see
how it $\mathfrak{B}$-changes along itself and to build quantities
describing intercommunication between the generated by $\varphi$ subspaces of
$\mathfrak{M}$. For example, assuming $\phi=\psi=P$, where $P\neq
id_{\mathfrak{M}}$ is a projection: $P\circ P=P$, this expression reduces to $$
\mathcal{A}(\mathfrak{B};P)(\mathbf{x},\mathbf{y})=
P(\mathfrak{B}(\mathbf{x},\mathbf{y}))+\mathfrak{B}(P(\mathbf{x}),P(\mathbf{y}))-
P(\mathfrak{B}(\mathbf{x},P(\mathbf{y})))-P(\mathfrak{B}(P(\mathbf{x}),\mathbf{y})) \ .
$$
Adding and subtracting now
$P\Big[\mathfrak{B}\big(P(\mathbf{x}),P(\mathbf{y})\big)\Big]$, after some
elementary transformations we obtain ($id$ is the identity in $\mathfrak{M}$).
$$
\mathcal{A}(\mathfrak{B};P)(\mathbf{x},\mathbf{y})=
P\Big[\mathfrak{B}\big[(id-P)(\mathbf{x}),(id-P)(\mathbf{y})\big]\Big]+
(id-P)\Big[\mathfrak{B}\big[P(\mathbf{x}),P(\mathbf{y}\big]\Big] .
$$
Recalling that $P$ and $(id-P)$ project on two subspaces of $\mathfrak{M}$,
the direct sum of which generates $\mathfrak{M}$,
and naming $P$ as {\it vertical} projection denoted by $V$, then
$(id-P)$, denoted by $H$, gets naturally the name {\it horizontal}
projection. So the above expression gets the final form of
$$
\mathcal{A}(\mathfrak{B};P)(\mathbf{x},\mathbf{y})=
V\Big[\mathfrak{B}\big[H(\mathbf{x}),H(\mathbf{y})\big]\Big]
+H\Big[\mathfrak{B}\big[V(\mathbf{x}),V(\mathbf{y})\big]\Big]
$$
$$
=\mathcal{R}_{P}(\mathfrak{B};\mathbf{x},\mathbf{y})+
\bar{\mathcal{R}}_{P}(\mathfrak{B};\mathbf{x},\mathbf{y}).
$$
As it is seen, the first term on the right,
$\mathcal{R}_{P}(\mathfrak{B};\mathbf{x},\mathbf{y})$, which may be called {\it
$\mathfrak{B}$-algebraic curvature of $P$}
\index{$\mathfrak{B}$-algebraic curvature}, measures the vertical component of
the $\mathfrak{B}$-image of the horizontal projections of
$(\mathbf{x},\mathbf{y})$, and then the second term
$\bar{\mathcal{R}}_{P}(\mathfrak{B};\mathbf{x},\mathbf{y})$, acquiring the name
of  $\mathfrak{B}$-{\it algebraic cocurvature of $P$}
\index{$\mathfrak{B}$-algebraic cocurvature}, measures the horizontal
component of the $\mathfrak{B}$-image of the vertical projections of
$(\mathbf{x},\mathbf{y})$. Hence, the curvature and cocurvature measure the
mutual $P$-influence between $V(\mathfrak{M})\subset\mathfrak{M}$ and
$H(\mathfrak{M})=(id-V)(\mathfrak{M})\subset\mathfrak{M}$ generated by the
restriction of $\mathfrak{B}$ to each of the two subspaces of
$\mathfrak{M}=P(\mathfrak{M})\oplus(id-P)(\mathfrak{M})$ binar
$\mathfrak{B}$-coupling of elements.

Recalling the above {\bf Remark}, for the case $\Lambda^{1}(M,TM)$ and the
corresponding {\it Fr$\ddot{o}$licher-Nijenhuis} bracket:
$\mathfrak{B}\rightarrow [\,,]_{(F,N)}$, we couild work out the
corresponding Bianchi identities (see [7, Sec.16] for details).

If $\Phi=\alpha^i\otimes e_i$ and $\Psi=\beta^j\otimes k_j$ are skew
symmetric forms on $E$ with values correspondingly in the linear spaces $W_1$
and $W_2$, and $\varphi: W_1\times W_2\rightarrow W_3$ is a bilinear mapping
valued in the linear space $W_3$ we can define a $W_3$-valued form
$\varphi(\Phi,\Psi)$ on $E$ according to $$
\varphi(\Phi,\Psi)=\varphi(\alpha^i\otimes e_i,\beta^j\otimes k_j)=
\alpha^i\wedge\beta^j\otimes\varphi(e_i,k_j).
$$
For example, if $W_3$ is just $E\wedge E$ and $\varphi$ is the exterior
product $E\times E\rightarrow E\wedge E$, we obtain
$\varphi(\Phi,\Psi)=\alpha^i\wedge\beta^j\otimes e_i\wedge k_j$, while if
$\varphi$ is the symmetric product $E\times E\rightarrow E\vee E$ we obtain
$\varphi(\Phi,\Psi)=\alpha^i\wedge\beta^j\otimes e_i\vee k_j$. As an
illustration, if $W_3$ is 2-dimensional, then in the exterior case we obtain
$\varphi(\Phi,\Psi)=\alpha^1\wedge\beta^2\otimes e_1\wedge e_2$ and in the
symmetric case we obtain
$$
\varphi(\Phi,\Psi)=\alpha^1\wedge\beta^1\otimes
e_1\vee e_1+\alpha^2\wedge\beta^2\otimes e_2\vee
e_2+(\alpha^1\wedge\beta^2+\alpha^2\wedge\beta^1)\otimes e_1\vee e_2.
$$

Note that if $\varphi$ is a linear mapping in $E$ with the corresponding
$\varphi^*$ in $E^*$, $\psi$ is a linear mapping in $W$, and
$\Phi=\alpha^i\otimes e_i$ is a $W$-valued skew symmetric form on $E$, it may
happen that $\varphi^*(\alpha^i)\otimes\psi(e_i)=\alpha^i\otimes e_i$. In such a
case $\Phi$ is called $(\varphi,\psi)-$ {\it equivariant} \index{equivariant
form}.

Finally, if $(\mathcal{A},+,.)$ is a graded algebra and $\mathcal{F}:
\mathcal{A}\rightarrow \mathcal{A}$ is NOT a derivation in $\mathcal{A}$,
i.e., $\mathcal{F}(a.b)\neq\mathcal{F}(a).b+\varepsilon_p\, a.\mathcal{F}(b)$ in
general, then the combination

$$
\{a,b\}=\mathcal{F}(a).b+\varepsilon_p\, a.\mathcal{F}(b)
-\mathcal{F}(a.b), \ a,b\in \mathcal{A}\, ,
$$
where $\varepsilon_p$ is the pairity of $a\in\mathcal{A}$, is called Leibniz
bracket of $\mathcal{F}$(for details see: arXiv : gr-qc/0306102, or,
J.Math.Phys.{\bf 45}(6),p.2405).

\section{Basic Examples of Algebraic Structures}
\subsection{Determinants.} A {\it determinant} function $\Delta$ in a
$n$-dimensional linear space $E$ is a skew symmetric n-linear function from $E$
to the scalars $\Gamma$ . Hence,
$$
\Delta(x_1,x_2,\dots,x_n)=
\sum_{\sigma\in S^n}\varepsilon_{\sigma}x_1^{\sigma(1)}
x_2^{\sigma(2)}\dots x_n^{\sigma(n)}.
$$
Clearly, $\Delta(x_1,\dots,x_n)$ will be not zero only if all $x_i,i=1,\dots,n$
are linearly independent.

If $\{e_i\}$ and $\{\varepsilon^j\}$ are two dual bases then the element
$\omega=\varepsilon^1\wedge\varepsilon^2\wedge\dots\wedge\varepsilon^{n}\in\Lambda^n(E^*)$
defines determinant function in $E$. So, the duality relation
$\langle\varepsilon^i,e_j\rangle=\delta^i_j$ yields
$$
\omega(e_1,\dots,e_n)=\langle\varepsilon^1\wedge\varepsilon^2\wedge\dots\wedge\varepsilon^{n},
e_1\wedge e_2\wedge\dots\wedge e_n\rangle=1.1.\dots.1=1.
$$
If $\varphi$ is a linear mapping in $E$ then
$\omega(\varphi(x_1),\dots,\varphi(x_2))=\gamma.\omega(x_1,\dots,x_n)$ and
$\gamma$ is called the {\it determinant} of $\varphi$ and is denoted by
$\mathrm{det}\varphi$. If $\varphi=\lambda.id$ then
$\mathrm{det}(\lambda.id)=\lambda^n$, so, $\mathrm{det}(id)=1$. Also,
$\mathrm{det}(\varphi\circ\psi)=(\mathrm{det}\varphi).(\mathrm{det}\psi)$, so if
$\psi$ is a linear isomorphism, then
$\mathrm{det}(\psi\circ\varphi\circ\psi^{-1})=\mathrm{det}(\varphi)$.
 If $\varphi^*$ is dual to $\varphi$ then
$\mathrm{det}\varphi^*=\mathrm{det}\varphi$.

The solutions of the equation $\mathrm{det}(\varphi-\lambda.id)=0$ are called
{\it eigen} values of $\varphi$. We obtain $$
\mathrm{det}(\varphi-\lambda.id)=\sum_{p=0}^n\alpha_p\lambda^{n-p},
$$
where the
coefficients $\alpha_p$ are expressed through the principal minors of the
representative matrix $\varphi^i_j$, and these coefficients are invariants:
$$
\alpha_p(\varphi)=\alpha_p(\psi\circ\varphi\circ\psi^{-1}),
$$
where $\psi$ is a linear isomorphism.

A linear mapping is an isomorphism iff $\mathrm{det}\varphi\neq 0$.

Any two determinant functions in $E$ may differ from each other just by a
scalar: $\Delta_2=\lambda\Delta_1$. Two determinant functions are called
equivalent if $\lambda>0$, so it is said that each class defines an {\it
orientation} in $E$ \index{orientation}. A basis $\{e_i\}$ in $E$ is called
{\it positive} with respect to $\Delta$ if $\Delta(e_1,\dots,e_n)>0$. Any even
permutation
$\sigma:(1,2,\dots,n)\rightarrow(\sigma(1),\sigma(2),\dots,\sigma(n))$ respects
the orientation chosen. An isomorphism $\varphi$ is orientation preserving if
$\Delta$ and $\Delta\circ\varphi$ have the same orientation.

If $\Delta$ is a determinant function in $E$ then the equality
$$
\sum_{i=1}^n\Delta(x_1,\dots,\varphi(x_i),\dots,x_n)=
\alpha.\Delta(x_1,\dots,x_n),
$$
defines the scalar $\alpha$, called the {\it trace} of $\varphi$, denoted by
$tr\,\varphi$. It satisfies
$$
tr(\lambda.\varphi+\mu.\psi)=\lambda.tr\varphi+\mu.tr\psi \ \ \text{and} \ \
tr(\psi\circ\varphi)=tr(\varphi\circ\psi).
$$
With respect to any couple of dual bases $\{\varepsilon^i\},\{e_j\}:
\langle\varepsilon^i,e_j\rangle=\delta^i_j$, we
obtain $tr\varphi=\sum_{i}\langle\varepsilon^i,\varphi(e_i)\rangle$.

In case $A$ is an antisymmetric 2-form on 4-dimensional space, then
$$
A\wedge A=(A_{12}A_{34}+A_{13}A_{42}+A_{14}A_{23})\,e^1\wedge e^2\wedge
e^3\wedge e^4=\sqrt{det(A_{ij})}\,e^1\wedge e^2\wedge e^3\wedge e^4,
$$ where $det(A)=(A_{12}A_{34}+A_{13}A_{42}+A_{14}A_{23})^2\geqq 0$.

\subsection{Metrics, Pseudo-metrics, Symplectic forms}
 1. {\bf Euclidean metrics.}
A metric (metric tensor, inner product) $g$ in a real linear space $E$ is
every element of $S^2(E^*)$ satisfying additionally the conditions for
nondegeneracy: $g(x,y)=0$ if for every $y\in E$ it follows $x=0$, and {\it
positivity}: $g(x,x)>0$ for any nonzero $x\in E$. In such a case $E$ is
called {\it inner product space}. In finite dimensional case if $\{e_i\}$ is a
basis in $E$ then $g$ is completely determined by its values
$g_{ij}=g(e_i,e_j)$. Then the nondegeneracy condition means
$\mathrm{det}\parallel g_{ij}\parallel\neq 0$. Having introduced a metric $g$
in $E$ we define a {\it norm} $|x|$ of $x\in E$ by $|x|=\sqrt{g(x,x)}$. A
vector $x$ is called {\it unit} if $|x|=1$. It follows that $g(x,y)$ can be
expressed in terms of the norm: $g(x,y)=\frac12(|x+y|^2-|x|^2-|y|^2)$.

Two vectors $x,y$ in $E$ are called {\it orthogonal} (with respect to $g$) if
$g(x,y)=0$. Correspondingly, two subspaces $E_1$ and $E_2$ of $E$ are called
orthogonal if any two vectors $x_1\in E_1$ and $x_2\in E_2$ are orthogonal.

\noindent
{\bf Remark}. Further, when no misunderstanding will take place, we are going
to write just $(x,y)$ instead of $g(x,y)$.

There is a basic inequality in every inner product space,
called Schwarz-inequality: $(x,y)^2\leqq|x|^2|y|^2$. This allows to introduce
a real number $\alpha$ by $0\leqq\alpha\leqq \pi$ and an angle between two
vectors according to $\cos\alpha=(x,y)/|x||y|$. Clearly, $(x,y)=0$ leads to
$\cos\alpha=0$ and $\alpha=\frac{\pi}{2}$. The {\it cosine theorem} asserts
$|x-y|^2=|x|^2+|y|^2-2|x||y|\cos\alpha$, and the {\it triangle} inequality
asserts $|x+y|\leqq|x|+|y|$.

A basis $\{e_i\}$ in $E$ is called {\it orthogonal} if $(e_i,e_j)=0$, and if
additionally each $e_i$ is unit, then the basis is called {\it orthonormal}.
Linear isomorphisms that transform orthonormal basis to orthonormal basis are
called {\it orthogonal} and satisfy $\varphi^*=\varphi^{-1}$. A vector $y_x$ is
called {\it orthogonal} projection of the vector $x$ into the subspace $E_1$ if
$y_x=\sum_\nu(x,k_\nu)k_\nu$, where $\{k_\nu\}$ is an orthonormal basis in
$E_1\subset E$. It follows $|y_x|\leqq|x|$. The number $|x-y_x|$ is called the
{\it distance} of $x$ from $E_1$. If the linearly
independent vectors $(x_1,x_2,\dots,x_p), \linebreak 1<p<n$, generate the
subspace $E_1\subset E$ and $\{\varepsilon^{\nu}\}$ is an orthonormal basis in
$E_1^*$ then the number $|\langle
\varepsilon^1\wedge\dots\wedge\varepsilon^p,x_1\wedge\dots\wedge x_p\rangle|$
is called {\it volume} of the $p-$dimensional parallelepiped spanned by the
vectors $(x_1,\dots,x_p)$.

The metric $(\,,)$ defines isomorphism $\tilde{g}$ between $E$ and $E^*$
 according to $\langle\tilde{g}(x),y\rangle=(x,y)$. We obtain
$$
\langle\tilde{g}(x),y\rangle=(x,y)=(y,x)=\langle\tilde{g}(y),x\rangle=
\langle x,\tilde{g}(y)\rangle.
$$
Thus, $\tilde{g}^*=\tilde{g}$. It follows that the matrices $(e_i,e_j)$ and
$(\varepsilon^i,\varepsilon^j)$ are inverse to each other, i.e.
$g_{ij}g^{jk}=\delta_i^k$. This allows to raise and lower indices:
$A_{ijmkk}g^{kl}=A_{ijm}^l$, $A^{ijmk}g_{kl}=A^{ijm}_l$.

 Finally, a linear mapping $\varphi: E\rightarrow E$ is called {\it isometry}
\index{isometry} if $\varphi^*(g)=g$. We give some properties of isometries:

1. All isometries in $E$ preserve the norm: $|\varphi(x)|=|x|$.

2. All isometries transform orthonormal basis to orthonormal basis.

3. All isometries $\varphi$ satisfy: $\mathrm{det}(\varphi)=\pm 1$.

4. All isometries have eigen values $\lambda=\pm 1$, and not all isometries
have eigen vectors.

5. Every isometry $\varphi$ in an odd dimensional space has at least one
positive eigen value equal, of course, to $1$. The corresponding eigen vector
$z$ is invariant with respect to $\varphi: \varphi(z)=z$.

6. Every isometry $\varphi$ in $E$ satisfies $tr(\varphi)\leqq(dimE)$.

7. For every isometry in $E$ there exists an orthogonal decomposition of $E$
into subspaces of dimension $1$ and $2$.

8. All isometries of $E^n$ form a group of dimension $\frac12n(n+1)$
and NOT a linear space.

If $\{e_i\}$ is orthonormal basis with respect to $g$ then $g$ has
components $g^{e}_{ii}=1, g_{ij}=0$ for $i\neq j$, so, $\det\parallel
g_{ij}\parallel=1$. If $\{k_i\}$ is any other basis with
$g_{ij}^{k}=A^m_iA^n_jg^{e}_{mn}$ then $\det\parallel g^k_{ij}\parallel=
(\det\parallel A^m_j\parallel)^2>0$ and the two volume forms
$\omega(e)$ and $\omega(k)$ are connected by
$\omega(k)=\sqrt{\det\parallel g^k_{ij}\parallel}\,\omega(e)$.

As it is seen from the above properties the isometries define a
group $O(n)$, this group has two components: $S^{+}(n)$ and $S^{-}(n)$. The
elements of $S^{+}(n)$ have determinants equal to $1$ and are called {\it
proper}, and the elements of $S^{-}(n)$ have determinants equal to $(-1)$ and
are called {\it improper}. So, the proper isometries preserve the orientation,
and the improper isometries change the orientation (defined by an orthonormal
basis). In the $3-$dimensional case every proper isometry $\varphi$ has unique
$1-$dimensional eigen (hence, invariant) subspace with respect to which the
isometry is reduced to $2-$dimensional rotation, the corresponding rotation
angle $\theta$ is defined by $\cos(\theta)=\frac12(tr\varphi-1)$.

2. {\bf Pseudo-Euclidean metrics.}
 These are nondegenerate bilinear forms $\eta$ on $E$ admitting positive, zero
and negative values when calculated on the same vector. If $\eta(x,x)>0$ then
$x$ is called {\it time-like}, if $\eta(x,x)<0$ then $x$ is called {\it
space-like} and if $\eta(x,x)=0$ then $x$ is called {\it isotropic/light-like}.
The set of all isotropic vectors form the {\it light-cone}. Each isotropic
vector is orthogonal to itself. A basis $\{e_\nu\}$ is called {\it orthonormal}
if $\eta_{\mu\mu}=\eta(e_\mu,e_\mu )=\pm 1$. The number of minuses is called
{\it index} of $\eta$, and the difference $pluses-minuses$ is called
{\it signature} of $\eta$, but sometimes the signature is denoted just by
$sign(\eta)=(-,-,...,-,+,+,...,+)$, and the number of minuses, or pluses, is
preliminary clear.

If the number of minuses is $(n-1)$ the following properties hold:

a/. Two time-like vectors are never orthogonal,

b/. A time-like vector is never orthogonal to an isotropic vector,

c/. Two isotropic vectors are orthogonal only if they are linearly dependent.

A 4-dimensional pseudo-Euclidean space with signature $(-,-,-,+)$ is called
Minkowski space $M=(\mathbb{R}^4,\eta)$. The corresponding isometries are
called Lorentz transformations, they form a 6-dimensional group.Together with
the translations we get the 10-dimensional Poincare group. A proper Lorentz
transformation possesses always at least one eigenvector on the light-cone.
If $F$ is a 2-form on $M$ and $G$ is a 3-form on $M$ then
$$
\eta(F,F)=\sum_{\mu<\nu}F_{\mu\nu}F^{\mu\nu}=
\frac12\sum F_{\mu\nu}F^{\mu\nu},
$$
$$
\eta(G,G)=\sum_{\mu<\nu<\sigma}G_{\mu\nu\sigma}G^{\mu\nu\sigma}=
\frac{1}{3!}\sum G_{\mu\nu\sigma}G^{\mu\nu\sigma}, \ \ \
\mu,\nu,\sigma=1,...,4.
$$

%\newpage
3. {\bf Some Structures associated with an exterior form.} A non-zero $p-$form
$\beta$ on $E$ is decomposable if there exist $p$ linearly independent
one-forms $\alpha_1,\alpha_2,\dots,\alpha_p$ such that
$\beta=\alpha_1\wedge\alpha_2\wedge\dots\wedge\alpha_p$. Inversely, the 1-forms
$\alpha_1,\alpha_2,\dots,\alpha_p$ are linearly independent if
$\alpha_1\wedge\alpha_2\wedge\dots\wedge\alpha_p\neq 0$.

With every p-form $\alpha$ a subspace $E_{\alpha}\subset E$ is associated:
$E_{\alpha}\subset E$ is generated by those $x\in E$ satisfying $i(x)\alpha=0$.
Clearly, if $\alpha$ is 1-form, then the corresponding $E_{\alpha}\subset E$ is
a hyperplain in $E$.

The subspace $E_{\alpha}\subset E$ can be defined by subspace
$E^*_{\alpha}\subset E^*$ of 1-forms which annihilate the whole
$E_{\alpha}\subset E$. So, $E^*_{\alpha}\subset E^*$ is called {\it associated
to $\alpha$ system}. If $\alpha$ is a 2-form on $E$ and $\{e_i\}$ is a basis in
$E$ then the 1-forms $i(e_i)\alpha$ generate the {\it associated to $\alpha$
system}. The dimension of $E^*(\alpha)$ is called {\it rank} of $\alpha$.
Obviously, the rank of $\alpha$ is equal to the codimension of $E(\alpha)$. If
$\alpha$ is decomposable, then its rank is equal to the number of 1-forms that
represent $\alpha$, e.g. if $rank(\alpha)$ is $p$, then
$\alpha=\alpha_1\wedge\dots\wedge\alpha_p$. A nonzero $(n-1)$-form has rank
$(n-1)$, and a nonzero $(n-2)$-form may have rank equal to $n$, or to
$(n-2)$.

For any 2-form $\alpha$ there exist even number $2s$ of 1-forms
$\varepsilon^{i}, i=1,2,\dots,2s\leqq n$, such that
$\alpha=\varepsilon^1\wedge\varepsilon^2+\dots
+\varepsilon^{2s-1}\wedge\varepsilon^{2s}$, 2-forms may have only even rank.

A symplectic structure on $E$ is introduced by a 2-form $\alpha$ of rank
$dimE$, so in such a case $dimE=2m$. Hence, the product
$\alpha\wedge\dots\wedge\alpha$, where $\alpha$ is multiplied by itself
$m$-times, defines a volume form on $E^{2m}$, and hence, an orientation called
{\it canonical} (with respect to $\alpha$). Moreover, $\alpha$ defines
isomorphism between $E$ and $E^*$ according to $x\rightarrow i(x)\alpha, x\in
E$.

A linear mapping $\varphi:E^{2m}\rightarrow E^{2m}$ satisfying
$\varphi^*\alpha=\alpha$ is called symplectic isomorphism. All
symplectomorphisms in $E$ form a $m(2m+1)$-dimensional group
$Sp(E^{2m},\alpha)$, and each element in $Sp(E^{2m},\alpha)$ has determinant
equal to 1, so symplectomorphisms preserve the canonical orientation.

There are two bases $\{\varepsilon^i\}, i=1,2,...,2m$ and
$\{\tilde{\varepsilon}^i\}, i=1,2,...,2m$
of $E^{2m}$ being canonical in some sense with respect to
the symplectic 2-form $\alpha$.  Making use of the first one, we get
$$
\alpha=\varepsilon^1\wedge\varepsilon^2+\dots
+\varepsilon^{2m-1}\wedge\varepsilon^{2m},
$$
The 1-forms $\{\varepsilon^i\}, i=1,2,\dots,2m$ form a basis in $E^*$ and
the dual to this basis $\{e_i\}$ in $E$ is usually called {\it symplectic basis}.

In the second basis we get
$$
\alpha=\tilde{\varepsilon}^1\wedge\tilde{\varepsilon}^{m+1}+\dots
+\tilde{\varepsilon}^{m}\wedge\tilde{\varepsilon}^{2m},
$$
and this basis is usually employed in the frame of symplectic mechanics on the
cotangent bundle of a manifold.

\chapter{Manifolds and Bundles}
\section{Topological and Smooth Manifolds.}
{\it We continue under the assumption that all linear spaces to be considered
are real, finite dimensional and endowed with the standard topology, so the
concepts of differentability and smoothness can be introduced and used.}
\vskip 0.2cm
1. {\bf Topological manifold.} Let $M$ denote a topological space with a
countable set $\Sigma_{\alpha}$ of open sets $U_{\alpha}$, so $\Sigma_{\alpha}$
covers the whole $M$. We say that $M$ is a $n$-dimensional topological manifold
\index{topological manifold}
if every $U_{\alpha}\subset M$ is homeomorphic to a an open set in
$\mathbb{R}^n$. So, the couple  $(U_{\alpha};\varphi_\alpha:U_\alpha\rightarrow
V\subset\mathbb{R}^n)$ is called a {\it local chart} on $M$ and
$(\Sigma_{\alpha},\varphi_{\alpha})$ is called an {\it atlas} \index{atlas}
on $M$. An atlas
is called {\it maximal} if it includes all possible local charts. Under these
conditions $M$ is called topological manifold.

Since $\varphi_{\alpha}$ are homeomorphisms between open sets then if we denote
the possible intersection by $U_{\alpha\beta}=U_{\alpha}\cap U_{\beta}$, a
homeomorphism
$$
\varphi_{\alpha\beta}:\varphi_\beta(U_{\alpha\beta})\rightarrow\varphi_\alpha(U_{\alpha\beta})
$$
is defined by $\varphi_{\alpha\beta}=\varphi_\alpha\circ\varphi_\beta^{-1}$,
called {\it identification map}, or {\it transition function} for $U_\alpha$
and $U_\beta$. Clearly, the inverse of $\varphi_{\alpha\beta}$ is
$\varphi_{\beta\alpha}$.

2. {\bf The Derivative.} If $E$ and $F$ are two real, finite dimensional vector
spaces, $U\subset E$ is an open subset and $\varphi:E\rightarrow F$ is a map,
then $\varphi$ is called differentiable at $x_o\in U$ if there is a linear map
$\psi:E\rightarrow F$ such that
$$
lim_{t\rightarrow 0}\frac{\varphi(x_o+th)-\varphi(x_o)}{t}=\psi_{x_o}(h), \
h\in E,
$$
where $t$ is a real external parameter.
If this is true for every $x_o\in U$ then $\varphi$ is called differentiable
map in $U$, and the map $\varphi':U\rightarrow L(E,F)$ defined by
$x_o\rightarrow\varphi'(x_o)$, where $\varphi'(x_o)(h)=\psi_{x_o}(h)$  is called
the {\it derivative} of $\varphi$.

Now, $L(E,F)$ is also real finite dimensional vector space, so
$\varphi':U\rightarrow L(E,F)$ can be tested for differentability, and if it
is differentiable in $U$ its derivative is denoted by $\varphi''$. So, a
$k^{th}$ derivative of $\varphi$ on $U$ could be considered to exist, and if it
exists then $\varphi$ is called to be $C^{(k)}-map$. If this process is
infinite with respect to $k$, then $\varphi$ is said to be of
$C^{\infty}-class$, or {\it smooth}.

If $\varphi:U\rightarrow V\subset F$ has smooth inverse, then $\varphi$ is
called a diffeomorphism between $U\subset E$ and $V\subset F$.

If it turns out that $\varphi'$ is continuous and $\varphi'(x_o):E\rightarrow F$
is a linear isomorphism, then the {\it inverse function theorem} states that
there are open sets $U$ of $x_o$ and $V$ of $\varphi(x_o)\subset F$ such that
$\varphi$ restricts to diffeomorphism between $U$ and $V$.

If all identification maps of an maximal atlas on $M$ are smooth then we say
that on $M$ is defined {\it smooth structure} \index{smooth structure}
 and $(M^n,U_\alpha)$ is called a
smooth n-dimensional manifold. Further under manifold we shall understand
always smooth manifold.

3. {\bf Smooth maps.} Let $M$ and $N$ be two manifolds with corresponding
atlases $(U_\alpha,u_\alpha)$ and $(V_i,v_i)$ and $\varphi:M\rightarrow N$ be a
continuous map such that $U_\alpha\cap\varphi^{-1}(V_i)$ is not
empty. We obtain a continuous map $$
\varphi_{i\alpha}:u_\alpha(U_\alpha\cap\varphi^{-1}(V_i))\rightarrow v_i(V_i).
$$
defined by $\varphi_{i\alpha}:v_i\circ\varphi\circ u_\alpha^{-1}$. Then
$\varphi$ is said to be smooth if all $\varphi_{i\alpha}$ are smooth. The
composition of two smooth maps $\varphi:M\rightarrow N$ and
$\psi:N\rightarrow P$ is obviously smooth.

A smooth map $\varphi:M\rightarrow N$ is called {\it diffeomorphism} if it has
smooth inverse. Correspondingly, two manifolds $M$ and $N$ are called
diffeomorphic if there exists a diffeomorphism $\varphi:M\rightarrow N$. The
set of diffeomorphisms of a manifold $M$ form a group $Diff(M)$ with respect to
the composition of two diffeomorphisms.

If $M^m$ and $N^n$ are smooth manifolds with corresponding atlases as above,
then the product $M\times N$ becomes a manifold of dimension $(m+n)$ and atlas
$(U_\alpha\times V_i,u_\alpha\times v_i)$ and the two projections $\pi_M:M\times
N\rightarrow M$ and $\pi_N:M\times N\rightarrow N$ are smooth.

The smooth maps $f:M\rightarrow \mathbb{R}$ from a manifold to the real numbers
are called smooth functions and the set of all such functions will be denoted
by $\mathcal{J}(M)$. The set $\mathcal{J}(M)$ is an (infinite dimensional)
algebra with respect to the real numbers and a (one dimensional) module with
respect to itself, where the function $f_1:M\rightarrow 1$ defines a basis.

A smooth path on $M$ is a smooth map $\varphi:\mathbb{R}\rightarrow M$. A
manifold is called smoothly path connected if for any two points $a$ and $b$ of
$M$ there is a smooth path $\varphi$ such that $\varphi(0)=a$ and
$\varphi(1)=b$. Clearly, if $M$ is connected as a topological space, then it is
a smoothly path-connected.

Every smooth map $\varphi:M\rightarrow N$ defines algebra homomorphism
$\varphi^*:\mathcal{J}(N)\rightarrow\mathcal{J}(M)$ through
$\varphi^*(f)=f\circ\varphi$, $f\in\mathcal{J}(N)$. The surjectivity of
$\varphi$ leads to injectivity of $\varphi^*$, and the composition
$\varphi\circ\psi$ leads to $(\varphi\circ\psi)^*=\psi^*\circ\varphi^*$.

The {\it carrier} or {\it support} of a smooth function $f$ on $M$ is the
closure of the set$\{x\in M|f(x)\neq 0\}$. A smooth function may have finite
carrier with respect to a submanifold $S\subset M$.

4. {\bf Local coordinates.} Let $(U_\alpha,u_\alpha)$ be a local chart on
$M^n$ and $\{\varepsilon^i\}$ be a basis in $\mathbb{R}^{n*}$. Then the
compositions $\varepsilon^i\circ u_\alpha$ are called {\it coordinate
functions}, or just {\it local coordinates} for $U_\alpha$ and are denoted
usually by $x^i|_{U_\alpha}=\varepsilon^i\circ u_\alpha.$ If the point $p\in M$
lays in the intersection $U_\alpha\cap U_\beta$, it is endowed with two
coordinates, say $x^i(p)$ and $y^i(p)$, and we have the smooth functions
$x^i=f^i(y^j)$, defining diffeomorphism between the corresponding regions of
$\mathbb{R}^n$. Correspondingly, a map $\varphi:M\rightarrow N$ can be
represented in corresponding local coordinates $x^i$ on $M$ and $y^j$ on $N$ by
$y^j=\varphi^j(x^i)$, i.e.
$$
\varepsilon^j\circ
v_\beta=\varepsilon^j\circ\varphi_{\alpha\beta}\circ\varepsilon^i\circ
u_\alpha,  \ \  i=1,2,\dots,m, \ \ j=1,2,\dots,n ,
$$
and $\varphi_{\alpha\beta}$ maps $u_\alpha(U_\alpha)\subset\mathbb{R}^m$ into
$v_\beta(V_\beta)\subset\mathbb{R}^n$.

\section{Smooth bundles and vector bundles.\\ Sections}

{\bf 2.2.1. Local product property.} Let $E$ and $B$ be smooth manifolds and
$\pi:E\rightarrow B$ be a smooth map. We say that $\pi$ has the {\it local
product property} with respect to the manifold $F$ if there is an open covering
$\{U_\alpha\}$ of $B$ and a family of diffeomorphisms
$$
\psi_\alpha:U_\alpha\times F\rightarrow \pi^{-1}(U_\alpha)
$$
such that
$$
\pi\circ\psi_{\alpha}(x,y)=x, \ \ \ x\in U_\alpha,\ \ \ y\in F.
$$
In such a case $\pi$ is obviously surjective, and the system
$\{U_\alpha,\psi_\alpha\}$ is called {\it local decomposition} of $\pi$.

{\bf 2.2.2. Smooth fiber bundle.} The
four-tuple $(E,\pi,B,F)$ is called {\it smooth fiber bundle}, \index{fiber
bundle} and any local decomposition of $\pi$ is called {\it coordinate
representation} for the fiber bundle. The manifold $E$ is called {\it the total
space}, $B$ is called {\it the base space}, for each $x\in B$ the set
$F_x=\pi^{-1}(x)$ is called the fiber over $x\in B$ and $F$ is called {\it
standard fiber}. Clearly, $E$ is a disjoint union of the fibers.

A smooth {\it cross-section}, or just a {\it section}, of the fiber bundle
$(E,\pi,B,F)$ is a smooth map $\sigma:B\rightarrow E$ with property
$\pi\circ\sigma=id_{B}$. Clearly, $\sigma(B)$ is diffeomorphic image of $B$.

Having a coordinate representation $(U_\alpha,\psi_\alpha)$ we obtain
bijections $\psi_{\alpha,x}:F\rightarrow F_x$ defined by
$\psi_{\alpha,x}(y)=\psi_\alpha(x,y), y\in F$.

If $x\in B$ lays in the intersection $U_{\alpha\beta}=U_\alpha\cap U_\beta$ we
obtain a map $\psi_{\beta,x}^{-1}\circ\psi_{\alpha,x}:F\rightarrow F$, which is
a diffeomorphism. The functions
$g_{\beta\alpha}(x)=\psi_{\beta,x}^{-1}\circ\psi_{\alpha,x}$ are in $Diff(F)$
and are called {\it transition functions} for the bundle with respect to
$(U_\alpha,\psi_\alpha)$.

Let $(E',\pi',B',F')$ be another fiber bundle. Then a map $\varphi:E\rightarrow
E'$ is called {\it fiber preserving}, or {\it homomorphism of bundles} if
whenever $\pi(z_1)=\pi(z_2)$ for $(z_1,z_2)\in E$ then
$\pi'\circ\varphi(z_1)=\pi'\circ\varphi(z_2)$. So, we get a map
$\varphi_B:B\rightarrow B'$ requiring $\pi'\circ\varphi=\varphi_B\circ\pi$,
and $\varphi_B$ is smooth.

There may exist various fiber bundles on the same base space $B$. If
$\xi=(E,\pi,B,F)$ and $\xi'=(E',\pi',B,F')$ are two such bundles  then a bundle
map $\varphi:\xi\rightarrow \xi'$ is called {\it strong bundle map} if the
induced map in $B$ is the identity $i_B$ of $B$.

If $A\subset B$ then $(\pi^{-1}(A),\pi,A,F)$ can be endowed with a bundle
structure, called {\it restriction} of $(E,\pi,B,F)$ on $A$.

Finally, a fiber bundle is called {\it trivial} if the bundle space $E$ is
diffeomorphic to the direct product $B\times F$ and $\pi(B\times F)=B$.

 \vskip 0.3cm Roughly speaking, a
smooth fiber bundle $(E,\pi,B,F)$ is a disjoint union of diffeomorphic images
of the same manifold $F$, which union is parametrized smoothly by the points of
a manifold $B$, and the atlases of $E$ are made to respect this intrinsic
structure of $E$ by the requirement that, locally, $E$ is diffeomorphic to the
direct product $U_\alpha\times F$.
\vskip 0.3cm
{\bf 2.2.3. Vector bundles.}
A {\it vector bundle} \index{vector bundle}
is a quadruple $(E,\pi,B,F)$, where

1. $(E,\pi,B,F)$ is a smooth bundle,

2. the spaces $F$ and $F_x=\pi^{-1}(x), \ x\in B$, are real finite
dimensional vector spaces,

3. there is a coordinate representation $(U_\alpha,\psi_\alpha)$ such that the
maps $\psi_{\alpha,x}:F\rightarrow F_x$ are linear isomorphisms.

The dimension of $F$ is called {\it rank} of the vector bundle, and the
required in p.3 coordinate representation is called {\it vector coordinate
representation}. The induced by the coordinate representation
$(U_\alpha,\psi_\alpha)$ maps $g_{\alpha\beta}:U_{\alpha\beta}\rightarrow
GL(F)$, given by $g_{\alpha\beta}(x)=\psi_{\alpha,x}^{-1}\circ\psi_{\beta,x}$
are smooth and satisfy $g_{\gamma\beta}(x)g_{\beta\alpha}(x)=g_{\gamma\alpha}(x)$
for each $x\in U_\gamma\cap U_\beta\cap U_\alpha$.

If $(U_\alpha,\psi_{\alpha})$ define a coordinate representation for $\eta$ then
$(U_\alpha,\varphi_\alpha)$, where
$\varphi_{\alpha,x}=(\psi^*_{\alpha,x})^{-1}$, define a coordinate
representation of the dual vector bundle $\eta^*=(E^*,\pi^*,B,F^*)$ where $E^*$
is the union of all $(F_x)^*$, $\pi^*$ is obvious, $F^*$ is the dual space of
$F$.

A subbundle of a vector bundle $(E,\pi,B,F)$ is any vector bundle
$(E',\pi,B,F')$ with a standard fiber $F'$ which is a vector subspace of the
vector space $F$, the vector spaces $F_x$ are vector subspaces of $F_x$, and
the induced inclusion map $E'\rightarrow E$ is smooth.

If $\eta$ and $\eta'$ are vector bundles and $\varphi:\eta\rightarrow \eta'$ is
a bundle map with $\psi:B\rightarrow B"$ the induced map between the base
spaces, let $(U_\alpha,\psi_\alpha)$ and $(V_i,\chi_i)$ be coordinate
representations for $\eta$ and $\eta'$. We get smooth maps
$$
\varphi_{i\alpha}:\psi^{-1}(V_i)\cap U_\alpha\rightarrow L(F,F'),\ \
\varphi_{i\alpha}(x)=\chi^{-1}_{i\psi(x)}\circ\varphi_x\circ\psi_{\alpha,x}.
$$
These smooth maps are called {\it mapping transformations} for $\varphi$ with
respect to the two coordinate representations.

A bundle map $\varphi:\eta\rightarrow\eta'$ is an isomorphism iff the induced
$\psi:B\rightarrow B'$ is diffeomorphism and $\varphi_x:F_x\rightarrow
F'_{\psi(x)}$ is a linear isomorphism.

We consider now strong bundle maps $\varphi$ between two vector bundles
$\eta=(E,\pi,B,F)$ and $\eta'=(E',\pi',B,F')$, recalling that these two bundles
have the same base space $B$ and that each $\varphi_x: x\in B$ is a linear map
between the linear spaces $F_x$ and $F'_x$ over the same point of the base.

1. If $\varphi:\eta\rightarrow\eta'$ and $\psi:\eta\rightarrow\eta'$ are such
strong bundle maps then their sum $\varphi+\psi$ is defined by
$(\varphi+\psi)(z)=\varphi(z)+\psi(z), \ z\in E$.

2. If $f\in\mathcal{J}(B)$ and $\varphi$ is a bundle map then a strong bundle
map is defined by $f\varphi$ by $(f\varphi)(z)=f(x)\varphi(z), \
x=\pi(z)=\pi'(z)$.

These two properties of the strong bundle maps between vector bundles make the
set of all such maps a module with respect to the algebra $\mathcal{J}(B)$ of
smooth functions on the base space $B$. Since these are in fact the bundle
homomorphisms between $\eta$ and $\eta'$ this set is denoted by
$Hom(\eta,\eta')$. Clearly, if we have three vector bundles
$\eta,\eta',\eta''$ on the same base then the composition of
$\varphi:\eta\rightarrow\eta'$ and  $\psi:\eta'\rightarrow\eta''$ is defined
and the map $(\varphi,\psi)\rightarrow\psi\circ\varphi$ is a
$\mathcal{J}(B)$-bilinear map $Hom(\eta,\eta')\times
Hom(\eta',\eta'')\rightarrow Hom(\eta,\eta'')$.

If $\eta_1,\eta_2,\dots,\eta_p;\eta$ are vector bundles on the same base, then the
above property of strong bundle maps is extended to all multilinear strong
bundle maps $\Phi(\eta_1,\dots.\eta_p)\rightarrow\eta$, again a
$\mathcal{J}(B)$-module is defined by the same rule. This is important when we
consider dual bundles, tensor product of vector bundles; tensor, skew-symmetric
and symmetric powers of a vector bundle, as well as the induced bundles of
linear and multilinear mappings between vector bundles; also bundles
with additional algebraic structure like exterior and symmetric algebra bundles
over a vector bundle.

{\bf 2.2.4. Sections of vector bundles.} A section of a vector bundle
$\eta=(E,\pi,B,F)$ is a map $\sigma:B\rightarrow E$ with the property
$\pi\circ\sigma=id_{B}$. Since every fiber $F_x$ is a vector space, clearly,
every vector bundle admits the zero section $o(x)=0_x\in F_x, \ x\in B$. Every
section has carrier $carr(\sigma)$ defined as the closure of $\{x\in
B|\sigma(x)\neq 0_x\}$.

The set of sections $Sec(\eta)$ is a $\mathcal{J}(B)$-module:
$(f\sigma)(x)=f(x)\sigma(x)$ and
$(\sigma_1+\sigma_2)(x)=\sigma_1(x)+\sigma_2(x)$. Since locally a vector bundle
is diffeomorphic to $U_\alpha\times F$, then if $F$ is $r$-dimensional, the
sections over $U_\alpha\subset B$ acquire a basis
$\sigma_{(\alpha,1)},\dots,\sigma_{(\alpha,r)}$ according to
$\sigma_{(\alpha,i)}(x)=\psi_{\alpha,x}(e_i)$, where $\{e_i\}$ is a basis of the
standard fiber $F$.

We consider now mappings of sections under bundle maps.

1. Let $\eta=(E,\pi,B,F)$ and $\eta'=(E',\pi',B',F')$ be two vector bundles
with corresponding dual bundles $\eta^*$ and $\eta'^*$, and let
$\varphi:\eta\rightarrow\eta'$ be a bundle map inducing $\psi:B\rightarrow B'$.
Now $\varphi_x:F_x\rightarrow F'_{\psi(x)}$ induces the dual map
$\varphi_x^*:F'^*_{\psi(x)}\rightarrow F_x^*$. If now $\sigma'$ is a section in
$\eta'^*$ then a section $\varphi^*(\sigma)$ in $\eta^*$ is defined by
$\varphi^*(\sigma')(x)=\varphi_x^*(\sigma'(\psi(x)))$. The so defined map
between the sections of $\eta'^*$ and $\eta^*$ is in fact a morphism of
the modules of sections of $\eta'^*$ and $\eta^*$: $\varphi^*(\rho_1+\rho_2)=
\varphi^*(\rho_1)+\varphi^*(\rho_2)$ and
$\varphi^*(f\rho)=\psi^*f.\varphi^*\rho$, where $f\in\mathcal{J}(B')$ and
$\rho_1,\rho_2$ are sections of $\eta'^*$.

2. If $\varphi: \eta\rightarrow\eta'$ restricts to isomorphisms in the fibers
then every section $\sigma$ of $\eta'$ is carried to section
$\varphi^{\#}\sigma$ in $\eta$ according to
$\varphi^{\#}\sigma(x)=\varphi_x^{-1}(\sigma(\psi(x)))$. The map $\varphi^\#$
is also morphism of $\mathcal{J}(B)$ modules $Sec(\eta)$ and $Sec(\eta')$.

3. If $\eta$ and $\eta'$ have the same base and $\varphi$ is a strong bundle
map then the sections $\sigma$ of $\eta$ are carried to sections
$\varphi_{*}\sigma$ in $\eta'$ according to
$\varphi_{*}\sigma(x)=\varphi_x(\sigma(x))$. Clearly, $\varphi_{*}$ is also
morphism of the $\mathcal{J}(B)$ modules $Sec(\eta)$ and $Sec(\eta')$.

\section{Vector Bundles with Additional Structure}

{\bf 2.3.1. Orientable vector bundles.} Let $\eta=(E,\pi,B,F)$ be a vector
bundle of rank $r$ and $\eta^{*}$ be its dual bundle. Then we have the exterior
powers $\Lambda^p(\eta)$ of $\eta$, in particular, $\Lambda^r(\eta)$, which is
of rank $1$. It is said that $\eta$ is orientable if $\Lambda^r(\eta)$ admits
nonzero section $\Delta\in Sec(\Lambda^r(\eta)): \Delta(x)\neq 0, x\in B$.

Any two nonzero sections of $\Lambda^r(\eta)$ differ from each other by a
nonzero element $f$ of $\mathcal{J}(B): \Delta_2=f\Delta_1$. We say that two
such determinant functions are equivalent if the corresponding
$f\in\mathcal{J}(B)$ is positive: $f(x)>0, \ x\in B$. Each class of equivalence
defines orientation in $\eta$. If $B$ is a connected space then $\eta$ admits
just two orientations.

Important result: $\eta$ is orientable iff it admits coordinate representation
$(U_\alpha,\varphi_\alpha)$ whose coordinate transformations
$g_{\alpha\beta}(x)=\varphi_{\alpha,x}^{-1}\circ\varphi_{\beta,x}$ have
positive determinant.

{\bf 2.3.2. Riemannian and pseudo-Riemannian vector bundles.} A pseudo-Riemannian
metric in $\eta$ is an element $g\in Sec(\vee^2(\eta^*))$ such that $g(x)$ is
nondegenerate for every $x\in B$. Under this condition $(\eta,g)$ is called
pseudo-Riemannian vector bundle \index{pseudo-Riemannian vector bundle}
. If $g(x)$ is positive definite for each $x\in
B$ then $\eta$ is called Riemannian vector bundle.

A section $\sigma\in Sec(\eta)$ in a pseudo-riemannian vector bundle is called
normed if $g(x;\sigma(x),\sigma(x))=1, \ x\in B$.

The metric $g$ in $\eta$ defines duality, i.e. linear isomorphism, between
$\eta$ and $\eta^*$. Also, $\eta^*$ acquires metric represented by the
inverse matrix of the matrix of $g(x), x\in B$.

If $\eta$ is riemannian then all tensor, exterior and symmetric powers of
$\eta$ are also Riemannian.

Two pseudo-Riemannian vector bundles are called isometric if there is a bundle
map $\varphi$ between them such that each $\varphi_x$ is isometry, i.e. it
preserves the metric. Clearly, if the standard fiber $F$ is euclidean space and
$\eta$ is riemannian, then there is a coordinate representation
$(U_\alpha,\varphi_\alpha)$ for $\eta$ for which the maps
$\varphi_{\alpha,x}:F\rightarrow F_x$ are isometries. Such coordinate
representation is called {\it Riemannian}.

If two riemannian bundles are
isomorphic then there is an isomorphism which is isometry.

Every pseudo-Riemannian vector bundle is orientable.

Since $F$ and every fiber $F_x$ in a Riemannian bundle are euclidean spaces,
then there is a unit sphere $S_x\subset F_x$. It is possible to construct a
smooth bundle over the base space $B$ with a standard fiber the unit sphere in
$F$.

We recall a theorem from vector bundle theory, which
establishes some important properties of those vector bundles which admit
pseudoriemannian structure. The theorem says that if a vector bundle
$\Sigma$ with a base manifold $B$ and standard fiber $V$
admits pseudoriemannian structure $g$ of signature $(p,q), p+q=dimV$, then it
is always possible to introduce in this bundle a riemannian structure $h$ and
a linear automorphism $\varphi$ of the bundle, such that two subbundles
$\Sigma^+$ and $\Sigma^-$ may be defined with the following properties:

\begin{enumerate}
\item $g(\Sigma^+,\Sigma^+)=h(\Sigma^+,\Sigma^+)$,
\item $g(\Sigma^-,\Sigma^-)=-h(\Sigma^-,\Sigma^-)$,
\item $g(\Sigma^+,\Sigma^-)=h(\Sigma^+,\Sigma^-)=0$.
\end{enumerate}
The automorphism $\varphi$ is defined by
\[
g_x(u_x,v_x)=h_x(\varphi(u_x),v_x),\quad u_x,v_x\in V_x, \ \ x\in B.
\]
In components we have
\[
g_{ij}=h_{ik}\varphi^k_j \rightarrow \ \varphi^k_i=g_{im}h^{mk}.
\]
In the tangent bundle case this theorem allows to separate a subbundle of the
tangent bundle if the manifold admits pseudoriemannian metric.

{\bf Structure theorem}: For every vector bundle $\eta$ there exists a vector
bundle $\eta'$ over the same base such that the Whitney sum $\eta\oplus\eta'$
is a trivial bundle. The proof of this assertion makes use of the following
lemma: $Sec(\eta)$ is a finitely generated $\mathcal{J}(B)$-module. This theorem
enables us to create isomorphisms between the spaces $Hom(\eta,\eta')$ and
$Hom_{B}(Sec(\eta),Sec(\eta'))$, and this result can be extended to the
multilinear cases by corresponding extension of the isomorphism.

This structure theorem allows also {\it unit tensor}
$\mathfrak{t}\in Sec(\eta^*)\otimes_B Sec(\eta)$  for $(\eta,\eta^*)$
to be introduced: there are finitely many sections $\sigma^*_i\in Sec(\eta^*)$
and $\sigma_i\in Sec(\eta)$ such that
$$
\mathfrak{t}=\sum_i\sigma^*_i\otimes_B\sigma_i,\ \ \
\sigma=\sum_i\langle\sigma^*_i,\sigma\rangle\sigma_i, \ \ \sigma\in Sec(\eta).
$$

\section{Tangent and Cotangent Bundles}
{\bf 2.4.1. Tangent space.} Let $M$ be a n-dimensional real smooth manifold with
$\mathcal{J}(M)$ be the algebra of real valued smooth functions on $M$. We shall
define the concept of {\it tangent vector} which is one of the basic concepts
in differential geometry.

{\bf Definition}. A tangent vector of $M$ at the point $a\in M$ is a linear map
$\xi: \mathcal{J}(M)\rightarrow \mathbb{R}$ satisfying
$$
\xi(f.g)=\xi(f)g(a)+f(a)\xi(g), \ \ \ f,g\in\mathcal{J}(M).
$$
All tangent vectors at the point $a\in M$ form a real vector space denoted by
$T_a(M)$ with respect to the following rules:
$$
(\lambda\,\xi+\mu\,\eta)\,f=\lambda\,\xi(f)+\mu\,\eta(f), \ \ \
\lambda,\mu\in\mathbb{R}, \ \ \xi,\eta\in T_a(M), \ \ f\in\mathcal{J}(M).
$$
The so defined linear space $T_a(M)$ is called the {\it tangent space} of the
manifold $M$ at the point $a\in M$.

If $g$ is the constant function $g(x)=\lambda$ then the linearity of $\xi$
requires  $\xi(g.f)=\xi(\lambda\,f)=\lambda\xi(f)$, and the above property
requires $\xi(g.f) =\xi(\lambda)f+\lambda\xi(f)$, so, $\xi(\lambda)=0$: {\it any
tangent vector maps constant functions to zero.}

{\bf 2.4.2. Derivative of a smooth map}. Recall that a map $\varphi$ between
manifolds induces a homomorphism $\varphi^*$ between the corresponding algebras
of real valued functions. If $\varphi:M\rightarrow N$ then
$\varphi^*:\mathcal{J}(N)\rightarrow\mathcal{J}(M)$ according to
$$
(\varphi^*f)(x)=f(\varphi(x)), \ \ f\in\mathcal{J}(N), \ \ x\in M.
$$
Let $\xi\in T_a(M)$ and consider the composition $\xi\circ\varphi^*$. Clearly,
$\xi\circ\varphi^*$ is a linear map from $\mathcal{J}(N)$ to $\mathbb{R}$.
Let's see how it acts on product of functions from $\mathcal{J}(N)$:
$$
(\xi\circ\varphi^*)(fg)=\xi(\varphi^*f.\varphi^*g)=
\xi(\varphi^*f).g(\varphi(a))+f(\varphi(a)).\xi(\varphi^*g).
$$
We see that $\xi\circ\varphi^*$ is a tangent vector in $T_{\varphi(a)}(N)$.
Hence, the correspondence $\xi\rightarrow\xi\circ\varphi^*$ defines a
$\mathbb{R}$-linear map from $T_a(M)$ to $T_{\varphi(a)}(N)$. This map is
called the derivative of $\varphi$ at $a\in M$, and denoted by $d\varphi_a$.
Thus, $d\varphi_a(\xi)(g)=\xi(\varphi^*g)$.

If we have composition of maps $\psi\circ\varphi$, then
$d(\psi\circ\varphi)_a=d\psi_{\varphi(a)}\circ d\varphi_a$ for $a\in M$.

The derivative of the identity of $M$ at $a\in M$ is the identity of $T_a(M)$.

The derivative $d\varphi_a$ of a diffeomorphism $\varphi:M\rightarrow N$
is a linear isomorphism $T_a(M)\rightarrow T_{\varphi(a)}(N)$ with inverse
linear isomorphism $d\varphi^{-1}_{\varphi(a)}:T_{\varphi(a)}(N)\rightarrow
T_a(M)$.

If $U\subset M$ is open and $j:U\rightarrow M$ is the inclusion map, then
$dj_a:T_a(U)\rightarrow T_a(M)$ is linear isomorphism.

There is a canonical linear isomorphism between the vector space $E$ and any
$T_a(U)$, where $U$ is an open subset of $E$. It is given by
$h\rightarrow\xi_h$, where $\xi_h(f)=f'(a;h)$, i.e. every element $h\in E$ we
consider as defining directional derivative of functions along itself.

The dimension of every $T_a(M)$ is equal to the dimension of $M$.

The constant map $\varphi:M\rightarrow N$ has zero derivative, conversely, if
$d\varphi_a=0, a\in M$, and the manifold is connected then $d\varphi_a=0, a\in
M$.

{\bf 2.4.3. Tangent and cotangent bundles.} \index{tangent/cotangent bundles}
If $M$ is a manifold consider the
disjoint union $$ T_M=\bigcup_{a\in M}T_a(M) \ \  \text{and the projection} \ \
\pi_M:T_M\rightarrow M: \pi_M(\xi)=a, \ \xi\in T_a(M).
$$
Then the quadruple $\tau_M=(T_M,\pi_M,M,\mathbb{R}^n)$ is a vector bundle
over $M$ with standard fiber $\mathbb{R}^n$ and fiber over a point of $x\in M$
given by $T_x(M)$. This bundle is called tangent bundle of $M$. If
$(U_\alpha,u_\alpha)$ is a coordinate atlas, $U_\alpha\cap U_\beta$ is not
empty and $u_{\beta\alpha}=u_\beta\circ u^{-1}_\alpha$, then the
corresponding transition functions
$$
\psi_{\beta\alpha}:U_\alpha\cap U_\beta\times\mathbb{R}^n\rightarrow
U_\alpha\cap U_\beta\times\mathbb{R}^n
$$
are given by
$$
\psi_{\beta\alpha}(x,h)=(x,(du_{\beta\alpha})_{u_\alpha(x)}(h)), \ x\in M, \
h\in\mathbb{R}^n.
$$

The above considered derivative $d\varphi_a:T_a(M)\rightarrow T_{\varphi(a)}N$
of a map $\varphi:M\rightarrow N$ is naturally extended to a set map
$d\varphi:T_M\rightarrow T_N$ just by $d\varphi(\xi)=(d\varphi)_x(\xi)$ for
each $x\in M$ and $\xi\in T_x(M)$. It follows that considered as a map between
bundles $\tau_M$ and $\tau_N$ it is a homomorphism of bundles, i.e. a bundle
map between vector bundles. In particular, the derivative $dj$ of the inclusion
map $j:U\rightarrow M$ induces such a (strong) bundle map from $\tau_{U}$ to the
restriction of $\tau_M$ to $U$.

Since $\tau_M$ is a vector bundle then the corresponding dual bundle,
called {\it cotangent bundle} and denoted by $\tau^*_M$, is also defined. The
standard fiber of $\tau^*_M$ is again $\mathbb{R}^n$, and the fiber over $x\in
M$ is the dual space $T_x^*(M)$ of $T_x(M)$.
\vskip 0.3cm
{\bf 2.4.4. Local properties of smooth maps.}
Let $\varphi:M^n\rightarrow N^r$ be
a smooth map and $a\in M^n$. Then $\varphi$ is called:
\vskip 0.2cm
{\it Local diffeomorphism} at a point $a\in M$ if
$(d\varphi)_a:T_a(M)\rightarrow T_{\varphi(a)}(N)$ is linear isomorphism. Then
there are neighborhoods $U\subset M^n$ of $a\in M^n$ and $V\subset N^r$ of
$b=\varphi(a)\in N^r$ such that $\varphi$ maps $U$ diffeomorphically to $V$.
\vskip 0.2cm
{\it
Immersion} at a point $a\in M$ if $(d\varphi)_a:T_a(M)\rightarrow
T_{\varphi(a)}(N)$ is injective. Then there are neighborhoods $U\subset M^n$
of $a\in M^n$, $V\subset N^r$ of $b=\varphi(a)\in N^r$, $W\subset
\mathbb{R}^{r-n}$ and a diffeomorphism $\psi: U\times W\rightarrow V$ such that
$\varphi(x)=\psi(x,0), x\in U$. Also, there is a smooth map
$\theta: V\rightarrow U$ such that $\theta\circ\varphi_U=id_U$.
\vskip 0.2cm
{\it Submersion} at a point $a\in M$ if $(d\varphi)_a:T_a(M)\rightarrow
T_{\varphi(a)}(N)$ is surjective. Then there are neighborhoods $U\subset M^n$
of $a\in M^n$, $V\subset N^r$ of $b=\varphi(a)\in N^r$, $W\subset
\mathbb{R}^{n-r}$ and a diffeomorphism $\psi: V\times W\rightarrow U$ such that
$\varphi(x)=\pi_{V}\psi(x), x\in U$, and $\pi_V:V\times W\rightarrow V$. Also,
there is a smooth map $\theta: V\rightarrow U$ such that
$\varphi\circ\theta=id_V$.
\vskip 0.2cm

If the above properties hold at every $a\in M$ then $\varphi$ is called local
respectively diffeomorphism, immersion, submersion of $M$ into $N$. If
$\varphi$ is smooth bijective and all $(d\varphi)_a$ are injective then
$\varphi$ is a diffeomorphism.

An {\it imbedded manifold} is a pair $(N,\varphi)$ such that the derivative
$d\varphi:T_N\rightarrow T_M$ is injective, so $dim(N)\leq dim(M)$. In this
case $\varphi(N)\subset M$ acquires smooth structure in which $\varphi$ is a
diffeomorphism. If the topology of $\varphi(N)$ is induced by that of $M$, then
$(N,\varphi)$ is called {\it submanifold}. Clearly, if $N\subset M$ and
$\varphi$ is the inclusion map, then $N$ is a submanifold.

{\it Rank} of $\varphi:M\rightarrow N$ at $a\in M$ is the rank of the linear
map $(d\varphi)_a:T_a(M)\rightarrow T_{\varphi(a)}(N)$. If $y\in N$ and the
rank of $\varphi$ is constant in $\varphi^{-1}(y)$, say
$rank(\varphi)_{\varphi^{-1}(y)}=k<dim(N)$, then $\varphi^{-1}(y)$ is a
submanifold of $M$. In such a situation one can choose local coordinates around
$a\in M$ and $\varphi(a)\in N$ such that in the corresponding
neighborhoods
$$
\varphi(x_1,x_2,\dots,x_{dim(M)})=(y_1,y_2,\dots,y_k,0,\dots,0).
$$ If
$\varphi:M\rightarrow M$ satisfies $\varphi\circ\varphi=\varphi$ then
$\varphi(M)$ is also submanifold of $M$.

A manifold $M$ is called {\it parallelizable} if its tangent bundle is trivial,
i.e. if $\tau(M^n)=M^n\times \mathbb{R}^n$.

A number of smooth functions $(f_1,f_2,\dots,f_p), p\leq n$ on a manifold $M^n$
are called {\it independent} at a point $x\in M$ if their differentials
$(df_1)_x,\dots,(df_p)_x$ are linearly independent as elements of $T^*_x(M)$.

Any manifold of dimension $n$ can be embedded into $\mathbb{R}^{2n+1}$ and in
$\mathbb{R}^{2n}$

The smooth mappings $\varphi_1:M_1\rightarrow N$ and $\varphi_2:M_2\rightarrow
N$ are called transversal at $y\in(\varphi_1(M_1)\cap\varphi_2(M_2))$ if
$dimT_y(N)=dim(d\varphi_1)_{x_1}+dim(d\varphi_2)_{x_2}$ whenever
$\varphi_1(x_1)=\varphi_2(x_2)=y$.

{\bf 2.4.5. The Inclusion map.} If $U\subset M$ is an open subset
of the manifold $M$ and $j: U\rightarrow M$ is the inclusion map then the
linear map $(dj)_x: T_x(U)\rightarrow T_x(M)$ is a linear isomorphism.

Let now $U$ be an open subset of the vector space $F$. Then every $a\in F$
defines a linear isomorphism $\lambda_a: F\rightarrow T_a(U)$ by means of the
derivative operation as follows. As we know, the tangent vectors on a
manifold act as derivations in the algebra of smooth real valued functions on
the manifold. If $f$ is such a smooth real valued function on $U$ then its
derivative $f'_a(h), h\in F$, at the point $a\in U$ defines an element
 $\xi_h\in T_a(U)$ according to
$$
(\xi_h)_a(f)=f'_a(h)=\langle df_a,h\rangle, \ \ a\in U, \ \ h\in F.
$$
The corresponding Leibniz rule is easily verified. The so defined map
$\lambda_a: h\rightarrow \xi_h$ is a linear isomorphism between $F$ and $T_a(U)$.

In order to extend this isomorphism to $T_a(F)$ we make the following
consideration. Let $j: F\rightarrow F$ be the inclusion map considered as
identity map, so, $(dj)_a$ is the identity $T_a(F)\leftrightarrow T_a(F)$, i.e.
the differential of the identity is the identity of the corresponding tangent
space. Hence, the composition $(dj)_a\circ\lambda_a: F\rightarrow T_a(F)$ is a
linear isomorphism between $F$ and $T_a(F)$. Roughly speaking, the identity map
in a vector space $F$ leads to identification  of $F$ and $T_a(F)$ for each
$a\in F$. For example, if $\varphi$ is a projection in $F$ then the restriction
$\varphi|_{\varphi(F)}$ of $\varphi$ to its image $\varphi(F)$ is the identity
map for $\varphi(F)$ and the elements of $\varphi(F)$ generate the
corresponding linear isomorphisms.

%\newpage
\section{Vector fields}
{\bf 2.5.1. Definition.}

A vector field $X$ \index{vector field} on a manifold $M^n$ is a smooth section
of the tangent bundle $\tau(M)$ of $M$, so, $X: M\rightarrow T_M$ is smooth and
$\pi\circ X=id_M$. If $(U_\alpha,u_\alpha)$ is an atlas of $M$ then with a
vector field $X$ on $M$ are associated the functions $X_\alpha:
U_\alpha\rightarrow \mathbb{R}^n$, such that
$$
X_\alpha(x)=d(u_\alpha u_\beta^{-1})_{u_\beta(x)}X_\beta(x), \
x\in U_\alpha\cap U_\beta.
$$
Further the set of vector fields on $M$ will be denoted by $\mathfrak{X}(M)$.
\vskip 0.2cm
\noindent
{\bf 2.5.2. Properties.}
Here are some elementary properties of $X\in\mathfrak{X}(M)$.

1. $\mathfrak{X}(M)$ is an infinite dimensional vector space over the real
numbers $\mathbb{R}$.

2. $\mathfrak{X}(M)$ is a module over the algebra of smooth functions
$\mathcal{J}(M)$.

3. $\mathfrak{X}(M)|_{U_\alpha}$ is finitely generated.

4. If $M$ is parallelizable then $\mathfrak{X}(M)$ has finite basis.

5. A vector field $X$ on $M$ can be restricted to $U\subset M$ and then the
restriction $X|_U$ satisfies $(fX)|_U(x)=f(x)X(x), \ x\in U$.

6. If $(U_\alpha,u_\alpha)$ is a local chart on $M$ inducing local coordinates
$(x^1,x^2,\dots,x^n)$ then the partial derivatives
$$
\frac{\partial}{\partial x^1}, \ \frac{\partial}{\partial x^2}, \ \dots, \
\frac{\partial}{\partial x^n}
$$
are (local) vector fields on $U\subset M$. They form a basis for $\tau(U)$
called {\it holonomic}. So, a vector field on $U$ can be represented by
(summation over the repeated indices)
$$
X=X^\sigma\frac{\partial}{\partial
x^\sigma}, \ \ \sigma=1,2,\dots,n.
$$

6. If $M$ is smooth, then the set of vector fields $\mathfrak{X}(M)$ is
isomorphic to the derivations in $\mathcal{J}(M)$, so every $X\in
\mathfrak{X}(M)$ maps $\mathcal{J}(M)$ into $\mathcal{J}(M)$ linearly with
respect to $\mathbb{R}$ and generates derivation: $$
X(\lambda\,f)=\lambda\,X(f), \ \ \ X(fg)=X(f)g+fX(g) ,
$$
where $\lambda\in \mathbb{R}$ and $f,g$ are two (smooth) functions on $M$. Such
an isomorphism does NOT hold in the nonsmooth case.

7. If $\varphi:M\rightarrow N$ is a diffeomorphism and $X$ is a vector
field on $M$, then $(d\varphi)\circ X\circ\varphi^{-1}$ is a vector field on
$N$. If $\varphi:M\rightarrow M$ then $(d\varphi)\circ X\circ\varphi^{-1}$ is
denoted usually by $\varphi_*X$. Locally,
$$
(\varphi_*X)(x)=(d\varphi)_{\varphi^{-1}(x)}X_{\varphi^{-1}(x)}, \ \ \ x\in N.
$$
If locally $\varphi:M\rightarrow M$ is given by
$y^\nu=\varphi^\nu(x^\sigma)$, and we consider this as {\bf change of
coordinates}, then the representation of $X$ with respect to the new coordinates
$(y^1,\dots,y^n)$ is computed as follows:

$1^o$. Represent $X$ locally as $X=X^\sigma(x^\nu)\frac{\partial}{\partial
x^\sigma}$.

$2^o$. Compute consecutively $X(y^1(x^\sigma))$, $X(y^2(x^\sigma))$, ...,
$X(y^n(x^\sigma))$.

$3^o$. Replace $x^\sigma$ according to $x^\sigma=(\varphi^{-1})^\sigma(y^\nu)$
in the expressions obtained.

$4^o$. The so obtained functions $\tilde{X}^\sigma(y^\nu)$ are the components of
$X$ with respect to the new local basis $\frac{\partial}{\partial y^{\nu}}$.
So, in coordinates $(y^\nu)$ we obtain
$X=\tilde{X}^\sigma(y^\nu)\frac{\partial}{\partial y^\sigma}$.

8. Let $\varphi: M\rightarrow N$ be a smooth map. Two vector fields $X\in
\mathfrak{X}(M)$ and $Y\in \mathfrak{X}(N)$ are called $\varphi-related$ if:
$Y(\varphi(x))=(d\varphi)_x(X(x)), \ \ x\in M$. Accordingly, $X$ and $Y$ are
$\varphi-related$ iff: $\varphi^*(Y(g))=X(\varphi^*g), \ g\in\mathcal{J}(N)$.
\vskip 0.2cm
\noindent
{\bf 2.5.3. Lie product of vector fields (Commutator).} \index{commutator of
vector fields} In view of the isomorphism $\mathfrak{X}(M)\cong
Der\mathcal{J}(M)$ and the fact that $Der\mathcal{J}(M)$ is a Lie algebra over
$\mathbb{R}$ with respect to $$
[\theta_1,\theta_2]=\theta_1\circ\theta_2-\theta_2\circ\theta_1, \ \
\theta_1,\theta_2\in Der\mathcal{J}(M) $$ we obtain a Lie product structure
with respect to $\mathbb{R}$ in $\mathfrak{X}(M)$: $$ [X,Y](f)=X(Y(f))-Y(X(f)),
\ \ f\in\mathcal{J}(M). $$ The local expression for $[X,Y]$ with respect to the
local coordinates $(x^1,x^2,\dots,x^n)$ in the coordinate frame
$\{\frac{\partial}{\partial x^\nu}\}$ is the following:
$$
X=X^\nu\frac{\partial}{\partial x^{\nu}}, \ \
Y=Y^\sigma\frac{\partial}{\partial x^\sigma}: \ \
[X,Y]=\left(X^\nu\frac{\partial Y^\sigma}{\partial x^\nu}- Y^\nu\frac{\partial
X^\sigma}{\partial x^\nu}\right)\frac{\partial}{\partial x^\sigma}.
$$

This Lie bracket is bilinear with respect to $\mathbb{R}$ and satisfies the
relations $(X,Y,Z\in \mathfrak{X}(M))$:
$$
[X,Y+Z]=[X,Y]+[X,Z];
$$
$$
[X,Y]=-[Y,X];
$$
$$
[X,fY]=(Xf)Y+f[X,Y],
$$
$$
[[X,Y],Z]+[[Y,Z],X]+[[Z,X],Y]=0  \ \ \text{the Jacobi identity}.
$$

If $X,Y\in \mathfrak{X}(M)$ and $X_1,Y_1\in \mathfrak{X}(N)$ are
correspondingly $\varphi-related$, $\lambda,\mu\in \mathbb{R}$ and $f\in
\mathcal{J}(N)$, then: $\lambda X+\mu Y$ is $\varphi-related$ to $\lambda
X_1+\mu Y_1$; $(\varphi^*f).X$ is $\varphi-related$ to $f.X_1$ and $[X,Y]$ is
$\varphi-related$ to $[X_1,Y_1]$.

If $\varphi: M\rightarrow N$ is a diffeomorphism, then $\varphi_*$ is an
isomorphism of Lie algebras, in particular,
$$
\varphi_*[X,Y]=[\varphi_*X,\varphi_*Y], \ \ X,Y\in \mathfrak{X}(M).
$$
%\newpage
{\bf 2.5.4. The flow of a vector field}.

Let $X$ be a vector field on the $n$-dimensional manifold $M$ and the map
$c:I\rightarrow M$, where $I=(t_o,t_1)$ is an open interval in $\mathbb{R}$,
defines a smooth curve in $M$. Then if $X^a$ are the components of $X$ with
respect to the local coordinates $(x^1, ..., x^n)$ and the equality
$c'(t)=X(c(t))$ holds for every $t\in I$, or in local coordinates,
\[
\frac{dc}{dt}\left(\frac{d}{dt}\right)=
\frac{dx^a}{dt}\frac{\partial}{\partial x^a}=
X^a(x^b(t))\frac{\partial}{\partial x^a} ,
\]
$c(t)$ is called {\it integral curve/orbit} of the vector field $X$ through the
point $c(t_o)\in M$. The product $\mathbb{R}\times M$ is
considered and the following important theorem for uniqueness and existence of a
solution is proved: For every point $p\in M$ and point $\tau \in \mathbb{R}$
there exist a vicinity $U$ of $p$, a positive number $\varepsilon $ and a
smooth map $\Phi:(\tau-\varepsilon,\tau +\varepsilon)\times U\rightarrow M$,
$\Phi:(t,x)\rightarrow \varphi_t (x)$, such that for every point $x\in U$ the
following conditions are met: $\varphi_\tau (x)=x,\
t\rightarrow \varphi_t (x)$ is an integral curve of $X$, passing through the
point $x\in M$; besides, if two such integral curves of $X$ have at least one
common point, they coincide. Moreover, if $(t',x),\ (t+t',x)$ and
$(t,\varphi_{t'}(x))$  are points of a vicinity $U'$ of $\{0\}\times \mathbb{R}$
in $\mathbb{R}\times M$, we have
$\varphi_{t+t'}(x)=\varphi_t (\varphi_{t'}(x))$. This last relation gives the
local group action: for every $t\in I$ we have the local diffeomorphism
$\varphi_t :U\rightarrow \varphi_t (U)$. So, through every point of $M$ there
passes only one integral curve of $X$ and in this way the manifold $M$ is
foliated to non-crossing integral curves - 1-dimensional manifolds, and these
1-dimensional manifolds define all trajectories of the defined by the vector
field $X$ system of ordinary differential equations.

If the vicinity $U'$ coincides with the whole
$\mathbb{R}\times M$ then the group $\varphi_t$ is called {\it global}, and it
satisfies: $\varphi_o=id_M$, $\varphi_{t+t'}=\varphi_t\circ\varphi_{t'}$,
$\varphi_{-t}=(\varphi_t)^{-1}$. In such a case the corresponding vector field
is called {\it complete}.

On compact manifolds all vector fields are complete.

If $X$ is not complete it is possible to find positive function $f>0$ on $M$
such that the field $fX$ is complete.

Around every nonsingular point $y$ for
$X$: $X(y)\neq 0$, there exists a coordinate system $(x_1,x_2,\dots,x_n)$ such
that locally $X=\frac{\partial}{\partial x_1}$.

Simplifying, we can say that every vector field $X$ defines 1-parameter group
$\varphi_t$ of local diffeomorphisms of $M$: for each couple $(t_o,t)$, varying
the "initial conditions" $x^a(t_o)$ inside an open set $U\subset M$, we can
define the diffeomorphic image $\varphi_t(U)$ of $U$, and this is true for any
$U\subset M$. This one-parameter group of local diffeomorphisms is called the
{\it flow} defined by $X\in\mathfrak{X}(M)$, and characterizes the
dynamical nature of the concept of vector field.

If $X$ and $Y$ are $\phi-related$, then
$\varphi_t^Y\circ\phi=\phi\circ\varphi_t^X$. If $[X,Y]=0$, then the
corresponding flows commute:
$\varphi_t^Y\circ\varphi_t^X=\varphi_t^X\circ\varphi_t^Y$. Also,
$(\varphi_{-t}^X)_*(Y)=Y$.

Finally, if $X$ is tangent to some submanifold $N\subset M$ then its orbit
through a point in $N$ lays entirely in $N$.

Let now look at the situation inversely: If $\psi_t$ is 1-parameter group of
diffeomorphisms of the smooth manifold $M$ then does there exist a vector field
on $M$ such that its flow $\varphi_t^X$ to coincide with the given $\psi_t$?
The answer is positive: there exists a vector field $X$ on $M$ such that
$\varphi_t^X=\psi_t$. The corresponding to $\psi_t$ vector field is defined by
the following relation:
$$
(Xf)(x)=lim_{t\rightarrow 0}\frac{f(\psi_t(x))-f(x)}{t}, \ \ f\in\mathcal{J}(M).
$$
The group $\psi_t$ defines a curve on $M$: $t\rightarrow \psi_t(x)$, passing
through the point $x=\psi_0(x)\in M$, and the corresponding vector field $X$ is
tangent to this curve, e.g. $X(\psi_t(x))$ is tangent to $t\rightarrow
\psi_t(x)$ at the point $\psi_t(x)$.

Now, if the function $f\in \mathcal{J}(M)$ is invariant with respect to the group
$\varphi_t$: $\varphi_t^*f=f$ then $f(\varphi_t(x))=f(x)$ for any $x\in M$, so
the generated by $\varphi_t$ vector field kills $f$: $X(f)=0$. This means that
$f$ is constant on every integral curve of $X$, therefore it is called {\it
first integral} of $X$.

\section{Covector fields}
{\bf 2.6.1. One-forms.}  Recall the cotangent bundle $\tau^*(M)$ of the manifold
$M$. The sections of $\tau^*(M)$ are called {\it covariant vector
fields, co-vector fields} or {\it one-forms}. Hence, if $\alpha$ is one-form
then $\alpha(x), x\in M$, is an element of $T^*_x(M)$.

The one-forms define a $\mathcal{J}(M)$-module over $M$: $(f,\alpha)\rightarrow
f\alpha$. This module will be denoted further by $\Lambda^1(M)$. The duality
between $T_x(M)$ and $T^*_x(M)$ induces duality between  $\Lambda^1(M)$ and
$\mathfrak{X}(M)$: $\Lambda^1(M)\times\mathfrak{X}(M)\rightarrow \mathcal{J}(M)$
given by
$$
\langle\alpha,X\rangle(x)=\langle\alpha(x),X(x)\rangle, \ \ \alpha\in
\Lambda^1(M), \ \ X\in\mathfrak{X}(M), \ \ x\in M.
$$
Clearly, $\Lambda^1(M)$ is isomorphic to
$Hom_M(\mathfrak{X}(M),\mathcal{J}(M))$.

If $\varphi: M\rightarrow N$ is a smooth map with corresponding bundle map
$d\varphi: \tau_M\rightarrow \tau_N$, then $(d\varphi)^*:
Sec(\tau^*_N)\rightarrow Sec(\tau^*_M)$, and $(d\varphi)^*$ is usually denoted
just by $\varphi^*: \Lambda^1(N)\rightarrow\Lambda^1(M)$. Explicitly,
$$
(\varphi^*\alpha)(x;\xi)=\alpha(\varphi(x);d\varphi_x(\xi)), \ \ x\in M, \ \
\xi\in T_x(M).
$$
Also,
$$
\varphi^*(f\alpha+g\beta)=(\varphi^*f).\varphi^*\alpha+(\varphi^*g).\varphi^*\beta,
 \ \ f,g\in \mathcal{J}(N), \ \ \alpha,\beta\in\Lambda^1(N).
$$

{\bf 2.6.2. The gradient of a function.}  Every $f\in\mathcal{J}(M)$ determines a
homomorphism $\varphi_f: \mathfrak{X}(M)\rightarrow \mathcal{J}(M)$ according
to $\varphi_f: X\rightarrow X(f)$. So, there is unique one-form
$df\in\Lambda^1(M)$ such that $X(f)=\langle df,X\rangle$, and $df$ is called
{\it gradient} of $f\in\mathcal{J}(M)$. We obtain
$$
d(\lambda\,f+\mu\,g)=\lambda\,df+\mu\,dg; \ \ d(f.g)=g.df+f.dg,
\ \ \lambda,\mu\in \mathbb{R}, \ \ f,g\in\mathcal{J}(M).
$$
If $\varphi: M\rightarrow N$ is a smooth map then
$\varphi^*(df)=d(\varphi^*f), \ f\in\mathcal{J}(N)$. If $f$ is a constant
function then $df=0$, conversely, if $df=0$ and $M$ is connected, then
$f=const$ on $M$.

The following result is important: The $\mathcal{J}(M)$-module $\Lambda^1(M)$
is generated by gradients. Locally, the gradients $(dx^1,dx^2,\dots,dx^n)$ of
the coordinate functions on $U\subset M$ define basis of $\Lambda^1(M)|_U$, so
any one-form $\alpha$ has the local representation of the kind
$\alpha=\alpha_\mu(x^1,\dots,x^n) dx^\mu$.

If $\frac{\partial}{\partial x^1},\dots.\frac{\partial}{\partial x^n}$ is local
basis in $\mathfrak{X}(M)|_U$ then the duality yields
$$
\left\langle dx^\mu,\frac{\partial}{\partial x^\nu}\right\rangle=
\frac{\partial}{\partial x^\nu}(x^\mu)=\delta^\mu_\nu.
$$

\section{Tensor and Exterior algebras over a\\ manifold.}
{\bf 2.7.1. Tensor algebras.}
Having the
$\mathcal{J}(M)$-modules $Sec(\tau_M)$ and $Sec(\tau^*_M)$ on a manifold $M$
we can construct the tensor products
$$
Sec(\tau_M)\otimes_MSec(\tau_M)\otimes_M\dots\otimes_MSec(\tau_M),
$$
$$
Sec(\tau^*_M)\otimes_MSec(\tau^*_M)\otimes_M\dots\otimes_MSec(\tau^*_M)
$$
and the $(p+q)$-product $\otimes^p\mathfrak{X}(M)\otimes\otimes^q\Lambda^1(M)$
over $M$ in the point-wise way with respect to the base space.
Thus, for $\Phi\in \otimes^p\mathfrak{X}(M)$ and $\Psi\in\otimes^q\Lambda^1(M)$
we have
$$
(\Phi\otimes\Psi)(x)=\Phi(x)\otimes\Psi(x), \ x\in M .
$$
The duality between $\otimes^p\mathfrak{X}(M)$ and $\otimes^p\Lambda^1(M)$ is also
introduced in the same point-wise way:
$$
\langle\Phi,\Psi\rangle(x)=\langle\Phi(x),\Psi(x)\rangle, \ x\in M.
$$
So, a mixed tensor field of type $(p,q)$ is a section of
$\tau^p(M)\otimes\tau^*_q(M)$. In particular, the unit tensor field
$\mathfrak{t}$ over $M$ is defined by
$\mathfrak{t}(x;X_x,\alpha_x)=\langle\alpha_x,X_x\rangle, \ x\in M$. If
$\{\alpha^\mu\}$ and $\{X_\nu\}$ are local dual bases we can write
$\mathfrak{t}=\alpha^\sigma\otimes X_\sigma$, and in coordinate bases we have
$$
\mathfrak{t}=dx^\sigma\otimes\frac{\partial}{\partial x^\sigma}, \ \text{so}, \ \
\mathfrak{t}(X,\alpha)=\left(dx^\sigma\otimes\frac{\partial}{\partial
x^\sigma}\right)(X,\alpha)
$$
$$
= \langle
dx^\sigma,X\rangle.\left\langle\alpha,\frac{\partial}{\partial x^\sigma}\right\rangle=
X^\sigma\alpha_\sigma.
$$
Local coordinate bases for $\otimes^p\mathfrak{X}(M)$ are given by all products
$$
\frac{\partial}{\partial x^{k_1}}\otimes\dots\otimes\frac{\partial}{\partial
x^{k_p}}.
$$

\noindent
{\bf 2.7.2. Exterior algebra over $\mathfrak{X}(M)$.} The $\mathcal{J}(M)$-module
$\mathfrak{X}(M)$ naturally determines its exterior $p$-powers
$\mathfrak{X}^p(M)=
\mathfrak{X}(M)\wedge\mathfrak{X}(M)\wedge\dots\wedge\mathfrak{X}(M)$.
Then the direct sum (which is in fact Whitney sum of vector bundles)
$$
\sum_{p=0}^n\mathfrak{X}^p(M)=
\mathfrak{X}^0(M)\oplus_M\mathfrak{X}^1(M)\oplus_M\mathfrak{X}^2(M)
\oplus_M\dots\oplus_M\mathfrak{X}^n(M), \ \ \mathfrak{X}^0(M)=\mathcal{J}(M),
$$
is anticommutative graded algebra (with respect to the wedge-product)
called {\it exterior algebra of multivectorfields} over $M$.

An element of the kind $X_1\wedge X_2\wedge\dots\wedge X_p$
is called {\it decomposable}, it is not zero only if the $X_i, i=1,2,\dots,p$
are linearly independent. Thus, a nonzero decomposable $p$-vector field defines
at every point $x\in M$ a $p$-dimensional subspace of $T_x(M)$. Here, in the
same way as in the algebraic case, but making use of the Lie product of vector
fields, the Schouten bracket of a $p$-vector field
and a $q$-vector field is defined. If $\Phi=X_1\wedge X_2\wedge\dots\wedge X_p$
and $\Psi=Y_1\wedge Y_2\wedge\dots\wedge Y_q$ the bracket $[\Phi,\Psi]$ is a
$(p+q-1)$-vector field and is given by
$$
[\Phi,\Psi]=\sum_{i,j}(-1)^{i+j}[X_i,Y_j]\wedge X_1\wedge\dots\wedge\hat{X_i}\wedge\dots\wedge X_p
\wedge Y_1\wedge\dots\wedge\hat{Y_j}\wedge\dots\wedge Y_q,
$$
with the corresponding properties:
$$
[\Phi,\Psi\wedge\Omega]=[\Phi,\Psi]\wedge\Omega+(-1)^{(p-1)q}\Psi\wedge[\Phi,\Omega],
$$
$$
[\Omega\wedge\Phi,\Psi]=\Omega\wedge [\Phi,\Psi]\wedge
+(-1)^{(q-1)p}[\Omega,\Psi]\wedge\Phi.
$$
Since the coordinate basis elements
are of the kind
$$ \frac{\partial}{\partial
x^{k_1}}\wedge\dots\wedge\frac{\partial}{\partial x^{k_p}}, \ \ \text{with} \ \
k_1<k_2<\dots<k_p, $$ for a p-multivector field $\Phi$ we obtain $$
\Phi(x^1,\dots,x^n)=\sum_{\sigma_1<\dots<\sigma_p}
\Phi^{\sigma_1\sigma_2\dots\sigma_p}(x^1,\dots,x^n)
\frac{\partial}{\partial
x^{\sigma_1}}\wedge\frac{\partial}{\partial
x^{\sigma_2}}\wedge\dots\wedge\frac{\partial}{\partial x^{\sigma_p}}.
$$

{\bf 2.7.3. Exterior algebra over $\Lambda^1(M)$.} In the same algebraic way we
form the Whitney sum of the wedge powers of antisymmetric covector fields.
$$\bigwedge(M)=
\sum_{p=o}^n=
\Lambda^0(M)\oplus
\Lambda^1(M)\oplus\Lambda^2(M)\oplus\dots\oplus\Lambda^n(M)
$$
where $\Lambda^0(M)=\mathcal{J}(M)$. The wedge product, defined point-wise,
makes this space into anticommutative graded algebra. The elements of
$\Lambda^p(M)$ are called {\it differential p-forms} over $M$.
Explicitly we have for $\Phi\in\Lambda^p(M)$ and $\Psi\in\Lambda^q(M)$ that
$\Phi\wedge\Psi\in\Lambda^{p+q}(M)$ given by at $x\in M$
$$
(\Phi\wedge\Psi)(x;\xi_1,\dots,\xi_{p+q})
$$
$$
=\frac{1}{p!q!}\sum_\sigma\varepsilon_{\sigma}
\Phi(x;\xi_{\sigma(1)},\dots,\xi_{\sigma(p)})
\Psi(x;\xi_{\sigma(p+1)},\dots,\xi_{\sigma(p+q)}), \ p,q\geqslant 1.
$$
The property $\Phi\wedge\Psi=(-1)^{pq}\Psi\wedge\Phi$ follows, so the wedge
product of odd forms with itself yields zero:
$\alpha^{(2k+1)}\wedge\alpha^{(2k+1)}=0$.

If $\varphi:M\rightarrow N$ is a smooth map, then a p-form $\Psi$ on $N$
determines a p-form  $\varphi^*\Psi$ on $M$ according to
$$
(\varphi^*\Psi)(x;\xi_1,\dots,\xi_p)=\Psi(\varphi(x);(d\varphi)\xi_1,
\dots,(d\varphi)\xi_p), \ x\in M, \ \ \xi_1,\dots,\xi_p\in T_x(M).
$$

In coordinate basis for a p-form
$\Phi$ we obtain
$$
\Phi=\Phi_{\sigma_1\sigma_2\dots\sigma_p}dx^{\sigma_1}\wedge\dots\wedge\,
dx^{\sigma_p} , \ \ \text{with} \ \ \sigma_1<\dots<\sigma_p.
$$

If we consider the diffeomorphism $\varphi: M\rightarrow M$ as change of
coordinates: $(x_1,\dots,x_n)\rightarrow (y_1,\dots,y_n)$, then in the canonical
basis p-forms $(dx^{\sigma_1}\wedge\dots\wedge dx^{\sigma_p})$ we consider
$x^{\sigma}$ as functions of $(y_1,\dots,y_n)$, compute the differentials
$dx^{\sigma}$ through the corresponding derivatives of $x^\sigma(y^\rho)$ and
the new differentials $dy^{\sigma}$, and replace in the component functions
$x^{\sigma}$ by $x^{\sigma}(y^1,\dots,y^n)$.
\vskip 0.3cm
Consider now the tensor product $\Lambda^n(\tau^*_M)\otimes\mathfrak{X}^p(M)$
and denote the space of sections of this bundle by $\mathcal{S}_p(M)$. The
elements of $\mathcal{S}_p(M)$ are called $p-densities$.

Let $\Phi=X_1\wedge X_2\wedge\dots\wedge X_p$ be a nonzero decomposable
$p$-vector field, i.e. a decomposable section of the bundle $\mathfrak{X}^p(M)$, and
$\omega$ be a nonzero section of the 1-dimensional determinant bundle
$\Lambda^n(\tau^*_M)$. Then the $p$-density $\omega\otimes\Phi$
defines a $(n-p)$-differential form according to
$$
i(\Phi)\omega=i(X_p)\circ i(X_{p-1})\circ\dots\circ i(X_1)\omega.
$$
Hence, after extension by linearity, we obtain a map
$$
\mathfrak{P}:\mathcal{S}_p(M)\rightarrow Sec(\Lambda^{(n-p)}(M)) ,
$$
or
$$
\mathfrak{P}:Sec(\Lambda^n(\tau^*_M)\otimes\mathfrak{X}^p(M))\rightarrow
Sec(\Lambda^{(n-p)}(M)) ,
$$
which is an $\mathcal{J}(M)$-isomorphism. It could be
said that every nonzero section of $\Lambda^n(\tau^*_M)$ defines an isomorphism
between $Sec(\mathfrak{X}^p(M))$ and $Sec(\Lambda^{n-p}(M))$, this isomorphism
is called {\it Poincare} isomorphism.

\section{Calculus on manifolds} \index{calculus on manifolds}
{\bf 2.8.1. The substitution operator (interior product).}
This operator was introduced from pure
algebraic point of in sec.1.4.2. In the context of algebraic structure over a
manifold it is defined in the algebraic structure of the graded
algebra $\bigwedge(M)$. It is antiderivation of degree $(-1)$, and is defined by
the relation
$$
(i(X)\alpha)(X_1,\dots,X_{p-1})=\alpha(X,X_1,\dots,X_{p-1})
$$
for $X\in\mathfrak{X}(M)$ and $\alpha\in\Lambda^p(M)$.
It satisfies
$$
i(X)\circ i(Y)=-i(Y)\circ i(X), \ \ \text{and}
$$
$$
i(X)(\alpha\wedge\beta)=(i(X)\alpha)\wedge\beta+(-1)^p\wedge i(X)\beta, \ \
$$
for $\alpha\in\Lambda^p(M)$. It is extended
to $\mathcal{J}(M)$ by $i(X)f=0$ for any smooth function $f$, and for 1-forms
$\alpha$ it yields $i(X)\alpha=\langle\alpha,X\rangle$, which in case of
gradients turns to $i(X)df=\langle df,X\rangle=X(f)$.

{\bf 2.8.2. The Lie derivative in
$Sec(\tau_M),Sec(\tau^*_M),Sec(\tau_M\otimes\tau^*_M)$.} \index{Lie derivative}
 This is a very
important operator, it describes how a tensor object on a manifold changes
along a vector field $X\in\mathfrak{X}(M)$ taking into account also the local
changes of the very $X$. It is a derivation in the tensor algebra over $M$ of
degree $0$.

The idea is very simple. Every vector field $X$ generates a flow $\varphi_t$,
where $t\in U\subset\mathbb{R}$ is external to the manifold $M$ parameter.
Since for each $t$ the flow $\varphi_t$ is a local diffeomorphism, then every
tensor object $\mathbf{T}$ is being transformed to
$\varphi_*\mathbf{T}$ in the contravariant case and to
$\varphi^*\mathbf{T}$ in the covariant case. Also in the mixed case
$\mathbf{T}=A\otimes B$ where $A$ is covariant tensor field and $B$ is
contravariant tensor field we get $\mathbf{T}\rightarrow
\varphi^*(A)\otimes\varphi_*(B)$, and on a smooth function $f$ it acts as
$f\rightarrow \varphi^*f$. Therefore, at every point $x\in M$ we can compare
$T(x)$ with $(\varphi_*T)(x)/(\varphi^*T)(x)$ just by subtracting $T(x)$ from
$(\varphi_*T)(x)$ in the covariant case and subtracting $(\varphi_*T)(x)$ from
$T(x)$ in the contravariant case,
and then to consider how these differences behave at $t\rightarrow 0$.
The corresponding "limes" we call the {\it derivative} of $\mathbf{T}$
with respect to $X$. In general the Lie derivative of $\mathbf{T}$ is denoted
by $L_X\mathbf{T}$. Also, it is linear with respect to reals:
$L_X(\lambda\,T_1+\mu\,T_2)=\lambda L_XT_1+\mu L_XT_2$, moreover
$L_X(fT)=(L_Xf)T+fL_XT$.

In case of a function $f$ this procedure looks like this
$$
L_Xf(x)=lim_{t\rightarrow 0}\frac{\varphi^*_tf(x)-f(x)}{t}=
lim_{t\rightarrow 0}\frac{f(\varphi_t(x))-f(x)}{t}.
$$
\vskip 0.3cm
In case of a vector field $Y$ the Lie derivative $L_X(Y)$ reduces to the
commutator: $L_X(Y)=[X,Y]$, and $L_X(fY)=X(f)Y+f[X,Y]$.
\vskip 0.3cm
If $T=T_1\otimes T_2\otimes T_3\otimes\dots$, then from the derivation property
it follows $$ L_XT=(L_XT_1)\otimes T_2\otimes T_3\otimes\dots+
T_1\otimes L_XT_2\otimes T_3\otimes\dots+T_1\otimes T_2\otimes
L_XT_3\otimes\dots.
$$
\vskip 0.3cm
In coordinate basis vectors and one-forms for
$X=X^\sigma\frac{\partial}{\partial x_\sigma}$ we obtain:
$$
L_X\left(\frac{\partial}{\partial x_\mu}\right)=-
\frac{\partial X^\sigma}{\partial x_\mu}\frac{\partial}{\partial x_\sigma};
\ \ \ \
L_X(dx^\sigma)=\frac{\partial X^\sigma}{\partial x_\mu}dx^\mu.
$$
In view of this for contravariant and covariant tensors in components we obtain
respectively:
$$
(L_XT)^{\alpha\beta\gamma\dots}=
X^\mu\frac{\partial T^{\alpha\beta\gamma\dots}}{\partial x_\mu}-
T^{\mu\beta\gamma\dots}\frac{\partial X^\alpha}{\partial x_\mu}-
T^{\alpha\mu\gamma\dots}\frac{\partial X^\beta}{\partial x_\mu}-
T^{\alpha\beta\mu\dots}\frac{\partial X^\gamma}{\partial x_\mu}-\dots
$$
$$
(L_XT)_{\alpha\beta\gamma\dots}=
X^\mu\frac{\partial T_{\alpha\beta\gamma\dots}}{\partial x_\mu}+
T_{\sigma\beta\gamma\dots}\frac{\partial X^\sigma}{\partial x_\alpha}+
T_{\alpha\sigma\gamma\dots}\frac{\partial X^\sigma}{\partial x_\beta}+
T_{\alpha\beta\sigma\dots}\frac{\partial X^\sigma}{\partial x_\gamma}+\dots
$$
When restricted to $\bigwedge(M)$ the Lie derivative satisfies also:
$$
L_X(df)=dL_Xf=dX(f);
$$
$$ i([X,Y])=[L_X,i(Y)]=L_X\circ i(Y)-i(Y)\circ L_X;
$$
$$ L_X(\alpha\wedge\beta
)=(L_X\alpha)\wedge\beta+\alpha\wedge L_X\beta;
$$
$$ L_{[X,Y]}=[L_X,L_Y]=L_X\circ L_Y-L_Y\circ L_X;
$$
$$ L_{fX}\alpha=fL_X\alpha+df\wedge(i(X)\alpha). $$

A tensor field $\mathbf{T}$ on $M$ is called {\it invariant} with respect to
the vector field $X$ if $L_X(\mathbf{T})=0$. Because of the derivation property
of $L_X$ the set of all $X$-invariant tensor fields form a subalgebra of the
$\mathbb{R}$-algebra $\otimes(M)$.

{\bf 2.8.3. Exterior derivative in} $\bigwedge(M)$. The {\it exterior
derivative} in $\bigwedge(M)$ is the $\mathbb{R}$-linear map $\mathbf{d}:
\bigwedge(M)\rightarrow\bigwedge(M)$ homogeneous of degree $1$. On functions it
coincides with $d: \mathbf{d}f=df$, and on $p$-forms acts as follows:
$$
\mathbf{d}\alpha((X_0,\dots,X_p)=\sum_{k=0}^p(-1)^kX_k(\alpha(X_0,\dots,\hat{X}_k,\dots,X_p))
$$
$$
+\sum_{0\leqslant i<j\leqslant
p}(-1)^{i+j}\alpha([X_i,X_j],\dots,\hat{X}_i,\dots,\hat{X}_j,\dots,X_p), \
\alpha\in\Lambda^p(M), \ X_k\in\mathfrak{X}(M).
 $$
If $\alpha$ is one-form then
$$
\mathbf{d}\alpha(X,Y)=X(\langle \alpha,Y\rangle)-Y(\langle \alpha,X\rangle)-
\alpha([X,Y]).
$$
If $\alpha$ is of the kind:
$\alpha=fdx^{i_1}\wedge dx^{i_2}\wedge\dots\wedge dx^{i_p}$
in coordinate basis, we get
$$
\mathbf{d}\alpha=df\wedge dx^{i_1}\wedge dx^{i_2}\wedge\dots\wedge dx^{i_p}.
$$

The exterior derivative $\mathbf{d}$ has the following properties:
\vskip 0.3cm
	$1. \ L_X=i(X)\circ \mathbf{d}+\mathbf{d}\circ i(X)$

	$2. \ \mathbf{d}(\alpha\wedge \beta)=\mathbf{d}\alpha\wedge\beta+
(-1)^p\alpha\wedge\mathbf{d}\beta, \ \alpha\in\lambda^p(M)$,

	$3. \ \mathbf{d}\circ\mathbf{d}=0$,

	$4. \ L_X\circ\mathbf{d}=\mathbf{d}\circ L_X$,

	$5. \ \varphi^*\circ\mathbf{d}=\mathbf{d}\circ\varphi^*, \
\varphi-\text{smooth}$.
\vskip 0.3cm

If $\{e_\alpha\}$ and $\{\varepsilon^\beta\}$ are dual bases on open set of
$M$ correspondingly for vector fields and 1-forms, and
$[e_\alpha,e_\beta]=C^\sigma_{\alpha\beta}e_{\sigma}$ then the following
relations hold: $$
L_{e_{\nu}}\varepsilon^\beta=-C_{\nu\sigma}^\beta\varepsilon^\sigma; \ \
\mathbf{d}\varepsilon^\sigma=
-\frac12C^\sigma_{\alpha\beta}\varepsilon^\alpha\wedge\varepsilon^\beta.
$$

Recalling the algebraic isomorphism (Sec.1.4.2)
$$
D^p(x): \Lambda^n(M)\otimes\mathfrak{X}^p(M)(x)\rightarrow\Lambda^{n-p}(M)(x),
x\in M
$$
we can define the {\it divergence} operator (with respect to a
definite volume form $\omega$ on $M$)
$$
\delta_{\omega}=(-1)^pD_{n-p+1}\circ\mathbf{d}\circ D^p:
\mathfrak{X}^p(M)\rightarrow\mathfrak{X}^{p-1}(M).
$$
Clearly, $\delta_{\omega}\circ\delta_{\omega}=0$. Note that a similar
formula can be established for any $q$-graded (anti)derivation
$\mathfrak{D^q}\,:\Lambda^p(M)\rightarrow\Lambda^{p+q}(M)$ in $\Lambda(M)$.

Making use of this $\delta_{\omega}$,
of the above defined Schouten bracket inside the exterior algebra on
$\mathfrak{X}(M)$
and of the extended by linearity
insertion operator with respect to a
$q$-multivector field $\Phi$, the Lie derivative of differential forms is
naturally extended to $q-$multivector fields \index{Lie derivative along
p-vectors} $\Phi$ in the following two ways:
$$
L_\Phi\,\alpha:=\mathbf{d}\,i_{\Phi}\alpha-
(-1)^{deg\Phi}i_{\Phi}\,\mathbf{d}\alpha, \ \ \
L_{(\Phi,\omega)}\,\alpha:=i(\delta_{\omega}\Phi)\alpha-
(-1)^{deg\Phi}i_{\Phi}\,\mathbf{d}\alpha .
$$
If $L_\Phi\,\alpha=0$, then $\alpha$ is called invariant with respect to
$\Phi$. Clearly, if $\alpha$ is $\Phi$-invariant: $L_\Phi\,\alpha=0$, and not
$\Phi$-sensitive: $i(\Phi)\alpha=0$, then this invariance reduces to the
requirement that $\mathbf{d}\alpha$ is not $\Phi$-sensitive. The second above
relation can be similarly characterized in these terms. Additionally, if
$i(\Phi)\alpha$ is closed: $\mathbf{d}i(\Phi)\alpha=0$ and $L_\Phi\,\alpha\neq
0$ then $(\Phi, \alpha)$ may be called $\mathbf{d}$-{\it partners}.

The forms $L_\Phi\,\alpha$ and $L_{(\Phi,\omega)}\,\alpha$ are
 $(deg\,\alpha-q+1)$-forms. If $(\Phi,\Psi)$ are
correspondingly $r$ and $s$ multivector fields, making use
of the Schouten bracket inside the exterior algebra on
$\mathfrak{X}(M)$, we obtain (arXiv: math-ph/0202043v1):
$$ \mathbf{d}\,L_\Phi\alpha=(-1)^{(r-1)}L_\Phi\,\mathbf{d}\alpha, $$
$$
i_{[\Phi,\Psi]}\alpha=(-1)^{(r-1)s}L_{\Phi}\,i_{\Psi}\alpha-
i_{\Psi}\,L_{\Phi}\alpha ,
$$
$$
L_{[\Phi,\Psi]}\alpha=(-1)^{(r-1)(s-1)}L_{\Phi}\,L_{\Psi}\alpha-
L_{\Psi}\,L_{\Phi}\alpha ,
$$
$$
L_{\Phi\wedge\Psi}\alpha=(-1)^{s}i_{\Psi}\,L_{\Phi}\alpha+L_{\Psi}\,i_{\Phi}\alpha .
$$
If now $\Omega$ is a $m$-multivector field we have:
$$
i_{[\Phi,\Psi\wedge\Omega]}\alpha=(-1)^{(r-1)(s+m)}L_\Phi\,i_{[\Psi\wedge\Omega]}\alpha-
i_{[\Psi\wedge\Omega]}\,L_\Phi ,
$$
$$
i_{[\Phi\wedge\Psi,\Omega]}\alpha=(-1)^{(r+s-1)m}L_{\Phi\wedge\Psi}\,i_{\Omega}\alpha
-i_{\Omega}\,L_{\Phi\wedge\Psi}\alpha .
$$

\vskip 0.3cm
{\bf 2.8.4. Vector valued differential forms.}  Let $E$ be a real finite
dimensional vector space. We consider the skew symmetric $p$-linear maps
$\Omega_x$ from $T_x(M)$ to $E$, smoothly depending on $x\in M$:
$$
\Omega_x: T_x(M)\times T_x(M)\times\dots\times T_x(M)\rightarrow E.
$$
These objects are called $E$-valued differential forms on $M$ and will be
denoted by $\Lambda^p(M,E)$. They form a module over the smooth functions
$\mathcal{J}(M)$. The direct sum of these modules is denoted by
$\bigwedge(M,E)$

Clearly, we have the isomorphism $\bigwedge(M)\otimes
E\rightarrow\bigwedge(M,E): \alpha\otimes v\rightarrow\Omega$, where $v\in E$,
and
 $$ \Omega(x;\xi_1,\dots,\xi_p)=\alpha(x;\xi_1,\dots,\xi_p).v,
\ \ x\in M, \ \ v\in E, \ \ \xi_i\in T_x(M).
$$
The operators $i(X), L_X, \mathbf{d}$ are naturally
generalized to $\bigwedge(M)\otimes E$ by \linebreak $i(X)\otimes id_E, \ \
L_X\otimes id_E, \ \ \mathbf{d}\otimes id_E$. So the above stated properties of
these operators are naturally carried to $\bigwedge(M,E)$.

Also, with respect to a smooth map $M\rightarrow N$ we get
$\varphi^*(\alpha\otimes v)=(\varphi^*\alpha)\otimes v$.

Another property of $\bigwedge(M)\otimes E$ is that it is a graded module over
the algebra of differential forms $\bigwedge(M)$
on $M$: $\alpha\wedge(\beta\otimes
v)=(\alpha\wedge\beta)\otimes v$. So, if $\Phi\in\bigwedge(M)\otimes E$ and
$\alpha\in\Lambda^p(M)$, then
we have the relations
$$
i(X)(\alpha\wedge\Phi)=(i(X)\alpha)\wedge\Phi+(-1)^p\alpha\wedge i(X)\Phi,
$$
$$
L_X(\alpha\wedge\Phi)=(L_X\alpha)\wedge\Phi+\alpha\wedge L_X\Phi,
$$
$$
\mathbf{d}(\alpha\wedge\Phi)=\mathbf{d}\alpha\wedge\Phi+
(-1)^p\alpha\wedge\mathbf{d}\Phi.
$$
If $F$ is another linear space and $\psi:E\rightarrow F$ is a linear map then
the space $\bigwedge(M,E)$ is transformed to $\bigwedge(M,F)$ according to
$\psi_*(\Phi)=\psi\circ\Phi$, and the commutations
$$
\psi_*\circ
i(X)=i(X)\circ\psi_*, \ \ \psi_*\circ L_X=L_X\circ\psi_*, \ \
\psi_*\circ\mathbf{d}=\mathbf{d}\circ\psi_*
$$ are obvious.

Another interesting case is when $E$ is an algebra (multiplication denoted just
by "."). Then we can multiply the elements in $\bigwedge(M,E)$ according to
$(\alpha\otimes v).(\beta\otimes w)=\alpha\wedge\beta\otimes v.w; \ v,w\in E,
$ i.e.
$$
(\Phi.\Psi)(x;\xi_1,\dots,\xi_{p+q})=
\frac{1}{p!q!}\sum_{\sigma}\varepsilon_{\sigma}\Phi(x;\xi_{\sigma(1)},\dots,\xi_{\sigma(p)})
.\Psi(x;\xi_{\sigma(p+1)},\dots,\xi_{\sigma(p+q)})$$
where $\Phi\in\Lambda^p(M,E)$, $\Psi\in\Lambda^q(M,E)$, $x\in M$ and $\xi_i\in
T_x(M)$. If $E$ is a commutative algebra then $\Phi.\Psi=(-1)^{pq}\Psi.\Phi$,
if $E$ is skew-commutative then $\Phi.\Psi=(-1)^{pq+1}\Psi.\Phi$, finally, if
$E$ is a Lie algebra, then
$$
(-1)^{pq}(\Phi.\Psi).\Omega +(-1)^{rp}(\Omega.\Phi).\Psi
+(-1)^{qr}(\Psi.\Omega).\Phi=0,
$$ where $\Phi\in\Lambda^p(M,E),\ \Psi\in\Lambda^q(M,E),\
\Omega\in\Lambda^r(M,E)$. Clearly, for $\Phi\in\Lambda^p(M,E)$ we have
$$
0=(-1)^{p.p}\Big((\Phi.\Phi).\Phi+(\Phi.\Phi).\Phi+(\Phi.\Phi).\Phi\Big)=
3(-1)^{p.p}(\Phi.\Phi).\Phi,
$$
i.e. $(\Phi.\Phi).\Phi=0$.

In general if $\varphi: (E,F)\rightarrow G$ is a bilinear map, $\alpha^i\otimes
e_i$ and $\beta^j\otimes k_j$ are correspondingly from $\bigwedge(M,E)$ and
$\bigwedge(M,F)$, where $\{e_i\}$ is a basis in $E$ and $\{k_j\}$ is basis in
$F$, we get a $G$-valued form according to
$$
(\alpha^i\otimes e_i).(\beta^j\otimes
k_j)=\alpha^i\wedge\beta^j\otimes\varphi(e_i,k_j).
$$
For example, if $\varphi$ is respectively: symmetrized: $E\vee E$, and
antisymmetrized: $E\wedge E$, tensor product, we obtain
$$
\sum_{i=1}^{n}\alpha^i\wedge\beta^i\otimes (e_i\vee e_i)+
\sum_{i<j}(\alpha^i\wedge\beta^j+\alpha^j\wedge\beta^i)\otimes (e_i\vee e_j),
$$
$$
\sum_{i<j=1}^{n}(\alpha^i\wedge\beta^j-\alpha^j\wedge\beta^i)\otimes(e_i\wedge
e_j).
$$

We construct the $\varphi$-extended insertion operator
\index{$\varphi$-extended insertion operator}
on $M$. Let
$T=\mathfrak{t}^i\otimes e_i$ be a $E_1$-valued q-vector,
$\Phi=\alpha^j\otimes k_j$ be a $E_2$-valued p-form with $q\leq p$ and
$\varphi:E_1\times E_2\rightarrow F$ be a bilinear map. Now we define
$i^{\varphi}_T\Phi\in\Lambda^{p-q}(M,F)$:
$$
i^{\varphi}_T\Phi=
i_{\mathfrak{t}^i}\alpha^j\otimes\varphi(e_i,k_j), \ \ \
i=1,2,...,dim(E_1), \ j=1,2,...,dim(E_2).
$$
Hence, we can define the $\varphi$-extended Lie derivative
\index{$\varphi$-extended Lie derivative}
$$
\mathcal{L}^{\varphi}_T:
\Lambda^p(M,E_1)\times\mathfrak{X}^q(M,E_2)\rightarrow\Lambda^{p-q+1}(M,F)
$$
as follows
$$
\mathcal{L}^{\varphi}_T(\Phi)=
\mathbf{d}\circ i^{\varphi}_T\Phi-
(-1)^{deg(T).deg(\mathbf{d})}i^{\varphi}_T\circ\mathbf{d}\Phi.
 $$
Accordingly, $T$ will be called ({\it Lie, $\varphi$)-symmetry}
\index{Lie $\varphi$ - symmetry} of $\Phi$ if
$\mathcal{L}^{\varphi}_T(\Phi)=0$, and {\it algebraic} $\varphi$ - symmetry of
$\Phi$ if $i^{\varphi}_T\Phi=\mathfrak{C}$, where $\mathfrak{C}$ is a constant
element of $\Lambda^{p-q}(M,F)$, leading to $\mathbf{d}\circ
i^{\varphi}_T\Phi=0$. Surely, the case $\mathfrak{C}=0$ is admissible.

\section{Orientation and Integration on manifolds}

{\bf 2.9.1. Orientation of vector spaces.}
Let $(V,V^*)$ be a couple of two dual $n$-dimensional real
vector spaces.
%We say that two bases of $V$ have the same orientation, or they
%are consistently oriented, if for any linear isomorphism $\varphi$ of
%$V$, if $\varphi$ transforms $K$ into $E$ and
%the determinant $det(\varphi)$ is positive: $det(\varphi)>0$.

If $\omega$ is a basis in $\Lambda^n(V^*)$ then for all
$\lambda>0, \lambda\in\mathbb{R}$, the relation $\omega=\lambda.\omega$ defines
equivalence relation on the nonzero elements of  $\Lambda^n(V^*)$, so we have
just two classes of equivalence. It is said that the choice of each class of
equivalence defines orientation in $V$, and if the choice is done, it is said
that the space $V$ is oriented.
If $\{\varepsilon^1,\dots,\varepsilon^n\}$ is a basis of $V^*$ and $\varphi\in
GL(V)$, then the class of the nonzero $n$-form
$\omega=\varepsilon^1\wedge \varepsilon^2\wedge\dots\wedge\varepsilon^n$
defines an orientation in $V$ since
$\omega(\varphi(e_1),\dots,\varphi(e_n))=det(\varphi)\omega(e_1,\dots, e_n)$.
Clearly, the basis $\{e_1,\dots, e_n\}$ defines orientation in $V$. We say that
the linear isomorphism $\varphi$ of $V$ preserves the orientation in $V$ if for
every nonzero $\omega\in\Lambda^n(V^*)$ the $n$-forms $\omega$ and
$\varphi^*\omega$ define the same orientation in $V$.
The standard orientation in $\mathbb{R}^n$ is defined by the standard basis
$\{e_1=(1,0,\dots,0),e_2=(0,1,\dots,0),\dots,e_n=(0,0,\dots,1)\}$.
\vskip 0.3cm
{\bf 2.9.2. Orientation of manifolds.} Let $M$ be a $n$-dimensional real
(smooth) manifold, and $(U,\psi:U\rightarrow \mathbb{R}^n)$ be a local chart on
$U\subset M$. Let $\{e_1,\dots,e_n\}$ be a local frame on $U\subset M$,
so, at every point $x\in U$ we have a basis $\{e_1(x),\dots,e_n(x)\}$ of the
corresponding tangent space $T_x(M)$, so, the tangent space $T_x(M)$ is
oriented.
%We say that this local frame is
%{\it continuously} oriented if the orientation does not change from
%point to point inside $U\subset M$, and then the chart $(U,\psi)$ is also
%called continuously oriented.

Two intersecting local charts
$\left((U_\alpha,\varphi_\alpha);(U_\beta,\varphi_\beta)\right)$ on $M$ are
called orientationally consistent if the corresponding change of coordinates
$\varphi_\beta\circ\varphi_\alpha^{-1}$ preserves the
orientations of the corresponding tangent spaces, i.e. the corresponding
Jacobians of $d_x( \varphi_\beta\circ\varphi_\alpha^{-1})$ on
$(U_\alpha\cap U_\beta)$ are positive for each $x\in(U_\alpha\cap U_\beta)$.

The manifold $M$ is called orientable if there exists an atlas
$(U_\alpha,\varphi_\alpha)$ on $M$ such that every two intersecting local
charts are orientationally consistent. Thus, every orientable manifold can be
oriented by means of choosing appropriate atlas of local charts.

Hence, an orientable manifold admits a nonzero section $\Delta(x)\neq 0$
of the $1$-dimensional (co)bundle
$\Lambda^n(T^*M)$ usually called {\it orientation representing $n$-form},
the corresponding class is said to orient the manifold, and the elements of
the orientation class are called {\it positive} $n$-forms. All bases
$\{e_1(x),\dots,e_n(x)\}$ of a $T_x(M)$ are called positive with respect to
$\Delta(x)$ if $\Delta(x;e_1(x),\dots,e_n(x))>0$.

If $M$ is oriented with $\Delta$ then any restriction of $\Delta$ on $U\subset
M$ defines orientation of $U$.

In view of the above we can say that the manifold $M$ is orientable if its
tangent bundle $\tau(M)$ is orientable vector bundle.

All orientable manifolds $M$ have trivial (co)bundles $\Lambda^n(T^*M)$, and they
are called {\it parallelizable}. This means that these 1-dimensional (co)bundles
admit global nonzero sections.

If $M$ and $N$ are two $n$-dimensional oriented manifolds and
$\varphi:M\rightarrow N$ is a diffeomorphism, then $\varphi$ is called
orientation preserving if each $d\varphi_x:T_x(M)\rightarrow T_{\varphi(x)}(N)$
respects the introduced orientations of $T_x(M)$ and $T_{\varphi(x)}(N)$.
\vskip 0.3cm
{\bf 2.9.3. Manifolds with boundary.} An upper half space, denoted by
$\mathbb{H}^n$, of $\mathbb{R}^n$ is called the closed subset
$$
\mathbb{H}^n=\{(x_1,x_2,\dots,x_n)\in \mathbb{R}^n| x_n\geq 0\}.
$$
The subset $\{x\in\partial \mathbb{H}^n: x_n=0\}$ is called {\it
boundary} of $\mathbb{H}^n$ and is obviously isomorphic to $\mathbb{R}^{n-1}$.
Now, the subset $Int(\mathbb{H}^n)=\mathbb{H}^n-\partial\mathbb{H}^n$ is called
the {\it interior} of $\mathbb{H}^n$.

The boundary $\partial\mathbb{H}^n$  is endowed with the corresponding subset
topology and may be used as a model space instead of $\mathbb{R}^n$ for
constructing manifolds with boundary. The smoothness of a map $\varphi$ between open
subsets $U$ and $W$ of $\mathbb{H}^n$ can be defined if $\varphi$ has an
extension $\tilde{\varphi}$ to open subsets $\tilde{U}\supseteq U$ and
$\tilde{W}\supseteq W$ in $\mathbb{R}^n$. The so obtained manifolds are called
{\it manifolds with boundary}. Now, the boundary $\partial M$ of a
manifold-with-boundary consists of those points $z\in M$ for which there is a
chart $(U_\alpha,\psi_\alpha)$ such that $z\in U_\alpha$ and
$\psi_\alpha(z)\in\partial \mathbb{H}^n$. Correspondingly, the set of all
points living in $Int(M)=(M-\partial M)$ is called {\it interior} of $M$.
Clearly, the (standard) manifolds have empty boundaries, and the interior of a
manifold with boundary is a smooth manifold without boundary.

It can be shown that the boundary $\partial M$ of a $n$-dimensional manifold
with boundary admits a smooth structure of $(n-1)$-dimensional manifold.
\vskip 0.3cm
{\bf 2.9.4. Integration.}
A subset $U\subset M$ of a manifold is called {\it compact} if it is compact as
a topological space.

A tensor field $\mathfrak{t}$ on $M$ is said to have a compact carrier (or
support) $U\subset M$ if $\mathfrak{t}(x)\neq 0$ only on the closure of $U$.

We consider the space $\bigwedge_c(M)$ of differential forms on $M$ with compact
support, they form a graded ideal in the space of all differential forms on $M$
and $\bigwedge_c(M)$ is invariant with respect to the operators $i_X, L_X$, and
$\mathbf{d}$:
$$
i_X(\Lambda^p_c(M))\subset\Lambda^{p-1}_c(M), \ \
L_X(\Lambda^p_c(M))\subset \Lambda^p_c(M), \ \
\mathbf{d}(\Lambda^p_c(M))\subset\Lambda^{p+1}_c(M).
$$
If $U\subset\mathbb{R}^n$ and $\mathbf{d\mu}=dx^1dx^2\dots dx^n$ is the
Lebesque measure we have the integral of a $n$-form
$\alpha\in\Lambda^n_c(\mathbb{R}^n): \int_U\alpha\,\mathbf{d\mu}$, which is a
real number. Thus we have a linear form on the space
$\Lambda^n_c(\mathbb{R}^n)$.

Let $\mathcal{V}=\{V_i\}$ be locally finite open cover of $U\subset\mathbb{R}^n$
and the functions $\theta_i$ represent a partition of unity that is subordinate
to $\mathcal{V}$. Then we have $\int_U\alpha=\sum_i\int_{V_i}\theta_i\alpha$.
If $f$ is a smooth function on $U$ with compact carrier and
$\varphi:V\rightarrow U$ is an orientation preserving diffeomorphism:
$det(d\varphi)>0$, then
$$
\int_U f\,\mathbf{d\mu}=\int_V(f\circ
\varphi)\,det(\varphi)\mathbf{d\mu}, \ \ \text{i.e.}, \ \ \
\int_U\alpha=\int_V\varphi^*\alpha .
$$
Now, since every $n$-dimensional smooth
manifold is locally diffeomorphic to an open set in $\mathbb{R}^n$, we can
carry the integration of a $n$-form $\alpha$ on $M^n$ to integration of the
$n$-form $\varphi^*\alpha$, where $\varphi:U\rightarrow V$ realizes the
corresponding orientation preserving local diffeomorphism
$\varphi:(U\subset\mathbb{R}^n)\rightarrow(V\subset M^n)$. So, by definition,
if $\alpha\in\Lambda^n_c(M^n)$ $$ \int_M\,\alpha=\int_U\,\varphi^*\alpha . $$
This definition is easily made consistent (when needed) with the
case when the carrier of $\alpha$ is covered by a local open covering
$\mathcal{V}$ subordinate to the corresponding coordinate atlas of orientation
preserving charts.

If $M^n$ and $N^n$ are two oriented manifolds and $\varphi:M\rightarrow N$ is
an orientation preserving/reversing diffeomorphism then for
$\alpha\in\Lambda^n_c(N)$ we obtain
$$
\int_N\,\alpha=\int_M\,\varphi^*\alpha \ ,\ \ \ \
\int_N\,\alpha=-\int_M\,\varphi^*\alpha .
$$

If $M$ is an oriented manifold-with-boundary, $\partial M$ is its boundary
canonically imbedded in $M$ by $j:\partial M\rightarrow M$, and $\partial M$ is
endowed with the induced orientation, then for any differential form
$\alpha\in\Lambda^{n-1}_c(M)$ the following important relation (the Stokes
formula) holds:
$$
\int_M\mathbf{d}\alpha=\int_{\partial M}j^*\alpha .
$$
As a consequence from this formula, if $M$ has no boundary, then for every
$\alpha\in\Lambda^{n-1}_c(M)$ we obtain $\int_M\mathbf{d}\alpha=0.$

\section{Lie Groups and Lie group actions on\\ manifolds .}
{\bf 2.10.1. Lie groups.} A Lie group \index{Lie group}
is a set $G$ which carries algebraic and
topological (smooth) structures, which are compatible as follows:

	(i) The group multiplication $\mathfrak{f}: G\times G\rightarrow G$,
written
as $$
(a,b)\rightarrow ab , \ \ a,b\in G;
$$
is a smooth map.

	(ii) The inversion map $\mathfrak{u}: G\rightarrow G$, given by
$$
\mathfrak{u}: a\rightarrow a^{-1} , \ \ a\in G,
$$
is smooth. So, a Lie group is a smooth manifold, such that the above maps are
also smooth.

The unit element of $G$ will be denoted further by $e$.

If $G$ and $H$ are two Lie groups and $\varphi: G\rightarrow H$ is smooth and
satisfies $\varphi(ab)=\varphi(a)\varphi(b)$ then $\varphi$ is called
homomorphism of Lie groups. If $\varphi$ is additionally a diffeomorphism then
it is called isomorphism of Lie groups.

Each $a\in G$ defines smooth maps $L_a:G\rightarrow G$  and
$R_a:G\rightarrow G$ by
$$
L_a(x)=ax, \ \ \ R_a(x)=xa, \ \ x\in G
$$
called {\it left/right translations} by $a\in G$. Thus we have:
$$
L_a\circ L_b=L_{ab}, \ \ R_a\circ R_b=R_{ba}, \ \
L_a\circ R_b=R_b\circ L_a.
$$
So, $L_a$ and $R_b$ are diffeomorphisms with corresponding inverses
$(L_a)^{-1}=L_{a^-1}$ and $(R_b)^{-1}=R_{b^-1}$.

The derivatives $dL_a$ and $dR_b$ of $L_a$ and $R_b$ map $T(G)$ into $T(G)$. We
obtain the relations ($X_a$ ia a tangent vector at $a\in G$):
$$
(dL_a)_b(X_b)=X_{L_a(b)}=X_{ab}, \ \
dL_a\circ dL_b=dL_{ab},
$$
$$
 dR_a\circ dR_b=dR_{ba}, \ \ dL_a\circ
dR_b=dR_b\circ dL_a.
$$

If $\varphi: G\rightarrow H$ is a Lie group homomorphism, then
$$
\varphi\circ L_a=L_{\varphi(a)}\circ\varphi, \ \ \
\varphi\circ R_b=L_{\varphi(b)}\circ\varphi; \
$$
$$
d\varphi\circ dL_a=dL_{\varphi(a)}\circ d\varphi, \ \ \
d\varphi\circ dR_b=dL_{\varphi(b)}\circ d\varphi.
$$
In particular $(d\varphi)_x: T_x(G)\rightarrow T_{\varphi(x)}(H), x\in G$, is
injective/surjective if $(d\varphi)_e$ is injective/surjective.

The derivatives of the two algebraic operations $\mathfrak{f}$ and
$\mathfrak{u}$ are given by
$$
d\mathfrak{f}: T_G\times T_G\rightarrow T_G:
$$
$$
d\mathfrak{f}(X,Y)=dR_b(X)+dL_a(Y), \ X\in T_a(G),
Y\in T_b(G).
$$
$$
d\mathfrak{u}: T_G\rightarrow T_G:
$$
$$
d\mathfrak{u}(X)=-(dL_{a^{-1}})_e\circ (dR_{a^{-1}})_a(X)=
-(dR_{a^{-1}})_e\circ (dL_{a^{-1}})_a(X), \ X\in T_a(G).
$$

{\bf 2.10.2. Vector fields and differential forms.}
 Some of the vector fields on a Lie
group $G$ form the corresponding Lie algebra $\mathfrak{X}(G)$ of vector
fields. The left and right translations by a $a\in G$ induce automorphisms
$(L_a)_*$ and $(R_a)_*$ of $\mathfrak{X}(G)$. A vector field
$X\in\mathfrak{X}(G)$ is called {\it left invariant} if $(L_a)_*X=X$, i.e. if
$(dL_a)_x(X(x))=X(ax)$. All left invariant vector fields on $G$ form a
subalgebra $\mathfrak{X}_L(G)\subset\mathfrak{X}(G)$ since every $(L_a)_*$
preserves the Lie bracket in $\mathfrak{X}(G)$.

Every $a\in G$ defines a linear isomorphism between $T_e(G)$ and $T_a(G)$ by
$(dL_a)_e$. So, every left invariant vector field on $G$ is determined by
unique element of $h\in T_e(G)$. We obtain an isomorphism between the Lie
algebra of left invariant vector fields $\mathfrak{X}_L(G)$ and the tangent
space $T_e(G)$ at the unit element $e\in G$. An isomorphism between
$\mathfrak{X}(G)$ and $\mathfrak{X}_L(G)\otimes\mathcal{J}(G)$ is also
induced.

Each $h\in T_e(G)$ defines unique $X_h\in\mathfrak{X}_L(G)$.

The Lie algebra of $GL(V)$, $V$ is a vector space, is the set of all linear
maps $L_{V}$ in $V$. A left-invariant vector field $X_{\alpha}$ on $GL(V)$,
defined by $\alpha\in L_{V}$, at the point $\sigma\in GL(V)$, is the couple
$(\sigma, \sigma\circ \alpha)$.

 In what follows the Lie algebra $\mathfrak{X}_L(G)$ will be denoted just
by $\mathfrak{g}$ when it is clear the connection with the corresponding Lie
group.

In the same way a Lie subalgebra $\mathfrak{X}_R(G)$ of right-invariant vector
fields on $G$ can be constructed. Now, if $X\in\mathfrak{X}_L(G)$ and
$Y\in\mathfrak{X}_R(G)$ then the corresponding Lie bracket is zero:
$[X,Y]=0$, so, the corresponding flows commute.

Since the inversion map $\mathfrak{u}$ has the property
$\mathfrak{u}\circ\mathfrak{u}=id_G$, it is a
diffeomorphism of $G$, satisfying the following relations:
$$
\mathfrak{u}\circ L_a=R_{a^{-1}}\circ\mathfrak{u}, \ \
d\mathfrak{u}\circ dL_a=dR_{a^{-1}}\circ d\mathfrak{u} ,
$$
and $\mathfrak{u}_{*}$ restricts to isomorphism between $\mathfrak{X}_{L}(G)$ and
$\mathfrak{X}_{R}(G)$ given by $\mathfrak{u}_{*}(X_h)=-Y_h$. Also, for $h,k\in
T_e(G)$ we get $[X_h,X_k](e)= -[Y_h,Y_k](e)$.
\vskip 0.3cm
If $\alpha$ is a
differential form on $G$ then it is called {\it lelt/right invariant} if
$$
L_a^*\alpha=\alpha, \ \ R_a^*\alpha=\alpha, \ \ a\in G .
$$
Let's write this for a 1-form in a more detail:
$$
\langle(L_a^*\alpha)_b,(X_b)\rangle=\langle\alpha_{L_a(b)},(dL_a)_b(X_b)\rangle=
\langle\alpha_{ab},(dL_a)_b(X_b)\rangle=\langle\alpha_b,X_b\rangle.
$$
For $a=b^{-1}$ we obtain
$$
\langle\alpha_b,X_b\rangle=
\langle\alpha_e, (dL_b)^{-1}_b(X_b)\rangle.
$$
Clearly, this relation says that every left-invariant 1-form is completely
determined by its value at the unity $e\in G$. Moreover, since
$(dL_b)^{-1}_b(X_b)\in \mathfrak{g}$, it follows that if $\alpha$ is
left-invariant 1-form and $X$ is left-invariant vector field then
$\langle\alpha,X\rangle=\langle\alpha_e,X_e\rangle=const$. Hence, denoting
the left invariant 1-forms by $\Lambda^1_L(G)$ we obtain the isomorphism of
$\Lambda^1_L(G)$ with $T^*_e(G)=\mathfrak{g^*}$.

Recalling the action of the exterior derivative on 1-forms and in view of the
above observation for the constancy of $\langle\alpha,X\rangle$ for
left-invariant objects we obtain
$$
\mathbf{d}\alpha(X,Y)=-\langle\alpha,[X,Y]\rangle, \ \ \alpha\in\mathfrak{g^*},
 \ \ X,Y\in\mathfrak{g}.
 $$

Let now $G$ be $r$-dimensional, so $\mathfrak{g}$ and
$\mathfrak{g}^*$ are also $r$-dimensional. Let $\{X_1,\dots,X_r\}$
and $\{\omega^1,\dots,\omega^r\}$ be two dual bases of $\mathfrak{g}$ and
$\mathfrak{g^*}$ respectively. We have
$$
[X_i,X_j]=\sum_{i,j=1}^r C_{ij}^k\,X_k,
$$
where $C_{ij}^k=-C_{ji}^k$ are constants, called {\it structure constants} for
$G$. It is easily obtained that
$$
\mathbf{d}\omega^k=-\sum_{i<j=1}^r\,C^k_{ij}\omega^i\wedge\omega^j.
$$

These remarks allow to introduce the so called
$\mathfrak{g}$-valued Maurer-Cartan 1-form $\omega_{MC}$ on $G$. It is defined
by
$$
(\omega_{MC})_a=(dL_{a^{-1}})_a, \ a\in G.
$$
Clearly, $(\omega_{MC})_e=id_{\mathfrak{g}}$, so,
$\omega_{MC}=\omega^i\otimes X_i$, where
$\langle\omega^i,X_j\rangle=\delta^i_j$.
 This 1-form satisfies the following
relations: $$ R_a^*\omega_{MC}=Ad(a^{-1})\circ\omega_{MC}, \ \
\mathbf{d}\omega_{MC}+\frac12[\omega_{MC},\omega_{MC}]=0,
$$
where $[\omega_{MC},\omega_{MC}]=\omega^i\wedge\omega^j\otimes[X_i,X_j], i<j$.

\vskip 0.3cm
{\bf 2.10.3. Representations.} Let $W$ be a finite dimensional vector space and
$GL(W)$ be its linear group of automorphisms. Then $GL(W)$ is a Lie group and
the space $L_W$ of {\bf all} linear transformations of $W$ is the corresponding
Lie algebra. Let $G$ be a Lie group. Then, a homomorphism $$ P: G\rightarrow
GL(W) $$ is called a {\it representation} of $G$ in $W$.

The derivative of $P$ at $e\in
G$: $P':\mathfrak{g}\rightarrow L_W$ is a Lie algebra homomorphism. This Lie
algebra homomorphism $P'$ is called a {\it representation of
$\mathfrak{g}$ in $W$}.

The representation $P$ is called {\it faithful} if $Ker(P')=0\in\mathfrak{g}$.

The subspace $W_I\subset W$ consisting of all elements $x\in W$ that satisfy
$$
P(a)(x)=x, \ a\in G,
$$
is called {\it invariant subspace} of $P$.

Similarly, the invariant subspace $W_o$ for $P'$ is given by
$$
W_o=\{x\in W| P'(h)x=0\}, \ h\in\mathfrak{g}.
$$
These two invariant subspaces satisfy $W_I\subset W_o$, and if $G$ is connected
then $W_I=W_o$.

A representation $P$ of $G$ in $W$ generates representation $P^\natural$ of $G$
in the dual space $W^*$, called {\it contragradient to} $P$, according to
$$
P^\natural(x)=[(P(x))^{-1}]^*, \ x\in G.
$$
Accordingly, a representation $P'^\natural$ of $\mathfrak{g}$ in $W^*$ that is
dual to $P'$ is given by
$$
(P')^{\natural}(h)=-(P(h))^*, \ h\in\mathfrak{g}.
$$

Each $a\in G$ defines inner automorphism $\theta_a$ of $G$ by
$$
\theta_a(x)=a\,x\,a^{-1}, \ x\in G.
$$
Since $\theta_a=L_a\circ R_{a^{-1}}$ the derivative $\theta'_a$ of $\theta_a$ is
given by
$$
\theta'_a=dL_a\circ dR_a^{-1}=dR_a^{-1}\circ dL_a, \ a\in G,
$$
and is denoted by $Ad(a)$, Obviously, $Ad(a)$ is an automorphism of
$\mathfrak{g}$. This representation of $G$ in $\mathfrak{g}$ is called {\it
adjoint representation}. The corresponding coadjoint representation is given by
$[Ad(a^{-1})]^*$.

Now each $h\in\mathfrak{g}$ induces representation $ad$ of the Lie algebra
$\mathfrak{g}$ in the linear space $\mathfrak{g}$ according to
$$
ad(h)(k)=[h,k], \ h,k\in\mathfrak{g} .
$$
It can be shown that $ad$ is the derivative of $Ad$.

Finally we note that all these representations induce representations and
derivations in the tensor, exterior and symmetric algebras over $\mathfrak{g}$
and $\mathfrak{g}^*$, as well as, over the tensor, exterior and symmetric
algebras built on the corresponding vector space $W$ where the representation
is initially defined.
\vskip 0.3 cm

{\bf 2.10.4. Action of a Lie group on a manifold.} \index{action of a Lie
group}

This important subsection will be represented as divided to four parts. As
usual, all manifolds are assumed to be finite dimensional and smooth.
\vskip 0.3cm
{\bf 2.10.4.1 Definition and basic properties.}  Let $M$ be a
$n$-dimensional real manifold and $G$ be a $r$-dimensional Lie group. Then a
right action of $G$ on $M$ is called every smooth map
$$
\Phi: M\times G\rightarrow M, \ \text{denoted by} \ \ (z,a)\rightarrow z.a
$$
satisfying the following condition:
$$
z.(a\,b)=(z.a).b, \ \ \text{and} \ \ z.e=z, \ \ z\in M, \ a,b\in G .
$$
The action is called {\it transitive} if every two points of $M$ can be
transformed to each other by an element of $G$, i.e. if $z_1,z_2\in M$, then
there exists an element $a\in G$ such that $z_1.a=z_2$. Clearly, then
$z_1=z_2.a^{-1}$.

The action $\Phi$ defines two partial maps:

$1^o$. Each (fixed) $a\in G$ determines a
diffeomorphism $\mathcal{R}_a: M\rightarrow M$ according to the action:
$\mathcal{R}_a(z)=\Phi(z,a)=z.a, \ z\in
M$. Clearly, $(\mathcal{R}_a)^{-1}=\mathcal{R}_{a^{-1}}$.

$2^o$. Each (fixed) $z\in M$ determines a smooth map $\phi_z: G\rightarrow M$
given by $\phi_z(a)=\Phi(z,a), \ a\in G$.

Recalling the notations $L_a, \ \ R_a, \ \theta_a(b)=a.b.a^{-1}$ from the
preceding (sub)section we obtain the following relations fulfilled:
$$
\mathcal{R}_b\circ\phi_z=\phi_z\circ R_b, \ \
\phi_{z.b}=\phi_z\circ L_b= \mathcal{R}_b\circ\phi_z\circ\theta_b, \ \
\mathcal{R}_a\circ \phi_z=\phi_{z.a}\circ\theta^{-1}_a.
$$
Let now $\hat{\Phi}: N\times G$ be an action of $G$ on the manifold $N$. Then a
smooth map $\varphi: M\rightarrow N$ is called {\it equivariant} with respect
to $\Phi$ and $\hat{\Phi}$ if
$$
\varphi\circ\Phi(z,a)=\hat{\Phi}\circ(\varphi\times id_G)(z,a), \ \ z\in M, \
a\in G.
$$
The so defined equivariance is equivalent to the following relations:
$$
\varphi(z.a)=\varphi(z).a, \ z\in M, \ a\in G;
$$
$$
\varphi\circ\mathcal{R}_a=\hat{\mathcal{R}_a}\circ \varphi,  \ a\in G ;
$$
$$
\varphi\circ\phi_z=\hat{\phi}_{\varphi(z)}, \ z\in M.
$$

Now, a left action of $G$ on $M$ is called every smooth map
$$
\Phi: G\times M\rightarrow M, \ \text{denoted by} \ \ (a,z)\rightarrow a.z
$$
satisfying the following condition:
$$
(a\,b).z=a.(b.z), \ \ \text{and} \ \ e.z=z, \ \ z\in M, \ a,b\in G .
$$
The corresponding {\it equivariance} condition looks like
$$
\varphi(a.z)=a.\varphi(z), \ \ a\in G, \ z\in M.
$$

Every representation $P$ defines a left action of $G$ in a vector
space $W$ by $a.w=P(a).w, \ a\in G, \ w\in W$.

Every action of $G$ on a manifold $M$ defines an action of $G$ on $T(M)$ by
$X.a=d\mathcal{R}_a(X), \ X\in T(M), \ a\in G$.
For a vector field $X$ and $a\in G$ we obtain
$$
(\mathcal{R}_a)_*X(z)=d\mathcal{R}_a(X(z.a^{-1})).
$$

A subset $S\subset M$ is called {\it stable} with respect to the action $\Phi:
M\times G\rightarrow M$ if $z.a\in S$ for each $a\in G$.
\vskip 0.3 cm
{\bf 2.10.4.2 Orbits of an action.} An {\it isotropy subgroup} $G_z$ of $G$ with
respect to the point $z\in M$ is given by all elements $a\in G$ such that
$z.a=z , \ a\in G$. If for each $z\in M$ the corresponding isotropy subgroup is
reduced to the unit element $e\in G$ then the action is called {\it free}.

The Lie algebra $\mathfrak{g}_z$ of $G_z$ is given by :
$\mathfrak{g}_z=Ker(d\phi_z)_e$.

An {\it orbit} of $G$ with respect to the point $z\in M$ is called the subset
$z.G$, i.e. the images of $z$ when transformed by all elements of $G$. The
orbits through different $z\in M$ are nonintersectable, or coincide with each
other. A transitive action makes the whole $M$ in an orbit. The isotropy group
of the point $z.a$ is equal to $a^{-1}G_z\,a$.
\vskip 0.3 cm {\bf 2.10.4.3 Induced vector
fields.} Consider the map $\phi_z: G\rightarrow M$. Its derivative
$(d\phi_z)_e$ at $e\in G$ maps $\mathfrak{g}$ into $T_z(M)$. So, each
$h\in\mathfrak{g}$ determines a tangent vector $Z_h(z)\in T_z(M)$.

Let's now fix $h\in \mathfrak{g}$ and vary $z\in M$. In this way we obtain a
vector field
$$
Z_h(z)=(d\phi_z)_e(h), \ \ z\in M,
$$
called {\it fundamental vector field generated by $h$}. Clearly, every function
$f: M\rightarrow \mathfrak{g}$ generates vector field $Z_f$ on $M$ according to
$Z_f(z)=Z_{f(z)}$.

Differentiating the formula $\mathcal{R}_a\circ \phi_z=\phi_{z.a}\circ
\theta^{-1}_a$ we obtain
$$
(\mathcal{R}_a)_*(Z_h)(z)=Z_{Ad\,a^{-1}(h)}(z.a), \ h\in\mathfrak{g},
\ a\in G, \ z\in M.
$$
Since the differential of $Ad$ is the Lie bracket in the Lie algebra
$\mathfrak{g}$ of $G$, the obtained map $\mathfrak{g}\rightarrow
\mathfrak{X}(M)$ given by $h\rightarrow Z_h$ is a homomorphism of Lie algebras:
$$
Z_{[h,k]}=[Z_h,Z_k].
$$

A vector field $X$ on $M$ is called invariant with respect to the action $\Phi$
of $G$ on $M$ if $(\mathcal{R}_a)_*X=X, \ a\in G$. The set
$\mathfrak{X}^{I}(M)$ of all invariant vector fields is a subalgebra of
$\mathfrak{X}(M)$. The invariant vector fields are generated by those $h\in
\mathfrak{g}$ satisfying $Ad\,a(h)=h, \ a\in G$. For connected Lie groups this
is equivalent to $[h,k]=0, \ k\in\mathfrak{g}$, i.e. when $h$ is in the center
of $\mathfrak{g}$.

The Lie bracket of a fundamental and invariant vector fields is zero.

When the action of $G$ on $M$ is free, then:

	-$M$ gets a fiber bundle structure with $G$ as a standard fiber.

	-the fundamental vector fields have no zeros, and they are tangent to
the fibers

In this case the correspondence $f\rightarrow Z_f$ defines isomorphism between
the set $\mathcal{J}(M,\mathfrak{g})$ of $\mathfrak{g}$-valued functions and
the sections of this fiber bundle.
\vskip 0.3 cm
{\bf 2.10.4.4  Differential forms under Lie group action.} Since every
$\mathcal{R}_a$ is a diffeomorphism of $M$ then $(\mathcal{R}_a)^*$ is an
automorphism of the graded algebra $\bigwedge(M)$. It follows
$$
(\mathcal{R}_{ab})^*=(\mathcal{R}_a)^*\circ(\mathcal{R}_b)^*, \ \
(\mathcal{R}_e)^*=id_{\bigwedge(M)}, \ \ a,b\in G.
$$
Other important relations are the following:
$$
i_X\circ(\mathcal{R}_a)^*=(\mathcal{R}_a)^*\circ\,i_{(\mathcal{R}_a)_*(X)};
$$
$$
L_X\circ(\mathcal{R}_a)^*=(\mathcal{R}_a)_*\circ L_{(\mathcal{R}_a)_*(X)}; \
$$
$$
(\mathcal{R}_a)^*\circ\,\mathbf{d}=\mathbf{d}\circ(\mathcal{R}_a)^*.
$$

A differential form $\alpha\in\bigwedge(M)$ is called {\it invariant} with
respect to the action of $G$ if for any $a\in G$ we have
$(\mathcal{R}_a)^*\alpha=\alpha$.

The following properties hold:

	- The invariant differential forms form a graded subalgebra \newline
$\bigwedge_I(M)\subset\bigwedge(M)$;

	- The invariant functions $\mathcal{J}_I(M)$ form a subalgebra of
$\mathcal{J}(M)$;

	- The invariant vector fields $\mathfrak{X}(M)$ form a module over
$\mathcal{J}_I(M)$;

	- The subalgebra $\bigwedge_I(M)$ is stable under the exterior
derivative $\mathbf{d}$;

	- If $X\in\mathfrak{X}(M)$ is $G$-invariant then $\bigwedge_I(M)$ is
stable under $i_X$ and $L_X$.

A differential form $\alpha$ is called {\it horizontal} with respect to the
action of $G$ if $i(Z_h)\alpha=0, \ h\in\mathfrak{g}$. The horizontal forms on
$M$ form a graded subalgebra of $\bigwedge(M)$, it is stable under
$L_{Z_h},h\in \mathfrak{g}$,
but it is not stable under the exterior derivative $\mathbf{d}$.

All invariant differential forms satisfy $L_{Z_h}\alpha=0, h\in\mathfrak{g}$,
they form a subalgebra which is stable under the exterior derivative
$\mathbf{d}$.

The intersection of horizontal and invariant differential forms is stable under
$\mathbf{d}$. In general, the set of horizontal forms is a subset of the set of
invariant forms, and on connected manifolds these two sets coincide.

If $G$ acts also on another manifold $N$ and $\varphi: M\rightarrow N$ is
equivariant and smooth map, i.e. $\varphi $ commutes with the two actions:
$\hat{\Phi}\circ\varphi=\varphi\circ\Phi $, then every two fundamental vector
fields are $\varphi $-related. Moreover:
$$
\varphi^*\circ i(Z_h^N)=i(Z_h^M)\circ\varphi^*, \ \
\varphi^*\circ L_{Z_h}^N=L_{Z_h}^M\circ\varphi^*.
$$
Also, since $\mathbf{d}$ commutes with $\varphi^*$, then $\varphi^*$ restricts
to a homomorphism \linebreak $\bigwedge_I(N)\rightarrow\bigwedge_I(M)$.

Let now $P$ be a (linear) representation of $G$ in the vector space $W$.
Consider the $W$-valued differential forms $\bigwedge(M,W)$. Then, if
$\Psi\in\bigwedge(M,W)$, the composition $P(a)\circ\Psi$ is well defined.
We obtain a left action of $G$ in $\bigwedge(M,W)$ according to
$$
a.\Psi=(P(a)\circ\phi_a^*)(\Psi)=(\phi_a^*\otimes P(a))\Psi, \
\Psi\in\bigwedge(M,W), \ a\in G.
$$
Clearly, $\mathbf{d}(a.\Psi)=a.\mathbf{d}\Psi$.

A $W$-valued differential form $\Psi$ on $M$ is called {\it P-equivariant} if
$a.\Psi=\Psi$, which is equivalent to $\phi_a^*(\Psi)=P(a^{-1})\circ\Psi$.

Recalling the induced representation $P'$ of $\mathfrak{g}$ in $W$ we find the
relation
$$
L_{Z_h}\Psi=-P'(h)\circ\Psi, \ \ h\in\mathfrak{g}.
$$
In particular, the adjoint representation gives
$$
\phi_a^*\Psi=(Ad\,a^{-1})\circ\Psi, \ \text{and} \ \
L_{Z_h}\Psi=-ad(h)\circ\Psi , \ h\in\mathfrak{g}.
$$
Also, the corresponding contragradient representation $P^{\natural}$ of $G$ in
$W^*$ gives the left action of $G$ in $\bigwedge(M,W^*)$

Finally, in view of further use (Sec.6.1.1), we'd like to specially note that
every spherically symmetric, i.e. $SO(3)$-invariant, with respect to the origin
of the space $\mathbb{R}^3$ differential 2-form $\Omega$ looks like in
corresponding spherical coordinates $(r,\theta,\varphi)$ as
$\Omega=f(r)\mathrm{sin}\theta\,d\theta\wedge d\varphi$. Therefore, the only
spherically symmetric representative of the cohomological class of the space
$\Sigma=\mathbb{R}^3\setminus\{0\}$ looks as
$\Omega_0=Const.\,\mathrm{sin}\theta\,d\theta\wedge d\varphi$.

\chapter{Integrability, Curvature, Connections}
\section{Distributions on manifolds. Morphisms and Symmetries}
{\bf 3.1.1 Integrability conditions.} \index{integrability}
The problem for integration of a system of partial differential equations of
the kind
$$
\frac{\partial y^a}{\partial x^i}=f^a_i(x^k,y^b),       %64%
\ i,k=1,...,p\,;\ a,b=1,...,q,
$$
where $f^a_i(x^k,y^b)$ are given functions, obeying some definite smoothness
conditions, has contributed to the formulation of a number of concepts,
which in turn have become generators of ideas and research directions, and
most of them have shown an wide applicability in many branches of
mathematics and mathematical physics.  A particular case of the above system
(nonlinear in general) of equations is when there is only one independent
variable, i.e.  when all $x^i$ are reduced to $x^1$, which is usually denoted
by $t$ and the system acquires the form
$$
\frac{dy^a}{dt}=f^a (y^b(t),t),\ a,b=1,...,q.                 %65%
$$

We recall now some of the concepts used in considering the integrability
problems for these equations, making use of the geometric language of
manifold theory. Let $X$ be a vector field on the $q$-dimensional manifold $M$
and the map $c:I\rightarrow M$, where $I=(t_o,t_1)$ is an open interval in
${\mathbb{R}}$, defines a smooth curve in $M$. Then if $X^a, a=1,2,\dots,q$,
are the components of $X$ with respect to the local coordinates $(y^1, ...,
y^q)$ and the equality $c'(t)=X(c(t))$ holds for every $t\in I$, or in local
coordinates,
$$
\frac{dy^a}{dt}=X^a (y^b(t)),
$$
$c(t)$ is an {\it integral
curve} of the vector field $X$ through the point $c(t_o)$, i.e. $c(t)$ defines
a 1-dimensional manifold such that $X$ is tangent to it at every point $c(t)$.

As it is seen, the difference between the above two systems of ODE is in the
additional dependence of the right side of first one on the independent variable
$t$.  Mathematics approaches such situations in an unified way as follows.
The product ${\mathbb{R}}\times M$ is considered and the important theorem for
uniqueness and existence of a solution is proved: For every point $p\in M$
and point $\tau \in {\mathbb{R}}$ there exist a vicinity $U$ of $p$, a positive
number $\varepsilon $ and a smooth map $\Phi:(\tau-\varepsilon,\tau
+\varepsilon)\times U\rightarrow M$, $\Phi:(t,y)\rightarrow \varphi_t (y)$,
such that for every point $y\in U$ the following conditions are met:
$\varphi_\tau (y)=y,\ t\rightarrow \varphi_t (y)$ is an integral curve of
$X$, passing through the point $y\in M$; besides, if two such integral curves
of $X$ have at least one common point, they coincide. Moreover, if $(t',y),\
(t+t',y)$ and $(t,\varphi(y))$  are points of a vicinity $U'$ of $\{0\}\times
{\mathbb{R}}$ in ${\mathbb{R}}\times M$, we have $\varphi_{t+t'}(y)=\varphi_t
(\varphi_{t'}(y))$. This last relation gives the local group action: for
every $t\in I$ we have the local diffeomorphism $\varphi_t :U\rightarrow
\varphi_t (U)$. So, through every point of $M$ there passes only one
trajectory of $X$ and in this way the manifold $M$ is foliated to
non-crossing trajectories - 1-dimensional manifolds, and these 1-dimensional
manifolds define all trajectories of the defined by the vector field $X$
system of ODE. This fibering of $M$ to nonintersecting submanifolds, the
union of which gives the whole manifold $M$, together with considering $t$
and $y(t)$ as 1-d submanifolds of the same manifold, is the leading idea in
treating systems of partial differential equations, where the number
of the independent variables is more than 1, but finite. For example, if we
consider two linearly independent vector fields on $M$, then through every point
of $M$ two trajectories will pass and the question: when a 2-dimensional
surface, passing through a given point can be built, and such that the
representatives of the two vector fields at every point of this 2-surface to be
tangent to the surface, naturally arises. The answer to this problem in the
case of more than one independent variables is given by corresponding
integrability conditions.

For simplicity, further we consider regions of the space
${\mathbb{R}}^p \times {\mathbb{R}}^q$, but this is not essentially important since
the integrability conditions are local statements, so the results will hold for
any $(p+q)$-dimensional manifold.

Let $U$ be a region in ${\mathbb{R}}^p\times{\mathbb{R}}^q$, and
$(x^1,...,x^p,y^1=x^{p+1},...,y^q =x^{p+q})$ are the canonical coordinates.
We set the question: for which points $(x_o,y_o)$ of $U$ the above written
system of
equations  has a solution $y^a=\varphi ^a(x^i)$, defined for points $x$,
sufficiently close to $x_o$ and satisfying the initial condition
$\varphi (x_o)=y_o$? The answer to this question is: for this to happen it is
necessary and sufficient the functions $f_i^a$ on the right hand side
to satisfy the following conditions:
$$
\frac{\partial f^a_i}{\partial x^j}(x,y)+
\frac{\partial f^a_i}{\partial y^b}(x,y).f^b_j(x,y)=
\frac{\partial f^a_j}{\partial x^i}(x,y)+                           %(66)%
\frac{\partial f^a_j}{\partial y^b}(x,y).f^b_i(x,y).
$$
This relation is obtained as a consequence of two basic steps: first,
equalizing the mixed partial derivatives of $y^a$ with respect to $x^i$ and
$x^j$, second, replacing the obtained first derivatives of $y^a$ with
respect to $x^i$ on the right hand side of the system again from the system.
If the functions $f^a_i$ satisfy the above equations  for each point of the
region $U$ the system is called {\it completely integrable} on $U$. In order to
give a coordinate free formulation of the situation mathematics comes to the
concept of distribution.
\vskip 0.3cm
{\bf 3.1.2. Distributions and co-distributions.} \index{distributions,
co-distributions} Let $M$ be an arbitrary $n=p+q$ dimensional manifold. At
every point $x\in M$ the tangent space $T_x(M)$ is defined. The union of all
these spaces with respect to the points of $M$ defines the {\it tangent
bundle}. On the other hand, the union of the co-tangent spaces $T^*_x(M)$
defines the {\it co-tangent bundle}. At every point now of $M$ we separate a
$p$ dimensional subspace $\Delta_x(M)$ of $T_x(M)$ in a smooth way, i.e. the
map $x\rightarrow \Delta_x$ ix smooth. If this is done we say that a
$p$-dimensional {\it distribution} $\Delta$ on $M$ is defined. Clearly, a
distribution defines a subbundle of the tangent bundle of $M$, and the sections
of this subbundle define a module with respect to the algebra $\mathcal{J}(M)$.

From the elementary linear algebra we know that every $p$-dimensional subspace
$\Delta_x$ of $T_x(M)$ defines unique $(n-p)=q$ dimensional subspace
$\tilde{\Delta}_x$ of the dual to $T_x(M)$ space $T^*_x(M)$, such that all
elements of $\tilde{\Delta}_x$ annihilate (i.e. send to zero) all elements of
$\Delta_x$. In this way we get a $\Delta$-insensitive $q$-dimensional {\it
co-distribution} $\tilde{\Delta}_x$ on $M$. We consider those vector fields,
the representatives of which at every point are elements of the distribution
$\Delta$, and those 1-forms, the representatives of which at every point are
elements of the co-distribution $\tilde{\Delta}_x$. We note that, every system
of $p$ independent and non-vanishing vector fields, belonging to $\Delta$,
may define $\Delta$ equally well, and in this case we call such a system a {\it
differential $p$-system} ${\cal P}$ on $M$.  The corresponding system
$\tilde{\cal P}$ of $q$ independent 1-forms is called $q$-dimensional {\it
Pfaff system}. Clearly, if $\alpha \in \tilde{\cal P}$ and $X\in {\cal P}$,
then $\langle\alpha,X\rangle=0$.

This allows to look at distributions as represented by a {\it nonvanishing
decomposable} $p$-vector(s), or by a {\it nonvanishing} $(n-p)$-differential
form(s). If $\mathcal{P}\Leftrightarrow\{X_1,X_2,\dots,X_p\}$ then the
$p$-vector field $\mathbb{P}=X_1\wedge X_2\wedge\dots\wedge X_p\neq 0$ defines
the distribution since at every point the representatives of $X_i, i=1,...,p$
define the corresponding subspace $\Delta_x\subset T_x(M)$. An appropriate
decomposable nonvanishing and not $\mathcal{P}$-attractive $(n-p)$-form
$\Omega=\alpha^{p+1}\wedge\alpha^{p+2}\wedge\dots\wedge\alpha^{n}$, i.e. such
that $i(\mathcal{P})\Omega=0$, also defines $\Delta_x(M)$ through its
restriction to $x\in M$.

A derivative of a distribution defined by the vector fields
$(X_1,X_2,\dots,X_p)$ is a new distribution $\mathcal{P}'$ defined by the given
$X_i, i=1,2,\dots,p$ plus all Lie brackets $[X_i,X_j], i<j=1,2,\dots,p$. In the
same way higher derivatives of a given distribution can be defined. The
corresponding SN-bracket $[\Delta, \Delta']$ presents how $\Delta$ changes
along $\Delta'$.

It deserves noting, that the above definition of a distribution $\Delta$ on a
manifold $M$ allows definite freedom in choosing appropriate local bases of
$\Delta$, but, on the other hand, it requires basis independence of all
essential statesments concerning $\Delta$.
\vskip 0.3cm
{\bf 3.1.3. Morphisms of distributions.} Let
now $M$ and $N$ be two manifolds and $\Delta(M)$ and $\Delta(N)$ be two
distributions on $M$ and $N$ respectively. Let $\varphi: M\rightarrow N$ be a
smooth map. If $X_x\in\Delta_x(M)$, we consider its image $(d\varphi)_x(X_x)\in
T_{\varphi(x)}(N)$. If for every $x\in M$ every image of elements of
$\Delta_x(M)$ is in $\Delta_{\varphi(x)}(N)$ we say that the coupe
$(\varphi,d\varphi)$ realize a morphism $\Delta(M)\rightarrow\Delta(N)$.

If $\varphi: M\rightarrow N$ is a diffeomorphism and $\Delta(M)$ and
$\Delta(N)$ have the same dimension then the image $Im(\Delta(M))=
(\varphi,d\varphi)|_{\Delta(M)}$ of $\Delta(M)$ is a well defined distribution
on $N$.

Correspondingly, every diffeomorphism $\varphi: M\rightarrow M$ sends a
distribution on $M$ into another (in general) distribution on $M$.

\vskip 0.3cm
\section{Integral manifolds, symmetries and\\ curvature of distributions}
{\bf 3.2.1. Integral manifolds.} \index{integrability of distributions}
The concept of {\it integral manifold} for a $p$-dimen\-sional distribution, or
differential system, is introduced as follows. Namely, we call an {\it integral
manifold} through the point $x\in M$ for the $p$-dimensional differential
system ${\cal P}$, or for the $p$-dimensional distribution $\Delta^p(M)$, to
which ${\cal P}$ belongs, any $(q\leq p)$-dimensional submanifold $V^q$ of $M$
if the tangent spaces at every point of $V^q$ are subspaces of the same
dimension of the corresponding subspaces of the distribution $\Delta^p(M)$ at
this point. If $\mathcal{P}$ admits at least one integral manifold of dimension
$r: 1<r\leq p$, then it is called {\it integrable}.

An integral manifold of $\Delta^p(M)$ is called {\it maximal}, if its
dimension $q$ satisfies $q\leq p$ and there are NO other integral manifolds of
dimension $r>q$.

 If through every point of $M$ there passes an
integral manifold for ${\cal P}$ of dimension equal to $dim\,\Delta^p(M)$, then
${\cal P}$ is called {\it completely integrable}. In this
case the various integral manifolds do NOT intersect and we say that the
manifold $M$ foliates to corresponding to $\Delta^p(M)$ lists.

A smooth function $f\in \mathcal{J}(M)$ is called {\it first integral} for
$\mathcal{P}$ if $df\in\tilde{\cal P}$.

It can be shown that $\mathcal{P}$ is completely integrable if and only if it
has $n-p=dim(\tilde{\cal P})$ functionally independent first integrals,
i.e. locally $\tilde{\cal P}(df_1,df_2,\dots,df_{n-p})$.
\vskip 0.2cm
\noindent
{\bf Remark}: Our further considerations will be connected mainly with
completely integrable distributions, unless the oposite is specially mentioned.
\vskip 0.3cm

{\bf 3.2.2. Symmetries of distributions.} A diffeomorphism $\varphi:
M\rightarrow M$ is called a {\it symmetry} of the distribution $\Delta$ on $M$
if \index{symmetries of distributions}
$$
(d\varphi)_x(\Delta_x)=\Delta_{\varphi(x)}, \ x\in M.
$$
If $\Delta$ is defined by the linearly independent 1-forms
$\{\alpha^1,\alpha^2,\dots,\alpha^{n-p}\}$ then we obtain the transformed
by $\varphi$ 1-forms
$\{\varphi^*\alpha^1,\varphi^*\alpha^2,\dots,\varphi^*\alpha^{n-p}\}$, which
are also linearly independent, so, we have the relations
$$
\varphi^*(\alpha^m)=A^m_n\alpha^n,
$$
where the matrix $A^m_n$ is non-degenerate at every point $x\in M$. These last
relations may be written as follows:
$$
\varphi^*(\alpha^m)\wedge\alpha^1\wedge\alpha^2\wedge\dots\wedge\alpha^{n-p}=0
, \ \ m=1,2,...,(n-p). $$
i.e. without making use of the matrix $A^m_n$.

A vector field $X\in\mathfrak{X}(M)$ is called {\it infinitesimal}
symmetry of the distribution $\Delta$ if the corresponding flow
$\varphi_t$ is a symmetry of $\Delta$. This is equivalent to say that
$$
L_X(\mathcal{P})\subset\mathcal{P}, \ \ \text{or}, \ \
L_X(\tilde{\cal P})\subset\tilde{\cal P}.
$$
In other words, the Lie derivative $L_XZ$ of every $Z\in\mathcal{P}$ is again
in $\mathcal{P}$ and the Lie derivative $L_X\alpha$ of every
$\alpha\in\tilde{\cal P}$ is again in $\tilde{\cal P}$, so, $\mathcal{P}$ and
$\tilde{\cal P}$ are locally $X$-attractive, or $X$ is
$(\mathcal{P},\tilde{\cal P})$-sensitive.

Also, a $q-$vector $\Psi=Y_1\wedge ... \wedge Y_q$ is a local symmetry of
$\mathcal{P}$ if every linear combination over $(Y_1,\dots,Y_q)$ is a
symmetry of $\mathcal{P}$. If $\Psi=Y_1\wedge ... \wedge Y_q$ represents
distribution $\Delta_1$, then $\Delta_1$ is a local symmetry of $\Delta$ if
every linear combination $Z=\Sigma_{i=1}^p f^iY_i$ is a local symmetry of
$\mathcal{P}$. In a similar way these definitions are extended for
codistributions.

Clearly, the set of infinitesimal symmetries of the distribution $\Delta$ is a
Lie algebra (over $\mathbb{R}$), i.e.  if $X,Y$ are infinitesimal symmetries
of $\Delta$, then $X+Y$, $\lambda\,X$, $\lambda\,Y$, $\lambda\in \mathbb{R}$,
and $[X,Y]$ are also infinitesimal symmetries of $\Delta$.

There are two naturally identified subsets of all (i.e. infinitesimal)
symmetries of $\Delta$. The first subset, denoted by $Char(\mathcal{P})$,
includes those vector fields which live in $\mathcal{P}$, and the second subset
includes all the rest. Since $X\in Char(\mathcal{P})$ can be represented as
a linear combination of elements of $\mathcal{P}$ it is tangent to every
integral manifold of $\Delta$. From the general relation $[L_X,i_Y]=i_{[X,Y]}$
it follows that the set $Char(\mathcal{P})$ is an ideal of the Lie algebra of
all local symmetries of $\Delta$. Also, the set $Char(\mathcal{P})$ is a module
over $\mathcal{J}(M)$, and if $\Delta$ is 1-dimensional then
$Char(\Delta)=\Delta$.

The other subset of local symmetries, denoted by $Shuf(\mathcal{P})$, live
entirely outside $\mathcal{P}$ and are called {\it shuffling} symmetries of
$\Delta$, the corresponding flows transform an integral manifold of $\Delta$ to
another integral manifold of $\Delta$, i.e. they shuffle the lists of the
corresponding foliation. The vector fields that represent these local
symmetries have constant coefficients along every list. For example, on
principle bundles, the fundamental, i.e vertical, vector fields are shuffling
symmetries of any horizontal distribution, and the basic/projectable vector
fields are shuffling symmetries of the vertical distribution. Another example,
consider the 2-manifold $(\mathbb{R}^2-\{0\})$ with standard coordinates
$(x,y)$ and define a distribution by the 1-form $\alpha=xdx+ydy$, which can
also be defined by the vector field $X=-y\partial_x+x\partial_y$. The integral
manifolds are the circles $x^2+y^2=const$. The vector field
$Z=x\partial_x+y\partial_y$  is a shuffling symmetry since $X\wedge Z\neq 0$,
$\langle\alpha,Z\rangle\neq 0$ and $[X,Z]=0$.

\vskip 0.3cm
{\bf 3.2.3. Curvature of distributions.}  \index{curvature of distributions}
The concept of {\it curvature} of a
distribution $\Delta^p(M)$ on a manifold $M$ is a local measure of the
integrability properties of $\Delta^p(M)$, i.e. when $\Delta^p(M)$ admits a
$p-$dimensional integral manifold. This problem is solved by the following
theorem of Frobenius, which can be formulated as follows:

{\bf Frobenius theorem}:  \index{Frobenius theorem}
A distribution $\Delta^p(M)$ is completely integrable iff every Lie bracket
$[X,Y]$ of two vector fields in $\Delta^p(M)$ stays in $\Delta^p(M)$.

If $\{X_i, i=1,...,p\}$ are constituents of $\Delta^p(M)$ then this theorem
says that $\Delta^p(M)$ is integrable only if all Lie brackets $[X_i,X_j],
i<j=1,...,p$ can be represented as $\mathcal{J}(M)$-linear combinations of
$X_i, i=1,...,p$ :
$$
[X_i,X_j]=f_{ij}^kX_k, \ \ i<j,k=1,...,p, \ \ f_{ij}^k\in\mathcal{J}(M).
$$
Therefore, if at least one of the following exterior products
$$
[X_i,X_j]\wedge X_1\wedge X_2\wedge\dots\wedge X_p, \ i,j=1,...,p
$$
is different from zero, then $\Delta^p(M)$ is {\it not completely integrable}.

Hence, integrability means that the differential system defined by
$\Delta^p(M)$ is, in fact, an algebra with respect to the Lie bracket, and
nonintegrability means that the differential system defined by $\Delta^p(M)$ is
just a $\mathcal{J}(M)$-module, and is not an algebra with respect to the Lie
bracket.

The following idea comes now to mind from this view on
integrability/nonin\-tegrability of distributions. For a mathematical model of a
time-stable continuous physical system $\Sigma$ that consists of several
time-recognizable and interacting subsystems $\Sigma_i$ to choose an
appropriate distribution $\Delta$ on a pseudoriemannian manifold, to every
subsystem $\Sigma_i$ to juxtapose an appropriate subdistribution
$\Delta_i\subset\Delta$, and to every couple $(\Sigma_i,\Sigma_j,i\neq j)$ of
interacting subsystems to juxtapose an appropriate couple of subdistributions
$(\Delta_i,\Delta_j, i\neq j)$ such, that some Lie brackets of sections of
$\Delta_i$ to live in the module of sections of $\Delta_j$, and, vice versa,
some Lie brackets of sections of $\Delta_j$ to live in the module of
sections of $\Delta_i$. So, the subspace linear structure of $\Delta_i$ and
$\Delta_j$ to be formally responsible for the time-recognizability of
$\Sigma_i$ and $\Sigma_j$, and their available Lie bracket intercommunication
to be formally responsible for the available local physical interaction between
$\Sigma_i$ and $\Sigma_j$.

\vskip 0.4cm
{\bf The following important result holds}:
\vskip 0.2cm
If  $\Delta^p(M)$ is completely
integrable and $X\in Shuf(\Delta^p(M))$ is nowhere zero, then the distribution
$\Delta^p(M)\oplus X$ is also completely integrable. This observation we
consider as a suggestive one for theoretical physics in the following sense.
%\vskip 0.4cm

Let the distribution $\Delta^p(M)$ be completely integrable.
Clearly, since the elements of $Shuf(\Delta^p(M))$ generate flows that
transform an integral manifold of $\Delta^p(M)$ to another integral manifold of
$\Delta^p(M)$, in theoretical physics, if the distribution $\Delta^p(M)$
is meant to represent a propagating in space spatially finite and time-stable
physical object, then  every element $\zeta\in Shuf(\Delta^p(M))$ with
unremovable time component (in Minkowski space-time these are the time-like and
the isotropic vector fields) is appropriate to define admissible dynamical
behaviour, i.e. propagation, of the physical object represented by
$\Delta^p(M)$.

%\vskip 0.4cm
Let's see now how the above integrability criterion looks in terms of the
co-distribution $\Delta_{n-p}^*$. Recall that if
$\{\alpha^{p+1},\alpha^{p+2},\dots,\alpha^{n}\}$ are constituents of
$\Delta_{n-p}^*$, then
$$
\langle\alpha^m,X_i\rangle=0, \ \ m=(p+1),...,n, \
i=1,2,...,p.
$$

{\bf Remark}. Note that the linear maps defined by all $\alpha^m\otimes X_i$,
are boundary maps: $(\alpha^m\otimes X_i)\circ(\alpha^m\otimes X_i)=0$.
\vskip 0.3cm
Further, since all $\alpha^m$ are 1-forms, then we obtain
$$
\mathbf{d}\alpha^m(X_i,X_j)=-\alpha^m([X_i,X_j]), \ m=(p+1),...,n.
$$
Thus, if $\Delta^p(M)$ is completely
integrable then the restriction of every $\mathbf{d}\alpha^m$ to $\Delta^p(M)$
has to be zero, i.e., all $\mathbf{d}\alpha^m$ will not be {\it
attractive} for products $Y\wedge Z$, where $Y,Z\in\Delta^p(M)$. In terms of
the constituents of $\Delta_{n-p}^*(M)$ the complete integrability of
$\Delta^p(M)$ reads
 $$
\mathbf{d}\alpha^m\wedge\alpha^{p+1}\wedge\alpha^{p+2}\dots\wedge\alpha^{n}=0,
 \ \ m=(p+1),...,n.
$$

Recalling relations in Sec.1.4.2, on every coordinate chart subset $U\subset M$
we can introduce transversal to $\Delta^p(U)$ distribution $\Delta^{n-p}(U)$,
i.e. linearly independent and nonvanishing vector fields $\{X_{p+1}, X_{p+2},...
X_n\}$, and transversal to $\Delta^*_{n-p}(U)$ codistribution $\Delta_p^*(U)$,
i.e. linearly independent and nonvanishing 1-forms
$\{\alpha^1,\alpha^2,...,\alpha^p\}$ such that the following direct sum
representations and duality relations to hold at every $x\in U$:
 $$
T_x(M)=\Delta^p_x(M)\oplus\Delta^{n-p}_x(M), \ \
\ T^*_x(M)=(\Delta^*_{p})_x(M)\oplus(\Delta^*_{n-p})_x(M),
$$
$$
\langle\alpha^i,X_j\rangle=\delta^i_j,
\ \ \langle\alpha^m,X_s\rangle=\delta^m_s,
$$
$$
\langle\alpha^i,X_m\rangle=\langle\alpha^m,X_j\rangle=0,\ \
i,j=1,2,...,p\,; \ \ m,s=p+1,...,n.
$$
The nonintegrability of $\Delta^p(U)$ requires NONexistence of functions
$f_{ij}^k$ such that
$$
[X_i,X_j]-f_{ij}^k(x)X_k=0, \ \ x\in M, \ \
i<j,k=1,...,p \ ,
$$
or, equivalently, NONexistence of functions $f^m_{sl}$, such that
$$
 \mathbf{d}\alpha^m+f^m_{sl}(x)\alpha^s\wedge\alpha^l=0, \ \ x\in
(M), \ \ m,(s<l)=p+1,\dots,n.
$$

As an example, consider the manifold $\mathbb{R}^3$ (with the corresponding
identification of forms and vector fields by the euclidean metric $g$) and a
distribution of codimension 1 defined by the 1-form $\alpha$. The integrability
condition looks like $\mathbf{d}\alpha\wedge\alpha=0$. In terms of vector
analysis this condition is equivalent to $\vec{A}.curl\vec{A}=0,
\alpha=\tilde{g}(\vec{A})$. Hence, the nonintegrability of the distribution
defined by $\alpha$ requires that $\vec{A}$ to be
non-orthogonal to $curl\vec{A}$, we recall that in hydrodynamics
$\vec{A}.curl\vec{A}$ is called {\it local helicity} \index{helicity} of the
vector field $\vec{A}$.

Let $\Delta^{n-1}(M)$ be a completely integrable distribution on $M^n$. Then
we have the corresponding 1-dimensional completely
integrable codistribution, so, there exists nonvanishing 1-form $\omega$ on
$M^n$ satisfying $\mathbf{d}\omega\wedge\omega=0$. The corresponding Pfaff
system is defined up to a nonvanishing function: $f\omega, f(x)\neq 0, x\in M$.
Obviously, $f\omega$ also satisfies $\mathbf{d}(f\omega)\wedge f\omega=0$.
From this last equation it follows that there is 1-form $\theta$ such, that
$\mathbf{d}\omega=\theta\wedge\omega$. Now, the Godbillon-Vey theorem says that
the 3-form $\Gamma=\mathbf{d}\theta\wedge\theta$ is closed:
$$
\mathbf{d}\Gamma=\mathbf{d}(\mathbf{d}\theta\wedge\theta)=0.
$$
Moreover, varying $\theta$ and $\omega$ in an admissible way:
$$\theta\rightarrow
(\theta+g\omega); \ \ \omega\rightarrow f\omega, \ g\in \mathcal{J}(M),
$$
leads to adding an exact 3-form to $\Gamma$, so we have a cohomological class
$\Gamma$ defined entirely by the integrable 1-dimensional Pfaff system. From
physical point of view this could happen to be very important, because finding
appropriate completely integrable 1-dimensional Pfaff system on Minkowski
space-time  will give a conservation law when the restriction of $\Gamma$ to
$\mathbb{R}^3$ is not zero, which conservation law will depend entirely on the
integrability properties of the physical system considered.

In order to come to the explicit expressins for the curvature and cocurvature
forms we note that $\alpha^i\otimes X_i$ and $\alpha^m\otimes X_m$ are two
projections in $T(M)$ such, that the unit tensor $\mathfrak{t}_o$ on $M$ is
given by
$$
\mathbf{t}_o=\sum_{j=1}^n(\alpha^j\otimes X_j)=
\sum_{i=1}^p(\alpha^i\otimes X_i)+\sum_{m=p+1}^n(\alpha^m\otimes X_m).
$$
Now, since the coupling between two vector fields is given by the Lie bracket
$[\,, ]$ and in view of the relations
$$
i_{[X_i,X_j]}(\alpha^m\otimes X_m)=\alpha^m([X_i,X_j])X_m
$$
$$=
-\mathbf{d}\alpha^m(X_i,X_j)X_m=-i_{X_i\wedge X_j}(\mathbf{d}\alpha^m\otimes
X_m)
$$
it is natural to define the objects
$$
\Omega=-\mathbf{d}\alpha^m\otimes X_m, \ m=p+1,...,n
$$
and
$$
\tilde{\Omega}=-\mathbf{d}\alpha^i\otimes X_i, \ i=1,2,...,p . $$
The restriction $\Omega|_{\Delta^p(U)}$
of $\Omega$ to $\Delta^p(U)$ will be called {\it curvature
form} for $\Delta^p(U)$, and the restriction
$\tilde{\Omega}|_{\tilde{\Delta}^{n-p}(U)}$ of $\tilde{\Omega}$ to
$\tilde{\Delta}^{n-p}(U)$ will be called {\it co-curvature form} for
$\Delta^p(U)$, or just {\it curvature form} for $\tilde{\Delta}^{n-p}(U)$
\index{curvature form}.

Clearly, $\Omega$ selects those Lie brackets in $\mathcal{P}(U)$ which "stick
out" of $\Delta^p(U)$, and so, having NON-zero projections in
$\tilde{\Delta}^{n-p}(U)$, and $\tilde{\Omega}$ selects those Lie brackets in
$\tilde{\mathcal{P}}(U)$ which "stick out" of $\tilde{\Delta}^{n-p}(U)$ and
having NON-zero projections in $\Delta^p(U)$. So, nonintegrability of
$\Delta^p(U)$ means $\Omega$ is attractive for some elements of
$[\mathcal{P}(U),\mathcal{P}(U)]$, and nonintegrability of $\Delta^{n-p}(U)$
means $\tilde{\Omega}$ is attractive for some elements of
$[\tilde{\mathcal{P}}(U),\tilde{\mathcal{P}}(U)]$.

It is important to note that a completely integrable distribution $\Delta^p$ on
$M$ may contain many non-integrable subdistributions $\Delta_1^{p_1},
\Delta_2^{p_2}, ... \ , \ p_1,p_2,...<p$. Clearly, the corresponding curvature
forms $\Omega^{p_1}_1, \Omega^{p_2}_2, ...$ of these subdistributions do NOT
take values outside $\Delta^p$, but, for example, $\Omega^{p_i}_i$ may take
values in $\Delta_j^{p_j}, i\neq j$, now $\Omega^{p_j}_j$ may take values in
$\Delta_k^{p_k}$, so, some of the values of $\Omega^{p_i}_i$ may be
retransferred to $\Delta_i^{p_i}$, and so on. So, generally speaking, the
initial completely integrable distribution may consist of many nonintegrable and
intercommunicating by their curvature forms, subdistributions, which deserve to
be called {\it interacting partners} \index{interacting partners}.

Such a picture of available "intercommunication" between subdistributions of a
higher dimensional {\it completely integrable} distribution by means of their
curvature forms suggests the idea to try this geometrical "intercommunication"
as an appropriate mathematical "picture" of local physical interaction among
physical systems. Hence, if a time-stable continuous physical system having
dynamical structure may be mathematically represented by an integrable
distribution $\Delta$, and if it is built of relatively time-stable and
continuously recognizable subsystems, existing through some permanent
energy-momentum inter-exchange so that these subsystems are representable by
corresponding nonintegrable subdistributions of $\Delta$, then the
corresponding curvature forms may be interpreted as "internal interacting
agents". The nonzero flows of the values of these internal interacting agents
through the volume forms of the corresponding co-subdistributions appear as
natural formal measures of the local energy-momentum exchanges.

Following the above direct sum representation  we are going to give the
corresponding formal expressions in the simple case of two interacting
distributions. We have the two sets of nonvanishing vector fields
$\{X_i, i=1,2,...,p\}$ and
$\{X_m, m=p+1, ...,n\}$, as well as two codistributions defined by the two
corresponding sets of nonvanishing
1-forms $\{\alpha^i, i=1,2,...,p\}$ and $\{\alpha^m,
m=p+1, ...,n\}:
%\langle\alpha^i,X_j\rangle=\delta^i_j, \langle\alpha^m,X_s\rangle=\delta^m_s,
\langle\alpha^i,X_s\rangle=0, \langle\alpha^m,X_j\rangle=0$, So, we have the two
representing multivectors $$ \mathcal{P}_{(1,p)}=X_1\wedge X_2\wedge ... \wedge
X_p, \ \ \mathcal{P}_{(p+1,n)}=X_{p+1}\wedge X_{p+2}\wedge ... \wedge X_n, $$
and the two differential forms (considered here as corresponding volume forms)
$$ \omega_{(1,p)}=\alpha^1\wedge\alpha^2\wedge ... \wedge\alpha^p, \ \
\omega_{(p+1,n)}=\alpha^{p+1}\wedge\alpha^{p+2}\wedge ... \wedge\alpha^n .
$$
Differentiating $\omega_{1,p}$ and $\omega_{p+1,n}$, we obtain
$$
\mathbf{d}\omega_{1,p}=
\sum_{k=1}^{p}(-1)^{k-1}\mathbf{d}\alpha^k\wedge\alpha^1\wedge
... \wedge\hat{\alpha}^k\wedge... \wedge\alpha^p,
$$
$$
\mathbf{d}\omega_{p+1,n}=
\sum_{k=1}^{n-p}(-1)^{k-1}\mathbf{d}\alpha^{p+k}\wedge\alpha^{p+1}\wedge
... \wedge\hat{\alpha}^{p+k}\wedge... \wedge\alpha^n,
$$
where the hat means "omission" as usually. Now the curvature of
$\Delta^{(1,p)}$ is measured by the values of $(-1)\mathbf{d}\alpha^m\otimes
X_m$ on the 2-dimensional subdistributions $\{X_i,X_j\}$, and the curvature of
$\Delta^{(p+1,n)}$ is measured by the values of $(-1)\mathbf{d}\alpha^i\otimes
X_i$ on the 2-dimensional subdistributions $\{X_m,X_s\}$. So, for each couple
$(i\neq j)$ and $m\neq s$ the flows of the vector fields
$$
\Omega(X_i,X_j)=-\mathbf{d}\alpha^m(X_i,X_j)X_m=\alpha^m([X_i,X_j])X_m,
$$
$$
\tilde{\Omega}(X_m,X_s)=-\mathbf{d}\alpha^i(X_m,X_s)X_i=\alpha^i([X_m,X_s])X_i
$$
across the corresponding volume forms $\omega_{p+1,n}$ and $\omega_{1,p}$,
respectively, are
$$
\mathbb{D}_{(1,p)}^{(p+1,n)}=
\sum_{k=1}^{n-p}(-1)^{k}\alpha^{p+k}([X_i,X_j])\alpha^{p+1}\wedge
... \wedge\hat{\alpha}^{p+k}\wedge... \wedge\alpha^n,
$$
$$
\mathbb{D}_{(p+1,n)}^{(1,p)}=
\sum_{k=1}^{p}(-1)^{k}\alpha^{k}([X_m,X_s])\alpha^{1}\wedge
... \wedge\hat{\alpha}^{k}\wedge... \wedge\alpha^p .
$$
We call them {\it Curvature Interaction (CI) operators}
\index{CI-operators}. Note that the NONzero values of these operators
guarantee the correspnding nonintegrabilities.

 Summing up on all $(i,j), i<j$,
and on all $(m,s), m<s$, we obtain formal expressions of the {\it total local
flow} of the corresponding quantity that the distribution $\Delta^{(1,p)}$
transfers to $\Delta^{(p+1,n)}$, and that the distribution $\Delta^{(p+1,n)}$
transfers to $\Delta^{(1,p)}$, respectively, by means of their curvature forms.
These operators suggest also to say that nonintegrability of $\Delta^{(1,p)}$
means that $\mathbf{d}\omega_{p+1,n}$ is {\it attractive} for (at least some
of) the 2-dimensional subdistributions of $\Delta^{(1,p)}$, and
nonintegrability of $\Delta^{(p+1,n)}$ means that $\mathbf{d}\omega_{(1,p)}$ is
{\it attractive} for (at least some of) the 2-dimensional subdistributions of
$\Delta^{(p+1,n)}$. On the other hand, the two expressions, which we call {\it
internal balance operators} \index{internal balance operator}, generated by
$(X_i\wedge X_j)$ and $(X_m\wedge X_s)$
$$
i(X_i\wedge X_j)\mathbf{d}\omega_{(1,p)}, \ \
i(X_m\wedge X_s)\mathbf{d}\omega_{(p+1,n)}
$$
will measure for each $i<j; m<s$
the internal for each of the two distributions
local exchange that take place inside $\Delta^{(1,p)}$ and $\Delta^{(p+1,n)}$
correspondingly, and the corresponding sums for $i<j; m<s$ will measure the
total internal exchanges taking place inside each of $\Delta^{(1,p)}$ and
$\Delta^{(p+1,n)}$ .

Recalling the Lie derivative of a $p$-form with respect to a
$q$-vector field (Sec.2.8.3) and the above relations, connecting
the representatives of these distributions and codistributions, we observe that
$$
 L_{X_i\wedge X_j}\omega_{(p+1,n)} =-i(X_i\wedge
X_j)\mathbf{d}\omega_{(p+1,n)}, \ \
L_{X_m\wedge X_s}\omega_{(1,p)}
=-i(X_m\wedge X_s)\mathbf{d}\omega_{(1,p)},
$$
These last relations suggest a dynamical interpretation of the above relations.
%In view of this we are allowed to make use of the corresponding terminology,
For example, we can say that the two distributions $\Delta^{(1,p)}$ and
$\Delta^{(p+1,n)}$ are in a {\it local dynamical equilibrium}
\index{ local dynamical equilibrium} if $p=n-p$ and
 the following relations hold (i,j=1,...,p\,; \ \ m,s=p+1,...,n):
$$
L_{X_i\wedge X_j}\omega_{(1,p)}=
\mathbf{d}\langle i_{X_i\wedge X_j},\omega_{1,p}\rangle,
\ \ \text{i.e.}, \ \ i_{X_i\wedge X_j}\mathbf{d}\omega_{1,p}=0,
$$
$$
L_{X_m\wedge X_s}\omega_{(p+1,n)}=
\mathbf{d}\langle i_{X_m\wedge X_s},\omega_{p+1,n}\rangle,
\ \ \text{i.e.}, \ \ i_{X_m\wedge X_s}\mathbf{d}\omega_{p+1,n}=0,
$$
$$
\mathbb{D}_{(p+1,n)}^{(1,p)}=-\mathbb{D}_{(1,p)}^{(p+1,n)}.
$$

The first two equations require recognizability of each of the two
distributions during evolution when dynamical equilibrium, required by the
third equation, is guaranteed. So,  if a spatially
finite, time stable and space propagating physical object is mathematically
represented by an integrable distribution $\Delta^{p+1}=\{X_1,...,X_p,Z\}$,
being extension of the integrable distribution $\Delta^{p}=\{X_1,...,X_p\}$
along a local shuffling symmetry $Z, [X_i,Z]\in\Delta^{p}$, then every new
recognizable 2-dimensional subdistribution of the kind $\{X_i,Z\}, i=1,...,p$,
should be either integrable, or in a state of dynamical equilibrium with
appropriate partner(s) inside $\Delta^p$. In such a case the evolution along a
symmetry admits a natural dynamical interpretation.

This idea we are going to work out on the example of photon-like objects in
Part IV of this book.

\section{Projections, Nonlinear connections, \\Curvature and Cocurvature}

{\bf 3.3.1. Projections in a linear space.}
The projections are linear maps $P$ in a linear space $W^n$ (under linear space
we mean here {\it module over a ring}, or {\it vector space over a field})
sending all elements of $W^n$ to some subspace $P(W^n)\subset W^n$, such that
$P\circ P=P$. We assume further $P\neq id_{W^n}$. Let
$(e_1,\dots,e_p,\dots,e_n)$ and
$(\varepsilon^1,\dots,\varepsilon^p,\dots,\varepsilon^n)$ be two dual bases:
$<\varepsilon^\mu,e_\nu>=\delta^\mu_\nu, \ \mu,\nu=1,\dots,n$, and let $N_i^a,
\ i=1,\dots,p\,; \ a=p+1,\dots,n$ be the corresponding to $P$ $[p\times(n-p)]$
matrix of rank $(n-p)$. We define another couple of dual bases (Berwald bases):
$$
k_\mu=(e_i+N_i^ae_a,\,e_a)\,; \ \
\omega^\nu=(\varepsilon^i,\,\varepsilon^b-N^b_j\varepsilon^j),\ \
\ j=1,\dots,p\,; \ a,b=p+1,\dots,n.
$$
Clearly, the (sub)basis $(\varepsilon^b-N^b_j\varepsilon^j)$
annihilates the (sub)basis $(e_i+N_i^ae_a)$, as well as, the (sub)basis
$(\varepsilon^i-N^i_b\varepsilon^b)$ annihilates the (sub)basis
$(e_a+N_a^je_j)$:
$$
\langle\varepsilon^b-N^b_j\varepsilon^j,e_i+N_i^ae_a\rangle=0, \ \
 \langle\varepsilon^i-N^i_b\varepsilon^b,e_a+N_a^je_j\rangle=0.
$$
Now the identity map $id_{W^n}=\omega^\nu\otimes k_\nu$ acquires the form
$$
id_{W^n}=\omega^\nu\otimes k_\nu=
\omega^i\otimes k_i+\omega^b\otimes k_b=
\varepsilon^i\otimes(e_i+N_i^ae_a)+
(\varepsilon^b-N^b_j\varepsilon^j)\otimes e_b.
$$

We obtain two vertical projections: $P^{p+1,n}_V$ and $P^{1,p}_V$.
The first one is
$$
P^{p+1,n}_V=(\varepsilon^b-N^b_j\varepsilon^j)\otimes e_b,
$$
and the projection property $P_V\circ P_V=P_V$ is readily verified. Clearly,
this $P_V$ projects onto the subspace generated by $\{e_{p+1},\dots,e_n\}$.
So the image $P_V(x)$ of any vector $x\in W^n$ acquires the form
$P_V(x)=A^a\,e_a, \ a=p+1,\dots,n$, in particular, $P^{p+1,n}_V(e_i)=-N^a_ie_a$,
$P^{p+1,n}_V(e_a)=e_a$.

If we want the image space to be generated by the first $p$ basis vectors
$\{e_1,\dots,e_p\}$ the projection should look like
$$
P^{1,p}_V=(\varepsilon^i-N^i_a\,\varepsilon^a)\otimes e_i,
$$
and the image of $x\in W^n$ will in general looks like
$P^{1,p}_V(x)=A^ie_i, \ i=1,2,\dots,p$. In particular, $P^{1,p}_V(e_i)=e_i$ and
$P^{1,p}_V(e_a)=-N^i_ae_i$.
\vskip 0.3cm

\noindent
{\bf Remark}: Here $"i"$ numbers the
rows, and $"a"$ numbers the columns and $P_V$ acts from the right on the basis
$\{e_1,e_2,\dots,e_p\}$. Also, the index "V" means here "vertical".
\vskip 0.3cm
The corresponding  horizontal
projections, denoted by $P_H$, are defined for any of the above
two cases by $P_H=id_{W^n}-P_V$. The projection property $P_H\circ P_H=P_H$ is
also readily verified. If the image space of $P_V$ is $\{e_{p+1},\dots,e_n\}$,
then for the corresponding $P^{1,p}_H$ we obtain
$$
P^{1,p}_H=\varepsilon^i\otimes(e_i+N_i^ae_a), \ \
P_H(e_i)=e_i+N^a_ie_a, \ \ P_H(e_a)=0,
$$
and if the image space of $P_V$ is
$\{e_1,\dots,e_p\}$, then
$$
P^{p+1,n}_H=\varepsilon^a\otimes(e_a+N^i_ae_i), \ \
P_H(e_i)=0, \ \ P_H(e_a)=e_a+N^i_ae_i.
$$
It is seen that
$$ KerP_V=ImP_H, \
\text{and} \ \ KerP_H=ImP_V.
$$
Hence,
$$
W^n=KerP_V\oplus ImP_V=KerP_H\oplus
ImP_H.
$$

The projection $P$ acquires the corresponding forms:
$$
P=P_{V}^{p+1,n}\oplus P_{H}^{1,p}=
\Big[(\varepsilon^b-N^b_j\varepsilon^j)\otimes e_b\Big]\oplus
\Big[\varepsilon^i\otimes(e_i+N_i^ae_a)\Big],
$$
$$
P=P_{V}^{1,p}\oplus P_{H}^{p+1,n}=
\Big[(\varepsilon^i-N^i_a\varepsilon^a)\otimes e_i\Big]\oplus
\Big[\varepsilon^a\otimes(e_a+N_a^ie_i)\Big].
$$

Finally we note that with every involution $\varphi, \ \varphi\circ\varphi=id$,
two projections are associated: $P^{-}_{\varphi}=\frac12(id-\varphi)$ and
$P^{+}_{\varphi}=\frac12(id+\varphi)$.
\vskip 0.3cm

{\bf 3.3.2. Projections in tangent bundles.} \index{Projections in tangent
bundles} Let's now turn to manifolds. Recalling the concept of distribution on
a manifold $M$ we see that every distribution $\Delta^p(M)$ in $T_M$ defines a
projection $P_x; T_x(M)\rightarrow T_x(M), x\in M$, and inversely, every
section $P$ of $T_M\otimes T_M^*$, i.e. $P\in \Lambda^1(M,T_M)$, with constant
kernel: $Ker(P_x)=p<n$ , defines a $p$-dimensional distribution $\Delta^p(M)$ on
$M$ and corresponding $(n-p)$-codistribution $\Delta^*_{n-p}(M)$.

Let $P^{(1,p)}_V(x)$ projects on the subspace generated by
$\{\frac{\partial}{\partial x^1},\dots,\frac{\partial}{\partial x^p}\}$, then
the vertical subspace at $x\in M$ is the image of $P^{(1,p)}_V(x)$, or the
kernel of $P^{(1,p)}_H(x)$, and all $dx^a, a=p+1,p+2,...,n$ are horizontal.
The local horizontal vector fields $X_a$ and
vertical 1-forms $\alpha^i$ are  given by
$$
X_a=\frac{\partial}{\partial x^a}+
N^i_a(x)\frac{\partial}{\partial x^i}, \
\ \ \ \alpha^i=dx^i-N^i_a(x)dx^a.
$$
The corresponding projections are
$$
P^{(1,p)}_V(x)=\alpha^i\otimes \frac{\partial}{\partial x^i}=
\Big[dx^i-N^i_a(x)dx^a\Big]\otimes\frac{\partial}{\partial x^i},
$$
$$
P^{(1,p)}_H(x)=dx^a\otimes X_a=
dx^a\otimes\left(\frac{\partial}{\partial x^a}+
N_a^j(x)\frac{\partial}{\partial x^j}\right)
$$
Clearly, $P^{(1,p)}_H(x)$ projects on the subspace generated by
$\left(\frac{\partial}{\partial x^a}+N_a^j\frac{\partial}{\partial
x^j}\right)$, so, the horizontal subspace at $x\in M$ is the image of
$P^{(1,p)}_H(x)$, or the kernel of $P^{(1,p)}_V(x)$,
and we have $\langle\alpha^i,X_a\rangle=0$.

If we want to call the subspace generated by
$\{\frac{\partial}{\partial x^{p+1}},\dots,\frac{\partial}{\partial x^n}\}$
vertical, then all $dx^i, i=1,2,...,p$, are horizontal,
the local horizontal vector fields $X_i$ and vertical 1-forms $\alpha^a$ are
given by
$$
X_i=\frac{\partial}{\partial x^i}+ N^a_i(x)\frac{\partial}{\partial
x^a}, \ \ \ \ \alpha^a=dx^a-N^a_i(x)dx^i.
$$
The corresponding projections will look like
$$
P^{(p+1,n)}_V(x)=\alpha^a\otimes \frac{\partial}{\partial x^a}=
\Big[dx^a-N^a_i(x)dx^i\Big]\otimes
\frac{\partial}{\partial x^a},
$$
$$
P^{(p+1,n)}_H(x)=dx^i\otimes X_i=dx^i\otimes
\left(\frac{\partial}{\partial x^i}+
N_i^a(x)\frac{\partial}{\partial x^a}\right).
$$
In both cases, of course, we have $P=P_V\oplus P_H$.

{\bf Definition.} The projections in $T_M$ of constant rank are called {\it
nonlinear}, or {\it general}, connections. \index{nonlinear connection}

According to Sec.1.4.3 every projection generates curvature $\mathcal{R}$
and co-curvature $\bar{\mathcal{R}}$
forms, where in our case here the binary map $\mathfrak{B}$ is given by the Lie
bracket of vector fields since $P$ takes values in $\mathfrak{X}(M)$.
Explicitly we have \index{curvature/cocurvature of nonlinear connection}
$$
\mathcal{A}([\,,];P)(X,Y)=
P_V\Big(\big[P_H(X),P_H(Y)\big]\Big)
+P_H\Big(\big[P_V(X),P_V(Y)\big]\Big)
$$
$$
=\mathcal{R}_{P}([\,,];X,Y)+
\tilde{\mathcal{R}}_{P}([\,,];X,Y).
$$
Recalling the introduced curvature and co-curvature of a distribution as
introduced in the previous subsection and identifying $KerP_V$ with
$\Delta^p(M)$,  and $ImP_V$ with $\Delta^{n-p}(M)$ we can write
$$
\mathcal{R}_{P}=\Omega|_{\Delta^p(M)}, \ \
\tilde{\mathcal{R}}_{P}=\Omega|_{\Delta^{n-p}(M)}.
$$

We are going now to see how these curvature forms look locally in terms of the
projection components $N^i_a(x)$, or $N^a_i(x)$. In the first case we have to
compute $\Omega^j_{ab}=\alpha^j([X_a,X_b]), a<b$, and in the second case we
have to compute \linebreak $\Omega^a_{ij}=\alpha^a([X_i,X_j]), \ i<j$, here and
further $i,j,k$ run from $1$ to $p$, while $a,b$ run from $p+1$ to $n$ and for
clarity we write just $N^i_a$ instead of $N^i_a(x)$.

We obtain
$$
\Omega^j_{ab}=\frac{\partial N^j_b}{\partial x^a}- \frac{\partial
N^j_a}{\partial x^b}+N^k_a\frac{\partial N^j_b}{\partial x^k}-
N_b^k\frac{\partial N^j_a}{\partial x^k},
$$
$$ \Omega^a_{ij}=\frac{\partial
N^a_j}{\partial x^i}- \frac{\partial N^a_i}{\partial x^j}+N^b_i\frac{\partial
N^a_j}{\partial x^b}- N_j^b\frac{\partial N^a_i}{\partial x^b}.
$$

If we'd like to obtain the two curvature forms directly from the corresponding
vertical 1-forms $\alpha^i, i=1,2,...,p$, or from $\alpha^a, a=p+1,...,n$, then
we have to compute the corresponding horizontal projections of
$\mathbf{d}\alpha^i$ and $\mathbf{d}\alpha^a$.

First, consider the case $\alpha^i=dx^i-N^i_adx^a$ are vertical 1-forms, so
$dx^i, i=1,2,...,p$ have non-zero vertical and horizontal projections. For
$\mathbf{d}\alpha^i$ we obtain
$$
\mathbf{d}\alpha^i=-\mathbf{d}(N^i_adx^a)= -\frac{\partial
N^i_a}{\partial x^j}dx^j\wedge dx^a- \frac{\partial N^i_a}{\partial
x^b}dx^b\wedge dx^a.
$$
Now, in this case $dx^a$ are horizontal:
$\langle dx^a,\partial_{x^i}\rangle=0$,
so we have to find the horizontal projection
of $dx^j$. Since the restriction of
$dx^j$ to $\Delta^{n-p}(M)$ is equal to $N^j_a(x^i(x^b),x^b)\,dx^a$, for the
horizontal projection of $\mathbf{d}\alpha^i$ we obtain
$$
\mathbf{d}\alpha^i|_{\Delta^{n-p}(M)}=
-\left(\frac{\partial N^i_a}{\partial x^b}-
\frac{\partial N^i_b}{\partial x^a}+
\frac{\partial N^i_a}{\partial x^j}N^j_b-
\frac{\partial N^i_b}{\partial x^j}N^j_a\right)_{x^i(x^a)}dx^b\wedge dx^a, \
b<a . $$

In the same way for the case $\alpha^a=dx^a-N^a_idx^i, i=1,2,...,p$ are
vertical and the horizontal projections of $dx^a$ are $N^a_i(x^j,x^b(x^j))dx^i$
we obtain
$$
\mathbf{d}\alpha^a|_{\Delta^{p}(M)}= -\left(\frac{\partial
N^a_i}{\partial x^j}- \frac{\partial N^a_j}{\partial x^i}+ \frac{\partial
N^a_i}{\partial x^b}N^b_j- \frac{\partial N^a_j}{\partial
x^b}N^b_i\right)_{x^a(x^i)}dx^j\wedge dx^i, \ j<i .
$$
Hence, for $\Omega|_{\Delta^p(M)}$
and $\Omega|_{\Delta^{n-p}}$ we obtain respectively:
$$
\Omega|_{\Delta^p(M)}=-\left(\mathbf{d}\alpha^a|_{\Delta^p(M)}\right)
\otimes\frac{\partial}{\partial x^a}
$$
$$
=\left(\frac{\partial N^a_i}{\partial x^j}-
\frac{\partial N^a_j}{\partial x^i}+
\frac{\partial N^a_i}{\partial x^b}N^b_j-
\frac{\partial N^a_j}{\partial x^b}N^b_i\right)_{x^a(x^i)}
dx^j\wedge dx^i\otimes\frac{\partial}{\partial x^a}, \ j<i .
$$
$$
\Omega|_{\Delta^{n-p}(M)}=-\left(\mathbf{d}\alpha^i|_{\Delta^{n-p}(M)}\right)
\otimes\frac{\partial}{\partial x^i}
$$
$$
=\left(\frac{\partial N^i_a}{\partial x^b}-
\frac{\partial N^i_b}{\partial x^a}+
\frac{\partial N^i_a}{\partial x^j}N^j_b-
\frac{\partial N^i_b}{\partial x^j}N^j_a\right)_{x^i(x^b)}dx^b
\wedge dx^a\otimes\frac{\partial}{\partial x^i}, \ b<a .
$$

\section{Connections and Curvature on Smooth\\ Bundles}

{\bf 3.4.1. Tangent Structure of Smooth Bundles.}

Let $(E,\pi,M,F)$ be a smooth bundle with $dim(M)=n$, $dim(F)=r$. The
derivative $d\pi$ of the projection map $\pi$ is a bundle map between the
tangent bundles of $E$ and $M$: $d\pi: \tau_E\rightarrow \tau_M$. For every
$z\in E$ we have $d\pi_z: T_z(E)\rightarrow T_{\pi(z)}(M)$. Since $\pi$ is
surjective then $Ker(d\pi_z)$ is not empty. The elements in $Ker(d\pi_z)$ are
called {\it vertical vectors} and the subspace $V_z=Ker(d\pi_z)\subset T_z(E)$ is
called {\it vertical subspace} at $z\in E$. From the surjectivity of each
$d\pi_z, z\in E$, it follows that the dimension of $V_z, z\in E$ is equal to the
dimension of $F$. Now the union $V_E=\bigcup_{z\in E}V_z$ acquires a structure
of subbundle of the tangent bundle of $E$: $V_E\subset \tau_E$, and is called
{\it vertical} subbundle of $\tau_E$. The dimension of $V_E$ is $(n+2r)$.

The sections of $V_E$ are called {\it vertical} vector fields. If $Z\in
Sec(V_E)$, clearly $\pi_*Z=0$. It follows that the Lie bracket of two vertical
vector fields is again a vertical vector field:
$\pi_*[Z_1,Z_2]=[\pi_*Z_1,\pi_*Z_2]=0$, so the vertical vector fields on $E$
form a Lie subalgebra $\mathfrak{X}_V(E)$ of the Lie algebra
$\mathfrak{X}(E)$ of all vector fields on $E$. It follows that

	$1^o$. $\mathfrak{X}_V(E)$ is finitely generated module over
$\mathcal{J}(E)$;

	$2^o$. the vertical subbundle defines a completely  integrable distribution.

Since $V_z(E)$ is a subspace of $T_z(E)$ then there exist other subspaces
$H_z(E)\subset T_z(E)$ such that $T_z(E)=V_z(E)\oplus H_z(E)=Ker(d\pi_z)\oplus
H_z(E), z\in E$. The
subspaces $H_z(E), z\in E$ are called {\it horizontal}. If at every $z\in E$
such a subspace is chosen then the union $H_E=\bigcup_{z\in E}H_z$ is defined
and $H_E$ is a subbundle of $\tau_E$, called {\it horizontal} subbundle.
Clearly $\tau(E)=V_E\oplus H_E$, and the dimension of $H_E$ is equal to
$(2n+r)$. The sections of $H_E$ are called {\it horizontal} vector fields, they
form a submodule $\mathfrak{X}_H$
of $\mathfrak{X}(E)=Sec(\tau_E)$ but NOT a Lie subalgebra of
$\mathfrak{X}(E)$. Every horizontal vector field $Z_H$ on $E$ is projectable,
i.e. there is a vector field $X$ on $M$ such that $\pi_*Z_H=X$.

The surjectivity of $d\pi_z$ leads to the fact that $d\pi_z|_{H_z(E)}$ is a
linear isomorphism between $H_z(E)$ and $T_{\pi(z)}(M)$. So, $H_E$ is
isomorphic to the pullback of $\tau(M)$ via $\pi_*$. We obtain that every choice
of $H_E$ leads to the decomposition of the vector fields on $E$: $Z=Z_V+Z_H,
Z_V\in \mathfrak{X}_V, Z_H\in \mathfrak{X}_H$.

Finally it deserves noting that $V_E$ is connected only with the bundle
structure of $(E,\pi,M,F)$, while $H_E$ depends on our choice, and if chosen,
the corresponding n-dimensional distribution should be checked about
integrability.

 \vskip 0.3cm
{\bf 3.4.2. Cotangent Structure of Smooth Bundles.}

The corresponding cotangent structure of $(E,\pi,M,F)$ inherits the following
specific properties.

A differential form $\phi\in \bigwedge(E)$ is called {\it horizontal} if it
is annihilated by the vertical vector fields, i.e. $i(Z_V)\phi=0,
Z_V\in\mathfrak{X}_V$.  Clearly, in view of the antiderivation properties of
$i(Z)$ the horizontal differential forms form a graded algebra denoted by
$\bigwedge_H(E)$, and this graded algebra depends only on the bundle structure
of $E$. The corresponding n-dimensional codistribution of 1-forms
$\Lambda^1_H(E)$ is, of course, completely integrable.

If a horizontal subbundle $H_E$ is defined, then a differential form
$\phi\in\bigwedge(E)$ is called {\it vertical} if every horizontal vector field
$Z_H$ annihilates $\phi$: $i(Z_H)\phi=0$. The corresponding graded algebra
$\bigwedge_V(E)$ is called {\it vertical} subalgebra of $\bigwedge(E)$, it
depends on the choice of $H_E$. The obtained r-dimensional codistribution of
vertical 1-forms $\Lambda^1_V(E)$ should be checked about integrability.

 The following basic property holds: The anticommutative tensor product of the
algebras $\bigwedge_V(E)$ and $\bigwedge_H(E)$: $\phi\otimes\psi\rightarrow
\phi\wedge\psi$ defines the (pointwise) isomorphism
$\bigwedge_H(E)\otimes_E\bigwedge_V(E)\leftrightarrow \bigwedge(E)$ of graded
algebras.

\vskip 0.3cm
{\bf 3.4.3. Connections and Curvature on Smooth Bundles.}

As we mentioned the above properties of the tangent and cotangent
structure of a smooth bundle $(E,\pi,M,F)$ obviously define a
completely integrable vertical distribution $\Delta_V(E)$ in $T(E)$, and the
corresponding completely integrable horizontal codistribution $\Delta_H^*(E)$
in $T^*(E)$. Hence, every choice of $H_E$ sets the problem of the
integrability of the corresponding distribution $\Delta_H$. Usually, only such
distributions are considered which are additional to the naturally existing
vertical distributions. In terms of projections $\mathcal{P}$
this means that the general
connections considered on smooth bundles are required to be compatible with the
existing specific bundle structure of $T_E$, i.e. $Im\mathcal{P}_z=V_z(E)$.

From local point of view, let $(x^i,y^a)=(x^1,...,x^n;y^1,...,y^r)$ be local
coordinates on $U\times\pi^{-1}(U)$, where $U\subset M$ is a coordinate
neighborhood on $M$, and the connection, i.e. the horizontal distribution, is
defined by the projection $\mathcal{P}$:
$\mathcal{P}_z(N^a_i(z)): T_z(E)\rightarrow V_z(E)$,
where $N^a_i$ are the components of $\mathcal{P}$,
so that the image of $\mathcal{P}_z$ coincides with $V_z(E), z\in E$.
Let $X_i, i=1,2,...,n$ represent the horizontal distribution, and $\alpha^a,
a=1,2,...,r$ represent the corresponding codistribution, so
$\langle\alpha^a,X_i\rangle=0$. If the curvature is not zero, then:

	$\mathbf{1^o}$. At least one of the expressions
$$
[X_i,X_j]\wedge X_1\wedge X_2\wedge...\wedge X_n, \ \ \ X_i\in\mathfrak{X}_H(M),
\ i,j=1,2,...,n $$ is NOT equal to zero.

	$\mathbf{2^o}$. At least one of the expressions
$$
\mathbf{d}\alpha^a\wedge\alpha^1\wedge\alpha^2\wedge...\wedge\alpha^r, \ \ \
\alpha^a\in \Lambda^1_V(E), \ a=1,2,...,r
$$
is NOT equal to zero.

	$\mathbf{3^o}$. The corresponding curvature form
$$
\Omega(H_E)=-\left(\mathbf{d}\alpha^a|_{H_E}\right)
\otimes\frac{\partial}{\partial y^a}=
$$
$$
=\left(\frac{\partial N^a_i}{\partial x^j}-
\frac{\partial N^a_j}{\partial x^i}+
\frac{\partial N^a_i}{\partial y^b}N^b_j-
\frac{\partial N^a_j}{\partial y^b}N^b_i\right)_{y^b(x^k)}
dx^j
\wedge dx^i\otimes\frac{\partial}{\partial y^a}, \ j<i .
$$
is NOT equal to zero.
%\newpage

\section{Connections and Curvature on Principal Bundles.}

{\bf 3.5.1. Principal Bundles.}

 Let $G$ be a $r$-dimensional Lie group with corresponding Lie algebra
$\mathfrak{g}$ and $B$ be a $n$-dimensional manifold.  A smooth bundle
$\mathcal{P}=(P,\pi,B,G)$ satisfying the following conditions

	1. $\mathcal{R}: P\times G\rightarrow P$ is a smooth right action of
$G$ on $P$.

	2. There is a coordinate representation $(U_\alpha,\psi_\alpha)$ of
$\mathcal{P}$ such that
$$
\psi(x,ab)=\psi(x,a).b, \ \ x\in B, \ \ a,b\in G
$$
is called a {\it principal bundle with a structure group} $G$ and the group
action $\mathcal{R}$ is called {\it principal action} of $G$ on $P$.
\index{principal bundle}
The following properties of $\mathcal{P}$ are obvious:

	-$\pi(z.a)=\pi(z)\rightarrow d\pi\circ d\mathcal{R}_a=d\pi,  \ \ z\in
P, \ \ a\in G$,

	-the action of $G$ on $P$ is {\bf free},

	-the orbit through $z\in P$ is the fiber through $z\in P$,

	-the fibers $G_x=\pi^{-1}(x), x\in B$ are submanifolds of $P$.

\noindent
If $\hat{\mathcal{P}}=(\hat{P},\hat{\pi},\hat{B},G)$ is a second principal
bundle with a principal action $\hat{\mathcal{R}}$ then a smooth map $\varphi:
P\rightarrow \hat{P}$ is called a {\it homomorphism of principal bundles} if
$\varphi$ is $G$-equivariant with respect to the two actions of $G$. The induced
map $\psi: B\rightarrow \hat{B}$ satisfies $\hat{\pi}\circ\psi=\psi\circ\pi$.
Also, every $\varphi_x: G_x\rightarrow G_{\psi(x)}$ satisfies
$\varphi_x(z.a)=\varphi_x(z).a, \ \ z\in G_x, \ \ x\in B$.

Every local section $\sigma: U\rightarrow P, U\subset B$ of $\mathcal{P}$
defines isomorphism between the trivial bundle $U\times G$ and the restriction
of $\mathcal{P}$ to $U$, so if $\mathcal{P}$ admits a (global) section, it is
trivial.

If $V\subset B$ is another open subset in $B$ such that the intersection
$U\cap V$ is not empty and $\tau$ is a section over $V$, then there is {\it
unique} smooth map $g_{UV}:U\cap V\rightarrow G$ such that
$\varphi(x,g_{UV}(x))=\tau(x)$, and $g_{UV}(x)\in G_x$ can be determined by the
equation $\tau(x)=\sigma(x).g_{UV}(x), x\in U\cap V$.
%\newpage
{\bf 3.5.2. Vector fields on principal bundles.}

Recall that a free action of a Lie group $G$ on a manifold $P$ defines
corresponding fundamental subbundle $F_P$ of $T_P$ with a fiber over $z\in P$
the corresponding image of the Lie algebra $\mathfrak{g}$ over $z$. The
important observation is that $F_P$ coincides with the vertical subbundle
$V_P$ of a principal bundle $(P,\pi,B,G)$, so the map
$$
P\times\mathfrak{g}\rightarrow V_P
$$
is a bundle isomorphism.

There are two aspects of this isomorphism deserving to be mentioned.

The first aspect is to  consider it as isomorphism
between $\mathcal{J}(P)\otimes\mathfrak{g}$ and  $\mathfrak{X}_VP$ according to
the map $f\otimes h\rightarrow f.Z_h, \ f\in\,\mathcal{J}(P), \ \
h\in\mathfrak{g}$.
The second view is to consider the $\mathfrak{g}$-valued
functions on $P$, i.e. the space $\mathcal{J}(P;\mathfrak{g})$, and the map
$\mathcal{J}(P;\mathfrak{g})\rightarrow\mathfrak{X}_V(P)$
defined by $(\phi\rightarrow Z_{\phi})$, where $Z_{\phi}(z)=Z_{\phi(z)}(z)$.

Another important class of vector fields on the bundle space $P$ consists of
vector fields satisfying the condition $(d\mathcal{R}_a)_zZ_z=Z_{z.a}, a\in
G$, they are called $G$-{\it invariant}, or just - {\it invariant} and are
denoted by $\mathfrak{X}^{I}(P)$. These vector fields are {\it projectable},
i.e. for every invariant vector field $Z\in \mathfrak{X}^{I}(P)$ there is a
vector field $X\in\mathfrak{X}(B)$ such that
$$
\pi_*Z=X, \ \ Z\in\mathfrak{X}^{I}(P), \ \ X\in \mathfrak{X}(B).
$$
Thus, $\pi_*|_{\mathfrak{X}^{I}(P)}$ is surjective with kernel the intersection
$\mathfrak{X}^{I}(P)\cap\mathfrak{X}_V(P)$. It follows that the Lie bracket
between invariant and vertical vector fields is vertical because
$$
\pi_*[Z,Y]=[\pi_*Z,\pi_*Y]=[\pi_*Z,0]=0, \ \ Z\in\mathfrak{X}^{I}(P), \ \
Y\in\mathfrak{X}_V(P).
$$
Also,

	-the Lie bracket between two projectable vector fields is projectable,

	-the module of vector fields on $\mathcal{P}$ is generated by the
vertical and projectable vector fields,

	-the vector field $Z$ on $P$ is projectable iff
$(Z-(\mathcal{R}_a)_*Z)$ is vertical for every $a\in G$.

%\vskip 0.3cm
\newpage
{\bf 3.5.3. Differential forms on a principal bundle.}

A differential form $\alpha\in\bigwedge(P)$ is called {\it invariant} if
$$
\mathcal{R}^*\alpha=\alpha, \ a\in G.
$$
All invariant differential forms on $P$ form a $\mathbb{R}$-algebra denoted
by $\bigwedge_I(P)$.

A differential form $\alpha\in\bigwedge(P)$ is called {\it horizontal} if
$$
i(Z_V)\alpha=0, \ Z_V\in \mathfrak{X}_V(P),
$$
i.e. if $\alpha$ is horizontal with respect to the action of $G$ on $P$.
All horizontal differential forms on $P$ form a $\mathbb{R}$-algebra
denoted by $\bigwedge_H(P)$.

The following result holds:
\vskip 0.3cm
\noindent
{\bf The homomorphism $\pi^*:
\bigwedge(B)\rightarrow\bigwedge(P)$ is injective. The image of $\pi^*$
consists of those elements of $\bigwedge(P)$ which are horizontal and
invariant}.
\vskip 0.3cm
The differential forms that are both invariant and horizontal form a subalgebra
$\bigwedge_B(P)\subset\bigwedge(P)$, so, $\pi^*$ can be considered as
isomorphism between $\bigwedge(B)$ and $\bigwedge_B(P)$.

If $\hat{\mathcal{P}}$ is another principal bundle with the same group $G$ and
$(\varphi,\psi): \mathcal{P}\rightarrow\hat{\mathcal{P}}$ is a homomorphism then,
because of the equivariance of $\varphi$, the fundamental vector fields on
$\mathcal{P}$ and $\hat{\mathcal{P}}$, generated by the same
$h\in\mathfrak{g}$, are $\varphi$-related.
$$
\varphi_*Z_h=\hat{Z}_h, \ \ h\in\mathfrak{g}.
$$

This leads to the following commutation relations ($h\in\mathfrak{g}, a\in G$):
$$
\varphi^*\circ L_{\hat{Z}_h}=L_{Z_h}\circ\varphi^*; \ \
\varphi^*\circ i(\hat{Z_h})=i(Z_h)\circ\varphi^*; \ \
\varphi^*\circ\hat{\mathcal{R}_a}^*=\mathcal{R}_a^*\circ\varphi^* .
$$
%\newpage
{\bf 3.5.4. Vector-valued differential forms on a principal bundle.}

Recall that if $W$ is a vector space then the $W$-valued differential forms
on $P$ form a $\bigwedge(P)$ graded module, usually denoted by
$\bigwedge(P,W)$ which is isomorphic to $\bigwedge(P)\otimes W$. Every such
differential form $\Phi$ is written down as
$$
\Phi=\alpha^m\otimes e_m, \ \ m=1,2,...,dimW ,
$$
where $\{e_m\}$ form a basis of $W$ and
$\alpha^m\in\bigwedge(P)$. If $\beta$ is a $p$-form on $P$ then the product
$(\beta.\Phi)$ is given by
$$
\beta.\Phi=\beta.(\alpha^m\otimes e_m)=(\beta\wedge\alpha^m)\otimes e_m.
$$
The operators $\mathbf{d}$, $i(Z)$, $L_Z$ and $\mathcal{R}_a^*, a\in G$, are
extended to $\bigwedge(P,W)$ according to
$$
\mathbf{d}(\alpha^m\otimes e_m)=(\mathbf{d}\alpha^m)\otimes e_m; \ \
i(Z)(\alpha^m\otimes e_m)=(i(Z)\alpha^m)\otimes e_m;
$$
$$
L_Z(\alpha^m\otimes e_m)=(L_Z\alpha^m)\otimes e_m; \ \
\mathcal{R}_a^*(\alpha^m\otimes e_m)=(\mathcal{R}_a^*\alpha^m)\otimes e_m.
$$
A $W$-valued form $\Phi$ is called {\it horizontal} if $i(Z_h)\Phi=0, h\in
\mathfrak{g}$. Horizontal forms form a graded subspace of $\bigwedge(P,W)$
denoted by $\bigwedge_H(P,W)$ which is isomorphic to $\Lambda_H(P)\otimes W$.

If $\rho: G\rightarrow W$ is a representation, and $\rho':
\mathfrak{g}\rightarrow W$ is the corresponding derived representation then a
$W$-valued form on $P$ is called {\it $\rho$-equivariant} if
$$
\mathcal{R}_a^*\Phi=\rho(a^{-1})\circ\Phi, \ \ a\in G,
$$
and if $G$ is connected this is equivalent to
$$
L_{Z_h}\Phi=-\rho'(h)\circ\Phi, \ \ h\in\mathfrak{g}.
$$
The $G$-invariant $W$-valued forms are denoted by $\Lambda_{I}(P)\otimes W$.
The forms that are at the same time horizontal and invariant are called {\it
basic} and are denoted by $\bigwedge_B(P,W)$.

Every linear map $\varphi: W\rightarrow W_1$ generates a map
$\varphi_*: (\bigwedge(P)\otimes W)\rightarrow \bigwedge(P)\otimes W_1$ given
by
$$
\varphi_*(\alpha^m\otimes e_m)=\alpha^m\otimes\varphi(e_m).
$$
This rule is extended to multilinear maps $\varphi: W_1\times
W_2\times...\times W_p\rightarrow W$ according to
$$
\varphi_*(\alpha_1^{m_1}\otimes e_{m_1},\alpha_2^{m_2}\otimes e_{m_2},...,
\alpha_p^{m_p}\otimes e_{m_p})=
$$
$$
(\alpha_1^{m_1}\wedge\alpha_2^{m_2}\wedge...\wedge\alpha_p^{m_p})\otimes
\varphi(e_{m_1},e_{m_2},...,e_{m_p}),
$$
where $m_i$ is the dimension of $W_i$ and
$\{e_{m_i}\}, i=1,2,...,p_i$ is a basis of $W_i$. In particular, if $\varphi$
is: tensor product, symmetrized tensor product, exterior product, we obtain
respectively $$
(\alpha_1^{m_1}\wedge\alpha_2^{m_2}\wedge...\wedge\alpha_p^{m_p})\otimes
(e_{m_1}\otimes e_{m_2}\otimes...\otimes\, e_{m_p}),
$$
$$
(\alpha_1^{m_1}\wedge\alpha_2^{m_2}\wedge...\wedge\alpha_p^{m_p})\otimes
(e_{m_1}\vee e_{m_2}\vee...\vee e_{m_p}),
$$
$$
(\alpha_1^{m_1}\wedge\alpha_2^{m_2}\wedge...\wedge\alpha_p^{m_p})\otimes
(e_{m_1}\wedge e_{m_2}\wedge...\wedge e_{m_p}).
$$
Every representation $\rho: G\rightarrow W$ generates the bilinear map $\rho':
\mathfrak{g}\times W\rightarrow W$ given by $(h,w)\rightarrow\rho'(h)w$. For
the corresponding map of differential forms we obtain
$$
(\Phi,\Psi)\rightarrow\rho'(\Phi,\Psi)=
\alpha^i\wedge\beta^m\otimes(\rho'(E_i)(e_m).
$$
If $\rho=Ad$, then $\rho'(h)(k)=[h,k]$, so if $\Phi=\alpha^i\otimes E_i$ and
$\Psi=\beta^j\otimes E_j$ are $\mathfrak{g}$-valued we obtain
$$ (\Phi,\Psi)\rightarrow
ad(\Phi,\Psi)\equiv[\Phi,\Psi]= \alpha^i\wedge\beta^j\otimes(\rho'(E_i)(E_j)=
\alpha^i\wedge\beta^j\otimes[E_i,E_j].
$$
It follows from this relation that if $\Phi\in\Lambda^1(P,\mathfrak{g})$ then
\begin{eqnarray*}
&&[\Phi,\Phi](X,Y)=[\alpha^i\otimes E_i, \alpha^j\otimes E_j](X,Y)\\
&&=(\alpha^i(X)\alpha^j(Y)-\alpha^i(Y)\alpha^j(X))[E_i,E_j]\\
&&=\alpha^i(X)\alpha^j(Y)[E_i,E_j]-\alpha^j(X)\alpha^i(Y)[E_i,E_j]\\
&&=\alpha^i(X)\alpha^j(Y)[E_i,E_j]+\alpha^j(X)\alpha^i(Y)[E_j,E_i]\\
&&=\alpha^i(X)\alpha^j(Y)[E_i,E_j]+\alpha^i(X)\alpha^j(Y)[E_i,E_j]\\
&&=2(\alpha^i(X)\alpha^j(Y))[E_i,E_j]=2(\alpha^i\wedge\alpha^j)(X,Y)[E_i,E_j]\\
&&=2[\Phi(X),\Phi(Y)].
\end{eqnarray*}

For the case $\rho'([\Phi,\Psi],\Omega)$, where $(\Phi,\Psi)$ are
$\mathfrak{g}$-valued and $\Omega$ is $W$-valued, in view of the relation
$$
\rho'([h,k])=\rho'(h)\circ\rho'(k)-\rho'(k)\circ\rho'(h)
$$
we obtain
$$
\rho'([\Phi,\Psi],\Omega)=\rho'(\Phi,\rho'(\Psi,\Omega))-
(-1)^{pq}\rho'(\Psi,\rho'(\Phi,\Omega)),
$$
where $\Phi\in\Lambda^p(P,\mathfrak{g})$ and
$\Psi\in\Lambda^q(P,\mathfrak{g})$. Clearly, if $\Phi=\Psi$ and $(p=q)$ is
even number we obtain $\rho'([\Phi,\Phi],\Omega)=0$, and if $(p=q)$ is
odd, then
$$
\rho'([\Phi,\Phi],\Omega)=2\rho'(\Phi,\rho'(\Phi,\Omega)).
$$
%\newpage
{\bf 3.5.5. Principal connections}

Note that from the above mentioned relation $d\pi\circ d\mathcal{R}_a=0, a\in G$
it follows that the vertical subbundle $V_P$ of $\mathcal{P}$ is stable under
the action of the group $G$. This suggests to introduce connections on
a principal bundle $\mathcal{P}$ as follows:
\vskip 0.3cm
{\bf Definition.} A principal connection on $\mathcal{P}$ is every bundle map
$(\Gamma,id_P): T_P\rightarrow T_P$ which satisfies the following conditions:

	${\bf 1^o.}$ $\Gamma\circ\Gamma=\Gamma$, so $\Gamma_z$ is a projection
in every tangent space $T_z(P), z\in P$;

	${\bf 2^o.}$ $\Gamma(T_z(P))=V_z, \ z\in P$, so, $\Gamma$ projects
every tangent space $T_z(P)$ on the vertical subspace $V_z\subset T_z(P)$ ;

	${\bf 3^o.}$ $d\mathcal{R}_a\circ\Gamma=\Gamma\circ d\mathcal{R}_a, \
a\in G$, i.e. $\Gamma$ is equivariant with respect to the action of $G$ on $P$.
\index{principal connection}
\vskip 0.3cm
We see that the difference between connections on smooth bundles and
connections on principal bundles is in the additional compatibility condition
with the action of $G$ on $P$. Therefore, the above equivariance condition
guarantees additionally that every choice of horizontal subspaces $H_z(P)$ at
every $z\in P$, i.e. such that $T_z(P)=V_z(P)\oplus H_z(P)$, leads to stability
of the corresponding horizontal subdistribution
$\mathbf{H}_P\subset\mathbf{\tau}(P)$.

The corresponding to $\Gamma$ horizontal projection $H_z$ at $z\in P$ is, of
course, given by
$$
H_z=id_{T_z(P)}-V_z, \ z\in P,
$$
and this decomposes the tangent bundle $\tau(P)$ to the direct sum
$$
\tau(P)=\mathbf{V}_{P}\oplus\mathbf{H}_{P}.
$$
We obtain the module decomposition of the vector fields on $P$:
$$
\mathfrak{X}(P)=\mathfrak{X}_V(P)\oplus\mathfrak{X}_H(P),
$$
and this decomposition commutes with the action of $G$:
$$
(\mathcal{R}_a)_*\circ V_*=V_*\circ(\mathcal{R}_a)_*, \ \
(\mathcal{R}_a)_*\circ H_*=H_*\circ(\mathcal{R}_a)_*, \ \ a\in G.
$$
This decomposition leads to the decomposition of invariant vector fields on
$P$:
$$
\mathfrak{X}^{I}(P)=\mathfrak{X}^{I}_{H}(P)\oplus\mathfrak{X}^{I}_V(P),
$$
where
$$
\mathfrak{X}^{I}_{H}(P)=\mathfrak{X}^{I}(P)\cap\mathfrak{X}_{H}(P) \  \
\text{and} \ \
\mathfrak{X}^{I}_{V}(P)=\mathfrak{X}^{I}(P)\cap\mathfrak{X}_{V}(P).
$$

Recalling the surjective homomorphism $\pi_*:
\mathfrak{X}^{I}_{H}(P)\rightarrow\mathfrak{X}(B)$ and that
$Ker(\pi_*)=\mathfrak{X}^{I}(P)\cap\mathfrak{X}_{V}(P)$, we obtain that the
restriction of $\pi_*$ to $\mathfrak{X}^{I}_{H}(P)$ is an isomorphism. The
inverse isomorphism $\chi: \mathfrak{X}(B)\rightarrow\mathfrak{X}^{I}_{H}(P)$
is called {\it horizontal lifting isomorphism} for the connection $\Gamma$. The
following important relation holds:
$$
\chi([X,Y])=H_*([\chi(X),\chi(Y)]), \ \ X,Y\in\mathfrak{X}(B).
$$
\vskip 0.3 cm
{\bf 3.5.6. The Connection Form.}

Recall the bundle isomorphism $P\times\mathfrak{g}\leftrightarrow
\mathbf{V}_P$, it is given by
$$
(z,h)\leftrightarrow Z_h(z)=(d\Phi_z)_e(h), \ \ h\in\mathfrak{g}, z\in P.
$$
Since $(d\Phi_z)_e$ is a linear isomorphism for each $z\in P$ we can consider
its inverse
$$
(d\Phi_z)_e^{-1}: V_z(P)\rightarrow\mathfrak{g}.
$$
Now we compose $(d\Phi_z)_e^{-1}$ with the projection $\Gamma$, so
at each $z\in P$ we get a map from $T_z(P)$ to the Lie algebra $\mathfrak{g}$.
\vskip 0.3 cm
{\bf Definition.} The $\mathfrak{g}$-valued differential 1-form $\omega$
on $P$ defined by
$$
\omega(z;Z_z)=(d\phi_z)_e^{-1}\circ\Gamma_z(z,Z_z), \ \ Z_z\in T_z(P), \ z\in P
$$
is called {\it the connection form} for $\Gamma$.
\vskip 0.3 cm
Clearly, $\omega(Z)=0$ iff $Z$ is horizontal.
\vskip 0.3cm
The connection form $\omega$ has the following two properties:

	1. $i(Z_h)\omega=h, \ \ h\in \mathfrak{g}$,

	2. $\mathcal{R}_a^*\omega=Ad(a^{-1})\circ\omega, \ \ a\in G$.

The first property is almost obvious, since
$$
\omega(Z_h)=\omega((d\phi_z)_e(h))=(d\phi_z)_e^{-1}\circ
\Gamma(Z_h)=(d\phi_z)_e^{-1}(Z_h)=h, \ h\in \mathfrak{g}.
$$
The second property, which  means that $\omega$ is $Ad$ equivariant, follows
from differentiating the relation
$\mathcal{R}_a\circ\phi_z=\phi_{z.a}\circ\theta_{a^{-1}}$ and from the
equivariance of $\Gamma$ with respect to the action of the differential of each
$\mathcal{R}_a, \ a\in G$.

From property 2. it follow also that locally we have
$$
L_{Z_h}\omega=-ad(h)\circ\omega, \ \ h\in\mathfrak{g}.
$$

Another interesting fact is that the Lie bracket of a fundamental and
horizontal vector fields is horizontal. In fact, if $Y$ is a horizontal
vector field, then from the last relation and from the general relation
$$
i([X,Y])=[L_X,i(Y)]=L_X\circ i(Y)-i(Y)\circ L_X, \ X,Y\in\mathfrak{X}(P),
$$
we obtain
$$
-i(Y)L_{Z_h}\omega=-ad(h)\circ\omega(Y)=0
$$
$$
=(L_{Z_h}\circ i(Y)-i(Y)\circ
L_{Z_h})\omega=i([Z_h,Y])\omega.
$$
\vskip 0.3 cm
{\bf 3.5.7. The Covariant Exterior Derivative with respect to principal
connection.} \index{covariant exterior derivative on principal bundle}

Having a principal connection $\Gamma$ on $\mathcal{P}$ with connection form
$\omega$ we have in every tangent space the horizontal projection
$H_z=id_{T_z(P)}-\Gamma_z$. The corresponding dual linear map $H_z^*$ is a
projection in $T_z^*$, and these projections are extended to the whole tensor
algebra on $P$. Moreover, $H^*$ is naturally extended to the space of
$W$-valued differential forms $\bigwedge(P,W)$ on $P$, where $W$ is a finite
dimensional vector space. If $\Phi\in\bigwedge(P,W)$, this extended projection
is defined by $$
(H^*\Phi)(z;Z_1,Z_2,...,Z_p)=\Phi(z;H_z(Z_1),H_z(Z_2),...,H_z(Z_p)), \ Z_i\in
T_z(P), $$ and carries the following properties:

	1. $H^*$ projects on the space of horizontal $W$-valued differential
forms;

	2. $H^*(\alpha\wedge\Phi)=(H^*\alpha)\wedge (H^*\Phi), \ \
\alpha\in\bigwedge(P)$;

	3. $H^*\omega=0$;

	4. $H^*\circ\mathcal{R}_a^*=\mathcal{R}_a^*\circ H^*, \ a\in G$;

	5. $H^*\circ L_{Z_h}=L_{Z_h}\circ H^*, \ \ h\in \mathfrak{g}$;

	6. $H^*\circ\pi^*=\pi^*$.
\vskip 0.3cm
{\bf Definition.} The operator
$$
\nabla:=H^*\circ\mathbf{d}:
\Lambda^p(P,W)\rightarrow\Lambda_H^{p+1}(P,W)
$$
is called {\it covariant exterior derivative}.
\vskip 0.3cm
The covariant exterior derivative $\nabla$ has the following properties:

	1. $\nabla(\alpha\wedge\Phi) =\nabla\alpha\wedge H^*\Phi+
(-1)^pH^*\alpha\wedge\nabla\Phi, \ \alpha\in\Lambda^p(P),
\ \Phi\in\bigwedge(P,W)$;

	2. $i(Z_h)\circ\nabla=0, \ h\in\mathfrak{g}$;

	3. $\nabla\circ\mathcal{R}_a^*=\mathcal{R}_a^*\circ\nabla, \ \ a\in G$;

	4. $\nabla\circ L_{Z_h}=L_{Z_h}\circ\nabla, \ \ h\in\mathfrak{g}$;

	5. $\nabla\circ\pi^*=\mathbf{d}\circ\pi^*$.

From 5. it follows that $\nabla$ reduces to $\nabla_H:
\bigwedge_H(P,W)\rightarrow\bigwedge_H(P,W)$.

Note that in general $\nabla\circ\nabla\neq 0$.

Note also, that restricted to the algebra of smooth functions $\mathcal{J}(P)$
the operator $\nabla$ satisfies:
$$
\nabla\,f=\mathbf{d}f; \ \ \nabla(f.g)=f.\nabla(g)+g.\nabla(f).
$$

Consider now the basic $W$-valued differential forms on $P$, and let
$\rho$ be a representation of
$G$ in $W$ with $\rho'$ the corresponding representation of the Lie algebra
$\mathfrak{g}$ in $W$. Let $\nabla$ be the covariant exterior derivative with
respect to the connection $\omega$. Then the following relation holds:
$$
\nabla(\Phi)=\mathbf{d}\Phi+\rho'(\omega,\Phi).
$$
If $\rho$ is the adjoint representation with $\rho'$ its derivative then the
above relation takes the form
$$
\nabla(\Phi)=\mathbf{d}\Phi+[\omega,\Phi],
$$
where $\Phi$ is basic and $\mathfrak{g}$-valued.

These last formulae suggest that with every connection $\omega$ on $P$, and
representations $\rho$ and corresponding $\rho'$ of $(G,\mathfrak{g})$ in a
finite dimensional vector space $W$ we can associate the operator
$$
\mathfrak{D}:\Lambda^p(P,W)\rightarrow\Lambda^{p+1}(P,W): \ \
\mathfrak{D}(\Phi)=\mathbf{d}\Phi+\rho'(\omega,\Phi),
$$
and on $\Lambda_B(P,\mathfrak{g})$ $\mathfrak{D}$ reduces to $\nabla$. This
operator has the following important property. Let $\langle\,,\rangle$ be a
bilinear map in $W$ which is invariant under the derivative $\rho'$ of the
representation considered
(e.g. scalar product), and denote by $\langle\langle\Phi,\Psi\rangle\rangle$
the corresponding map of differential forms:
$$
\langle\langle\Phi,\Psi\rangle\rangle=
\langle\langle\alpha^i\otimes e_i,\beta^j\otimes e_j\rangle\rangle=
\alpha^i\wedge\beta^j\langle e_i,e_j\rangle.
$$
 The invariance of $\langle\,,\rangle$ leads to
$$
\mathbf{d}\langle\langle\Phi,\Psi\rangle\rangle=
\langle\langle\mathfrak{D}\Phi,\Psi\rangle\rangle+
(-1)^p\langle\langle\Phi,\mathfrak{D}\Psi\rangle\rangle, \ \
\Phi\in\Lambda^p(P,W).
$$

\vskip 0.3cm

{\bf 3.5.8. Curvature of a Principal Connection.}
 \index{curvature of a principal connection}

Let $\omega$ and $\nabla$ be the connection form and the covariant exterior
derivative on a principal bundle $\mathcal{P}=(P,\pi,B,G)$ defined by the
principal connection $\Gamma$. In terms of the NF-bracket
 the curvature $\mathcal{R}$ of $\Gamma$
should look like $\mathcal{R}=\frac12[\Gamma,\Gamma]$. The values of
$\mathcal{R}$ are in the vertical subbundle $T_V(P)\subset T(P)$. So, the
composition $(d\phi_z)^{-1}\circ\mathcal{R}$ defines a $\mathfrak{g}$-valued
2-form on $P$.

{\bf Definition.} The $\mathfrak{g}$-valued 2-form
$\Omega\in\Lambda^2(P,\mathfrak{g})$ on $P$ defined by
$$
\Omega:=\nabla\omega
$$
is called {\it curvature form} of the connection $\Gamma$.

As we shall see this definition of curvature of a principal
connection is the negative of $(d\phi_z)^{-1}\circ\mathcal{R}$,
i.e.
$$
\Omega=-(d\phi_z)^{-1}\circ\mathcal{R}.
$$
The curvature form
$\Omega$ enjoys the following properties:

	1. $\Omega$ is horizontal;

	2. $\Omega$ is equivariant:
$\mathcal{R}_a^*\Omega=Ad(a^{-1})\circ\Omega$. The proof is based on the
commutativity of $\nabla$ and $\mathcal{R}_a^*, \ a\in G$. From this property it
follows the relation: $L_{Z_h}\Omega=-ad(h)\circ\Omega, \ h\in\mathfrak{g}$.

	3. If $(Y_1,Y_2)$ are horizontal vector fields, then
$$
Z_{\Omega(Y_1,Y_2)}=-\Gamma_*[Y_1,Y_2].
$$

	4. If $\chi$ is the lifting isomorphism for $\Gamma$ then
$$
[\chi(X_1),\chi(X_2)]=\chi([X_1,X_2])-Z_{\Omega(\chi(X_1),\chi(X_2))},
\ X_1,X_2\in\mathfrak{X}(B).
$$
From property 4. it follows that the curvature form $\Omega$ is zero only if
the Lie bracket of every two horizontal vector fields is horizontal, i.e. if
the horizontal distribution defined by $\Gamma$ is completely integrable.

	5. The structure equation of Maurer-Cartan:
$$
\Omega=\mathbf{d}\omega+ \frac12[\omega,\omega].
$$
To proof this we consider the three cases:

	5.a the vector fields $X,Y$ are horizontal: then
$[\omega,\omega](X,Y)=0$. So,
 \begin{eqnarray*}
\Omega(X,Y)&=&(H^*\circ\mathbf{d}\omega)(X,Y)=\mathbf{d}\omega\circ H_*(X,Y)=
\mathbf{d}\omega(X,Y)\\
&=&X(\omega(Y))-Y(\omega(X))-\omega([X,Y])=-\omega([X,Y]).
\end{eqnarray*}
On the other hand, according to the general formula for curvature of a
distribution we must have
$$
\mathcal{R}(X,Y)=\frac12[\Gamma,\Gamma](X,Y)=
\Gamma([H_*X,H_*Y])=\Gamma([X-\Gamma(X),Y-\Gamma(Y)])
$$
$$
=\Gamma([X,Y]-[X,\Gamma(Y)]-[\Gamma(X),Y]+[\Gamma(X),\Gamma(Y)]).
$$
But $[X,\Gamma(Y)]$ and $[\Gamma(X),Y]$ are horizontal so $\Gamma$ sends them
to zero, also $\Gamma(X)=\Gamma(Y)=0$ since $X,Y$ are horizontal. We obtain
$$
\mathcal{R}(X,Y)=\Gamma([X,Y])=Z_{\omega([X,Y])}
$$
So,
$$
\Omega(X,Y)=-(d\phi_z)^{-1}\circ\mathcal{R}(X,Y)=
(\mathbf{d}\omega+\frac12[\omega,\omega])(X,Y)=-\omega([X,Y]).
$$

	5.b: $(X=Z_h,Y=Z_k)$ are vertical:
then
$$
\Omega(Z_h,Z_k)=0
$$
and
$$
-(d\phi_z)^{-1}\circ\frac12[\Gamma,\Gamma](Z_h-\Gamma(Z_h),Z_k-\Gamma(Z_k)]=
-(d\phi_z)^{-1}\circ\frac12[\Gamma,\Gamma](0,0)=0.
$$
Also,
\begin{eqnarray*}
\mathbf{d}\omega(Z_h,Z_k)&+&\frac12[\omega,\omega](Z_h,Z_k)\\
&=&Z_h\omega(Z_k)-Z_k\omega(Z_h)-\omega([Z_h,Z_k])+[\omega(Z_h),\omega(Z_k)]\\
&=&Z_h(k)-Z_k(h)-\omega([Z_h,Z_k])+[h,k]\\
&=&-\omega(Z_{[h,k]})+[h,k]=-[h,k]+[h,k]=0.
\end{eqnarray*}

	5.c: $X$-horizontal, $Y=Z_h$-vertical:
$$
\Omega(X,Z_h)=0,\ \ [\omega(X),\omega(Z_h)]=0,
$$
$$
\mathbf{d}\omega(X,Z_h)=-\omega([X,Z_h])=0, \text{since} \ \
[X,Z_h] \ \text{is horizontal}.
$$
Also,
$$
\mathcal{R}(X,Z_h)=\Gamma([X-\Gamma(X),Z_h])=\Gamma([X,Z_h])=0
$$
since $[X,Z_h]$ is horizontal.

	6. The Bianchi identity; $\nabla\Omega=0$:
$$
\nabla\Omega=\nabla(\mathbf{d}\omega+ \frac12[\omega,\omega])=
H^*\mathbf{d}(\mathbf{d}\omega)+\frac12H^*\mathbf{d}[\omega,\omega]
$$
$$
=H^*[\mathbf{d}\omega,\omega]=[H^*\mathbf{d}\omega,H^*\omega]=0.
$$
In terms of $\mathcal{R}$ this Bianchi identity looks like:
$[\Gamma,\mathcal{R}]=0$. To prove it, from the graded Jacobi
identity applied to $\Gamma$ and $\mathcal{R}=\frac12[\Gamma,\Gamma]$,
we obtain
$$
[\Gamma,\mathcal{R}]=\frac12[\Gamma,[\Gamma,\Gamma]]=
\frac12[[\Gamma,\Gamma],\Gamma]+\frac12(-1)^{1.1}[\Gamma,[\Gamma,\Gamma]],
$$
i.e.
$$
2[\Gamma,\mathcal{R}]=\frac12[[\Gamma,\Gamma],\Gamma].
$$
On the other hand, from the corresponding commutator we obtain
$$
\frac12[[\Gamma,\Gamma],\Gamma]=-\frac12[\Gamma,[\Gamma,\Gamma]].
$$
Hence
$$
2[\Gamma,\mathcal{R}]=-[\Gamma,\mathcal{R}]=0.
$$
	7. If $\Phi\in\Lambda_B(P,W)$, and $\rho$ is a representation of $G$ in
$W$,  then
$$
\nabla(\nabla \Phi)=\rho'(\Omega,\Phi).
$$
Recalling the operator $\mathfrak{D}$ (Sec.3.5.7) we get the formula
$$
\mathfrak{D}\omega=\Omega+\frac12[\omega,\omega].
$$

Finally, if $\mathcal{P}=(B\times G,\pi,B,G)$ is a trivial  principal bundle,
then with every connection form $\omega$ on $\mathcal{P}$ it is possible to
associate a $\mathfrak{g}$-valued 1-form $\theta$ on $B$ according to
$$
\omega(x,e;X_x,h)=h+\theta(x;X_x), \ \ x\in B, h\in\mathfrak{g} ,
$$
where $e$ is
the unit element of $G$. The horizontal projection $H_{(x,a)}$ in this case is
given by
$$
H_{(x,a)}(X_x,Y_a)=(X_x,-(dR_a)_e\theta(x;X_x)),
$$
where $x\in B, \ a\in G,\ X_x\in T_x(B),\ Y_a\in T_a(G)$.

%\newpage

\section{Linear connections.}

Linear connections in vector bundles are special kind of first order
differential operators acting on the crossections of the bundles. They can
induce various differential operators on the sections of the corresponding
tensor algebras, therefore, we first shall consider briefly some of the
properties of the induced tensor bundles and their crossections.
%\newpage

\vskip 0.3cm
{\bf 3.6.1. Vector bundle valued differential forms}.
\index{vector bundle valued differential forms}
\vskip 0.2cm
Let $\eta=(E,\pi,B,F)$ be a vector bundle with crossections $Sec(\eta)$. We
consider all smooth screw-symmetric multilinear maps $\Phi_x: T_x(B)\rightarrow
F_x, x\in B$, they form a linear space. The union $\bigcup_{x}\Phi_x, x\in B$
defines a new bundle space over $B$ with obvious projection $\pi:
\Phi_x\rightarrow x$, denoted by $\bigwedge(B,\eta)$, and its crossections are
denoted by $\Lambda(B,\eta)=\Sigma_{p=0}^{n}\Lambda^p(B,\eta), \ \
\Lambda^0(B,\eta)=Sec(\eta)$. The elements of $\Lambda(B,\eta)$ are called
$\eta$-{\it valued differential forms} on $B$.

According to the usual isomorphisms we get the isomorphism
$$
\Lambda(B,\eta)\leftrightarrow \Lambda(B)\otimes_B Sec(\eta).
$$
So, $\Lambda(B,\eta)$ is a $\Lambda(B)$-graded module with multiplication (in
the decomposable case)
 $$
\alpha\wedge\Phi=\alpha\wedge(\beta\otimes\sigma)=(\alpha\wedge\beta)\otimes\sigma,
 \ \ \alpha,\beta\in\Lambda(B), \ \ \sigma\in Sec(\eta).
$$
The substitution operator in $\Lambda(B)$ with respect to a vector field
$X\in\mathfrak{X}(B)$ is naturally extended to $\Lambda(B,\eta)$ according to
$$
i(X)\sigma=0, \ \ \sigma\in Sec(\eta), \ \
i(X)\Phi=i(X)(\alpha\otimes\sigma)=(i(X)\alpha)\otimes\sigma.
$$

Recalling from Sec.2.2.4 the bundle maps $\varphi^*, \varphi^{\#}, \varphi_*$
generated by the bundle map
$\varphi: \eta=(E,\pi,B,F)\rightarrow \eta'=(E',\pi',B',F')$
and the induced map
$\psi: B\rightarrow B'$ between the two base-spaces,
we obtain their natural extension to the corresponding bundle-valued
differential forms (everywhere $\sigma$ denotes corresponding section of $\eta$
or $\eta^*$):
$$
\varphi^*(\sigma)(x)=\varphi_x^*(\sigma(\psi(x)), \ \
\varphi^*\Phi(x; h_1,...,h_p)=
\varphi^*_x(\Phi_{\psi(x)}(d\psi(h_1),...,d\psi(h_p)),
$$
so, $\varphi^*(\alpha\wedge\Phi)=(\psi^*\alpha)\wedge\varphi^*\Phi$.
$$
(\varphi^{\#}\sigma)(x)=\varphi_x^{-1}(\sigma(\psi(x)), \ \
(\varphi^{\#}\Phi)(x,h_1,...,h_p)=
\varphi_x^{-1}(\psi(x);d\psi(h_1),...,d\psi(h_p).
$$
Also, we obtain
$\varphi^{\#}(\alpha\wedge\Phi)=\psi^*\alpha\wedge\varphi^{\#}\Phi$.

Finally, when the two vector bundles are on the same base $B$ and the induced
$\psi$ reduces to the identity map of $B$
$$
(\varphi_*\Phi)(x;h_1,...,h_p)=\varphi(\Phi(x;h_1,...,h_p)).
$$
In the above formulae $h_i, i=1,2,...,p$ are elements of the corresponding
tangent spaces.

If $\eta_i=(E^i,\pi^i,B,F^i), i=1,2,...,p$ and $\eta'=(E',\pi,B,F')$ are vector
bundles on the same base $B$ then every $m$-linear map $\mathcal{A}_x$
$$
\mathcal{A}_x: F^1_x\times F^2_x\times...\times F^p_x\rightarrow F'_x
$$
determines a map
$$
\mathcal{A}_*: \Lambda^p(B,\eta_1)\times...\times\Lambda^p(B,\eta_p)\rightarrow
\Lambda^{mp}(B,\eta'),
$$
and in the case of decomposable forms $\Phi^i=\alpha^i\otimes\sigma^i$
(no summation on $i$) reduces to
$$
\mathcal{A}_*(\alpha^1\otimes\sigma^1,\alpha^2\otimes\sigma^2,...,\alpha^p\otimes\sigma^p)=
(\alpha^1\wedge\alpha^2\wedge...\wedge\alpha^p)\otimes\mathcal{A}_*(\sigma^1,\sigma^2,...,\sigma^p),
$$
where $\alpha^i$ are $p$-forms on $B$ and $\sigma^i\in Sec(\eta_i)$. Thus we
have
$$
\mathcal{A}_*(\alpha\wedge\Phi_1,\Phi_2,...,\Phi_m)=
\alpha\wedge\mathcal{A}_*(\Phi_1,\Phi_2,...,\Phi_m),
$$
and  if $\alpha$ is a $p$-form on $B$ then
$$
\mathcal{A}_*(\Phi_1,...,\Phi_p,\alpha\wedge\Phi_{p+1},...,\Phi_m)=
(-1)^{qp}\alpha\wedge\mathcal{A}_*(\Phi_1,...,\Phi_m),
$$
where $q$ is equal to the sum of the degrees of $\Phi_1,\Phi_2,...,\Phi_p$.

Here are some examples (we shall consider just decomposable bundle-valued
forms).

	1. Dual vector bundles, the bilinear map given by $\langle \,,\rangle$.
$$
\mathcal{A}_*(\alpha\otimes\sigma,\beta\otimes\sigma^*)=
\alpha\wedge\beta\langle\sigma^*,\sigma\rangle.
$$
\vskip 0.3cm
	2. If $(\varphi,\psi)$ defines isomorphism between $\eta$ and $\eta'$,
 $\eta^*$ and $\eta'^*$ being the corresponding dual bundles, then
$$
\mathcal{A}_*(\varphi^*(\alpha\otimes\sigma^*),\varphi^{\#}(\beta\otimes\sigma))=
\psi^*\circ\mathcal{A}_*(\alpha\otimes\sigma^*,\beta\otimes\sigma)
=\psi^*(\alpha\wedge\beta)\langle\sigma^*,\sigma\rangle.
$$
\vskip 0.3cm
	3. $\mathcal{A}: \eta\times\eta\rightarrow\eta$ is bilinear. Then if
$\mathcal{A}$ is symmetric, we obtain for $\alpha\in\Lambda^p(B)$ and
$\beta\in\Lambda^q(B)$
$$
\mathcal{A}_*(\alpha\otimes\sigma,\beta\otimes\tau)=
\alpha\wedge\beta\otimes\mathcal{A}(\sigma,\tau)
$$
$$=
(-1)^{pq}\beta\wedge\alpha\otimes\mathcal{A}(\sigma,\tau)=
(-1)^{pq}\beta\wedge\alpha\otimes\mathcal{A}(\tau,\sigma)=
(-1)^{pq}\mathcal{A}_*(\beta\otimes\tau,\alpha\otimes\sigma).
$$
If $\mathcal{A}$ is antisymmetric, then
$$
\mathcal{A}_*(\alpha\otimes\sigma,\beta\otimes\tau)=
(-1)^{pq+1}\mathcal{A}_*(\beta\otimes\tau,\alpha\otimes\sigma).
$$

	4. With every vector bundle $\eta=(E,\pi,B,F)$ can be associated the
bundle $L_{\eta}=(\mathcal{E},\pi,B,L_F)$, where the standard fiber $L_F$ is
the set of all linear transformations $E\rightarrow E$. The compositions of
linear transformations $\varphi_x\circ\psi_x, x\in B$ is a bilinear map, so,
the corresponding map $\mathcal{A}_*:
(\Lambda(B,L_{\eta}),(\Lambda(B,L_{\eta}))\rightarrow \Lambda(B,L_{\eta})$ of
$L_{\eta}$ -valued differential forms naturally arises. It looks like (at the
point $x\in B$) $$
\mathcal{A}_*(\alpha_x\otimes\varphi_x,\beta_x\otimes\psi_x)=
\alpha_x\wedge\beta_x\otimes\mathcal{A}_*(\varphi_x,\psi_x)=
\alpha_x\wedge\beta_x\otimes (\varphi_x\circ\psi_x).
$$
On the other hand, there is a bilinear map $(L_\eta,\eta)\rightarrow\eta$
defined by the action of a linear map on its argument:
$(\varphi,y)\rightarrow\varphi(y)$. The corresponding map of $L_{\eta}$-valued
and $\eta$-valued differential forms looks like
$$
\mathcal{A}_*(\alpha_x\otimes\varphi_x,\beta_x\otimes\sigma_x)=
\alpha_x\wedge\beta_x\otimes\mathcal{A}(\varphi_x,\sigma_x)=
\alpha_x\wedge\beta_x\otimes\varphi_x(\sigma_x).
$$
Also, if $\varphi_x, x\in B$, is a linear isomorphism then we obtain the map
$$
\psi_x\rightarrow\varphi_x\circ\psi_x\circ\varphi_x^{-1},
$$
 which in turn defines
corresponding map of bundle valued differential forms. The same thing happens
with respect to tensor powers of $\eta$, with respect to the exterior powers
of $\eta$ and with respect to the symmetric powers of $\eta$. The
corresponding explicit expressions for differential forms are easy to deduce, so
we shall not reproduce them here.
\vskip 0.5cm
%\newpage
{\bf 3.6.2. Definition and elementary properties of linear connections}.
\index{linear connection}
\indent
Let $\eta=(E,\pi,B,F)$ be a vector bundle with $Sec(\eta)$ - the
corresponding $\mathcal{J}(B)$-module of sections, and $(e_1,e_2,...,e_r)$-be
the (local) basis sections, so, locally, every section of $\eta$ gets the
representation $\sigma=\sigma^i\,e_i, \ \sigma^i\in\mathcal{J}(B)$.

{\bf Definition}.
A linear connection in $\eta$ is a differential operator
$$
\nabla:
Sec(\eta)\rightarrow\Lambda^1(B,\eta),
$$
satisfying the following conditions:
\begin{align}
	1.&\nabla(\lambda\,\sigma+\mu\,\tau)=\lambda\nabla(\sigma)+\mu\nabla(\tau), \ \
\lambda,\mu \in\mathbb{R}, \ \ \sigma,\tau\in Sec(\eta);\\
	2.&\nabla(f\sigma)=df\otimes\sigma+f\nabla(\sigma),
\ \ f\in\mathcal{J}(B),\ \ \sigma\in Sec(\eta).
\end{align}

{\bf Note}: The term $df\otimes\sigma$ does NOT depend on $\nabla$, so it is the
same for all linear connections in $\eta$.

Now, since $\nabla (e_i)$ must
be 1-form with values in $\eta$, we may write
$$
\nabla(e_i)= \Gamma_{\mu i}^jdx^\mu\otimes e_j,
$$
where $\Gamma_{\mu i}^j$ are called components of $\nabla$ (in the
corresponding bases).

{\bf Conclusion}: The components  $\Gamma_{\mu i}^j$ are arbitrary
and determine $\nabla$ completely.

{\bf Definition}: A section $\sigma\in Sec(\eta)$ is called {\it parallel} with
respect to $\nabla$ if $\nabla(\sigma)=0$.

Let now $X\in \mathfrak{X}(B)$ be a vector field on $B$. The above property
$2.$ allows to form the interior product $i(X)\nabla(\sigma)$:
$$
i(X)(\nabla(\sigma^ie_i))=(i(X)d\sigma^i)\otimes e_i+
(i(X)dx^\mu)\Gamma_{\mu i}^j\varepsilon^i\otimes e_j
$$
$$
=\big(X(\sigma^i)+\sigma^iX^\mu\Gamma_{\mu i}^j\big)e_j.
$$

{\bf Note}: The operator $\nabla_X:=i(X)\circ\nabla$ does NOT depend on the
derivatives of $X$, so $\nabla_X\sigma(x), x\in B$ depends only on the
representative $X_x$ of $X$ at the point $x\in B$, i.e.
$$
\nabla_X\sigma(x)=\nabla_{X_x}\sigma.
$$

If  $\Gamma_{\mu i}^j$ are components of a linear connection in $\eta$ and
$\Psi\in\Lambda^1(B,L_\eta)$ is a $\Lambda^1(B,L_\eta)$ valued 1-form, the map
$$
\sigma\rightarrow\nabla(\sigma)+\mathcal{A}_*(\Psi,\sigma)
$$
is another linear connection in $\eta$. Conversely, every two linear connections
$\nabla_1$ and $\nabla_2$ define an element $\Psi\in\Lambda^1(B,L_\eta)$ by
$\Psi=\nabla_1-\nabla_2$.

The linear connections in the tangent bundle $\tau(B)$ are usually called
linear connections in $B$. With every linear connection in $B$ is associated
the so called {\it torsion} 2-form $S$, valued in $\tau(B)$, according to
$$
S(X,Y):=\nabla_X(Y)-\nabla_Y(X)-[X,Y].
$$

The components of a linear connection $\nabla$ in $B$ in canonical (coordinate)
bases are given by
$$
\nabla\left(\frac{\partial}{\partial x^\mu}\right)=
\Gamma_{\mu\nu}^\sigma\,dx^\nu\otimes\frac{\partial}{\partial x^\sigma}.
$$
So, in coordinate bases the components of the torsion form are given by
$$
S_{\mu\nu}^\sigma=\Gamma_{\mu\nu}^\sigma-\Gamma_{\nu\mu}^\sigma.
$$

The linear connections in $B$ with zero torsion forms are called {\it
symmetric}.

Two connections $\nabla_1$ and $\nabla_2$ in $B$ are called {\it conjugate}
if $S_{\nabla_1}=-S_{\nabla_2}$.

If $\nabla$ is a linear connection in $\eta$ then there is unique linear
connection $\nabla^*$ in the dual bundle $\eta^*$, satisfying the condition :
$$
\mathcal{A}_*(\nabla^*\sigma^*,\sigma)+
\mathcal{A}_*(\sigma^*,\nabla\sigma)=d\langle\sigma^*,\sigma\rangle.
$$
In components this looks like
$$
\mathcal{A}_*(\nabla^*(\varepsilon^i),e_j)+
\mathcal{A}_*(\varepsilon^i,\nabla(e_j))=d\langle\varepsilon^i,e_j\rangle
$$
where $\{\varepsilon^i\}$ and $\{e_j\}$ are dual bases, so, the righthand side
is zero. In fact
$$ \mathcal{A}_*(\nabla^*(\varepsilon^i),e_j)=
\langle(\Gamma^*)^i_{\mu k}dx^\mu\otimes\varepsilon^k,e_j\rangle=
(\Gamma^*)^i_{\mu k}dx^\mu\delta^k_j=(\Gamma^*)^i_{\mu j}dx^\mu.
$$
Also
$$
\mathcal{A}_*(\varepsilon^i,\nabla(e_j))=
\langle\varepsilon^i,\Gamma^k_{\mu j}dx^\mu\otimes e_k\rangle=
\Gamma^k_{\mu j}dx^\mu\delta^i_k=\Gamma^i_{\mu j}dx^\mu.
$$
So, $(\Gamma^*)^i_{\mu j}=-\Gamma^i_{\mu j}$. This allows to induce linear
connection $\hat{\nabla}$ in the bundle $L_{\eta}$ of linear maps
$(\varphi,id_B)$ according to
$$
(\hat{\nabla}\varphi)(\sigma)=\nabla(\varphi(\sigma))-\varphi(\nabla(\sigma)),
$$
where $\nabla$ is linear connection in $\eta$.

If $\nabla^{\eta}$ and $\nabla^{\zeta}$ are linear connections in the bundles
$\eta$ and $\zeta$, then connection $\nabla^{\eta\otimes\zeta}$ in the tensor
product bundle $\eta\otimes\zeta$ is induced according to
$$
\nabla^{\eta\otimes\zeta}(\sigma\otimes\tau)=
\mathcal{A}_*(\nabla^{\eta}\sigma,\tau)+\mathcal{A}_*(\sigma,\nabla^{\zeta}\tau).
$$
So, if $X$ is a vector field on $B$ we get
$$
\nabla^{\eta\otimes\zeta}_X(\sigma\otimes\tau)=
\nabla^{\eta}_X(\sigma)\otimes\tau+\sigma\otimes\nabla^{\zeta}_X(\tau),
$$
where $\sigma$ and $\tau$ are corresponding sections.

In the same way every linear connection $\nabla$ in $\eta$ induces linear
connections in each $\otimes^p\eta$, in $\wedge^p\eta$ and in $\vee^p\eta$,
$p=1,2,...n$. In particular, in view of the isomorphism between $L_\eta$ and
$\eta\otimes\eta^*$, the induced connection in the tensor product
$\eta\otimes\eta^*$ operates well enough in $L_\eta$.
\vskip 0.3cm
{\bf 3.6.3. Change of the local basis in} $Sec(\eta)$.

Let $\{e_i\}$ and $\{\hat{e}_j\}$ be two local bases in $Sec(\eta)$
corresponding to two intersecting local trivializations $U\times F$ and
$V\times F$, $U\subset B$, $V\subset B$ and $U\cap V$ is not empty,
of our vector bundle
$\eta=(E,\pi,B,F)$. We want to see how the connection components
$\Gamma_{\mu i}^j$ change when passing from $\{e_i\}$ to $\{\hat{e}_j\}$
at the point $x\in U\cap V$.

We can write
$$
\hat{e}_i(x)=\varphi_i^j(x)e_j(x), \ \ det||\varphi_i^j(x)||\neq 0,\ \
x\in U\cap V.
$$
According to the action of the linear connection $\Gamma$ we can write
$$
\nabla(e_i)=\Gamma_{\mu i}^jdx^\mu\otimes e_j, \ \
\nabla(\hat{e}_i)=\hat{\Gamma}_{\mu i}^jdx^\mu\otimes\hat{e}_j.
$$
So, in view of the above transformation for the bases (we omit writing argument
$x$ for clarity)
 $$
\nabla(\hat{e}_i)=\nabla(\varphi_i^je_j)=d\varphi_i^k\otimes e_k+
\varphi_i^k\nabla(e_k)
$$
 $$ =\frac{\partial \varphi_i^k}{\partial
x^\mu}dx^\mu\otimes e_k+ \varphi_i^j\Gamma_{\mu j}^kdx^\mu\otimes e_k $$
$$
=\left(\varphi_i^j\Gamma_{\mu j}^k+
\frac{\partial \varphi_i^k}{\partial x^\mu}\right)dx^\mu\otimes e_k
=\hat{\Gamma}_{\mu i}^kdx^\mu\otimes\hat{e}_k
$$
$$
=\hat{\Gamma}_{\mu i}^kdx^\mu\otimes\varphi_k^je_j=
\hat{\Gamma}_{\mu i}^j\varphi_j^kdx^\mu\otimes e_k.
$$
It follows
$$
\hat{\Gamma}_{\mu i}^j\varphi_j^k=\varphi_i^j\Gamma_{\mu j}^k+
\frac{\partial \varphi_i^k}{\partial x^\mu}.
$$
Multiplying by $(\varphi_k^m)^{-1}$ and restoring the $x$-dependence we obtain
finally $$
\hat{\Gamma}_{\mu i}^m(x)=\varphi_i^j(x)\Gamma_{\mu
j}^k(x)(\varphi_k^m(x))^{-1}+ (\varphi_k^m(x))^{-1}\frac{\partial
\varphi_i^k}{\partial x^\mu}(x),
$$
so, $\Gamma_{\mu i}^j$ do not define a vector bundle 1-form on $M$.
\vskip 0.3cm
{\bf 3.6.4. Linear connections in $\eta$ and connections in} $\tau(\eta)$.
\indent

We are going to show how a linear connection in $\eta=(E,\pi,B,F)$ defines a
connection, i.e. a projection operator, in $\tau(\eta)$.

 Recall the relations obtained in Sec.3.3.2 and define locally the
corresponding projections in the tangent bundle of a manifold, and the
identification of a linear space $F$ with the tangent space of $F$ at $h\in F$.
In our context now we consider the differential $dj_{\sigma(x)}, \ x\in B$ of
the inclusion map $j_{\sigma(x)}: F_x\rightarrow E$ as given in Sec.2.4.5.

Let $\sigma$ be a section in $\eta$, so, locally,
$\sigma(x)=(x,\sigma^a(x)e_a(x))$. Its differential $d\sigma_x$ sends
$T_x(B)$ to $T_{\sigma(x)}(E)$. If $\nabla$ is a linear connection in $\eta$
then we can form the expression
$$
\psi_x:= d\sigma_x-dj_{\sigma(x)}\circ\nabla\sigma
$$
Let $\{x^\mu\}, \ \mu=1,2,...,n,$ be local coordinates on $U\subset B$, and
$\{e_a\}, a=n+1,n+2,...,n+r,$ give basis local sections in the restriction
of $\eta$ on $U\subset B$. If $\Gamma^a_{\mu b}$ are the corresponding
components of $\nabla$ we obtain (omitting the dependence on $x$ wherever
possible)
$$
\psi_x=-\sigma^b\Gamma_{\mu b}^a\,dx^\mu\otimes
dj_{\sigma(x)}(e_a)= -\sigma^b\Gamma_{\mu b}^a\,dx^\mu\otimes e_a,
$$
where we have identified $dj_{\sigma(x)}(e_a(x))$ with $e_a(x)$.

Now, let $\{\varepsilon^a\}$ be the dual basis for $\{e_a\}$, correspondingly
identified with the dual to $dj_{\sigma(x)}(e_a(x))$. The corresponding
vertical $V^{\nabla}$ and horizontal $H^{\nabla}$ projections in $\tau(E)$
are given by:
$$
V^{\nabla}(\sigma)=(\varepsilon^a+\sigma^b\Gamma_{\mu b}^adx^\mu)\otimes e_a, \
\
H^{\nabla}(\sigma)=
dx^\mu\otimes\left(\frac{\partial}{\partial x^\mu}-
\sigma^b\Gamma_{\mu b}^ae_a\right).
$$
It is easily verified that $V^{\nabla}(\sigma)\circ H^{\nabla}(\sigma)=0$.

Vice versa, if we know the vertical projection $V$, we can define
the covariant derivative of every $\sigma\in Sec(\eta)$. For
$\sigma\in Sec(\eta)$ the linear connection map is given by  the combination
$$
(dj_{\sigma(x)})^{-1}\circ V_{\sigma(x)}\circ (d\sigma)_x.
$$
In fact, respecting the above mentioned identifications and the explicit
expression for $d\sigma_x$
$$
d\sigma_x=dx^\nu\otimes\frac{\partial}{\partial x^\nu}+
\frac{\partial \sigma^a}{\partial x^\nu}dx^\nu\otimes e_a,
$$
we obtain
$$
(d\sigma)_x\left(\frac{\partial}{\partial x^\mu}\right)=
\frac{\partial}{\partial x^\mu}+
\frac{\partial \sigma^a}{\partial x^\mu}e_a.
$$
Now,
$$
V_{\sigma(x)}\left(\frac{\partial}{\partial x^\mu}+
\frac{\partial \sigma^a}{\partial x^\mu}e_a\right)=
(\varepsilon^a+\sigma^b\Gamma_{\mu b}^adx^\mu)\otimes e_a
\left(\frac{\partial}{\partial x^\mu}+
\frac{\partial \sigma^b}{\partial x^\mu}e_b\right)
$$
$$
=\left(\frac{\partial \sigma^b}{\partial x^\mu}+
\sigma^a\Gamma_{\mu a}^bdx^\mu\right)e_b.
$$
Following our identification convention we come to
$$
(dj_{\sigma(x)})^{-1}\circ V_{\sigma(x)}\circ (d\sigma)_x
\left(\frac{\partial}{\partial x^\mu}\right)=
\left(\frac{\partial \sigma^b}{\partial x^\mu}+
\sigma^a\Gamma_{\mu a}^bdx^\mu\right)e_b.
$$

\section{Curvature of Linear Connections}
  \index{curvature of linear connection}
{\bf 3.7.1. Covariant exterior derivative with respect to a linear\\
connection}.

 Let's recall how the Leibnitz differential has been developed so far.

	$d(f.g)=df.g+f.dg, \ \ f,g\in\mathcal{J}(M)$,

	$d(f\alpha)=df\wedge\alpha+f\,\mathbf{d}\alpha, \ \
\alpha\in\Lambda^p(M)$,

	$d(f^ie_i)=df^i\otimes e_i, \ \ f^ie_i\in\mathcal{J}(M,F)$,

	$\mathbf{d}(\alpha\otimes(f^ie_i))=\mathbf{d}\alpha\otimes
f^ie_i+(-1)^p\alpha\wedge df^i\otimes e_i$,

	$\mathbf{d}(\alpha\wedge(\beta^i\otimes e_i))=
(\mathbf{d}\alpha\wedge\beta^i)\otimes e_i+
(-1)^p(\alpha\wedge\mathbf{d}\beta^i)\otimes e_i, \ \ \beta^i\in\Lambda(M)$

	$D_{\varphi}(\alpha^i\otimes
e_i,\beta^j\otimes k_j)=
\mathbf{d}(\alpha^i\wedge\beta^j)\otimes\varphi(e_i,k_j)$

We are going now to extend the exterior derivative $\mathbf{d}$ in $\Lambda(M)$
to exterior covariant derivative $\mathbf{D}$ in vector bundle
valued differential forms $\Lambda(M,\eta)$, where we make
use of our usual notation for the vector bundle $\eta=(E,\pi,M,F)$ and for the
linear connection $\nabla$. We recall that $\Lambda(M,\eta)$ is a module with
respect to $\mathcal{J}(M)$ and a graded module with respect to the graded
algebra $\Lambda(M)$.

Recall the defining property for $\nabla$:
$$
\nabla(f\sigma)=df\otimes\sigma+f\nabla \sigma, \ \ f\in\mathcal{J}(M), \ \
\sigma\in Sec(\eta)=\Lambda^0(M,\eta).
$$
Let now $\Phi\in\Lambda^p(M,\eta)$ be decomposable: $\Phi=\alpha\otimes\sigma$.
We define $\mathbf{D}\Phi$ as follows:
$$
\mathbf{D}\Phi=\mathbf{D}(\alpha\otimes\sigma):=\mathbf{d}\alpha\otimes\sigma+
(-1)^p\alpha\wedge\nabla(\sigma).
$$
If in this relation $\alpha$ is 1-form on $B$ then we obtain
$$
\mathbf{D}(\Phi)(X,Y)=\mathbf{D}_X\Phi(Y)-\mathbf{D}_Y\Phi(X)-\Phi([X,Y]).
$$
Since $\mathbf{D}(f\alpha\otimes\sigma)=\mathbf{D}(\alpha\otimes f\sigma)$ the
definition is correct. If $\{e_i\}$ is a local basis for $Sec(\eta)$ then
$$
\mathbf{D}\Phi=\mathbf{D}(\alpha^i\otimes e_i)=
\mathbf{d}\alpha^i\otimes
e_i+(-1)^p\alpha^i\wedge\Gamma_{\mu i}^jdx^{\mu}\otimes e_j.
$$

We note the following important property of $\mathbf{D}$:
$$
\mathbf{D}\circ\mathbf{D}(\alpha\wedge\Phi)=
\alpha\wedge\mathbf{D}\circ\mathbf{D}\Phi.
$$
The proof is elementary, in fact
$$
\mathbf{D}\circ\mathbf{D}(\alpha\wedge\Phi)=
\mathbf{D}(\mathbf{d}\alpha\wedge\Phi+(-1)^p\alpha\wedge\mathbf{D}\Phi)
$$
$$
=(\mathbf{d}\mathbf{d}\alpha)\wedge\Phi+(-1)^{p+1}\mathbf{d}\alpha\wedge\mathbf{D}\Phi
+(-1)^p\mathbf{d}\alpha\wedge\mathbf{D}\Phi
+(-1)^{2p}\alpha\wedge\mathbf{D}\circ\mathbf{D}\Phi
$$
$$
=\alpha\wedge\mathbf{D}\circ\mathbf{D}\Phi.
$$
Therefore,
$$
\mathbf{D}\circ\mathbf{D}(f\Phi)=f\,\mathbf{D}\circ\mathbf{D}\Phi,
$$
i.e. $\mathbf{D}\circ\mathbf{D}$ is $\mathcal{J}(M)$-linear, and in this way it
defines an element of
$$
Sec(\eta^*)\otimes\Lambda^2(M,\eta)=
Sec(\eta^*)\otimes\Lambda^2(T^*M)\otimes Sec(\eta)
$$
$$
=\Lambda^2(T^*M)\otimes (Sec(\eta^*)\otimes Sec(\eta))=
\Lambda^2(T^*M;\eta^*\otimes\eta).
$$

The exterior covariant derivative allows the usual Lie derivative $L_X$
in $\Lambda^p(M)$ to be extended to {\it covariant Lie derivative}
\index{covariant Lie derivative}
with respect
to a $q$-vector $Z\in\mathfrak{X}^q(M), q\leq p$,
denoted by $\mathcal{L}^{\nabla}_Z:
\Lambda^p(M,\eta)\rightarrow\Lambda^{(p-q+1)}(M,\eta)$ according to
$$
\mathcal{L}^{\nabla}_Z\Phi=\mathbf{D}\circ i_Z\Phi-
(-1)^{deg(Z).deg(\mathbf{D})}i_Z\circ\mathbf{D}\Phi, \ \ deg(\mathbf{D})=1.
$$
The corresponding generalization with respect to a bilinear map
$\varphi:\eta_1\times\eta_2\rightarrow\Sigma$, where $\Sigma$ is another
vector bundle on $M$, $Z=Z^i\otimes e_i$ is a $\eta_1$-valued $q$-vector,
$\Phi=\alpha^j\otimes k_j$ is a $\eta_2$-valued $p$-form, will look like
($deg(\mathbf{D})=1$)
\index{$\varphi$-extended covariant Lie derivative}
$$
\mathcal{L}^{(\nabla,\nabla',\varphi)}_Z\Phi=\mathbf{D'}\circ
i_Z^{\varphi}\Phi- (-1)^{deg(Z)}i_Z^{\varphi}\circ\mathbf{D}\Phi,
$$
where $\mathbf{D'}$ is defined by a linear connection $\nabla'$ in $\Sigma$.
So, $Z$ may be called $(\nabla,\nabla',\varphi)$ - symmetry of $\Phi$ if
$\mathcal{L}^{(\nabla,\nabla',\varphi)}_Z\Phi=0$.

\vskip 0.3cm

{\bf 3.7.2. Curvature of a linear connection}

	{\bf Definition}. The element $\mathcal{R}:=\mathbf{D}\circ\mathbf{D}$
is called {\it curvature} of the linear connection $\nabla$.
\index{curvature of linear connection}
The following relation holds:
$$
\mathcal{R}(X,Y)=\nabla_{X}\nabla_{Y}-\nabla_{Y}\nabla_{X}-\nabla_{[X,Y]}, \ \
X,Y\in\mathfrak{X}(M).
$$
We obtain also
$$
\mathcal{R}(fX,Y)=\mathcal{R}(X,fY)=f\,\mathcal{R}(X,Y).
$$
So, $\mathcal{R}: (X,Y,\sigma)\rightarrow\mathcal{R}(X,Y,\sigma)$ is 3-linear
map.

In components:
$$\mathcal{R}=\mathcal{R}_{\mu\nu,i}^jdx^\mu\wedge
dx^\nu\otimes\varepsilon^i\otimes e_j, \ \ \mu,\nu=1,2,...,dim(B), \
i,j=1,2,...,dim(F).
 $$

As an example let's compute the curvature of the connection
$$
\nabla=\mathbf{d}+\Psi, \
\Psi=\Psi_{\mu i}^jdx^\mu\otimes \varepsilon^i\otimes e_j
$$
in the trivial bundle $\eta=B\times F$. Omitting the index $\mu$ we obtain
$$
\mathcal{A}_*(\Psi,\Psi)=\Psi^k_j\wedge\Psi^n_m\otimes(\varepsilon^j\otimes
e_k)\circ(\varepsilon^m\otimes e_n)
$$
$$
=\Psi^k_j\wedge\Psi^n_m\delta^j_n\otimes(\varepsilon^m\otimes e_k)=
\Psi^k_j\wedge\Psi^j_m\otimes(\varepsilon^m\otimes e_k).
$$
Hence, $\mathcal{A}_*(\Psi,\Psi)(e_i)=\Psi^k_j\wedge\Psi^j_i\otimes e_k$.

Computing $\nabla(e_i)=\mathbf{D}(e_i)$ we obtain $\Psi^k_i\otimes e_k$, since
$e_i$ are constant basis vectors in $F$. Now for $\mathbf{D}(\mathbf{D}(e_i))$
we obtain
$$
\mathbf{D}(\mathbf{D}(e_i))=\mathbf{D}(\Psi^k_i\otimes e_k)=
\mathbf{d}\Psi^k_i\otimes e_k+(-1)^1\Psi^k_i\wedge \nabla(e_k)
$$
$$
=\mathbf{d}\Psi^k_i\otimes e_k-\Psi^k_i\wedge\Psi^m_k\otimes e_m=
(\mathbf{d}\Psi^j_i+\Psi^j_k\wedge\Psi^k_i)\otimes e_j.
$$
Therefore, for the curvature $\mathcal{R}$ we obtain
$$
\mathcal{R}=\mathbf{d}\Psi+\mathcal{A}_*(\Psi,\Psi).
$$

We are going to prove the Bianchi identity satisfied by the curvature 2-form.
Recall from Sec.3.6.2. that a linear connection $\nabla$ in $\eta$ induces
linear connection $\hat{\nabla}$ in $\eta^*\otimes\eta$, i.e. in $L_\eta$,
according to
$$
(\hat{\nabla}(\varphi))(\sigma)=\nabla(\varphi(\sigma))-\varphi(\nabla(\sigma))
,\ \ \varphi\in Sec(L_\eta),
$$
and corresponding exterior covariant derivative
$\hat{\mathbf{D}}$ in $\Lambda(B,L_\eta)$ according to
$$
(\hat{\mathbf{D}}(\alpha\otimes\varphi))(\sigma)=
\mathbf{D}(\alpha\otimes\varphi)(\sigma))-
\mathcal{A}_*(\alpha\otimes\varphi,\mathbf{D}\sigma).
$$

The curvature 2-form $\mathcal{R}$, defined by $\nabla$ in $\eta$, takes values
in $L_\eta$, so, we can ask how much is $\hat{\nabla}\mathcal{R}=
\hat{\mathbf{D}}(\mathcal{R})$. Recalling that $(\mathbf{D}\circ\mathbf{D})$ is
linear with respect to $\Lambda(B)$, we obtain for a decomposable
$\eta$-valued form $\alpha\otimes\sigma, \ \alpha\in\Lambda(B),
\sigma\in Sec(\eta)$:
$$
(\hat{\mathbf{D}}(\mathcal{R}))(\alpha\otimes\sigma)=
\mathbf{D}(\mathbf{D}\circ\mathbf{D}(\alpha\otimes\sigma))-
\mathbf{D}\circ\mathbf{D}(\mathbf{D}(\alpha\otimes\sigma))
$$
$$
=\mathbf{D}(\alpha\wedge\mathbf{D}\circ\mathbf{D}(\sigma))-
\mathbf{D}(\alpha\wedge\mathbf{D}\circ\mathbf{D}(\sigma))=0.
$$

Finally, consider the tangent bundle $\tau(B)$ and a linear connection $\nabla$
in $\tau(B)$. In every tangent bundle there is a canonical $\tau(B)$-valued
1-form $\omega$, defined by $\omega(X)=X, X\in\mathfrak{X}(B)$. We obtain
$$
\mathbf{D}(\omega)(X,Y)=\nabla_{X}\omega(Y)-\nabla_{Y}\omega(X)-\omega([X,Y])
$$
$$
=\nabla_XY-\nabla_YX-[X,Y]=S(X,Y),
$$
where $S$ is the torsion of $\nabla$. So, we can write
$$
\mathbf{D}\omega=S, \ \ \text{and} \ \
\mathbf{D}\circ\mathbf{D}\omega=\mathcal{A}_*(\mathcal{R},\omega).
$$
We note also, that if we consider $\omega$ as an element in $L_{\tau(B)}$, in
fact the identity map in $\tau(B)$, then with respect to the induced connection
$\hat{\nabla}$ in $L_{\tau(B)}$ we easily obtain
$$
\hat{\nabla}\omega=\hat{\nabla}(id_{\tau(B)})=0.
$$
\vskip 0.3cm

{\bf 3.7.3. Generalized parallelism}
\index{generalized parallelism}
\indent

We give here a more general view on the concept of parallelism.
Recall that a section $\sigma$ in a vector bundle is called parallel with
respect to the linear connection $\nabla$ if $\nabla\sigma=0$.

 We start with the algebraic structure to be used further in the
bundle picture. The basic concepts to be used are the {\it tensor product}
$\otimes$ of two linear spaces (we shall use the same term {\it linear space}
for a vector space over a field, and for a module over a ring, and from the
context it will be clear which case is considered) and {\it bilinear maps}.

Let
$(U_1,V_1)$, $(U_2,V_2)$ and $(U_3,V_3)$ be three couples of linear spaces, and
let
$$
\Phi:  U_1\times U_2\rightarrow U_3 \ \ \text{and} \ \
\varphi: V_1\times V_2\rightarrow V_3
$$
be two bilinear maps. Then we can form
the tensor products
$$
U_1\otimes V_1, \ \  U_2\otimes V_2, \ \ U_3\otimes V_3 .
$$
Consider now the elements
$$
(u_1\otimes v_1)\in U_1\otimes V_1 \ \ \text{and} \ \
(u_2\otimes v_2)\in U_2\otimes V_2.
$$
Apply now the given bilinear maps $\Phi,\varphi$ as follows:
$$
(\Phi,\varphi)(u_1\otimes v_1,u_2\otimes v_2)=
\Phi(u_1,u_2)\otimes\varphi(v_1,v_2).
$$
The obtained element $\Phi(u_1,u_2)\otimes\varphi(v_1,v_2)$ is in
$U_3\otimes V_3$.

We give now the corresponding bundle picture. Let $M$ be a smooth
n-dimensional real manifold. We assume that the following vector bundles over
$M$ are constructed:
 $\xi_i, \eta_i$, with standard fibers $U_i,V_i$ and sets of
sections $Sec(\xi_i), Sec(\eta_i), i=1,2,3$.

Assume the two bundle maps are given: $(\Phi,id_M):
\xi_1\times\xi_2\rightarrow\xi_3$ and $(\varphi,id_M):
\eta_1\times\eta_2\rightarrow\eta_3$. Then if $\sigma_1$ and $\sigma_2$ are
sections of $\xi_1$ and $\xi_2$ respectively, and $\tau_1$ and $\tau_2$ are
sections of $\eta_1$ and $\eta_2$ respectively, we can form an element of
$Sec(\xi_3\otimes\eta_3)$:
%\begin{equation}
$$
(\Phi_x,\varphi_x)(\sigma_1(x)\otimes\tau_1(x),\sigma_2(x)\otimes\tau_2(x))
$$
$$
=
\Phi_x(\sigma_1(x),\sigma_2(x))\otimes\varphi_x(\tau_1(x),\tau_2(x)), \ \ x\in
M.
$$
%\end{equation}

Let now $\tilde\xi$ be a new vector bundle on $M$ and $\sigma_2\in
Sec(\xi_2)$ be obtained by the action of the differential operator
$$
D:Sec(\tilde\xi)\rightarrow Sec(\xi_2)
$$
on a section $\tilde\sigma$ of
$\tilde\xi$, so we can form the section (instead of $\sigma_1$ we write just
$\sigma$)
$$\Phi(\sigma,D\tilde\sigma)\otimes\varphi(\tau_1,\tau_2)\in
Sec(\xi_3\otimes\eta_3).
$$ We give now the following
\vskip 0.4cm
\noindent
{\bf Definition}:
The section $\tilde\sigma$ will be called
$(\Phi,\varphi;D)$-{\it parallel} with respect to $\sigma$ if
$$
%\begin{equation}
(\Phi,\varphi;D)(\sigma\otimes \tau_1, \tilde\sigma\otimes \tau_2)=
(\Phi,\varphi)(\sigma\otimes \tau_1, D\tilde\sigma\otimes \tau_2)
$$
$$=
\Phi(\sigma,D\tilde\sigma)\otimes\varphi(\tau_1,\tau_2)=0.
$$
%\end{equation}
\vskip 0.4cm
The map $\Phi$
"projects" the "changes" $D\tilde\sigma$ of the section $\tilde\sigma$ on the
 section $\sigma$ ($\sigma$ may depend on $\tilde\sigma$), and $\varphi$
"works" usually on the (local) bases of the bundles where $\sigma$ and
$D\tilde\sigma$ take values.

Here are two examples.

	1. Let $\xi_1=\tau(M)=\tilde\xi$,
$\xi_2=\Lambda^1(M)\otimes_M\xi_1$,
and $\eta_1=\eta_2=M\times\mathbb{R}$. Also, $\Phi(X,\nabla\sigma)=\nabla_X
\sigma$ and $\varphi(f,g)=f.g$. Hence, we obtain
$$
(\Phi,\varphi;\nabla)(X\otimes 1, \sigma\otimes 1)=           %5%
(\Phi,\varphi)(X\otimes 1,(\nabla \sigma)\otimes 1) =\nabla_X \sigma\otimes
1=\nabla_X \sigma,
$$
and the section $\sigma$ is called shortly $\nabla$-{\it
parallel} with respect to $X$ if $\nabla_X\sigma=0$.

	2. Consider the case
 $\tilde\xi$ being the bundle of $\mathfrak{g}$-valued exterior $p$-forms
on $M$ with the available differential operator {\it exterior derivative}
$\mathbf{d}:
\Lambda^p(M,\mathfrak{g})\rightarrow\Lambda^{p+1}(M,\mathfrak{g})$,
where $\mathfrak{g}$ is a Lie algebra.
Let  "$\Phi$" be the exterior product in  $\Lambda^p(M)$,
and "$\varphi$" be the Lie bracket $[\,,]$ in $\mathfrak{g}$.
Let finally
$$
\xi_1=\Lambda^p(M), \ \ \tilde{\xi}=\Lambda^r(M), \ \ \xi_2=\Lambda^{q}(M), \ \
\eta_1=\eta_2=M\times\mathfrak{g},
$$
$$
\sigma_1=\alpha^i\otimes E_i\in\Lambda^p(M,\mathfrak{g}), \ \
\tilde{\sigma}=\beta^j\otimes E_j\in \Lambda^{r}(M,\mathfrak{g})
$$
where $\{E_i\}$ is a basis of $\mathfrak{g}$, and a summation over the repeated
indices is understood.

In this case our definition acts as follows:
$$
(\wedge,[\,,];\mathbf{d})(\alpha^i\otimes E_i, \beta^j\otimes E_j)=
\alpha^i\wedge\mathbf{d}\beta^j\otimes [E_i,E_j]=0,
$$
meaning that $\beta^j\otimes E_j$ is $(\wedge,[\,,];\mathbf{d})$-parallel with
respect to $\alpha^i\otimes E_i$.

%\newpage
{\bf 3.7.4. Riemannian connections and curvature in a vector bundle}

Recall that a vector bundle $\eta=(B,\pi,F,B)$ is called {\it riemannian} if a
symmetric nondegenerate bilinear form $g_x: F_x\times F_x\rightarrow
\mathbb{R}$, smoothly dependent on $x\in B$ is introduced, so, $g\in
Sec(\bigvee^2\eta^*)$.

Let now $\nabla$ is a linear connection in $\eta$ (further $\nabla$ will
denote all induced connections). The covariant derivative of $g$ lives in the
space $\Lambda^1(B,\bigvee^2\eta^*)$, and we have
$$
d(g(\sigma,\tau))=(\nabla g)(\sigma,\tau)+g(\nabla \sigma,\tau)+g(\sigma,\nabla
\tau), \ \ \sigma,\tau\in Sec(\eta).
$$
\vskip 0.3cm
{\bf Remark:} Following our previous notation we should write e.g.
$\mathcal{A}_*(g;\nabla\sigma,\tau)$, but in order to simplify notation we
write just $g(\nabla \sigma,\tau)$, since it is clear what is meant here. This
simplified notation will be use throughout this subsection.
\vskip 0.3cm
{\bf Definition}. The connection $\nabla$ is called {\it riemannian} if $g$ is
constant (parallel) with respect to $\nabla$, i.e. if $\nabla g=0$.
\index{riemannian connection and curvature}
Thus, for riemannian connections we can write
$$
d(g(\sigma,\tau))=g(\nabla\sigma,\tau)+g(\sigma,\nabla\tau).
$$

Clearly, all induced riemannian metrics are constant with respect to the
corresponding induced linear connections.

If $\sigma\in Sec(\eta)$ satisfies $g(\sigma,\sigma)=1$ then
$d(g(\sigma,\sigma))=0$, so
$$
g(\nabla\sigma,\sigma)+g(\sigma,\nabla\sigma)=2g(\sigma,\nabla\sigma)=0.
$$
If $\{e_i\}$ are local basis vectors in $\eta$ we obtain
$$
d(g(e_i,e_j))=g(\Gamma_i^ke_k,e_j)+g(e_i,\Gamma_j^ke_k)=
\Gamma_i^ke_kg(e_k,e_j)+\Gamma_j^ke_kg(e_i,e_k)
$$
$$
=
(\Gamma_{\mu i}^kg_{kj}+\Gamma_{\mu j}^kg_{ik})dx^\mu=
\frac{\partial g_{ij}}{\partial x^\mu}dx^\mu.
$$

Since $g(\sigma,\tau)$ is a function on $B$ its differential
$d(g(\sigma,\tau))$ is an exact 1-form, so $d\circ dg(\sigma,\tau)=0$. On the
other hand
$$
d\circ d(g(\sigma,\tau))=d(g(\nabla\sigma,\tau))+d(g(\sigma,\nabla\tau))
$$
$$
=g(\nabla\circ\nabla\sigma,\tau)+2g(\nabla\sigma,\nabla\tau)+
g(\sigma,\nabla\circ\nabla\tau).
$$
Making use again of the local basis vectors $\{e_i\}$ we get
$$
g(\nabla e_i,\nabla e_j)=g(\Gamma_i^k\otimes e_k,\Gamma_j^m\otimes e_m)=
\Gamma_i^k\wedge\Gamma_j^mg(e_k,e_m)=0.
$$
Therefore
$$
g(\nabla\circ\nabla e_i,e_j)+g(e_i,\nabla\circ\nabla e_j)=
g(\mathcal{R}_i^ke_k,e_j)+g(e_i,\mathcal{R}_j^ke_k)
$$
$$
=\mathcal{R}_i^kg_{kj}+\mathcal{R}_j^kg_{ik}=
\mathcal{R}_{ij}+\mathcal{R}_{ji}=0.
$$
Hence, the Riemann curvature tensor $\mathcal{R}_{\mu\nu ik}$ satisfies
$$
\mathcal{R}_{\mu\nu, ik}=-\mathcal{R}_{\mu\nu, ki}=
-\mathcal{R}_{\nu\mu, ik}=\mathcal{R}_{\nu\mu, ki},
$$
$$
\ \ \mu,\nu=1,2,...,dim(B),\ \ i,k=1,2,...,dim(F).
$$
\vskip 0.3cm
%\newpage
{\bf 3.7.5. Riemannian connections and curvature in a tangent bundle}

The above obtained relation connecting $g$ in the vector bundle $\eta$ with the
components of $\nabla, \nabla(g)=0$, reduces in the case $\eta=\tau(B)$ to
$$
\Gamma_{\mu\nu}^{\alpha}g_{\alpha\beta}+\Gamma_{\mu\beta}^{\alpha}g_{\nu\alpha}=
\frac{\partial g_{\nu\beta}}{\partial x^\mu}.
$$
Now if the torsion of $\nabla$ is zero, i.e.
$\Gamma_{\mu\nu}^{\alpha}=\Gamma_{\nu\mu}^{\alpha}$, then
$\Gamma_{\mu\nu}^{\alpha}$ can be represented in terms of the derivatives of
$g_{\mu\nu}$. The solution looks like
$$
\Gamma_{\mu\nu}^{\alpha}=\frac12g^{\alpha\sigma}\left(
\frac{\partial g_{\sigma\nu}}{\partial x^\mu}+
\frac{\partial g_{\mu\sigma}}{\partial x^\nu}-
\frac{\partial g_{\mu\nu}}{\partial x^\sigma}\right), \ \
g^{\mu\sigma}g_{\sigma\nu}=\delta_\nu^\mu.
$$

The following result (Ricci lemma) is important:

If $S\in\Lambda^2(B,\tau(B))$
then there is just one riemannian connection $\nabla$ in $\tau(B)$ with torsion
equal to $S$.

The riemannian connection with torsion $S=0$, i.e. with symmetric connection
coefficients: $\Gamma_{\mu\nu}^{\alpha}=\Gamma_{\nu\mu}^{\alpha}$,
is called {\it Levi-Civita} connection.

The Riemann curvature tensor defined by the Levi-Civita connection acquires the
following additional properties:
$$
\mathcal{R}_{\mu\nu,\alpha\beta}=\mathcal{R}_{\alpha\beta,\mu\nu} \ \,
$$
$$
\mathcal{R}_{\mu\nu,\alpha\beta}+\mathcal{R}_{\nu\alpha,\mu\beta}+
\mathcal{R}_{\alpha\mu,\nu\beta}=0 \ .
$$
The first of these two properties can be interpreted as a symmetric linear map
$$
\mathcal{R}:
\mathfrak{X}(M)\wedge\mathfrak{X}(M)\rightarrow\mathfrak{X}(M)\wedge\mathfrak{X}(M).
$$
The Bianchi identity looks in components as follows: $$
\nabla_\sigma\mathcal{R}_{\mu\nu,\alpha\beta}+
\nabla_\mu\mathcal{R}_{\nu\sigma,\alpha\beta}+
\nabla_\nu\mathcal{R}_{\sigma\mu,\alpha\beta}=0.
$$
The number $N$ of algebraically independent components of $\mathcal{R}$ in this
case is $$ N=\frac{n^2(n^2-1)}{12}, \ \ n=dim(B). $$

Recall now the divergence operator $\delta$ from Sec.2.8.3. Choosing
a volume form $\omega_o$ on our (pseudo)riemannian manifold $(M,g)$
(note our notation: $(-1)^{ind(g)}=\kappa_{g}$, where $ind(g)$ denotes
the number of minuses of the signature of $g$), according to
$$
\omega_o=\sqrt{|det(g_{\mu\nu})|}dx^1\wedge...\wedge dx^n,
$$
we orient $M$ by $\omega_o$ and
make use of the induced by $g$ isomorphism
$\tilde{g}^p\,:\mathfrak{X}^p(M)\rightarrow\Lambda^p(M)$ to define the
isomorphism $*$ between $\Lambda^p(M)$ and $\Lambda^{n-p}(M)$ according to
$$
*_p:=\kappa_gD^p\circ(\tilde{g}^p)^{-1}:
\Lambda^p(M)\rightarrow\Lambda^{n-p}(M).
$$
We obtain now the $g$-modified Poincare isomorphism
 $D^p_{g}=\kappa_g*_{p}\circ\,(\tilde{g}^p)$, and note that the quantity
$i_{\Phi}(D^p_g\Phi), \Phi\in\mathfrak{X}^p(M)$, is NOT now
always zero.

Another approach to defining the $*$-operator is based on the relation
$$ \alpha\wedge*\beta=\kappa_gg(\alpha,\beta)\omega_o , $$ where $\alpha$ and
$\beta$ are $p$-forms. We note that, the coefficient $\kappa_g$ is introduced
here for convenience in view of our future work with Minkowski space-time
$(M,\eta)$ with signature $sign(\eta)=(-,-,-,+)$.
\index{divergence operator, laplacian operator}
Having the $*$-operator, we can define the corresponding
divergence operator $\delta$ and laplacian operator $\Delta$ according to
$$
\delta:=(-1)^p*^{-1}\mathbf{d}\,* :\Lambda^p(M)\rightarrow\Lambda^{p-1}(M), \ \
$$
$$
\Delta=\mathbf{d}\delta+\delta\mathbf{d}: \
\Lambda^p(M)\rightarrow\Lambda^{p}(M).
$$
Having in view this definition of
 $*$ it can be shown that on $p$-forms we obtain
$$
\delta_p=\kappa_g(-1)^{(pn+n+1)}\,*\,\mathbf{d}\,*_p.
$$
If $\alpha$ is a p-form, in components we obtain
$$
(\delta\alpha)_{\nu_{1}\nu_{2}...\nu_{p-1}}=
-\nabla_\sigma\alpha^{\sigma}\,_{\nu_{1}\nu_{2}...\nu_{p-1}} ,
$$
where $\nabla$ is the Levi-Civita covariant derivative.

If the vector field $X$ is an infinitesimal isometry : $L_{X}g=0$, then $X$
satisfies the equations
$$
\Delta(\tilde{g}(X))=2\mathcal{R}_{\mu\nu}X^\mu
dx^\nu; \ \ \delta(\tilde{g}(X))=0,
$$
where $\mathcal{R}_{\mu\nu}=\mathcal{R}^\sigma\,_{\mu,\sigma\nu}$
is the corresponding Ricci tensor. Conversely, if the vector field $X$
satisfies these equations it is local isometry.

From the above relations it follows that if the Ricci tensor is zero,
$\mathcal{R}_{\mu\nu}=0$, which is the case of vacuum gravitational fields in
General relativity, then every local isometry $X$ defines closed $(n-2)$-form.
In fact, in such a case $$ \Delta(\tilde{g}(X))=\delta\,\mathbf{d}\tilde{g}(X)=
-\kappa_{g}*\,\mathbf{d}\,*\mathbf{d}\tilde{g}(X)=
2\mathcal{R}_{\mu\nu}X^\mu dx^\nu=0.
$$
So, the $(n-2)$-form $\alpha=*\,\mathbf{d}\tilde{g}(X)$ satisfies
$\mathbf{d}\alpha=\mathbf{d}*\,\mathbf{d}\tilde{g}(X)=0$.

Finally we note that if a $p$-form $\alpha$ on a Riemannean manifold satisfies
the equation $\alpha\wedge *\mathbf{d}\alpha=0$, which is equivalent to
$i(\tilde{g}^{-1}(\alpha))\mathbf{d}\alpha=0$, it was called {\it autoclosed}
(see paper No.7 in the List of studies of the authors, p.377).

\vskip 1.3cm
{\bf Literature}
\vskip 0.2cm
\addcontentsline{toc}{subsection}{{\bf Literature}}
\vskip 0.4cm
1. {\bf N. Bourbaki}, {\it Algebra I: Chapters I-III}, Springer, 1989

2. {\bf W.H. Greub}, {\it Linear Algebra}, third edition, Springer, 1967

3. {\bf W.H. Greub}, {\it Multilinear Algebra}, second edition, Springer, 1978

4. {\bf S. Kobayashi and K. Nomizu}, {\it Foundations of Differential
Geometry}, Vols. I-II, Interscience, New York, 1963 and 1969.

5. {\bf W.H. Greub, S.Halperin, R.Vanstone}, {\it Connections, Curvature, and
Cohomology}, Vols. I-II, Academic Press, 1972-1973

6. {\bf C. Godbillon},  {\it Geometrie differentielle et mecanique
analytiqe}, Hermann, Paris (1969)

7. {\bf P. Michor}, {\it Topics in Differential Geometry}, AMS, 2008

8. {\bf S. Vacaru}, et al., arXiv/gr-qc/0508023v2

9. {\bf H. Cartan}, {\it Calcul differentiel. Formes differentielles}, Herman,
Paris, 1967

10. {\bf A. Kushner, V. Lychagin, V. Rubtsov,}, {\it Contact Geometry and
Non-linear Differential Equations}, Cambridge University Press 2007

11. {\bf S. S. Chern, W. H. Chen, K. S. Lam,}, {\it Lectures on Differential
Geometry}, World Scientific, Reprinted 2000.

12. {\bf I. Tamura}, {\it Topology of Foliations: An Introduction},
Amer.Math.Soc, 1992.

13. {\bf S. Kobayashi}, {\it Transformation groups in differential geometry},
Springer - Verlag, Berlin-Heidelberg-New York, 1972

14. {\bf Michael Forger, Cornelius  Paufler, Hartmann R¨omer},
{\it The Poisson Bracket for Poisson Forms
in Multisymplectic Field Theory}, arXiv, math-ph/0202043v1.

15. {\bf Charles-Michel Marle}, {\it The Schouten-Nijenhuis bracket and
interior products},  Journal of Geometry and Physics, {\bf 23}, 350-359, 1997.

\part{Basics of classical mechanics and vacuum\\ electrodynamics}

\chapter{General Notions about Physical Objects and Interactions}
\section{The concept of Physical Object}
When we speak about physical objects, e.g. classical particles, solid bodies,
elementary particles, fields, etc., we always suppose that some {\it definite
properties of the object under consideration do not change during its
time-evolution under the influence of the existing environment}.  The
availability of such time-stable features of any physical object makes it {\it
recognizable} among the other physical objects, on one hand, and  guarantees
its proper {\it identification} during its existence in time, on the other
hand. Without such an availability of constant in time properties (features),
which are due to the object's resistance and surviving abilities, we could
hardly speak about objects and knowledge at all. So, for example, two classical
mass particles together with their own gravitational fields survive under the
mutual influence of their gravitational fields through changing their states of
motion: change of state compensates the consequences of the violated dynamical
equilibrium that each of the two particles had been established with the
physical environment before the two gravitational fields have begun
perturbating each other.

The above view implies that three kinds of quantities will be necessary to
describe as fully as possible the existence and the evolution of a given
physical object: \index{physical object}

	1. {\bf Proper (identifying) characteristics}, i.e. quantities which
do NOT change during the entire existence of the object. The availability of
such quantities allows to distinguish a physical object among the other ones.

	2. {\bf Kinematical characteristics}, i.e. quantities, which describe
the allowed space-time evolution, where "allowed" means {\it consistent with
the constancy of the identifying characteristics}.

	3. {\bf Dynamical characteristics}, i.e. quantities which are
functions (explicit or implicit) of the {\it proper} and of the {\it
kinematical} characteristics.

Some of the dynamical characteristics must have the following two important
properties:  they are in a definite degree {\bf universal}, i.e. a class of
physical objects (may be all physical objects) carry nonzero value of them (e.g.
energy-momentum), and they are {\bf conservative}, i.e. they may just be
transferred from one physical object to another (in various forms) but no loss
is allowed.

Hence, the evolution of a physical object subject to bearable/acceptable
exterior influence (perturbation), coming from the existing environment,
has three aspects:

	1. {\it constancy of the proper (identifying) characteristics},

	2. {\it allowed kinematical evolution},

	3. {\it exchange of dynamical quantities with the physical
environment}.

Moreover, if the physical object under study is space-extended (continuous) and
demonstrates internal structure and dynamics, i.e. available interaction of
time-stable subsystems, it should be described by a many-component mathematical
object, e.g., vector valued differential form, therefore, we must consider this
{\it internal exchange} of some dynamical characteristics among the various
subsystems of the object as {\it essential} feature, determining in a definite
extent object's integral appearance. For example, the relativistically
described electromagnetic field respects two subsystems, formally represented
by two differential 2-forms $(F,*F)$, so, internal energy-momentum
exchange between $F$ and $*F$ should be considered as possible, and to be
appropriately taken into account.

The above features suggest that the dynamical equations, describing locally
the evolution of the object, may come from giving an explicit form of the
quantities controlling the local internal and external exchange processes
i.e. from writing down corresponding local balance equations.  Hence, denoting
the local quantities that describe the external exchange processes by $Q_i,
i=1,2,\dots$, the object should be considered to be $Q_i$-free,
$i=1,2,\dots$, if the corresponding integral values are time constant,  which
can be achieved only if $Q_i$ obey differential equations presenting
appropriately (implicitly or explicitly) corresponding local versions of the
conservation laws (continuity equations). In case of absence of external
exchange similar equations should describe corresponding {\it internal}
exchange processes.  The corresponding evolution in this latter case may be
called {\it proper evolution}.
\vskip 0.2cm
Summarizing, we may assume the rule that the available changes of a
physical field system, or a recognizable subsystem of a given field system, must
be refered/related somehow to the very system, or to some of the susbsystems,
in order to evaluate their significance:

	-if the value of the refered quantity is zero, then the changes are
admissible and the system/subsystem keeps its identity;

	-if the mutually refered quantities among subsystems establish dynamical
equilibrium, i.e., each subsystem gains as much as it loses, then the whole
system keeps its identity;

	-if the value of the refered quantity is not zero, then the identity of
the system is partially, or fully, lost, so, our system undergoes essential changes
leading to becoming subsystem of another system, or to destruction, giving
birth to new system(s).

In trying to formalize these views it seems appropriate to give some initial
explicit formulations of some most basic features (properties) of what we call
physical object, which features would  lead us to a, more or less,
adequate theoretical notion of our intuitive notion of a physical object.
Anyway, the following properties of the theoretical concept "physical object"
we consider as necessary:

      1. It can be created during finite period(s) of time.

      2. It can be destroyed during finite period(s) of time.

      3. It occupies finite 3-volume at any moment of its existence, so it has
spatial structure and may be considered as a system consisting of two or more
interconnected subsystems.

      4. It has a definite stability to withstand definite external
disturbances.

      5. It has definite conservation properties.

      6. It necessarily carries sufficiently universal
measurable quantities, e.g., energy-momentum.

      7. It exists in an appropriate environment (called usually vacuum), which
 provides all necessary existence needs. Figuratively speaking, every
physical object lives in a dynamical equilibrium with the outside world, which
dynamical equilibrium may be realized in various regimes.

      8. It can be detected by other physical objects through allowed
exchange of appropriate physical quantities, e.g., energy-momentum exchange.

      9. It  may combine/coexist through interaction with other appropriate
physical objects to form new objects/systems of higher level structure. In
doing this it may keep its identity and can be recognized and identified
throughout the existence of the system as its constituent/subsystem.

      10. Its destruction gives necessarily birth to new objects, and this
process respects definite rules of conservation. In particular, the available
interaction energy among its subsystems may transform entirely or partly to
kinetic one, and carried away by the newly created objects/systems.

\vskip 0.3cm
%\noindent {\bf Remark}.
The property {\it to be spatially finite} we consider as a very essential one.
So, the above features do NOT allow the classical "material
points" and "infinite classical fields" (e.g.  plane waves) to be considered as
appropriate theoretical notions, since the point-likeness excludes any
structure and forbids destruction, and the spatially infinite fields, even if
they carry finite energy, they cannot be finite-time-created. Hence, the
Born-Infeld "principle of finiteness" [1] stating that {\it "a satisfactory
theory should avoid letting physical quantities become infinite"} may be
strengthened as follows: \vskip 0.3cm {\bf All real physical objects are
spatially finite entities and NO infinite values of the physical quantities
carried by them are allowed}.

\vskip 0.2cm
Clearly, together with the purely qualitative
features physical objects carry important quantitatively described physical
properties, and any external interaction may be considered as an exchange of
the corresponding quantities provided both the object and the corresponding
environment carry them. Hence, the more universal is a physical quantity the
more useful for us it is, and this moment determines the exclusively important
role of energy-momentum, which modern physics considers as the most universal
one, i.e., it is more or less assumed that:

{\bf All physical objects necessarily carry energy-momentum and most of them
are able in a definite extent to lose and gain energy-momentum.}

\vskip 0.3 cm
The above notes clearly say that we make use of the term "physical object"
when we consider it from integral point of view, i.e. when its stability
against external perturbations is guaranteed. We make use of the term "physical
system" when time-stable interacting subsystems are possible to be
recognized/identified, and the behavior of the system as a whole, i.e.
considered from outside, we try to consider as seriously dependent on its
internal dynamical structure, i.e. on an available stable interaction of its
time-recognizable/time-identifiable subsystems. Therefore we shall follow the
rule:
\vskip 0.2cm
{\bf Physical recognizability of time-stable subsystems of a physical
system requires corresponding mathematical recognizability in the theory}.
\vskip 0.2cm
From pure formal
point of view a description of the evolution of a given continuous physical
object/system must include obligatory two mathematical objects:
\vskip 0.3cm
{\bf 1.} The mathematical object $\Psi$, having in general various
vector components, is meant to represent as fully as possible the wholeness and
integrity of the object under consideration: when subject to appropriate
operators, $\Psi$ must reproduce explicitly all important information about the
{\it structure, admissible changes} and admissible {\it dynamical evolution} of
the physical object;
\vskip 0.3cm
{\bf 2.} The mathematical object $D\Psi$ which represents the {\it admissible
recognizable} changes, \index{admissible changes} where $D$ is appropriately
chosen differential operator acting mainly on the kinematical and dynamical
characteristics. The object $D\Psi$ must have tensor nature in order to be used
to define appropriate physical quantities. If $D$ does not depend on $\Psi$ and
its derivatives, and, so, on the corresponding proper characteristics of
$\Psi$, the relation $D\Psi=0$ then would mean that those kinematical and
dynamical properties of $\Psi$, which feel the action of $D$, are constant with
respect to $D$, so the evolution prescribed by $D\Psi=0$, would have "constant"
character and would not say much about possible changes of those
characteristics, which do not feel $D$. From principle point of  view, it does
not seem so important if the changes $D\Psi$ are zero, or not zero. The {\it
significantly important} point is that the {\it changes $D\Psi$ are
admissible}, and appropriate combinations of $D\Psi$ and $\Psi$ may
represent quantitavely corresponding changes of physically important quantities,
e.g., energy-momentum changes, while the very $\Psi$ is not obliged, in general,
to represent phisical quantities.

The changes are admissible in the following two cases:

{\bf First}, when related to
the very object $\Psi$ through some "projection" $P$ upon $\Psi$, the
"projections" $P(D\Psi,\Psi)$ vanish, and then the object may be called {\it
free} (with respect to those characteristics which feel $D$);

{\bf Second}, when the
projections $P(D\Psi,\Psi)$ do not vanish, but the object still survives, then
the object is called {\it not free} (with respect to the same characteristics).

In the first case the corresponding admissible changes $D\Psi$ have to be
considered as having an {\it intrinsic} for the object nature, and they should
be generated by some {\it necessary} for the very existence of the object {\it
internal} energy-momentum redistribution during evolution (recall point 6 of
the above stated 10 properties of a physical object). In the second case the
admissible changes $D\Psi$, in addition to the intrinsic factors, depend also
on external factors, so, the corresponding projections $P(D\Psi,\Psi)$ should
describe explicitly or implicitly some energy-momentum exchange with the
environment.

These views correspond in some sense to mathematics, where we always meet
coupling between {\it mathematical structure} and corresponding set of
transformations, or {\it group of invariance} of this structure.

Following this line of considerations we come to a conclusion that every
description of a free physical object must include some mathematical
expression of the kind $\mathbb{F}(\Psi,D\Psi;S)=0$, specifying (through the
additional quantities $S$) what and how changes, and specifying also what is
projected and how it is projected.  If the object is not free but survives
when subject to the external influence, then it is very important the
quantity $\mathbb{F}(\Psi,D\Psi;S)\neq 0$ to have, as much as possible,
universal character and to present a change of a conservative quantity, so
that this same quantity to be expressible through the characteristics
$\mathcal{F}$ of the external object(s).  Hence, specifying differentially
some conservation/balance properties of the object under consideration, and
specifying at every space-time point the corresponding admissible exchange
processes with the environment through an equation of the form
$$
\mathbb{F}(\Psi,D\Psi;S)=
\mathbb{G}(\mathcal{F},d\mathcal{F};\Psi,D\Psi,...)
$$
we obtain corresponding equations of motion being consistent with the
corresponding integral conservation properties.

This notion of a (finite continuous) physical object sets the problem to try
to consider and understand its integral characteristics through its local
dynamical characteristics, i.e., as we mentioned above, to try to understand its
nature and integral appearance as determined and caused by its dynamical
structure. The integral appearance of the local features may take various
forms, in particular, it might influence the spatial structure of the object.
In view of this, the propagational behavior of the object as a whole,
considered in terms of its local translational and rotational components of
propagation, which, in turn, should be related to the internal energy-momentum
redistribution during propagation, could be stably consistent only with some
distinguished and time-recognizable spatial structures. We note that available
local rotational components of propagation does NOT always produce integral
rotation of the object, but if they are available, time stable, and consistent
with the translational components of propagation, some specific conserved
quantity should exist.  And if the rotational component of propagation shows
some consistent with the object's spatial structure periodicity, clearly, the
corresponding frequency may be used to introduce such a quantity.

Finally we note that the rotational component of propagation of
a (continuous finite) physical object may be of two different origins: {\it
relative} and {\it intrinsic}. In the "relative" case the corresponding
physical quantity, called {\it angular momentum}, depends on the
choice of some {\it external} to the system factors, usually these are {\it
relative axis} and {\it relative point}. In the "intrinsic" case the
rotational component (if it is not zero) is meant to carry intrinsic
information about the internal dynamics of the object, considered now as
a system, so, the corresponding physical quantity, usually called
{\it spin}, should NOT depend on any external factors as far as the object
survives.
\vskip 0.3cm
{\bf References}
\vskip 0.2cm
\addcontentsline{toc}{subsection}{{\bf References}}
1. {\bf M.Born, L.Infeld}, {\it Proc.Roy.Soc.}, A 144 (425), 1934

\section{The concept of Interaction}

We recall that this idea of {\it change-conservation} nature of a physical
object has been used firstly by Newton in his momentum balance equation
$$
\dot{\mathbf{p}}=\mathbf{F}, \ \ \mathbf{p}=m\mathbf{v},
$$
which is the restriction of the nonlinear partial differential system
$$
\nabla_{\mathbf{p}}\mathbf{p}=\mathrm{m}\mathbf{F}, \ \text{or} \ \
\mathbf{p}^i\nabla_i \mathbf{p}^j=\mathrm{m}\mathbf{F}^j
$$
on some trajectory, under the assumption to make use of time as parameter on
the trajectory. This Newton's system of equations just says that there are
physical objects in Nature which admit the "point-like" approximation, and
which can exchange energy-momentum with "the rest of the world" but keep
unchanged their identification properties, and this allows these objects to be
recognized in space-time and studied as a whole, i.e., as point-like ones.

In macrophysics, as a rule, the external influences are such that they do NOT
destroy the system, and in microphysics a full restructuring is allowed: the
old ingredients of the system may fully vanish, or fully transform to new ones,
(e.g., the photon destruction) provided
energy-momentum  and may some other conservation laws hold. The essential point
in this second case is that {\it it always results in appearing of relatively
stable objects, carrying energy-momentum and some other particular physically
measurable quantities, demonstrating in such a way their recognizability}. This
conclusion emphasizes once again the importance of having an adequate notion of
what is called physical object, and of the rules that regulate the admissible
transformations of a set of objects to a set of new objects.

As it becomes clear from the above, we consider physical objects as permanently
interacting entities with the rest of the world according to their individual
structure and exchange abilities. A physical object we call {\it free} if the
exchange with the outside world is optimal in the sense that all intrinsic
exchanges among its subsystems guarantee its time-existence as an individual
recognizable entity. If the optimal regime of exchange with the outside world
is violated by an external factor, the object is no more free. In trying to
protect itself from destruction, the object may attempt two initiatives:
to appropriately change the rates of the internal exchange processes, and,
in order to restore its dynamical equilibrium with the physical
environment, to change its state as a whole.

Classical physics recognizes two kinds of interaction of a physical object with
the outside world: direct and indirect. The basic examples of direct kind of
interaction come from mechanics, where we consider as admissible one material
body to push another material body without any kind of intermediates. Since in
the point-like approximation this does not lead to destruction of any of the
two bodies this process is regulated by the momentum conservation law. If the
two bodies begin changing their state of motion when they are distant from each
other, physics introduces the concept of field, continuous physical object
associated with every one of the bodies, and the mutual throughout space
influence between these two fields is considered as responsible for violating
the dynamical equilibrium of each of the bodies with the outside world, which
causes the observed change of state, e.g., appearance of acceleration,
of the two bodies.

We note that two continuous systems may interact, i.e. exchange
energy-momentum, and in this way considered as recognizable subsystems of a
larger system, in two ways: with {\it available interaction energy} and without
available interaction energy. In the first case we have to define locally the
interaction energy density, which may be {\it positive} or {\it negative}, and
this "hidden", or "potential" energy must always be taken into account in the
energy balance relations. In the second case there is no interaction energy, so
if the system is isolated, then any energy-momentum loss of
one of the subsystems is gained by the other. In particular, if each of
the two subsystems keeps its integral energy unchanged, then the two subsystems
are in dynamical equilibrium: {\it each one gains as much as it loses}. We
shall see that electromagnetic photon-like objects make use namely of this
second way of interaction: in order to keep their nature, the electro-magnetic
and the magneto-electric components (mathematically expressible correspondingly
by $F$ and $*F$) {\it carry always the same stress-energy-momentum}, so, the
local energy-momentum exchange between them during propagation may take place
only in equal quantities, i.e., $F$ and $*F$ appear as partners  living always
in dynamical equilibrium.

\section{Further remarks on objects and \\interaction}

Modern science seeks and aims at a good adequacy between the real objects and
the corresponding mathematical model objects.  So, the mathematical model
objects $\Psi$ must necessarily be spatially finite, and even temporally
finite if the physical object considered has by its intrinsic nature finite
life-time.  This most probably means that $\Psi$ must satisfy
partial differential equation(s), together with its derivatives it should be
able to  define in a consistent way the {\it interaction instruments}, the {\it
admissible} changes and the {\it conservation} properties of the object under
consideration. Hence, talking about physical objects we shall mean the
following: \begin{center} \hfill\fbox{
    \begin{minipage}{0.97\textwidth}
\begin{center}
\vskip 0.3cm
{\bf Physical objects are time-stable spatially finite entities which have a
well established internal and external dynamical balance between change and
conservation, and this balance is kept by a permanent and strictly fixed
interaction with the environment.}
 \end{center}
\vskip 0.3cm
\end{minipage}} \hfill \end{center}

As an idealized (mathematical) example of an object as outlined above, let's
consider a spatial region  $D$ of one-step piece of a helical cylinder with
some proper (or internal) diameter $r_o$, and let $D$ be winded around some
straightline axis $Z$. Let at some (initial) moment $t_o$ our mathematical
model-object $\Psi$ be different from zero only inside $D$.  Let now at
$t>t_o$ the object $\Psi$, i.e. the region $D$, begin moving as a whole along
the helical cylinder with some constant along $Z$ (translational) velocity
$c$ in such a way that every point of $D$ follows its own (helical)
trajectory around $Z$ and never crosses the (helical) trajectory of any other
point of $D$.  Obviously, the rotational component of propagation is
available, but the object does NOT rotate as a whole.  Moreover, since the
translational velocity $c$ along $Z$ is constant, the spatial periodicity
$\lambda$, i.e.,  the height of $D$ along $Z$, should be proportional to the
time periodicity $T$, and for the corresponding frequency $\nu=1/T$ we obtain
$\nu=c/\lambda$. Clearly, this is an idealized example of an object with a
space-time compatible dynamical structure, so, any physical interpretation
would require to have explicitly defined $D$, $\Psi$, corresponding dynamical
equations, local and integral conserved quantities, $\lambda$, $c$ and,
probably, some other parameters.
\index{sense of equality sign}

We repeat now what the sign of equality $"="$ means:
\begin{center}
\hfill\fbox{
    \begin{minipage}{0.97\textwidth}
\begin{center}
\vskip 0.3cm
{\bf On the two sides of the equality sign stays the SAME
element/quantity, which can be defined in different terms.}
\end{center}
\vskip 0.3cm
\end{minipage}} \hfill \end{center}

One of the basic in our view lessons that we more or
less have been taught is that any detection and further study of a physical
object {\bf requires} some energy-momentum exchange.
So, {\it every physical object necessarily carries energy-momentum, every
quantity of energy-momentum needs a carrier,
and every interaction between two physical objects has an energy-momentum
exchange aspect}. The second lesson concerning any interaction is that, beyond
its {\it universality}, {\it energy-momentum is conserved} quantity, so NO loss
of it is allowed: it may only pass from one object to another. This means also,
that an {\it annihilation} process may cause {\it creation} process(es), and
the full energy-momentum that has been carried by the annihilated objects, must
be carried away by the created ones. {\bf Energy-momentum always needs
carriers, as well as, every physical object always carries energy-momentum}.
Hence, the energy-momentum exchange abilities of any physical object realize
its protection against dangerous external influence on one side, and reveal its
intrinsic nature, on the other side. Therefore, our knowledge about the entire
complex of properties of a physical object relies on getting information
about its abilities in this respect and finding out corresponding quantities
describing quantitatively these abilities.

The above views make us think and assume the standpoint
that the most reliable dynamical equations,
describing locally or integrally the time-evolution of a physical system, should
express energy-momentum balance relations.

\section{Symmetries, Conservative Quantities and Isometries}

The conservation properties of an object manifest themselves through
corresponding symmetry properties, and these physical symmetry properties
appear as mathematical symmetries of the corresponding equations
$\mathbb{F}(\Psi,D\Psi;Q)=0$ in the theory. Usually, responsible for
these symmetries are some new (additional) mathematical objects defining the
explicit form of the equation(s), e.g., the Minkowski pseudometric tensor
$\eta$ in the relativistic mechanics and relativistic field theory, the
symplectic 2-form $\omega$ in the Hamilton mechanics, etc.  Knowing such
symmetries we are able to find new solutions from the available ones, and in
some cases to describe even the whole set of solutions.  That's why the Lie
derivative operator (together with its generalizations and prolongations)
and the integrability conditions for the
corresponding equation(s) $\mathbb{F}(\Psi,D\Psi;Q)=0$ play a very
essential and hardly overestimated role in theoretical physics. Of course,
before to start searching for symmetries of an equation, or of a mathematical
object $\Psi$ which is considered as a model of some physical object, we must
have done some preliminary work of specifying the mathematical nature of
$\Psi$, and the necessary information may come only from an initial data
analysis of appropriately set and carried out experiments.

The mathematical concept of symmetry has many faces and admits various
formulations and generalizations. The simplest case is a symmetry of a
real valued function $f: M\rightarrow \mathbb{R}$, where $M$ is a
manifold, with respect to a map $\varphi: M\rightarrow M$: $\varphi$ is
a symmetry (or a symmetry transformation) of $f$ if $f(\varphi(x))=f(x), x\in
M$. If $\varphi_t, t\in(0,1)\subset\mathbb{R}$ is 1-parameter group of
diffeomorphisms of $M$, then the symmetry $f(\varphi_{t}(x))=f(x)$ may be
locally expressed through the Lie derivative $L_X(f)=0$, where the vector
field $X$ on $M$ generates $\varphi_t$. If $T$ is an arbitrary tensor field
on $M$ then the Lie derivative is naturally extended to act on $T$ and we
call $T$ symmetric, or {\it invariant} with respect to $X$, or with respect
to the corresponding (local, in general) 1-parameter group of diffeomorphisms
of $M$, if $L_{X}T=0$. In this way the Lie derivative represents an
universal tool to search symmetries of tensor fields on $M$ with respect to
the diffeomorphisms of $M$. Unfortunately, this universality of $L_X$ does
not naturally extend to sections of arbitrary vector bundles on $M$, where we
need additional structures in order to introduce some notion of symmetry or
invariance. \index{symmetry, isometries, conservation}

In classical field theory integral conserved quantities, i.e.
time-independent integral characteristics of the system considered, are usually
constructed by means of a symmetric second rank tensor, called
stress-energy-momentum tensor of the corresponding physical system,
$Q_{\mu\nu}$, with zero divergence $\nabla_\nu Q_\mu^\nu=0$, by making use of
isometries, i.e., symmetries of the metric tensor, in the following way. The
local symmetries of the metric tensor are also called {\it Killing} vector
fields. The equation $L_X g=0$, where $g$ is given, looks as follows \[
(L_Xg)_{\mu\nu}=\nabla_\mu X_\nu+\nabla_\nu X_\mu=0,
\]
where $\nabla$ is the corresponding to $g$ Levi-Civita connection.
If now $Q_{\mu\nu}$ is a conservative tensor field, i.e.
$\nabla_\nu Q_\mu^\nu=0$, and $X$ is a local isometry, we obtain
\[
\nabla_\nu (Q_\mu^\nu X^\mu)=(\nabla_\nu Q_\mu^\nu)X^\mu+
Q^{\mu\nu}\nabla_\nu X_\mu =Q^{\mu\nu}\nabla_\nu X_\mu.
\]
Because of the symmetry of $Q$, in the sum $Q^{\mu\nu}\nabla_\mu X_\nu$ only
the symmetric part of $\nabla_\mu X_\nu$ may contribute, but this symmetric
part is zero since $X$ is a local isometry. In this way with every local
isometry $X$ of the metric the 1-form $Q_{\mu\nu}X^\mu dx^\nu$ is associated,
and this 1-form has zero divergence. This  means that the 3-form
$*(Q_{\mu\nu}X^\mu dx^\nu)$ is closed: $\mathbf{d}*(Q_{\mu\nu}X^\mu dx^\nu)=0$,
so according to the Stokes theorem, the
integral over $\mathbb{R}^3$ of the restriction of this 3-form to $\mathbb{R}^3$
will not depend on time. Of course, from physical point of view,
these considerations make sense only for finite valued such 3-integrals, which
corresponds to the natural view that physical objects are {\it spatially
finite}, so the corresponding field functions should be spatially finite too.

The role of a local isometry here is two-sided: quantitative and qualitative,
namely, besides its use to define quantitatively the conserved quantity, the
nature of the corresponding Killing vector field determines the nature of the
corresponding conserved quantity. For example, on Minkowski space-time the four
translations along the standard coordinates define the three integral momentums
and energy:
$$
\int_{\mathbb{R}^3}j^*(*(Q_{\mu,i}dx^\mu)), \ \
\int_{\mathbb{R}^3}j^*(*(Q_{\mu,4}dx^\mu)),
$$
where $j:\mathbb{R}^3\rightarrow\mathbb{R}^4$ is the canonical embedding
$(x,y,z)\rightarrow(x,y,z,0)$.

In general, in order to compute an integral conserved quantity for a continuous
physical system on Minkowski space-time we always need a closed differential
3-form $\alpha: \mathbf{d}\alpha=0$, such that its restriction $j^*\alpha$
 on $\mathbb{R}^3$ to be different from zero, and the integral
$\int{j^*\alpha}$ over $\mathbb{R}^3$ to be finite.

\subsection{A note on Angular momentum and Helicity}
In classical mechanics on $(\mathbb{R}^{3},g)$, $g$-is the euclidian metric,
one of the important constants of motion of a particle is the so called {\it
angular momentum}. This can be traced, for example, in celestial mechanics
where the external field is assumed to be spherically symmetric, and the center
of symmetry of the external field is chosen for reference point, i.e. for
center of spherical coordinate system $(r,\theta,\varphi)$. Introducing the so
called {\it radius-vector} $\mathbf{r}$ with respect to the symmetry center and
denoting the momentum vector of a particle by $\mathbf{p}$, the angular
momentum $\mathfrak{m}$ of the particle is defined by $$
\mathfrak{m}=\mathbf{r}\times\mathbf{p},
$$
where it is assumed that $\mathbf{r}$ is parallely transported from the origin
to the point
where the particle is at the moment considered. So, at this point we have two
vector fields, which in canonical coordinates look like
$$
\mathbf{r}=x\frac{\partial}{\partial x}+y\frac{\partial}{\partial y}+
z\frac{\partial}{\partial z},  \ \ \
\mathbf{p}=\mathbf{p}^1\frac{\partial}{\partial x}
	   +\mathbf{p}^2\frac{\partial}{\partial y}+
\mathbf{p}^3\frac{\partial}{\partial z}\cdot
$$
It is shown now that under static external fields the quantity
$\mathfrak{m}=\mathbf{r}\times\mathbf{p}$ is conserved, i.e. the particle
moves inside a fixed plane passing through the symmetry center, and
$|\mathfrak{m}|$ keeps its value. So, any other point that is out of this plane
will define another such plane through the symmetry center.

Our purpose now is to find the most general condition on a vector field $X$,
such that its representative at a given point and $\mathbf{r}$ at this point to
define unique such plane through the center $(0,0,0)$, and to find the
corresponding equivalent condition in terms of the {\it helicity} concept. We
shall work in standard spherical coordinates $(r,\theta,\varphi)$ given by $$
r=\sqrt{x^2+y^2+z^2}, \ \ \theta=arccos\frac{z}{\sqrt{x^2+y^2+z^2}},\ \
\varphi=arctg\frac{y}{x},
$$
where $(x,y,z)$ are that canonical cartesian coordinates on the
manifold $\mathbb{R}^3$. In these coordinates we obtain
$$
dr=\frac{xdx+ydy+zdz}{\sqrt{x^2+y^2+z^2}}, \ \ \text{so}, \ \ \
rdr=xdx+ydy+zdz,
$$
therefore, since the metric has components $(1,r^2,r^2\mathrm{sin}^2\theta)$, we
have
$$ \mathbf{r}=\tilde{g}(rdr)=r\frac{\partial}{\partial r}\cdot
$$

We are looking now for conditions on a vector field $X(r,\theta,\varphi)$, such
that any trajectory of $X$ to lie in a plane passing through the zero point
$(0,0,0)$, through a given initial point $p_o$ outside the zero point, and of
course, through the straight line passing through these two points.

The above conditions imply that the two vector fields $(\mathbf{r},X)$ must be
tangent to the plane we are searching for whatever the initial point $p_o$ is.
This means that this 2-plane must be {\it integral} surface for the
2-dimensional distribution defined by $(\mathbf{r},X)$, hence, the Lie bracket
$[\mathbf{r},X]$ must also be tangent to this plane. Therefore, the condition
we are searching for is equivalent to the Frobenius integrability relation
$$
[\mathbf{r},X]\wedge\mathbf{r}\wedge X=0
$$
meaning that these three vector fields are linearly dependent, so, each
of them can be represented as a linear combination of the other two.

We obtain consecutively:
\begin{eqnarray*}
\left[r\frac{\partial}{\partial r},X\right]&=&
\left(r\frac{\partial X^1}{\partial r}-X^1\right)\frac{\partial}{\partial r}+
r\frac{\partial X^2}{\partial r}\frac{\partial}{\partial \theta}
+r\frac{\partial X^3}{\partial r}\frac{\partial}{\partial \varphi}\\
r\frac{\partial}{\partial r}\wedge X&=&rX^2
\frac{\partial}{\partial r}\wedge \frac{\partial}{\partial \theta}+
rX^3\frac{\partial}{\partial r}\wedge\frac{\partial}{\partial\varphi}.
\end{eqnarray*}
Therefore,
$$
[\mathbf{r},X]\wedge\mathbf{r}\wedge X=
r^2\left(-X^3\frac{\partial X^2}{\partial r}+
X^2\frac{\partial X^3}{\partial r}\right)\frac{\partial}{\partial r}\wedge
\frac{\partial}{\partial \theta}\wedge\frac{\partial}{\partial \varphi}.
$$
Thus, the condition $[\mathbf{r},X]\wedge\mathbf{r}\wedge X=0$ is equivalent to
$$
X^3\frac{\partial X^2}{\partial r}-X^2\frac{\partial X^3}{\partial r}=0, \ \
i.e. \ \ \frac{\partial}{\partial r}\left(ln\frac{X^3}{X^2}\right)=0.
$$
In view of this we can assume
$$
X^1=X^1(r,\theta,\varphi), \ \ X^2=f(r)\phi(\theta,\varphi), \ \
X^3=f(r)\psi(\theta,\varphi).
$$

We give now the helicity-form of this condition, i.e. we have to find 1-form
$\alpha$ on $\mathbb{R}^3$, such that the equation
$\mathbf{d}\alpha\wedge\alpha=0$ to be equivalent to the above condition:
$$
\mathbf{d}\alpha\wedge\alpha=0 \ \Leftrightarrow \
[\mathbf{r},X]\wedge\mathbf{r}\wedge X=0.
$$
An arbitrary such 1-form $\alpha=hdr+fd\theta+gd\varphi$ gives
$$
\mathbf{d}\alpha\wedge\alpha=\left[g\left(-\frac{\partial h}{\partial \theta}+
\frac{\partial f}{\partial r}\right)-f\left(-\frac{\partial h}{\partial\varphi}+
\frac{\partial g}{\partial r}\right)+h\left(-\frac{\partial f}{\partial\varphi}+
\frac{\partial g}{\partial \theta}\right)\right]
dr\wedge d\theta\wedge d\varphi.
$$
Choosing $h=0, f=X^2, g=X^3$ and putting the result equal to zero we obtain
$$
\mathbf{d}\alpha\wedge\alpha=\left(X^3\,\frac{\partial X^2}{\partial r}-
X^2\frac{\partial X^3}{\partial r}\right)dr\wedge d\theta\wedge d\varphi=0,
$$
which is equivalent to $\frac{\partial}{\partial
r}\left(ln\,(X^3/X^2)\right)=0$.

\chapter{Classical Mechanics and Classical Fields}

\section{Classical mechanics}
{\it Our aim in this section of the chapter is to point out those
moments of classical mechanics, which lead to the conclusion that, from
theoretical point of view, the potential approach to description of interaction
between point-like particles and external fields has to be reconsidered}.

\subsection{Symplectic View on Mechanics}
The symplectic formulation of classical mechanics \index{symplectic mechanics}
presents a rigorous
geometric formulation of the hamiltonian formulation of dynamical equations of
a particle in an external field, where the external field is represented as a
rule by a scalar (real or complex) function, together with  the corresponding
conservation laws [1]. The basic mathematical object in this approach is the
concept of {\it symplectic manifold}\index{symplectic manifold}.

{\bf Definition}: An even dimensional smooth manifold $M^{2n}$ is called {\it
symplectic} if a closed nondegenerate 2-form $\omega\in \Lambda^{2}(M^{2n}),
\mathbf{d}\omega=0, det||\omega_{\mu\nu}(x)||\neq 0, x\in M^{2n}$, is defined.

So, every tangent space $T_x(M,\omega)$ is a symplectic 2n-dimensional vector
space. Also, $(M^{2n},\omega)$ is orientable, and the orientation is given by
the volume form $(-1)^{n(n-1)/2}\omega^{n}$.

The nondegeneracy of $\omega$ allows to introduce $\omega^{-1}$ such that
$\omega_{\mu\sigma}(\omega^{-1})^{\sigma\nu}=\delta_\mu^\nu$, so, the cotangent
space at $x\in M^{2n}$ becomes also symplectic vector space. Thus, a linear
isomorphism $\tilde{\omega}$ between the tangent and cotangent spaces is at
hand: locally we have, if $X\in T_x(M)$ then
$\tilde{\omega}(X)_\mu=\omega_{\nu\mu}X^\nu$ and if $\alpha\in T_x^{*}(M)$ then
$\tilde{\omega}(\alpha)^{\mu}=(\omega^{-1})^{\nu\mu}\alpha_\nu$. Hence, we
obtain an isomorphism between the vector fields $\mathfrak{X}(M)$
and 1-forms $\Lambda^1(M)$ according to $X\rightarrow i(X)\omega$. Clearly, to
the Lie bracket $[X,Y]$ corresponds unique 1-form $i([X,Y])\omega$.

A diffeomorphism $\varphi:(M^{2n},\omega)\rightarrow (M^{2n},\omega)$ is called
{\it symplectic} if $\varphi^*\omega=\omega$.

The following result holds: On every symplectic manifold $(M,\omega)$ there is
a local coordinate system $(y^1,...,y^{2n})$ such that $\omega$ is represented
locally as
$$
\omega=dy^1\wedge dy^{n+1}+...+dy^{n}\wedge dy^{2n}.
$$

The cotangent bundle $(M,\pi,\mathbb{R}^n)$ of every manifold is a symplectic
manifold, where the symplectic 2-form $\omega_{x,\alpha_x}, x\in M$ at the point
$(x,\alpha_x)\in T_\alpha(T^*(M))$ is defined by the differential of the
canonically defined 1-form $\mathbf{\lambda}$ on $T^*(M)$ according to
$\mathbf{\lambda}(X_{\alpha}):= \alpha(\pi_*(X_{\alpha})$. Now, the symplectic
2-form is given by $\mathbf{d}\mathbf{\lambda}$.
If $(q^\mu,p^\mu=\frac{\partial}{\partial q^{\mu}}$
are local coordinates on $T^*(U\subset M)$ then, usually, $\mathbf{d}\lambda$
is chosen to be represented locally as
$\mathbf{d}\lambda=\sum_{\mu}dp^\mu\wedge dq^\mu$.

If $X_\alpha$ corresponds to $\alpha$ and $X_\beta$ corresponds to $\beta$
then the bracket of $\alpha$ and $\beta$ is the 1-form $[\alpha,\beta]$ defined
by $[\alpha,\beta]=i([X_\alpha,X_\beta])\omega$. It is easily verified that if
$\alpha$ and $\beta$ are closed, then the bracket $[\alpha,\beta]$ is an exact
1-form:
$$
[\alpha,\beta]=-\mathbf{d}(\omega(X_\alpha,X_\beta)).
$$
This equality allows to introduce bracket in the algebra of smooth functions on
a symplectic manifold. In fact, if $f,g\in\mathcal{J}(M)$, then the bracket
$(f,g)$ of $f$ and $g$ is defined as $-\mathbf{d}\omega(X_{df},X_{dg})$,
so we have the relations
$$
(f,g)=-\omega(X_{df},X_{dg})=X_{df}(g)=-X_{dg}(f),\ \ \
\mathbf{d}(f,g)=(df,dg).
$$
Locally, if $\omega$ is represented as $\omega=dp^\mu\wedge dq^\mu$ and
$\alpha=a_\mu dq^\mu+b_\mu dp^\mu$, then
$$
X_\alpha=-b_\mu\frac{\partial}{\partial q^\mu}+a_\mu\frac{\partial}{\partial
p^\mu}, \ \ \
(f,g)=\frac{\partial f}{\partial q^\mu}\frac{\partial g}{\partial p^\mu}-
\frac{\partial g}{\partial q^\mu}\frac{\partial f}{\partial p^\mu}.
$$
From the above it is seen that $f$ will be a first integral of $X_{dg}$ and $g$
will be a first integral of $X_{df}$ if $(f,g)=0$, which motivates the
introduction of the concept of {\it involution}: two 1-forms $(\alpha,\beta)$ on
a symplectic manifold are called to be in involution if
$\omega(X_\alpha,X_\beta)=0$. Hence, two functions on a symplectic manifold are
in involution if their differentials are in involution.

	{\bf Definition}: The vector fields $X$ on a symplectic
manifolds which correspond to closed 1-forms, i.e., $\mathbf{d}i(X)\omega=0$,
are called {\it hamiltonian systems}. If $i(X)\omega$ is exact, i.e.,
$i(X)\omega=-dH$, then $H$ is called {\it hamiltonian} for $X$.

The above implies that $X$ is a hamiltonian system iff the Lie
derivative of $\omega$ along $X$ vanishes: $L_X\omega=0$.
Clearly the
hamiltonian for $X$ is a first integral of $X$, called {\it integral energy} of
the corresponding dynamical system, i.e., it keeps the same value along any
fixed trajectory of $X$. We could also say that every local symmetry $X$ of the
symplectic 2-form $\omega$, i.e., $L_X\omega=0$, generates first integral of
$X$. Therefore, if the Lie group $G$ acts as a group of symmetries of $\omega$
then a map $M\rightarrow \mathfrak{g}^*$ can be constructed unifying all
integrals corresponding to the local symmetries of $\omega$ generated by the
corresponding to the action fundamental vector fields on $M$.

If $\alpha\in\Lambda^1(M)$ is closed, and nonvanishing on the open subset
$U\subset M$, then $\mathbf{d}\alpha\wedge\alpha=0$, so, $\alpha$ generates
1-dimensional completely integrable Pfaff system. Denoting its
$(2n-1)$-dimensional integral manifold by $(N,h)$, where $h$ denotes the
corresponding embedding: $h:N\rightarrow M$, we have the following properties:
\begin{itemize}
\item
$X_\alpha$ is tangent to $h(N)\subset M$.
\item
$h^*\omega$ is entirely described by the coordinates on $N$.
\end{itemize}
Locally, in coordinates $(q^\mu,p^\mu)$, if $H$ is a hamiltonian, then the
corresponding dynamical equations look as follows \index{hamilton equations}:
$$
\frac{dq^\mu}{dt}=\frac{\partial H}{\partial p^\mu}, \ \ \
\frac{dp^\mu}{dt}=-\frac{\partial H}{\partial q^\mu},
$$
where $t$ is a parameter along the trajectories and has nothing to do {\it in
general} with the physical time parameter.

 The following result due to E.Cartan \index{Cartan's theorem}
deserves to be specially noted. If $(M,\omega)$ is symplectic manifold and
$H$ is a function on $M\times\mathbb{R}$, then on the (odd dimensional) manifold
$(M\times\mathbb{R})$ there exists unique vector field $Y$ having the
properties: $$
	 Y_{x,t}=X_t(x)+\frac{\partial}{\partial t}, \ \ \
	 i(Y)(p_1^*\omega-dH\wedge dt)=0,
$$
where $(t)$ is a coordinate on $\mathbb{R}$, and
$p_1:M\times\mathbb{R}\rightarrow M$ is the projection on $M$. If $H$
does not depend on $t$, then $X_t$ also does not depend on $t$ and coincides
with the hamiltonian vector field defined by $dH$ on $M$. Moreover, the
relation $Y(H)=\frac{\partial H}{\partial t}$ always holds, so in general the
function $H$ depends on $t$ and is not a first integral of $Y$.

Finally it deserves noting that symplectic geometry gives just an appropriate
scheme: every smooth function $f$ defines through its symplectic gradient a
vector field $X_f$, and $f$ is constant along each trajectory of $X_f$, but
finding appropriate from physical point of view such functions, i.e.
hamiltonians, is not its engagement, it must come from physics. This goes along
also with the role of the variational principle in theoretical physics, where
the optimal nature of this principle leaves the choice of lagrangians free of
answer. \newpage

\subsection{Contact Structures and Invariant Forms}
Let $M$ be a manifold.

	{\bf Definition}. A {\it contact structure} on $M$ is defined by any
$(n-1)$-dimensional distribution $\Delta^{(n-1)}$ such that the corresponding
curvature form $\Omega$ is nondegenerate [2] \index{contact structure}.

Since this $\Delta^{(n-1)}$ can be defined by a nonvanishing 1-form $\alpha$,
$\alpha(x)\neq 0, x\in M$, such that $\langle\alpha,\Delta^{(n-1)}\rangle=0$,
this is equivalent to say that $\mathbf{d}\alpha$ restricted on
$\Delta^{(n-1)}$ is nondegenerate. There is unique vector field $X$ on $M$
satisfying $\langle\alpha(x),X(x)\rangle=1, x\in M$, $X$ is nonvanishing on
$M$, and $\Omega=\mathbf{d}\alpha\otimes X$.

The nondegeneracy of $\Omega$ on $\Delta^{(n-1)}$ requires the nondegeneracy of
$\mathbf{d}\alpha$ on $\Delta^{(n-1)}=Ker(\alpha)$, so, every tangent space
$T_x(M)$ admits the splitting
$$
T_x(M)=Ker_x(\alpha)\oplus
Ker_x(\mathbf{d}\alpha), x\in M.
$$
Therefore, $(\mathbf{d}\alpha)_x$ is a
symplectic structure in $Ker_x(\alpha)$, and the integral manifold
of $\Delta^{(n-1)}$ becomes symplectic manifold. We conclude that our initial
manifold $M$ has odd dimension, say $dim(M)=2n+1$. Moreover, $M$ becomes
orientable, and the orientation is given by the volume form
$\alpha\wedge(\mathbf{d}\alpha)^{n-1}$.

It is also important to note that $i_X\mathbf{d}\alpha=0$. In fact, around
every point $x\in M$ there exists a local coordinate system $(y,x^1,...,x^n)$
such that in this coordinate system we have
$$
X=\frac{\partial}{\partial y}, \ \ \
\alpha=dy+\beta_j(x^i)dx^j, \ \ i,j=1,2,...,2n.
$$
Obviously, $i_X\mathbf{d}\alpha=0$, and $L_X\alpha=L_X\mathbf{d}\alpha=0$.
Moreover, for every differentiable function $f$ on $M$ it is easily obtained
that $L_{fX}\alpha=df$ and $L_{fX}\mathbf{d}\alpha=0$.

We continue to consider the general concept of invariance of geometric objects
on a manifold and its application to conservation laws in mechanics.

{\bf Definition}. A tensor field $\mathfrak{T}$ on a manifold $M^n$ (further we
shall assume that $M$ is $n$-dimensional) is called invariant with respect to a
vector field $X\in\mathfrak{X}(M)$ if its Lie derivative $L_X\mathfrak{T}$ with
respect to $X$ vanishes: $L_X\mathfrak{T}=0$.

So if $\varphi_t$ is the flow of $X$ the invariance of $\mathfrak{T}$ means
that $\varphi_t^*\mathfrak{T}=\mathfrak{T}$.

Clearly, every function $f$ satisfying $L_Xf=0$ is a first integral of $X$,
i.e., $f$ does not change along any integral line of $X$.

Since $L_X$ is a derivation in the exterior algebra of differential forms on
 $M$: $L_X(\alpha\wedge\beta)=L_X\alpha\wedge\beta+\alpha\wedge L_X\beta$, all
invariant forms with respect to $X$ form a subalgebra of $\Lambda(M)$.

Locally, the invariance of $\alpha\in\Lambda(M)$ with respect to $X$ means that
around some point $x\in M$ the field $\alpha$ does not depend on
one of the corresponding local coordinates.

For example, the symplectic structure $\omega$ is invariant with respect to any
hamiltonian vector field, and the generated by a contact structure forms
$\alpha$ and $\mathbf{d}\alpha$ are invariant with respect to the unique
corresponding vector field $X, \langle\alpha,X\rangle=1$.

If $\omega\in\Lambda^n(M)$ and $X\in\mathfrak{X}(M)$ have compact support then
we can form the integral $$ I(t)=\int_{M}\varphi_t^*\omega<\infty. $$ Now, the
invariance of $\omega$ with respect to $X$ leads to $$
\frac{dI(t)}{dt}=\int_{M}\varphi_t^*L_X\omega=0.
$$
Hence, the integral has constant value along any trajectory of $X$. Vice versa,
if the above integral keeps its value with respect to any differentiable map
$h:M\rightarrow M$, then the integral
$$
\int_{M}(\varphi_t\circ h)^*\omega
$$
does not depends on $t$, so, $\omega$ is invariant with respect to $X$.

If the invariant with respect to $X$ differential form $\alpha$ satisfies in
addition $i_X\alpha=0$, then $\alpha$ is called {\it absolute integral
invariant} \index{absolute integral invariant} of $X$. Examples for such forms
are the reductions of the symplectic forms to any integral manifold of the
corresponding Pfaff system, and the corresponding to a contact structure 2-form
$\mathbf{d}\alpha$ with respect to the generated vector field $X,
\langle\alpha,X\rangle=1$. Of course, the corresponding integral $I(t)$ also
does not depend on $t$.

This concept of invariance of a differential form with respect to a vector field
$X$ admits the following two extensions.

A differential form $\alpha$ on $M$ is
called {\it relative integral invariant} \index{relative integral invariant}
with respect to $X$ if
$i_X(\mathbf{d}\alpha)=0$. As an example, we recall that the 1-form $\alpha$
defined by a contact stricture, satisfies $i_X\mathbf{d}\alpha=0,
\langle\alpha,X\rangle=1$.

A differential form $\alpha$ on $M$ is called {\it integral invariance
relation} \index{integral invariance relation} for the vector field $X$ if
$i_X\alpha=0$.

The above concepts are used in continuous mechanics also, where one tries to
compute how much of the energy carried by the vector field $X$ on $\mathbb{R}^3$
passes throughout an infinitesimal 2-volume defined by some 2-form, and how much
energy $X$ has lost locally during this process.

Finally we note that denoting the parameter along the trajectories of the
vector fields considered by $t$, we do NOT engage ourselves with the
presumption to interpret physically this parameter as time.

\subsection{Analytical Mechanics}
The geometric view on Analytical mechanics [1] \index{analytical mechanics}
(AnMech) makes use of the
tangent bundle $T(M)$ of a manifold, in particular, on the tangent bundle of
the manifold $\mathbb{R}^3$. So, the basic space of AnMech is $T(M)$, it is
always even dimensional, so a symplectic approach is possible in general. If
the local coordinates on $U\subset M$ are denoted by $(q^1,...,q^n)$ and the
projection $T(M)\rightarrow M$ is denoted by $p_M$, then these coordinates
canonically induce coordinates on $p_M^{-1}(U)\subset T(M)$ according to
$(q^\mu\circ p_M, \dot{q}^\mu=dq^\mu)$. Now the projection $p_{T(M)}$ of $T(T(M))$ to
$T(M)$ is given by $(q^\mu,\dot{q}^\mu,dq^\mu,d\dot{q}^\mu)\rightarrow
(q^\mu,\dot{q}^\mu)$, and the differential of $p_M$ is given by
$(q^\mu,\dot{q}^\mu,dq^\mu,d\dot{q}^\mu)\rightarrow (q^\mu,dq^\mu)$.

On the other hand the vector bundle $p_M^*(T(M))$ is always available and its
projection $\pi$ to $T(M)$ is given locally by
$\pi:(q^\mu,\dot{q}^\mu,d\dot{q}^\mu)\rightarrow (q^\mu,\dot{q}^\mu)$.

The following morphisms now can be defined:
$$
\pi': p_M^*(T(M))\rightarrow
T(M):  \ \pi'(q^\mu,\dot{q}^\mu,dq^\mu)\rightarrow(q^\mu,dq^\mu),
$$
$$
H: p_M^*(T(M))\rightarrow T(T(M)): \ \
H(q^\mu,\dot{q}^\mu,dq^\mu)\rightarrow (q^\mu,\dot{q}^\mu,0,dq^\mu),
$$
$$
K: T(T(M))\rightarrow p_M^*(T(M)): \ \
K(q^\mu,\dot{q}^\mu,dq^\mu,d\dot{q}^\mu)\rightarrow (q^\mu,\dot{q}^\mu,dq^\mu) .
$$
From the above local relations is seen that the composition $v=H\circ K$
generates a inear map of every tangent space of $T(TM)$,
called {\it vertical endomorphismsm}, \index{vertical endomorphism}
it satisfies $v\circ v=0$. Locally, we get
$$
v=dq^{\mu}\otimes\frac{\partial}{\partial \dot{q}^\mu}
\rightarrow \
v\left(f^\mu\frac{\partial}{\partial q^\mu}+
g^\mu\frac{\partial}{\partial \dot{q}^\mu}\right)=
f^\mu\frac{\partial}{\partial \dot{q}^\mu} \ \cdot
$$
Clearly, if $V=\dot{q}^\mu\frac{\partial}{\partial\dot{q}^\mu}$ is the Liouville
vector field on $T(M)$ \index{Liouville vector field}
then a vector field $X$ on $T(M)$ defines differential equation of
second order if $v(X)=V$.

The dual map $v^*: \Lambda(T(M))\rightarrow \Lambda(T(M))$ to $v$ is called
vertical operator \index{vertical operator}, and at every point of $T(TM))$ is
given by $$ v^*(f)=f, \ \ v^*(dq^\mu)=0, \ \ v^*(d\dot{q}^\mu)=dq^\mu, \ \
f\in\mathcal{J}(T(M)) . $$ So we can write $$ v^*(f_\mu dq^\mu +g_\mu
d\dot{q}^\mu)=g_\mu dq^\mu . $$ From the above expressions it is seen that the
endomorphism $v$ transforms the $p_M^*(T(M))$-part of a tangent vector to
$T(M)$ to "vertical" one, i.e. to tangent to the fibers of $T(T(M))$.
Accordingly, the dual map $v^*$ transforms the "vertical" part of a
differential form over $T(M)$, i.e. the one developed along $d\dot{q}^\mu $, to
a $p_M^*(T(M))$-part, i.e. to a one developed along $dq^\mu$.

Similarly, we can consider the "horizontal" endomorphism
$$
v^t=d\dot{q}^{\mu}\otimes\frac{\partial}{\partial q^\mu}
\rightarrow \
v^t\left(f^\mu\frac{\partial}{\partial q^\mu}+
g^\mu\frac{\partial}{\partial \dot{q}^\mu}\right)=
g^\mu\frac{\partial}{\partial q^\mu} \ \cdot
$$
Of course, the components of these "horizontal/vertical" parts of the
objects may depend on all coordinates $(q^\mu,\dot{q}^\mu)$ of $T(M)$. So, the
components of a form living in the image of $v^*$ will look like $$
\alpha_{\mu_1...\mu_p}(q^\mu,\dot{q}^\mu)dq^{\mu_1}\wedge dq^{\mu_2}\wedge ...\wedge
dq^{\mu_p}, \ \ \mu_1<\mu_{2}<...<\mu_p.
$$
Such forms on $T(M)$ are called {\it semibasic} \index{semibasic}
 which comes from the fact that
every form $\alpha$ on $M$ generates a form $p_M^*\alpha$ on $T(M)$ of this
kind. Every semibasic form annihilates the vectors living in the image of the
endomorphism $v$.

The endomorphism $v$ transforms every p-form $\alpha$ on $T(M)$ to a new p-form
$\mathbf{i}_v\alpha$ according to
$$
\mathbf{i}_v\alpha(X_1,X_2,...,X_p)=\sum_{k}\alpha(X_1,...,vX_k,...X_p).
$$
Since, assuming $\mathbf{i}_{v}f=0$, we obtain that $\mathbf{i}_v$ is
derivation (of degree zero) in $\Lambda(T(M))$, we can consider the commutator
$$ \mathbf{d}_v=[\mathbf{i}_v,\mathbf{d}]=
\mathbf{i}_v\circ\mathbf{d}-\mathbf{d}\circ\mathbf{i}_v,
$$
which is obviously an (anti)derivation of degree 1. Locally we have
$$
\mathbf{d}_vf=\frac{\partial f}{\partial \dot{q}^\mu}dq^\mu, \ \
\mathbf{d}_v(dq^\mu)=0, \ \ \mathbf{d}_v(d\dot{q}^\mu)=0 .
$$
Clearly, $\mathbf{d}_v\circ p_M^*=0$.

Now, in these terms, a mechanical system $\mathcal{M}$ is defined by the triple
$(M,T,\alpha)$, where $M$ is a manifold called configuration space, $T$ is a
differentiable function on $T(M)$ and $\alpha$ is a semibasic 1-form on
$T(M))$. In physical terms, $T$ is called {\it kinetic} energy and $\alpha$ is
called {\it force-generating field} of $\mathcal{M}$. The exact 2-form
$\omega:=\mathbf{d}\mathbf{d}_vT$ is called {\it fundamental form} of the
mechanical system $\mathcal{M}$. If $\omega$ is nondegenerate, which we shall
further assume, then $\mathcal{M}$ is called {\it regular}, so, in such a case
we have a symplectic 2-form on $T(M)$. Locally we have
$$
\mathbf{d}_vT=\frac{\partial T}{\partial \dot{q}^\mu}dq^\mu, \ \
\mathbf{d}\mathbf{d}_vT=
\frac{\partial^2 T}{\partial q^\mu\partial \dot{q}^\nu}dq^\mu\wedge dq^\nu+
\frac{\partial^2 T}{\partial\dot{q}^\mu\partial\dot{q}^\nu}d\dot{q}^\mu\wedge dq^\nu .
$$
Clearly, the regularity of $\mathcal{M}$ is equivalent to
$$
det\left(\frac{\partial^2 T}
{\partial\dot{q}^\mu\partial\dot{q}^\nu}\right)\neq 0.
$$
Since $\omega$ is nonsingular we have the one to one correspondence between
vector fields and 1-forms. We consider the 1-form
$$
\beta:=\mathbf{d}\left(T-\dot{q}^\mu\frac{\partial
T}{\partial\dot{q}^\mu}\right)+\alpha.
$$
The corresponding vector field $X$ defined by
$$
i_X\omega=\beta=
\mathbf{d}\left(T-\dot{q}^\mu\frac{\partial T}{\partial\dot{q}^\mu}\right) +\alpha
$$
determines a dynamical system on the symplectic manifold $T(M)$. Moreover, this
dynamical system is described in fact by a second order differential equation on
$M$.
Since
$$ \alpha(X)=X\left(\dot{q}^\mu\frac{\partial
T}{\partial\dot{q}^\mu}-T\right)
$$
we obtain that the function
$$
\left(\dot{q}^\mu\frac{\partial T}{\partial\dot{q}^\mu}-T\right)
$$
satisfies the following relation along any trajectory $c: (a,b)\rightarrow
T(M)$ of $X$:
$$ \int_{(a,b)}c^*\alpha=
\left(\dot{q}^\mu\frac{\partial T}{\partial\dot{q}^\mu}-T\right)_{c(a)}^{c(b)} .
$$
We easily obtain in these terms that the Lagrange equations for the trajectories
of $X$ look like:
$$
X^\mu=\frac{d}{dt}\left(\frac{\partial
T}{\partial\dot{q}^\mu}\right)- \frac{\partial T}{\partial q^\mu}.
$$

The mechanical system $(\mathcal{M},T,\alpha)$ is called {\it conservative} if
the semibasic 1-form $\alpha$ is closed: $\mathbf{d}\alpha=0$. Under this
condition the corresponding vector field $X$ is hamiltonian and corresponds to
the 1-form $\beta$ as given above, so, $\beta$ is a first integral of $X$. If
there is a function $U$ on $M$ such that
$\beta=\mathbf{d}(p_M^*U )$, then the mechanical system is called {\it
lagrangian}. All lagrangian mechanical systems are conservative. The
corresponding hamiltonian $H$ looks like
$$
H=\frac{\partial T}{\partial\dot{q}^\mu}\dot{q}^\mu-T-U.
$$
We easily obtain the relation $i_X\omega=-dH$, and that $H$ is first integral
of $X$.

The function $T+p^*_M(U)$ is called {\it lagrangian} of the
corresponding mechanical system. Introducing the function $L$ according to
$$
H=\frac{\partial L}{\partial\dot{q}^\mu}\dot{q}^\mu-L ,
$$
we get a new look at the Lagrange equations :
$
\frac{d}{dt}\left(\frac{\partial
L}{\partial\dot{q}^\mu}\right)- \frac{\partial L}{\partial q^\mu}=0.
$

\subsection{The role of potential function in mechanics}

The considerations in the last three subsections suggest that no coordinate on
the basic manifold should be physically interpreted in general as a
time-coordinate, otherwise, the hamiltonian function $H$ will NOT be a
conservative quantity, so, the presumption that it could be interpreted as
energy of an isolated physical system fails. This observation leads to the
undesirable conclusion, that in case of the standard classical symplectic
manifold $T^*(\mathbb{R}^3)$ where $H=\frac12mv^2+U(x,y,z)$, the particle can
not change its kinetic energy since the so called potential $U(x,y,z)$ is
traditionally interpreted as a static {\it field object}, i.e. as a
continuously distributed physical object. Therefore, all its physically
meaningful characteristics, in particular its differential $\mathbf{d}U$, or
$grad\,U$, been usually understood as the real energy-momentum transfer agent
called {\it force}, shall NOT change with time at any point. It deserves
noting, also, that its change from point to point, in particular along a
trajectory, does NOT allow to conclude that the particle moving along a fixed
trajectory should change its energy-momentum at the expense of the field $U$,
otherwise, according to the universal energy conservation law, the field $U$
{\it must} change its energy, which requires its explicit and essential time
dependence.

Recalling that the energy of the particle is a quadratic function of its
momentum, we also meet with something very strange: the particle changes its
momentum at the expense of a static external field. This strongly contradicts
the momentum conservation law since every static field has zero momentum.

Traditionally, the systems considered in classical mechanics consist of two
interacting subsystems: the point-like particles identified by their masses
$m$ and their behaviour, on one hand, and the so called external field
identified by its ability to change the state of the particles through changing
their kinetic energy, on the other hand. Let's note that this understanding
presumes that the field carries energy. Now the problem is that any static
physical object, no matter is it particle or field, can not interact with other
objects since from local as well as from integral viewpoints its energy does not
change with time, so, the particles presented can not exchange energy-momentum
with the physical field represented mathematically by the static function
$U(x,y,z)$ or $\mathbf{d}U$, where $(x,y,z)$ are supposed to be coordinates in
the spatial region under consideration. This contradicts the observable fact
that particles change their energy and momentum, although this external field
has zero momentum, so, the question {\it where the particles take energy and
momentum from} stays unanswered.

All this suggests that the physical interpretation of the
potential function in classical mechanics as {\it representing a continuous
physical field object}, and so as {\it necessarily carrying an energy-momentum
resource} to be at disposal for exchange with other physical objects (like
particles), has to be reconsidered.

This problem is quite clearly seen when we try to understand, for example, what
happens in the case of two $R$-distant charged particles with charges $(q,Q)$.
The potential there is given by $qQ/R$, and the usual view is based on the
assumption that each of the particles moves in the field of the other one, so
that the force field is written as $q\mathbf{E}_{Q}$, or as $Q\mathbf{E}_q$.
Now, the Gauss theorem requires existence of field every time when a charge is
available, so, where is the field associated with the charge $q$ in the first
force field, and with charge $Q$ in the second force field? Moreover, if there
are two fields in the space away from the two particles, no answer to the
natural question "do and how these physical fields of the {\it same nature}
interact" can be found in the literature.

It seems to us that $U$ has rather integral sense, a sense of {\it integral
interaction energy},  i.e., carrying information about the system through
presenting the integral interaction energy of the system considered in terms of
the configurational parameters $(Q,q,R)$ at a given moment of the system of
particles+fields. Such a view would suggest that $U$ should have sense only at
the spatial points being occupied by the particles at every moment of time.
However, if this is so, then how to
understand the values of $U$ outside these occupied points it is not quite
clear. In particular, how to understand the concept of force as $grad\,U\neq
0$, i.e., as constructed by the three partial derivatives of $U$, which requires
$U$ to be well defined and differentiable function outside the trajectories of
the particles?

Our vision towards a correct theoretical view on these problems is that from
theoretical point of view the point-like notion about particles has to be left
off/abandoned, and the theoretical concept of field object has to be duly
clarified and correspondingly respected throughout all theoretical studies. The
very basic principles of physics say that any real interaction in Nature
requires direct energy-momentum exchange between the interacting objects,
including vanishing of some of the initially presented ones provided energy
conservation. So, every adequate view on physical interaction among particles
in mechanics must take care to explain what happens in the external space, in
the space out of the points/regions occupied by the particles.
\vskip 0.2cm
{\bf Literature}
\vskip 0.2cm
\addcontentsline{toc}{subsection}{{\bf Literature}}
\vskip 0.1cm
1. {\bf C. Godbillon},  {\it Geometrie differentielle et mecanique
analytiqe}, Hermann, Paris (1969)
\vskip 0.1cm

2. {\bf A.Kushner, V.Lychagin, V.Rubtsov}, {\it Contact Geometry and
Non-linear Differential Equations}, Cambridge University Press 2007

\section{Stress and Strain}
\subsection{Preliminary notes}
The concepts of {\it stress} \index{stress} and {\it strain} \index{strain}
arise in classical
mechanics of continuous physical media as a first step towards finding an
adequate approach to describing possible propagation and interaction, i.e.
energy-momentum exchange, between the propagating in the corresponding medium
physical field $F$, on one hand, and the medium, considered as physical object
that is able to detect $F$ and to keep its identity during this interaction, on
the other hand. The {\it stress} concept is engaged mainly with the notion "$F$
{\it acts upon the medium}", and the {\it strain} concept is engaged mainly
with the notion "{\it the medium feels that it is acted upon, but it deforms in
order to survive}". Hence, there is action-reaction process but no destructions
are allowed. It should be noted also, that here, as in all classical mechanics,
time, as far it is involved in the process, is external parameter not depending
on the described physical processes.

It should be noted, that the further assumed in continuum mechanics linear
relation between the stress and strain tensors reduces the initial ambition for
dynamical understanding of what locally happens inside the material, to
parametrization of the admissible static configurations of the material subject
to the external field $F$. The introduced Lame coefficients as characteristics
of the ability of the attacked material not to be destroyed by the action of
$F$, give in fact the necessary knowledge from application point of view.

Hence, {\it no local dynamical} understanding of the process of passing
from one admissible configuration of the material to another admissible one,
i.e. how the admissible local energy-momentum exchange is realized, is achieved.
Therefore, we could not get directly in such a way some  knowledge about the
propagation of the attacking field $F$ inside the material from local point of
view. Moreover, no understanding of the problem "what really happens locally
inside each of the two interacting objects" seems to has been really obtained.

If we want to understand dynamically and describe mathematically the
propagation with no destruction of a finite continuous object in appropriate
vacuum, i.e. in a media admitting this propagation by means of keeping a
{\it permanent dynamical equilibrium} with the object so that we may speak about
"propagation in vacuum", then, we have to set the question: "how an
external observer would understand the propagational appearance of a continuous
object as determined by its internal dynamical structure?"

This problem requires new visions/insight and, therefore, new mathematical
objects in order to extend appropriately the old concepts of {\it stress
tensor} and {\it strain tensor}, which, more or less, are based on an elementary
notion about action and reaction. And this is due to the fact that the whole
dynamics and surviving of our propagating object, been in {\it permanent
dynamical equilibrium} with the environment, should be understood
and explained {\it only} in terms of its interacting subsystems, which
subsystems may be not able to exist separately, but only as interacting ones.

One of the greatest, from our point of view, achievements of Maxwell is mainly
in this direction: he found the way to attack this problem by constructing a
stress tensor for a {\it vacuum electromagnetic field}, although not directly,
but through making use of the hypothesis that the electromagnetic field is well
defined in the media of continuously distributed electric charges.
Nevertheless, writing down this tensor only in terms of the electric and
magnetic fields, he showed the right way in approaching this problem.

Our purpose in this section is to try to look at Maxwell's results as
appropriate tools suggesting how to define corresponding quantities in terms of
which to understand the internal dynamics of propagating time-stable and
spatially finite field objects. The main idea will be based on the
understanding that all the necessary surface and volume elements across which
the internal energy-momentum exchange should take place have to be constructed
out of the very field (vector) components, and not of concepts connected with
the external world. In some sense we shall try to peep into the field object's
internal structure, considered as static, but thought of as being in dynamical
equilibrium. The line of extension of the classical quantities will be directed
by the theoretical comprehension that our propagating, finite and time-stable
object carries smart enough dynamical structure in order to present itself to
the rest of the world as an appropriately self-organized dynamical system
knowing how to deserve the qualification of "free object".

\subsection{Stress}
From formal point of view the most natural mathematical object to be considered
as stress generating, seems to be any vector field, because every vector
field defined on an arbitrary manifold $M$ generates 1-parameter family
$\varphi_t$ of (local in general) diffeomorphisms of $M$. Therefore, having
defined a vector field $X$ on $M$ we can consider for each $t\in\mathbb{R}$ the
corresponding diffeomorphic image $\varphi_t(U)$ of any region $U\subset M$.
Hence, if the spatially finite object under consideration occupies for each
$t$ the finite region $U_t\subset \mathbb{R}^3$, carries dynamical structure
and propagates in the 3-space, should stay identical to itself during
propagation, the available 1-parameter group of diffeomorphisms can be
considered as the appropriate mathematical tool to describe its propagation as
a whole, as well as from local point of view, since every admissible change
described by $\varphi_t$ has unique opposite, described by $\varphi_{-t}$,
i.e. nothing essential has been lost irreversibly and the object stays
identical to itself.

Let now $X$ be a vector field on the euclidean space $(\mathbb{R}^3,g)$, where
$g$ is the euclidean metric in $T\mathbb{R}^3$ having in the canonical global
coordinates $(x^1,x^2,x^3=x,y,z)$ components $g_{11}=g_{22}=g_{33}=1$ and
$g_{12}=g_{13}=g_{23}=0$. The induced euclidean metric in $T^*\mathbb{R}^3$ has
in the dual bases the same components and will be denoted further by the same
letter $g$. The corresponding isomorphisms between the tangent and cotangent
spaces and their tensor, exterior and symmetric products will be denoted by
the same letter $\tilde{g}$, so
$$
\tilde{g}\left(\frac{\partial}{\partial x^i}\right)=
g^{ik}\left(\frac{\partial}{\partial x^k}\right)=dx^i,
\ \ (\tilde{g})^{-1}(dx^i)=\frac{\partial}{\partial x^i} \ \  \cdots
$$
Having another vector field  $Y$ on $\mathbb{R}^3$ we can form their scalar
product
$$
g(X,Y)\equiv X.Y=g_{ij}X^iY^j=X_iY^i=X_1Y^1+X_2Y^2+X_3Y^3.
$$
According to classical vector analysis on $\mathbb{R}^3$ for the differential
of the function $g(X,Y)$ we obtain
$$
\mathbf{d}g(X,Y)=(X.\nabla)Y+(Y.\nabla)X+X\times\mathrm{rot}\,(Y)+
Y\times\mathrm{rot}\,(X),
$$
where in our coordinates
$$
X.\nabla=\nabla_{X}=X^i\frac{\partial}{\partial x^i}, \ \ \
(X.\nabla)Y=\nabla_{X}Y=
X^i\frac{\partial Y^j}{\partial x^i}\frac{\partial}{\partial x^j},
$$
"$\times$" is the usual vector product, and
$$
(\mathrm{rot}X)^i=\left(\frac{\partial X^3}{\partial x^2}-\frac{\partial
X^2}{\partial x^3}, \ \frac{\partial X^1}{\partial x^3}-\frac{\partial
X^3}{\partial x^1}, \ \frac{\partial X^2}{\partial x^1}-
\frac{\partial X^1}{\partial x^2}\right)\cdot
$$
The Hodge $*$-operator acts in these coordinates as follows:
$$
*dx=dy\wedge dz, \ \ *dy=-dx\wedge dz, \ \ *dz=dx\wedge dy,
$$
$$
*(dx\wedge dy)=dz, \ \ *(dx\wedge dz)=-dy, \ \ *(dy\wedge dz)=dx,
$$
$$
*(dx\wedge dy\wedge dz)=1, \ \ \ *1=dx\wedge dy\wedge dz .
$$
{\bf Corollary}. The following relation holds:
$$
\mathrm{rot}X=(\tilde{g})^{-1}\, *\,\mathbf{d}\,\tilde{g}(X), \ \ \text{or} \ \
\tilde{g}(\mathrm{rot}X)=*\,\mathbf{d}\,\tilde{g}(X).
$$
Assume now that in the above expression for $\mathbf{d}g(X,Y)$ we put $X=Y$. We
obtain
$$
\frac12\mathbf{d}(X^2)=\frac12\mathbf{d}g(X,X)=
X\times\mathrm{rot}X+(X.\nabla)X=X\times\mathrm{rot}X+\nabla_XX.
$$
In components, the last term on the right reads
\[ (\nabla_X
X)^j=X^i\nabla_i X^j=\nabla_i(X^iX^j)-X^j\nabla_iX^i=
\nabla_i(X^iX^j)-X^j\mathrm{div}\,X ,
\]
where
$$
\mathrm{div}X=*L_X\left(dx\wedge dy\wedge dz\right)=
*\left(\frac{\partial X^i}{\partial x^i}dx\wedge dy\wedge dz\right)=
\frac{\partial X^i}{\partial x^i}.
$$
Substituting into the preceding relation, replacing
$\mathbf{d}(X^2)$ by $(\nabla_i\delta^i_jX^2)dx^j$ and making some elementary
transformations we obtain
\[
\nabla_i\left(X^iX^j-\frac12
g^{ij}X^2\right)= \big [(\mathrm{rot}\,X)\times X+X\mathrm{div}\,X\big ]^j.
\]
The symmetric 2-tensor
$$
M^{ij}=X^iX^j-\frac12\,g^{ij}X^2=
\frac12\Big
[X^iX^j+\big(\tilde{g}^{-1}*\tilde{g}(X)\big)^{ik}
\big(*\tilde{g}(X)\big)_{k}\,^j\Big ]
$$
we shall call further Maxwell stress tensor \index{Maxwell stress tensor}
generated by the (arbitrary) vector
field $X\in\mathfrak{X}(\mathbb{R}^3)$. Clearly, when we raise and lower indices
in canonical coordinates with $g$ we shall have the following component
relations:
$$
M_{ij}=M_i^j=M^{ij}
$$ which does not mean, of course, that we equalize
quantities being elements of different linear spaces.

We note now the easily verified relation between the vector product "$\times$"
and the wedge product in the space of 1-forms on $\mathbb{R}^3$:
$$
X\times Y=(\tilde{g})^{-1}\,(*\,(\tilde{g}(X)\wedge\tilde{g}(Y)))
$$
$$
=(\tilde{g})^{-1}\,\circ i(X\wedge Y)(dx\wedge dy\wedge dz), \ \
X,Y\in\mathfrak{X}(\mathbb{R}^3) .
$$
We shall prove now the following

	{\bf Proposition}. If $\alpha=\tilde{g}(X)$ then the following relation
holds:
$$
\tilde{g}(\mathrm{rot}\,X\times X)=i(X)\mathbf{d}\alpha=
-*(\alpha\wedge *\mathbf{d}\alpha) .
$$
{\it Proof}.
$$\tilde{g}(\mathrm{rot}\,X\times X)=
\tilde{g}\circ(\tilde{g})^{-1}\,*(\tilde{g}(\mathrm{rot}X)\wedge\tilde{g}(X))=
-*(\alpha\wedge *\mathbf{d}\alpha).
$$
For the component of $i(X)\mathbf{d}\alpha$ before $dx$ we obtain
$$
-X^2\left(\frac{\partial \alpha_2}{\partial x^1}-\frac{\partial
\alpha_1}{\partial x^2}\right)-X^3\left(\frac{\partial \alpha_3}{\partial
x^1}- \frac{\partial \alpha_1}{\partial x^3}\right),
$$
and the same quantity is easily obtained for the component of
$[-*(\alpha\wedge *\mathbf{d}\alpha)]$ before $dx$. The same is true for the
components of the two 1-forms before $dy$ and $dz$. The proposition is proved.

Hence, $\mathbf{d}\alpha=\mathbf{d}\tilde{g}(X)$ is the 2-form across which the
vector field $X$ will drag the points of the finite region
$U\subset\mathbb{R}^3$, and local interaction will take place if
$i(X)\mathbf{d}\alpha$ is  not zero.

As for the second term
$X\mathrm{div}X$ of the divergence $\nabla_jM_i^j$, since
$\mathbf{d}*\alpha=\mathrm{div}X(dx\wedge dy\wedge dz)$ we easily obtain
$$
i(\tilde{g}^{-1}(*\alpha))\mathbf{d}*\alpha=\tilde{g}(X)\,\mathrm{div}X .
$$
Hence, additionally,
the 2-vector $\tilde{g}^{-1}(*\alpha)$ will drag the points of $U\subset
\mathbb{R}^3$ across the 3-form $\mathbf{d}*\alpha$.

We can write now
$$
\nabla_iM^i_jdx^j=\big[i(X)\mathbf{d}\alpha+
i(\tilde{g}^{-1}(*\alpha))\mathbf{d}*\alpha\big]_jdx^j .
$$

Physically this may be understood in the sense, that our physical field is
represented by two objects: $\alpha=\tilde{g}(X)$ and $*\alpha$, and each of
them interacts with itself only, so that there is no stress exchange between
$\alpha$ and $*\alpha$, as well as, there is no available local nonzero
interaction stress between
$\alpha$ and $*\alpha$. The static nature of the situation and the naturally
isolated two terms in $\nabla_iM^i_jdx^j$ suggest the following equations to
hold
$$ i(X)\mathbf{d}\alpha=0, \ \
i(\tilde{g}^{-1}(*\alpha))\mathbf{d}*\alpha=0, \ \ \text{i.e.} \ \
X\times\mathrm{rot}X=0, \ \mathrm{div}X=0,
$$
the solutions of which are known as Beltrami vector fields
\index{Beltrami vector fields}.

In the above consideration the role of the euclidean metric $g$ was somehow
overlooked since no derivatives of $g$ appeared explicitly. We are going now to
come to these equations paying due respect to $g$ and showing its significance
in the problem of constructing conservative quantities, a point that will be of
crucial importance when we consider time-dependent and propagating in
the 3-space  real field objects.

First, we recall the invariance of any definite integral on
$U\subset\mathbb{R}^3$ with respect to orientation preserving diffeomorphisms.
Let $\omega$ be a 3-form on $\mathbb{R}^3$ with compact support
$U\subset\mathbb{R}^3$. Then the invariance of the integral with respect to the
diffeomorphism $\varphi$ means
$$
 \int_U\varphi^*\omega=\int_{\varphi(U)}\omega.
$$
If the 1-parameter group of diffeomorphisms $\varphi_t$ is generated by the
vector field  $Y\in\mathfrak{X}(\mathbb{R}^3)$, then for each
$t\in [0,1]\subset\mathbb{R}$ the quantity
$$
I(t)=\int_{U}\varphi_t^*\omega
$$
is well defined, and  the following relation holds:
$$
\frac{dI(t)}{dt}=\int_{U}\varphi_t^*L_Y\omega .
$$

We shall work further in this part of the section with the 1-form
$\alpha=\tilde{g}(X)$, where the vector field $X$ represents our finite field
object as in the previous part of the section, and generates stress according to
the Maxwell stress tensor $M^i_j(X)$. We want to see how the 3-form
$\alpha\wedge *\alpha=i(X)\alpha\,dx\wedge dy\wedge
dz=\langle\alpha,X\rangle\omega$ changes along an arbitrary vector field $Y\neq
X$, so we have to find the corresponding Lie derivative. $$ L_Y(\alpha\wedge
*\alpha)=(L_Y\alpha)\wedge *\alpha+\alpha\wedge L_Y(*\alpha) $$ $$ =
(L_Y\alpha)\wedge *\alpha+\alpha\wedge *L_Y\alpha+\alpha\wedge[L_Y,*_1]\alpha=
2(L_Y\alpha)\wedge *\alpha+\alpha\wedge[L_Y,*_1]\alpha  , $$ where $[L_Y,*_1]$
is the commutator $L_Y\circ\,*_1-*_1\circ L_Y$, and the index of $*$ denotes
the degree of the form it is applied to. Further we get $$ (L_Y\alpha)\wedge
*\alpha=(i_Y\mathbf{d}\alpha)\wedge*\alpha-<\alpha,Y>.\mathbf{d}*\alpha+
\mathbf{d}(<\alpha,Y>.*\alpha).
$$
Noting that $L_Y(\alpha\wedge *\alpha)=\mathbf{d}(\alpha^2i_Y\omega)$
and the relation between the exterior derivative
$\mathbf{d}$ and the coderivative $\delta$ in euclidean case for p-forms, given
by
$$
(-1)^{p(n-p)}\delta_p\circ *_{n-p}=*_{(n-p+1)}\circ \mathbf{d}_{(n-p)}
$$
we obtain consecutively
$$
\mathbf{d}\left(\frac12\alpha_i\alpha^ii_Y\omega-<\alpha,Y>*\alpha\right)=
-[\alpha^i(\mathbf{d}\alpha)_{ij}+\alpha_j\,\mathrm{div}X]Y^j\omega
+\frac12\alpha\wedge[L_Y,*_1]\alpha ,
$$
$$
\delta\circ
*_2\left(\frac12\alpha_i\alpha^i\,i_Y\omega-<\alpha,Y>*\alpha\right)=
-[\alpha^i(\mathbf{d}\alpha)_{ij}+\alpha_j\,\mathrm{div}X]Y^j
+\frac12\alpha\wedge[L_Y,*_1]\alpha ,
$$
$$
*_2<\alpha,Y>*\alpha=\alpha<\alpha,Y>,
$$
$$
\delta\left(\frac12\alpha_i\alpha^ig_{jk}-\alpha_j\alpha_k)Y^jdx^k\right)=
-\nabla_j\left[\left(\frac12\alpha_i\alpha^i\delta^j_k-
\alpha_k\alpha^j\right)Y^k\right] .
$$
So we get
$$
\nabla_j\left[\left(\frac12\alpha_i\alpha^i\delta^j_k-
\alpha_k\alpha^j\right)Y^k\right]=
[\alpha^i(\mathbf{d}\alpha)_{ij}+\alpha_j\,\mathrm{div}X]Y^j+
\frac12*\big(\alpha\wedge[L_Y,*_1]\alpha\big) .
$$
Hence, if $\alpha\wedge[L_Y,*_1]\alpha =0$, then
the equations for our 1-form $\alpha=\tilde{g}(X)$ are
$$
\alpha^i(\mathbf{d}\alpha)_{ij}=0, \ \ \alpha_j\,\mathrm{div}X=0 , \ \
i,j=1,2,3
$$
and the co-closed 1-form $M_{ij}Y^idx^j$ defines the closed
2-form $*(M_{ij}Y^idx^j)$, so, its integral over a closed 2-surface that
separates the 3-volume where $\alpha$ is different from zero, gives a
conservative quantity and the nature of this conservative quantity is connected
with the nature of the vector field $Y$.
Of course, the term "conservative" here is, more or less,
trivial, since everything is static, no propagation of the field $\alpha$ takes
place.

Finally we note that if the vector fields $X_1,X_2, ...,X_p$ describe
non-interac\-ting physical objects, we can form the sum of their Maxwell stress
tensors
$$
M^{ij}(X_1)+M^{ij}(X_2)+...+M^{ij}(X_p)=
\sum_{k=1}^p\left[(X_k)^i(X_k)^j-\frac12(X_k)^2g^{ij}\right] ,
$$
which is not equal to the Maxwell stress tensor of their sum
$M^{ij}(X_1+...+X_p)$.

As for solutions of the above equations, we see two classes of solutions:

	1. Linear, i.e. those generated by a function $f$ satisfying the
Laplace equation $\Delta f=0$  with $\alpha=\mathbf{d}f$ .

	2. Those, satisfying
$\mathbf{d}\alpha\neq 0$ but $\alpha^i\mathbf{d}\alpha_{ij}=0$ and
$\mathbf{d}*\alpha=0$. Note that for such solutions the determinant
$det||(\mathbf{d}\alpha)_{ij}||, i,j=1,2,3$, is always zero, so the
homogeneous linear system
$$
\alpha^i(x,y,z)\mathbf{d}\alpha_{ij}(x,y,z)=0
$$
allows to express explicitly $X$ through the derivatives of its components.

Of course, all these solutions are static, so they can not serve as models of
spatially propagating finite physical objects with dynamical structure.

\subsection{Strain}
The concept of {\it strain}
is introduced in studying {\it elastic} materials subject to external forces of
different nature: mechanical, electromagnetic, etc. In nonrelativistic
continuum physics the local representatives of the external forces in this
context are usually characterized in terms of {\it stresses}, considered briefly
above. Since the force means energy-momentum transfer leading to corresponding
mutual energy-momentum change of the interacting objects, then according to the
energy-momentum conservation law the material must react somehow to the
external influence in accordance with its structure and reaction abilities.
The classical strain describes mainly the abilities of the material to bear
force-action from outside through deformation, i.e. through changing its
shape, or, configuration. The term {\it elastic} \index{elastic}
now means that any two
allowed configurations can be deformed to each other without appearance of
holes and breakings, in particular, if the material considered has
deformed from configuration $C_1$ to configuration $C_2$, it is able to
return smoothly to its initial configuration $C_1$.

The general geometrical description starts with the assumption
that an elastic material is a continuum $\mathbb{B}\subset\mathbb{R}^3$ which
can {\it smoothly deform} inside the space $\mathbb{R}^3$, so, it can be
endowed with differentiable structure, i.e. having an elastic material is
formally equivalent to have a smooth real 3-dimensional submanifold
$\mathbb{B}\subset\mathbb{R}^3$. The deformations are formally considered as
smooth maps $\varphi: \mathbb{B}\rightarrow\mathbb{R}^3$. The spaces
$\mathbb{B}$ and $\mathbb{R}^3$ are endowed with riemannian metrics
$\mathbf{G}$ and $g$ respectively (and corresponding riemannian co-metrics
$\mathbf{G}^{-1}$ and $g^{-1}$), and induced isomorphisms $\tilde{\mathbf{G}}$
and $\tilde{g}$ between the corresponding tangent and cotangent spaces . Now,
an important combination is the difference $$
\mathbf{E}_x:=\frac12(\varphi^*g-\mathbf{G})_x:
T_x\mathbb{B}\times T_x\mathbb{B}\rightarrow \mathbb{R} ,
$$
where $\varphi^*g$ denotes the induced on $\mathbb{B}$ metric from the metric
$g$ (usually euclidean) on $\mathbb{R}^3$.

We could look at the problem from a more general formal point of view as
follows. The mathematical counterparts of the allowed
deformations are the diffeomorphisms $\varphi$ of a riemannian
manifold $(M,g)$, and every $\varphi(M)$ represents a possible configuration of
the material considered. But some diffeomorphisms may not lead to
deformation (i.e. to shape changes), so, a criterion must be introduced to
separate those diffeomorphisms which should be considered as essential. For
such a criterion is chosen the distance change: {\it if the distance between
any two fixed points does not change during the action of the external force
field, then we say that there is no deformation}. Now, every essential
diffeomorphism $\varphi$ must transform the metric $g$ to some new metric
$\varphi^*g$, such that $g\neq\varphi^*g$. The naturally arising tensor field
$e=(\varphi^*g-g)\neq 0$ appears as a measure of the physical abilities of the
material to withstand external force actions. It deserves to be noted that if
$g$ does not induce nonzero riemannian curvature, then $\varphi^*g$ also does
not induce riemannian curvature for any diffeomorphism $\varphi$ of $M$.

Since the external force is assumed to act locally and the material
considered gets the corresponding to the external force field final
configuration in a smooth way, i.e., passing smoothly through a family of
allowed configurations, we may introduce a localization of the above scheme, such
that the isometry diffeomorphisms to be eliminated. This is done by means
of introducing 1-parameter group $\varphi_t, t\in [a,b]\subset\mathbb{R}$
of local diffeomorphisms, so, $\varphi_a(M)$ and $\varphi_b(M)$ denote
correspondingly the initial and final configurations. Now $\varphi_t$
generates a family of metrics $\varphi_t^*\,g$, and a corresponding family
of tensors $e_t=\varphi_t^*g-g$. According to the local analysis every local
1-parameter group of diffeomorphisms is generated by a vector field on $M$. Let
the vector field $X$ generate $\varphi_t$. Then the quantity
$$
\frac12\,L_{X}g:=\frac12\,\lim_{t\rightarrow 0}\frac{\varphi_t^*\,g-g}{t} \ ,
$$
i.e. one half of the {\it Lie derivative} of $g$ along $X$, is called
{\it infinitesimal strain tensor}, or just {\it deformation tensor}, or
{\it deformation velocity tensor} \index{deformation tensor}. Clearly, the
tensor $L_Xg$ is different from zero only if $X$ is not local isometry, i.e., if
$g$ is locally $X$-attractive. In local coordinates we have in general $$
(L_Xg)_{ij}=X^k\frac{\partial
g_{ij}}{\partial x^k}+ g_{ik}\frac{\partial X^k}{\partial x^j}+
g_{jk}\frac{\partial X^k}{\partial x^i},
$$
so, in the euclidean case in standard coordinates, where $g_{ij}$ are
constants, we get
$$
(L_Xg)_{ij}= g_{ik}\frac{\partial X^k}{\partial x^j}+
g_{jk}\frac{\partial X^k}{\partial x^i}= \frac{\partial X_i}{\partial x^j}+
\frac{\partial X_j}{\partial x^i} \cdot
$$
If we assume the mentioned above
linear relation between stress and strain to hold also between our generated by
the arbitrary vector field $X$ stress tensor $M(X)=X\otimes X-\frac12g^{-1}X^2$
and the corresponding infinitesimal strain tensor $L_Xg$, it would look like $$
M(X)_{ij}=C_{ij}^{kl}(L_Xg)_{kl}, \ \ \ \text{or}, \ \ \
(L_Xg)_{kl}=S_{kl}^{ij}M(X)_{ij}.
$$

Considering $X$ as formal image of a physical field, we note, that the tensors
$C/S$ connect a quadratic function of its components $X^i$ with a linear
function of the derivatives of its components, i.e., the value of field at a
point depends on the values of the field around this point. This goes along
somehow with the static nature of the field, but still there is no time change
and no propagation in space, i.e., there is no dynamics. In our view, in these
terms, a {\it dynamics suggesting relation should connect the divergence of
$M(X)$ with some projection of} $L_Xg$, for example, along the expected
direction of propagation.

There is another moment to be pointed out. In our view, an internal dynamics
may take place only between/among time-recognizable subsystems, which we
mathematically understand as coordinate-free recognized/identi\-fied
objects. The components of a vector field do not satisfy such a condition: by a
coordinate change we could nullify a given component of a vector field at a
given space point. This suggests that to any real object, existing through a
permanent internal dynamics, in the theory should correspond a mathematical
object with vector components. Thus, to every recognizable subsystem of the
physical object considered the theory should juxtapose a coordinate free
mathematical object. Such mathematical objects are, for example, the vector
valued differential forms. This is the reason why we
 paid special attention to $\alpha$ and $*\alpha$ above.

\section{Some Formal Relativity}

The relativistic approach in mechanics and field theory appeared
as a necessary generalization of classical approach in order to
incorporate formally and appropriately in the theories the time aspect of all
real physical processes. According to it {\it time is not external and depending
on nothing theoretical parameter, on the contrary, a physical process acquires
time characteristics when related to another physical process}. In order
to compare real time periods this approach assumes the frame invariance of the
speed of light in vacuum $c$ and introduces the concept of 4-dimensional
space-time in the theory as a basic manifold in which every physical process
must be considered. Moreover, a basic assumption in this approach is that no
physical frame object should propagate translationally in the 3-space with speed
greater than or equal to the speed of light $c$ in vacuum\index{relativistic
time}.

A basic mathematical object that controls all this is the space-time
pseudometric $\eta$, defined on the mathematical manifold $\mathbb{R}^4$, and
having components in the canonical coordinates
$$
x^\mu=(x,y,z,\xi=ct), \ \ \ \mu=1,2,3,4
$$
as
$$
\eta_{11}=\eta_{22}=\eta_{33}=-\eta_{44}=-1,  \ \ \eta_{\mu\nu}=0 \ \ \
\text{for} \ \ \ \mu\neq \nu, \ \ \ ind(\eta)=3.
$$
Hence, the space-time $(\mathbb{R}^4,\eta)$ is a flat
4-dimensional pseudo-riemannian manifold, usually called Minkowski
space-time, which will be denoted further by $M^4$, or just by $M$ when no
misunderstanding may happen.

All diffeomorphisms $\varphi:M\rightarrow M$ that respect the canonical values
of $\eta$ and are not translations are called Lorentz transformations. These
transformations are linear, they admit physical interpretation as relatively
moving frames with respect to each other with constant translational velocity
$v<c$, and these frames are usually called {\it inertial} \index{inertial
frame}. The classical principle of inertia says now, that all physical
processes proceed in the same way with respect to any inertial frame.

Of course, the concept of inertial frame does not require to work only in
canonical coordinates, it only allows to make use of such coordinates. For,
example, in any inertial frame we can pass to coordinates
$(r,\theta,\varphi,\xi)$, where $(r,\theta,\varphi)$ are the standard spherical
coordinates on $\mathbb{R}^3$. Passing to canonical coordinates is guaranteed
by the existing possibility to separate the time coordinate in an invariant
way. In fact, making use of any euclidean metric $h$ on $\mathbb{R}^4$ we can
form the linear map $\psi_\mu^\nu=\eta_{\mu\sigma}h^{\sigma\nu}$ which has just
one time-like eigen direction.

If the frame we work in is not inertial we can not pass to canonical
coordinates, i.e. to coordinates where $\eta$ acquires its canonical values.
This does not mean, however, that we can not pass to another noninertial frame
where $\eta$ will have the same components as in the preceding noninertial
frame.

A vector field $X\in\mathfrak{X}(M)$ is called geodesic, or autoparallel, if it
satisfies the nonlinear equation $\nabla_XX=0$, where $\nabla$ is defined by the
$\eta$-determined Levi-Civita connection in $TM$.  Clearly, all mass bodies
of an inertial frame are parallely transported along the geodesic
trajectories of the same geodesic vector field.

A natural extension of the concept of inertial frame reads: if $\Gamma$ is a
linear connection in $TM$ then every $\Gamma$-geodesic vector field defines a
$\Gamma$-inertial frame, corresponding classes of $\Gamma$-inertial frames and
corresponding principle of inertia, which principle differs seriously, of
course, from the $\eta$-defined principle of inertia.

Since the $\eta$-metric is not positively definite the sets of vector fields
$X$ and one-forms $\alpha$ on $M$ admit time-like ones:
$$
X^2=\eta(X,X)>0, \ \ \alpha^2=\eta^{-1}(\alpha,\alpha)>0;
$$
space-like ones:
$$
X^2=\eta(X,X)<0, \ \ \alpha^2=\eta^{-1}(\alpha,\alpha)<0;
$$
and isotropic (frequently called null) ones:
$$
X^2=\eta(X,X)=0, \ \ \alpha^2=\eta^{-1}(\alpha,\alpha)=0.
$$
The time-like vector fields are physically interpreted as momentum vector fields
of mass particles, the isotropic vector fields are physically interpreted
as momentum vector fields of massless (i.e. photon-like) particles, and the
space-like ones are interpreted as stress generating.

It deserves to be noted that these classes are NOT corresponding subspaces.

Since these are invariant properties, we obtain three classes of curves on
$M$: time-like, space-like and isotropic. The above property of vector fields
and one-forms is correspondingly extended to the whole tensor algebra on $M$,
i.e., we can talk about space-like, time-like and isotropic symmetric tensor
fields, differential forms, etc.

The flow of a geodesic time-like or isotropic vector field started from a region
$U\subset M$ fills up a subset called {\it geodesic tube}. If a geodesic tube
is filled up by a spatially finite physical object then this object is called
{\it free}.

The manifold $M$ has natural $\eta$-defined volume form
$$
\omega_o=\sqrt{|det(\eta_{\mu\nu})|}dx^1\wedge dx^2\wedge dx^3\wedge dx^4=
dx\wedge dy\wedge dz\wedge d\xi.
$$
So, we can introduce the Poincare isomorphisms
$D^p: \mathfrak{X}^p(M)\otimes\omega_o\rightarrow\Lambda^{4-p}(M)$. For
decomposable $p$-vectors we obtain:
$$ D^1(X\otimes\omega_o)=i(X)\omega_o; \ \
D^2(X\wedge Y\otimes\omega_o)=i(Y)\circ i(X)\omega_o;
$$
$$
D^3(X\wedge Y\wedge Z\otimes\omega_o)=i(Z)\circ i(Y)\circ i(X)\omega_o.
$$
For example, in canonical coordinates we obtain
$$
D^1(X\otimes\omega_o)=X^1dy\wedge dz\wedge d\xi-
X^2dx\wedge dz\wedge d\xi+X^3dx\wedge dy\wedge d\xi-
X^4dx\wedge dy\wedge dz.
$$
\begin{align*}
D^2(X\wedge Y\otimes\omega_o)&=
(X^3Y^4-X^4Y^3)dx\wedge dy+(X^4Y^2-X^2Y^4)dx\wedge dz\\
&+(X^1Y^4-X^4Y^1)dy\wedge dz+(X^2Y^3-X^3Y^2)dx\wedge d\xi\\
&+(X^3Y^1-X^1Y^3)dy\wedge d\xi+(X^1Y^2-X^2Y^1)dz\wedge d\xi.
\end{align*}

The Hodge $*$-operator \index{Hodge $*$-operator on Minkowski space}
defined by $\eta$ acts as follows in canonical
coordinates:
$$
 \alpha\wedge *\beta=\beta\wedge *\alpha=
(-1)^{ind(\eta)}\eta (\alpha,\beta)\omega_o,\ \ \alpha,\beta\in\Lambda^p(M),
\ \ \ ind(\eta)=3.
$$
$$
 \alpha\wedge\beta=-\alpha\wedge *(*\beta)=
-(-1)^{ind(\eta)}\eta (\alpha,\beta)\omega_o=\eta (\alpha,\beta)\omega_o
,\ \ \alpha,\beta\in\Lambda^2(M).
$$
$$
*_{(4-p)}*_p=(-1)^{ind(\eta)+p(4-p)}id,\
(*^{-1})_p=(-1)^{ind(\eta)+p(4-p)}*_p,\ *\omega_o=1,\
*1=(-1)^{ind(\eta)}\omega_o,
$$
$$
 \begin{array}{ll}
    *dx=dy\wedge dz\wedge d\xi                 &*dx\wedge dy\wedge dz=d\xi
\\ {*}dy=-dx\wedge dz\wedge d\xi               &*dx\wedge dy\wedge d\xi=dz
\\ {*}dz=dx\wedge\ dy\wedge d\xi               &*dx\wedge dz\wedge d\xi=-dy
\\ {*}d\xi=dx\wedge dy\wedge dz                &*dy\wedge dz\wedge d\xi=dx
\end{array}
$$
$$
\begin{array}{ll}
*dx\wedge dy=-dz\wedge d\xi          &*dx\wedge d\xi=dy\wedge dz \\
{*}dx\wedge dz=dy\wedge d\xi         &*dy\wedge d\xi=-dx\wedge dz \\
{*}dy\wedge dz=-dx\wedge d\xi         &*dz\wedge d\xi=dx\wedge dy.
\end{array}
$$
We give the corresponding relations for the euclidean case where $ind(\eta)=0$.
$$
 \begin{array}{ll}
    *dx=dy\wedge dz\wedge d\xi                 &*dx\wedge dy\wedge dz=d\xi
\\ {*}dy=-dx\wedge dz\wedge d\xi               &*dx\wedge dy\wedge d\xi=-dz
\\ {*}dz=dx\wedge\ dy\wedge d\xi               &*dx\wedge dz\wedge d\xi=dy
\\ {*}d\xi=-dx\wedge dy\wedge dz                &*dy\wedge dz\wedge d\xi=-dx
\end{array}
$$
$$
\begin{array}{ll}
*dx\wedge dy=dz\wedge d\xi          &*dx\wedge d\xi=dy\wedge dz \\
{*}dx\wedge dz=-dy\wedge d\xi         &*dy\wedge d\xi=-dx\wedge dz \\
{*}dy\wedge dz=dx\wedge d\xi         &*dz\wedge d\xi=dx\wedge dy.
\end{array}
$$
We continue with pseudoeuclidean case and specially note the following property
of $*_2$: $$ *_2\circ *_2=-id_{\Lambda^2(M)}, $$ which means that $*_2$ is a
complex structure in the space $\Lambda^2(M)$. The matrix $J$ of $*_2$ in this
basis looks like ($J$ acts from the left on the basis 2-forms)
 $$
 J=\begin{Vmatrix} 0  & 0  & 0  & 0
& 0 & -1\\ 0  & 0  & 0  & 0  & 1  & 0 \\ 0  & 0  & 0  & -1 & 0  & 0 \\ 0  & 0
& 1 & 0  & 0  & 0 \\ 0  & -1 & 0  & 0  & 0  & 0 \\ 1  & 0  & 0  & 0  & 0  & 0
\\ \end{Vmatrix},
$$
 We see that
$$
\eta(*(dx^\mu\wedge dx^\nu),dx^\mu\wedge dx^\nu)=0,   \ \mu<\nu,
$$
but this does not mean that every $*$-corresponding
2-forms $(F,*F)\in\Lambda^2(M)$ are always orthogonal to each other. Also, if we
order the above canonical basis elements as $$ (dx\wedge dy,\,dx\wedge dz,\,
dy\wedge dz,\,dx\wedge d\xi,\, dy\wedge d\xi,\, dz\wedge d\xi), $$ we get the
following signature of the induced by $\eta$ metric in $\Lambda^2(M)$: $$
sign(\eta_{\Lambda^2(M)})=(+,+,+,-,-,-). $$

 The Poincare isomorphism $D_p:
\Lambda^p(M)\rightarrow\mathfrak{X}^{4-p}(M)$ is connected to the isomorphism
 $(\tilde{\eta}^{-1})_p: \Lambda^p(M)\rightarrow\mathfrak{X}^{4-p}(M)$ defined by
the metric $\eta$ as follows:
$$
(\tilde{\eta}^{-1})_p=(-1)^{p-1}D_{4-p}\circ *_p, \ \ p=1,2,3.
$$
For example, if $\omega^o$ is the basis 4-vector in $\mathfrak{X}^4(M)$ dual to
$\omega_o$ so that $<\omega_o,\omega^o>=1$, we obtain
$$
D_3\circ *_1(dx)=D_3(dy\wedge dz\wedge d\xi)=
i(d\xi)\circ i(dz)\circ i(dy)\omega^o
$$
$$
=i(d\xi)\circ i(dz)\circ i(dy)\left(\frac{\partial}{\partial x}\wedge
\frac{\partial}{\partial y}\wedge \frac{\partial}{\partial z}\wedge
\frac{\partial}{\partial \xi}\right)=-\frac{\partial}{\partial x}=
\tilde{\eta}^{-1}(dx),
$$
$$
D_2\circ *(dx\wedge d\xi)=D_2(dy\wedge dz)=i(dz)\circ i(dy)\omega^o=
\frac{\partial}{\partial x}\wedge\frac{\partial}{\partial \xi}=-
\tilde{\eta}^{-1}(dx\wedge d\xi).
$$
We are going now to find the local symmetries of the Hodge $*$-operator, i.e
the vector fields $X$ on $M$ satisfying $[L_X,*]=0$ . First we recall the
relation
$$
L_X(\alpha\wedge *\beta)=(L_X\alpha)\wedge *\beta+
\alpha\wedge [L_X,*]\beta+\alpha\wedge*L_X\beta ,
$$
where $\alpha,\beta$ are arbitrary forms on $M$.
On the other hand we obtain
\begin{align*}
& L_{X}\left(\alpha\wedge *\beta\right)=
-(L_X\eta)(\alpha,\beta)\omega_o - \eta(L_X\alpha,\beta)\omega_o-
\eta(\alpha,L_X\beta)\omega_o-
\eta(\alpha,\beta)L_X\omega_o \\
&=-(L_X\eta)(\alpha,\beta)\omega_o+
\left(L_X \alpha\right)\wedge *\beta
+\alpha\wedge *L_X\beta -\eta(\alpha,\beta){\rm div}X.
\omega_o \\
&=-(L_X\eta)(\alpha,\beta)\omega_o+
L_{X}\left(\alpha\wedge *\beta\right)-
\alpha\wedge\left[L_X,*\right]\beta-
\eta(\alpha,\beta){\rm div}X.\omega_o.
\end{align*}
Since $\alpha$ and $\beta$ are arbitrary $p$-forms from this relation it
follows that $\left[L_X,*\right]=0$ iff
$$
L_X\eta=-{\rm div}X.\eta,\quad p=1,2,3,4.                %26%
$$
From this relation we obtain the following (independent) equations for the
components of any local symmetry $X$ of the Hodge $*$.
\begin{alignat*}{2}
 2\left(\frac{\partial X^1}{\partial x}+\frac{\partial X^2}{\partial y}\right)
& ={\rm div}X, & \qquad
2\left(\frac{\partial X^1}{\partial x}+\frac{\partial X^1}{\partial
\xi}\right)
& ={\rm div}X,  \qquad \\
 2\left(\frac{\partial X^1}{\partial x}+\frac{\partial X^3}{\partial z}\right)
& ={\rm div}X, & \qquad
 2\left(\frac{\partial X^2}{\partial y}+\frac{\partial X^4}{\partial
\xi}\right)
& ={\rm div}X,  \qquad  \\
 2\left(\frac{\partial X^2}{\partial y}+\frac{\partial X^3}{\partial z}\right)
& ={\rm div}X, & \qquad
 2\left(\frac{\partial X^3}{\partial z}+\frac{\partial X^4}{\partial
\xi}\right)
& ={\rm div}X,  \qquad \\
\left(\frac{\partial X^2}{\partial x}+\frac{\partial X^1}{\partial y}\right)
& =0, & \qquad
\left(\frac{\partial X^4}{\partial x}-\frac{\partial X^1}{\partial \xi}\right)
& =0,  \qquad  \\
\left(\frac{\partial X^3}{\partial x}+\frac{\partial X^1}{\partial z}\right)
& =0, & \qquad
\left(\frac{\partial X^4}{\partial y}-\frac{\partial X^2}{\partial \xi}\right)
& =0,  \qquad \\
\left(\frac{\partial X^3}{\partial y}+\frac{\partial X^2}{\partial z}\right)
& =0, & \qquad
\left(\frac{\partial X^4}{\partial z}-\frac{\partial X^3}{\partial \xi}\right)
&=0.
\end{alignat*}

These equations have the following solutions:

1. Translations:
$$
X=\frac{\partial}{\partial x},\quad X=\frac{\partial}{\partial y},\quad
X=\frac{\partial}{\partial z},\quad X=\frac{\partial}{\partial \xi},
$$
as well as any linear combination with {\it constant} coefficients of
these four vector fields;
\vskip 0.4cm
2. Spatial rotations:
$$
X=y\frac{\partial}{\partial x}-x\frac{\partial}{\partial y},\quad
X=z\frac{\partial}{\partial y}-y\frac{\partial}{\partial z},\quad
X=x\frac{\partial}{\partial z}-z\frac{\partial}{\partial x};
$$
\vskip 0.4cm
3. Space-time rotations:
$$
X=x\frac{\partial}{\partial \xi}+\xi\frac{\partial}{\partial x},\quad
X=y\frac{\partial}{\partial \xi}+\xi\frac{\partial}{\partial y},\quad
X=z\frac{\partial}{\partial \xi}+\xi\frac{\partial}{\partial z}
$$
\vskip 0.4cm
4. Dilatations:
$$
X=x\frac{\partial}{\partial x}+y\frac{\partial}{\partial y}+
z\frac{\partial}{\partial z}+\xi\frac{\partial}{\partial \xi},\quad \text{or}
\quad X=x^\mu\frac{\partial}{\partial x^\mu};
$$
\vskip 0.4cm
5. Conformal (with respect to $\eta$) vector fields:
$$
X_\mu=\left(\eta_{\alpha\beta} x^\alpha x^\beta\right)
\frac{\partial}{\partial x^\mu}-2\eta_{\mu\nu}x^\nu\left(x^\sigma
\frac{\partial}{\partial x^\sigma}\right),\quad \mu=1,\dots, 4.
$$

Let's consider the flows generated by the above vector
fields.

1. The translation vector fields generate flows as follows:
$$
x^{\mu'}=x^\mu+a^\mu, \ \ a^\mu \ \text{are 4 constants}.
$$
\vskip 0.4cm
2. The spatial rotations generate "rotational" flows inside the three planes
$(x,y)$, $(x,z)$ and $(y,z)$ as follows:
\begin{alignat*}{3}
x'&=x\,\cos(s)+y\,\sin(s)&,\quad x'&=x\,\cos(s)+z\,\sin(s)&,\quad
y'&=y\,\cos(s)+z\,\sin(s) \\
y'&=-x\,\sin(s)+y\,\cos(s)&,\quad
z'&=-x\,\sin(s)+z\,\cos(s)&\quad z'&=-y\,\sin(s)+z\,\cos(s).
\end{alignat*}

\vskip 0.4cm
3. The space-time rotations generate the following flows:
\begin{alignat*}{3}
x'& =x\,{\rm ch}(s)+\xi\,{\rm sh}(s),&\quad y'& =y\,{\rm ch}(s)+\xi\,{\rm
sh}(s),&\quad z'& =z\,{\rm ch}(s)+\xi\,{\rm sh}(s), \\
\xi'&=x\,{\rm sh}(s)+\xi\,{\rm ch}(s),&\quad\xi'&=y\,{\rm sh}(s)+\xi\,{\rm
ch}(s),&\quad\xi' &=z\,{\rm sh}(s)+\xi\,{\rm ch}(s).
\end{alignat*}
Let's concentrate for a while on the flow in the plane
$(x,\xi)$.  It is obtained by solving the equations
$$
\frac{dx}{ds}=\xi,\quad \frac{d\xi}{ds}=x.
$$
Let $x_{s=0}=x_\circ,\ \ \xi_{s=0}=\xi_\circ$. Then the solution is
$$
x=x_\circ {\rm ch}(s)+\xi_\circ {\rm sh}(s)=
\frac{x_\circ +{\rm th}(s)\xi_\circ}{\sqrt{1-{\rm th}^2(s)}}=
\frac{x_\circ +\beta ct_\circ}{\sqrt{1-\beta^2}},
$$
$$
\xi=x_\circ {\rm sh}(s)+\xi_\circ {\rm ch}(s)=
\frac{x_\circ {\rm th}(s)+ct_\circ}{\sqrt{1-{\rm th}^2(s)}}=
\frac{x_\circ\beta+ct_\circ}{\sqrt{1-\beta^2}},
$$
where $\beta^2={\rm th}^2(s)\le 1$, and $s$ is fixed. The standard physical
interpretation of these relations is that the frame $(x_\circ, \xi_\circ)$
moves with respect to the frame $(x,\xi)$ along the common axis $x\equiv
x_\circ$ with the velocity $v=\beta c$, and since $|\beta|\le 1$ then $|v|\le
c$. It is important to have in mind that this interpretation requires that $c$
has the same value in all such frames.
\vskip 0.4cm
4. The dilatation vector field generates the flow:
$$
x^{\mu'}=ax^\mu, \ \ a=exp(s)=\text{const}.
$$
\vskip 0.2cm
5. The conformal (with respect to $\eta$) vector fields generate
the nonlinear flows
$$
x^{\mu'}=\frac{x^\mu+d^\mu\eta(x,x)}
{1+2\eta(d,x) + \eta(x,x).\eta(d,d)}\ \ ,
$$
where $d$ is a 4-vector and its four components $d^\mu$ are the four
constants-parameters of the special conformal transformations. Note that these
transformations may be considered as coordinate transformations only if the
corresponding denominators are different from zero.

These symmetry considerations show undoubtedly some analogy with the symplectic
mechanics: the canonical $(q,p)$-transformations defined as symmetries of the
symplectic 2-form on $T^*(\mathbb{R}^3)$ determine symmetries of the hamilton
equations; in the same way, the transformations of $\mathbb{R}^4$, defined as
(or generated by) symmetries of the Hodge $*$-operator, are possible symmetry
generators of the equations where it participates.

\section{Classical fields - general notions}
\subsection{Wave fields}
From physical point of view when we talk about {\it waves} \index{wave} we mean
{\it propagation of some disturbance, or perturbation, in a given medium}. It is also
assumed that the perturbation does not alter the characteristic properties of
the medium, and the time-evolution of the perturbation depends on the medium
properties as well as on the specificities of the very perturbation. The waves
are divided to 2 classes: {\it elementary} ({\it linear}) and {\it
intrinsically coordinated} ({\it nonlinear}). The elementary waves are
observed in homogeneous media and are generated by perturbing the
equilibrium state of the medium by means of small quantities of
external energy and momentum. The important properties of linear waves come
from the condition, that during the propagation of the initial disturbance
throughout the medium the structure of the medium does not change
irreversibly, and the various such propagating perturbations do not interact
with each other substantially. From mathematical point of view this means
that the corresponding evolution equations, which are partial differential
equations, describing such phenomena, are linear, so any linear combination
with constant coefficients of solutions gives again a solution. In other
words, the set of solutions of such equations forms a real (finite or
infinite dimensional) vector space. Clearly, any attemt to use these waves as
models of real spatially finite time-stable physical objects having dynamical
structure should be made with great attention.

The intrinsically coordinated, or nonlinear, waves disturb more deeply the
medium structure, but the corresponding changes of the medium structure stay
reversible. When subject to several such perturbations, the medium responses
to the various disturbances are different in general, so the medium
reorganization requires more complicated intrinsic coordination. All this
demonstrates itself in various ways, depending on the medium properties and
the initial perturbation. What we observe from outside is, that some important
properties of the initial perturbations are changed in result of the
interaction. In some cases we observe a time-stable coordination among the
responding reactions of the medium and if the corresponding formation is
finite and time recognizable, we may consider it as a new object. If this
object keeps its energy and momentum we frequently call the corresponding
medium {\it vacuum} (with respect to the object).
 Clearly, such objects can exist only in appropriate media. In
such cases, studying the objects, we get some information about the medium
itself.  From mathematical point of view these waves are described by
nonlinear equations, so that a linear combination of solutions is not, as a
rule, a new solution. The huge variety of various such cases could hardly be
looked at from a single point of view, except when some most general features
are under consideration.

It is important to note, that in both cases, linear and nonlinear, the
perturbations are bearable for the medium in the sense, that they do not
destroy it. We are not going to consider here unbearable perturbations.

One common for every kind of waves characteristic is the {\it
polarization} \index{polarization}. The polarization determines the relation
between the direction of propagation (at some point of the medium) and the
direction of deviation from the equilibrium state of the medium point
considered. If these two directions are parallel we say that the polarization
is {\it longitudinal}, and if these directions are not parallel (e.g.
orthogonal) we say that the polarization is {\it transverse}. In general the
polarization depends on the space-time point, i.e., it is a local
characteristic. When the wave passes through some region of the medium, the
points inside this medium commit some displacements along some (usually closed)
trajectories. If these trajectories are straight lines we say that the
polarization is {\it linear}, if they are circles we say the polarization is
circular, etc. It is important to note that the polarization is an intrinsic
property of the system wave-medium, therefore it is a very important
characteristic for the corresponding theory. In particular, the mathematical
character of the object (scalar, tensor, spinor, differential form, etc.),
describing the wave, depends substantially on it. If the wave is linear, and
the corresponding equation admits solutions with various polarizations, then
summing up solutions with appropriate polarizations we can obtain a solution
with a desired polarization.

Other common characteristics of the waves are the {\it propagation velocity},
determining the energy transfer from point to point of the medium, and
{\it the phase surface} \index{phase surface}, built of all points, being in
the same state with respect to the equilibrium state at a given moment.

\subsection{Solitary waves and solitons}
The concepts of {\it solitary wave} and {\it soliton} \index{solitary
wave, soliton}
 appeared in physics as a
nonlinear elaboration - physical and mathematical - of the general notion for
propagating excitation in a medium.  The following features will be mentioned:
\vskip 0.4cm
I. PHYSICAL.
\vskip 0.4cm
1. The medium is homogeneous, isotropic and has
definite properties of elasticity. \vskip 0.3cm 2. The excitation does not
destroy the medium. \vskip 0.3cm
	3. The excitation is physically finite and flexible:
\vskip 0.3cm
-at every moment it is concentrated in a comparatively small volume of the
medium,

-it carries finite quantities of energy-momentum and
of any other physical quantity too,

-it keeps its spatial stress-strain structure, and may have appropriate
time-periodical dynamical structure, \vskip 0.3cm
	4. The excitation is time-stable, i.e. at lack of external
perturbations its evolution does not lead to a self-ruin.
In particular, the spatial shape of the excitation does not (significantly)
change during its propagation.
\vskip 0.3cm
The above 4 features outline the physical notion of a {\it solitary wave}. A
solitary wave becomes a {\it soliton} if it has in addition the following
property of stability:
\vskip 0.3cm
	5. The excitation survives when collides with another excitation of the
same nature.
\vskip 0.3cm
We make some comments on the features 1-5.

Feature 1 requires homogeneity and some elastic properties of the medium,
which means that it is capable to bear the excitation, and every region of it,
subject to the excitation, i.e. dragged out of its natural (equilibrium) state,
is capable to recover entirely after the excitation leaves that region.

Feature 2 puts limitations on the excitations considered in view of the medium
properties:  they should not destroy the medium.

Feature 3 is very important, since it requires {\it finite} nature of the
excitations, it enables them  to represent some initial level self-organized
physical objects with dynamical structure, so that these objects "feel good"
in this medium. This finite nature assumption admits only such excitations
which may be {\it created} and {\it destroyed} ; no point like and/or
physically infinite excitations are admitted.  The excitation interacts
permanently with the medium and if time periodicity is available it can be
interpreted as a measure of this interaction.

Feature 4 guarantees the very existence of the excitation in this medium, and
the shape keeping during propagation allows its recognition and
identification when observed from outside. This feature 4 carries in some
sense the first Newton's principle from mechanics of particles to
dynamics of continuous finite objects, it implies conservation of
energy-momentum and of other characteristic quantities of the excitation.

The last feature 5 is frequently not taken in view, especially when one
considers single excitations. But in presence of many excitations in a given
region it allows only such kind of interactions between/among the excitations,
which do not destroy them, so that the excitations get out of the
interaction region (almost) the same. This feature is some continuous version of
the elastic collisions of particles.
\vskip 0.4cm
II. MATHEMATICAL
\vskip 0.4cm
1. The excitation defining functions $\Phi^a$ are components of {\it one}
mathematical object (usually a section of a vector/tensor bundle), and for
the soliton case it depends most frequently on one spatial and one time
independent variables.
\vskip 0.3cm 2.
The components $\Phi^a$ satisfy some
system of nonlinear partial differential equations (except the case of (1+1)
linear wave equation), and admit some "running wave" dynamics as a whole.
Compatible with its spatial structure internal dynamics is not excluded as a
rule.
\vskip 0.3cm
3. There are many, even infinite many in some cases, conservation laws.
\vskip 0.3cm
4. The components $\Phi^a$ are usually localized functions with respect to the spatial
coordinate, and the conservative quantities are finite. \vskip 0.3cm
5. The multisoliton solutions, describing elastic interaction (collision), tend
to many single soliton solutions at $t\rightarrow \infty$. \vskip 0.4cm
Comments: \vskip 0.3cm 1. Feature 1 introduces some notion of {\it integrity}:
{\it one excitation - one mathematical object}, although having many
algebraically independent but differentially interrelated (through the
equations) components $\Phi^a$. \vskip 0.3cm 2. Usually, the system of PDE is
of {\it evolution} kind : the initial stress configuration is kept during the
evolution. The "running wave" character of the evolution may result in bringing
Galilei/Lorentz invariance in correspondence to the physical feature 4. The
nonlinearity of the equations is meant to guarantee the spatially localized
(finite) nature of the solutions. \vskip 0.3cm 3. The infinite many
conversation laws frequently lead to complete integrability of the equations.
\vskip 0.3cm
4. The spatially localized $\Phi^a$ represents the finite nature of the excitation.
\vskip 0.3cm 5.
The asymptotic behaviour at $t\rightarrow\infty$ of a
multisoliton solution mathematically represents the elastic character of the
allowed interactions, and so it takes care of the stability of the physical
objects being modelled.

\vskip 0.4cm
The above physical/mathematical features are not always strictly accounted
for in the literature. For example, the word {\it soliton} is frequently used
for a {\it solitary wave} excitation.  Another example, in optics
the soliton behavior is described by the corresponding amplitude of the
solution.

We would like to note the following. As we mentioned above,
one usually makes use of this soliton terminology for spatially {\it localized},
i.e.  going to zero just at spatial infinity, but {\it not} spatially {\it
finite} $\Phi^a$, i.e. when the spatial support of $\Phi^a$ is a compact set.
In fact, all soliton solutions of the well known KdV, SG, NLS equations are
spatially localized and {\it not} spatially finite. This feature may
motivate, from theoretical point of view, some lack of satisfaction since
the creation of any soliton-like excitation would require infinite time in view
of the finite speed of propagation of any physical signal: there is no way to
go to infinity during finite time intervals. So, soliton solutions are rather
approximations than entirely correct models of real physical excitations, and,
apart from their $(1+1)$-dimensionality, they could hardly serve as adequate
enough models of real physical objects. Nevertheless, they may be used in the
frame of accuracy required by the corresponding application. It should be noted
however that the corresponding mathematics, developed during the past half
century, opens new directions and insights in the mathematical comprehension of
the physical world.

Finally, it is curious, that the
{\it linear} (1+1) wave equation admits {\it spatially finite} solitary wave
and even soliton solutions of arbitrary spatial shape.

\subsection{Dynamical equations and conservation laws}
Following our earlier considerations (Sec.4.1) we assume further that
{\it dynamical equations must relate dynamical quantities}. Recall that
the dynamical quantities, considered as characteristics of physical objects,
must depend on their proper/identifying characteristics, on one hand, and on
their kinematical characteristics, which describe their space-time
evolution abilities, on the other hand.
The important moment is that these two kind of
properties {\it must be compatible}, i.e., {\it consistent with each other},
so, the first problem to face when we want to build a
theoretical description of a spatially structured physical object is to point
out the corresponding to its structure stress-strain characteristics, and {\it
the time evolution of the corresponding dynamical quantities must keep the
initial stress-strain structure recognizable}.

Looking back in time we see that the introduced about century and
a half ago energy-momentum quantities appear to be the most reliable and the
most universal ones. Besides their dynamical nature these quantities have the
exclusively important property to be {\it conservative}. The importance of this
property consists in its theoretical power: it allows to write down dynamical
equations according to the principle:
\begin{center}
\hfill\fbox{
    \begin{minipage}{0.97\textwidth}
\begin{center}
\vskip 0.3cm
{\bf If some quantity of energy-momentum
is lost by the physical object $A$, the same quantity of energy-momentum must
be\\
gained by another physical object $B$}.
\end{center}
\vskip 0.3cm
\end{minipage}} \hfill \end{center}

Accepting this principle as universally valid we must find the needed
characteristics of the two objects in terms of which to express the
energy-momentum lost by the object $A$ and gained by the object $B$, and to
write them on the two sides of the equality sign "=".

When we apply this principle to spatially finite objects having dynamical
structure and propagating in the space as a whole we must have available its
corresponding {\it local} version. Every dynamical structure requires at least
two time-recognizable and interacting subsystems, so that the corresponding
local energy-momentum exchange to be appropriately understood and formally
well defined. Our view is based on the understanding that the time stability of
the dynamical structure rests on the time stability of the corresponding
internal exchange process(es). Moreover, these internal local exchange
processes must be strong enough in order to withstand the possible harm-causing
disturbances coming from the outside world.

We are going now to consider and comment some examples of field theories as given
in many textbooks and monographs.
\vskip 0.3cm

{\bf 1. Scalar field $\Phi$.} \index{scalar field}

Let $(M=\mathbb{R}^4,\eta)$ denote the Minkowski space-time manifold with
$sign(\eta)=(-,-,-,+)$, standard coordinates, volume form and corresponding
Hodge $*$-operator as given in Sec.5.4. From formal point of view the usual
approach passes through defining lagrangian $L$ and corresponding action
integral $\mathcal{A}$ (in standard coordinates) according to:

$$ L=\mathbf{d}\Phi\wedge
*\mathbf{d}\Phi=-\eta(\mathbf{d}\Phi,\mathbf{d}\Phi)\omega_o=
-(\mathbf{d}\Phi)_\mu(\mathbf{d}\Phi)^\mu dx\wedge dy\wedge dz\wedge d\xi.
$$
$$
\mathcal{A}=\int_{\mathcal{D}}L
=-\int_{\mathcal{D}}(\mathbf{d}\Phi)_{\mu}(\mathbf{d}\Phi)^\mu\omega_o ,
$$
where $\Phi(x,y,z,\xi)$,
$\mathcal{D}$ is an appropriately defined space-time region. Now the
principle for minimal (or stationary) action  requires $\delta\mathcal{A}=0$,
where $\delta\mathcal{A}$ is computed with respect to the variation of the
field $\Phi$. The commutation between $\mathbf{d}$ and $\delta$ leads to the
equation
$$
\eta^{\mu\nu}\frac{\partial \Phi}{\partial x^\mu\partial x^\nu}=0, \ \ \
\text{i.e.} \ \ \ \mathbf{d}*\mathbf{d}\Phi=0 .
$$

This is the well known D'Alembert wave equation.
We are interested in the following: {\it Does this  equation
admit spatially finite and time-stable solutions, so that such solutions to
serve as models of propagating as a whole and spatially finite real objects?}.
The positive answer to this question would be a serious virtue from the point
of view of its adequacy as model equation for an important class of real
objects, while the negative answer would make us searching for new equations,
having solutions with the desired properties. This problem has been essentially
solved in the 19-th century, and because of its importance we shall give some
explanatory comments.

Before to go to the general solution of the corresponding Cauchy problem
\index{Cauchy problem for wave equation} we
give the following suggesting consideration. Since we aim to describe free
time-stable {\it spatially finite} propagating as whole in the 3-space with the
speed of light $c$ objects, the simplest solution would look like $\Phi(x,y,z\pm
ct)$, where, $\Phi$ has to be spatially finite, and since the object is free, it
may be assumed that the propagation shall follow some straight line direction
in the 3-space, so, it is assumed this direction to be the $z$-coordinate.
Substituting this function in the equation we see that the second derivatives
along $z$ and along $\xi=ct$ cancel each other, so, with respect to $(x,y)$ the
function $\Phi$ must satisfy the equation $\Phi_{xx}+\Phi_{yy}=0$. From
harmonic function theory, however, is known that such finite and smooth functions
do not exist.

Let's consider the general case. We are interested in the Caushy problem, i.e. in
the behaviour of $\Phi$ at $t>0$, if at $t=0$ the function $\Phi$ satisfies the
following initial conditions
$$
 \Phi|_{t=0}=f(x,y,z),\quad
\frac{\partial \Phi}{\partial t}\biggl|_{t=0}=F(x,y,z).
$$
Further we assume that the functions
$f(x,y,z)$ and $F(x,y,z)$ are finite, i.e. they are different from zero in some
finite connected region $D\subset\mathbb{R}^3$, which corresponds to the above
introduced concept of a real object. Besides, we assume also that $f$ is
continuously differentiable up to third order, and $F$ is continuously
differentiable up to the second order.
Under these conditions an unique solution
$\Phi(x,y,z,t)$ of the above wave equation is defined, and it is expressed by
the initial conditions $f$ and $F$ by the following formula (called sometimes
Poisson's formula):
$$ \Phi(x,y,z,t)=\frac{1}{4\pi c}\left\{\frac{\partial}{\partial
t}\Biggl[\int_ {S_{ct}}\frac{f(P)}{r}d\sigma_r
\Biggr]+\int_{S_{ct}}\frac{F(P)}{r} d\sigma_r \right\}, $$ where $P$ is a point
on the sphere $S$ centered at the point $(x,y,z)$ and a radius $r=ct$,
$d\sigma_r$ is the surface element on $S_{r=ct}$.

The above formula shows the following. In order to get the solution
at the point $(x,y,z)$, being at an arbitrary position {\it outside} the
spatially finite and usually topologically trivial region $D$, where the
initial condition, defined by the two functions $f$ and $F$, is concentrated,
it is necessary and sufficient to integrate these initial conditions over a
2-sphere $S$, centered at $(x,y,z)$ and having a radius of  $r=ct$. Clearly,
the solution will be different from zero only if the sphere $S_{r=ct}$ crosses
the region $D$ at the moment $t>0$. Consequently, if $r_1=ct_1$ is the shortest
distance from $(x,y,z)$ to $D$, and $r_2=ct_2$ is the longest distance from
$(x,y,z)$ to $D$, then the solution will be different from zero only inside the
interval $(t_1,t_2)$.

From another point of view this means the following. The initially
concentrated perturbation in the region $D$ begins to expand radially , it
comes to the point $(x,y,z)$ at the moment $t>0$, makes it "vibrate" (
i.e. our devices show the availability of a field) during the time interval
$\Delta t=t_2-t_1$, after this the point goes back to its initial condition
and our devices find no more the field. Through every point out of $D$ there
will pass a wave, and its forefront reaches the point $(x,y,z)$ at the moment
$t_1$ while its backfront leaves the same point at the moment $t_2$.
Figuratively speaking, the initial condition "blows up" radially and goes to
infinity with the velocity of light.

This mathematical result shows that {\it every}  spatially finite initial
condition for this equation generates time-unstable solution, so this equation
{\it has no} smooth enough time-dependent solutions, which could be used as
models of time stable real objects. As for the static solutions, as it was
mentioned earlier, such solutions also can not describe real time-stable finite
physical objects having dynamical structure.

Although the above action integral does not give direct physically sensible
solutions for the scalar field $\Phi$, its optimal nature should not be
neglected at all. In fact, it is suggestive in the following sense.

First, it shows how in an optimal way we can come to a null 1-form
$\mathbf{d}\Phi, \eta(\mathbf{d}\Phi,\mathbf{d}\Phi)=0$, or null vector field
$\tilde{\eta}(\mathbf{d}\Phi)$, on $M$ in a quite general way, and to associate
with such a field corresponding dynamical quantities.

Second, it directly leads to a 1-dimensional completely integrable Pfaff system
defined by $\mathbf{d}\Phi: \mathbf{d}(\mathbf{d}\Phi)\wedge\mathbf{d}\Phi=0$.

Third, it suggests how to construct a completely integrable 3-dimensional
distribution $(X_1,X_2,X_3)$ on $M$, where
$\langle\mathbf{d}\Phi,X_i\rangle=0$, and shows the way how to choose one of
the three vector fields representing this distribution, namely, we can assume
$X_3=\tilde{\eta}^{-1}(\mathbf{d}\Phi)$, since
$\langle\mathbf{d}\Phi,\tilde{\eta}^{-1}(\mathbf{d}\Phi)\rangle=0$.

Finally, choosing the couple $\{X_1,X_2\}$ to define space-like completely
integrable distribution orthogonal to the spatial projection of the null vector
$\tilde{\eta}^{-1}(\mathbf{d}\Phi)$ we can easily come to a 3-dimensional
distribution of special kind:

-$\{X_1,X_2\}$ to be integrable,

-$\{X_1,\tilde{\eta}^{-1}(\mathbf{d}\Phi\}$ and $\{X_2,\tilde{\eta}^{-1}(\mathbf{d}\Phi\}$
to be NONintegrable.

\noindent
Hence, if we put in $\{X_1,X_2\}$ the essential physical information and
require $\tilde{\eta}^{-1}(\mathbf{d}\Phi)$ to be a local symmetry of $\{X_1,X_2\}$,
then the 3-dimensional completely integrable distribution
$\{X_1,X_2,\tilde{\eta}^{-1}(\mathbf{d}\Phi)\}$ may represent a propagating along
$\tilde{\eta}^{-1}(\mathbf{d}\Phi)$ physical object consisting of two recognizable
subsystems mathematically represented by the subdistributions
$\{X_1,\tilde{\eta}^{-1}(\mathbf{d}\Phi)\}$ and
$\{X_2,\tilde{\eta}^{-1}(\mathbf{d}\Phi)\}$, and dynamical structure, represented by
the corresponding two curvature forms measuring the nonintegrability of
$\{X_1,\tilde{\eta}^{-1}(\mathbf{d}\Phi)\}$ and
$\{X_2,\tilde{\eta}^{-1}(\mathbf{d}\Phi)\}$. If such a suggestion seems
realistic in some cases, it should not be underestimated from theoretical point
of view.

\vskip 0.3cm
{\bf 2. Vector bundle valued differential $p-$forms}
\vskip 0.3cm
Here we are going to consider two vector bundles: $\mathbf{V},dim\mathbf{V}=r$,
and $\mathbf{W}$, on a  manifold $M, dimM=n$. The field will be represented by a
$\mathbf{V}$-valued differential $p-$form $\Phi$ on $M$, so we can say
that $\Phi$ has $r$ recognizable subobjects, which are meant to represent the
recognizable interacting subsystems of our physical object. Note that every
such subobject is a $p-$form, so, it can be locally interpreted as a volume
form through which something, e.g. a $q-$vector denoted further by $Z$, may/will
flow.

Locally, $Z=Z^i\otimes e_i$ is represented by
$$
Z=Z^{\sigma_1\sigma_2\dots\sigma_q;i}(x^1,...,x^n) \frac{\partial}{\partial
x^{\sigma_1}}\wedge\frac{\partial}{\partial
x^{\sigma_2}}\wedge\dots\wedge\frac{\partial}{\partial x^{\sigma_q}}
\otimes e_i(x^1,...,x^n)
, \ \ \sigma_1<...<\sigma_q ,
$$ and $\Phi=\alpha^i\otimes e_i$ is represented by
$$
\Phi=\Phi_{\mu_1\mu_2...\mu_p}^{i}(x^1,...,x^n)
dx^{\mu_1}\wedge...\wedge dx^{\mu_p}\otimes
e_i(x^1,...,x^n) , \ \mu_1<...<\mu_p \ ,
$$
where $i=1,2,...,r$,
$\{e_1,...,e_r\}$ is a local basis in $Sec(\mathbf{V})$. A linear
connection $\nabla$ in $\mathbf{V}$ is given with exterior covariant derivative
$\mathbf{D}$ in the $\Lambda(M)$-module of $\mathbf{V}$-valued differential
forms denoted further by $\Lambda(M,\mathbf{V})$. Additionally, a bilinear
map $\varphi_x: \mathbf{V}_x\times\mathbf{V}_x\rightarrow\mathbf{W}_x$ for
every $x\in M$ is given, determining the algebraic couplings between the
subobjects of $\Phi$. From physical point of view we could say that $\varphi$
determines which couples of subsystems of a larger complicated physical system
interact.

Recalling now the concept of {\it attractiveness/sensitivity} between vector
valued vectors and forms (Sec.1.4.2), we are ready  to see how $\Phi$
changes along a $\mathbf{V}$-valued $q$-vector $Z$ on $M$, with $q\leq p$,
computing the covariant $\varphi$-extended Lie derivative
$\mathcal{L}^{(\nabla,\nabla',\varphi)}_Z(\Phi)$ of $\Phi$ with respect to $Z$
as given at the end of Sec.3.7.1.

 Let $\{r_m,m=1,2,...,s\}$
be a local basis of Sec.$\mathbf{W}$, $\Gamma$ and $\Gamma'$ be
 linear connections in $\mathbf{V}$ and $\mathbf{W}$ respectively, with
corresponding exterior covariant derivatives $\mathbf{D}$ and $\mathbf{D'}$.
The indices $k,j$ will take values $\{1,2,...,dim\mathbf{V}\}$, the indices
$m,l$ will take values $\{1,2,...,dim\mathbf{W}\}$, and the greek indices
$\mu,\nu$ will take values $\{1,2,...,dim(M)\}$. Denoting
$\varphi(e_k,e_i)=A_{ki}^{m}r_m$ we obtain
\begin{eqnarray*}
\mathcal{L}^{\varphi}_{Z}\Phi&=&\mathbf{D'}i_Z^{\varphi}\Phi-
(-1)^{deg(Z)}i_Z^{\varphi}\mathbf{D}\Phi,\\
i_Z^{\varphi}\mathbf{D}\Phi&=&A_{kj}^{m}\left[i_{Z^k}\mathbf{d}\alpha^j+
(-1)^pi_{Z^k}(\alpha^i\wedge\Gamma_{i\mu}^jdx^\mu)\right]\otimes r_m,\\
\mathbf{D'}i_Z^{\varphi}\Phi&=&\left[\mathbf{d}(A_{ki}^{m}i_{Z^k}\alpha^i)+
(-1)^{(p-q)}A_{ki}^{l}(i_{Z^k}\alpha^i\wedge
\Gamma^{'m}_{l\mu}dx^\mu)\right]\otimes r_m .
\end{eqnarray*}

We consider now the case when $Z$ and $\Phi$ are valued in the same
vector space $\mathbb{V}$ with basis $\{e_i, i=1,2,...,r\}$. So,
$\varphi$ is bilinear in $\mathbb{V}$:
$$
\varphi=\varphi_{ij}^ke_k\otimes \varepsilon^i\otimes \varepsilon^j, \ \ \
\langle\varepsilon^i,e_j\rangle=\delta^i_{j},
$$
 and the covariant exterior
derivatives reduce to the usual exterior derivative: $(\mathbf{D,D'})\rightarrow
\mathbf{d}$ on $M$, and the $\varphi$-extended Lie derivative becomes
$$
\mathcal{L}^{\varphi}_{Z}\Phi=\mathbf{d}\,i_Z^{\varphi}\Phi-
(-1)^{deg(Z)}i_Z^{\varphi}\,\mathbf{d}\Phi.
$$
Note that if the $\varphi$-modulated flow of $Z$ across $\Phi$ is a closed form:
$\mathbf{d}i_Z^{\varphi}\Phi=0$, then $\mathcal{L}^{\varphi}_{Z}\Phi$ reduces
to the $\varphi$-modulated flow of $Z$ across $\mathbf{d}\Phi$.

Clearly, if $\Phi$ is $Z$-attractive, then
at least one of the corresponding coefficients $\varphi_{ij}^k$ should
be different from zero.

Specializing the bilinear map $\varphi$ as tensor product, symmetrized tensor
product and antisymmetrized tensor product, we obtain respectively:
$$
\mathcal{L}_Z^{\otimes}\Phi=\mathbf{d}\,i_Z^{\otimes}\Phi-
(-1)^{degZ}i_Z^{\otimes}\,\mathbf{d}\Phi,
$$
$$
\mathcal{L}_Z^{\vee}\Phi=\mathbf{d}\,i_Z^{\vee}\Phi-
(-1)^{degZ}i_Z^{\vee}\,\mathbf{d}\Phi,
$$
$$
\mathcal{L}_Z^{\wedge}\Phi=\mathbf{d}\,i_Z^{\wedge}\Phi-
(-1)^{degZ}i_Z^{\wedge}\,\mathbf{d}\Phi.
$$

We shall give explicitly the cases of symmetrized and antisymmetrized tensor
product, $dim\mathbb{V}=2$, $Z=Z_1\otimes e_1+Z_2\otimes e_2$ will
be a $\mathbb{V}$-valued 2-vector, so $degZ=2$, and
$\Phi=\alpha^1\otimes e_1+\alpha^2\otimes e_2$ will be a $\mathbb{V}$-valued
2-form.
\begin{eqnarray*}
\mathcal{L}_Z^{\vee}\Phi&=&
\left[\mathbf{d}\langle\alpha^1,Z^1\rangle-
i_{Z^{1}}\mathbf{d}\alpha^1\right]\otimes e_1\vee e_1+
\left[\mathbf{d}\langle\alpha^2,Z^2\rangle-
i_{Z^{2}}\mathbf{d}\alpha^2\rangle\right]\otimes e_2\vee e_2\\
&+&\left[\mathbf{d}\langle\alpha^2,Z^1\rangle-i_{Z^{1}}\mathbf{d}\alpha^2+
\mathbf{d}\langle\alpha^1,Z^2\rangle-
i_{Z^{2}}\mathbf{d}\alpha^1\right]\otimes e_1\vee e_2,\\
\mathcal{L}_Z^{\wedge}\Phi&=&\left[\mathbf{d}\left(\langle \alpha^2,Z_1\rangle-
\langle \alpha^1,Z_2\rangle\right)-\left(i_{Z_1}\mathbf{d}\alpha^2-
i_{Z_2}\mathbf{d}\alpha^1\right)\right]\otimes e_1\wedge e_2.
\end{eqnarray*}

We specialize now the above symmetrized expression for the case of Minkowski
space-time, $\mathbb{V}$ will be real 2-dimensional, also, let's denote
the 2-form $\alpha^1$ as $F$ and choose $\alpha^2=*F$, finally let
$Z^1=\tilde{\eta}^{-1}(F)$ and $Z^2=\tilde{\eta}^{-1}(*F)$. So,
$$
\Phi=F\otimes e_1+*F\otimes e_2, \ \ \
Z=\tilde{\eta}^{-1}(F)\otimes e_1+\tilde{\eta}^{-1}(*F)\otimes e_2.
$$
The local $"\vee"$-change of this $\Phi$ along the so defined $Z$ looks
as follows (everywhere $\mu<\nu<\sigma$): \begin{eqnarray*}
\mathcal{L}_{Z}^{\vee}\Phi&=&
\left[\mathbf{d}(F^{\mu\nu}F_{\mu\nu})-F^{\mu\nu}(\mathbf{d}F)_{\mu\nu\sigma}dx^\sigma
\right]\otimes e_1\vee e_1\\
&+&\left\{\mathbf{d}[(*F)^{\mu\nu}(*F)_{\mu\nu}]-(*F)^{\mu\nu}(\mathbf{d}*F)_{\mu\nu\sigma}dx^\sigma
\right\}\otimes e_2\vee e_2\\
&+&\{2\mathbf{d}[F^{\mu\nu}(*F)_{\mu\nu}]-F^{\mu\nu}\mathbf{d}(*F)_{\mu\nu\sigma}dx^\sigma-
(*F)^{\mu\nu}\mathbf{d}F_{\mu\nu\sigma}dx^\sigma
\}\otimes e_1\vee e_2.
\end{eqnarray*}
If $\mathcal{L}_{Z}^{\vee}\Phi=0$, i.e., if $Z$ is $(\vee,\eta)$-symmetry of
$\Phi$, this mutual interaction between $F$ and $*F$ is rather "partnership".
In fact, then we have the equations
\begin{eqnarray*}
&&\mathbf{d}(F^{\mu\nu}F_{\mu\nu})-
F^{\mu\nu}(\mathbf{d}F)_{\mu\nu\sigma}dx^\sigma=0,\\
&&\mathbf{d}[(*F)^{\mu\nu}(*F)_{\mu\nu}]-
(*F)^{\mu\nu}(\mathbf{d}*F)_{\mu\nu\sigma}dx^\sigma=0\\
&&2\mathbf{d}[F^{\mu\nu}(*F)_{\mu\nu}]-F^{\mu\nu}\mathbf{d}(*F)_{\mu\nu\sigma}dx^\sigma-
(*F)^{\mu\nu}\mathbf{d}F_{\mu\nu\sigma}dx^\sigma=0.
\end{eqnarray*}
Since in the pseudoeuclidean case
$F^{\mu\nu}F_{\mu\nu}=-(*F)^{\mu\nu}(*F)_{\mu\nu}$, from the first two
equations follows
$$
F^{\mu\nu}(\mathbf{d}F)_{\mu\nu\sigma}+
(*F)^{\mu\nu}(\mathbf{d}*F)_{\mu\nu\sigma}=0,
$$
which, under electromagnetic interpretation of $F$ would mean, that the
divergence of the standard electromagnetic energy-momentum tensor
$$
Q^\nu_\mu=-\frac12[F_{\mu\sigma}F^{\nu\sigma}+
(*F)_{\mu\sigma}(*F)^{\nu\sigma}]
$$
is equal to zero.

In the null-field case, where $F^{\mu\nu}F_{\mu\nu}=0, \
F^{\mu\nu}(*F)_{\mu\nu}=0$, the above equations become equivalent to
$$
L_{\bar{F}}F=0, \ \ L_{\bar{*F}}*F=0, \ \ L_{\bar{F}}*F+L_{\bar{*F}}F=0 .
$$
where $\bar{F}, \bar{*F}$ denote the metric-corresponding bi-vectors, i.e., the
above $(Z^1,Z^2)$. Clearly, the third equation $L_{\bar{F}}*F+L_{\bar{*F}}F=0$
represents the idea for "partnership" : what $F$ loses is gained by $*F$, and
vice versa : $L_{\bar{*F}}F=-L_{\bar{F}}*F$.

Formally, since $deg({\mathbf{d}})=1$, we could write generalized equations in
the form
$$
\mathcal{L}^{\varphi}_{Z}\Phi-\mathbf{d}\,i_{Z}^{\varphi}\Phi=
(-1)^{deg\,Z}i^{\varphi}_{Z}\mathbf{d}\Phi=0.
$$

\section{A note on Finiteness and Infinities}

Let's go back to the variational formulation of the description of a physical
object/system through the action integral and lagrangian 4-form. Usually, the
domain of integration looks like $\mathbb{R}^3\times (t_2-t_1), t_2>t_1$. In
order this definite integral to be well defined it is necessary the lagrangian
4-form to have compact support, or to go rapidly enough to zero at spatial
infinity. Since the lagrangian function is built in terms of the corresponding
field functions and their derivatives we must be careful enough not to admit
omissions in this respect: the lagrangian 4-form must not have singularities,
on one hand, and must give non-zero value to the action integral, on the other
hand. Unfortunately, this mathematically important detail is not always fully
respected: for example, the widely used classical action integral for
individual plane electromagnetic waves is zero since the lagrangian
function $(\mathbf{B}^2- \mathbf{E}^2)$, been given relativistically by
$\frac12F^{\mu\nu}F_{\mu\nu}$, is equal to zero. Hence, the local and the
integral action of these solutions is zero, and, somehow, this does not seem to
generate any concern in the theoretical physics community.

Another example of the kind comes from variational formulation of classical
mechanics. The lagrangian for a particle moving in an external field described
by the potential function $U=U(x,y,z)$ is given by the expression $L=T-U$,
where $T$ is the kinetic energy of the particle. The corresponding equation of
motion requires change of $T$, so according to the energy conservation law
this change of $T$ must lead to corresponding change in $U$, but $U$ is static
and does not admit changes with time. Nevertheless, the variational formulation
leads to conservation of $(T+U): \frac{d}{dt}(T+U)=0$, so, since $U$ does not
change with time where the particle takes energy from? The fact that $U$ may
change from point to point along the trajectory of the particle does not save
the situation. This sets very seriously the question: what essentially is the
real/true sense of $U$? Is it an integral energy of the external field, or
something else. For example, in the electrostatic case is it
possible to treat $U$ as local/integral energy of the field created by the
central charge? Moreover, where is the field of the charged particle
moving in external field, this question is strongly motivated by the
undisputable fact that around every charge there is always region where
the field strength  of the other charge is much weaker?

Assume that we want to compute
the integral energy of a continuous free physical object, and this is motivated
by the energy conservation law: since our object is free its integral energy
must be an important integral characteristic whatever happens inside among the
object's subsystems. According to Sec.4.4 we have to build some 3-form of the
kind $wdx\wedge dy\wedge dz$, where $w$ represents the energy density of the
object, and then to work out the integration over the 3-volume occupied by the
object at some fixed moment.

Clearly, the very formulation of this task presumes its
mathematical correctness: since this is definite integral the energy density $w$
{\it must be integrable function}, i.e. the integration to produce a {\it
finite} number. Otherwise, in case of infinite value of this integral, as it,
for example, is in the case of plane wave solutions of the Maxwell free
equations, why should we trust the equations giving such "infinite energy"
solutions. Clearly, such "infinite energy" plane waves can not be created.
Recalling now that these "infinite energy" electromagnetic plane wave solutions
generate, according to the action integral, zero local and integral action, on
one hand, and infinite integral action according to the formula: \begin{center}
 "integral action"="integral energy"$\times$"time
interval"=$"\infty"\times$"time interval", \end{center} on the other hand,
additionally complicates the situation.

We shall mention other two strange, from our point of view, moments coming from
gauge theory.

Every gauge field in theoretical physics is, in fact, a connection 1-form on a
principal bundle been projected on the base space of the bundle by means of a
local section of the bundle. As it is assumed in these theories, the
stress-energy-momentum tensor $T_{\mu\nu}$ of the field is given as a bilinear
tensor function of the corresponding curvature:
$T_{\mu\nu}(R_{\alpha\beta}^i)$. This should imply, in accordance with the rest
of theoretical physics, that the gauge field $A_{\mu}^i$ can exert influence on
the dynamical behavior of other physical objects only through the curvature
components. Nevertheless, possible direct influence by means of $A_{\mu}^i$ is
widely accepted (covariant derivatives, Bohm-Aharonov effect, etc.). May be
this is just kinematical and not dynamical exert. In fact, the existence of
normal frames (i.e. frames with respect to which the connection components are
zero at a point, or along a not-selfintersecting curve) in the module of
sections in every vector bundle no matter if there is an action of the gauge
group in the standard fiber of the bundle, seriously suggests such an
interpretation.

The other moment is connected with the so called self-dual fields $\Phi=*\Phi$,
and corresponding instanton solutions. Two characteristic properties of these
solutions are:

{\bf 1.} they cannot exist on pseudoeuclidean $4$-dimensional manifolds,

{\bf 2.} they have necessarily zero stress-energy-momentum tensor.
\vskip 0.2cm
So, although their interesting mathematical properties, it is still not clear
if instanton-like physical objects can be physically detected and studied, or
to be just of methaphysical interest, at least for now.

Turning now to General Relativity we see that the situation here is similar and
in some sense more serious. Almost a century past from the appearance of this
theory and we {\it still have no theoretically good enough expression for the
energy density of the free gravitational fields} although the theory
generically needs the energy-momentum tensor of the other (nongravitational)
fields, and requires this energy-momentum tensor to be locally conservative,
i.e. to have zero divergence, although this does not lead to integral
conservation when there are no isometries. Moreover, every free gravitational
field in this theory requires zero Ricci scalar curvature
$\mathbf{R}=g^{\mu\nu}\mathbf{R}_{\mu\nu}=0$, and, consequently, zero
local action and zero integral action $$
\mathcal{A}_{grav}=\int{\mathbf{R}\sqrt{|det(g_{\mu\nu})|}dx\wedge dy\wedge
dz\wedge d\xi}=0.
$$
Another strange looking moment is that
the trajectories of a particle, in fact of a planet, do NOT depend at all on
the planet's real characteristics as a physical object and on the possible
interaction of the planet's own gravitational field with the external one, for
example, the Sun-Earth system. All this makes it difficult to understand why,
on one hand, the predicted trajectories are so true, and on the other hand,
where the electric energy-momentum produced in a water-power electric station
comes from.

We'd like to note specially, that what quantity we shall talk about,
"energy" or some else, is not so essential. The important moment is that we
{\it must have corresponding quantities} in terms of which to describe the
interaction between physical objects and to understand the stability of a
physical object built of time-recognizable interacting subsystems. The point is
that this quantity "energy" has proved its flexibility and ability to
appropriate generalization when needed, that's why it is traditionally used as
preferable kinematical and interaction measure. Moreover, its conservative
nature allows to write down corresponding balance relations. In terms of such
balance relations we can talk about {\it dynamical equilibrium}, which is a
basic concept in trying to understand how physical objects/systems succeed to
keep their identity under the permanent internal interaction among their
subsystems, on one hand, and under the permanent attacks from the external
world, on the other hand. In view of this, it seems not reasonable to leave off
such kind of concepts as theoretical tools, our view is that we must learn how
to appropriately reformulate them in every new physical situation in order to
extend appropriately the existing harmony among concepts used in the seriously
tested theoretical constructions.

\chapter{Classical Vacuum Electrodynamics}
\section{Basic Concepts}
\subsection{Analysis of the Coulomb Law} \index{Coulomb law analysis}
This section presents an attempt to reconsider the long stayed problem of
locally performed momentum exchange between a vacuum spherically symmetric
solution of Maxwell equations, which carries NO momentum and necessarily
conserves its energy, and another charged mass-particle
which changes its mechanical momentum, so: {\it where this mechanical momentum
change goes to, or comes from, if the whole system is assumed to be isolated?}
The main steps we follow are:

	{\bf 1.} To pay {\bf equal respect to both} presenting fields connected
to the two charges.

	{\bf 2.} Then to introduce a reasonable notion about
{\bf local interaction} between the two
fields at a given moment $t_o$ in a way allowing to compute the corresponding {\bf
integral interaction energy} $U(t_o)$ as a function of corresponding
configurational parameters: masses, charges, distances.

	{\bf 3.} Then, following the rule that {\bf any isolated system tends
to configurations with less values} of $U$, the Coulomb force picture to appear
as a quasistatic {\bf decreasing} of $U: U(t_o)\rightarrow U(t)<U(t_o), t>t_o$
only through appropriate changes of some of the configurational parameter(s).

	{\bf 4.} Corresponding changes of the kinetic mechanical energies of the
two particles to be considered as responsible for carrying away the decrease of
$U$. \vskip 0.3cm {\bf 1. The Problem}

Usually, theoreticians introduce the Coulomb force field starting with some
field interpretation of the Coulomb force law: $f=\frac{qQ}{r^2}$, where $q$
and $Q$ are the charges of two small bodies (usually considered as
point-particles) and $r$ is the euclidean distance between them. Two fields
$\mathbf{E}_Q$ and $\mathbf{E}_q$, considered as generated correspondingly by
each of the two charges $Q$ and $q$, in corresponding spherical coordinates
$(r,\theta,\varphi)$ and $(\bar{r},\bar{\theta},\bar{\varphi})$ connected with
each charge $Q$ and $q$, are defined by the relations

$$
\mathbf{E}_Q=\frac{f}{q}\frac{\partial}{\partial r}
= \frac{Q}{r^2}\frac{\partial}{\partial r}
 \ \
\ \text{and} \ \ \ \mathbf{E}_q=\frac{f}{Q}\frac{\partial}{\partial\bar{r}}=
\frac{q}{\bar{r}^2}\frac{\partial}{\partial\bar{r}}\cdot
$$

We mention now the following:
\begin{itemize}
\item	 $\mathbf{E}_Q$ and $\mathbf{E}_q$ are considered as vector fields
(or identified by the euclidean metric 1-forms),
i.e. local objects, defined outside the regions ocupied by two mass source
objects of charge magnitudes $Q$ and $q$. \item
 These fields are interpreted as force-fields in the sense of classical
mechanics,
acting directly on other mass particles carrying unit charges, hence, the force
acting on the $(m_q)$-particle is $q\mathbf{E}_Q$ and the force acting on the
$(m_Q)$-particle is $Q\mathbf{E}_q$.

\item
The mechanical behaviour of the $m_q-$particle in the
reference frame connected with the $Q$-particle is defined by the Newton law
$\dot{\mathbf{p}_{q}}=q\,\mathbf{E}_{Q}$, where $\mathbf{p}_{q}$ is the
mechanical momentum of the $m_q$-particle in this reference frame (clearly,
$\mathbf{p}_{Q}=0$ in this frame). In the same way, the $m_Q$-particle with
respect to the $m_q$-particle frame, satisfies the corresponding dynamical
equation: $\dot{\mathbf{p}_{Q}}=Q\,\mathbf{E}_{q}$.
\item Close to the
$q/Q$-particle the field generated by the charge $q/Q$ is much stronger than
the one generated by the $Q/q$-particle, so neglecting the proper field of any
of the two charges seems not correct. \end{itemize} Looking closer to the
sitution we note that each of the dynamical equations $$
\dot{\mathbf{p}_{q}}=q\,\mathbf{E}_{Q}, \ \ \
 \dot{\mathbf{p}_{Q}}=Q\,\mathbf{E}_{q}
$$
presupposes that the change of the mechanical momentum of
the $q/Q$-particle comes from (or goes to) the corresponding change of the
momentum/energy carried by the field $\mathbf{E}_{Q}/\mathbf{E}_q$ in
accordance with the (presupposed by the local nature of the above differential
equations) local character of momentum/energy transfer and with the universal
momentum/energy conservation law. However, such a justification assumes that
each of the fields $\mathbf{E}_{Q}/\mathbf{E}_q$ carries {\it non-zero}
momentum. How much is this field momentum? This question requires corresponding
definition of the field momentum, which directs our attention to the theory
based on Maxwell vacuum equations. Maxwell theory, however, gives at least
three objections to this understanding of the physical situation:
 \vskip0.2cm
$\mathbf{1^o}$.
Each of the fields $\mathbf{E}_{Q}/\mathbf{E_q}$, considered as local object,
i.e. vector field, is STATIC, it satisfies Maxwell vacuum equations outside
its source:\linebreak $\mathrm{rot}\,\mathbf{E}=0,\ \mathrm{div}\,\mathbf{E}=0$,
and according to the theory, every such solution field conserves its energy,
and carries NO momentum with respect to the proper frame of its generator;

$\mathbf{2^o}$. The static nature of the field forbids any
time-changes of any field characteristic;

$\mathbf{3^o}$. The field momentum density in Maxwell theory is proportional to
the Poynting vector, so, neither the electric field $\mathbf{E}$ nor the
magnetic field $\mathbf{B}$ which is missing in the static case, are allowed to
carry momentum separately.
\vskip 0.2cm
In general, every vacuum solution of Maxwell equations conserves its energy,
momentum and angular momentum, so, {\bf NO vacuum solution
$(\mathbf{E},\mathbf{B})$ should be allowed to participate directly as force
generating agent in the expression
$q\mathbf{E}+\frac{q}{c}\mathbf{v}\times\mathbf{B}$}.

We see that from theoretical point of view
for vacuum fields the usual setting "charged particle in external vacuum
field" does not work: any such field conserves its energy, momentum and
angular momentum, therefore, the "test particle" does NOT have any chance
to gain directly from the external field energy-momentum in a local way as
supposed by the above differential equations.

In view of the above, how to understand the
experimentally observed Coulomb force law from theoretical point of view in the
frame of Maxwell theory?

%{\bf 2. Possible solution.}

In an attempt to answer the above stated question we make the following
considerations. First, some clarifications concerning the structure and
admissible changes of the physical situation. We have two mass particles
carring electric charges $q$ and $Q$. The two masses "generate" two
gravitattional fields which are further neglected as physical field factors.
The two charges "generate" two electric fields: $\omega_q$ denoted further just
by $\omega$, and $\Omega_Q$ denoted further just by $\Omega$. The whole system
is isolated and time stable, so, the two fields and the two particles
considered as inertia carrying mechanical objects, exist consistently with each
other.

\vskip 0.3cm \noindent {\bf Remark}.\hskip 0.1cm We put the
term "generate" in commas no ocasionally, but intentionally, because in this
case the charge-field configuration we consider as the real one, i.e., no
charged particle can exist without such a field, and no such a field can exist
without charged particle, so, both the
charge and the field aspects of the situation should be paid equal respect.
\vskip 0.3cm \noindent
Since we consider electrostatic situation, no magnetic fields are
assumed to be present. The admissible changes, by assumption, do NOT lead to
destruction of any of the objects. Paying now due rspect to the Gauss theorem
we have to assume that each of the two fields is NOT defind inside the small
regions ocupied by the two sources. Therefore, at the moment $t_o$, the two
fields $\omega$ and $\Omega$ are defined on the topologically non-trivial space
$\Sigma=\mathbb{R}^3\diagdown(W_q\cup W_Q)$, where $W_q$ and $W_Q$ are the two
small nonintersecting regions, treated further as two $R$-distant balls with
boundaries $S^2_q$ and $S^2_Q$, occupied by the two particles. The first
question to answer is: how to specify the mathematical nature of $\omega$ and
$\Omega$ at this moment ?
\vskip 0.3cm
{\bf 2. The model structure}
\vskip 0.2cm
The nontrivial topology of $\Sigma$, which {\bf must be kept unchanged} at
$t>t_o$, suggests to choose as local representatives of the two physical
fields, namely, the de Rham representatives $\Theta_Q$ and $\Theta_q$ of the
two cohomological classes. In view of the expected spherical symmetry of the
fields far enough from the charges we may assume
$\Theta_Q=\Omega+\mathbf{d}\alpha_1$, $\Theta_q=\omega+\mathbf{d}\alpha_2$, and
spherical symmetry of $\Omega$ and $\omega$ with
$\mathbf{d}\Omega=\mathbf{d}\omega=0$.

In carrying out this idea we introduce two spherical
coordinate systems $(r,\theta,\varphi)$ and
$(\bar{r},\bar{\theta},\bar{\varphi})$,  originating at the centers of
$W_Q$ and $W_q$ respectively, so, recalling the end of Sec.2.10,
the spherical symmetry {\it away from} the two
centers leads to assume $\mathbf{d}\alpha_1= \mathbf{d}\alpha_2=0$, therefore,
in regions away from the two centers we may assume
$$
\Theta_Q=\Omega(r,\theta)=h(r)\sin\theta\mathbf{d}\theta\wedge\mathbf{d}\varphi,
 \ \ \ \Theta_q=\omega(\bar{r},\bar{\theta})=
\bar{h}(\bar{r})\sin\bar{\theta}\mathbf{d}\bar{\theta}\wedge\mathbf{d}\bar{\varphi}.
$$
Being representatives of corresponding cohomology classes, $\Omega$ and $\omega$
must satisfy $\mathbf{d}\Omega=0$ and $\mathbf{d}\omega=0$, so,
$h(r)=Const$ and $\bar{h}(\bar{r})=const$. We denote $Const=Q$ and $const=q$.
Now, the euclidean Hodge star operator $*$ and the euclidean identification
$\tilde{g}$ of vectors and covectors give
$$
\mathbf{E}_{Q}=\tilde{g}^{-1}(*\Omega(r,\theta))=\tilde{g}^{-1}
\left(\frac{Q}{r^2}dr\right)
 =\frac{Q}{r^2}\frac{\partial}{\partial r}, \ \ \
$$
$$
\mathbf{E}_{q}=\tilde{g}^{-1}(*\omega(\bar{r},\bar{\theta}))=
\tilde{g}^{-1}\left(\frac{q}{\bar{r}^2}d\bar{r}\right)
 =\frac{q}{\bar{r}^2}\frac{\partial}{\partial\bar{r}}\cdot
$$

Going further we note that at every point of $\Sigma$ the real stress is
built of two physical fields of the {\it same physical nature},
therefore, the resulted stress should depend on the
local mutual influence/interaction between the two stress generating fields.
The point is how to model mathematically this local interaction of the two
fields? At this moment the Maxwell stress tensor
$$
M^{j}_{i}=\mathbf{E}_i\mathbf{E}^j-\frac12\mathbf{E}^2\delta_i^j
=\frac14(*\mathbf{E})_{mn}\,(*\mathbf{E})^{mn}\delta_i^j-
(*\mathbf{E})_{im}\,(*\mathbf{E})^{jm}
$$
helps us as follows.

Identifying the vector fields and 1-forms on $\Sigma$ by means of $\tilde{g}$
and omitting $\tilde{g}$, for the two Maxwell stress tensors we have
$$
M_q\equiv M(\mathbf{E_{q}})=\mathbf{E}_{q}\otimes
\mathbf{E}_{q}-\frac12\,\mathbf{E}^2_{q}\,id_{T\Sigma},
$$
$$
M_Q\equiv M(\mathbf{E}_{Q})=\mathbf{E}_{Q}\otimes
\mathbf{E}_{Q}-\frac12\,\mathbf{E}^2_{Q}\,id_{T\Sigma}.
$$

Mathematically, each of these two tensors can be considered as a quadratic
map from the vector fields on $\Sigma$ to $(1,1)$-tensors, i.e. to the linear
maps in the linear space of vector fields. Each of our two fields generates
such $(1,1)$-tensor field:  $M(\omega)$ and $M(\Omega)$. Let's assume that the
real field $\mathfrak{F}$ on $\Sigma$ that generates the real stress in
$\Sigma$ is the sum of $\mathbf{E}_{q}$ and $\mathbf{E}_{Q}$:
$\mathfrak{F}=\mathbf{E}_{q}+\mathbf{E}_{Q}$. Recall now that every quadratic
map $\Phi$ between two linear spaces generates a bilinear map $T_{\Phi}$
according to $T_{\Phi}(x,y)=\Phi(x+y)-\Phi(x)-\Phi(y)$, where $(x,y)$ are
corresponding variables. So, in our case we can define corresponding bilinear
map denoted by $\mathbb{T}$.  We obtain
$$
\mathbb{T}(\mathfrak{F})=
\mathbb{T}(\mathbf{E}_{q}+\mathbf{E}_{Q})-\mathbb{T}(\mathbf{E}_{q})-\mathbb{T}(\mathbf{E}_{Q})=
\mathbf{E}_{q}\otimes\mathbf{E}_{Q}+
\mathbf{E}_{Q}\otimes\mathbf{E}_{q}-\mathbf{E}_{Q}.\mathbf{E}_{q}\,id_{T\Sigma}.
$$
In components we have correspondingly (here "bar" means vector)
\begin{eqnarray*}
(M_{q})_i^j&=&(E_{q})_i(\bar{E}_{q})^j-\frac12(\mathbf{\bar{E}}_{q})^2\delta_i^j=
\frac14\omega_{mn}\,\omega^{mn}\delta_i^j-\omega_{im}\,\omega^{jm},\\
(M_{Q})_i^j&=&(E_{Q})_i(\bar{E}_{Q})^j-
\frac12(\mathbf{\bar{E}}_{Q})^2\delta_i^j=
\frac14\Omega_{mn}\,\Omega^{mn}\delta_i^j-\Omega_{im}\,\Omega^{jm},\\
\mathbb{T}_i^j(\mathfrak{F})&=&
(E_{q})_i(\bar{E}_{Q})^j+(E_{Q})_i(\bar{E}_{q})^j-
\bar{\mathbf{E}}_{q}.\bar{\mathbf{E}}_{Q}\,\delta_i^j.
\end{eqnarray*}

The tensor field $(-\frac{1}{4\pi}\mathbb{T})$ may be called
{\it interaction stress tensor}\index{interaction stress tensor}.
In the sections of the bundle $T^*\Sigma\otimes T\Sigma$
we have the {\it trace} form $tr$, and on $\Sigma$ we have the
standard volume form $\omega_o=dx\wedge dy\wedge dz$. So we can form the object
$tr\otimes\omega_o$.

By definition, the two quantities $w$ and $U$ defined by
$$
w=(tr\otimes\omega_o)(-\frac{1}{4\pi}\mathbb{T})=
-\frac{1}{4\pi}\langle tr,\mathbb{T}\rangle\omega_o=
\frac{1}{4\pi}\bar{\mathbf{E}}_{q}.\bar{\mathbf{E}}_{Q}\omega_o,
\ \ \ \text{and} \ \ \
U=\int_{\Sigma}w
$$
will be called {\it interaction energy density}\index{interaction energy
density} and {\it interaction energy} for
$\omega$ and $\Omega$. Clearly, $w$ represents the mutual flow of
$\mathbf{E}_q$ and $\mathbf{E}_Q$, and $w$ and $U$ may acquire positive and
negative values.
 \newpage
{\bf 3. Introducing admissible changes} \vskip 0.2cm Further we
shall follow the rule: \begin{center} \hfill\fbox{
    \begin{minipage}{0.97\textwidth}
\begin{center}
\vskip 0.3cm
{\bf An isolated (quasistatic) physical
system of this kind tends to configurations with less values of the
integral interaction energy.}
\end{center}
\vskip 0.3cm
\end{minipage}} \hfill \end{center}

Hence, an {\it intrinsically induced drifting} should require time-decreasing
$\delta U$ of $U$ {\bf with respect to the configuration parametres} $(Q,q,R)$.
Since by assumption the topology of $\Sigma$ must not change, $Q$ and $q$
must keep their values. So, the only confgurational parameter allowed to
change in time is $R$, therefore,
$\frac{\delta U}{\delta t}=\frac{\delta
U}{\delta R}\frac{\delta R}{\delta t}<0$, where $t$ denotes time.

In order to compute $U$ we compute first $w$ and obtain
$$
w=\frac{1}{4\pi}\bar{\mathbf{E}}_{q}.\bar{\mathbf{E}}_{Q}\,\omega_o=
\frac{1}{8\pi}\,(\Omega\wedge *\omega+ \omega\wedge *\Omega)=
-\frac{1}{8\pi}\left[\mathbf{d}\left(\frac{q}{\bar{r}}\Omega\right)+
\mathbf{d}\left(\frac{Q}{r}\,\omega\right)\right].
$$
\vskip 0.2cm
{\bf Remark.}
The above expression for the local
interaction energy $w$ clearly shows that $w$ has maximum value on the straight
line connecting the two charges. This suggests that the attracting/repelling
force could be expected to be directed along the same line since the derivatve
of the interaction energy is expectable to take its maximum value namely along
the same line. \vskip 0.2cm

Since $w$ is an {\it exact} 2-form we can make use of the Stokes theorem, so,
the integral over $\Sigma$ is transformed to 2-dimensional surface integral
over the boundary $\partial\Sigma$ of $\Sigma$:
 $$
\partial\Sigma=S^2_{(r,\bar{r})=\infty}\cup S^2_q\cup S^2_{Q}.
$$
On
$S^2_{\infty}$ the corresponding integrals have zero values. So, in the induced
on $\partial \Sigma$ orientation, and denoting by $R_q$ and $R_Q$ the radiuses
of $S^2_q$ and $S^2_Q$ respectively, we have
$$
U=\frac{q}{2}\frac{1}{4\pi
R^2_Q}\int_{S^2_q\cup S^2_Q} \frac{R^2_Q
\,Q\,\mathrm{sin}\theta\,d\theta\wedge\,d\varphi}{\bar{r}}+
\frac{Q}{2}\frac{1}{4\pi R^2_q}\int_{S^2_q\cup S^2_Q}
\frac{R^2_q\,q\,\mathrm{sin}\bar{\theta}\,d\bar{\theta}\wedge
d\bar{\varphi}}{r}\ .
$$
 On $S^2_q$ we have $\bar{r}=const$ and
$\int_{S^2_q}\Omega=0$. Similarly, on $S^2_Q$ we have $r=const$ and
$\int_{S^2_Q}\omega=0$. Notice further that $\frac{1}{r}$ is a harmonic
function, so, at every point $p\in \Sigma$ it can be represented by its avarege
value on the corresponding 2-sphere centered at $p$. Now, the first integral
reduces to integral over the 2-sphere $S^2_Q$ and it is equal to
$\frac{qQ}{2R}$, similarly, the second integral reduces to integral over the
2-sphere $S^2_q$ and has the same value, $\frac{qQ}{2R}$, where $R$ is the
euclidean distance between the centers of the two small spheres. Thus, the
computation gives finally $U=\frac{qQ}{R}$, where $(q,Q,R)$ are treated as
configuration parameters.

Now, according to the above mentioned rule that $\delta_t U<0$, and that $q$
and $Q$ do not change, we obtain:

	1. for the  case $q.Q>0$ we shall have
$\delta_t U=-\frac{qQ}{R^2}\delta_t R<0$, so $\delta_t R>0$, i.e.
{\bf repulsion} should be expected;

	2. for the case $q.Q<0$ we shall have
$\delta_t U=-\frac{qQ}{R^2}\delta_t R<0$, so $\delta_t R<0$, i.e. {\bf
atraction} should be expected.

The above consideration clearly suggests the conclusion:

 {\bf The Coulomb force law originates from available local interaction
between the two fields $\mathbf{E}_{Q}$ and $\mathbf{E}_{q}$ under quasistatic
time-changes of the integral interaction energy $U$ leading to
minimization of} $U$.
\vskip 0.2cm
In fact, if $U$ changes with time, then the change
$\delta_t U=-\frac{qQ}{R^2}\delta_t R$ must be carried over
to the mechanical kinetic energies of the two particles:
$$
\delta_t\left(\frac{\mathbf{p_1^2}}{2m_1}+\frac{\mathbf{p_2^2}}{2m_2}\right)=-
\delta_t U ,
$$
since, by assumption, there are no other energy consuming factors in the system
considered. So, the Coulomb force can be understood as an {\it integral}
characteristic of the system, therefore its field, i.e., spatially local,
interpretation may be reconsidered. On the other hand, in the corresponding
spherical coordinates, $(Q.*\omega)$ and $(q.*\Omega)$ look very much as
$\delta U$, but this {\it first-sight resemblance should not mislead us}. The
difference is quite serious: $(Q.*\omega)$ and $(q.*\Omega)$ are 1-forms, local
objects on $\Sigma$ by definition, while $\frac{qQ}{R^2}\delta R$ is not local
(with respect to the coordinates) object, $R$ is not the coordinate $r$ and,
contrary to $dr$, $\delta R$ is not 1-form on $\Sigma$. We may allow ourselves
to call $(\omega)$ and $(\Omega)$, or $(*\omega)$ and $(*\Omega)$, Coulomb
fields but NOT Coulomb force fields because {\bf they can NOT generate any
direct local change of momentum}, since as we mentioned earlier, {\bf these
fields conserve their energy, and their static nature requires zero momentum}.
The local force of stress nature is given in the theory by the divergence of
the Maxwell stress tensor which is a nonliner object, namely, a bilinear
combination of the field components and their derivatives.

The Coulomb force gets an admissible interpretation as an
integral characteristic of the system describing some integral tendences to
minimization of the integral interaction energy $U$ of the two fields. Surely,
$\omega$ and $\Omega$ carry some local physical information but in a quite
indirect manner: except spherical symmetry (which, of course, is not specific
only for electric fields) any of these two local objects {\it can not clarifiy
the physical nature of the local changes in the space when charged particles
are around}. In other words, from local point of view, we could not identify
$\omega$ and $\Omega$ as electric fields. Any topologically nontrivial region
of the kind "$\mathbb{R}^3$ minus a point" generates such fields, so, the
electric nature of the field can be proved only by means of additional
procedures concerning the integral structure of the system.

Also, the topological interpretation of $\omega$ and $\Omega$
suggests that the description is rather integral than local: although
$\omega$ and $\Omega$ are local objects, in fact they are just specially chosen
representatives of integral characteristics of the physical system considered:
they specify the topology of the space where the two fields are defined. For
another example, the Newton gravitation force law looks the same except the
different interpretation of the corresponding topological numbers as masses,
assuming only positive values. Following the same argument, the Newton
gravitation force law is of integral nature and shows corresponding tendences
except that the masses are always positive numbers and repulsion is not
allowed: $U<0$. But this integral difference says too little about the local
physical nature of the two physically different field structures. \vskip 0.3cm

The above consideration makes us think that, from theoretical point of view,
the Maxwell stress tensor field is the right object in terms of which local
force (in fact stress) fields must be defined, namely, through computing its
divergence. If the field is free then this divergence is zero and represents
physically admissible local nonhomogenities, and any additional conditions must
be consistent with this zero divergence. In the static case for just one field
$\omega$ we shall have
$\mathbf{d}\omega=0,\ \mathbf{d}*\omega=0$, so,  the divergence must be
zero:
$$ \nabla_iM^i_jdx^j=\Big[(*\omega)^{i}(\mathbf{d}*\omega)_{ik}
+\frac12\,\omega^{ij}(\mathbf{d}\omega)_{ijk}\Big]dx^k=
(\mathrm{rot}\,\mathbf{E})\times\mathbf{E}+\mathbf{E}\,\mathrm{div}\mathbf{E}=0
,
$$
where
$$ \frac12\,\omega^{ij}(\mathbf{d}\omega)_{ijk}dx^k=
\mathbf{E}\,\mathrm{div}\mathbf{E},
$$
$$
(*\omega)^{i}(\mathbf{d}*\omega)_{ik}dx^k=
(\mathrm{rot}\,\mathbf{E})\times\mathbf{E}, \ (x^1,x^2,x^3)=(x,y,z),
$$
and vectors and covectors are identified through the euclidean metric.

The situation seriously changes when we are going to consider independent,
self-consistent time-dependent, time-stable spatially finite and
propagating in space physical objects of electromagnetic nature, namely, we have
no such topologically motivated suggestions to choose adequate mathematical
objects been able to represent appropriately the corresponding physical
stresses. Hence, the mathematical model must be created on the basis of
assumptions of quite different nature, for example: requirements for definite
and appropriately defined integrability properties representing the object's
time stability; experimentally proved and traditionally assumed straight-line
propagation of the energy-density; orthogonality of the electric and magnetic
components of the field suggesting absence of local interaction energy between
the electric and magnetic components or their new versions; some notion for
internal energy redistribution during time-evolution, etc. In our view, in such
cases, the eigen and other algebraic properties of the corresponding
stress-energy-momentum tensor field should play a basic role, so,
not the very fields, but their  stress
tensor fields and the corresponding mutual stress tensors seem to be the right
objects in terms of which local force fields should be defined. If there is
just one free field then the divergence of the corresponding stress tensor is
zero, representing in this way the physically admissible local spatial
nonhomogenities. Any additional conditions must be consistent with this zero
divergence. In the special case of electric field
$\mathbf{E}=\tilde{g}^{-1}(*\omega)$ considered, this divergence is given
above. Note that, even in this very special {\it static} case, the field
demonstrates structure consisting of two recognizable subsystems, formally
represented by the differential forms $(\omega,*\omega)$, while represented
through $\mathbf{E}$, such a 2-component nature is not easily recognizable.
Such a 2-component nature of the field may demonstrate itself as a
characteristic property in the above mentioned time dependent free field case,
allowing corresponding time-stabalization through an internal
$(\omega\leftrightarrow *\omega)$ energy-momentum exchange.

\subsection{Interconnecting electric $\mathbf{E}$ and magnetic
$\mathbf{B}$ fields}
Following the development of experiment, theoretical and
mathemetical physics of the second half of the 19th century make some serious
steps in building corresponding model of electromagnetic phenomena. Among these
we mention the following (further we consider only the vacuum case):

	{\bf 1.} The electromagnetic phenomena in vacuum have field nature.

	{\bf 2.} The mathematical description is based on a couple of two
differentially time-interconnectd vector fields $(\mathbf{E,B})$ defined on
the space $(g,\mathbb{R}^3)\times(\mathbb{R}=TIME)$, and having no {\it
TIME}-directed components.

	{\bf 3.} The time-change of each of $(\mathbf{E,B})$ is generated by the
spatial nonhomogenity of the other.

	{\bf 4.} The infinitesimal flow of each of the vector fields
$(\mathbf{E,B})$ does not change the infinitesimal 3-volume element
$\omega_o=dx\wedge dy\wedge dz=\sqrt{det(g_{ij})}dx^1\wedge dx^2\wedge dx^3$.

The leading idea in finding how each time-dependence is connected to the
corresponding spatial nonhomogenity is based classically on the assumption that
{\it the time-change of the flow of each of the two vector fields
$(\mathbf{E,B})$ across a finite 2-surface $S$ with boundary $\partial S$ is
proportional up to a sign to the line-integral of the $g$-corresponding to the
other vector field 1-form on the the boundary $\partial S$}. The necessary
coefficient $c$ is understood as propagation velocity. The sign choice comes
from correspondence with the experiment. Finally, time is external parameter
not supposed to be involved in coordinate transformations.

In modern notation these assumptions lead to the followling relations. First,
the absolute character of the time parameter allows to write down the time
changes of $(\mathbf{E,B})$ simply as
$$
\mathbf{E}\rightarrow\frac{\partial \mathbf{E}}{\partial t}=\frac{\partial
\mathbf{E}^i}{\partial t}\frac{\partial}{\partial x^i} \ \ \
\mathbf{B}\rightarrow \frac{\partial \mathbf{B}}{\partial t}=\frac{\partial
\mathbf{B}^i}{\partial t}\frac{\partial}{\partial x^i}\cdot
$$
Now the
corresponding local flows should be represented by corresponding differential
2-forms, and in terms of the Poincare isomorphism and the Hodge $*$, they are
given by reducing the 2-forms
 $$
i\left(\frac{\partial \mathbf{E}}{\partial t}\right)\omega_o
=\frac{\partial}{\partial t}i(\mathbf{E})\omega_o=
\frac{\partial}{\partial t}*\tilde{g}(\mathbf{E}),  \ \ \
\ i\left(\frac{\partial \mathbf{B}}{\partial t}\right)\omega_o
=\frac{\partial}{\partial t}i(\mathbf{B})\omega_o=
\frac{\partial}{\partial t}*\tilde{g}(\mathbf{B})
$$
on the 2-surface $S$ considered, so, if $\varphi:S\rightarrow\mathbb{R}^3$ is the
imbedding, then the two integrands are
$$
\varphi^*\left(i\left(\frac{\partial \mathbf{E}}{\partial t}\right)\omega_o\right), \
\ \varphi^*\left(i\left(\frac{\partial \mathbf{B}}{\partial t}\right)\omega_o\right).
$$

The two boundary line integrals are transformed to surface integrals according
to the Stokes theorem, so, the two integrands written down as differential
2-forms (as it should be) are just
$\varphi^*(\mathbf{d}\tilde{g}(\mathbf{E}))$ and
$\varphi^*(\mathbf{d}\tilde{g}(\mathbf{B}))$, where $\tilde{g}(\mathbf{E})$ and
$\tilde{g}(\mathbf{B})$ are the euclidean 1-form images of the two vector
fields. Hence, determining the signs in correspondence with the experiment, the
equations read (we omit $\varphi^*$ since arbitrary not physical
2-surfaces are considered)
$$
 i\left(\frac{\partial \mathbf{E}}{\partial
t}\right)\omega_o=c\,\mathbf{d}\tilde{g}(\mathbf{B}), \ \
i\left(\frac{\partial
\mathbf{B}}{\partial t}\right)\omega_o=-c\,\mathbf{d}\tilde{g}(\mathbf{E}), \ \
L_{\mathbf{E}}\omega_o=0, \ \ L_{\mathbf{B}}\omega_o=0.
$$
\vskip 0.3cm
\noindent {\bf Remark}. In order not to have misunderstanding, we note that the
usual notations we meet in some electrodynamics textbooks :
$\mathbf{E}.\vec{\mathbf{ds}}$ and $\mathbf{B}.\vec{\mathbf{ds}}$, which are
interpreted as local flows of $\mathbf{E}$ and $\mathbf{B}$ across a finite
2-surface and given by scalar products, are not quite appropriate, since,
formally speaking, scalar product of two vectors does not produce differential
2-form. Moreover, the 2-surface $S$ plays here just a helping role, it does not
participate as a physical exchange partner, and since $\mathbf{E}$ and
$\mathbf{B}$ do not represent real energy flows, the equations obtained have
not local energy-momentum balance sense.

Another note that should be made here is the silent assumption that the
electric and magnetic components are {\it time recognaizable during
propagation} in the 3-space, so, the 2-dimensional distribution defined by
$(\mathbf{E},\mathbf{B})$ admits local shuffling symmetry, i.e., a vector field
$Z$ such that $ \mathbf{E}\wedge\mathbf{B}\wedge Z\neq 0$, and the Lie brackets
$[\mathbf{E},Z]$ and $[\mathbf{B},Z]$ are lineary representable by $\mathbf{E}$
and $\mathbf{B}$. Since the distribution $(\mathbf{E},\mathbf{B},Z)$ on
$\mathbb{R}^3$ is integrable by dimension considerations, this would suggest to
consider the integrability properties of the three 2-dimensional
subdistributions $(\mathbf{E},\mathbf{B})$, $(\mathbf{E},Z)$ and
$(\mathbf{B},Z)$ in order to obtain explicit expressions for the expected
interaction between $\mathbf{E}$ and $\mathbf{B}$ through the corresponding
curvature forms, as it is clearly suggested by the above stated assumptions
$\mathbf{(1-4)}$.

\section{Maxwell Equations: Nonrelativistic and Relativistic forms}
\subsection{General features}
In standard coordinates $(x,y,z)$ the above equations acquire the form
$$
{\rm rot}\,\mathbf{E}+
\frac 1c      \frac{\partial {\mathbf{B}}} {\partial t}=0
, \quad {\rm div}\,\mathbf{B}=0,             %1%
$$
$$
{\rm rot}\,\mathbf{B} -
	\frac 1c \frac{\partial {\mathbf{E}}} {\partial t}=0 ,
\quad {\rm div}\,\mathbf{E}=0,
$$
further related to as Maxwell vacuum equations (MVE)\index{Maxwell vacuum
equations}.

First we note, that because of the linearity of these equations if
$({\mathbf{E}}_i,\mathbf{B}_i), i=1,2,...$  are a collection of solutions,
then every couple of linear combinations of the form
$$
{\mathbf{E}}=a_i{\mathbf{E}}_i,\ \mathbf{B}=a_i\mathbf{B}_i   %3%
$$
(sum over the repeated $i=1,2,...$) with arbitrary constants $(a_i)$
gives a new solution.

Following Sec.5.3.2
the local dynamical characteristics are represented by the Maxwell stress
tensor $M$ given as sum of the $\mathbf{E}$-stress and $\mathbf{B}$-stress by
$$
M=M(\mathbf{E})+M(\mathbf{B})=
\mathbf{E}\otimes\mathbf{E}-\frac12\mathbf{E}^2.(g^{-1})+
\mathbf{B}\otimes\mathbf{B}-\frac12\mathbf{B}^2.(g^{-1})
$$
$$
=\mathbf{E}\otimes\mathbf{E}+\mathbf{B}\otimes\mathbf{B}-
\frac12(\mathbf{E}^2+\mathbf{B}^2).(g^{-1}),
$$
or in components (recall the identification of
contravariant and covariant tensor fields by $\tilde{g}$ and
$\tilde{g}^{-1}$)
$$
M^i_j=E^iE_j+B^iB_j-\frac12(\mathbf{E}^2+\mathbf{B}^2)\delta^i_j,
$$
and by the Poynting vector
$$
\mathbf{S}=\frac1c\,(\mathbf{E}\times\mathbf{B}).
$$
The energy density ${\bf w}$ of the field is defined by
$$
{\bf w}=-tr(M)=\frac12(\mathbf{E}^2+\mathbf{B}^2)
$$
and the Poynting vector $\mathbf{S}$ defines two important features: the
direction of propagation at each point $(x,y,z)$ and the local momentum of the
field.

If the field is free then
$$
\nabla_iM^i_j=0, \ \ \ \text{and} \ \
\frac{\partial\bf{w}}{\partial t}=-\mathrm{div}(\mathbf{S}).
$$

These definitions tacitly introduce the following important features of the
dynamical behavior of the field:
\vskip 0.3cm
	{\bf 1.} The stress-energy tensor of the electromagnetic field is a sum
of the stress-energy tensors of the electric and magnetic component-fields, so,
{\it there is NO mutual  $(\mathbf{E},\mathbf{B})$-interaction stress and
$(\mathbf{E},\mathbf{B})$-interaction energy}.
\vskip 0.3cm
	{\bf 2.} Non of the two component-fields  $(\mathbf{E},\mathbf{B})$ is
able to carry separately local momentum, although the two components are
time-recognizable.
\vskip 0.3cm

	{\bf 3.} The direction of propagation of the field is orthogonal to
each of of the two component-fields  $(\mathbf{E},\mathbf{B})$.
\vskip 0.3cm

\subsection{Nonrelativistic duality} \index{nonrelativistic duality}
The important observation
made by clever men at the end of 19th century, is that the substitution $$
{\mathbf{E}}\rightarrow -\mathbf{B},\quad
\mathbf{B}\rightarrow {\mathbf{E}}      %4%
$$
transforms the first couple of the equations into the
second couple, and, vice versa, the second couple is transformed into
the first one. This symmetry transformation
is called {\it special duality transformation}, or
SD-transformation.  It clearly shows that the electric and magnetic
components of the vacuum electromagnetic field are interchangeable and the
interchange transforms solution into solution. This feature of the
electromagnetic field reveals its {\it dual} nature.

It is important to note that the SD-transformation does not
change the energy density ${\bf w}=\frac12({\mathbf{E}}^2+\mathbf{B}^2)$,
the Poynting vector $\mathbf{S}=\frac1c({\mathbf{E}}\times\mathbf{B})$ , and the
(nonlinear) Poynting relation
$$
\frac1c\frac {\partial}{\partial t} \frac {{\mathbf{E}}^2+\mathbf{B}^2}{2}=
-{\rm div}\,\mathbf{({\mathbf{E}}\times\mathbf{B})}.
$$
Hence, from energy-momentum point of view two dual
solutions are indistinguishable.

Note that this substitution may be considered as a transformation of the
following kind:
$$
(\mathbf{E},\mathbf{B})
	%\left\| \matrix{0   &1\cr -1  &0\cr} \right\|=      %5%
	\begin{Vmatrix}
0   &1
\\
-1  &0
	\end{Vmatrix}
=
(-\mathbf{B},\mathbf{E}).
$$
The following question now arises naturally: do there exist constants
$(a,b,m,n)$, such that the linear combinations
$$
{\mathbf{E}}'=a{\mathbf{E}}+m\mathbf{B},\
\mathbf{B}'=b{\mathbf{E}}+n\mathbf{B},       %6%
$$
or in a matrix form
$$
({\mathbf{E}}',\mathbf{B}')=({\mathbf{E}},\mathbf{B})
%\left\|\matrix{a   &b\cr m  &n\cr} \right\|
	\begin{Vmatrix}
a  & b
\\
m  & n                                                    %7%
	\end{Vmatrix}
=
(a{\mathbf{E}}+m\mathbf{B},b{\mathbf{E}}+n\mathbf{B}),                         %8%
$$
form again a vacuum solution?
Substituting $\mathbf{E}'$ and $\mathbf{B'}$ into Maxwell's vacuum
equations we see that the answer to this question is affirmative iff $m=-b,
n=a$, i.e. iff the corresponding matrix $S$ is of the form
$$
A=
%\left\|\matrix{a   &b\cr -b  &a\cr} \right\|
	\begin{Vmatrix}
a       & b
\\
-b      & a                                   %8%
	\end{Vmatrix}
.
$$
The new solution
will have now energy density ${\bf w}'$ and momentum density $\mathbf{S}'$ as
follows:
\[
{\bf w}'=\frac12\left({\mathbf{E}'}^2 +
			{\mathbf{B}'}^2\right)=
\frac {1}{2}(a^2 + b^2)\biggl({\mathbf{E}}^2 + \mathbf{B}^2\biggr),
\]
\[\mathbf{S}'=(a^2+b^2)\frac1c\,{\mathbf{E}}\times\mathbf{B}.
\]
Obviously, the new and the old solutions will have the same energy and
momentum if $a^2+b^2 =1$, i.e. if the matrix $A$ is {\it unimodular}.
In this case we may put
$a=\cos \alpha$ and $b=\sin \alpha$, where $\alpha=const$,
so the transformation becomes
$$
%\begin{equation}
\tilde{\mathbf{E}}={\mathbf{E}}\cos \alpha -
				\mathbf{B}\sin \alpha,\        %9%
\tilde{\mathbf{B}}={\mathbf{E}}\sin \alpha + \mathbf{B}\cos
\alpha.
%\end{equation}
$$
This transformation is known as {\it electromagnetic duality transformation},
or D-transformation. Note that the energy density and the Poynting vector stay
the same even if the above parameter $\alpha$ depends on the coordinates.

From physical point of view a basic feature of the D-transformation is,
that the difference between the electric and magnetic fields becomes
non-essential: we may superpose the electric and the magnetic vectors, i.e.
vector-components, of a general electromagnetic field to obtain new
solutions.  From mathematical point of view we see that Maxwell vacuum equations
, besides the usual linearity mentioned above, admit also
"cross"-linearity, i.e.  linear combinations of ${\mathbf{E}}$ and $\mathbf{B}$ of a
definite kind determine new solutions.

On the other hand, any linear map $\phi: \mathbb{R}^2\rightarrow\mathbb{R}^2$,
having in the canonical basis of $\mathbb{R}^2$ a matrix $A$, is a symmetry of
the canonical complex structure $\mathcal{I}$ of $\mathbb{R}^2$; we recall that
if the canonical basis of $\mathbb{R}^2$ is denoted by
$(\varepsilon^1,\varepsilon^2)$ then ${\cal I}$ is defined by ${\cal
I}(\varepsilon^1)=\varepsilon^2$, ${\cal I}(\varepsilon^2)=-\varepsilon^1$, so
$A$ must satisfy $A.{\cal I}.A^{-1}={\cal I}$. We note also that
$\mathcal{I}(\varepsilon^1\wedge\varepsilon^2)=\varepsilon^1\wedge\varepsilon^2$,
so, $\mathcal{I}$ is unimodular, i.e. it keeps the volume
$\varepsilon^1\wedge\varepsilon^2$ unchanged. Hence, {\bf the
electromagnetic D-transformations coincide with the unimodular symmetries of
the canonical complex structure} ${\cal I}$ {\bf of} $\mathbb{R}^2$.  This
important in our view remark clearly points out that the canonical complex
structure ${\cal I}$ in $\mathbb{R}^2$ should be considered as an {\bf essential
element} of classical electromagnetic theory, so we should in no way neglect
it.

Finally we note that D-transformations change the two well known
invariants $I_1=(\mathbf{B}^2-{\mathbf{E}}^2)$ and $I_2=2{\mathbf{E}}.\mathbf{B}$ in the
following way:
\begin{align*}
& \tilde{I_1}=\tilde{\mathbf{B}}^2-\tilde{\mathbf{E}}^2=
(\mathbf{B}^2-{\mathbf{E}}^2)\cos 2\alpha+
			2{\mathbf{E}}.\mathbf{B}\sin 2\alpha=
I_1\cos 2\alpha+I_2\sin 2\alpha, \\               %10%
&\tilde{I_2}= 2\tilde{\mathbf{E}}.\tilde{\mathbf{B}}=
({\mathbf{E}}^2-\mathbf{B}^2)\sin 2\alpha+
			2{\mathbf{E}}.\mathbf{B}\cos 2\alpha=
-I_1\sin 2\alpha+I_2\cos 2\alpha.                  %11%
\end{align*}
It is seen that even the SD-transformation, where $\alpha=\pi/2$, changes
these two invariants: $I_1\rightarrow -I_1, \ I_2\rightarrow -I_2$.
This shows that if these two invariants define which solutions should be
called {\it different}, then by making an arbitrary dual transformation we
will always produce different solutions, no matter if these solutions carry
the same energy-momentum or not.
In general we always have
\[
\tilde{I_1}^2+\tilde{I_2}^2=I_1^2+I_2^2,
\]
i.e. the sum of the squared invariants is a D-invariant.

These notices are in accordance with the above made assumption, that the
electromagnetic field, considered as  {\it one physical object}, has {\it two
physically distinguishable interrelated vector components}, so the adequate
mathematical model-object must have two vector components and must admit
2-dimensional linear transformations of its components, which should be closely
related to the invariance properties of the energy-momentum characteristics of
the field. Therefore, in the frame of this nonrelativistic 3-dimensional form
of the theory it seems reasonable to assume the following:
 \vskip 0.5cm
{\it
The electromagnetic field is mathematically represented on $\mathbb{R}^3$ by an
$\mathbb{R}^2$-valued differential 1-form $\omega$, such that in the canonical
basis $(\varepsilon^1,\varepsilon^2)$ in $\mathbb{R}^2$ the 1-form $\omega$
looks as follows}
$$ \omega=\tilde{g}(\mathbf{E})\otimes \varepsilon^1 +
\tilde{g}(\mathbf{B})\otimes \varepsilon^2.                %12%
$$
\vskip 0.5cm
\noindent We recall that we identify the vector fields and 1-forms on
$\mathbb{R}^3$ through the euclidean metric and we write, e.g.,
$*({\mathbf{E}}\wedge\mathbf{B})= {\mathbf{E}}\times\mathbf{B}$. Also, we
identify $(\mathbb{R}^2)^*$ with $\mathbb{R}^2$ through the euclidean metric.
\vskip 0.5cm Now we have to present Maxwell vacuum equations  correspondingly,
i.e. in terms of $\mathbb{R}^2$-valued objects.

The above assumption requires a general covariance with respect to
transformations in $\mathbb{R}^2$, so, the complex structure $\mathcal{I}$
has to be introduced explicitly in the equations.  In order to do this we
recall that the linear map $\mathcal{I}:\mathbb{R}^2\rightarrow\mathbb{R}^2$
induces a map
\[
{\mathcal{I}_*:\omega\rightarrow \mathcal{I}_*(\omega)=
{\mathbf{E}}\otimes \mathcal{I}(\varepsilon^1)+ \mathbf{B}\otimes
\mathcal{I}(\varepsilon^2)= -\mathbf{B}\otimes \varepsilon^1+
\mathbf{E}}\otimes \varepsilon^2.
\]
We recall also that every operator
$\mathcal{D}$ in the set of differential forms is naturally extended to
vector-valued differential forms according to the rule
$\mathcal{D}\rightarrow \mathcal{D}\times id$,
and $id$ is usually omitted.  Having in mind the
identification of vector fields and 1-forms through the euclidean metric we
introduce now $\mathcal{I}$ in Maxwell's equations through $\omega$
in the following way ($\delta$ is the coderivative):
%\begin{equation}
$$
*\mathbf{d}\omega-\frac {1}{c} \frac {\partial }{\partial t}{\cal I}_*
(\omega)=0,\quad \delta\omega=0.                          %13%
%\end{equation}
$$
Two other equivalent forms are:
$$
{\bf d}\omega-*\frac {1}{c} \frac
{\partial }{\partial t}{\cal I}_* (\omega)=0,\quad \delta \omega =0,
$$
$$
*{\bf d}{\cal I}_*(\omega)+
\frac {1}{c} \frac {\partial }{\partial t}\omega=0, \quad \delta \omega =0.
$$
\noindent
In order to verify the equivalence to Maxwell vacuum equations
we compute the marked operations.  We obtain
$$
*{\bf d}\omega-
\frac {1}{c} \frac {\partial }{\partial t}{\cal I}_*(\omega)=
\left({\rm rot}\,\mathbf{E}+\frac1c\frac{\partial \mathbf{B}}{\partial
t}\right) \otimes\varepsilon^1+ \left({\rm rot}\,\mathbf{B}-
\frac1c\frac{\partial {\mathbf{E}}}{\partial t}\right)\otimes\varepsilon^2,
$$
The second equation $\delta \omega=0$ is, obviously,
equivalent to
$$
{\rm div}\,{\mathbf{E}}\otimes \varepsilon^1 +
{\rm div}\,\mathbf{B}\otimes \varepsilon^2=0
$$
since $\delta=-{\rm div}$.

We shall emphasize once again that according to our general assumption
the field $\omega$ will have different representations in the different bases
of $\mathbb{R}^2$.
Changing the basis $(\varepsilon^1,\varepsilon^2)$ to any other basis
$(\varepsilon^{1'},\varepsilon^{2'})$ by a linear map $\varphi$,
means, of course, that the field $\omega$ changes to
$\varphi_*\omega$ and the complex structure ${\cal I}$
changes to $\varphi{\cal I}\varphi^{-1}$. In some sense this means that we
have two fields now: $\omega$ and ${\cal I}$, but ${\cal I}$ is given
beforehand. So, in the new basis
the $\mathcal{I}$-dependent equations will look like
$$
*{\bf d}\varphi_*\omega-
\frac1c\frac{\partial }{\partial t}
(\varphi{\cal I}\varphi^{-1})_*(\varphi_*\omega)=0.
$$
If $\varphi$ is a symmetry of ${\cal I}:
\varphi{\cal I}\varphi^{-1}={\cal I}$, then we transform just $\omega$ to
$\varphi_*\omega$.

In order to write down the Poynting energy-momentum balance relation we
recall the product of vector-valued differential forms. Let
$\Phi= \Phi^a\otimes e_a$ and $\Psi= \Psi^b\otimes k_b$ be two differential
forms on some manifold with values in the vector spaces $V_1$ and $V_2$
with bases $\{e_a\}, a=1,...,n$ and $\{k_b\}, b=1,...,m$, respectively.
Let $f:V_1\times V_2\rightarrow W$ be a bilinear map valued in a third vector
space $W$.  Then a new differential form, denoted by $f(\Phi,\Psi)$, on the
same manifold and valued in $W$ is defined by
$$
f(\Phi,\Psi)=\Phi^a\wedge \Psi^b\otimes f(e_a,k_b).
$$
Clearly, if the original forms are $p$ and $q$ respectively,
then the product is a $(p+q)$-form.

Assume now that $V_1=V_2=\mathbb{R}^2$ and the bilinear map is the
exterior product:
$\wedge:\mathbb{R}^2\times \mathbb{R}^2\rightarrow
\Lambda^2(\mathbb{R}^2)$.

Let's compute the expression $\wedge(\omega,{\bf d}\omega)$.
\begin{align*}
& \wedge(\omega,{\bf d}\omega)=\wedge({\mathbf{E}}\otimes \varepsilon^1+
\mathbf{B}\otimes \varepsilon^2,{\bf d}{\mathbf{E}}\otimes \varepsilon^1+
{\bf d}\mathbf{B}\otimes \varepsilon^2)
\\
&=
({\mathbf{E}}\wedge{\bf d}\mathbf{B}-
\mathbf{B}\wedge{\bf d}{\mathbf{E}})\otimes
\varepsilon^1\wedge\varepsilon^2
\\
& =-{\bf
d}({\mathbf{E}}\wedge\mathbf{B})\otimes\varepsilon^1\wedge\varepsilon^2=
-{\bf d}(**({\mathbf{E}}\wedge\mathbf{B}))\otimes\varepsilon^1\wedge\varepsilon^2=
*\delta({\mathbf{E}}\times\mathbf{B})\otimes\varepsilon^1\wedge\varepsilon^2
\\
& =-*{\rm
div}({\mathbf{E}}\times\mathbf{B})\otimes\varepsilon^1\wedge\varepsilon^2=
-{\rm div}({\mathbf{E}}\times\mathbf{B})dx\wedge dy\wedge dz\otimes
\varepsilon^1\wedge\varepsilon^2.
\end{align*}
Following the same rules we obtain
$$
\wedge\left(\omega,*\frac1c\frac{\partial }{\partial t}
{\cal I}_*\omega\right)=
\frac1c\frac{\partial }{\partial t} \frac{{\mathbf{E}}^2+
\mathbf{B}^2}{2}dx\wedge dy\wedge
dz\otimes\varepsilon^1\wedge\varepsilon^2,
$$
So, the Poynting energy-momentum
balance relation is given by
$$
\wedge\left(\omega, {\bf d}\omega-*\frac1c\frac{\partial }{\partial t}
{\cal I}_*\omega\right)=0.                                    %14%
$$
Since the orthonormal 2-form $\varepsilon^1\wedge\varepsilon^2$ is invariant
with respect to rotations (and even with respect to unimodular
transformations in $\mathbb{R}^2$) we have the duality invariance of the
above energy-momentum quantities and relations.

Note the following simple forms of the energy density
$$
\frac{1}{2}*\wedge\left(\omega,*{\cal I}_*\omega\right)=
\frac{{\mathbf{E}}^2+\mathbf{B}^2}{2}\varepsilon^1\wedge\varepsilon^2,
$$
and of the Poynting vector,
$$
\frac{1}{2}*\wedge(\omega,\omega)=
{\mathbf{E}}\times\mathbf{B}
\otimes\varepsilon^1\wedge\varepsilon^2.
$$

We make the following remark. Physically, we can get information
about the very fields just studying their physical appearence, i.e. by
studying the stress-energy properties of the field: these properties
demonstrate themselves through allowed losing and gaining stress-energy, i.e.
interaction with other physical objects but keeping the field identity.

From formal point of view, frequently, these
propertieis are assumed to be expressed by quadratic functions of the very field
functions. In our case here important such characteristics are the above
mentioned two quantitis $I_1=(\mathbf{B}^2-\mathbf{E}^2),
I_2=2\mathbf{E}.\mathbf{B}$, which transform under the $A$-action on the field
$(\mathbf{E},\mathbf{B})$ according to
$$
I_1'=(a^2-b^2)\,I_1+2ab\,I_2,\quad
I_2'=-2ab\,I_1+(a^2-b^2)\,I_2,
$$
and the determinant of this transformation is
$(a^2+b^2)^2\neq 0$. This shows that the only case when these two invariants do
not change under the action of $A$ is when they are zero, the so called "null
field". We can say that NO non-null field can be transformed to a null field by
means of duality transformation, and, conversely, NO null field can be
transformed to a non-null field in this way. The following two corollaries hold:

{\bf Corollary}: If any couple inside
$(I_1,I_2,I_1',I_2')$ is zero, then the other couple is also zero.

{\bf Corollary}: If $a\neq 0, b\neq 0$, and $(\mathbf{E},\mathbf{B})$ and
$(\mathbf{E'}=a\mathbf{E}-b\mathbf{B},\mathbf{B'}=b\mathbf{E}+a\mathbf{B})$
satisfy
$$
{\rm rot}\,\mathbf{X}+
\frac 1c      \frac{\partial {\mathbf{Y}}} {\partial t}=0
, \quad {\rm div}\,\mathbf{Y}=0,             %1%
$$
then they satisfy also
$$
{\rm rot}\,\mathbf{Y} -
	\frac 1c \frac{\partial {\mathbf{X}}} {\partial t}=0 ,
\quad {\rm div}\,\mathbf{X}=0.
$$

We consider now the problem: is there a joint action of the matrices of the
kind $A$ on $(\mathbf{E},\mathbf{B})$ and on the bases
$(\varepsilon^1,\varepsilon^2)$ such that the field $\omega$ to stay the same,
i.e., $\omega$ to be correspondingly equivariant? The answer is positive.
In fact,
consider the new basis $(e_1,e_2)$ given by
$$
e_1=\frac{1}{a^2+b^2}(a\varepsilon^1-b\varepsilon^2),\quad
e_2=\frac{1}{a^2+b^2}(b\varepsilon^1+a\varepsilon^2).
$$
Accordingly, $A$
transforms the basis through right action by means of
$(A^{-1})^*=A/det(A)$.  Then the "new" field
$\omega'(\mathbf{E}',\mathbf{B}')$ is, in fact, the "old" field
$\omega(\mathbf{E},\mathbf{B})$:
$$
\omega'=\mathbf{E}'\otimes e_1+\mathbf{B}'\otimes e_2
= (a\mathbf{E}-b\mathbf{B})\otimes
\frac{a\varepsilon^1-b\varepsilon^2}{a^2+b^2}+
(b\mathbf{E}+a\mathbf{B})\otimes
\frac{b\varepsilon^1+a\varepsilon^2}{a^2+b^2}
$$
$$
=\mathbf{E}\otimes \varepsilon^1+\mathbf{B}\otimes \varepsilon^2=\omega,
$$
i.e., the "new" field $\omega'(\mathbf{E}',\mathbf{B}')$, represented in
the new basis $(e_1,e_2)$ coincides with the "old" field
$\omega(\mathbf{E},\mathbf{B})$, represented in the old basis
$(\varepsilon^1,\varepsilon^2)$.

\subsection{Amplitude and Phase of a vacuum field}

If the component-fields $(\mathbf{E},\mathbf{B})$ are lineary independent,
the triple $(\mathbf{E},\mathbf{B},\mathbf{E}\times\mathbf{B})$
defines a basis
of the tangent (or cotangent) space at every point, where the field is
different from zero.
We denote the corresponding frame by ${\cal R'}$, so we can write
${\cal R'}=(\mathbf{E},-\varepsilon\mathbf{B},
-\varepsilon\mathbf{E}\times\mathbf{B})$, where $\varepsilon=
-1$ corresponds to motion from $-\infty$ to $+\infty$ and
$\varepsilon=1$ corresponds to motion from $+\infty$ to $-\infty$.

 Since the physical dimension of the third vector
$\mathbf{E}\times\mathbf{B}$ is different from that of the first two, we introduce the
factor $\alpha$
$$
\alpha=\frac{1}{\sqrt{\frac{\mathbf{E}^2+\mathbf{B}^2}{2}}}.
$$
Making use of $\alpha$, we introduce the so called {\it electromagnetic frame}
\index{electromagnetic frame}:
$$
{\cal R}=\left[\alpha\mathbf{E},-\varepsilon\alpha\mathbf{B},
-\varepsilon\alpha^2\mathbf{E}\times\mathbf{B}\right].
$$
Hence, at every point we've got two frames: ${\cal R}$, and the coordinate frame
${\cal R}_0=\Bigl[{\partial_x},{\partial_y},{\partial_z}\Bigr]$, as well as the
corresponding co-frames ${\cal R}^*$ and ${\cal R}_0^*=(dx,dy,dz)$.  We
are interested in the invariants of the corresponding transformation matrix
${\cal M}:{\cal R}_0\rightarrow{\cal R}$.  It is defined by the relation
${\cal R}_0.{\cal M}={\cal R}$.  So, we obtain
$$
\mathcal{M}=\begin{vmatrix}
\alpha E^1&-\varepsilon\alpha
B^1&-\varepsilon\alpha^2(\mathbf{E}\times\mathbf{B})^1\\ \alpha
E^2&-\varepsilon\alpha B^2&-\varepsilon\alpha^2(\mathbf{E}\times\mathbf{B})^2\\
\alpha E^3&-\varepsilon\alpha
B^3&-\varepsilon\alpha^2(\mathbf{E}\times\mathbf{B})^3 \end{vmatrix}.
$$

We shall try to express the physically important concepts of {\it amplitude}
and {\it phase} as functions of the invariants of this matrix $\mathcal{M}$.
\index{invariance of amplitude and phase}
So, in all
cases , where this is possible, the invariant character of the so defined phase
and amplitude will be out of doubt.  As it is well known, in general, every
square $(n\times n)$-matrix ${\cal L}$ has $n$ invariants $J_1,J_2,...,J_n$,
where $J_k$ is the sum of all principle minors of order $k$. The invariant
$J_1({\cal L})={\cal L}_{11}+...+{\cal L}_{nn}$ is the sum of all elements on
the principle diagonal, and the invariant $J_n=det({\cal L})$ is the
determinant of the matrix. In our case $n=3$, so for the invariant $J_2$ we get
$$
J_2=\det\left| \begin{matrix} m_{11} &m_{12}\\ m_{21}
&m_{22}\end{matrix}\right| + \det\left| \begin{matrix} m_{11} &m_{13}\\ m_{31}
&m_{33}\end{matrix}\right| + \det\left| \begin{matrix} m_{22} &m_{23}\\ m_{32}
&m_{33} \end{matrix}\right|.
$$
Denoting $E=|\mathbf{E}|, \mathbf{E}=(E^1,E^2,E^3)$ and
$B=|\mathbf{B}|, \mathbf{B}=(B^1,B^2,B^3)$, for the invariants we get
\begin{eqnarray*}
J_1({\cal M})&=&\alpha E^1-\varepsilon\alpha B^2-\varepsilon\alpha^2
(\mathbf{E}\times\mathbf{B})^3,\\
J_2({\cal M})&=&-\varepsilon\alpha^2
(\mathbf{E}\times\mathbf{B})^3+\varepsilon
\alpha^3\Bigl[\mathbf{E}\times(\mathbf{E}\times\mathbf{B})\Bigr]^2+
\alpha^3\Bigl[\mathbf{B}\times(\mathbf{E}\times\mathbf{B})\Bigr]^1,\\
J_3({\cal M})&=&\alpha^4 (\mathbf{E}\times \mathbf{B}).(\mathbf{E}\times
\mathbf{B}),
\end{eqnarray*}
obviously, the invariants are physically dimensionless.

Now, if the
couple $(\mathbf{E},\mathbf{B})$ represents the field, we introduce the matrix
${\cal M}({\cal R})$ of the frame
${\cal R}=\Bigl[\alpha\mathbf{E}, -\varepsilon\alpha\mathbf{B},
-\varepsilon\alpha^2(\mathbf{E}\times\mathbf{B})\Bigr]$
and define the amplitude ${\cal A}$ of the field, considered to have the
physical dimension of energy-density, by
$$
{\cal A}(\mathbf{E},\mathbf{B}) =\sqrt{\alpha^{-4}
J_3({\cal M})}=|\mathbf{E}\times\mathbf{B}|.
$$

When the inequality
$$
\frac 12\biggl|J_1({\cal M})-1\biggr|\leq 1,
$$
holds, then the function $arccos$ is defined on the expression on the left.
In these cases, by definition, the phase $\theta$ of the field
$(\mathbf{E},\mathbf{B})$ we define by
$$
\theta=arccos\Biggl[\frac 12\Bigl[J_1({\cal M})-1\Bigr]\Biggr].
$$

Let's now see when the basis ${\cal R}$ is normed, i.e. when
$$
 |\alpha\mathbf{E}|=1,\ |\alpha\mathbf{B}|=1,\
\alpha^2|(\mathbf{E}\times \mathbf{B})|=1.
$$
From the first two equations it obviously follows
$|\mathbf{E}|=|\mathbf{B}|$, and from the third equation it follows
$\mathbf{E}.\mathbf{B}=0$. In fact, writing down the third equation in the
form
$$
|\mathbf{E}|^2-2|\mathbf{E}||\mathbf{B}||sin\beta|+|\mathbf{B}|^2=0
$$
where $\beta$ is the angle between $\mathbf{E}$ and $\mathbf{B}$
and expressing $|\mathbf{E}|$ as a function of $|\mathbf{B}|$, through solving
this quadratic equation with respect to $|\mathbf{E}|$, we obtain
$$
0<|\mathbf{E}|_{1,2}=|\mathbf{B}||sin\beta|\pm |\mathbf{B}|\sqrt{sin^2\beta-1}.
$$
This inequality is possible only if $|sin^2\beta|=1$, i.e. when $\mathbf{E}$ and
$\mathbf{B}$ are, in addition, orthogonal. Moreover, both relations
$|\mathbf{E}|=|\mathbf{B}|,\ \mathbf{E}.\mathbf{B}=0$
follow from the third equation only: $\alpha^2|\mathbf{E}\times\mathbf{B}|=1$.
 So, the normed
character of ${\cal R}$ leads to its {\it orthonormal} character, consequently,
$\det{\cal M}({\cal R})=1$. Vice versa, the requirement
$\det{\cal M}({\cal R})=1$ leads to the orthonormal character of ${\cal R}$.

{\bf Corollary}.
The condition $\det{\cal M}({\cal R})=1$ requires
 null character of the field: $\mathbf{B}^2=\mathbf{E}^2,\
\mathbf{E}.\mathbf{B}=0$.

The relations obtained suggest to define and consider the following 4-linear
map:
$R:\mathbb{R}^3\times\mathbb{R}^3\times\mathbb{R}^3\times\mathbb{R}^3
\rightarrow\mathbb{R}$.
$$
R(x,y,v,w)=\det\left| \begin{matrix} x_1 &y_1 &(v\times w)_1\\ x_2 &y_2
&(v\times w)_2\\ x_3 &y_3 &(v\times w)_3 \end{matrix}\right|.
$$
Making use of the vector algebra in $\mathbb{R}^3$ we
come to the following relations:
$$
 R(x,y,v,w)=(x\times y).(v\times w)=\Bigl[y\times(v\times w)\Bigr].x=
\Bigl[(v\times w)\times x\Bigr].y,
$$
$$
 R(x,y,v,w)=-R(y,x,v,w),
$$
$$
 R(x,y,v,w)=-R(x,y,w,v),
$$
$$
 R(x,y,v,w)+R(x,v,w,y)+R(x,w,y,v)=0,
$$
$$
 R(x,y,v,w)=R(v,w,x,y),
$$
$$
 R(x,y,x,y)=(x\times y)^2.
$$
We note that this 4-linear map has all algebraic properties of the Riemannian
curvature tensor, therefore in the frame of this section, we shall call it
{\it $\mathfrak{det}$-algebraic curvature}
\index{$\mathfrak{det}$-algebraic curvature}. For the corresponding 2-dimensional
curvature $K(x,y)$, determined by the two vectors $(x,y)$ we obtain
$$
K(x,y)=\frac {R(x,y,x,y)}{x^2 y^2 -(x.y)^2}=\frac{(x\times y)^2}{x^2
y^2(1-cos^2 (x,y))}= \frac{x^2 y^2 sin^2(x,y)}{x^2 y^2 sin^2(x,y)}=1.
$$
Let
$(e_1,e_2,e_3)$ be a basis. We compute the corresponding Ricci tensor $R_{ik}$
and the scalar curvature ${\bf R}$.
\begin{eqnarray*}
&&R_{ijkl}=R(e_i,e_j,e_k,e_l)=(e_i\times
e_j).(e_k\times e_l),\\
&&R_{ik}=\sum_l {R_i^l}_{kl}=(e_i \times e_1).(e_k
\times e_1)+ (e_i \times e_2).(e_k \times e_2)+(e_i \times e_3).(e_k \times
e_3),\\
&&{\bf R}=\sum_i R^i_i =2\Bigl[(e_1 \times e_2)^2 +(e_1 \times e_3)^2
+ (e_2 \times e_3)^2\Bigr].
\end{eqnarray*}
For our electromagnetic frame ${\cal R}$ we obtain the
following non-zero components:
\begin{eqnarray*}
R_{12,12}&=&4\frac{|\mathbf{E}|^2.|\mathbf{B}|^2}{(|\mathbf{E}|^2+
|\mathbf{B}|^2)^2}sin^2\beta,\\
R_{13,13}&=&R_{12,12}.\frac{2\mathbf{E}^2}{\mathbf{E}^2+\mathbf{B}^2},\\
R_{23,23}&=&R_{12,12}.\frac{2\mathbf{E}^2}{\mathbf{E}^2+\mathbf{B}^2},
\end{eqnarray*}
and for the scalar curvature we get
$$
{\bf R}(E,B)=24\frac{\mathbf{E}^2
\mathbf{B}^2}{(\mathbf{E}^2+\mathbf{B}^2)^2}sin^2\beta.
$$

After this short retreat let's go back to the  quantities {\it phase} and
{\it amplitude}. The above mathematical consideration suggests to try to
relate these two concepts with the notion of curvature in purely formal
sense, namely as a 2-form with values in the bundle $L_{T({\cal R}^3)}$ of
linear maps in the tangent bundle. Most generally, a 2-form $R$ with values
in the bundle $L_{T({\cal R}^3)}$ looks as follows
$$
R=\frac 12 R_{ijl}^{k} dx^i \wedge dx^j \otimes \frac{\partial}
{\partial x^k}\otimes dx^l.
$$
We have to determine the coefficients $R_{ijl}^{k}$,
i.e. we have to define a $(3\times 3)$-matrix $\mathfrak{R}$ of 2-forms. We
define this matrix in the following way:
$$
\mathfrak{R}=\left|
\begin{matrix}
\alpha E_1 dy\wedge dz  &-\varepsilon\alpha B_1 dy\wedge dz  &-\varepsilon\alpha^2 (E\times B)_1
dy\wedge dz\\
\alpha E_2 dz\wedge dx  &-\varepsilon\alpha B_2 dz\wedge dx  &-\varepsilon\alpha^2
(E\times B)_2 dz\wedge dx\\
\alpha E_3 dx\wedge dy  &-\varepsilon\alpha B_3 dx\wedge dy
&-\varepsilon\alpha^2 (E\times B)_3 dx\wedge dy
\end{matrix}\right|.
$$
The columns of this matrix
are the 2-forms $*\tilde{g}(\mathbf{E}),\ *\tilde{g}(\mathbf{B}),\
*\tilde{g}(\mathbf{E}\times\mathbf{B})$, multiplied by the factor $\alpha$
at some degree in order to obtain physically dimensionless quantities.

We are going to represent the amplitude and the phase of the field
$(\mathbf{E},\mathbf{B})$,
making use of this matrix. In order to get the same above given value for
the amplitude ${\cal A}$ of the field in these terms we can define it by
$$
{\cal A}=\frac{1}{\alpha^2}\sqrt{\mathfrak{R}_{ijkl}\mathfrak{R}^{ijkl}-2}=
$$
$$
\frac{1}{\alpha^2}\sqrt{\alpha^2(\mathbf{E}^2+\mathbf{B}^2)+
\alpha^4(\mathbf{E}\times\mathbf{B}).(\mathbf{E}\times\mathbf{B})-2}=
|\mathbf{E}\times\mathbf{B}|.
$$
In order to define the phase we first consider the 2-form
$tr\circ\mathfrak{R}$. We get
$$
tr\circ\mathfrak{R}=\alpha E_1dy\wedge dz
-\varepsilon\alpha B_2 dz\wedge dx -
\varepsilon\alpha ^2 (\mathbf{E}\times\mathbf{B})_3 dx\wedge dy.
$$
The square of this 2-form is
$$
(tr\circ\mathfrak{R})^2=\alpha
^2\Biggl[(E_1)^2+(B_2)^2+
\alpha ^2\bigl[(\mathbf{E}\times\mathbf{B})_3\bigr]^2\Biggr].
$$
Now the phase $\theta$ of the field should be defined by $$
\theta=arccos\Biggl[\pm\sqrt{\biggl|\frac
{(tr\circ\mathfrak{R})^2-1}{2}\biggr|}\ \Biggr] $$ whenever the right-hand
expression is well defined.

The above definitions are motivated by the null-field case, where the
corresponding orthonormal electromagnetic frame looks
like ($u$,$p$ are two functions)
$$
\alpha\mathbf{E}=\left(\frac{u}{\sqrt{u^2+p^2}},\frac{p}{\sqrt{u^2+p^2}},0\right), \ \
\alpha\mathbf{B}=\left(\frac{\varepsilon p}{\sqrt{u^2+p^2}},
\frac{-\varepsilon u}{\sqrt{u^2+p^2}},0\right),
$$
$$
\ \ \alpha^2\mathbf{E}\times\mathbf{B}=(0,0,-\varepsilon), \ \ \
J_1=\frac{2u}{\sqrt{u^2+p^2}}+1.
$$ So,
for this case for the phase $\theta$ we obtain
$\theta=arccos\frac{u}{\sqrt{u^2+p^2}}\cdot$

We obtain that every null electromognetic field generates an isometry
linear map in every tangent and cotangent space where it is well defined, with
respect to the euclidean metric there. In fact the
corresponding linear transformation
$$ \varphi:\left(\frac{\partial}{\partial
x},\frac{\partial}{\partial y},\frac{\partial}{\partial z}\right)\rightarrow
\mathcal{R}
$$
in this case is orthogonal, has determinant $det(\varphi)=1$, and its trace
satisfies $|tr(\varphi)|\leq 3$. As is well known, the euclidean isometries in
an odd dimensional space have at least one positive eigen value, which should
be equal to 1. In the 3-dimensional case the corresponding 1-dimensional eigen
subspace is unique and, of course, invariant. So,
the isometry $\varphi$ reduces to 2-dimensional rotation, and the
corresponding rotation angle $\theta$ satisfies
$$
\cos(\theta)=\frac12[tr(\varphi)-1]=\frac{u}{\sqrt{u^2+p^2}}\cdot
$$
In our case, this invariant subspace is defined, obviously, by
$\alpha^2\mathbf{E}\times\mathbf{B}$, so, the allowed rotation is in the
2-dimensional subspace defined by the couple $(\alpha\mathbf{E},\alpha\mathbf{B})$.

\subsection{Relativistic form of Maxwell equations.}
The basic difference between the nonrelativistic and relativistic formulations
of the vacuum classical electrodynamics consists in the formal interpretation of
the time variable: while in nonrelativistic formulation the time variable $t$ has
{\it absolute}, i.e. not dependent on the frame changes, charachter, in the
relativistic formulation it becomes $\xi=ct$, a
{\it coordinate} variable, so, it is
treated formally in the same way as  the space coordinates $(x,y,z)$. Hence,
the corresponding base manifold becomes 4-dimensional, and endowed with
pseudo-euclidean metric. This pseudo-euclidean nature of the metric represents
the understanding that no motions of physical mass objects with velocities $v$
greater than or equal to the velocity of light in vacuum "$c$" are possible. So
the corresponding isometries naturally depend on a parameter $\beta^2=v^2/c^2$
allowing frame changes with $\beta<1$ only, which is consistent with the
assumption that frames consist of mass objects. This time coordinate we
shall further denote by $\xi=ct$.

This new look at the processes in nature requires new formulation of the
equations in the theory: the 3-dimensional relations in the old theory must be
made consistent with the new mathematical strcture, called {\it Minkowski
space-time}, which we shall denote here by $M$. Moreover, the
respect paid to the objectivity of the physical processes of interaction in
Nature requires coordinate free formulation of the corresponding dynamical
equations in the theory. In view of this we approach the problem as follows.

{\bf First}, we shall have in view that the free electromagnetic field has {\it
six scalar componenets}, i.e. it requires six functions of the space-time
coordinates in general for a full characterization, in standard coordinates
they are the classical components of the electric and magnetic vector feilds
(or 1-form fields).

{\bf Second}, we shall pay due respect to the classical hypotesis that the
matematical structure to be used must originate from the very Minkowski
space-time, since the time now is built in it. Formally, we have to look for a
mathematical object constructed form the tangent and cotangent objects on $M$.

{\bf Third}, classical Maxwell equations require inter-dependence between the
electric and magnetic components, so, the new formulation must take care of
this. Also, the classical hypothesis that the flows of the electric and magnetic
vector fields do not change the 3-volumes and the course of time should be
carefully reconsidered.

{\bf Fourth}, the classical understanding that the electric and magnetic
interacting partners exist consistently {\it without available non-zero
interaction energy}, as we see this in the corresponding Maxwell stress-energy
tensor, has to be correspondingly respected.

{\bf Fifth}, the classical Poynting dynamical relation between the field energy
and momentum densities requires new formal identification of the interacting
and time recognizable partners because neither the electric nor the magnetic
prerelativistic components are able separatly to carry local momentum.
Therefore, the new, relativistic, components must be identified in a manner
consistent with the understanding that each component must be able to carry,
besides energy, also momentum.

Finally, the new formulation should respect also the independence of the
Minkowski pseudo-matric and all objects constructed out of it on the field.

In view of the above mentioned features, presupposing that each new
relativistic time-recognizable component must depend on both, the electric and
magnetic classical components, we identify them as follows. In general, each
one must have six components, so, on a 4-dimensional manifold such objects are
the differential 2-forms and the 2-multivector fields. In view of the
\vskip 0.2cm
	-strong invariance of the exterior derivative "$\mathbf{d}$",

	-natural consideration of cotangent objects defined on submanifolds
of $M$ as objects on the whole $M$,

	-at least local volume-form interpretation of a differential $p$-form on
the coresponding $p$-dimensional submanifold,
\vskip 0.2cm
\noindent
we choose the first option - the differential 2-forms.

Further in the section we consider $\mathbb{R}^3$, parametrized by $(x,y,z)$, as
a submanifold of $\mathbb{R}^4$, where $\mathbb{R}^3$ is endowed with the
euclidean metric $g$, which is the "minus" of the corresponding induced by the
Minkowski pseudometric $\eta$ on $\mathbb{R}^4$, and $sign\,\eta=(-,-,-,+)$.
The volume forms in standard coordinates are correspondingly
$\omega_o^3=dx\wedge dy\wedge dz$ and $\omega_o=dx\wedge dy\wedge dz\wedge
d\xi$.

Let $\mathbf{E}$ and $\mathbf{B}$ denote the electric and magnetic
3-vectors and $\tilde{g}(\mathbf{E})$,  $\tilde{g}(\mathbf{B})$ denote
the corresponding euclidean 1-forms. We define the new
mathematical identification of the relativistic time-recognizable
and interacting  components (or, partners) of the free electromagnetic field by
means of the following two differential 2-forms:
$$
F=i(\mathbf{B})\omega_o^3+\tilde{g}(\mathbf{E})\wedge d\xi, \ \ \
G=i(\mathbf{E})\omega_o^3-\tilde{g}(\mathbf{B})\wedge d\xi.
$$
So, the component identification is:
$$
F_{ij}=(i(\mathbf{B})\omega_o^3)_{ij}=(B_3,-B_2,B_1), \ \
F_{i4}=(E_1,E_2,E_3), \ \ \  i<j=1,2,3;
$$
$$
G_{ij}=(i(\mathbf{E})\omega_o^3)_{ij}=(E_3,-E_2,E_1), \ \
G_{i4}=(-B_1,-B_2,-B_3), \ \ \  i<j=1,2,3 .
$$
Explicitly
$$
F=B_3dx\wedge dy-B_2dx\wedge dz+B_1dy\wedge
dz+E_1dx\wedge d\xi+E_2dy\wedge d\xi+E_3dz\wedge d\xi
$$
$$
G=E_3dx\wedge
dy-E_2dx\wedge dz+E_1dy\wedge dz-B_1dx\wedge d\xi-B_2dy\wedge d\xi-B_3dz\wedge
d\xi.
$$
Also, recalling the isomorphism $\tilde{\eta}: T(M)\rightarrow T^*(M)$ and its
opposite $\tilde{\eta}^{-1}$, then  $\mathbf{E}$ and $\mathbf{B}$ may be
considered as vector fields on $M$ as follows:
$$
\mathbf{E}=-\tilde{\eta}^{-1}\circ
i\left(\frac{\partial}{\partial \xi}\right)F, \ \
\mathbf{B}=\tilde{\eta}^{-1}\circ i\left(\frac{\partial}{\partial
\xi}\right)G .
$$
As it is seen, we may identify the vector field components of
$(\mathbf{E},\mathbf{B})$ considered as vector fields on $\mathbb{R}^3$ and on
$M$, so, $\tilde{\eta}(\mathbf{E})=-\tilde{g}(\mathbf{E})$ and
$\tilde{\eta}(\mathbf{B})=-\tilde{g}(\mathbf{B})$.
\vskip 0.2cm
{\bf Proposition}. The two differential 2-forms $F$ and $G$ are interconnected
by the $\eta$-generated Hodge $*$-operator according to $G=*F$.
\vskip 0.2cm
{\it Proof}. It easily verified making use of the action of $*$-operator on the
canonical basis 2-forms of $\Lambda^2(M)$ (Sec.5.4).
\vskip 0.2cm
Now, the basic equations for $F$ and $G=*F$ on regions where they are well
defined, i.e., away from charges and other continuous physical objects,
were given by H.Minkowski ({\it www.minkowskiinstitute.org}), and in modern
notation they read $$ \mathbf{d}F=0, \ \ \ \mathbf{d}*F=0. $$ So, the
time-recognizable two subsystems of a free electromagnetic field are
relativistically described by two closed differential 2-forms on Minkowski
space-time. Since the components of $F$ and $*F$ are just differently arranged
components of $\mathbf{E}$ and $\mathbf{B}$ it is naturally to expect that the
space-time change of each will cause change of the other, provided the above
equations hold. Locally, we can always find two interdependent 1-forms
$\mathcal{A}$ and $\mathcal{A}^*$ such that $$ F=\mathbf{d}\mathcal{A}, \ \ \
*F=\mathbf{d}\mathcal{A}^* . $$ This possibility should not be considered as
too arbitrary and not physically motivated. Any physically justified relations
come from experimantal observations and, therefore, shall be formulated in
terms of {\it admissible stress-energy-momentum changes}, i.e. changes which do
NOT lead to destruction of the field.

According to the properties of the $*$-operator (Sec.5.4) we always have the
relation $*_{(4-p)}\circ *_p=-(-1)^{p(4-p)}id$. So, for $p=2$ we obtain
$*_2\circ *_2=-id_{\Lambda^2(M)}$. In order always to have in mind some
elementary properties of the Hodge $*$ we shall give them in a form of three
corollaries: \vskip 0.2cm
{\bf Corollary.} The Hodge $*$-operator on Minkowski space-time when
reduced to the space $\Lambda^2(M)$ of differential 2-forms generates complex
structure.
\vskip 0.2cm
{\bf Corollary.} The relativistic equations for a free electromagnetic field
are invariant with respect to the transformation $F\rightarrow *F$.
\vskip 0.2cm
{\bf Corollary.} The relativistic equations for a free electromagnetic field
are conformally invariant.
\vskip 0.2cm
\noindent
This third corollary holds since $*_2$ is conformally invariant, i.e. a
conformal change of $\eta$ through an arbitraray function : $\eta\rightarrow
exp(f).\eta$ gives the same $*_2$.

The required space-time recognizability of $F$ and $*F$ and the conformal
invariance of $*_2$ allow to look at the mathematical identificaion of a free
electromagnetic field in the frame of relativistic consideration as a
$\mathbb{R}^2$-valued equivariant differential 2-form $\Omega$ in the
following sense. Define the object $\Omega$ by
$$
\Omega=F\otimes e_1+(*F)\otimes e_2,
$$
where $(e_1,e_2)$ are the standard basis vectors in $\mathbb{R}^2$. Since
$\mathbb{R}^2$ carries the standard complex structure
$J:\mathbb{R}^2\rightarrow\mathbb{R}^2$, given by
$$
J(e_1)=e_2, \ \ J(e_2)=-e_1,
$$
we obtain
$$
(*,J)\Omega=(*F)\otimes J(e_1)+(**F)\otimes J(e_2)=
*F\otimes e_2+(-F)\otimes (-e_1)=\Omega,
$$
i.e. $\Omega$ is $(*,J)$-equivarant: $(*,J)\Omega =\Omega$.
This property can also be written as $*\Omega=J^{-1}\circ\Omega$. In terms of
$\Omega$ the above two equations may be written as
$$
\mathbf{d}\Omega=0.
$$

The recognizability of the two components of the field is guaranteed as
follows. Note, that the equation $aF+b*F=0$ requires $a=b=0$. In fact,
if $a\neq 0$ then $F=-\frac{b}{a}*F$. From $aF+b*F=0$ we get $a*F-bF=0$ and
substituting $F$, we obtain $(a^2+b^2)*F=0$, which is possible only if
$a=b=0$ since $*F\neq 0$ and $*$ is isomorphism. In other words, $F$ and $*F$
are lineary independent.

Let now $(k_1,k_2)$ be another basis of $\mathbb{R}^2$ and let's consider the
2-form $\Psi=G\otimes k_1+*G\otimes k_2$. We express $(k_1,k_2)$ through
$(e_1,e_2)$ and obtain
$$
G\otimes k_1+*G\otimes k_2=G\otimes(ae_1+me_2)+*G\otimes (be_1+ne_2)=
$$
$$
=(aG+b*G)\otimes e_1+(mG+n*G)\otimes e_2=(aG+b*G)\otimes e_1+
*(aG+b*G)\otimes e_2.
$$
Consequently, $mG+n*G=a*G-bG$, i.e. $(b+m)G+(n-a)*G=0$, which
requires $m=-b,\ n=a$, i.e., the transformation matrix $A$ is
$$
A=\left| \begin{matrix} a &b\\
	               -b &a
\end{matrix}\right|.
$$
This matrix is a symmetry of $J: J=A.J.A^{-1}$, so the class of admissible bases
in $\mathbb{R}^2$ must satisfy the condition to be an orbit of the group of
invariance of $J$ through the canonical basis $(e_1,e_2)$.

Additional requirement comes from physics as follows.
In order to come to it,
we recall the product of 2 vector valued differential forms. If
$\Phi$ and $\Psi$ are respectively $p$ and $q$ forms on the same manifold $N$,
taking values in the vector spaces $W_1$ and $W_2$ with corresponding bases
$(e_1,...,e_m)$ and $(k_1,...,k_n)$, and
$\varphi :W_1\times W_2\rightarrow W_3$ is a bilinear map into the vector
space $W_3$, then a $(p+q)$-form $\varphi\left(\Phi,\Psi\right)$ on $N$ with
values in $W_3$ is defined by
$$
\varphi\left(\Phi,\Psi \right)=\sum_{i,j}\Phi^i\wedge \Psi^j \otimes \varphi(e_i,k_j).
$$
In particular, if $W_1=W_2$ and $W_3=\mathbb{R}$, and the bilinear map is
scalar (inner) product $g$, we get
$$
\varphi\left(\Phi,\Psi \right)=\sum_{i,j}\Phi^i\wedge \Psi^j g_{ij}.
$$

We introduce now the stress-energy-momentum tensor ${\bf Q}$ of the field,
considered as a symmetric 2-form on $M$ as follows:
$$
\left(\mathbf{Q}\otimes e_1\wedge e_2\right)(X,Y)=
\frac{1}{2}*\wedge\Bigl(i_X\Omega,*i_YJ_*\Omega\Bigr),
$$
where $X$
and $Y$ are two arbitrary vector fields, $i_X$ is the interior product by the
vector field $X$, and $J_*\Omega=F\otimes J(e_1)+(*F)\otimes J(e_2)$.
We obtain

$$
i_X\Omega=X^\mu F_{\mu\nu}dx^\nu\otimes e_1+
X^\mu (*F)_{\mu\nu}dx^\nu\otimes e_2,\
$$

$$
*i_YJ_*\Omega= *\Bigl[Y^\mu F_{\mu\nu}dx^\nu\Bigr]\otimes e_2-
*\Bigl[Y^\mu (*F)_{\mu\nu}dx^\nu\Bigr]\otimes e_1.
$$

$$
\wedge\Bigl(i_X\Omega,*i_Y{\cal I}_*\Omega\Bigr)=
-X^\mu Y^\nu\Bigl[F_{\mu\sigma}F_\nu^\sigma +
(*F)_{\mu\sigma}(*F)_\nu^\sigma\Bigr]dx\wedge dy\wedge dz\wedge
d\xi \otimes e_1\wedge e_2.
$$
Finally,
$$
\frac12*\wedge(i_X\Omega,*i_YJ_*\Omega)=
X^\mu Y^\nu\Bigl[-\frac12F_{\mu\sigma}F_\nu^\sigma -
\frac12(*F)_{\mu\sigma}(*F)_\nu^\sigma\Bigr]
e_1\wedge e_2.
$$
The symmetric tensor in the brackets is by definition the
stress-energy-momentum tensor of the field.

{\bf Remark:} Here and further the greek indices run form 1 to 4.

Let now $\Omega$ be of the above kind, $Q_{\mu\nu}$ be the stress-energy-momentum
tensor introduced above,
and $g$ be the canonical euclidean inner product in $\mathbb{R}^2$. Then
the class of bases in $\mathbb{R}^2$ we shall use will be required to satisfy also the
following equation
$$ Q_{\mu\nu}X^\mu Y^\nu=\frac12
*g\Bigl(i(X)\Omega,*i(Y)\Omega\Bigr).
$$
We develop the right hand side of this
equation and obtain
$$
\frac12 *g\Bigl(i(X)\Omega,*i(Y)\Omega\Bigr)
$$
$$
=\frac12 *g\Bigl(i(X)F\otimes e_1+i(X)*F\otimes e_2,*i(Y)F\otimes
e_1+*i(Y)*F\otimes e_2\Bigr)
$$
$$
=\frac12*\biggl[\Bigl(i(X)F\wedge*i(Y)F\Bigr)g(e_1,e_1)+
\Bigl(i(X)F\wedge*i(Y)*F\Bigr)g(e_1,e_2)+
$$
$$
+\Bigl(i(X)*F\wedge*i(Y)F\Bigr)g(e_2,e_1)+
\Bigl(i(X)*F\wedge*i(Y)*F\Bigr)g(e_2,e_2)\biggr]
$$
$$
=-\frac12 X^\mu Y^\nu\biggl[F_{\mu\sigma}F_\nu^\sigma g(e_1,e_1)+
(*F)_{\mu\sigma}(*F)_\nu^\sigma g(e_2,e_2)+
$$
$$
+\Bigl(F_{\mu\sigma}(*F)_\nu^\sigma+(*F)_{\mu\sigma}F_\nu^\sigma\Bigr)g(e_1,e_2)\biggr]=
-\frac12 X^\mu Y^\nu\biggl[F_{\mu\sigma}F_\nu^\sigma +
(*F)_{\mu\sigma}(*F)_\nu^\sigma\biggr].
$$
In order this relation to hold it is necessary to have
$$
g(e_1,e_1)=1,\ g(e_2,e_2)=1,\ g(e_1,e_2)=0,
$$
i.e., we have to use {\it orthonormal} bases in $\mathbb{R}^2$. If, however, by
some reasons, we make use of nonorthonormal basis, we see that interaction
stress-energy-momentum between the two components $F$ and $*F$ of the form
$$
F_{\mu\sigma}(*F)_\nu^\sigma+(*F)_{\mu\sigma}F_\nu^\sigma=
2\,F_{\mu\sigma}(*F)_\nu^\sigma=2\,F_{\nu\sigma}(*F)_\mu^\sigma
$$
$$=\frac12F_{\alpha\beta}(*F)^{\alpha\beta}\eta_{\mu\nu}=
2\,\mathbf{E}.\mathbf{B}\eta_{\mu\nu}
$$
will appear and has to be taken into account.

\section{Conserved Quantities}
\subsection{Some relations on Minkowski space-time}

	{\bf 1.} If $\alpha$ is 1-form on $M$ and $F$ is a 2-form on $M$ the
the following relation holds:
$$
*(\alpha\wedge *F)=-\alpha^\mu F_{\mu\nu}dx^\nu.
$$
We have:
$$
\alpha=\alpha_1dx+\alpha_2dy+\alpha_3dz+\alpha_4d\xi.
$$
$$
*F=-F_{12}dz\wedge d\xi+F_{13}dy\wedge d\xi-F_{23}dx\wedge d\xi+
F_{14}dy\wedge dz-F_{24}dx\wedge dz+F_{34}dx\wedge dy.
$$
We obtain:
\begin{eqnarray*}
\alpha\wedge *F
&=&(\alpha_1F_{14}+\alpha_2F_{24}+\alpha_3F_{34})dx\wedge
dy\wedge dz\\&+&(\alpha_1F_{13}+\alpha_2F_{23}+\alpha_4F_{34})dx\wedge dy\wedge
d\xi\\&+&(-\alpha_1F_{12}+\alpha_3F_{23}-\alpha_4F_{24})dx\wedge dz\wedge
d\xi\\&+&(-\alpha_2F_{12}-\alpha_3F_{13}+\alpha_4F_{14})dy\wedge dz\wedge
d\xi
\end{eqnarray*}
\begin{eqnarray*}
*(\alpha\wedge *F)&=&(-\alpha^2F_{21}-\alpha^3F_{31}-\alpha^4F_{41})dx\\&+&
(-\alpha^1F_{12}-\alpha^3F_{32}-\alpha^4F_{42})dy\\&+&
(-\alpha^1F_{13}-\alpha^2F_{23}-\alpha^4F_{43})dz\\&+&
(-\alpha^1F_{14}-\alpha^2F_{24}-\alpha^3F_{34})d\xi=-\alpha^\mu
F_{\mu\nu}dx^\nu. \end{eqnarray*}
{\bf Remark}: In the euclidean case we obtain
$*(\alpha\wedge *F)=\alpha^\mu F_{\mu\nu}dx^{\nu}$.
\vskip 0.2cm
{\bf 2.} Let now $F$ be a 2-form and $G$ be a 3-form, so $*G$ will be a 1-form.
Making use of the above relation we obtain:
$$
*(F\wedge *G)=*(*G\wedge F)
=-*(*G\wedge **F)=(*G)^\nu(*F)_{\nu\mu}dx^\mu. $$ On the other hand, since $$
*G=G_{123}d\xi+G_{124}dz-G_{134}dy+G_{234}dx'
$$
we obtain
\begin{eqnarray*}
*(F\wedge *G)&=&(F^{23}G_{231}+F^{24}G_{241}+F^{34}G_{341})dx\\
&+&(F^{34}G_{342}+F^{14}G_{142}+F^{13}G_{132})dy\\
&+&(F^{12}G_{123}+F^{14}G_{143}+F^{24}G_{243})dz\\
&+&(F^{12}G_{124}+F^{13}G_{134}+F^{23}G_{234})dz=
\frac12F^{\mu\nu}G_{\mu\nu\sigma}dx^\sigma.
\end{eqnarray*}
\vskip 0.2cm
{\bf 3.} If $G,F$ are 2-forms, $G=*F$, and recalling that the coderivative
$\delta$ in the case of Minkowski space-time satisfies
$\delta=*\,\mathbf{d}\,*$ we obtain
\begin{eqnarray*}
*(F\wedge *\mathbf{d}F)&=&-*(F\wedge *\mathbf{d}**F)
=-*(F\wedge\delta*F)=-*(\delta *F\wedge F)\\
&=&*(\delta *F\wedge **F)
=-(\delta *F)^\nu (*F)_{\nu\mu}dx^\mu=
\frac12F^{\mu\nu}(\mathbf{d}F)_{\mu\nu\sigma}dx^\sigma.
\end{eqnarray*}
Also, replacing $F$ with $*F$ in this relation we obtain
$$
*(*F\wedge *\mathbf{d}*F)=-(\delta F)^\nu F_{\nu\mu}dx^\mu=
\frac12(*F)^{\mu\nu}(\mathbf{d}*F)_{\mu\nu\sigma}dx^\sigma.
$$
Finally, we also have
$$
*(\delta F\wedge F)=(\delta
F)^\nu(*F)_{\nu\mu}dx^\mu=\frac12F^{\mu\nu}(\mathbf{d}*F)_{\mu\nu\sigma}dx^\sigma,
$$
$$
*(\delta *F\wedge *F)=-(\delta *F)^\nu F_{\nu\mu}dx^\mu=
-\frac12(*F)^{\mu\nu}(\mathbf{d}F)_{\mu\nu\sigma}dx^\sigma .
$$
Composing the coderivative operator
$\delta_p=(-1)^{p}*^{-1}\mathbf{d}*_p$ from the left
and from the right correspondingly with $*_{p-1}$ and $*_{p}^{-1}$ we obtain
the following commutation relations (the metric here is denoted by $g$):
$$ *_{p-1}\circ
\delta_p=(-1)^{ind(g)+p}\mathbf{d}\circ *_p, $$ $$ \delta\circ
*_p=(-1)^{p+1}*_{p+1}\circ\,\mathbf{d}_p\rightarrow \delta\circ
*_2=-*\circ\,\mathbf{d}_2. $$ \vskip 0.3cm
	{\bf 4.} We give now the corresponding 3-dimensional form of these
relations, assuming that the two relativistic 2-forms $(F,*F)$ are expressed respectively
by the space-like vectors: $(\mathbf{E},\mathbf{B})$
%\newpage
\begin{eqnarray*}
\delta F&=&\left[-\left(\mathrm{rot}\mathbf{B}-\frac{\partial
\mathbf{E}}{\partial \xi}\right), \mathrm{div}\mathbf{E}\right],\\
\delta *F&=&\left[-\left(\mathrm{rot}\mathbf{E}+\frac{\partial
\mathbf{B}}{\partial \xi}\right), -\mathrm{div}\mathbf{B}\right],\\
F_{\mu\nu}(\delta F)^\nu dx^\mu
&=&\frac12(*F)^{\mu\nu}(\mathbf{d}*F)_{\mu\nu\sigma}dx^\sigma\\
&=&\left[\left(\mathrm{rot}\mathbf{B}-\frac{\partial \mathbf{E}}{\partial
\xi}\right)\times\mathbf{B}+ \mathbf{E}\,\mathrm{div}\mathbf{E},
\,-\mathbf{E}.\left(\mathrm{rot}\mathbf{B}-\frac{\partial \mathbf{E}}{\partial
\xi}\right)\right],\\
(*F)_{\mu\nu}(\delta *F)^\nu dx^\mu
&=&\frac12F^{\mu\nu}(\mathbf{d}F)_{\mu\nu\sigma}dx^\sigma\\
&=&\left[\left(\mathrm{rot}\mathbf{E}+\frac{\partial \mathbf{B}}{\partial
\xi}\right)\times\mathbf{E}
+ \mathbf{B}\,\mathrm{div}\mathbf{B},
\,\mathbf{B}.\left(\mathrm{rot}\mathbf{E}+\frac{\partial \mathbf{B}}{\partial
\xi}\right)\right]\\
F_{\mu\nu}(\delta *F)^\nu dx^\mu
&=&-\frac12(*F)^{\mu\nu}(\mathbf{d}F)_{\mu\nu\sigma}dx^\sigma\\
&=&\left[\left(\mathrm{rot}\mathbf{E}+\frac{\partial \mathbf{B}}{\partial
\xi}\right)\times\mathbf{B}- \mathbf{E}\,\mathrm{div}\mathbf{B},
\,-\mathbf{E}.\left(\mathrm{rot}\mathbf{E}+\frac{\partial \mathbf{B}}{\partial
\xi}\right)\right]\\
(*F)_{\mu\nu}(\delta F)^\nu dx^\mu
&=&-\frac12F^{\mu\nu}(\mathbf{d}*F)_{\mu\nu\sigma}dx^\sigma\\
&=&\left[\left(\mathrm{rot}\mathbf{B}-\frac{\partial \mathbf{E}}{\partial
\xi}\right)\times\mathbf{E}-\mathbf{B}\,\mathrm{div}\mathbf{E},
\,\mathbf{B}.\left(\mathrm{rot}\mathbf{B}-\frac{\partial \mathbf{E}}{\partial
\xi}\right)\right] .
\end{eqnarray*}
\vskip 0.3cm
	{\bf 5.} We give some additional relations.
Let again $F$ and $G$ be two 2-forms on Minkowski space-time, then we recall
the relations
$$
F\wedge *G=-\frac12F_{\mu\nu}G^{\mu\nu}\omega_o
=-\frac12G_{\mu\nu}F^{\mu\nu}\omega_o=-\frac12F_{\mu\nu}G^{\mu\nu}\,dx\wedge
dy\wedge dz\wedge d\xi.
$$
$$
F\wedge G=-F\wedge **G=\frac12F_{\mu\nu}(*G)^{\mu\nu}\omega_o.
$$
If $X$ is an arbitrary vector field we obtain
$$
i_X(F\wedge *G)=i_XF\wedge *G+F\wedge
i_X*G=-\frac12F_{\mu\nu}G^{\mu\nu}i_X\omega_o.
$$
\begin{eqnarray*}
*i_X(F\wedge *G)&=&*(i_XF\wedge *G)-*(i_X*G\wedge**F)\\
&=&-X^\sigma F_\sigma\,^\nu
G_{\nu\mu}dx^\mu+X^\sigma(*G)_\sigma\,^\nu(*F)_{\nu\mu}dx^\mu\\
&=&-X^\sigma\Big[F_\sigma\,^\nu
G_{\nu\mu}-(*G)_\sigma\,^\nu(*F)_{\nu\mu}\Big]dx^\mu=
-\frac12F_{\mu\nu}G^{\mu\nu}*i_X\omega_o.
\end{eqnarray*}
\begin{eqnarray*}
*i_X\omega_o&=&X^1dx+X^2dy+X^3dz-X^4d\xi\\
&=&-X_1dx-X_2dy-X_3dz-X_4d\xi=-X_\mu dx^\mu=-\eta_{\mu\nu}X^\mu dx^\nu.
\end{eqnarray*}
{\bf Remark}: For the euclidean case we obtain also
$*i_X\omega_o=-\eta_{\mu\nu}X^\mu dx^\nu$.

So, the antisymmetries $F_{\mu\nu}=-F_{\nu\mu}$ and $G_{\mu\nu}=-G_{\nu\mu}$
lead to
$$
*i_X(F\wedge *G)=X^\sigma\big[
F_\sigma\,^\nu G_{\mu\nu}-(*G)_\sigma\,^\nu (*F)_{\mu\nu}\big]dx^\mu.
$$
Since $X$ is arbitrary we obtain the well known identity
$$
\frac12F_{\alpha\beta}G^{\alpha\beta}\eta_{\mu\nu}=
F_\mu\,^\sigma G_{\nu\sigma}-(*G)_\mu^\sigma\,(*F)_{\nu\sigma}, \ \ \text{or}
$$
$$
\frac12F_{\alpha\beta}G^{\alpha\beta}\delta_\mu^\nu=
F_{\mu\sigma} G^{\nu\sigma}-(*G)_{\mu\sigma}(*F)^{\nu\sigma}.
$$

Substituting $G=F$ and $G=*F$ we obtain
$$
\frac12F_{\alpha\beta}F^{\alpha\beta}\eta_{\mu\nu}=
F_\mu\,^\sigma F_{\nu\sigma}-(*F)_\mu\,^\sigma (*F)_{\nu\sigma}=
(\mathbf{B}^2-\mathbf{E}^2)\eta_{\mu\nu}=I_1\eta_{\mu\nu}
$$
$$
\frac12F_{\alpha\beta}(*F)^{\alpha\beta}\eta_{\mu\nu}=
2F_{\mu\sigma}(*F)_\nu\,^\sigma=2F_{\nu\sigma}(*F)_\mu\,^\sigma
=2\mathbf{E}.\mathbf{B}\eta_{\mu\nu}=I_2\eta_{\mu\nu}.
$$
\vskip 0.3cm
	{\bf 6.}
Finally, recalling the stress-energy-momentum tensor $Q_{\mu\nu}$
\index{eigen relations for energy tensor}
 for the field
$(F,*F)$, and making use of the above relations we easily obtain the
important and useful Rainich relation (see the proof in Sec.8.1):
$$
Q_{\mu\sigma}Q^{\nu\sigma}=\frac14\left[\left(\frac12F_{\alpha\beta}F^{\alpha\beta}\right)^2+
\left(\frac12F_{\alpha\beta}(*F)^{\alpha\beta}\right)^2\right]\delta_\mu^\nu, \
\ \ \rightarrow \ \ \ Q_{\mu\nu}Q^{\mu\nu}=I_1^2+I_2^2.
$$
From these Rainich relations it directly follows that the eigen values
of the stress-energy-momentum $Q_\mu^\nu$ are
$$
\lambda_{1,2}=\pm\frac12\sqrt{I_1^2+I_2^2}.
$$

For the eigen value equations for $F$ and $*F$
$$
det||F_\mu^\nu-\lambda\delta_\mu^\nu||=0, \ \
det||(*F)_\mu^\nu-\lambda^*\delta_\mu^\nu||=0
$$
we obtain correspondingly
$$
\lambda^4+I_1\lambda^2-\frac14I_2^2=0, \ \ \
(\lambda^*)^4-I_1(\lambda^*)^2-\frac14I_2^2=0.
$$
The eigen values are
$$
\lambda_{1,2}=\pm \sqrt{-\frac12 I_1 +\frac12 \sqrt{I_1^2+I_2^2}} ,\quad
\lambda_{3,4}=\pm \sqrt{-\frac12 I_1 -\frac12 \sqrt{I_1^2+I_2^2}} ,
$$
$$
\lambda^*_{1,2}=\pm \sqrt{\frac12 I_1 +\frac12 \sqrt{I_1^2+I_2^2}} ,\quad
\lambda^*_{3,4}=\pm \sqrt{\frac12 I_1 -\frac12 \sqrt{I_1^2+I_2^2}} .
$$
If we denote for a while the eigen values of $Q_\mu^\nu$ by $\gamma$, the
following relations between $(\gamma,\lambda)$ and $(\gamma,\lambda^*)$ exist:
$$
\gamma=\left[\frac12I_1+\lambda^2\right], \ \
\gamma=\left[-\frac12I_1+(\lambda^*)^2\right] .
$$
The formula at the end of Sec.1.5.1 gives the relations:
$$
det\|F_{\mu\nu}\|=det\|(*F)_{\mu\nu}\|=\frac14 (I_2)^2,\quad
det\|(F\pm*F)_{\mu\nu}\|=(I_1)^2 .
$$

{\subsection {Conservation and dynamics}
We are going to consider here what conservation laws one may obtain if the
field is mathematically identified by a vector bundle valued differential form
on the base space under the following conditions.

	1. The base space $M^n$ is endowed with a (pseudo)riemannian metric
$g$, and the corresponding Levi-Civita covariant derivative $\nabla$.

	2. The vector bundle, denoted by $\tau$, is real, $r$-dimensional and
is endowed with riemannian metric $\langle\,,\rangle$, and the corresponding
exterior covariant derivative $\mathbf{D}$ and coderivative $\mathcal{D}$.

The $\tau$-valued differential p-forms will be denoted by
$\Lambda^p(M,\tau)=\Lambda^p(M)\otimes Sec(\tau))$, and the
$\mathcal{J}(M)$-module $Sec(\tau)$ may be denoted sometimes by
$\Lambda^0(M,\tau)$.

Making use of the notations $\Phi,\Psi\in \Lambda(M,\tau)$,
$\alpha\in\Lambda^p(M); \sigma,\rho\in Sec(\tau);
X\in\mathfrak{X}(M)$, and $\omega_o$ denotes the $g$-generated volume form on
the base manifold $M$, we recall the relations:
\begin{eqnarray*}
&&\langle\langle
\Phi,\Psi\rangle\rangle=
\langle\langle\alpha\otimes\sigma,\beta\otimes\rho\rangle\rangle=
\alpha\wedge\beta\langle\sigma,\rho\rangle, \ \langle\langle
\Phi,*\Phi\rangle\rangle \in\Lambda^n(M),\\
&&\mathbf{d}\langle\rho,\sigma\rangle=
\langle\langle\mathbf{D}\sigma,\rho\rangle\rangle +
\langle\langle\sigma,\mathbf{D}\rho\rangle\rangle,\\
&&\mathbf{D}(\alpha\otimes\sigma)=\mathbf{d}\alpha\otimes\sigma+
(-1)^p\alpha\wedge\mathbf{D}\sigma, \ \ \alpha\in\Lambda^p(M).
\end{eqnarray*}
The covariant coderivative $\mathcal{D}$ and the covariant Lie derivative
$\mathcal{L}_X$ are given by
$$
\mathcal{D}_p=(-1)^{ind(g)+np+n+1}*\mathbf{D}*_p=(-1)^p*^{-1}\mathbf{D}*_p,
$$
$$
\mathcal{L}_X\Phi=\mathbf{D}i_X\phi+i_X\mathbf{D}\Phi,
$$
where $i_X\Phi=i_X(\alpha\otimes\sigma)=(i_X\alpha)\otimes\sigma$, and
$*\Phi=*(\alpha\otimes\sigma)=(*\alpha)\otimes\sigma$.
Clearly,
$$
\mathcal{L}_X\Phi=\mathcal{L}_X(\alpha\otimes\sigma)=(L_X\alpha)\otimes\sigma+
\alpha\otimes\mathbf{D}_X\sigma.
$$
Moreover,
$$
L_X\langle\langle\Phi,\Psi\rangle\rangle=
\langle\langle\mathcal{L}_X\Phi,\Psi\rangle\rangle+
\langle\langle\Phi,\mathcal{L}_X\Psi\rangle\rangle.
$$

Let now our field be represented by $\Phi\in\Lambda^p(M,\tau)$. We note
that now $\mathbf{D}$ and $\mathcal{D}$ will respect the same commutation
relations with the riemannian Hodge-$*$. We consider a lagrangian
$\mathfrak{L}$ representing the $\langle\,,\rangle$-flow of $\tilde{g}(\Phi)$
across $*\Phi$:
$$
\mathfrak{L}=\langle\langle\Phi,*\Phi\rangle\rangle
\in\Lambda^n(M).
$$
Let our field propagate along the (arbitrary) vector field
$X\in\mathfrak{X}(M)$. We want to see how the lagrangian $\mathfrak{L}$ changes
along $X$. \begin{eqnarray*}
L_X\langle\langle\Phi,*\Phi\rangle\rangle&=&\langle\langle\mathcal{L}_X\Phi,*\Phi\rangle\rangle+
\langle\langle\Phi,\mathcal{L}_X*\Phi\rangle\rangle\\
&=&\langle\langle\mathcal{L}_X\Phi,*\Phi\rangle\rangle+
\langle\langle\Phi,*\mathcal{L}_X\Phi\rangle\rangle+
\langle\langle\Phi,[\mathcal{L}_X,*]\Phi\rangle\rangle\\
&=&2\langle\langle\mathcal{L}_X\Phi,*\Phi\rangle\rangle+
\langle\langle\Phi,[\mathcal{L}_X,*]\Phi\rangle\rangle,
\end{eqnarray*}
where $[\mathcal{L}_X,*]=\mathcal{L}_X\circ *-*\circ\mathcal{L}_X$. Further we
obtain
\begin{eqnarray*}
\langle\langle\mathcal{L}_X\Phi,*\Phi\rangle\rangle&=&
\langle\langle i_X\mathbf{D}\Phi+\mathbf{D}i_X\Phi,*\Phi\rangle\rangle\\
&=&\langle\langle i_X\mathbf{D}\Phi,*\Phi\rangle\rangle+
\langle\langle\mathbf{D}i_X\Phi,*\Phi\rangle\rangle\\
&=&(-1)^p\langle\langle\mathbf{D}\Phi,i_X*\Phi\rangle\rangle+
\mathbf{d}\langle\langle i_X\Phi,*\Phi\rangle\rangle+
(-1)^p\langle\langle i_X\Phi,\mathbf{D}*\Phi\rangle\rangle.
\end{eqnarray*}
\begin{eqnarray*}
(-1)^p\langle\langle i_X\Phi,\mathbf{D}*\Phi\rangle\rangle&=&
(-1)^p\langle\langle i_X\Phi,**^{-1}\mathbf{D}*\Phi\rangle\rangle=
\langle\langle i_X\Phi,*\mathcal{D}\Phi\rangle\rangle,\\
(-1)^p\langle\langle\mathbf{D}\Phi,i_X*\Phi\rangle\rangle&=&
(-1)^{pn+n+1}\langle\langle i_X*\Phi,\mathbf{D}\Phi\rangle\rangle\\
&=&(-1)^{pn+n+1}\langle\langle
i_X*\Phi,\mathbf{D}*^{-1}*\Phi\rangle\rangle\\&=&
(-1)^{ind(g)+pn+n+1}\langle\langle i_X*\Phi,*\mathcal{D}*\Phi\rangle\rangle.
\end{eqnarray*}
On the other hand,
$L_X\langle\langle\Phi,*\Phi\rangle\rangle=
\mathbf{d}[(*\langle\langle\Phi,*\Phi\rangle\rangle)i_X\omega_o]$, so,
%We have obtained the identity:
$$
\mathbf{d}\left[\frac12(*\langle\langle\Phi,*\Phi\rangle\rangle)i_X\omega_o-
\langle\langle i_X\Phi,*\Phi\rangle\rangle\right]=
\frac12\langle\langle\Phi,[\mathcal{L}_X,*]\Phi\rangle\rangle
$$
$$
+(-1)^{ind(g)+pn+n+1}\langle\langle i_X*\Phi,*\mathcal{D}*\Phi\rangle\rangle+
\langle\langle i_X\Phi,*\mathcal{D}\Phi\rangle\rangle.
$$
On the left hand side of this identity stays an exact $n$-form, so if the
right hand side becomes zero, we can in principle have conserved
integral quantities provided the field functions and the other
participating object components generate integrable integrands. The very
physical interpretation could come from appropriate interpretation of the
vector field $X$.

The first requirment would be $[\mathcal{L}_X,*]=0$, and since this
requirment is equivalent to $[L_X,*]=0$, then in order with every local
symmetry of the corresponding Hodge $"*"$ to
associate a conserved quantity it is sufficient to require
$$
 \langle\langle i_X*\Phi,*\mathcal{D}*\Phi\rangle\rangle=0, \ \ \
\langle\langle i_X\Phi,*\mathcal{D}\Phi\rangle\rangle=0,
$$
i.e., the two componenets $\Phi$ and $*\Phi$ of the field to have the same
relation to the local symmetry of the Hodge $*$ proposed by the vector field
$X$.
 The component form of these equations is
$$
X^\mu(*\Phi)^a_{\mu\nu_1...\nu_{n-p-1}}(\mathcal{D}*\Phi)_a^{\nu_1...\nu_{n-p-1}}=0, \ \ \ \
X^\mu\Phi^a_{\mu\nu_1...\nu_{n-p-1}}\mathcal{D}\Phi_a^{\nu_1...\nu_{n-p-1}}=0.
$$
Equivalently, omitting the $X$-participation in the above expressions, these
equations can be written as follows ($\nu_1<\nu_2<....$ here and further):
\begin{eqnarray*}
i_{\tilde{g}(\mathcal{D}*\Phi)}(*\Phi)&=&0\leftrightarrow
i_{\tilde{g}(\Phi)}\mathbf{D}\Phi=0, \ \
\text{i.e.} \ \ \
\Phi_a^{\nu_1...\nu_p}(\mathbf{D}\Phi)^a_{\mu\nu_1...\nu_p}dx^\mu=0\\
i_{\tilde{g}(\mathcal{D}\Phi)}\Phi&=&0\leftrightarrow
i_{\tilde{g}(*\Phi)}(\mathbf{D}*\Phi)=0, \ \
\text{i.e.} \ \ \
(*\Phi)_a^{\nu_1...\nu_p}(\mathbf{D}*\Phi)^a_{\mu\nu_1...\nu_p}dx^\mu=0,
\end{eqnarray*}
where the summations with respect to the bundle index $a=1,2,...,r$ and the
base manifold indices $\nu_1<\nu_2<...<\nu_p$ are supposed to be made, also,
$\tilde{g}$ acts only on the base-form components of $\Phi$
to $p$-vector components like this:
$\tilde{g}(\Phi)=\tilde{g}(\alpha\otimes\sigma)=
(\tilde{g}(\alpha))\otimes\sigma$. The equivalent component form of these
equations in terms of the covariant coderivative $\mathcal{D}$ is
$$
\tilde{g}(\mathcal{D}*\Phi)_a^{\nu_1...\nu_{n-p-1}}
(*\Phi)^a_{\mu\nu_1...\nu_{n-p-1}}dx^\mu=0, \ \ \
\tilde{g}(\mathcal{D}\Phi)_a^{\nu_1...\nu_{p-1}}
\Phi^a_{\mu\nu_1...\nu_{p-1}}dx^\mu=0.
$$
Making use of the relations $*\,\mathbf{D}_{n-1}=(-1)^n\mathcal{D}\,*_{n-1}$,
$*i_X\omega_o=\pm g(X)$,  we obtain also
\begin{eqnarray*}
&&*\mathbf{d}\left[\frac12(*\langle\langle\Phi,*\Phi\rangle\rangle)i_X\omega_o-
\langle\langle i_X\Phi,*\Phi\rangle\rangle\right]=\\
&&(-1)^n\delta\Big[\frac12(*\langle\langle\Phi,*\Phi\rangle\rangle)\varepsilon
g(X)- *\langle\langle i_X\Phi,*\Phi\rangle\rangle\Big], \ \ \varepsilon=\pm 1.
\end{eqnarray*}
The sign of $\varepsilon$ depends on the dimension of $M$ as well as on the
signature of the metric, for $dim\,M=4$ we have $\varepsilon=-1$.

Let's consider, for example, the case of {\it euclidean}
metric $g$, so that at any point of $M$ we may assume all $g_{kk}=1$, also let
$dim(M)=N=4n, n=1,2,...<\infty $, $\Phi=*\Phi $, so
$deg\Phi=deg(*\Phi)=2n $
and $\omega_o=dx^1\wedge dx^2\wedge...\wedge dx^N $,
the so called {\it self-dual case}.
We obtain ("hat" means ommision)
\begin{eqnarray*}
*i_X\omega_o&=&*\left(\sum_{k=1}^N(-1)^{k+1}X^kdx^1\wedge
dx^2\wedge...\wedge\hat{dx^k}\wedge...\wedge dx^N\right)\\
&=&\sum_{k=1}^N(-1)^{k+1}(-1)^{N-k}X^kdx^k=
\sum_{k=1}^N(-1)^{N+1}X^kdx^k\\&=&
(-1)^{N+1}\sum_{k=1}^Ng_{kk}\left(X^k\frac{\partial}{\partial x^k}\right)
=(-1)^{N+1}\tilde{g}(X)=-\tilde{g}(X).
\end{eqnarray*}
Consider now the expression inside the brackets. For the first term we obtain
\begin{eqnarray*}
&&\frac12(*\langle\langle\Phi,*\Phi\rangle\rangle)(-\tilde{g}(X))=\frac12\left(*\langle\langle
\Phi,*\Phi\rangle\rangle\right)*i_X\omega_o
=\frac12*(i_X\langle\langle\Phi,*\Phi\rangle\rangle)\\
&=&\frac12\Big[\langle\langle i_X\Phi,*\Phi\rangle\rangle+
\langle\langle\Phi,i_X*\Phi\rangle\rangle\Big]=
\frac12\Big[\langle\langle i_X\Phi,*\Phi\rangle\rangle+
\langle\langle i_X\Phi,*\Phi\rangle\rangle\Big]\\
&=&\frac12*2\langle\langle i_X\Phi,*\Phi\rangle\rangle=
\langle\langle i_X\Phi,*\Phi\rangle\rangle.
\end{eqnarray*}
So, in this special case we see that the quantity inside the brackets
at the end of the previuos page, which is an analog of the energy-momentum
tensor in the pseudo-euclidean case, is zero by {\it pure algebraic reasons},
so it could hardly represent important characteristics of self-dual fields:
$\Phi=*\Phi$.

Another example, let $M^n$ be also of dimension $n=4k, k=1,2,...<\infty$, but
$ind(g)$ is {\bf odd}, as in relativistic theories where $n=4$ and $ind\,g=3$.
(Further we denote by the same letter the $\tilde{g}$-corresponding objects.)
Observe, that in such a case the restriction of the corresponding Hodge $*$ to
$2k$-forms satisfies $*\circ *_{2k}=(-1)^{ind(g)+2k(4k-2k)}id=-id$, so,
$*_{2k}$ defines a complex structure in $\Lambda^{2k}(M)(x), x\in M$. In such a
case it is impossible to have $\Phi\in\Lambda^{2k}(M)$ to be equal to $*\Phi$,
since the equality $\Phi=*\Phi$ leads to $\Phi=-\Phi$, i.e. $\Phi=0$).
Since $i_X\Phi$ is of odd degree $(2k-1)$ now, and in view of the
easily verified relations
$$
*\langle\langle\Phi,*\Phi\rangle\rangle=(-1)^{ind(g)}g(\Phi,\Phi)*\omega_o=
-\Phi_{\nu_1...\nu_{2k}}\Phi^{\nu_1...\nu_{2k}}, \ \ \nu_1<\nu_2<...<2k,
$$
$$
*\langle\langle i_X\Phi,*\Phi\rangle\rangle=-
X^\sigma\Phi_{\sigma}\,^{\nu_1...\nu_{2k-1}}\Phi_{\nu_1...\nu_{2k-1}\mu}dx^\mu=
X^\sigma\Phi_{\sigma}\,^{\nu_1...\nu_{2k-1}}\Phi_{\mu\nu_1...\nu_{2k-1}}dx^\mu
$$
we obtain  in components
$$
\delta\Big[\frac12\Phi^a_{\nu_1...\nu_{2k}}\Phi_a^{\nu_1...\nu_{2k}}X^\sigma
g_{\sigma\mu}dx^\mu-X^\sigma\Phi^a_{\sigma\nu_1...\nu_{2k-1}}
\Phi_{a\,\mu}^{\nu_1...\nu_{2k-1}}dx^\mu\Big].
$$
Denoting now
$$
T_{\mu\nu}\equiv\frac12\Phi^a_{\nu_1...\nu_{2k}}\Phi_a^{\nu_1...\nu_{2k}}g_{\mu\nu}-
\Phi_\mu^{a\,\nu_1...\nu_{2k-1}}\Phi_{a\,\nu\nu_1...\nu_{2k-1}},
$$
for the case $X$ is local isometry so that $\nabla_\mu X_\nu+\nabla_\nu
X_\mu=0$, i.e. $\nabla_\mu X_\nu$ is antisymmetric and $L_X(*)=0$, on the
solutions of the above equations, in view of the symmetry of $T_{\mu\nu}$, we
obtain
$$
X^{\mu}\nabla_{\nu}T_\mu^\nu=0, \ \ \ \text{i.e.} \ \ \
\mathbf{d}*(X^{\mu}T_{\mu\nu}dx^{\nu})=0,
$$
which is standard relation in classical field theories on Minkowski space-time.
So, on Minkowski space-time we can always construct such closed 3-forms
$*(X^\mu T_{\mu\nu}dx^\nu)$, and to interpret correspondingly the computed
integral conserved quantities, provided the spatial 3-integrals are finite.

Making use of the easily extension to
$2k$-forms of the above identity, proved in the previous section for 2-forms on
Minkowski space, ($\nu_1<\nu_2<...$)
$$
\frac12\Phi^a_{\nu_1...\nu_{2k}}\Phi_a^{\nu_1...\nu_{2k}}\delta_\alpha^\beta=
\Phi^a_{\alpha\nu_1...\nu_{2k-1}}\Phi_a^{\beta\nu_1...\nu_{2k-1}}-
(*\Phi)^a_{\alpha\nu_1...\nu_{2k-1}}(*\Phi)_a^{\beta\nu_1...\nu_{2k-1}},
$$
we can write
$$
T_\mu^\nu=-\frac12\Big[
\Phi^a_{\mu\nu_1...\nu_{2k-1}}\Phi_a^{\nu\nu_1...\nu_{2k-1}}+
(*\Phi)^a_{\mu\nu_1...\nu_{2k-1}}(*\Phi)_a^{\nu\nu_1...\nu_{2k-1}}\Big].
$$
Assume the metric $g$ does not depend on the field, then this form of
$T_\mu^\nu$ clearly suggests the following:

	1. Our field is mathematically represented by two
recognizable components: $\Phi$ and $*\Phi$.

	2. The full stress-energy-momentum is a sum of the
stress-energy-momentum carried by each of the two components $\Phi$ and $*\Phi$.

	3. There is NO internal interaction stress-energy-momentum.

	{\bf Conclusion:}
If the two components $\Phi$ and $*\Phi$ satisfy the above equations and
exchange energy-momentum at all, then the
exchange process must realize local dynamical equilibrium: {\it each of the
components must gain locally the same energy-momentum from the other as it
gives to it locally}.

In fact, the local energy-momentum changes of the two components are given by
the flows of the two components considered as vector bundle
valued $p$-multivector fields $\Phi$ and $*\Phi$ through the corresponding
$(p+1)$-forms $\mathbf{D}\Phi$ and $\mathbf{D}*\Phi$ and are given by the
1-forms $\Phi_a^{\nu_1...\nu_p}(\mathbf{D}\Phi)^a_{\mu\nu_1...\nu_p}dx^\mu$ and
$(*\Phi)_a^{\nu_1...\nu_p}(\mathbf{D}*\Phi)^a_{\mu\nu_1...\nu_p}dx^\mu$. The
zero values of these two flows say that the ballance between loss and gain
of each component is zero.

On the other hand, the components of $\Phi$ participate in $*\Phi$ too,
so, some kind of interaction is expectable, and the important
problem is how much is the corresponding inter-exchanged energy-momentum along
both directions, from $\Phi$ to $*\Phi$ and from $*\Phi$ to $\Phi$. A natural
answer to this question-problem is to consider the flows of each
component-field $\tilde{g}(\Phi)$ and $\tilde{g}(*\Phi)$
across the generated by the other component-field $\tau$-valued
$(p+1)$-forms, respectively, $\mathbf{D}\Phi$ and $\mathbf{D}*\Phi$,
which flows, making use of the bundle metric, shall be given by
$$
\Phi_a^{\nu_1...\nu_p}(\mathbf{D}*\Phi)^a_{\mu\nu_1...\nu_p}dx^\mu,
\ \ \ (*\Phi)_a^{\nu_1...\nu_p}(\mathbf{D}\Phi)^a_{\mu\nu_1...\nu_p}dx^\mu,
\ \ p=2k.
$$
Hence, if our base manifold has a distinguished time direction so that we could
speak about dynamical behaviour of the field considered, the possible
consistent system of dynamical equations for the field represented by
$(\Phi,*\Phi)$ could read ($p=2k, \nu_1<\nu_2<...<\nu_{2k+1}$)
\begin{eqnarray*}
&&\Phi_a^{\nu_1...\nu_p}(\mathbf{D}\Phi)^a_{\mu\nu_1...\nu_p}dx^\mu=0, \\
&&(*\Phi)_a^{\nu_1...\nu_p}(\mathbf{D}*\Phi)^a_{\mu\nu_1...\nu_p}dx^\mu=0,\\
&&\Phi_a^{\nu_1...\nu_p}(\mathbf{D}*\Phi)^a_{\mu\nu_1...\nu_p}dx^\mu+
(*\Phi)_a^{\nu_1...\nu_p}(\mathbf{D}\Phi)^a_{\mu\nu_1...\nu_p}dx^\mu=0.
\end{eqnarray*}
So, $\Phi$ and $*\Phi$ are $\mathbf{D}$-autoclosed.

\subsection
{External and internal local interaction through curvature forms}
 We are going to describe a possible formal approach to local physical
interaction taking place inside a spatially distributed physical object,
formally represented by a distribution. The basic idea of the corresponding
mathematical scheme is: the initial spatial stress-strain structure of the
object to be appropriately integrable, and the internal Frobenius curvature
forms to be identified as energy-momentum transfering agents between any two
interacting, i.e., energy-momentum exchanging and time-recognizable, subsystems.
So, the mathematical concept of {\it integrability of distributions} we are
going to physically interpret as dynamical equilibrium between the physical
system and the outside world, and in this sense, guaranteeing its time
stability. If such a dynamical equilibrium exists and is time stable, i.e., if
all existence needs of the system are provided and it can keep itself
ricognizable, we could speak about {\it isolated} or {\it free} system. On the
other hand, the energy-momentum exchange between any two subsystems of our
physical field we are going to mathematically interpret in terms of the
corresponding curvature forms that can be associated with {\it available
nonintegrability} of the corresponding subdistributions.

The first thing that we have to explain seems to be {\it why distributions}?
The answer is based on the {\it dynamical nature of a
vector field}, i.e., on its ability to generate flows, or families of local
diffeomorphisms, in other words, transformations which preserve all properties
of the corresponding manifold. In the physical world we detect energy-momentum
flows from one physical system to another, and any such flow we are going to
mathematically interpret as generated locally by an appropriate vector field.
So, an isolated (in the above sense) and time-evolved physical system appears to
us as an appropriately interconnected system of such energy-momentum flows,
and this time-stable interconnection among the flows guarantees the system's
recognizability and time stability. Therefore, we consider the mathematical
concept of distribution, or differential system, on an appropriate manifold as
a good mathematical concept to start with.

Let's sketch now the formal picture. We denote by $M^n$, or just by $M$, a
$n$-dimansional real manifold. Let the vector fields $\{X_1,X_2,...,X_p\}$
define a $p$-dimensional distribution $\Delta^p(M)$ on $M$, and the
corresponding dual codistribution $\Delta^*_p(M)$ be represented by the 1-forms
$\{\alpha^1,\alpha^2,...,\alpha^p\}$. So, at every point $x\in M$ we have two
dual spaces with corresponding bases $X_i(x), i=1,2,...,p$ and $\alpha^j(x),
j=1,2,...,p$.

On the other hand, let the $(n-p)$-dimensional system of vector fields
$\Delta^{n-p}(M)=\{Y_1,Y_2,...,Y_{n-p}\}$ be such that at every point $x\in M$
the following relation to hold: $T_x(M)=\Delta^p(x)\oplus\Delta^{n-p}(x), x\in
M$. In such a case we can write also
$T^*_x(M)=\Delta^*_p(x)\oplus\Delta_{n-p}^*(x), x\in M$,
where $\Delta_{n-p}^*(x)$ is
generated by $\{\beta^1(x),\beta^2(x),...,\beta^{n-p}(x)$. Hence, we obtain
another couple of dual spaces at $x\in M$, namely,
$\{Y_1(x),Y_2(x),...,Y_{n-p}(x)\}$ and
$\{\beta^1(x),\beta^2(x),...,\beta^{n-p}(x)\}$. These objects satisfy
$$
\langle\beta^m,X_i\rangle=0, \ \ \ \langle\alpha^i,Y_m\rangle=0\,, \ \ \
i=1,...,p\, , \ \ \ m=1,...,n-p.
$$

If now our physical system is represented by
$\Delta^p(M)=\{X_1,X_2,...,X_p\}$, we always can build the other three
distributions $\Delta^{n-p}(M)$, $\Delta^*_p(M)$ and $\Delta^*_{n-p}(M)$. The
corresponding curvature forms $\Omega_{X}$ and $\Omega_{Y}$ are given by
(Sec.3.2.3) \begin{eqnarray*} &&\Omega_{(X)}=-\mathbf{d}\beta^k\otimes Y_k, \ \
k=1,...,n-p \\ &&\Omega_{(Y)}=-\mathbf{d}\alpha^i\otimes X_i, \ \ i=1,...,p.
\end{eqnarray*}

The quantity
$$
\mathfrak{D}_{(1,p)}^{(p+1,n)}=\sum_{i<j=1}^p\,i_{X_i\wedge
X_j}\Omega_{(X)}=\sum_{i<j=1}^p\beta^k([X_i,X_j])Y_k,
$$
represents the sum of the {\it flow generators}
from $\Delta^p(M)$ to the (n-p)-dimensional "outside
world" $\Delta^{n-p}(M)$, and the quantity
$$
\mathfrak{D}^{(1,p)}_{(p+1,n)}=\sum_{k<l=1}^{n-p}\,i_{Y_k\wedge
Y_l}\Omega_{(Y)}=\sum_{k<l=1}^{n-p}\alpha^i([Y_k,Y_l])X_i
$$
represents the sum of the {\it flow generators} from the "outside world"
$\Delta^{n-p}(M)$ into $\Delta^p(M)$.

The introduced in Sec.3.2.3 {\it CI-operators},
representing the corresponding transfers of the quantities carried by
the  flow generators, for example energy-momentum, are given
by the flows of $\mathfrak{D}_{(1,p)}^{(p+1,n)}$ and
$\mathfrak{D}_{(p+1,n)}^{(1,p)}$ through the corresponding volume forms,
i.e. the corresponding interior products:
\begin{eqnarray*}
\mathbb{D}_{(1,p)}^{(p+1,n)}&\equiv&
i_{\mathfrak{D}_{(1,p)}^{(p+1,n)}}(\beta^1\wedge\beta^2\wedge...\wedge\beta^{(n-p)})\\
\mathbb{D}_{(p+1,n)}^{(1,p)}&\equiv&
i_{\mathfrak{D}_{(p+1,n)}^{(1,p)}}(\alpha^1\wedge\alpha^2\wedge...\wedge\alpha^p).
\end{eqnarray*}

We assume further that a dynamical equilibrium with the external world will
always hold. Our purpose now is to see what happens inside the physical system
$\Delta^p(M)$. Generalizing the classical concept "flow of a vector field
across a 2-surface", we introduce some terminology.

Let $Z$ be a $p$-multivector field and $\Phi$ be a $q$-differential form on
the manifold $M$, and let $p\leq q$.

	- the quantity $i_Z(\Phi)$ will be called {\it algebraic flow} of
$Z$ across $\Phi$, and if $i_Z\Phi\neq 0$ then $\Phi$ is $Z$-{\it attractive},
or, $Z$ is $\Phi$-{\it sensitive},

	- the quantity $i_Z(\mathbf{d}\Phi)$ will be called {\it
differential flow}, or {\it dynamical flow} of $Z$ across $\Phi$, and if
$i_Z\mathbf{d}\Phi\neq 0$ then $\mathbf{d}\Phi$ is $Z$-{\it attractive}, or, $Z$
is $\mathbf{d}\Phi$-{\it sensitive}.

	-the quantity $L_Z\Phi$ will be called {\it Lie flow} of $Z$ across
$\Phi$. If $L_Z\Phi=0$ then $Z$ will be called {\it symmetry} of $\Phi$,
and if $L_Z\Phi\neq 0$ then $\Phi$ is $(Lie,Z)$-{\it attractive}, or, $Z$ is
$(Lie,\Phi)$-{\it sensitive}.

Note the very suggestive relation between Lie flow and differential flow
(Sec.2.8.3): $$ L_Z\Phi-\mathbf{d}(i_Z\Phi)=-(-1)^{deg\,Z}i_Z\mathbf{d}\Phi. $$
These concepts are naturally extended to $E_1$-valued $p$-vectors and
$E_2$-valued differential forms with respect to a bilinear map
$\varphi: E_1\times E_2\rightarrow F$ (Sec.2.8.4).

Let now the two, may nontrivially intersected, distributions $\Delta^p_1$ and
$\Delta^p_2$ be represented by the two $p$-multivector fields $Z_1$ and $Z_2$
respectively, and the $p$-forms $\Phi^1$ and $\Phi^2$ represent the
corresponding codistributions, i.e. at every point $x\in M$ the space
$\Phi^1_x$ is the dual space to $(\Delta^p_1)_x$ and $\Phi^2_x$ is the dual
space to $(\Delta^p_2)_x$.
 So, we can form the expressions $i(Z_1)\Phi^1, i(Z_2)\Phi^2,i(Z_1)\Phi^2,
i(Z_2)\Phi^1$.

It seems convenient the general concept of $\varphi$-symmetry between two
distributions (Sec.2.8.4) to be called {\it dynamical equilibrium} between
two distributions $\Delta^p_1$ and $\Delta^p_2$ when $\varphi\rightarrow\vee$:
two distributions $\Delta^p_1$ and $\Delta^p_2$ will be called to be in dynamical
equilibrium, or {\it partners in equilibrium}, if
$$
\mathcal{L}^{\vee}_{Z_1\otimes e_1+Z_2\otimes e_2}(\Phi^1\otimes e_1+
\Phi^2\otimes e_2)=0,
$$
where $(e_1,e_2)$ is a basis in $\mathbb{R}^2$.

If the two distributions $\Delta^p_1$ and $\Delta^p_2$ satisfy
additionally the relations
$$
	 i(Z_1)\Phi^1=const, \quad i(Z_2)\Phi^2=const,  \quad
	 i(Z_1)\Phi^2=-i(Z_2)\Phi^1.
$$
we shall say that these two
distributions $\Delta^p_1$ and $\Delta^p_2$ are in {\it full equilibrium}.

We shall show further that among the distributions in dynamical equilibrium
there are many that are in full equilibrium, in particular, these are all
nonlinear solutions of the corresponding equations.

The system of $p$-dimensional distributions
$\Sigma^{p}=(\Delta_1^p,...,\Delta_k^p)$ will be
said to be in dynamical equilibrium if every distribution gains as much as it
loses locally during the exchange processes with all its partners.

If our manifold $M$ is endowed with a riemannian or pseudoriemannian metric $g$
then the Hodge star $*_{g}$ and the explicit isomorphisms $\tilde{g}$ between
distributions and codistributions are naturally to be in use.

For example, on
Minkowski space-time $M=(\mathbb{R}^4,\eta)$ every two {\it isotropic}
2-dimensional codistributions,  defined by the 2-forms $(\Phi,*\Phi)$ and the
corresponding 2-vectors
$(\bar{\Phi},\bar{*\Phi})=(\tilde{\eta}(\Phi),\tilde{\eta}(*\phi))$ will
be in full equilibrium if
$$ i(\bar{\Phi})*\Phi=const, \quad i(\bar{\Phi})\mathbf{d}\Phi=0, \quad
i(\bar{*\Phi})\mathbf{d}*\Phi=0, \quad
i(\bar{\Phi})\mathbf{d}*\Phi+i(\bar{*\Phi})\mathbf{d}\Phi=0.
$$

Note that we may
come to understand the dynamical behavior of the field
$\Omega=\Phi\otimes e_1+*\Phi\otimes e_2$ by
means of assuming that $\Omega$ keeps its identity along its
$\tilde{\eta}$-image $\bar{\Omega}$, i.e. assuming
$\mathcal{L}_{\bar{\Omega}}^{\vee}\Omega=0$. This suggests some analogy with,
e.g., the autoparallelisim of vector fields with respect to a given linear
connection, where the vector field $Z$ is projected on its own $\nabla$-change
$\nabla Z$, and the projection is $i_Z(\nabla Z)=\nabla_ZZ=0$. So, if $\nabla$ is
riemannian, then $\nabla_Z\tilde{g}(Z)=0$.

Four serious
differences with "action-variational" approach in field theory are seen:

	-{\it first}, the above $\varphi$-extended Lie derivative
does not make use of any additional local structure, like, for example, linear
connection,

	-{\it second}, it is applicable, in principle, without available
metric,

	-{\it third}, unlike the standard variational approach to field
equations, NO derivatives of the field functions are necessary for coming to
dynamical equations, but extension to corresponding jet-spaces is always
possible,

	-{\it fourth}, if metric presents, then explicit interaction
$\Phi\leftrightarrows *\Phi$ terms can be obtained.

In view of the further application of the above concepts and relations to real
systems we give some preliminary considerations coming from relativistic
physics, where $(M,g)=(\mathbb{R}^4,\eta), sign\,\eta=(-,-,-,+)$.

First, the physical system we are going to model
by an appropriate integrable distribution $\Delta^p(M)$, should be
allowed to {\it propagate in space} keeping its identity, as every real system
does, so, the distribution must admit (at least one) external/shuffling
symmetry along {\it time-like} or {\it null} vector field(s) to be additionally
introduced. Let $\bar{\zeta}$, with the $\eta$-corresponding local 1-form
$\zeta$, be such one, in which case $\Delta^p(M)$ may be called
$\eta$-{\it adaptable}. Hence, any vector $X\in\Delta^p(M)$ is naturally
extended to a two dimensional distribution $X\wedge\bar{\zeta}$, such that
$[X_i,\bar{\zeta}]\in \Delta^p(M)$, and the distribution
$\Delta^p(M)\wedge\bar{\zeta}$ is integrable.

Second, it seems naturally to require the integrability of the 1-dimensional
codistribution defined by $\zeta$ in view of the integrability of
$\Delta^p(M)$.

Third, the internal dynamics of the system is allowed to be carried out
{\it only in time}, so, the vector fields that are meant to generate the
internal dynamics of the system may fulfill such a function only through
allowed space-time propagation of the system, i.e. only through some coupling
with the vector field $\bar{\zeta}$. Since at every moment each of the vector
fields of $\Delta^p(M)$ must represent definite stress, we may assume that in
the Minkowski space-time case $\Delta^p(M)$ is {\it space-like}:
$\eta(X_i,X_i)<0$. Further, the time-coupling between each $\tilde{\eta}(X_i)$
and $\zeta$ defines the 2-forms $F^i:=\tilde{\eta}(X_i)\wedge\zeta$, and the
Hodge $*$-operator assigns the corresponding $(4-2)$-forms $*F^i$. So, it seems
natural to expect the 2-forms $(F^i,*F^i)$ to play essential role in describing
the {\it internal dynamics} \index{internal dynamics} of the system considered.

The time-recognizability of our physical system during
propagation and the assumed dynamical equilibrium with the outside world
require corresponding time-stability of its entire structure and
dynamics, so the integrability of
$\Delta^p(M)$ and  $\Delta^p(M)\wedge\bar{\zeta}$ should be considered as
natural, while the internal interaction
among the subsystems suggests available nonintegrability of most of the
2-dimensional subdistributions $F_i=X_i\wedge\bar{\zeta}$.

Also, the assumed time-stability and time-recognizability of any subsystem
$F_i$ during propagation suggests with each $F_i$ to associate a dimension in
an external vector space, such that the number of its dimensions to be equal to
the number of the time recognizable subsystems $F_i$. So, if $N$ is the number
of time-recognizable subsystems $F_i, i=1,2,...,N$, we can associate with our
system the quantity $\sum_{i=1}^NF_i\otimes e_i$, where $\{e_i\}, i=1,...,N$ is
a basis of an appropriate $N$-dimensional vector space, and the space
propagation and time-stability of the system to interpret formally
as requirement for dynamical equilibrium:
$$
\sum_{j>i=1}^N\mathcal{L}^{\varphi}_{F_i\otimes e_i+(*F)_j\otimes e_j}
\big(F^i\otimes e_i+(*F)^j\otimes e_j\big)=0, \ \ \
$$
under appropriate $\varphi$, e.g., $\varphi=\vee$. This relation can be
interpreted in the sense, that the internal interaction does not violate the
consistency and compatability of the subsystems of our system, on the contrary,
it {\it supports} these consistency and compatability, and guarantees the
surviving of each subsystem and the whole system, mathematicaly represented by
$\Delta^p(M)\wedge\bar{\zeta}$, by means of corresponding space-time
propagation along the shuffling local symmetry $\bar{\zeta}$.

From a more general viewpoint, if our system is decribed by $N_1$ $p\,$-vectors
$\Psi_i, i=1,2,...,N_1$, and $N_2$ $q\,$-forms $\Phi^j, j=1,2,...,N_2$,
$p\leq q$, satisfying $\mathbf{d}(i_{\Psi^j}\Phi_i)=0$, i.e.,
$\langle\Phi^j,\Psi_i\rangle$ are closed forms, then a possible extension
of the above equation on any manifold would read
$\sum_{i,j=1}^N\mathcal{L}^{\varphi}_{\Psi_i\otimes e_i}(\Phi^j\otimes e_j)=0$.

The connection of these $\varphi$-extended Lie derivatives with the internal
curvature forms $\Omega^i$ associated with each $\Phi_i/F_i$ will be further
used.

Fanally, these structures can be extended to vector bundles, where the
$p$-forms should be replaced by corresponding bundle-valued $p$-forms, and the
exterior derivative $\mathbf{d}$ should be replaced by corresponding covariant
exterior derivative $\mathbf{D}$.

 %\newpage

\section{The Gauge idea for field interaction and Maxwell equations}
 \subsection{The interaction in mechanics.}

The very idea for {\it field interaction}, i.e. {\it local interaction of two
continuous physical objects}, has proved to be a serious problem in theoretical
physics, and this is understandable. In mechanics, as introduced by the second
Newton law, interaction is represented in a very simple form, it just says that
two mechanical objects can keep their identities under mutual influence, if at
least one of them, changes its behavior as a whole, which theoretically is
noted by changing its relative velocity and all dynamical characteristics that
are functions of the velocity. The proper characteristics are those
that do not change under this influence, and so, the observer is allowed to
think that he continues observation of the same object(s). As an universal such
proper characteristic of a body has been assumed the {\it mass}, been
understood as a measure of its {\it inertial} properties. The inertial
properties of a body have been identified with its gravitational abilities to
feel external influence from other distant mass bodies, and, of its side, to
exert on the other distant mass bodies corresponding influence. In result, from
theoretical viewpoint, all these mass bodies that participate in the
interaction, keep their identities through changing their behavior, i.e. their
momentum, kinetic energy, etc. The principle of inertia defines a change of
behavior through identifying all states characterised by straight-line constant
velocity of the body as a whole. Also, the concept of inertial frame of
reference has been introduced as a system of bodies being in relative rest with
respect to each other and may moving as a whole along straight lines with
constant velocity.

From historical perspective, the most important theoretical quantity been used
to take care about availability of interaction in mechanics,
i.e. of changing its mechanical state of motion as a whole, has been the so
called {\it potential energy}. This quantity measures the final and integral
balance between energy losses and gains when a mechanical system suxessfully
withstands transitions between two admissible configurations. The corresponding
quantity characterising the total energy change has been named {\it work
against the external influence}, and the Newton force $\mathcal{F}$ measures
this work for a unit distance, so $\mathcal{F}$ acquires in modern terms the
mathematical sense of covariant tangent vector, or 1-form. The integral of this
1-form along the road-curve from infinity, where the mass body is considered to
be in inertal condition, to a given point where the external influence acts,
gives quantitatively the corresponding energy balance.

The further mathematical development of classical mechanics has been based on
the assumption that this {\it integral in nature} concept of potential energy,
can be universelized to the concept of {\it potential}, or potential function
$U(x,y,z)$, and considered as a {\it local} energy measure of external influence,
and its differential $\mathbf{d}U$, reduced on the trajectory, as a local force
field. This view, together with the assumption that the measured time can
always be used as a parameter along any trajectory, has been utilised further in
the lagrangian and hamiltonian formulations of mechanics.

We would specially note at this point two things.

{\bf First}, as far as the
potential depends only on the spatial coordinates $(x_a,y_a,z_a)$ of the mass
particles with masses $m_a, a=1,2,...$, it is rather {\it configurational
characteristic} of the mechanical system considered, and in {\it no way} a
local one. Hence, the {\it external} parameter "time" parametrizes family
of admissible configurations of the mechanical system considered.

{\bf Second}, if even we consider $U$ as function of the coordinates inside
the region not ocupied by the mass particles, it stays quite unknown how this
potential function should be defined in the various cases arising in practice,
in other words, the problem of understanding what really happens in the space
out of the volumes occupied by the bodies, stays unanswered. That's why the
theoretical concept of mechanical system consisting of interacting point-like
objects has been formulated, where the interaction is formulated in terms of
potential function depending on the coordinates of the points where the mass
points stay at a given moment. The total interaction energy is then defined as
a sum of the interaction energies of all couples of point like mass objects.
Now, stepping on one of these mass objects, i.e. choosing it as a reference
frame (with assuming absolute time parameter) we study the behavior of all the
rest point-like mass objects on the base of supposition that the potential
function for each observed mass point depends mainly on the distances to the
other mass points, and this dependence is speculatively treated as
{\it local}, i.e. the potential function has transformed from configurational
integral parameter to a field parameter.

Such a speculative transformation of the potential, together with the
approximation for point-like mass object, has braught the researchers to the
idea for 3-dimensional spherical symmetry of the potential function connected
with any two mass points. Relative to one of the points the other has to feel
influence through a spherically symmetric potential function $U(r)$,
this dependence must decrease with the distance treated now as coordinate on
$\mathbb{R}^3$, and this function should {\it not be defined} at the reference
mass point. So, from mathematical point of view,
the potential function $U(r)$ will be defined on topologically
nontrivial subspace of $\mathbb{R}^3$, around every mass point the
corresponding cohomological class is defined by the unique spherially symmetric
representative - the closed 2-form $\omega=const.sin\theta d\theta\wedge
d\varphi$, so the force acting on unit mass, or unit electric charge, becomes
$\mathcal{F}=*\omega=\frac{const}{r^2}dr$ and
$U(r)=\pm\,\frac{const}{r}+const$. The confugurational nature of such a
consideration presumes two kinds of interaction: repulsion and attraction, so,
the "-" sign of $U$ is chosen when attraction takes place, and the "+" sign of
$U$ is chosen when repulsion takes place by obvious reasons. We dare thinking
that the universality of this potential, proving its strength from classical
gravity and electricity through quantum mechanics and intra-nuclear
interaction, lies namely in its topological nature and applicability of the
point-like approximation concerninig sources.

The important point we'd like to specially note is that {\it the
interpretation neither of $U$ nor of $\mathbf{d}U$ as mathematical
images of physical field objects is posible while they do not depend on time,
because {\it static} means that all their characteistics do NOT change with
time, so they can NOT participate in any dynamical physical process connected
with energy exchange since the energy is conserved quantity, and, therefore,
the energy change of the distant to each other mass points has no where to come
from}: {\bf static physical fields can not act upon other physical objects by
means of transfering energy and whatever in view of their static nature}. So,
when such potentials are introduced in physical equations, they determine just
the admissible configurations of the system and, most probably, they can not
determine local energy-momentum exchange between/among recognizeble subsystems.

In conclusion, if we'd like to define local interaction making use of such
potential approach, the very contents of the concept of potential should be
appropriately modified.

\subsection{Field interaction in classical electrodynamics.}
\hskip 1.5cm
	{\bf 1. The Maxwell vacuum equations (MVE) case.}

Recall MVE from Sec.6.2.1:
$$
{\rm rot}\,\mathbf{E}+
\frac 1c      \frac{\partial {\mathbf{B}}} {\partial t}=0
, \quad {\rm div}\,\mathbf{B}=0,             %1%
$$
$$
{\rm rot}\,\mathbf{B} -
	\frac 1c \frac{\partial {\mathbf{E}}} {\partial t}=0 ,
\quad {\rm div}\,\mathbf{E}=0.
$$
These equations obviously imply that the vacuum electromagnetic field consists
of two locally recognizable subsystems/components represented by the
space-like vector fields $\mathbf{E}$ and $\mathbf{B}$, and that there is a
permanent mutual physical influence between $\mathbf{E}$ and $\mathbf{B}$,
which we understand physically as energy-momentum exchange between the
individualized $\mathbf{E}$ and $\mathbf{B}$ components: $\mathbf{E}$ acts upon
$\mathbf{B}$ and $\mathbf{B}$ acts upon $\mathbf{E}$. Now, according to the
above equations, each of these two components propagates and keeps its
individualization during propagation, so, $\mathbf{E}$ and $\mathbf{B}$ should
be able to carry energy (because each of them is assumed to be able to act
upon), and momentum (because each of them propagates in space as
individualizable system), separately. On the other hand, the energy concept of
the theory excludes nonzero interaction energy between these two components to
exist since the energy density $w$ is given by the sum of the energies carried
by $\mathbf{E}$ and $\mathbf{B}$: $w=\frac12(\mathbf{E}^2+\mathbf{B}^2)$, so,
how does the presumed by the equations energy exchange take place?

Further, the local momentum concept in the
theory is defined and experimentally proved quantitatively to be given by
$\frac1c(\mathbf{E}\times\mathbf{B})$, so neither of the assumed in the theory
electric and magnetic components is allowed to carry momentum separately, while
each component is allowed to carry energy separately. But the equations require
some kind of mutual influence, which we measure locally by local
energy-exchange. Recall now the null-field solutions, where the relations
$\mathbf{E}^2=\mathbf{B}^2$ and
$\mathbf{E}^2+\mathbf{B}^2=2|\mathbf{E}\times\mathbf{B}|$ always hold. The
assumed by the equations permanent space-time identification of $\mathbf{E}$
and $\mathbf{B}$ as propagating physical subsystems of the field obviously
implies besides energy also momentum exchange. So, how this implied internal
energy-momentum exchange between the two space-time recognizable subsystems,
mathematically identified as $\mathbf{E}$ and $\mathbf{B}$, is performed? May
be we have not made the right mathematical identification of the subsystems, or
may be we have to look, in analogy with mechancs, for some potential object?

The above remarks set the question : are these equations directly verifiable by
appropriate experiments? Our answer to this question rather "no", just because
we have not appropriate devices. What we are able to check directly is the
result of field's action upon some other physical object which we are able to
watch/observe. This turns our attention to the logic we usually meet in the
textbooks and even in monographs. Let's recall it, keeping in mind that the
time variable is of external nature in the nonrelativistic approach.

The basic concept introduced there is {\it integral flow of a vector field
across a 2-dimensional surface} $S$. This requires the nature of the integrand
to be differential 2-form on $\mathbb{R}^3$ having no singularities on the
2-surface, so that this definite integral to have well defined finite value.
With any vector field $Z$ on $\mathbb{R}^3$, endowed with euclidean metric $g$
and corresponding volume 3-form $\omega_o=dx\wedge dy\wedge dz$, we can
associate natarally two differential 2-forms: $i_Z\omega_o$ and
$*\tilde{g}(Z)$. It turns out that in this case these two differential forms
coincide: $$ i_Z\omega_o=Z^1dy\wedge dz-Z^2dx\wedge dz+Z^3dx\wedge
dy=*\tilde{g}(Z). $$ Hence, the flows of $\mathbf{E}$ and $\mathbf{B}$ across
the 2-surface $S$ are just $$ \int_{S}*\tilde{g}(\mathbf{E}), \ \ \
\int_{S}*\tilde{g}(\mathbf{B}). $$ The next step is to equilize the time
derivatives of these integrals to the integrals of
$\mathbf{d}\tilde{g}(\mathbf{B})$ and $(-\mathbf{d}\tilde{g}(\mathbf{E}))$
respectively ($\xi=ct$) $$ \frac{d}{d\xi}\int_{S}*\tilde{g}(\mathbf{E})=
\int_{S}\mathbf{d}\tilde{g}(\mathbf{B}), \
\
\frac{d}{d\xi}\int_{S}*\tilde{g}(\mathbf{B})=
-\int_{S}\mathbf{d}\tilde{g}(\mathbf{E}).
$$
The final step is to get free of the 2-surface $S$ on the assumption that $S$
is arbitrary and does NOT participate in the interaction, it just helps to
introduce dynamics, in fact linear equations, which presume interaction but do
not directly describe it in appropriate terms. So, we come to the equations $$
\frac{\partial}{\partial \xi}*\tilde{g}(\mathbf{E})=
\mathbf{d}\tilde{g}(\mathbf{B}), \ \ \frac{\partial}{\partial
\xi}*\tilde{g}(\mathbf{B})=- \mathbf{d}\tilde{g}(\mathbf{E}). $$ The assumption
that the 2-surface $S$ is of {\it no-physical nature} leads to the conclusion
that the so defined flow of a physical field across a mathemetical 2-surface is
not quite sensible from physical viewpoint since it cannot be observed and
verified. The flows of $\mathbf{E}$ and $\mathbf{B}$ {\bf must be across a
physical 2-surface}, in order to expect observable interaction between the
vector field and the 2-surface, considered as section of some physical object
been able to interact with the field flow. Otherwise, we must consider
energy-momentum flows, e.g. flow of the Poynting vector, across imaginable
2-surface.

The two scalar equations $\mathrm{div}\,\mathbf{E}=0, \
\mathrm{div}\,\mathbf{B}=0$ say geometrically that the volume form $\omega_o$
is not $(\mathbf{E},\mathbf{B})$-attractive, i.e. the 2-forms
$*\tilde{g}(\mathbf{B})$ and $*\tilde{g}(\mathbf{E})$ are closed:
$$
\mathbf{d}*\tilde{g}(\mathbf{B})=\mathbf{d}\,i_{\mathbf{B}}\omega_o=L_{\mathbf{B}}\omega_o=0,
\ \ \ \mathbf{d}*\tilde{g}(\mathbf{E})=\mathbf{d}\,i_{\mathbf{E}}\omega_o
=L_{\mathbf{E}}\omega_o=0,
$$
a supposition, seeming not sufficiently motivated in view of the fact that
direct experimental proof of the relations $L_{\mathbf{E}}g=0,
L_{\mathbf{B}}g=0$, where $g$ is the euclidean metric, for the general
case, are missing. We note that, these two equations clearly suggest to look
for two potential 1-forms $\alpha$, $\beta$, such, that
$\mathbf{d}\alpha=*\tilde{g}(\mathbf{E})$ and
$\mathbf{d}\beta=*\tilde{g}(\mathbf{B})$.

The above consideration suggests to look for additional theoretical motivation
for assuming MVE as appropriate local description of time dependent and space
propagating free electromagnetic fields.

In trying to overcome these theoretically motivated difficulties, the creatively
thinking men at the beginning of the last century made a very radical step
building a new viewpoint on MVE, called relativistic electrodynamics. They
introduced new point of view: adequate mathematical objects that represent such
two {\it interconnected, time-recognizable and spatially propagating
substructures} of the general vacuum field are NOT $\mathbf{E}$ and
$\mathbf{B}$, but two differential 2-forms $F_{(\mathbf{E},\mathbf{B})}$ and
$*F_{(-\mathbf{B},\mathbf{E})}$ on Minkowski space-time, so, from the new point
of view, any internal energy-momentum exchange should take place between $F$
and $*F$. However, by some reasons, the next radical step, leading to new
equations, was not made. Namely, the "new" field equations $\mathbf{d}F=0,
\mathbf{d}*F=0$, although in terms of $F$ and $*F$, keep the old viewpoint, and
in fact, coincide with the old equations and the problems connected with the
above mentioned internal energy-momentum exchange between the two new
components $F$ and $*F$ were not resolved: corresponding local energy-momentum
exchange expressions in terms of $F$, $*F$ and their derivatives, were not
appropriately introduced and used. Nevertheless, the new point of view brought
in a quite clear way the idea how to introduce potential object(s). We shall
consider this new and important step right after a glance at the so called
"quasi-vacuum" field equations, claiming successes in describing the field
evolution in space regions continuously filled with electrically charged mass
particles.

In terms of $\mathbf{E}$ and $\mathbf{B}$ these equations look like (we omit
dimensional constants)
$$
{\rm rot}\,\mathbf{E}+ \frac 1c      \frac{\partial
{\mathbf{B}}} {\partial t}=0 \, , \quad {\rm div}\,\mathbf{B}=\rho,
$$
$$ {\rm rot}\,\mathbf{B} -
	\frac 1c \frac{\partial {\mathbf{E}}} {\partial t}=\mathbf{j} \, ,
\quad {\rm div}\,\mathbf{E}=0,
$$
where, usually is assumed $\mathbf{j}=\rho\,\mathbf{v}$, $\rho(x,y,z,t)$ is the
so called "charge density", and $\mathbf{v}(x,y,z,t)$ is the velocity vector of
the charged mass particles filling a small volume around a space point at a
given moment of time. The used term of "quasi-vacuum" now means that no
mechanical collisions among the charged mass particles are allowed.

The main reason not to trust these equations is that they violate our creed,
according to which {\it on the two sides of "=" must stay the same quantity}.
If we ask which physical quantity may be represented equally well as
$\mathrm{rot}\mathbf{B}-\frac{\partial \mathbf{E}}{\partial \xi}$ and at the
same time as $\mathbf{j}$, no easy answer could be found. The same motive works
also for the equation $\mathrm{div}\mathbf{E}=\rho$. Of course, this does not
mean that these equations should not be used, if they work in various cases and
there are not better ones, let them be used. From theoretical viewpoint,
however, the introduced quantities have to be duly respected, so {\it quntities
of different physical nature should not be equalized}.
\vskip 0.3cm
	{\bf 2. The gauge view}

Let's go back now to the relativistic formulation of MVE:
$\mathbf{d}F=0, \mathbf{d}*F=0$. The suggestion for available two potential
1-forms could hardly be avoided. However, following the idea that the
"quasi-vacuum" equations should be kept and the appropriate form for this is
$\mathbf{d}*F=*\mathbf{J}$, where $\mathbf{J}=(\mathbf{j},\rho)$ is the
corresponding electric 4-current, the relativists have decided to choose the
following perspective:

	- there is just one potential 1-form $A$ on the
Minkowski space-time taking values in
the Lie algebra of the abelian group $U(1)$, and such that
$\mathbf{d}A=F$,

	- the interaction of the field with the charged mass
particles is performed in accordance with the principle of
"minimal coupling", i.e. by means of the induced through an appropriate
representation of $U(1)$ in $\mathbb{C}^4$ linear connection in a complex vector
bundle with a standard fiber $\mathbb{C}^4$ on the Minkowski space time.

In this way the 2-form $F$ became an image of a curvature form $\mathbf{F}$ of a
connection $\mathbf{A}$ on the pricipal bundle $\mathcal{P}=(M,U(1))$
through a section $\sigma:M\rightarrow\mathcal{P}$ of this bundle:
$A=\sigma^*\mathbf{A}, F=\sigma^*\mathbf{F}$. The sections $\Psi$ of the
$\mathbb{C}^4$-vector bundle over $M$ were called {\it spinors}, the Dirac
matrices $\gamma_\mu, \mu=1,2,3,4$ were introduced through the relation
$$
\gamma_\mu\gamma_\nu+\gamma_\nu\gamma_\mu=2\,\eta_{\mu\nu}id_{\mathbb{C}^4},
$$
and the 4-current $\mathbf{J}^\mu$ for an electron with charge $e$ appeared in
the form (after appropriate choice of the $\gamma$-matrices)
$$
\mathbf{J}^\mu\backsim e\bar{\Psi}\gamma^\mu\Psi, \ \ \rho\backsim
e\Psi^{+}\Psi,
$$
where $\Psi^{+}$ is the Hermit congugated of $\Psi$, and $\bar{\Psi}$ is the
Dirac congugated of $\Psi$.

This simple example brings us to the modern guage theory as the basic
theoretical tool in approaching microsystems. We give now just a brief
formal sketch of this approach since we are not going to make use of it
further.

Recall that each closed differential form is locally exact, i.e. if the
$p$-form is closed: $\mathbf{d}F=0$, then there are many $(p-1)$-forms
$A,A', A'', ...$ giving the same $F$ through exterior differentiation. In fact,
if $F=\mathbf{d}A$ and $\mathbf{d}A'=\mathbf{d}A''=...=0$, so that we can
locally assume $A'=\mathbf{d}B', A''=\mathbf{d}B'', ...$, where $B',B'',...$
are $(p-2)$-forms, then we obtain many "potentials"
$A,A+\mathbf{d}B',A+\mathbf{d}B'',...$ for $F$ :
$F=\mathbf{d}A=\mathbf{d}(A+\mathbf{d}B')=...$.

Let now $F$ be a 2-form on a manifold $M$, then $A$ is
1-form, so, $A',A'',...$ are functions $\psi$ on the corresponding manifold. In
such a case we could write
$$
A\rightarrow A^{\psi}=A+\mathbf{d}\psi.
$$
Considering $A$ as 1-form, taking values in the Lie algebra $u(1)=i\mathbb{R}$
of the Lie group $U(1)$, we can replace the real valued function $\psi$ with
the $U(1)$-valued function $g=e^{i\psi}$. Now, with the above identification of
$u(1)$ as $i\mathbb{R}$,  the above transformation can be written as
$$
iA\rightarrow iA^g=g^{-1}(iA)g+g^{-1}\mathbf{d}g, \ \ g\in\mathcal{J}(M,U(1)).
$$
Assuming now the Minkowski space-time as base space of principal bundle
with group $G=U(1)$, in view of the above, we obtain relativistic formulation
of electrodynamics in gauge terms if the 2-form $F$ satisfies additionally the
equation $\mathbf{d}*F=0$. Since the equation $\mathbf{d}F=0$ is now
concequence of the assumption that the potential $A$ is a projection on $M$ of
a connection on the principal bundle with $G=U(1)$, the desired equation is
equivalent to the requirement for extremum of the integral $\int_{M}F\wedge
*F$ with respect to variation of $A$. In this way we come to the so called gauge
formulation of vacuum classical electrodynamics.

Following the rules and concepts of the geometry of principal bundles, this
scheme is easily carried to connections on principal bundles on an arbitrary
(pseudo)riemannian manifold with a finite dimensional Lie group $G$. The
main additional requirement is to have well defined metric $h$ on
$G$ in order to have a metric on the bundle, so that, the integrand
$$
F^a\wedge *F^b\,h(E_a,E_b), \ \ a,b=1,2,...,dim G
$$
to be well defined. Of course, the very integral
$$
\int_MF^a\wedge *F^bh(E_a,E_b)\omega_o
$$
where $\omega_o$ is a volume form on $M$, and $\{E_a\}, a=1,2,...,dimG $ is a
basis of the Lie algebra of $G$, also should be well defined. In this general
situation with nonabelian Lie group $G$ if the connection form is
$\omega=\omega^a\otimes E_a$, then the curvature 2-form
$$
F=F^a\otimes
E_a=\mathbf{d}\omega^a\otimes E_a+\omega^a\wedge\omega^b\otimes [E_a,E_b]
$$
is already a nonlinear function of the components of $\omega $. If
$\mathbf{d}^{\omega}$ is the corresponding exterior differential, the variation
of the above action integral gives besides the Bianchi identity
$\mathbf{d}^{\omega}F=0$, the equation $\mathbf{d}^{\omega}*F=0$. Note that the
mentioned nonlinearity of these equations with respect to the components of
$\omega$ comes from the presumed nonabelian nature of $G$, and is not
explicitly connected with some physical understanding of local physical
interaction.

Usually, these equations are considered in terms of the projections
$\sigma^*\omega$ and $\sigma^*F$ of the connection and curvature forms on the
base space through the section $\sigma\in Sec{\mathcal{P}}$. The values of any
such sections are in the diffeomorphic image $G_x$ of $G$, so every $\sigma(x)$
can perform transformations in $G_x$ as well as in the Lie algebra
$\mathfrak{g}$ and its dual $\mathfrak{g}^*$, of $G$. These are the so called
local gauge transformatons.

The important property of the above action integral
is its invariance with respect to these transformations provided the metric on
$G$ is correspondingly invariant, which holds for the corresponding
Killing metric:
$$
\langle x,y\rangle=Tr[\mathrm{ad}(x)\circ\mathrm{ad}(y)] , \ \
x,y\in\mathfrak{g}. $$ The solutions of these equations are usually called
Yang-MIlls fields.

An extension of these fields are the so called Yang-Mills-Higgs fields. In
order to come to these fields we need a representation of the group $G$ in some
linear space, the natural example is the adjoint representation $Ad$, of
course, of $G$ in $\mathfrak{g}$ and in its dual $\mathfrak{g}^*$, but the
scheme works for any other representation, even for the case of action of $G$
on a manifold. If such a representaion $\rho: G\rightarrow GL(\mathbb{R}^m)$ is
given we also have the representation
$\rho': \mathfrak{g}\rightarrow GL(\mathbb{R}^m)$. Recall that with every such
representation a vector bundle on $M$ with a standard fiber $\mathbb{R}^m$ can
be associated. Now, the principal connection $\omega$ on the principal bundle
induces a linear connection in the associated vector bundle, so the sections of
this vector bundle and its tensor extensions can be differentiated covariantly
with respect to this induced linear connection. Now, introducing some metric
$\mathcal{M}$ in the vector bundle,
 the Yang-Mills action is extended as follows:
$$
\int_{M}[*(F\wedge *F)+const_1\mathcal{M}(\phi,\phi)+
const_2V(\mathcal{M}(\phi,\phi))]\omega_o,
$$
where $\phi$ is a section of the associated vector bundle, called usually
"matter" field, and $V(\mathcal{M}(\phi,\phi))$ is the so called "self
interaction" term. If the representation $\rho$ is the adjoint $\mathrm{Ad}$,
and $\mathcal{D}^\omega$ is the corresponding covariant coderivative, the
equations obtained are
$$
\mathcal{D}^\omega\,F+const_1[\phi,\nabla\,\phi]=0, \ \
\mathcal{D}^\nabla\circ
\nabla\phi+const_2V'(\mathcal{M}(\phi,\phi))\phi=0,
$$
where $\nabla$ is the induced by $\omega$ covariant derivative in the vector
bundle, $\mathcal{D}^\nabla$ is the covariant coderivative in the vector
bundle, $\mathcal{M}$ is here the Killing metric, $[\phi,\nabla\,\phi]$ is
induced by the Lie bracket in $\mathfrak{g}$, and $V'$ is the derivative of $V$
with respect to the appropriately squared $\phi$.

Finally, the folowing relations are identically satisfied:
$$
\mathbf{d}^\omega\,F=0, \ \ \
\mathbf{D}\circ\nabla\phi=[F,\phi],
$$
where $\mathbf{D}$ is the exterior derivative in the vector bundle valued
differential forms on the base manifold $M$.

\vskip 0.2cm
{\bf Literature}
\addcontentsline{toc}{subsection}{{\bf Literature}}
\vskip 0.2cm

1. {\bf K. Marathe}, {\it Topics in Physical Mathematics}, Springer-Verlag
London Limited 2010

2. {\bf B. Felsager}, {\it Geometry, Particles and Fields}, Odense University
Press, Second edition 1983, Copenhagen.

3. {\bf P. Deligne, D. Freed}, {\it Classical field theory}, Amer. Math. Soc.,
1999.

\part{Extended Electrodynamics}

\chapter{Extended Electrodynamics. Nonrelativistic approach}
{\it In this chapter we present in nonrelativistic terms our nonlinear approach
to vacuum electrodynamics, based on the understanding that the basic equations
must represent direct local energy-momentum balance relations in order to be
directly verifiable in principle, and so, trusted enough. We mention  three
references related somehow to our approach} [1],[2],[3].

\section{Maxwell Stress tensors}
Following our considerations and suggestions in Chapters 4,5 we begin with the
mentioned in Sec.5.2
well known differential relation satisfied by the square of
every vector field $V$ on the euclidean space $\mathbb{R}^3$. Our attention is
directed to the square of $V$ just because of the experimentally suggested
assumption that $\mathbf{E}^2,\mathbf{B}^2$ should measure the
energy-densities correspondingly of the electric and magnetic components.

Let $\mathbb{R}^3$ be related to the standard coordinates
$(x^i=x,y,z), i=1,2,3$; we denote by
$"\times"$ the vector product, and make use of the $\nabla$-operator:
$$
\frac12\nabla(V^2)=V\times
\mathrm{rot}\,V+(V.\nabla)V =V\times \mathrm{rot}\,V + \nabla_V V.
$$
Clearly, on the two sides of this relation stay well defined quantities, i.e.
quantities defined in a coordinate free way. The first term on the right hand
side of this identity accounts for the rotational component of the change of
$V$, and the second term accounts mainly for the translational component of the
change of $V$. Making use of component notation we write down the last term on
the right side as follows (summation over the repeated indices):
$$
(\nabla_V
V)^j=V^i\nabla_i V^j=\nabla_i(V^iV^j)-V^j\nabla_iV^i=
\nabla_i(V^iV^j)-V^j\mathrm{div}\,V .
$$
Substituting into the first identity,
and making some elementary transformations we obtain
$$\nabla_i(M_{V}^{ij})=
\nabla_i\left(V^iV^j-\frac12 \delta^{ij}V^2\right)= \big
[(\mathrm{rot}\,V)\times V+V\mathrm{div}\,V\big ]^j,
$$
where $\delta^{ij}=1$
for $i=j$, and $\delta^{ij}=0$ for $i\neq j$ are the euclidean metric
components. If now $W$ is another vector field it must satisfy the same
identity:
$$\nabla_i(M_{W}^{ij})=
\nabla_i\left(W^iW^j-\frac12 \delta^{ij}W^2\right)= \big
[(\mathrm{rot}\,W)\times W+W\mathrm{div}\,W\big ]^j.
$$
Summing up these two
identities we obtain the new identity
\setlength\arraycolsep{8pt}
\begin{eqnarray*}
\lefteqn{ \nabla_iM^{ij}_{(V,W)}\equiv \nabla_i\left(V^iV^j+W^iW^j-
\delta^{ij}\frac{V^2+W^2}{2}\right)={} } \nonumber\\ & & {}=\big
[(\mathrm{rot}\,V)\times V+ V\mathrm{div}\,V+(\mathrm{rot}\,W)\times
W+W\mathrm{div}\,W\big ]^j.
\end{eqnarray*}

Let now $(a(x,y,z),b(x,y,z))$ be two arbitrary functions on $\mathbb{R}^3$. We
consider the transformation
$$
(V,W)\rightarrow (V\,a-W\,b,V\,b+W\,a).
$$

	{\bf Corollary.}

The tensor $M_{(V,W)}$ transforms to $(a^2+b^2)M_{(V,W)}$.

	{\bf Corollary.}

The transformations
$(V,W)\rightarrow (V\,a-W\,b,V\,b+W\,a)$ do not change the eigen
directions structure, i.e the eigen (sub)spaces, of $M^{ij}_{(V,W)}$.

	{\bf Corollary.}

If $a=\mathrm{cos}\,\theta, b=\mathrm{sin}\,\theta$, where
$\theta=\theta(x,y,z)$ then the tensor $M_{(V,W)}$ stays invariant:
$$
M_{(V,W)}=M(V\mathrm{cos}\,\theta-W\mathrm{sin}\,\theta,
V\mathrm{sin}\,\theta+W\mathrm{cos}\,\theta).
$$

The expression inside the parenteses above, denoted by $M^{ij}$, looks formally
the same as the introduced by Maxwell tensor from physical considerations
concerned with the electromagnetic stress energy properties of continuous media
in presence of external electromagnetic field. Hence, any vector $V$, or any
couple of vectors $(V,W)$, defines such tensor which we denote by $M_V$, or
$M_{(V,W)}$, and call {\bf Maxwell stress tensor}. The term, "stress" in this
general mathematical setting is not topologically motivated as in the
considerations connected with the Coulomb case, but could be justified in the
following way. Every vector field on $\mathbb{R}^3$ generates corresponding
flow by means of the trajectories started from some domain
$U_{t=0}\subset\mathbb{R}^3$, where $t$ is an arbitrary parameter: at $t>0$ the
domain $U_{t=0}$ is diffeomorphically transformed to a new domain
$U_t\subset\mathbb{R}^3$. Having two vector fields on $\mathbb{R}^3$ we obtain
two {\it compatible} flows, so, the points of any domain
$U_{t=0}\subset\mathbb{R}^3$ are forced to accordingly move to new positions.

We emphasize the following moments: {\bf first}, the identity we started with
is purely mathematical and $t$ is {\it arbitrary} parameter, not time in
general; {\bf second}, on the two sides of this identity stay well defined
coordinate free quantities; {\bf third}, these tensors do not introduce
interaction stress: the full stress is the sum of the stresses generated by
each one of the couple $(V,W)$.

Physically, we say that the corresponding physical medium that occupies the spatial
region $U_o$ and is parametrized by the points of the mathematical subregion
$U_o\subset\mathbb{R}^3$, is subject to {\it compatible} and {\it admissible}
physical "stresses" generated by physical interactions mathematically described
by the vector fields $(V,W)$, and these physical stresses are
quantitatively described by the corresponding physical interpretation of the
tensor $M^{ij}$. Clearly, we could extend the couple $(V,W)$ to more vectors
$(V_1,V_2,...,V_p)$, but then the mentioned invariance properties of
$M_{(V,W)}$ may be lost, or appropriately extended.

We note that the stress tensor $M^{ij}$ appears as been subject to the
divergence operator, and if we interpret the components of $M^{ij}$ as physical
stresses, then the left hand side of the divergence acquires in general the
physical interpretation of force density. Of course, in the static situation as
it is given by the relation considered, no energy-momentum propagation is
possible, so at every point the local forces mutually compensate:
$\nabla_{i}M^{ij}=0$. If propagation is allowed then the force field may NOT be
zero: $\nabla_{i}M^{ij}\neq 0$, and we may identify the right hand side as a
{\bf real time-change} of appropriately defined momentum density $\mathbf{P}$.
So, assuming some expression for this momentum density $\mathbf{P}$ we are
ready to write down corresponding field equation of motion of Newton type
through equalizing the spatially directed force densities $\nabla_{i}M^{ij}$
with the momentum density changes along the time variable, i.e. equalizing
$\nabla_iM^{ij}$ with the $ct$-derivative of $\mathbf{P}$, where $c=const$ is
the translational propagation velocity of the momentum density flow of the
physical system considered. In order to find how to choose $\mathbf{P}$ in
case of free EM-field we have to turn to the intrinsic physical properties of
the field, so, it seems natural to turn to the eigen properties of $M^{ij}$,
since, clearly, namely $M^{ij}_{(\mathbf{E},\mathbf{B})}$ is assumed to carry
the physical properties of the field.

\section{Eigen properties of Maxwell stress tensor}
\index{maxwell stress tensor properties}
We consider $M^{ij}(\mathbf{E},\mathbf{B})$ at some point $p\in\mathbb{R}^3$
and assume that in general the vector fields $\mathbf{E}$ and $\mathbf{B}$ are
lineary independent, so $\mathbf{E}\times\mathbf{B}\neq 0$. Let the coordinate
system be chosen such that the coordinate plane $(x,y)$ to coincide with the
plane defined by $\mathbf{E}(p),\mathbf{B}(p)$. In this coordinate system
$\mathbf{E}=(E_1,E_2,0)$ and $\mathbf{B}=(B_1,B_2,0)$, so, identifying the
contravariant and covariant indices through the Euclidean metric $\delta^{ij}$
(so that $M^{ij}=M^i_j=M_{ij}$), we obtain the following nonzero components of
the stress tensor: $$ M^1_1=(E^1)^2+(B^1)^2-\frac12(\mathbf{E}^2+\mathbf{B}^2);
\ \ M^1_2=M^2_1=E^1\,E_2+B_1\,B^2; $$ $$
M^2_2=(E^2)^2+(B^2)^2-\frac12(\mathbf{E}^2+\mathbf{B}^2); \ \
M^3_3=-\frac12(\mathbf{E}^2+\mathbf{B}^2). $$ Since $M^1_1=-M^2_2$, the trace
of $M$ is $Tr(M)=-\frac12(\mathbf{E}^2+\mathbf{B}^2)$.

The eigen value equation acquires the simple
form
$$
\big[(M^1_1)^2-(\lambda)^2\big]+(M^1_2)^2\big](M^3_3-\lambda)=0.
$$
The corresponding eigen values are
$$
\lambda_1=-\frac12(\mathbf{E}^2+\mathbf{B}^2);\ \
\lambda_{2,3}=\pm\sqrt{(M^1_1)^2+(M^1_2)^2}= \pm\frac12\sqrt{(I_1)^2+(I_2)^2} ,
$$
where
$I_1=\mathbf{B}^2-\mathbf{E}^2,\, I_2=2\mathbf{E}.\mathbf{B}$.

The corresponding to
$\lambda_1$ eigen vector $Z_1$ must satisfy the equation
$$
\mathbf{E}(\mathbf{E}.Z_1)+\mathbf{B}(\mathbf{B}.Z_1)=0,
$$
 and since
$(\mathbf{E},\mathbf{B})$ are lineary independent, the two coefficients
$(\mathbf{E}.Z_1)$ and $(\mathbf{B}.Z_1)$ must be equal to
zero, therefore, $Z_1\neq 0$ must be orthogonal to $\mathbf{E}$ and
$\mathbf{B}$, i.e. $Z_1$ must be colinear to $\mathbf{E}\times\mathbf{B}$:

The other two eigen vectors $Z_{2,3}$
satisfy correspondingly the equations
$$ \mathbf{E}(\mathbf{E}.Z_{2,3})+
\mathbf{B}(\mathbf{B}.Z_{2,3})=\Big[\pm\frac12\sqrt{(I_1)^2+(I_2)^2}+
\frac12(\mathbf{E}^2+\mathbf{B}^2)\Big]Z_{2,3}.         \ \ \ \ \ \   (*)
$$
Taking into account the easily verified relation
$$
\frac14\Big[(I_1)^2+(I_2)^2\Big]=
\left(\frac{\mathbf{E}^2+\mathbf{B}^2}{2}\right)^2-
|\mathbf{E}\times\mathbf{B}|^2 ,
$$
so that
$$
\frac{\mathbf{E}^2+\mathbf{B}^2}{2}- |\mathbf{E}\times\mathbf{B}|\geq 0 \ ,
$$
we conclude that the coefficient before $Z_{2,3}$ on the right is always different
from zero, therefore, the eigen vectors $Z_{2,3}(p)$ lie in the plane defined
by $(\mathbf{E}(p),\mathbf{B}(p)), \ p\in \mathbb{R}^3$. In particular,
the above mentioned transformation properties of the
Maxwell stress tensor $M(V,W)\rightarrow (a^2+b^2)M(V,W)$ show that the
corresponding eigen directions do not change under the transformation
$(V,W)\rightarrow (V\,a-W\,b,V\,b+W\,a)$.
\vskip 0.2cm
The above consideration suggests: {\it the intrinsically allowed
dynamical abilities of the field are: translational
along $(\mathbf{E}\times\mathbf{B})$, and rotational inside the plane defined
by $(\mathbf{E},\mathbf{B})$, hence, we may expect finding field objects the
propagation of which shows intrinsic local consistency between rotation and
translation.}
\vskip 0.2cm
It is natural to ask now under what conditions the very $\mathbf{E}$ and
$\mathbf{B}$ may be eigen vectors of $M(\mathbf{E},\mathbf{B})$? Assuming
$\lambda_2=\frac12\sqrt{(I_1)^2+(I_2)^2}$ and $Z_2=\mathbf{E}$
in the above relation (*) and having in view that
$\mathbf{E}\times\mathbf{B}\neq 0$ we obtain that
$\mathbf{E}(\mathbf{E}^2)+\mathbf{B}(\mathbf{E}.\mathbf{B})$ must be
proportional to $\mathbf{E}$, so, $\mathbf{E}.\mathbf{B}=0$,
i.e. $I_2=0$. Moreover, substituting now $I_2=0$ in that same  relation we obtain
$$
\mathbf{E}^2=\frac12(\mathbf{B}^2-\mathbf{E}^2)+
\frac12(\mathbf{E}^2+\mathbf{B}^2)=\mathbf{B}^2, \ \ \text{i.e.}, \ \ I_1=0.
$$
The case
"-" sign before the square root, i.e. $\lambda_3=-\frac12\sqrt{(I_1)^2+(I_2)^2}$,
leads to analogical conclusions just the role of $\mathbf{E}$ and $\mathbf{B}$
is exchanged.

\vskip 0.2cm
{\bf Corollary}. $\mathbf{E}$ and $\mathbf{B}$ may be eigen
vectors of $M(\mathbf{E},\mathbf{B})$ only if
 $I_1=I_2=0$.
\vskip 0.2cm

The above notices suggest to consider in a more detail the case
$\lambda_2=-\lambda_3=0$ for the vacuum case. We shall show, making use of the
Lorentz transformation in 3-dimensional form that, if
these two relations do not hold then under $\mathbf{E}\times\mathbf{B}\neq 0$
the translational velocity of propagation
is less then the speed of light in vacuum $c$. Recall first the
transformation laws of the electric and magnetic vectors under Lorentz
transformation defined by the 3-velocity vector $\mathbf{v}$ and corresponding
parameter $\beta=v/c, v=|\mathbf{v}|$. If $\gamma$ denotes the factor
$1/\sqrt{1-\beta^2}$ then we have $$
\mathbf{E'}=\gamma\,\mathbf{E}+\frac{1-\gamma}{v^2}\mathbf{v}(\mathbf{E}.
\mathbf{v})+\frac{\gamma}{c}\mathbf{v}\times\mathbf{B} , $$ $$
\mathbf{B'}=\gamma\,\mathbf{B}+\frac{1-\gamma}{v^2}\mathbf{v}(\mathbf{B}.
\mathbf{v})-\frac{\gamma}{c}\mathbf{v}\times\mathbf{E} . $$

Assume first that $I_2=2\mathbf{E}.\mathbf{B}=0$, i.e. $\mathbf{E}$ and $\mathbf{B}$ are
orthogonal, so, in general, in some coordinate system we shall have
$\mathbf{E}\times\mathbf{B}\neq 0$ .

If $I_1>0$, i.e. $|\mathbf{E}|<|\mathbf{B}|$, we shall show that the
conditions $\mathbf{E'}=0, \mathbf{v}.\mathbf{B}=0, \infty>\gamma>0$ are
compatible. In fact, these assumptions lead to
$\gamma\,\mathbf{v}.\mathbf{E}+(1-\gamma)(\mathbf{E}.\mathbf{v})=0$, i.e.
$\mathbf{E}.\mathbf{v}=0$. Thus,
 $c|\mathbf{E}|=v|\mathbf{B}||\mathrm{sin}(\mathbf{v},\mathbf{B})|$, and since
$\mathbf{v}.\mathbf{B}=0$ then $|\mathrm{sin}(\mathbf{v},\mathbf{B})|=1$.
It follows that the speed
$v=c\frac{|\mathbf{E}|}{|\mathbf{B}|}<c$ is allowed.

If $I_1<0$, i.e. $|\mathbf{E}|>|\mathbf{B}|$, then the conditions
$\mathbf{B'}=0$ and $\mathbf{v}.\mathbf{E}=0$ analogically lead to the
conclusion that the speed $v=c\frac{|\mathbf{B}|}{|\mathbf{E}|}<c$ is allowed.

Assume now that $I_2=2\mathbf{E}.\mathbf{B}\neq 0$. We are looking for a reference frame
$K'$ such that $\mathbf{E'}\times\mathbf{B'}=0$, while in the reference frame $K$ we have
$\mathbf{E}\times\mathbf{B}\neq 0$. We choose the relative velocity $\mathbf{v}$ such
that $\mathbf{v}.\mathbf{E}= \mathbf{v}.\mathbf{B}=0$. Under these conditions the
equation $\mathbf{E'}\times\mathbf{B'}=0$ reduces to $$
\mathbf{E}\times\mathbf{B}+\frac{\mathbf{v}}{c}(\mathbf{E}^2+\mathbf{B}^2)=0 ,\ \
\text{so}, \ \ \frac{v}{c}=|\mathbf{E}\times\mathbf{B}|/(\mathbf{E}^2+\mathbf{B}^2). $$
Now, from the above mentioned inequality
$\mathbf{E}^2+\mathbf{B}^2-2|\mathbf{E}\times\mathbf{B}|\geq 0$ it follows that
$\frac{v}{c}<1$.

Physically, these considerations show that under nonzero $I_1$ and $I_2$ the
translational velocity of propagation of the field, and of the stress field
energy density of course, will be less than $c$. Hence, the only realistic
choice for the vacuum case (where this velocity is assumed by definition to be
equal to $c$), is $I_1=I_2=0$, which is equivalent to
$\mathbf{E}^2+\mathbf{B}^2=2|\mathbf{E}\times\mathbf{B}|$. Hence, assuming
$|Tr(M)|$ to be the stress energy density of the field, the names
"electromagnetic energy flux" for the quantity $c\mathbf{E}\times\mathbf{B}$,
and "momentum" for the quantity $\frac1c\mathbf{E}\times\mathbf{B}$, seem well
justified without turning to any dynamical field equations.

These considerations suggest also that if $I_1=0$, i.e.
$|\mathbf{E}|^2=|\mathbf{B}|^2$ during propagation, then the electric and
magnetic components of the field should carry always the same stress energy
density, so, a local mutual energy exchange between $\mathbf{E}$ and
$\mathbf{B}$ is not forbidden in general, but, if it takes place, it must be
{\it simultaneous} and in {\it equal quantities}. Hence, under zero invariants
$I_1=0$ and $I_2=2\mathbf{E}.\mathbf{B}=0$, internal energy redistribution
among possible subsystems of the field would be allowed but such an exchange
should occur {\it without available interaction energy} because the full energy
density is always equal to the sum of the energy densities carried by the
electric and magnetic components of the field. However, the required time
stability and propagation with velocity "c" of the field suggest/imply also
available internal momentum exchange since under these conditions the energy
density is always equal to
$|\mathbf{E}\times\mathbf{B}|$, and $\mathbf{E}$ and $\mathbf{B}$ can not carry
momentum separately.

The following question now arizes: is it {\it physically} allowed to interprit
each of the two vector fields  $\mathbf{E},\mathbf{B}$ as mathematical image of
a recognizable time-stable physical subsystem of the EM-field?

Trying to answer this question we note that the relation
$\mathbf{E}^2+\mathbf{B}^2=2|\mathbf{E}\times\mathbf{B}|$ and the required
time-recognizability during propagation (with velocity "c") of each subsystem
of the field suggest/imply also that {\it each of the two subsystems must be
able to carry locally momentum and to exchange locally momentum with the other
one}, since this relation means that the energy density is always strongly
proportional to the momentum density magnitude
$\frac1c|\mathbf{E}\times\mathbf{B}|$. Hence, the couple
$(\mathbf{E},\mathbf{B})$ is able to carry momentum, but neither of
$\mathbf{E},\mathbf{B}$ can carry momentum separately. Moreover, the important
observation here is that, verious combinations constructed out of the
constituents $\mathbf{E}$ and $\mathbf{B}$, e.g.,
$(\mathbf{E}\,cos\theta-\mathbf{B}\,sin\theta,
\mathbf{E}\,sin\theta+\mathbf{B}\,cos\theta)$, where $\theta(x,y,z;t)$ is a
functon, may be considered as possible representatives of the two recognizable
subsystems since they carry the same energy
$\frac12(\mathbf{E}^2+\mathbf{B}^2)$ and momentum
$\frac1c|\mathbf{E}\times\mathbf{B}|$ densities. Therefore, the suggestion by
Maxwell vacuum equations that the very $\mathbf{E}$ and $\mathbf{B}$ may be
considered as appropriate mathematical images of recognizable time-stable
subsystems of a time-dependent and space propagating electromagnetic field
object does NOT seem adequate and has to be reconsidered.

Hence, which combinations of $\mathbf{E}$ and $\mathbf{B}$ deserve to represent
mathematically the two subsystems of a time-dependent and space-propagating
electromagnetic field object?

In view of these considerations we assume the following understanding:
\vskip 0.3cm
{\bf Every real
EM-field is built of two recognizable subsystems, the mathematical images of
which are not the very $(\mathbf{E},\mathbf{B})$, but are
expressed in terms of $(\mathbf{E},\mathbf{B})$, both these
subsystems carry always the same quantity of energy-momentum, guaranteeing in
this way that the supposed internal energy-momentum exchange will also be
in equal quantities and simultanious}.

\section{Double field notion about time-dependent EM-fields}

In accordance with the above assumption the description of dynamical and
space-propagating behavior of the field will need two appropriate mathematical
objects to be constructed out of the two constituents
$(\mathbf{E},\mathbf{B})$. These two mathematical objects must meet the
required property that the two physical subsystems of the field carry always
the same quantity of energy-momentum, and that any possible internal
energy-momentum exchange between the two subsystems shall be {\it simultaneous}
and {\it in equal quantities}.

We are going to consider time dependent fields, and begin with noting once
again the assumption that the full stress tensor (and the energy density, in
particular) is a sum of the stress tensors carried separately by the two
subsystems. As we mentiond above, this does NOT mean that there is no energy
exchange between the two subsystems of the field.

Now, following the above stated idea we have to find two appropriate
mathematical images of the field which images are NOT  represented directly by
the electric $\mathbf{E}$ and magnetic $\mathbf{B}$ vectors, but are
constructed out of them. In terms of these two appropriate mathematical
representatives of the corresponding two partnering subsystems we must express
the mentioned  special kind of energy-momentum exchange, respecting in this way
the fact that {\it neither} of the two constituents
$(\mathbf{E},\mathbf{B})$ {\it is able to carry momentum separately}.

In view of the above we have to assume that the field keeps its identity through
adopting some special and appropriate dynamical behavior according to its
intrinsic capabilities. Since the corresponding dynamical/field behavior must
be consistent with the properties of the intrinsic stress-energy-momentum nature
of the field, we come to the conclusion that Maxwell stress tensor
$M(\mathbf{E},\mathbf{B})$ should play the basic role, and its zero-divergence
in the static case should suggest how to determine the appropriate
structure and allowed dynamical propagation.

Recall that any member of the family
$$
(\mathcal{E},\mathcal{B})=(\mathbf{E},\mathbf{B},\theta)=
(\mathbf{E}\,\mathrm{cos}\,\theta-
\mathbf{B}\,\mathrm{sin}\,\theta; \ \mathbf{E}\,\mathrm{sin}\,\theta+
\mathbf{B}\,\mathrm{cos}\,\theta), \ \ \theta=\theta(x,y,z;t),
$$
generates the same Maxwell stress tensor. So, the most natural assumption should
read like this:

{\bf Any member
$(\mathbf{E},\mathbf{B},\theta_1)$
of this $\theta$-family is looking for an energy-momentum exchanging partner
$(\mathbf{E},\mathbf{B},\theta_2)$ inside the family, and identifies itself
through appropriate (local) interaction with the partner found, defining in
this way corresponding dynamical behavior of the field}.

Simply speaking, a time-dependent EM-field is formally represented by two
members of the above $\theta$-family, and the coupling
$(\mathbf{E},\mathbf{B},\theta_1)
\leftrightarrow(\mathbf{E},\mathbf{B},\theta_2)$ is unique.

%Assuming for simplicity $\alpha_1=0$ then
%$(\mathcal{E},\mathcal{B},0)\rightarrow(\mathbf{E},\mathbf{B})$

Note that working with $\theta$-invariant quantities, e.g., $M^{ij}$ and
$\mathbf{E}\times\mathbf{B}$, we may consider the couple
$(\mathbf{E},\mathbf{B})$ as any member of the $\alpha$-family. In view of this
we shall make use of the local divergence of the Maxwell stress tensor and the
time derivative of the local momentum flow of the field in order to find
the corresponding partner-subsystem of $(\mathbf{E},\mathbf{B})$.

Further we shall call these two subsystems just partner-fields.

Recall the divergence \setlength\arraycolsep{8pt}
\begin{eqnarray*} \lefteqn{ \nabla_iM^{ij}\equiv
\nabla_i\left(\mathbf{E}^i\mathbf{E}^j+\mathbf{B}^i\mathbf{B}^j-
\delta^{ij}\frac{\mathbf{E}^2+\mathbf{B}^2}{2}\right)={} } \nonumber\\ & & {}= \big
[(\mathrm{rot}\,\mathbf{E})\times \mathbf{E}+ \mathbf{E}\mathrm{div}\,\mathbf{E}+
(\mathrm{rot}\,\mathbf{B})\times \mathbf{B}+ \mathbf{B}\mathrm{div}\,\mathbf{B}\big ]^j.
\end{eqnarray*}

As we mentioned, in the static case, i.e., when the vector fields
$(\mathbf{E},\mathbf{B})$ do not depend on the time "coordinate" $\xi=ct$, NO
propagation of field momentum density $\mathbf{P}$ should take place, so, at
every point, where $(\mathbf{E},\mathbf{B})\neq 0$, the  stress generated
forces must mutually compensate, i.e., the divergence $\nabla_iM^{ij}$ should be
equal to zero: $\nabla_iM^{ij}=0$. In this static case Maxwell vacuum equations
$$
\mathrm{rot}\,\mathbf{E}+\frac{\partial\mathbf{B}}{\partial \xi}=0,\quad
\mathrm{rot}\,\mathbf{B}-\frac{\partial\mathbf{E}}{\partial \xi}=0,\quad
\mathrm{div}\,\mathbf{E}=0,\quad \mathrm{div}\,\mathbf{B}=0 \ \ \
(\text{J.C.M.})
$$
give: $\mathrm{rot}\mathbf{E}=\mathrm{rot}\mathbf{B}=0;\,
\mathrm{div}\mathbf{E}=\mathrm{div}\mathbf{B}=0$, so, all static solutions to
Maxwell equations determine a sufficient, but NOT necessary, condition that
brings to zero the right hand side of the divergence through forcing each of
the four vectors there to get zero values.

In the non-static case, i.e. when $\frac{\partial\mathbf{E}}{\partial t}\neq 0;
\,\frac{\partial\mathbf{B}}{\partial t}\neq 0$, time change and propagation of
field momentum density should be expected, so, a full mutual compensation of the
generated by the Maxwell stresses at every spatial point local forces may NOT
be possible, which means $\nabla_iM^{ij}\neq 0$ in general. These local forces
generate time-dependent momentum inside the corresponding region.
Therefore, if we want to describe this physical process of field
energy-momentum density time change and spatial propagation we have to
introduce explicitly the dependence of the local momentum vector
field $\mathbf{P}$ on $(\mathbf{E},\mathbf{B})$, and to express the flow of the
electromagnetic energy-momentum across an arbitrary static finite 2-dimensional
surface $S$ in two ways: in terms of $\nabla_iM^{ij}\neq 0$ and in terms of the
time change of $\mathbf{P}(\mathbf{E},\mathbf{B})$, and then to appropriately
equilize them. Hence, we have to construct the corresponding two differential
2-forms to be integrated on $S$.

Note that compare to classical approach where the flows of
the very $\mathbf{E},\mathbf{B}$ through some 2-surface are considered, we
consider flows of quantities having direct stress-energy-momentum change sense.

In terms of $\mathfrak{F}^j=\nabla_iM^{ij}\neq 0$ the 2-form that is to be
integrated on $S$ is given by reducing
$i_{\mathfrak{F}}\omega_o=i_{\mathfrak{F}}(\sqrt{|g|}dx^1\wedge dx^2\wedge
dx^3)=*\tilde{g}(\mathfrak{F})$ on $S$,
where $i_{\mathfrak{F}}$ denotes the interior product
between the vector field $\mathfrak{F}$ and the volume 3-form $\omega_o$,
i.e. the local flow of $\mathfrak{F}$ across $\omega_o$, and
$*$ denotes the euclidean Hodge $*$. On the other hand, the momentum density
flow time change across $S$ should naturally be represented by
$\frac{d}{dt}\int_{S}*\tilde{g}(\mathbf{P}(\mathbf{E},\mathbf{B}))$
(recall that $t$ is considered as external parameter).
Restricting now $*\tilde{g}(\mathbf{P(\mathbf{E},\mathbf{B})})$ and
$*\tilde{g}(\mathfrak{F})$ on $S$ we get:
$$
\frac{d}{dt}\int_{S}*\tilde{g}(\mathbf{P(\mathbf{E},\mathbf{B})})=
\int_{S}*\tilde{g}(\mathfrak{F}).
$$
 \vskip 0.3cm
 The explicit expression for
$\mathbf{P}(\mathbf{E},\mathbf{B})$, paying due respect to J.Poynting, and
to J.J.Thomson, H.Poincare, M. Abraham, and in view of the huge, a century
and a half available experience, has to be introduced by the following
\vskip
0.4cm \noindent {\bf Assumption}: {\it The entire field momentum density is
given by $\mathbf{P}:=\frac1c\mathbf{E}\times\mathbf{B}$} .
 \vskip 0.4cm
According to this {\bf Assumption}, to the above interpretation of the relation
$\nabla_iM^{ij}\neq 0$, in view of the assumed by us local
energy-momentum exchange approach to description of the dynamics of the field,
in vector field terms and in canonical coordinates on $\mathbb{R}^3$ we come to
the following vector differential equation (the 2-surface $S$ is arbitrary and
static)
 $$
\frac{\partial}{\partial
\xi}\left(\mathbf{E}\times\mathbf{B}\right)=\mathfrak{F}, \ \ \ \xi\equiv ct,   \ \ \ \ \
\ \ (*)
$$
which is equivalent to
$$
\left(\mathrm{rot}\,\mathbf{E}+\frac{\partial\mathbf{B}}{\partial \xi}\right)\times
\mathbf{E}+ \mathbf{E}\mathrm{div}\,\mathbf{E}+
\left(\mathrm{rot}\,\mathbf{B}-\frac{\partial\mathbf{E}}{\partial \xi}\right)\times
\mathbf{B}+ \mathbf{B}\mathrm{div}\,\mathbf{B}=0.
$$
This last equation we write down in the following equivalent way:
$$
\left(\mathrm{rot}\,\mathbf{E}+\frac{\partial\mathbf{B}}{\partial \xi}\right)\times
\mathbf{E}+\mathbf{B}\mathrm{div}\,\mathbf{B}=
-\left[\left(\mathrm{rot}\,\mathbf{B}-\frac{\partial\mathbf{E}}{\partial
\xi}\right)\times\mathbf{B}+\mathbf{E}\mathrm{div}\,\mathbf{E}\right].\ \ \ \
(**)
 $$

The above relation (*) and its explicit forms we consider as mathematical
adequate in energy-momentum-change terms of the so called electric-magnetic and
magnetic-electric induction phenomena in the charge free case. We recall that
it is usually assumed these induction phenomena to be described in classical
electrodynamics by the following well known integral equations
$$
\frac{d}{d\xi}\int_{S}*\tilde{g}(\mathbf{B})|_S=-
\int_{S}*\tilde{g}(\mathrm{rot}\mathbf{E})|_S
\ \ \ \text{(the Faraday induction law)},
$$
$$
\frac{d}{d\xi}\int_{S}*\tilde{g}(\mathbf{E})|_S=
\int_{S}*\tilde{g}(\mathrm{rot}\mathbf{B})|_S
\ \ \ \text{(the Maxwell displacement current law)},
$$
where $(...)|_S$ means restriction of the corresponding 2-form to the 2-surface
$S$.

We would like to note that these last Faraday-Maxwell relations have NO
{\it direct} energy-momentum change-propagation (i.e. force flow) nature, so
they could not be experimentally verified in a {\it direct} way. Our feeling is
that, in fact, they are stronger than needed. So, on the corresponding
solutions of these equations we'll be able to write down {\it formally
adequate} energy-momentum change expressions, but the correspondence of these
expressions with the experiment will crucially depend on the nature of these
solutions. As is well known, the nature of the free solutions (with no
boundary conditions) to Maxwell vacuum equations with spatially finite and
smooth enough initial conditions requires strong time-instability (the
corresponding  theorem for the D'Alembert wave equation which each component of
$\mathbf{E}$ and $\mathbf{B}$ must necessarily satisfy). And time-stability of
time-dependent vacuum solutions usually requires spatial infinity (plane
waves), which is physically senseless. Making calculations with spatially
finite parts of these spatially infinite solutions may be practically
acceptable, but from theoretical viewpoint assuming these equations for {\it
basic} ones seems not acceptable since the relation "time stable physical
object - exact free solution" is strongly violated.

Before to go further we write down the right hand side bracket expression of
$(**)$ in the following two equivalent ways:
$$
\left[\left(\mathrm{rot}\,\mathbf{B}+\frac{\partial\mathbf{(-E)}}{\partial
\xi}\right)\times \mathbf{B}+ \mathbf{(-E)}\mathrm{div}\,\mathbf{(-E)}\right];\,%5%
$$
$$
\left[\left(\mathrm{rot}\,\mathbf{(-B)}+\frac{\partial\mathbf{E}}
{\partial\xi}\right)\times\mathbf{(-B)}+\mathbf{E}\mathrm{div}\,\mathbf{E}\right].
$$
These last two expressions can be considered as obtained from the left
hand side of the above relation $(**)$  under the substitutions
$(\mathbf{E},\mathbf{B})\rightarrow(\mathbf{B},\mathbf{-E})$ and
$(\mathbf{E},\mathbf{B})\rightarrow(\mathbf{-B},\mathbf{E})$ respectively.
Hence, the subsystem $(\mathbf{E},\mathbf{B})$ chooses as a partner
 $(\mathbf{-B},\mathbf{E})$, or $(\mathbf{B},\mathbf{-E})$.
We conclude that the subsystem
$(\mathbf{E},\mathbf{B},\alpha)$ will choose as partner-susbsystem
$(\mathbf{E},\mathbf{B},\alpha+\frac{\pi}{2})$ or
$(\mathbf{E},\mathbf{B},\alpha-\frac{\pi}{2})$.

We may summarize this nonrelativistic approach as follows: \index{notion of EM-field}
%\vskip 0.3cm

{\bf A real free field consists of two interacting subsystems
$(\Sigma_1,\Sigma_2)$, and each
subsystem is described by two partner-fields
inside the $\theta(x,y,z;t)$-family
$$
(\mathcal{E},\mathcal{B})=
(\mathbf{E}\,\mathrm{cos}\,\theta- \mathbf{B}\,\mathrm{sin}\,\theta; \
\mathbf{E}\,\mathrm{sin}\,\theta+ \mathbf{B}\,\mathrm{cos}\,\theta),
$$
$\Sigma_1=(\mathcal{E}_{\theta_{1}},\mathcal{B}_{\theta_{1}}),
\Sigma_2=(\mathcal{E}_{\theta_{2}},\mathcal{B}_{\theta_{2}})$, giving the same
Maxwell stress-energy tensor, so the full stress energy tensor is the sum:
$M(\Sigma_1,\Sigma_2)=\frac12M(\Sigma_1)+\frac12M(\Sigma_2)$. Each
partner-field has interacting electric and magnetic constituents, and each
partner-field is determined by the other through $(\pm\frac{\pi}{2})$ -
rotation-like transformation.  Both partner-fields carry the same
stress-energy-momentum : $M(\Sigma_1)=M(\Sigma_2)$, and the field propagates in
space through minimizing the relation} $I_1^2+I_2^2\geqslant 0$. {\bf The
intrinsic dynamics of a free real time-dependent field establishes and
maintains local energy-momentum exchange partnership between the two
partner-fields, and since these partner-fields carry always the same
stress-energy, the allowed exchange is necessarily simultaneous and in equal
quantities, so, each partner-field conserves its energy-momentum during
propagation}.

\section{Internal interaction and evolution in energy-momentum terms}

In order to find how much is the locally exchanged energy-momentum we are going
to interpret the above equation in accordance with the view on equations of
motion as stated in Sec.4.1. Our object of interest $\Phi$, representing the
{\it wholeness} and {\it integrity} of a real time dependent electromagnetic
field, is the couple
$\Big[(\mathbf{E},\mathbf{B});(\mathbf{-B},\mathbf{E})\Big]$ (the other case
$\Big[(\mathbf{E},\mathbf{B});(\mathbf{B},\mathbf{-E})\Big]$ is considered
analogically). In view of the above considerations our equations should
directly describe admissible energy-momentum exchange between these two
recognizable subsystems. Hence, we have to define the {\it admissible}
change-objects $D(\mathbf{E},\mathbf{B})$ and $D(\mathbf{-B},\mathbf{E})$
(having, of course, tensor nature) for each partner-field, their
self-"projections" and their mutual "projections", i.e. the corresponding
"change-field" flows in energy-momentum-change terms.

The explicit forms of {\it non-zero admissible changes} and {\it their
"projections" on the partner-fields}, are suggested by the developed form of
equation (**). Following this suggestion, the change object
$D(\mathbf{E},\mathbf{B})$ for the first partner-field
$(\mathbf{E},\mathbf{B})$ we naturally define as
$$
D(\mathbf{E},\mathbf{B}):=
\left(\mathrm{rot}\mathbf{E}+ \frac{\partial\mathbf{B}}{\partial \xi}; \,
\mathrm{div}\mathbf{B}\right).
$$
The corresponding "projection"
$\mathfrak{P}:D(\mathbf{E},\mathbf{B})\rightarrow(\mathbf{E},\mathbf{B})$
$$
\mathfrak{P}\left[D(\mathbf{E},\mathbf{B}); (\mathbf{E},\mathbf{B})\right]=
\mathfrak{P}\left[\left(\mathrm{rot}\mathbf{E}+ \frac{\partial\mathbf{B}}{\partial \xi};
\, \mathrm{div}\mathbf{B}\right); (\mathbf{E},\mathbf{B})\right]
$$
is suggested by the left hand side of the above energy-momentum
exchange expressions, so we define
it by : $$ \mathfrak{P}\left[\left(\mathrm{rot}\mathbf{E}+
\frac{\partial\mathbf{B}}{\partial \xi}; \, \mathrm{div}\mathbf{B}\right);
(\mathbf{E},\mathbf{B})\right]:=
\left(\mathrm{rot}\,\mathbf{E}+\frac{\partial\mathbf{B}}{\partial \xi}\right)\times
\mathbf{E}+\mathbf{B}\mathrm{div}\,\mathbf{B}.
$$
For the second partner-field
$(-\mathbf{B},\mathbf{E})$, following the same procedure we obtain:
$$
\mathfrak{P}\left[D(\mathbf{-B},\mathbf{E}); (\mathbf{-B},\mathbf{E})\right]=
\mathfrak{P}\left[\left(\mathrm{rot}\mathbf{(-B)}+ \frac{\partial\mathbf{E}}{\partial
\xi}; \, \mathrm{div}\mathbf{E}\right); (\mathbf{-B},\mathbf{E})\right]
$$
$$
=\left(\mathrm{rot}\,\mathbf{(-B)}+\frac{\partial\mathbf{E}}{\partial \xi}\right)\times
\mathbf{(-B)}+\mathbf{E}\mathrm{div}\,\mathbf{E}=
\left(\mathrm{rot}\,\mathbf{B}-\frac{\partial\mathbf{E}}{\partial \xi}\right)\times
\mathbf{B}+\mathbf{E}\mathrm{div}\,\mathbf{E}.
$$
Hence, relation (**) takes the form
$$
\mathfrak{P}\left[D(\mathbf{E},\mathbf{B});
(\mathbf{E},\mathbf{B})\right]+ \mathfrak{P}\left[D(\mathbf{-B},\mathbf{E});
(\mathbf{-B},\mathbf{E})\right]=0 .
$$

The accepted "two subsystem" view on a real time dependent electromagnetic field
allows in principle admissible energy-momentum exchange with the outside world
through any of the two partner-fields. The above calculations determine how
much each partner-field $(\mathbf{E},\mathbf{B})$, or
$(\mathbf{-B},\mathbf{E})$, is potentially able to give to some other physical
object without destroying itself, and these quantities are expressed in terms
of $\mathbf{E},\mathbf{B}$ and their derivatives only. In the case of free
field, since no energy-momentum is lost by the field, there are two
possibilities: first, there is NO energy-momentum exchange between the two
partner-fields, second, each of the partner-fields changes its energy-momentum
at the expense of the other through simultaneous and in equal quantity
exchanges. Such kind of mutual exchange is in correspondence with the
mathematical representatives of the two subsystems: the partner-fields
$(\mathbf{E},\mathbf{B})$ and $(\mathbf{-B},\mathbf{E})$ being members of the
above mentioned $\alpha(x,y,z;t)$-family, obviously carry the same energy
and momentum. If we denote by $\Delta_{11}$ and
by $\Delta_{22}$ the allowed energy-momentum changes of the two
partner-fields, by $\Delta_{12}$ the energy-momentum that the first
partner-field receives from the second partner-field, and by $\Delta_{21}$ the
energy-momentum that the second partner-field receives from the first
partner-field, then according to the energy-momentum local conservation law we
may write the following equations:
$$
\Delta_{11}=\Delta_{12}+\Delta_{21}; \ \
\Delta_{22}=-\left(\Delta_{21}+\Delta_{12}\right),
$$
which gives $\Delta_{11}+\Delta_{22}=0$.

We determine now how the mutual exchange between the two partner-fields
$(\mathbf{E},\mathbf{B})\rightleftarrows
(\mathbf{-B},\mathbf{E})$, or,
$(\mathbf{E},\mathbf{B})\rightleftarrows
(\mathbf{B},\mathbf{-E})$ is performed, i.e. the explicit expressions for
$\Delta_{12}$ and $\Delta_{21}$,
 keeping in mind that both subsystems carry equal energy-momentum densities.
The formal expressions are easy to obtain. In fact, in the case
$(\mathbf{E},\mathbf{B})\rightarrow (\mathbf{-B},\mathbf{E})$,
i.e. the quantity $\Delta_{21}$, we have to "project" the change object for the second
partner-field given by
$$
D(\mathbf{-B},\mathbf{E}):= \left(\mathrm{rot}\mathbf{(-B)}+
\frac{\partial\mathbf{E}}{\partial \xi}; \, \mathrm{div}\mathbf{E}\right)
$$
on the first partner-field $(\mathbf{E},\mathbf{B})$. We obtain:
$$
\Delta_{21}=
\left(\mathrm{rot}\,(\mathbf{-B})+\frac{\partial\mathbf{E}}{\partial \xi}\right)\times
\mathbf{E}+\mathbf{B}\mathrm{div}\,\mathbf{E}=
-\left(\mathrm{rot}\,\mathbf{B}-\frac{\partial\mathbf{E}}{\partial \xi}\right)\times
\mathbf{E}+\mathbf{B}\mathrm{div}\,\mathbf{E} \ .
$$
In the reverse case
$(\mathbf{-B},\mathbf{E})\rightarrow (\mathbf{E},\mathbf{B})$,
i.e. the quantity $\Delta_{12}$, we have to "project" the change-object for the
first partner-field $(\mathbf{E},\mathbf{B})$ given by $$
D(\mathbf{E},\mathbf{B}):= \left(\mathrm{rot}\mathbf{E}+
\frac{\partial\mathbf{B}}{\partial \xi}; \, \mathrm{div}\mathbf{B}\right)
$$
on the second partner-field $(\mathbf{-B},\mathbf{E})$. We obtain
$$
\Delta_{12}= \left(\mathrm{rot}\,\mathbf{E}+\frac{\partial\mathbf{B}}{\partial
\xi}\right)\times (\mathbf{-B})+\mathbf{E}\mathrm{div}\,\mathbf{B}=
-\left(\mathrm{rot}\,\mathbf{E}+\frac{\partial\mathbf{B}}{\partial
\xi}\right)\times\mathbf{B}+\mathbf{E}\mathrm{div}\,\mathbf{B}.
$$
So, the
internal local balance is governed by the equations
\vskip 0.2cm
\begin{eqnarray*}
&&\left(\mathrm{rot}\,\mathbf{E}+\frac{\partial\mathbf{B}}{\partial
\xi}\right)\times \mathbf{E}+\mathbf{B}\mathrm{div}\,\mathbf{B}\\
&&=-\left(\mathrm{rot}\,\mathbf{E}+\frac{\partial\mathbf{B}}{\partial
\xi}\right)\times \mathbf{B}+\mathbf{E}\mathrm{div}\,\mathbf{B}-
\left(\mathrm{rot}\,\mathbf{B}-\frac{\partial\mathbf{E}}{\partial \xi}\right)\times
\mathbf{E}+\mathbf{B}\mathrm{div}\,\mathbf{E}, \\
&&\left(\mathrm{rot}\,\mathbf{B}-\frac{\partial\mathbf{E}}{\partial
\xi}\right)\times \mathbf{B}+\mathbf{E}\mathrm{div}\,\mathbf{E}\\
&&=\left(\mathrm{rot}\,\mathbf{B}-\frac{\partial\mathbf{E}}{\partial
\xi}\right)\times \mathbf{E}-\mathbf{B}\mathrm{div}\,\mathbf{E}+
\left(\mathrm{rot}\,\mathbf{E}+\frac{\partial\mathbf{B}}{\partial \xi}\right)\times
\mathbf{B}-\mathbf{E}\mathrm{div}\,\mathbf{B}.
\end{eqnarray*}
\vskip 0.2cm
According to these equations the intrinsic dynamics of a free
electromagnetic field is described by two couples of vector fields,
$[(\mathbf{E},\mathbf{B}); (\mathbf{-B},\mathbf{E})]$, or
$[(\mathbf{E},\mathbf{B}); (\mathbf{B},\mathbf{-E})]$, and this intrinsic
dynamics could be interpreted as a direct energy-momentum exchange between two
appropriately individualized subsystems mathematically described by these two
partner-fields.

A further natural specilization of the above two vector equations
could be made if we recall that this internal energy-momentum exchange realizes
a {\it special kind of dynamical equilibrium} between the two partner-fields,
namely, {\it the two partner-fields necessarily carry always the same energy and
momentum}:
$$
M^{ij}(\mathbf{E},\mathbf{B},\alpha_1)=M^{ij}(\mathbf{E},\mathbf{B},\alpha_2),
\ \ \
\mathbf{P}(\mathbf{E},\mathbf{B},\alpha_1)=
\mathbf{P}(\mathbf{E},\mathbf{B},\alpha_2) ,
$$
so, each partner-field conserves its energy-momentum:
$\Delta_{11}=\Delta_{22}=0$.
In such a dynamical situation each partner-field
loses as much as it gains during any time period, so, the equations
reduce to
\begin{center}
\hfill\fbox{
    \begin{minipage}{0.97\textwidth}
\begin{center}
$$
\Delta_{11}\equiv\left(\mathrm{rot}\,\mathbf{E}+\frac{\partial\mathbf{B}}{\partial
\xi}\right)\times \mathbf{E}+\mathbf{B}\mathrm{div}\,\mathbf{B}=0,     %10%
$$
$$
\Delta_{22}\equiv
\left(\mathrm{rot}\,\mathbf{B}-\frac{\partial\mathbf{E}}{\partial      %11%
\xi}\right)\times \mathbf{B}+\mathbf{E}\mathrm{div}\,\mathbf{E}=0,
$$
$$
\Delta_{12}+\Delta_{21}\equiv\left(\mathrm{rot}\,\mathbf{E}+\frac{\partial\mathbf{B}}{\partial
\xi}\right)\times \mathbf{B}-\mathbf{E}\mathrm{div}\,\mathbf{B}+        %12%
\left(\mathrm{rot}\,\mathbf{B}-\frac{\partial\mathbf{E}}{\partial \xi}\right)\times
\mathbf{E}-\mathbf{B}\mathrm{div}\,\mathbf{E}=0.
$$
\end{center}
\vskip 0.3cm
\end{minipage}}
\hfill
\end{center}
The third equation fixes, namely, that {\it the exchange of energy-momentum
density between the two partner-fields is {\bf simultaneous} and in {\bf equal}
quantities}, i.e. the mutual balance is realized as {\bf permanent dynamical
equilibrium} between the two partner-fields:
$\mathbf{P}_{(\mathbf{E},\mathbf{B})}\rightleftarrows
\mathbf{P}_{(\mathbf{-B},\mathbf{E})}$, or,
$\mathbf{P}_{(\mathbf{E},\mathbf{B})}\rightleftarrows
\mathbf{P}_{(\mathbf{B},\mathbf{-E})}$.

Note that, this double-field viewpoint and the corresponding mutual
energy-momentum exchange described by the last equation are essentially new
moments. The left-hand sides of the first two equations also suggest how the
corresponding fields are able to exchange energy-momentum with other physical
systems. If such an exchange has been done, then the exchanged energy-momentum
quantities can be given in terms of the characteristics of the other physical
system (or in terms of the characteristics of the both systems) and to be
correspondingly equalized to the left hand sides of our three equations in
accordance with the local energy-momentum conservation law.

The above equations can be given the following form in terms of differential
forms. If $g$ is the euclidean metric let's introduce the following notations:
$$
\tilde{g}(\mathbf{E})=\eta, \ \ \tilde{g}(\mathbf{B})=\beta, \ \
\tilde{g}^{-1}(*\eta)=\bar{*\eta}, \ \ \tilde{g}^{-1}(*\beta)=\bar{*\beta}.
$$
Then we obtain (in terms of corresponding flows):
$$
\tilde{g}(\mathrm{rot}\mathbf{E}\times\mathbf{E})=i(\mathbf{E})\mathbf{d}\eta,\
\ \tilde{g}(\mathrm{rot}\mathbf{B}\times\mathbf{B})=i(\mathbf{B})\mathbf{d}\beta,
$$
$$
\tilde{g}(\mathrm{rot}\mathbf{E}\times\mathbf{B})=i(\mathbf{B})\mathbf{d}\eta, \ \
\tilde{g}(\mathrm{rot}\mathbf{B}\times\mathbf{E})=i(\mathbf{E})\mathbf{d}\beta
, $$ $$ \mathbf{E}\,\mathrm{div}(\mathbf{B})=i(\bar{*\eta})\mathbf{d}*\beta, \
\ \mathbf{B}\,\mathrm{div}(\mathbf{E})=i(\bar{*\beta})\mathbf{d}*\eta .
$$
Under these notations and relations the above three framed equations are
respectively equivalent to:
$$
i(\mathbf{E})\mathbf{d}\eta+i(\bar{*\beta})\mathbf{d}*\beta=
-*\left(\frac{\partial\beta}{\partial\xi}\wedge\eta\right)=
i\left(\frac{\partial\mathbf{B}}{\partial\xi}\right)*\eta ,
$$
$$
i(\mathbf{B})\mathbf{d}\beta+i(\bar{*\eta})\mathbf{d}*\eta=
*\left(\frac{\partial\eta}{\partial\xi}\wedge\beta\right)=-
i\left(\frac{\partial\mathbf{E}}{\partial\xi}\right)*\beta ,
$$
$$
i(\mathbf{B})\mathbf{d}\eta-i(\bar{*\eta})\mathbf{d}*\beta+
i(\mathbf{E})\mathbf{d}\beta-i(\bar{*\beta})\mathbf{d}*\eta=
*\left(\frac{\partial\eta}{\partial\xi}\wedge\eta-
\frac{\partial\beta}{\partial\xi}\wedge\beta\right) .
$$
Denoting now the Maxwell stress tensors of $\mathbf{E}$ and $\mathbf{B}$
correspondingly by $M(\mathbf{E})$ and $M(\mathbf{B})$, and introducing a new
stress tensor
$\mathbb{T}=M(\mathbf{E})+M(\mathbf{B})-M(\mathbf{E}+\mathbf{B})$ we obtain
$$
\mathrm{div}\Big[M(\mathbf{E})+M(\mathbf{B})\Big]=0, \ \
\mathrm{div}\mathbb{T}=\frac{\partial\mathbf{E}}{\partial\xi}\times\mathbf{E}-
\frac{\partial\mathbf{B}}{\partial\xi}\times\mathbf{B}.
$$

\section{Properties of the equations and their \\solutions}
Consider the second equation, $\Delta_{22}=0$,   and replace
$(\mathbf{E},\mathbf{B})$ acording to
$$
(\mathbf{E},\mathbf{B})
\rightarrow(a\mathbf{E}-b\mathbf{B},b\mathbf{E}+a\mathbf{E}),
$$
where $(a,b)$ are two constants.
After the corresponding computation we obtain
$$
a^2\left[
\left(\mathrm{rot}\,\mathbf{B}- \frac{\partial \mathbf{E}}{\partial
\xi}\right) \times\mathbf{B}+\mathbf{E}\,\mathrm{div}\,\mathbf{E}\right]+
b^2\left[\left(\mathrm{rot}\,\mathbf{E}+
\frac{\partial \mathbf{B}}{\partial \xi}\right)
\times\mathbf{E}+\mathbf{B}\,\mathrm{div}\,\mathbf{B}\right]+
$$
$$
+ab\left[
\left(\mathrm{rot}\,\mathbf{E}+
\frac{\partial \mathbf{B}}{\partial \xi}\right)
\times\mathbf{B}-\mathbf{E}\,\mathrm{div}\,\mathbf{B}+
\left(\mathrm{rot}\,\mathbf{B}-
\frac{\partial \mathbf{E}}{\partial \xi}\right)
\times\mathbf{E}-\mathbf{B}\,\mathrm{div}\,\mathbf{E}\right]=0.
$$
Since the constants $(a,b)$ are arbitrary the other two equations
follow. The same property holds with respect to any of the three equations.

	{\bf Corollary.} The system of the three equations is invariant with
respect to the transformation
$$
(\mathbf{E},\mathbf{B})
\rightarrow(a\mathbf{E}-b\mathbf{B},b\mathbf{E}+a\mathbf{E}).
$$
Writing down this transformation in the form
\begin{eqnarray*}
(\mathbf{E},\mathbf{B})\rightarrow &(\mathbf{E}',\mathbf{B}')          %7%
=(\mathbf{E},\mathbf{B}).\alpha(a,b)
=(\mathbf{E},\mathbf{B})\begin{Vmatrix} a & b \\ -b & a \end{Vmatrix}\\&
=(a\mathbf{E}-b\mathbf{B}, b\mathbf{E}+a\mathbf{B}), \
a=const, \ b=const,
\end{eqnarray*}
we get a "right action" of the matrix $\alpha$ on the solutions.
 The new solution $(\mathbf{E}',\mathbf{B}')$ has energy and momentum densities
equal to the old ones multiplied by $(a^2+b^2)$. Hence, the space of all
solutions factors over the action of the group of matrices of the kind
\[
\alpha(a,b)= \begin{Vmatrix} a & b \\ -b & a \end{Vmatrix},
\quad (a^2+b^2)\neq 0
\]
in the sense, that the corresponding classes are determined by the value of
$(a^2+b^2)$.

All such matrices with nonzero determinant form a group
with respect to the usual matrix product. The special property of this group is
that it represents the symmetries of the canonical complex structure in
$\mathbb{R}^2$.

Clearly, all solutions to Maxwell pure field equations are solutions to
our nonlinear equations, we shall call these solutions linear, and
will not further be interested of them, we shall concentrate our attention on
those solutions of our equations which satisfy the conditions
$$
\mathrm{rot}\,\mathbf{E}+\frac{\partial\mathbf{B}}{\partial \xi}\neq 0,\quad
\mathrm{rot}\,\mathbf{B}-\frac{\partial\mathbf{E}}{\partial \xi}\neq 0,\quad
\mathrm{div}\,\mathbf{E}\neq 0,\quad \mathrm{div}\,\mathbf{B}\neq 0.
$$
These solutions we call further nonlinear.

We consider now some properties of the nonlinear solutions.
\vskip 0.3cm
$\bf 1.$ Among the nonlinear solutions  there
are no constant ones.
\vskip 0.3cm
$\bf 2.$\ $\mathbf{E}.\mathbf{B}=0;$ This is obvious, no proof is needed.
\vskip 0.3cm
$\bf 3.$ The following relations are also obvious:
$$\left(\mathrm{rot}\,\mathbf{E}+ \frac{\partial\mathbf{B}}{\partial
\xi}\right).\mathbf{B}=0; \ \ \left(\mathrm{rot}\,\mathbf{B}-
\frac{\partial\mathbf{E}}{\partial \xi}\right).\mathbf{E}=0.
$$
\noindent
\vskip 0.3cm
{\bf 4.} It is elementary to see from the last two relations that the
classical Poynting energy-momentum balance equation follows.
 \vskip 0.3cm
{\bf 5.} $\mathbf{E}^2=\mathbf{B}^2$.

In order to prove this let's take the scalar product of the first
equation from the left by ${\mathbf B}$.  We obtain

$$
{\mathbf B}.\Biggl\{\left(\mathrm{rot}{\mathbf E}+
\frac{\partial {\mathbf B}}{\partial \xi}\right)
\times {\mathbf E}\Biggr\}+{\mathbf B}^2\mathrm{div}{\mathbf B}=0. \ \ \ \ \ (*)
$$
Now, multiplying the second equation from the left by ${\mathbf E}$ and having
in view $\mathbf{E}.\mathbf{B}=0$, we obtain
$$
{\mathbf E}.\Biggl\{\left(\mathrm{rot}{\mathbf E}+
\frac{\partial {\mathbf B}}{\partial \xi}\right) \times {\mathbf
B}\Biggr\}-{\mathbf E}^2\mathrm{div}{\mathbf B}=0.
$$
 This last relation is equivalent
to
$$
-{\mathbf B}.\Biggl\{\left(\mathrm{rot}{\mathbf E}+ \frac{\partial {\mathbf
B}}{\partial \xi}\right) \times {\mathbf E} \Biggr\}- {\mathbf
E}^2\mathrm{div}{\mathbf B}=0.\ \ \ \\ (**)
$$
Now, summing up $(*)$ and $(**)$, in view of $\mathrm{div}{\mathbf B}\neq 0$,
we come to the desired relation.

Properties {\bf 2.} and {\bf 5.} say that all nonlinear solutions are {\it
null fields}, i.e. the two well known relativistic invariants
$I_1=\mathbf{B}^2-\mathbf{E}^2$ and $I_2=2\mathbf{E}.\mathbf{B}$ of the field
are zero, and this property leads to optimisation of the inequality
$I_1^2+I_2^2\geqslant 0$ (recall the eigen properties of Maxwell stress tensor),
 which in turn guarantees $\alpha(x,y,z;t)$-invariance of $I_1=I_2=0$.

\vskip 0.3cm
{\bf 6.} The {\it helicity} property: \index{helicity property}
$$
\mathbf{B}.\left(\mathrm{rot}\,\mathbf{B}-
\frac{\partial\mathbf{E}}{\partial \xi}\right)- \mathbf{E}.
\left(\mathrm{rot}\,\mathbf{E}+ \frac{\partial\mathbf{B}}{\partial \xi}\right)=
\mathbf{B}.\mathrm{rot}\mathbf{B}-\mathbf{E}.\mathrm{rot}\mathbf{E}=0.
$$

\noindent To prove this property we first multiply (vector product) the third
equation from the right by $\mathbf{E}$, recall property {\bf 2}., then
multiply (scalar product) from the left by $\mathbf{E}$, recall again
$\mathbf{E}.\mathbf{B}=0$, then multiply from the right (scalar product) by
$\mathbf{B}$ and recall property {\bf 5.}

Property {\bf 6.} suggests the following consideration. If $\mathbf{V}$ is an
arbitrary vector field on $\mathbb{R}^3$ then the quantity
$\mathbf{V}.\mathrm{rot}\mathbf{V}$ is known as {\it local helicity} and its
integral over the whole (compact) region occupied by $\mathbf{V}$ is known as
{\it integral helicity}, or just as {\it helicity} of $\mathbf{V}$. Hence,
property {\bf 6.} says that the electric and magnetic constituents of a
nonlinear solution generate the same helicities. If we consider (through the
euclidean metric $g$) the 1-form $\tilde{g}(\mathbf{E})$ and denote by
$\mathbf{d}$ the exterior derivative on $\mathbb{R}^3$, then $$
\tilde{g}(\mathbf{E})\wedge\mathbf{d}\tilde{g}(\mathbf{E})=
\mathbf{E}.\mathrm{rot}\mathbf{E}\,dx\wedge dy\wedge dz,
$$
so, the zero helicity
says that the 1-form $\tilde{g}(\mathbf{E})$ defines a completely integrable
Pfaff system: $\tilde{g}(\mathbf{E})\wedge\mathbf{d}\tilde{g}(\mathbf{E})=0$.
The nonzero helicity
says that the 1-form $\tilde{g}(\mathbf{E})$ defines non-integrable
1-dimensional  Pfaff system, so the nonzero helicity defines corresponding
curvature. Therefore the equality between the $\mathbf{E}$-helicity and the
$\mathbf{B}$-helicity suggests to consider the corresponding integral
helicity
$$
\int_{\mathbb{R}^3}\tilde{g}(\mathbf{E})\wedge\mathbf{d}\tilde{g}(\mathbf{E}) =
\int_{\mathbb{R}^3}\tilde{g}(\mathbf{B})\wedge\mathbf{d}\tilde{g}(\mathbf{B})
$$
(when it takes finite nonzero values) as a measure of the spin properties
of the solution.

We specially note that the equality of the local helicities
defined by $\mathbf{E}$ and $\mathbf{B}$ holds also, as it is easily seen from
the above relation, for the solutions of the linear Maxwell vacuum equations,
but appropriate solutions giving well defined and time independent integral
helicities in this case are missing. The next property shows that our nonlinear
solutions admit such appropriate solutions giving finite constant integral
helicities.

 \vskip 0.3cm {\bf 7.}\ \ Example of nonlinear solution(s):
\begin{align*} &\mathbf{E}=\left[\phi(x,y,\xi\pm z)
\mathrm{cos}(-\kappa\frac{z}{\mathcal{L}_o}+const), \, \phi(x,y,\xi\pm
z)\mathrm{sin}(-\kappa\frac{z}{\mathcal{L}_o}+const),\,0\right];\\ &\mathbf{B}=\left[\pm
\phi(x,y,\xi\pm z)\, \mathrm{sin}(-\kappa\frac{z}{\mathcal{L}_o}+const),\,\mp
\phi(x,y,\xi\pm z) \mathrm{cos}(-\kappa\frac{z}{\mathcal{L}_o}+const),\,0\right],
\end{align*}
where $\phi(x,y,\xi\pm z)$ is an arbitrary positive function,
$0<\mathcal{L}_o<\infty$ is an arbitrary positive constant
with physical dimension of length, and $\kappa$ takes values $\pm1$
. Hence, we are allowed to choose the function $\phi$ to have compact
3d-support, and since the energy density of this solution is $\phi^2dx\wedge
dy\wedge dz$, then this solution will describe time-stable and space
propagating with the speed of light finite field objects carrying finite
integral energy.

Modifying now the corresponding helicity 3-forms to
$$
\frac{2\pi
\mathcal{L}_o^2}{c}\tilde{g}(\mathbf{E})\wedge\mathbf{d}\tilde{g}(\mathbf{E})=
\frac{2\pi
\mathcal{L}^2_o}{c}\tilde{g}(\mathbf{B})\wedge\mathbf{d}\tilde{g}(\mathbf{B}),
$$
then the corresponding 3d integral gives $\kappa TE$, where $\kappa=\pm 1$,
$T=2\pi\mathcal{L}_o/c$ and $E=\int{\phi^2}dx\wedge dy\wedge dz$ is the integral
energy of the solution.

\section{Scale factor and Planck's constant} We consider the vector
fields $$ \vec{\mathcal{F}}=\mathrm{rot}\,\mathbf{E}+ \frac{\partial
\mathbf{B}}{\partial \xi}+ \frac{\mathbf{E}\times\mathbf{B}}
{|\mathbf{E}\times\mathbf{B}|}\,\mathrm{div}\,\mathbf{B},   %44%
$$
$$
\vec{\mathcal{M}}=\mathrm{rot}\,\mathbf{B}-
\frac{\partial {\mathbf E}}{\partial \xi}-                   %45%
\frac{\mathbf{E}\times\mathbf{B}}
{|\mathbf{E}\times\mathbf{B}|}\,\mathrm{div}\,\mathbf{E},
$$
defined by a nonlinear solution.

It is obvious that on the solutions of Maxwell's vacuum equations
$\vec{\mathcal{F}}$ and $\vec{\mathcal{M}}$ are equal to zero. Note also that
under the transformation
$(\mathbf{E},\mathbf{B})\rightarrow (\mathbf{-B},\mathbf{E})$
we get
$\vec{\mathcal{F}}\rightarrow
-\vec{\mathcal{M}}$ and $\vec{\mathcal{M}}\rightarrow \vec{\mathcal{F}}$.

We shall consider now the relation between  $\vec{\cal F}$ and $\mathbf{E}$,
and between $\vec{\cal M}$ and $\mathbf{B}$ on the nonlinear solutions of our
equations assuming that $\vec{\cal F}\neq 0$ and $\vec{\cal M}\neq 0$.

Recalling $\mathbf{E}.\mathbf{B}=0$ we obtain
$$
({\mathbf E}\times {\mathbf
B})\times {\mathbf E}= -{\mathbf E}\times ({\mathbf E}\times {\mathbf B})=
-[{\mathbf E}({\mathbf E}.{\mathbf B})-
{\mathbf B}({\mathbf E}.{\mathbf E})]={\mathbf B}({\mathbf E}^2),
$$
and since $|{\mathbf E}\times {\mathbf B}|
=|\mathbf{E}||\mathbf{B}||\mathrm{sin}(\mathbf{E},\mathbf{B})|=
 {\mathbf E}^2={\mathbf B}^2$, we get
$$
\vec{\cal F}\times {\mathbf E}=
\left(\mathrm{rot}{\mathbf E}+\frac{\partial {\mathbf B}}{\partial \xi}\right)
\times {\mathbf E}+{\mathbf B}\mathrm{div}{\mathbf B}=0,
$$
according to our first nonlinear equation.

In the same way, in accordance with our second nonlinear equation,
we get $\vec{\cal M}\times {\mathbf B}=0$. In other words, on
the nonlinear solutions we obtain that $\vec{\cal F}$ is co-linear to
${\mathbf E}$ and $\vec{\cal M}$ is co-linear to ${\mathbf B}$.  Hence, we
can write the relations
$$
\vec{\cal F}=f_1.{\mathbf E},\ \ \vec{\cal M}=f_2.{\mathbf B},        %46%
$$
where $f_1$ and $f_2$ are two functions, and of course, the
interesting cases are $f_1\neq 0,\infty;\ f_2\neq 0,\infty$.
Note that the physical dimension of $f_1$ and $f_2$ is the
reciprocal to the dimension of coordinates, i.e.
$[f_1]=[f_2]= [length]^{-1}$.

Note also that $\vec{\cal F}$ and $\vec{\cal M}$ are mutually orthogonal:
$\vec{\cal F}.\vec{\cal M}=0$.

We shall prove now that $f_1=f_2$. In fact, making use of the same formula
for the double vector product, used above, we easily obtain
$$
\vec{\cal F}\times {\mathbf B}+\vec{\cal M}\times {\mathbf E}=
$$
$$
=\left(\mathrm{rot}{\mathbf E}+\frac{\partial {\mathbf B}}{\partial
\xi}\right)
\times {\mathbf B}+
\left(\mathrm{rot}{\mathbf B}-\frac{\partial {\mathbf E}}{\partial \xi}\right)
\times {\mathbf E}-{\mathbf E}\mathrm{div}{\mathbf B}
-{\mathbf B}\mathrm{div}{\mathbf E}=0,
$$
in accordance with our third nonlinear equation.
Therefore,
$$
\vec{\cal F}\times {\mathbf B}+\vec{\cal M}\times {\mathbf E}=
$$
$$
=f_1{\mathbf E}\times {\mathbf B}+f_2{\mathbf B}\times {\mathbf E}=
(f_1-f_2){\mathbf E}\times {\mathbf B}=0.
$$
The assertion follows.

The relation $|\vec{\cal F}|=|\vec{\cal M}|$ is now obvious.

Note that the two relations $|\vec{\cal F}|^2=|\vec{\cal M}|^2$ and
$\vec{\cal F}.\vec{\cal M}=0$  and the duality correspondence $(\vec{\cal
F},\vec{\cal M})\rightarrow (-\vec{\cal M},\vec{\cal F})$
 suggests to consider $\vec{\cal F}$ and $\vec{\cal M}$
as nonlinear analogs of $\mathbf{E}$ and $\mathbf{B}$ respectively.

These considerations suggest to introduce the quantity
$$
\mathcal{L}(\mathbf{E},\mathbf{B})=
\frac{1}{|f_1|}=\frac{1}{|f_2|}=
\frac{|\mathbf{E}|}{|\vec{\mathcal{F}}|}=                      %47%
\frac{|\mathbf{B}|}{|\vec{\mathcal{M}}|},
$$
which we call {\it scale factor} \index{scale factor}, and this quantity will
appear in various forms further. Note that the physical dimension of
$\mathcal{L}$ is {\it length}. Hence, every nonlinear solution defines its own
{\it scale factor} and, concequently, the nonlinear solutions factorize with
respect to $\mathcal{L}$. It seems natural to connect the constant
$\mathcal{L}_o$ in the above given family of solutions with the so introduced
scale factor. Assuming $\mathcal{L}=\mathcal{L}_o=const$, this could be done in
the following way.

A careful look at the solutions above shows that at a given moment, e.g. $t=0$,
the finite spatial support of the function $\phi$ is built of continuous sheaf
of nonintersecting helices along the coordinate $z$. Every such helix has a
special length parameter $b=\lambda/2\pi$ giving the straight-line advance
along the external straight-line axis (the coordinate $z$ in our case) for a
unit angle, and $\lambda$ is the $z$-distance between two equivalent points on
the same helix. So, we may put $\lambda=2\pi\mathcal{L}_o=const$, hence, the
$z$-size of the solution may, naturally, be bounded by $2\pi\mathcal{L}_o$.

Consider now a nonlinear solution with integral energy $E$ and scale factor
$\mathcal{L}_o=const$. Since this solution shall propagate in space with the
speed of light $c$, we may introduce corresponding time period
$T=2\pi\mathcal{L}_o/c$, and define the quntity $\mathfrak{h}=E.T$, having
physical dimension of "action". The temptation to separate a class of solutions
requiring $\mathfrak{h}$ to be equal to the Planck constant $h$ is great, isn't
it, especially if this $T$ can be associated with some helix-like real
periodicity during propagation?!
\vskip 0.2cm
{\bf References}
%\vskip 0.2cm
\addcontentsline{toc}{subsection}{{\bf References}}
\vskip 0.1cm
[1]. {\bf B. Lehnert, S. Roy}, {\it Extended Electromagnetic Theory}, World
Scientific, 1998.

[2]. {\bf B. Lehnert}, {\it A Revised Electromagnetic Theory with Fundamental
Applications}, Swedish Physic Arhive, 2008.

[3]. {\bf D. Funaro}, {\it Electromagnetism and the Structure of Matter},
Worldscientific, 2008; also: {\it From photons to atoms}, arXiv: gen-ph/1206.3110
(2012).

\chapter{Extended Electrodynamics. Relativistic approach}
\section{The Rainich identity}
We are going to sketch a proof of the important
Rainich identity, \index{Rainich identity} mentioned in
Sec.6.3.1, in view of its appropriate use in studying the eigen properties of
the electromagnetic energy-momentum tensor on Minkowski space-time
$M=(\mathbb{R}^4,\eta)$. We recall from Sec.6.3.1 the following relations
\begin{eqnarray*}
\frac12F_{\alpha\beta}F^{\alpha\beta}id_\mu^\nu&=&F_\mu\,^\sigma
F^\nu\,_\sigma-(*F)_\mu\,^\sigma (*F)^\nu\,_\sigma=
[F\circ F-(*F)\circ (*F)]_\mu^\nu\\
\frac14F_{\alpha\beta}(*F)^{\alpha\beta}id_\mu^\nu&=&
F_\mu\,^\sigma (*F)^\nu\,_\sigma=[F\circ(*F)]id_\mu^\nu=
[(*F)\circ F]_\mu^\nu \\
Q_\mu^\nu&=&-\frac12\left[F\circ F+(*F)\circ(*F)\right]_\mu^\nu=
\frac14F_{\alpha\beta}F^{\alpha\beta}id_\mu^\nu-F_\mu\,^\sigma F^\nu\,_\sigma.
\end{eqnarray*}
 Now for the composition $Q\circ Q$ we obtain
\begin{eqnarray*} Q\circ
Q&=&\frac14\Big[F\circ F\circ F\circ F+F\circ F\circ(*F)\circ(*F)\\
&+&(*F)\circ(*F)\circ F\circ F+(*F)\circ(*F)\circ(*F)\circ(*F)\Big]\\
&=&\frac14\Big[F\circ F\circ F\circ F+(*F)\circ(*F)\circ(*F)\circ(*F)+
2(*F)\circ(*F)\circ F\circ F\Big].
\end{eqnarray*}
Making use of the above identities we obtain
\begin{eqnarray*}
F\circ F\circ F\circ
F&=&\frac14(F.F)^2id+\frac{1}{16}(F.*F)^2id+\frac12(F.F)(*F)\circ(*F)\\
(*F)\circ(*F)\circ(*F)\circ(*F)&=&\frac{1}{16}(F.*F)^2id-\frac12(F.F)(*F\circ
*F)\\
2(*F)\circ(*F)\circ F\circ F&=&\frac18(F.*F)^2id,
\end{eqnarray*}
where $(F.F)=F_{\alpha\beta}F^{\alpha\beta}$ and
$(F.*F)=F_{\alpha\beta}(*F)^{\alpha\beta}$. Summing up we get to the Rainich
relation
$$ Q\circ
Q=\frac14\left[\left(\frac12F.F\right)^2+\left(\frac12F.*F\right)^2\right]id=
\frac14\left[I_1^2+I_2^2\right]id
$$
Clearly, since $tr(id)=4$, we obtain $$ Q_{\mu\nu}Q^{\mu\nu}=I_1^2+I_2^2. $$
Now the eigen relation $Q^\mu_\nu X^\nu=\lambda\,X^\mu$ gives the eigen values
$$
\lambda_{1,2}=\pm\frac12\sqrt{I_1^2+I_2^2}.
$$
We recall now that under the duality transformation
\begin{eqnarray*}
F'&=& F\mathrm{cos}\,\alpha-*F\mathrm{sin}\,\alpha\\
*F'&=& F\mathrm{sin}\,\alpha+*F\mathrm{cos}\,\alpha
\end{eqnarray*}
the two invariants $(I_1,I_2)$ keep their values only if they are zero:
$I_1=I_2=0$. Hence,
the only dually invariant eigen direction $\bar{\zeta}$ of the energy-momentum
tensor must satisfy $Q^\mu_\nu\bar{\zeta}^\nu=0$, where $Q$ must satisfy
$det||Q_\mu^\nu||=0$ and $Q\circ Q=0$, i.e. $Q$ becomes {\it boundary map}.
As we have mentioned earlier, under these conditions the field $(F,*F)$ is
usually called {\it null} field.

We would like specially to note the conformal invariance of the restriction
of the Hodge $*$ to 2-forms. In fact, $\eta'=f^2\eta, f(a)\neq 0, a\in M$,
and $\eta$ generate the same $*$:
$$
*'F=\frac 12 F_{\mu\nu}*'(dx^\mu\wedge dx^\nu)=
-\frac 12 F_{\mu\nu}\eta'^{\mu\sigma}\eta'^{\nu\tau}
\varepsilon_{\sigma\tau\alpha\beta}\sqrt{|det\,\eta'|}
dx^{\alpha}\wedge dx^{\beta}
$$
$$
=-\frac 12 F_{\mu\nu}f^{-4}\eta^{\mu\sigma}\eta^{\nu\tau}
\varepsilon_{\sigma \tau\alpha\beta}f^4 \sqrt{|det\,\eta|}
dx^{\alpha}\wedge dx^{\beta}=*F.
$$
It follows that the stress-energy-momentum tensor $Q_{\mu}^{\nu}$ transforms to
$f^{-4}Q_{\mu}^{\nu}$ under such conformal change of the metric $\eta$.
\vskip 0.3cm
{\bf Literature}
\addcontentsline{toc}{subsection}{{\bf Literature}}
\vskip 0.2cm
1. {\bf G. Rainich}, {\it Electrodynamics in general Relativity},
Trans.Amer.Math.Soc., {\bf 27} (106-136).

2. {\bf W. Misner, J. Wheeler}, {\it Classical Physics as geometry}, Ann.Phys.,
{\bf 2} (525-603)

3. {\bf J. Franca, J. Lopez-Bonilla}, {\it The Algebraic Rainich Conditions},
Progress in Physics, vol.3, July 2007.

\section{Some basic properties of null fields}

All null fields $(F,*F)$, by definition, satisfy  $Q_{\mu\nu}Q^{\mu\nu}=0$,
i.e.,
$$
I_1=\mathbf{B}^2-\mathbf{E}^2=0, \ \ \ I_2=2\mathbf{E}.\mathbf{B}=0.
$$
(For details see: J.L.Synge, {\bf Relativity: The Special Theory},
North-Holland, 1956, Ch.IX, \S\,7).
Since the field propagates freely, i.e., $\mathbf{E}\times\mathbf{B}$ defines a
sheaf of straight lines, the results of Sec.7.5 and the Rainich identity allow
to consider the mentioned in Sec.4.1 "appropriate initial stress-strain
spatial" structure that is to be kept during propagation along a straight line,
as {\it complete integrabilty} of the space-like 2-dimensional distribution
$(\mathbf{E},\mathbf{B})$, since in this null-field case this 2-dimensional
spatial distribution $(\mathbf{E},\mathbf{B})$ is easily verified to satisfy
$[\mathbf{E},\mathbf{B}]\wedge\mathbf{E}\wedge\mathbf{B}=0$. Thus we have
2-dimensional space-like foliation: any 2-dimensional spatial planes defined by
$(\mathbf{E},\mathbf{B})$ do NOT intersect.

This available 2-dimensional  space-like foliation
allows to choose a coframe as follows: $dx$ and $dy$
to determine coframe on each integral 2-dimensional plane of the
distribution $(\mathbf{E},\mathbf{B})$, $dz$ to be spatially orthogonal to
$dx$ and $dy$, i.e., to any integral manifold of $(\mathbf{E},\mathbf{B})$,
and $\xi=ct$ to denote the time coordinate.
Denoting further $\tilde{\eta}(\mathbf{E})\equiv A$,
$\tilde{\eta}(\mathbf{B})\equiv A^*$, the above zero values of the invariants
$I_1,I_2$ mean that in this coframe we have
$$ A=u\,dx+p\,dy, \ \
A^*=-\varepsilon\,p\,dx+\varepsilon\,u\,dy , \ \ \varepsilon=\pm 1.
$$
The corresponding frame looks like
$$
\bar{A}=-u\,\frac{\partial}{\partial x} -
p\,\frac{\partial}{\partial y}; \ \
\bar{A^*}=\varepsilon\,p\,\frac{\partial}{\partial x} -
\varepsilon\,u\,\frac{\partial}{\partial y}; \ \
\frac{\partial}{\partial z}; \ \ \frac{\partial}{\partial \xi}\cdot
$$
The eigen null direction $\bar{\zeta}$ is defined in this frame by
$$
\bar{\zeta}=-\varepsilon\frac{\partial}{\partial z} +          %1%
\frac{\partial}{\partial \xi}, \ \ \varepsilon=\pm 1,
$$
and the corresponding 1-form $\zeta=\tilde{\eta}(\bar{\zeta})$ looks like
$$
\zeta=\varepsilon\,dz+d\xi.
$$
The only non-zero componenets of $Q_{\mu}^{\nu}$ are
$$
Q_{4}^{4}=-Q_3^3=\varepsilon Q_3^4=-\varepsilon Q_4^3=
\frac12(|A|^2+|A^*|^2)=|A|^2=u^2+p^2.
$$
The two 2-forms $F$ and $*F$ look as follows
$$
F=A\wedge\zeta, \ \ \ *F=A^*\wedge\zeta.
$$
Since we consider free field, according to the relativity principle the
energy density of such field has to propagate along fixed straight line, and
since our field propagates along the Poynting vector, this frame/coframe and
the corresponding coordinate system $(x,y,z,\xi)$ may be used globally,
provided the relations studied to be written in coordinate free way. Further
this coordinate system will be called $\zeta$-adapted for short.

We note also the following specific properties of a {\it null} field:

	1. A null field is determined just by two functions, denoted here by
$u(x,y,z,\xi),p(x,y,z,\xi)$.

	2. The direction of translational propagation is determined {\it
intrinsically}, namely by a null eigen vector, and the trajectories of this
null eigen vector are null straight lines in Minkowski space-time.

	3. A null field is represented by two algebraically interconnected
through the Hodge $*$-operator and locally recognizable subfields $(F,*F)$
carrying always the same stress-energy-momentum:
$$
I_1=\frac12F_{\mu\nu}F^{\mu\nu}=0
\ \Rightarrow F_{\mu\sigma}F^{\nu\sigma}=(*F)_{\mu\sigma}(*F)^{\nu\sigma}.
 $$

	4. The eigen values of $F_\mu^\nu$ and $(*F)_\mu^\nu$ are also zero
(Sec.6.3.1).

	5. The following relations hold:
$$
i_{\bar{\zeta}}F=i_{\bar{\zeta}}(*F)=0, \
 \ i_{\bar{A}}(*F)=i_{\bar{A}^*}F=0.
$$
Hence, $\bar{\zeta}$ is eigen vector of $F$ and $*F$, $\bar{A}^*$ is eigen
vector of $F$, and $\bar{A}$ is eigen vector of $*F$.

Other two interesting properties of these $F=A\wedge\zeta$ and
$*F=A^*\wedge\zeta$ are the folowing. Consider the $TM$-valued 1-forms
$A\otimes\bar{\zeta}$ and $A^*\otimes\bar{\zeta}$ and compute the corresponding
{\it Fr$\ddot{o}$licher-Nijenhuis} brackets $[A\otimes\bar{\zeta},A^*\otimes\bar{\zeta}]=
[A^*\otimes\bar{\zeta},A\otimes\bar{\zeta}]$,
and $[A\otimes\bar{\zeta},A\otimes\bar{\zeta}]$ (see {\bf Imp.remark} in
Sec.1.3.4). We obtain
$$
[A\otimes\bar{\zeta},A^*\otimes\bar{\zeta}]
=-\frac12\varepsilon\big[(u^2+p^2)_{\xi}-\varepsilon(u^2+p^2)_z\big]
dx\wedge dy\otimes\bar{\zeta};
$$
$$
[A\otimes\bar{\zeta},A\otimes\bar{\zeta}]=
-[A^*\otimes\bar{\zeta},A^*\otimes\bar{\zeta}]
$$
$$
=[u(p_\xi-\varepsilon\,p_z)-
p(u_\xi-\varepsilon\,u_z)]dx\wedge dy\otimes\bar{\zeta}.
$$

The coressponding Schouten brackets $[\bar{F},\bar{F}]$ and
$[\bar{F},\bar{*F}]$ give
$$
[\bar{F},\bar{F}]=[\bar{A}\wedge\bar{\zeta},\bar{A}\wedge\bar{\zeta}]
$$
$$
=-\varepsilon\big[u(p_{\xi}-\varepsilon p_{z})-p(u_{\xi}-\varepsilon
u_{z})\big]\frac{\partial}{\partial x}\wedge
\frac{\partial}{\partial y}\wedge\frac{\partial}{\partial z}
$$
$$
+\big[u(p_{\xi}-\varepsilon p_{z})-p(u_{\xi}-\varepsilon
u_{z})\big]
\frac{\partial}{\partial x}\wedge
\frac{\partial}{\partial y}\wedge\frac{\partial}{\partial\xi},
$$
\vskip 0.3cm
$$
[\bar{F},\bar{*F}]=[\bar{A}\wedge\bar{\zeta},\bar{A^*}\wedge\bar{\zeta}]
$$
$$
=\frac12\big[(u^2+p^2)_\xi-\varepsilon(u^2+p^2)_z\big]
\frac{\partial}{\partial x}\wedge
\frac{\partial}{\partial y}\wedge\frac{\partial}{\partial z}
$$
$$
-\frac12\varepsilon\big[(u^2+p^2)_\xi-\varepsilon(u^2+p^2)_z\big]
\frac{\partial}{\partial x}\wedge
\frac{\partial}{\partial y}\wedge\frac{\partial}{\partial\xi}.
$$

\section{Basic Equations and their properties}
\subsection{Mathematical identification of the field}
In order to write down relativistic dynamical equations, describing the
space-time evolution of free electromagnetic field-object, we first need to
specify the mathematical object that we consider as mathematical image of the
electromagnetic field-object considered as physical object.

The physical object we are going to mathematically describe by means of
differential equations on Minkowski space-time, is characterized as follows:
\vskip 0.2cm
	{\bf 1.} It exists through a permanent space-time propagation with the
translational speed equal to the speed of light.
\vskip 0.2cm
	{\bf 2.} It has dynamical structure, represented by two recognizable
interacting subsystems and their $\eta$ co-images.
\vskip 0.2cm
	{\bf 3.} These two subsystems carry the same stress-energy-momentum.
\vskip 0.2cm
	{\bf 4.} These two subsystems are in a permanent local dynamical
equilibrium: making use of their $\eta$ co-images they permanently and directly
exchange energy-momentum in equal quantities without available local interaction
energy.

Following the rule in Sec.4.1, the available two space-time recognizable
subsystems $(\mathfrak{F},\mathfrak{*F})$
of the field we are going to mathematricaly identify by two
subdistributions in the tangent bundle of Minkowski space-time and their
$\eta$-codistributions, so that, {\it no admissible coordinate/frame change to
result in nullifying locally or globaly of any of these two components}.

The above four conditions suggest formally these two components and their
$\eta$ co-images to recognize each other in two ways: \vskip 0.2cm
	a/. Algebraically, i.e. there must exist a one-to-one algebraic map
between them.
\vskip 0.2cm
	b/. Differentially, i.e. there must exist appropriate relation using the
derivatives of the scalar components of the field vector components.
\vskip 0.2cm
Of course, these two kinds of contact between the two mathematical
representatives (we may call them vector components) should
 be {\it physically motivated}, i.e. they should reflect some physical
appearences of the field object carrying such dynamical structure.

From algebraic point of view we start with the observation that the exterior
powers of a vector space naturally separate lineary independent elements:
$x\wedge y$ is not zero only if $x\neq \lambda y$. So, if our physical object
of interest has two interacting components and each component has $p$
recognizable time-stable subsystems, it seems natural to turn to the exterior
algebras built over corresponding couple of dual linear spaces. This view is
supported also by the considered natural definitions and physical
interpretations as quantitative measures of local energy-momentum exchange of
the algebraic and differential flows of $p$-vectors accros $q$-forms as
given in Sec.6.3.3, where the introduced concepts of {\it attractivness} and
{\it sensitivity} were illustrated.

Let's recall also that choosing appropriate
$p$-vector $\Phi$ over a linear space $E^n$ as a starting mathematical model
object,  this $p$-vector defines a $p$-dimensional subspace $E^p_{\Phi}\subset
E^n$. Now, making use of the Poincare isomorphism (Sec.1.4.2) $D^p$ we can
determine the object $i(\Phi)\omega$ which defines a $(n-p)$-dimensional subspace
$D^p(E^{p}_{\Phi})\subset (E^n)^*$, where $(E^n)^*$ is the dual for $E^n$
space. Two more subspaces, namely, $(E^p_{\Phi})^*\subset (E^n)^*$ which is the
dual to $E^p_{\Phi}$, and $(D^p(E^{p}_{\Phi}))^*$, which is the dual to
$D^p(E^{p}_{\Phi})$, immediately appear.

Let's see now what Minkowski space-time manifold $M$ may offer in this direction.

The basic mathematical object on $M$ is its metric tensor $\eta$, it defines
the mathematical procedure that corresponds to the experimental procedure for
measuring space distance making use of light signals. In terms of $\eta$ we
algebraically define 4-volume on $M$ and appropriate linear isomorphisms in the
tensor algebra over $M$. Also, the
exterior algebra of differential forms can be equiped with
the $\eta$-defined linear isomorphism between $\Lambda^p(M)$ and
$\Lambda^{4-p}(M)$ by the so called Hodge $*_p$-operator. In view of the
pseudoeuclidean nature of $\eta$, we are going to make use of the Hodge-$*$ and
the $\tilde{\eta}$-isomorphisms which will serve as good substitutes of the
Poincare isomorphisms $D^p$.

In view of these remarks and of the considerations in Sec.3.7.5 and Sec.6.3.3 we
see that the most natural choice for mathematical images of the two physical
components should be corresponding representatives of the above defined four
subspaces when a $p$-vector is introduced. What we still have to determine in
our 4-dimensional case is $p=?$.

In order to find appropriate $p$ we first assume that the Hodge-$*$ defines the
required algebraic correspondence between the formal representatives of the two
subsystems, so, we have two choices: $p=1$ and $p=2$, since the choice $p=3$
reduces to $p=1$. In the case $p=1$ our couple looks as $(\alpha,*\alpha)$,
where $\alpha$ is 1-form. The required local dynamical equilibrium formally
means that the two differential flows
$i_{\tilde{\eta}(\alpha)}\mathbf{d}*\alpha$ and
$i_{\tilde{\eta}(*\alpha)}\mathbf{d}\alpha$ shall be nonzero in general, and
shall differ only in sign, i.e.,
$i_{\tilde{\eta}(\alpha)}\mathbf{d}*\alpha$ to be equal to
$-i_{\tilde{\eta}(*\alpha)}\mathbf{d}\alpha$,
but such relation is impossible since $-i_{\tilde{\eta}(*\alpha)}\mathbf{d}\alpha$
is a trivial zero: $\tilde{\eta}(*\alpha)$ is a 3-vector, and $\mathbf{d}\alpha$
is a 2-form, therefore, the case $p=1$ does not work.

The above simple consideration clearly shows that, following H.Minkowski,
we have to choose for mathematical identification of the field a
$\mathbb{R}^2$-valued differential 2-form $\Omega$ on $M$ and its $\eta$
co-image $\bar{\Omega}$ as follows: $$ \Omega=F\otimes e_1+*F\otimes e_2, \ \ \
\bar{\Omega}=\tilde{\eta}^{-1}(F)\otimes e_1+\tilde{\eta}^{-1}(*F)\otimes e_2 ,
$$
where $(F,*F)$ are 2-forms and $(e_1,e_2)$ is a basis of the vector space
$\mathbb{R}^2$. Therefore, the two balance and two interaction differential
flows will be, respectively, $$ i(\tilde{\eta}^{-1}(F)\mathbf{d}F, \ \ \
i(\tilde{\eta}^{-1}(*F))\mathbf{d}*F , $$ $$
i(\tilde{\eta}^{-1}(F)\mathbf{d}*F, \ \ \ i(\tilde{\eta}^{-1}(*F))\mathbf{d}F.
$$

\subsection{Dynamical equations}

Our field object now must survive through space-time propagation during which it
has to keep its structure, establishing and supporting internal dynamical
equilibrium between its two recognizable subsystems. Our mathematical
interpretation of this vision differs substantially from that of
Maxwell-Minkowski, simply speaking, it consists in considering $\bar{\Omega}$
as a $\vee$-extended {\it algebraic} and {\it Lie}-symmetry of
$\Omega$: (Sec.2.8.4; Sec.6.3.3):
$$
i^{\vee}_{\bar{\Omega}}\Omega=\mathfrak{C}\in \Lambda^2_{const}(M,\mathbb{R}^2)
, \ \ \, \text{i.e.}, \ \ \
\mathbf{d}i^{\vee}_{\bar{\Omega}}\Omega=0 ; \ \ \
\mathcal{L}^{\vee}_{\bar{\Omega}}\Omega=0.
$$
Explicitly, denoting
$\bar{F}=\tilde{\eta}^{-1}(F)$, and $\bar{*F}=\tilde{\eta}^{-1}(*F)$, the
required differential $\vee$-symmetry gives
\begin{eqnarray*}
\mathcal{L}_{\bar{\Omega}}^{\vee}\Omega&=&
\left[\mathbf{d}\langle F,\bar{F}\rangle-i_{\bar{F}}(\mathbf{d}F)
\right]\otimes e_1\vee e_1\\
&+&\left[\mathbf{d}\langle
*F,\bar{*F}\rangle-i_{\bar{*F}}(\mathbf{d}*F)\right]\otimes
e_2\vee e_2\\ &+&
\{2\mathbf{d}\langle F,\bar{*F}\rangle-[i_{\bar{*F}}(\mathbf{d}F)+
i_{\bar{F}}(\mathbf{d}*F)]\}\otimes e_1\vee e_2=0 .
\end{eqnarray*}
\vskip 0.2cm
{\bf Remark}. We have chosen the $"\vee"$-extension of the Lie derivative
paying due respect to the entire symmetry between the two components $F$ and
$*F$ and to the dynamical inter-equilibrium they keep during propagation.
\vskip 0.2cm
The equations we obtain are
\begin{eqnarray*}
&&\mathbf{d}\langle F,\bar{F}\rangle-i_{\bar{F}}(\mathbf{d}F)=0 ,
\\
&&\mathbf{d}\langle
*F,\bar{*F}\rangle-i_{\bar{*F}}(\mathbf{d}*F)=0 ,
\\
&&2\mathbf{d}\langle F,\bar{*F}\rangle-[i_{\bar{*F}}(\mathbf{d}F)+
i_{\bar{F}}(\mathbf{d}*F)]=0.
\end{eqnarray*}

Since in our case the formal identity
$\langle F,\bar{F}\rangle=-\langle *F,\bar{*F}\rangle$ always holds,
summing up the first two equations we obtain
$$
i_{\bar{F}}\mathbf{d}F+i_{\bar{*F}}\mathbf{d}*F=0, \ \ \text{i.e.} \ \ \
F^{\alpha\beta}(\mathbf{d}F)_{\alpha\beta\mu}+
(*F)^{\alpha\beta}(\mathbf{d}*F)_{\alpha\beta\mu}=0,
$$
which coincides with the zero divergence of the standard
and well trusted electromagnetic stress-energy-momentum tensor $Q_\mu^\nu$
(Sec.6.3.1): $$ \nabla_\nu
Q^\nu_\mu=-\nabla_\nu\Big[\frac12\big(F_{\mu\sigma}F^{\nu\sigma}+
(*F)_{\mu\sigma}(*F)^{\nu\sigma}\big)\Big]=0.
$$
From this explicit expression of $Q_\mu^\nu$ in terms of $F$ and $*F$
it is clearly seen that the full stress-energy-momentum of the field is the sum
of the stress-energy-momentum carried by $F$, i.e.
$\frac12F_{\mu\sigma}F^{\nu\sigma}$, and the stress-energy-momentum carried by
$(*F)$, i.e. $\frac12(*F)_{\mu\sigma}(*F)^{\nu\sigma}$.
Now, the algebraic symmetry equation
$$
i^{\vee}_{\bar{\Omega}}\Omega=
\frac12F_{\mu\sigma}F^{\mu\sigma}e_1\vee e_1+
\frac12(*F)_{\mu\sigma}(*F)^{\mu\sigma}e_2\vee e_2+
F_{\mu\sigma}(*F)^{\mu\sigma}e_1\vee e_2=\mathfrak{C}
$$
requires
$$
I_1=\frac12F_{\mu\sigma}F^{\mu\sigma}
=-\frac12(*F)_{\mu\sigma}(*F)^{\mu\sigma}=const, \ \
I_2=\frac12F_{\mu\sigma}(*F)^{\mu\sigma}=const.
$$

Hence, the above two equations, i.e., the $\vee$-extended algebraic and
Lie symmetry of $\Omega$ with respect to $\bar{\Omega}$, give:
\begin{eqnarray*}
&&(L_{\bar{F}}F)_\mu=-F^{\alpha\beta}(\mathbf{d}F)_{\alpha\beta\mu}=0,
\rightarrow F \ \ \text{is autoclosed},\\
&&(L_{\bar{*F}}*F)_\mu=-(*F)^{\alpha\beta}(\mathbf{d}*F)_{\alpha\beta\mu}=0,
\rightarrow *F \ \ \text{is autoclosed},\\
&&(L_{\bar{*F}}F)_\mu+(L_{\bar{F}}*F)_\mu=
-(*F)^{\alpha\beta}(\mathbf{d}F)_{\alpha\beta\mu}-
F^{\alpha\beta}(\mathbf{d}*F)_{\alpha\beta\mu}=0 , \ \ \alpha<\beta.
\end{eqnarray*}
The conformal invariance holds, and every solution $(F,*F)$ realizes the idea
for {\it local equilibrium} (Sec.6.3.3). In terms of the coderivative
$\delta=*\,\mathbf{d}\,*$ we get
\begin{eqnarray*}
&&F_{\mu\nu}(\delta F)^\nu=0,\\
&&(*F)_{\mu\nu}(\delta *F)^\nu=0,\\
&&(*F)_{\mu\nu}(\delta F)^\nu+F_{\mu\nu}(\delta *F)^\nu=0.
\end{eqnarray*}
The coordinate-free form of these
equations reads:
\begin{eqnarray*}
(*F)\wedge*\mathbf{d}*F &\equiv&\delta F\wedge *F=0,\\
F\wedge *\mathbf{d}F &\equiv& -F\wedge \delta *F=0,\\
F\wedge *\mathbf{d}*F+(*F)\wedge*\mathbf{d}F
&\equiv&\delta F\wedge F-\delta *F\wedge *F=0.
\end{eqnarray*}

Let us not forget that: {\it only {\bf time-stable}
null field solutions with {\bf finite} spatial support of the above equations
will be of interest further}.

We can come to these dynamical equations recalling
that vector valued differential forms can be multiplied according to the
rule
$$
\varphi(\Phi,\Psi)=\varphi(\alpha^i\otimes e_i).(\beta^j\otimes e_j)=\alpha^i\wedge
\beta^j\otimes \varphi(e_i,e_j),
$$
where $\{e_i\}$ is a basis of the vector space $W$, and
$\varphi:W\times W\rightarrow W$
is a bilinear map. So, we can form the expression
$\vee(\Omega,*\mathbf{d}\Omega)$ and to put it equal to zero, meaning that the
change $\mathbf{d}\Omega $
of the field $\Omega $ is not essential for the
surviving of $\Omega $, i.e. the change $\mathbf{d}\Omega $
is {\it admissible}.
The same result may be obtained if we replace the change operator $\mathbf{d}$
with the coderivative change operator $\delta $. Hence, using again the $"\vee"$
operator to fix the equilibrium between the two subsystems, the equation
$$
\vee(\delta\Omega,*\Omega)=0
$$
gives the above equations
$$
F_{\mu\nu}(\delta F)^\nu=0, \ \ (*F)_{\mu\nu}(\delta *F)^\nu=0, \ \
(*F)_{\mu\nu}(\delta F)^\nu +F_{\mu\nu}(\delta *F)^\nu=0.
$$

We shall consider now these equations more in detail, in particular, if
they admit spatially finite, time-stable and spin-carrying null solutions.

Clearly, these equations admit linear and nonlinear solutions. The
linear solutions satisfy $\delta F=\delta *F=0$, and coincide with the
solutions of Maxwell vacuum equations, so {\bf these solutions will be out of
consideration}.

We turn now to the nonlinear ones, i.e. to those satisfying $\delta
F\neq 0$, $\delta *F\neq 0$. The first important property reads: {\bf all
nonlinear solutions are null fields}, i.e. $\mathfrak{C}=0$.

It is clearly seen that the first two groups of these equations may be
considered as two linear homogeneous systems with respect to $(\delta F)^\mu $
and $(\delta*F)^\mu $ respectively. These
homogeneous systems may have non-zero solutions only if
$\mathrm{det}(F_{\mu\nu})=\mathrm{det}((*F)_{\mu\nu})=
(\frac12F_{\mu\nu}(*F)^{\mu\nu})^2=0$, i.e. if
$I_2=2{\bf E}.{\bf B}=0$.

Further, summing up these three systems of equations, we obtain
$$
(F+*F)_{\mu\nu}(\delta F+\delta *F)^\nu=0.
$$
Recall that a 2-form $A$ on $M$ satisfies $*A\wedge *A=-A\wedge A$, and
$$
A\wedge A=\sqrt{det(A_{\mu\nu})}\,\omega_o,
 \ \ A\wedge A=-A\wedge *(*A)=\frac12A_{\mu\nu}(*A)^{\mu\nu}\omega_o .
$$
If $(\delta F+\delta *F)^\nu\neq 0$, then
$$
(F+*F)\wedge(F+*F)=F\wedge F+2F\wedge *F-F\wedge F=
\sqrt{det[(F+*F)_{\mu\nu}]}\,\omega_o,
$$
so,
$$
0=\mathrm{det}||(F+*F)_{\mu\nu}||=
\left [-F_{\mu\nu}F^{\mu\nu}\right ]^2=(2I_1)^2.
$$
If $\delta F^\nu=-(\delta *F)^\nu\neq 0$, we sum up the first two systems
of equations and obtain $(*F-F)_{\mu\nu}(\delta *F)^\nu=0$. Consequently,
$$
0=\mathrm{det}||(*F-F)_{\mu\nu}||=
\left [F_{\mu\nu}F^{\mu\nu}\right ]^2=(2I_1)^2.
$$
This completes the proof. Hence, the corresponding $\mathfrak{C}$ is zero.

In view of this, from the identity (Sec.6.3.1)
$$
I_1=\frac12F_{\alpha\beta}F^{\alpha\beta}\delta^\nu_\mu=
F_{\mu\sigma}F^{\nu\sigma}-(*F)_{\mu\sigma}(*F)^{\nu\sigma}
$$
it follows that the two susystems carry the same stress-energy-momentum:
$F_{\mu\sigma}F^{\nu\sigma}=(*F)_{\mu\sigma}(*F)^{\nu\sigma}$.

In this way a permanent local dynamical equilibrium between
the two subsystems, represented by $F$ and $*F$, is established.

Recalling that all null fields have zero eigen values
we come to the following two corollaries:

{\bf Corollary}. The vector $\delta F^\mu$ is an eigen vector of $F_\mu^\nu$;
the vector $(\delta *F)^\mu$ is an eigen vector of $(*F)_\mu^\nu$.

{\bf Corollary}. The vectors $\delta F^\mu$ and $(\delta *F)^\mu$ are eigen
vectors of the energy tensor $Q^\mu_\nu$.

We continue studying the properties of the equations and their solutions
keeping in mind the two properties of the vectors in Minkowski space:

	1. There are NO mutually orthogonal time-like vectors in Minkowski
space.

	2. All eigen vectors of $F$ and $*F$ are eigen vectors of $Q_\mu^\nu$.

First we look at the invariance properties of our nonlinear system with respect
to the duality transformations:
$$
F\rightarrow \mathcal{F}=aF-b*F,\ \ *F\rightarrow*\mathcal{F}=bF+a*F,\ \
a,b\in \mathbb{R}.
$$
 We substitute and obtain:
$$
\delta \mathcal{F}\wedge *\mathcal{F}=a^2(\delta F\wedge *F)-
b^2(\delta *F\wedge F)+ab(\delta F\wedge F-\delta *F\wedge *F)
$$
$$
\delta *\mathcal{F}\wedge \mathcal{F}=a^2(\delta *F\wedge F)-
b^2(\delta F\wedge *F)+ab(\delta F\wedge F-\delta *F\wedge *F)
$$
$$
\delta\mathcal{F}\wedge\mathcal{F}-\delta *\mathcal{F}\wedge *\mathcal{F}=
(a^2-b^2)(\delta F\wedge F-\delta *F\wedge *F)-
2ab(\delta F\wedge *F+\delta *F\wedge F).
$$
It is seen that if $F$ defines a solution then $\mathcal{F}$ also defines a
solution. Conversely, if $\mathcal{F}$ defines a solution then subtracting the
second equation from the first and taking in view that $a^2+b^2\neq 0$ we
obtain $\delta F\wedge *F=(\delta *F)\wedge F$.
Now, the case $a^2=b^2$ directly leads to $\delta F\wedge F-\delta *F\wedge
*F=0$ and $\delta *F\wedge F=\delta F\wedge *F=0$. If $a^2\neq b^2$ then,
 since also
$\delta\mathcal{F}\wedge \mathcal{F}-\delta *\mathcal{F}\wedge *\mathcal{F}=0$,
the compatibility with the third equation leads to
$$
\delta F\wedge F-\delta *F\wedge *F=-\frac{a^2-b^2}{ab}\delta F\wedge *F,
$$
$$
\delta F\wedge F-\delta *F\wedge *F=\frac{4ab}{a^2-b^2}\delta F\wedge *F.
$$
These two equations require $4a^2b^2=-(a^2-b^2)^2$, which is impossible, so
$\delta F\wedge F-\delta *F\wedge *F=0$ and
$\delta F\wedge *F=\delta *F\wedge F=0$.
The duality invariance follows.

 We recall that in the {\it null field}
case $Q_\mu^\nu$ has just one isotropic eigen direction, defined by the
isotropic vector $\bar{\zeta}$, and all of its other eigen directions are
space-like.

We prove now that all nonlinear solutions satisfy the conditions
$$
(\delta F)_\mu (\delta *F)^\mu =0,\
\left|\delta F\right|=\left|\delta *F\right| .            %(35)%
$$
From $i(\tilde{\eta}^{-1}({\delta *F}))(\delta F\wedge *F)=0$, we get
$$
(\delta *F)^\mu(\delta F)_\mu(*F)-
\delta F\wedge (\delta *F)^\mu(*F)_{\mu\nu}dx^\nu=0.
$$
Because of the obvious nullification of the second term the first term will be
equal to zero (at non-zero $*F$) only if
$(\delta F)_\mu (\delta *F)^\mu=0$.

Further we form the interior product
$i(\tilde{\eta}^{-1}(\delta *F))(\delta F\wedge F-\delta *F\wedge *F)=0$ and
obtain
$$
(\delta *F)^\mu(\delta F)_\mu F-
\delta F\wedge (\delta *F)^\mu F_{\mu\nu}dx^\nu-
$$
$$
-(\delta *F)^2(*F)+\delta *F\wedge(\delta *F)^\mu (*F)_{\mu\nu}dx^\nu=0.
$$
Clearly, the first and the last terms are equal to zero. So, the inner
product by $\tilde{\eta}^{-1}(\delta F)$ gives
$$
-(\delta F)^2 (\delta *F)^\mu F_{\mu\nu}dx^\nu+
\left[(\delta F)^\mu (\delta *F)^\nu F_{\mu\nu}\right]\delta F
-(\delta *F)^2(\delta F)^\mu (*F)_{\mu\nu}dx^\nu=0.
$$
The second term of this equality is zero since $\delta F^\mu F_{\mu\nu}=0$.
Besides, our third equation $\delta F\wedge F-\delta *F\wedge *F=0$ means
$$
(\delta *F)^\mu F_{\mu\nu}dx^\nu=-(\delta F)^\mu
(*F)_{\mu\nu}dx^\nu.
$$ So,
$$
\left[(\delta F)^2-(\delta *F)^2\right](\delta
F)^\mu(*F)_{\mu\nu}dx^\nu=0.
$$
Now, since by supposition $(\delta
F)^\mu(*F)_{\mu\nu}dx^\nu\neq 0$, then the relation $\left|\delta
F\right|=\left|\delta *F\right|$ follows immediately.

It follows from this result and from the above corollaries that

1. $\delta F$ and  $\delta *F$ can NOT be time-like,

2. $\delta F$ and $\delta *F$ are {\bf simultaneously space-like}, i.e.
$(\delta F)^2=(\delta *F)^2<0$, or {\bf simultaneously isotropic}, i.e.
$|\delta F|=|\delta *F|=0$.  We note that in this last case the isotropic
vectors $\delta F$ and $\delta *F$ are also eigen vectors of $Q_\mu^\nu$, and
since $Q_\mu^\nu$ has just one isotropic eigen direction, which we denoted by
 $\bar{\zeta}$, we conclude that $\delta F, \delta*F$ and $\zeta$ should be
collinear. But, the requirement $\delta F\wedge F=\delta *F\wedge *F\neq 0$
requires non-null character of $\delta F$ and $\delta *F$, otherwise, since
$F$ and $*F$ can be represented respectively as $A\wedge\zeta$ and $A^*\wedge
\zeta$, if $\delta F$ and $\delta *F$ are colinear to $\zeta$, obviously,
$\delta F\wedge F=\delta *F\wedge *F=0$. In other words, the space-like
nature of $\delta F$ and $\delta *F$ guarantees {\bf nonzero equal
energy-momentum flows} between $F$ and $*F$, which, in view of the relation
$F_{\mu\sigma}F^{\nu\sigma}=(*F)_{\mu\sigma}(*F)^{\nu\sigma}$, establishes
permanent dynamical equilibrium between $F$ and $*F$.

 In our further study of the nonlinear solutions we shall make use
of the above mentioned representation in a $\zeta$-adapted frame:
$$
F=A\wedge\zeta,\ *F=A^*\wedge\zeta,
$$
where $A$ and $A^*$ are 1-forms,
$\bar{\zeta}^\mu=\eta^{\mu\nu}\zeta_\nu$. It follows that
$$
F\wedge\zeta=*F\wedge\zeta=0.
$$
\vskip 0.3cm
\noindent
{\bf Remark}. Further we are going to skip
the "bar" over $\zeta,\delta F, A, A^*$, and from the context it will be clear
the meaning: one-forms, or vector fields.
\vskip 0.3cm We establish
now some useful properties of these quantities. \vskip 0.3cm

	$1^o. \  A_\mu\zeta^\mu=A^*_\mu\zeta^\mu=0$

	$2^o. \  A^\sigma(*F)_{\sigma\mu}=0,\ \ (A^*)^\sigma F_{\sigma\mu}=0$

	$3^o. \  A^\sigma A^*_\sigma=0$,\ \ $A^2=(A^*)^2<0$.

In order to prove $1^o$ we note
\begin{align*}
& 0=I_1=\frac12 F_{\mu\nu}F^{\mu\nu}=\frac12(A_\mu\zeta_\nu-A_\nu\zeta_\mu)
(A^\mu\zeta^\nu-A^\nu\zeta^\mu)  \\
&=\frac12(2A_\mu A^\mu.\zeta_\nu\zeta^\nu-2(A_\mu\zeta^\mu)^2)=
-(A_\mu\zeta^\mu)^2.
\end{align*}
The second of $1^o$ is proved in the same way just replacing $F$ with $*F$
and $A$ with $A^*$.

To prove $2^o$ we notice
$$
0=*(A^*\wedge A^*\wedge\zeta)=*(A^*\wedge *F)=
-(A^*)^\sigma F_{\sigma\mu}dx^\mu.
$$
Similarly
$$
0=*(A\wedge A\wedge\zeta)=*(A\wedge F)=
A^\sigma (*F)_{\sigma\mu}dx^\mu.
$$
Hence, $A^*$ is an eigen vector of $F$ and $A$ is an eigen vector of $*F$.
The case $|A|=0$ is not considered since then $A$ is collinear to
$\zeta$ and $F=A\wedge \zeta=0$, so $*F=0$ too.

Now, $3^o$ follows from $2^o$ because
$$
0=-(A^*)^\sigma F_{\sigma\mu}=-(A^*)^\sigma(A_\sigma\zeta_\mu-
A_\mu\zeta_\sigma)
$$
$$
=-(A^*.A)\zeta_\mu+A_\mu(A^*.\zeta)=-(A^*.A)\zeta_\mu.
$$

Finally we express $Q_\mu^\nu$ in terms of $A$, or $A^*$, and $\zeta$. Since
$I_1=0$,i.e., $F_{\mu\sigma}F^{\nu\sigma}=(*F)_{\mu\sigma}(*F)^{\nu\sigma}$, we
have
$$
Q_\mu^\nu=-\Big[F_{\mu\sigma}F^{\nu\sigma}\Big]= -(A_\mu\zeta_\sigma-
A_\sigma\zeta_\mu)(A^\nu\zeta^\sigma-A^\sigma\zeta^\nu)=
$$
$$
-A^2\zeta_\mu\zeta^\nu=
-(A^*)^2\zeta_\mu\zeta^\nu.
$$
Hence, since $Q_4^4>0$ and $\zeta_4\zeta^4>0$ we obtain
$A^2=(A^*)^2<0$.

Noting that $\zeta$ and its normalized form
$Z=\zeta/\zeta^4$ define the same integral
lines, we show how the geodesic character of $\zeta$ follows
from the local conservation law $\nabla_\nu Q^{\mu\nu}=0$.  In fact,
$$
\nabla_\nu Q^{\mu\nu}=
\nabla_\nu\left(-A^2(\zeta^4)^2\,Z^\mu\,Z^\nu\right)=
-\Big[Z^\mu\nabla_\nu A^2(\zeta^4)^2\,Z^\nu+
A^2(\zeta^4)^2Z^\nu\nabla_\nu\,Z^{\mu}\Big]=0.
$$
This equation holds for any $\mu=1,2,3,4$. We consider it for $\mu=4$ and
recall that for $\mu=4$ we have $Z^4=1$. In our coordinates
$\nabla_\nu=\partial_\nu$, and we obtain that the second term becomes zero,
so $\nabla_\nu A^2(\zeta^4)^2\,Z^\nu=0$.  Therefore,
$Z^\nu\nabla_\nu\,Z^\mu=0$, which means that {\bf all the trajectories
of} $\zeta$ {\bf are parallel straight isotropic} lines. Hence, with
every nonlinear solution $F$ we are allowed to introduce $\zeta$-{\it adapted}
coordinate system by the requirement that the trajectories of $\zeta$ to be
parallel to the plane $(z,\xi)$. In such a coordinate system we may assume
$\zeta_\mu=(0,0,\varepsilon,1),\ \varepsilon=\pm 1$. From 3-dimensional point
of view this means that the field propagates along the coordinate $z$, and
$\varepsilon =-1$ implies propagation along $z$ from $-\infty$ to $+\infty$,
while $\varepsilon =+1$ implies propagation from $\infty$ to $-\infty$.

\vskip 0.4cm
We repeat once again: {\it The translational direction of propagation of any
null-field nonlinear solution is determined intrinsically}.

\vskip 0.4cm
We note that in Maxwell theory all time
stable null-field solutions are {\it spatially infinite}, otherwise they have
to blow-up radially according to the corresponding theorem for D'Alembert
equation. We shall see that Extended electrodynamics (EED) has no problems in this
respect, i.e. spatially finite and time stable null field solutions are allowed.

We express explicitly now $F, *F, A$ and $A^*$ in the corresponding $\zeta$-{\it
adapted} coordinate system making use of the relations $F=A\wedge\zeta,\
*F=A^*\wedge\zeta$, where $\zeta=\varepsilon dz+d\xi$, and of the above
established properties. We easily obtain:
$$
F_{12}=F_{34}=0,\ \ F_{13}=\varepsilon F_{14}, \ \
F_{23}=\varepsilon F_{24},
$$
$$
(*F)_{12}=(*F)_{34}=0,\ \
(*F)_{13}=\varepsilon (*F)_{14}=-F_{24}, \ \ (*F)_{23}=\varepsilon
(*F)_{24}=F_{14}.
$$
Moreover, from $A.\zeta=A^*.\zeta=0$ it follows that in an
$\zeta$-adapted coordinate system we obtain
\begin{eqnarray*}
&&A=A_1 dx+A_2 dy
+f.\zeta=(F_{14})dx+(F_{24})dy +f.\zeta \\ &&A^*=A^*_1 dx+A^*_2 dy
+f^*.\zeta=(-F_{23})dx+(F_{13})dy +f^*.\zeta\\ &&=-(\varepsilon
A_2)dx+(\varepsilon A_1)dy+f^*.\zeta,
\end{eqnarray*}
where $f$ and $f^*$ are two arbitrary functions.
\vskip 0.3cm
Having this in mind we prove the following:
\vskip 0.2cm
\noindent
All nonlinear solutions satisfy the relations:
$$
\zeta^\mu(\delta F)_\mu=0,\ \zeta^\mu(\delta *F)_\mu=0.       %(38)%
$$
\vskip 0.2cm
\noindent
 We form the interior product
 $i(\zeta)(\delta F\wedge *F)=0$, and recalling that $A^*.\zeta=0$, we get
$$
0=\left[\zeta^\mu
(\delta F)_\mu\right]*F-\delta F\wedge(\zeta)^\mu(*F)_{\mu\nu}dx^\nu=
\zeta^\mu(\delta F)_\mu *F.
$$
So, $\zeta^\mu(\delta F)_\mu=0$. Similarly, from
$(\delta *F)\wedge F=0$ we get $\zeta^\mu(\delta *F)_\mu=0$.

\vskip 0.3cm
We also note that the 3-form $\delta F\wedge F$ is isotropic: $(\delta
F\wedge F)^2=0$. In fact
$$
(\delta F\wedge F)^{\mu\nu\sigma}(\delta F\wedge F)_{\mu\nu\sigma}=
F^{\mu\nu}\big[(\delta F)^2 F_{\mu\nu}\big]-
F^{\mu\nu}\big[\delta F\wedge i(\delta F)F\big]_{\mu\nu}
$$
$$
=(\delta F)^2 F^{\mu\nu}F_{\mu\nu}=0, \
 \mu<\nu<\sigma.
$$

It deserves noting that there are no nonlinear spherically symmetric solutions,
i.e. if $F$ is a spherically symmetric solution then it is a solution
of Maxwell's equations. In fact, the most general spherically symmetric
2-form in spherical coordinates $(r,\theta,\varphi)$, originating at the
symmetry center, is $$ F=f(r,\xi)\,dr\wedge d\xi+
h(r,\xi)\,\mathrm{sin}\theta\, d\theta\wedge d\varphi. $$ Now the equation
$F\wedge *\mathbf{d}F=0$ requires $h(r,\xi)=const$, and the equation $\delta
F\wedge *F=0$ requires $f(r,\xi)=const/r^2$. It follows: $\mathbf{d}F=0,\
\delta F=0$.

Recall that at every point, where the field is different from zero, we have
in fact three coframes: the pseudoorthonormal ($\zeta$-adapted) coframe
$(dx,dy,dz,d\xi)$, the pseudoorthonormal coframe
$\chi^0=(\mathbf{A}=A/|A|,\varepsilon\mathbf{A^*}=A^*/|A^*|,
\mathbf{R}=-dz,\mathbf{S}=d\xi)$,
and the pseudoorthogonal frame $\chi=(A,\varepsilon A^*,{\bf R}, {\bf S})$.
The matrix $\chi_{\mu\nu}$ of  $\chi$ with respect to the coordinate coframe is
$$
 \chi_{\mu\nu}=\begin{Vmatrix} u  &-p  & 0  &0 \\ p  & u
& 0  &0 \\ 0  & 0  &-1  &0 \\ 0  & 0  & 0  &1 \end{Vmatrix}.
$$
We define now
the amplitude $\phi>0$ of the solution $(F,*F)$ by
$$
\phi=\sqrt{|det(\chi_{\mu\nu})|}.                              %(41)%
$$
Clearly, in an $\zeta$-adapted coordinate system
$$
\phi=\sqrt{u^2+p^2}=\sqrt{Q_4^4}=|A|.
$$
Each of these three coframes defines its own volume form:
\begin{eqnarray*}
&&\omega=dx\wedge dy\wedge dz\wedge d\xi ,\\
&&\omega_{\chi^{o}}=\mathbf{A}\wedge\mathbf{\varepsilon A}\wedge
\mathbf{R}\wedge\mathbf{S}=-\omega ,\\
&&\omega_{\chi}=A\wedge A^{*}\wedge\mathbf{R}\wedge\mathbf{S}=
-(u^2+p^2)\omega.
\end{eqnarray*}
We proceed further to define the {\it phase} of the nonlinear solution in
these terms. We shall need the matrix $\chi^0_{\mu\nu}$ of the frame $\chi^0$
with respect to the coordinate basis. We obtain

$$
\chi^0_{\mu\nu}=\begin{Vmatrix}
\frac{u}{\sqrt{u^2+p^2}}  &\frac{-p}{\sqrt{u^2+p^2}}  & 0  &0 \\
\frac{p}{\sqrt{u^2+p^2}}  &\frac{u}{\sqrt{u^2+p^2}}   & 0  &0 \\
0                         & 0                         &-1  &0 \\
0                         & 0                         & 0  &1
\end{Vmatrix}.
$$
The {\it trace} of this matrix is
$$
tr(\chi^0_{\mu\nu})=\frac{2u}{\sqrt{u^2+p^2}}.
$$
Obviously, the inequality $|\frac 12 tr(\chi^0_{\mu\nu})|\leq 1$ is
fulfilled. Now, the phase function $\varphi$ and the phase $\psi$ of the
nonlinear solution are defined by
$$
\varphi=\frac 12 tr(\chi^0_{\mu\nu}),\
\psi=\mathrm{arccos}(\varphi)=
\mathrm{arccos}\left(\frac 12 tr(\chi^0_{\mu\nu})\right) .
$$                                                 %(42)%

Making use of the amplitude function $\phi$, of the phase function $\varphi $
and of the phase $\psi$, we can write
$$
u=\phi.\varphi=\phi\,\mathrm{cos}\psi,\
\ p=\phi\,(\pm\sqrt{1-\varphi^2})=\phi\,\mathrm{sin}\psi.  %(43)%
$$

We consider now the 1-forms $(A,A^*)$ on a region
$U\subset M$, where $A\neq 0, \ A^*\neq 0$, and
since $A.A^*=0$, we have a 2-dimensional Pfaff system on
$U$. This 2-dimensional Pfaff system
$(A,A^*)$ is completely integrable, i.e. the following
equations hold:
$$
{\bf d}A\wedge A\wedge A^* =0,
\ {\bf d}A^*\wedge A\wedge A^*=0.
$$
In fact, $A\wedge A^*=(u^2+p^2)dx\wedge dy$, and in every
term of ${\bf d}A$ and ${\bf d}A^*$ at least one of the basis
covectors $dx$ and $dy$ will participate, so the above exterior products will
vanish.  The assertion is proved.
\vskip 0.3cm
\noindent
{\bf Remark}.
As we shall see further, this property of Frobenius integrability  is dually
invariant for the nonlinear solutions.
\vskip 0.3cm
\noindent
{\bf Remark}. These considerations stay in force also
for those time stable linear solutions, which have zero invariants
$I_1=I_2=0$.  But Maxwell's equations require $u$ and $p$ to be {\it infinite
running plane waves} in this case, so the corresponding amplitudes will NOT
depend on two of the spatial coordinates and the phase functions will be also
running waves.  As we'll see further, the phase functions for the nonlinear
solutions are arbitrary bounded functions.

We proceed further to give the relativistic definition of the earlier
introduced concept of {\it scale factor} $\mathcal{L}_o$ \index{scale factor}
for a given nonlinear
solution $F$ with $|\delta F|\neq 0$. First we recall a theorem from vector
bundle theory, which establishes some important properties of those vector
bundles which admit pseudoriemannian structure. The theorem says that if a
vector bundle $\Sigma$ with a base manifold $B$ and standard fiber $V$ admits
pseudoriemannian structure $g$ of signature $(p,q), p+q=dimV$, then it is
always possible to introduce in this bundle a riemannian structure $h$ and a
linear automorphism $\varphi$ of the bundle, such that two subbundles
$\Sigma^+$ and $\Sigma^-$ may be defined with the following properties:

\begin{enumerate}
\item $g(\Sigma^+,\Sigma^+)=h(\Sigma^+,\Sigma^+)$,
\item $g(\Sigma^-,\Sigma^-)=-h(\Sigma^-,\Sigma^-)$,
\item $g(\Sigma^+,\Sigma^-)=h(\Sigma^+,\Sigma^-)=0$.
\end{enumerate}
The automorphism $\varphi$ is defined by
$$
g_x(u_x,v_x)=h_x(\varphi(u_x),v_x),\quad u_x,v_x\in V_x, \ \ x\in B.
$$
In components we have
$$
g_{ij}=h_{ik}\varphi^k_j \rightarrow \ \varphi^k_i=g_{im}h^{mk}.
$$
In the tangent bundle case this theorem allows to separate a subbundle of the
tangent bundle if the manifold admits pseudoriemannian metric. In the simple
case of Minkowski space $(M,\eta),\  sign(\eta)=(-,-,-,+)$, introducing
the standard Euclidean metric $h$ on $M$ we may separate 1-dimensional
subbundle, i.e. a couple of vector fields $\pm X_o$, being eigen vectors
of $\varphi$, and to require $\eta(X_o,X_o)=1$. In our canonical coordinates
we obtain $X_o=\partial/\partial \xi$.

Now, the scale factor $\mathcal{L}_o$ for a nonlinear solution
\index{scale factor}
 $(F,*F)$ with
$|\delta F|\neq 0$, $|\delta *F|\neq 0$ is defined by
$$
\mathcal{L}_o=\frac{|i(X_o)F|}{|\delta F|}= \frac{|A|}{|\delta     %43%
F|}=\frac{|i(X_o)(*F)|}{|\delta *F|}=\frac{|A^*|}{|\delta*F|}\cdot
$$

\section{On the Homology defined by the null fields energy tensor}

\subsection{Introductory Remarks} \index{null fields homology}
From pure algebraic point of view we speak about homology (or, cohomology)
every time when we meet a linear map $D$ in a vector space $\mathbb{V}$ over
a field (e.g. $\mathbb{R}$, or $\mathbb{C}$), or in a module $\mathbb{W}$
over some ring, having the property $D\circ D=0$. Then we have two related
subspaces, $Ker(D)=\{x\in \mathbb{V}: D(x)=0\}$ and
$Im(D)=D(\mathbb{V})$. Since $Im(D)$ is a subspace of $Ker(D)$, we can
factorize, and the corresponding factor space $H(D,\mathbb{V})=Ker(D)/Im(D)$
is called the {\it homology space} for $D$. The dual linear map $D^*$ in the
dual space $\mathbb{V}^*$ has also the property $D^*\circ D^*=0$, so we
obtain the corresponding {\it cohomology space} $H^*(D^*,\mathbb{V}^*)$. In
such a situation the map $D$ (resp. $D^*$) is called {\bf boundary operator}
(resp {\bf coboundary operator}). The elements of $Ker(D)$ (resp. $Ker(D^*)$)
are called {\it cycles} (resp. {\it cocycles}), and the elements of $Im(D)$
(resp. $Im(D^*)$) are called {\it boundaries} (resp. {\it coboundaries}).

The basic property of a boundary operator $D$ is that every linear map
$\mathcal{B}:\mathbb{V}\rightarrow \mathbb{V}$ which commutes with $D$:
$D\circ \mathcal{B}=\mathcal{B}\circ D$, induces a linear map
$\mathcal{B}_*: H(D)\rightarrow H(D)$. So, a boundary operator realizes the
general idea of distinguishing some properties of a class of objects which
properties are important from a definite point of view, and to find those
transformations which keep invariant these properties.

The basic example for boundary operator used in theoretical physics is the
exterior derivative $\mathbf{d}$ (inducing the de Rham cohomology).
This operator acts in the space of differential forms over a manifold, e.g.,
the Euclidean space $(\mathbb{R}^3,g)$, or the Minkowski space-time
$M=(\mathbb{R}^4,\eta)$. The above mentioned basic property of every boundary
operator $D$ appears here as a commutation of $\mathbf{d}$ with the smooth
maps $\mathfrak{f}$ of the manifold considered:
$\mathbf{d}\circ\mathfrak{f}^*=\mathfrak{f}^*\circ \mathbf{d}$.  A well known
physical example for cocycles of $\mathbf{d}$ (which are called here {\it
closed differential forms}) comes when we consider a spherically symmetric
gravitational or electrostatic field generated by a point source.  Since the
field is defined only outside of the point-source, i.e. on the space
$N=\mathbb{R}^3-\{0\}$, then a natural object representing the field is a
closed differential 2-form $\omega, \mathbf{d}\omega=0$. In view of the obvious
spherical symmetry such a (spherically symmetric or $SO(3)$-invariant) closed
2-form is defined up to a constant coefficient $q$, and in standard spherical
coordinates originating at the point-source we obtain
$\omega=q\,\mathrm{sin}\,\theta\, d\theta\wedge d\varphi$. The Hodge star on
$N$ gives $*\omega=(q/r^2)dr$, which is usually called gravitational/electric
field generated by the point source $m/q$.  Now, the Stokes' theorem
establishes the corresponding charge $m/q$ as a topological invariant,
characterizing the nontrivial topology of the space $N$.

The above mentioned boundary operator $\mathbf{d}$ is a
differential operator. Pure algebraic boundary operators in a linear space
$V$ also exist and one way to introduce such operators is as
follows.  Let $V^*$ be the dual to $V$ space and $\langle\,,\rangle$ denote
the canonical conjugation:  $(x^*,x)\rightarrow\langle x^*,x\rangle$, where
$x^*\in V^*$ and $x\in V$.  Now fix $x\in V$ and let $x^*$ be such that
$\langle x^*,x\rangle =0$.  Consider now the decomposable element $x^*\otimes
x \in V^*\otimes V$. The isomorphism between $V^*\otimes V$ and $L(V,V)$
allows to consider $x^*\otimes x$ as a linear map
$\varphi_{(x^*,x)}:V\rightarrow V$ as follows:
$$
\varphi_{(x^*,x)}(y)=(x^*\otimes x)(y)=\langle x^*,y\rangle x, \ \ y\in V.
$$
Hence, the linear map $\varphi_{(x^*,x)}$ sends all elements of $V$ to the
1-dimensional space determined by the non-zero element $x\in V$. Clearly,
$\varphi_{(x^*,x)}$ is a boundary operator since
\begin{eqnarray*}
&&\varphi_{(x^*,x)}\circ\varphi_{(x^*,x)}(y)=
\varphi_{(x^*,x)}(\langle x^*,y\rangle x)\\
&&=
\langle x^*,y\rangle\varphi_{(x^*,x)}(x)=
\langle x^*,y\rangle\langle x^*,x\rangle x=0,\ \ y\in V.
\end{eqnarray*}
So, if $\{e_i\}$ and $\{\varepsilon^j\}$ are two dual bases in the
$n$-dimensional linear spaces $V$ and $V^*$ respectively, then every couple
$(\varepsilon^j,e_i), i\neq j$, determines the linear boundary operators
$\varphi_{(\varepsilon^j,e_i)}=\varepsilon^j\otimes e_i, i\neq j$.

If $g$ is an inner product in $V$ and the two elements $(x,y)$ are
$g$-orthogonal: $g(x,y)=0$, then denoting by $\tilde{g}$ the linear
isomorphism $\tilde{g}: V\rightarrow V^*$ (lowering indices) we can define
the linear map $\tilde{g}(x)\otimes y$, which is obviously a boundary
operator in $V$:  $$ (\tilde{g}(x)\otimes y)\circ(\tilde{g}(x)\otimes y)(z)=
\langle \tilde{g}(x),z\rangle(\tilde{g}(x)\otimes y)(y)=
\langle \tilde{g}(x),z\rangle g(x,y)(y)=0, \ z\in V.
$$
In particular, if $(V,\eta)$ is the Minkowski space, then every isotropic
vector $\zeta: \eta(\zeta,\zeta)=\zeta^2=0$ defines a boundary operator
$\tilde{\eta}(\zeta)\otimes\bar{\zeta}=\zeta\otimes\bar{\zeta}$. In fact,
$$
(\zeta\otimes\bar{\zeta})\circ
(\zeta\otimes\bar{\zeta})(x)=
\langle \zeta,x\rangle \langle\zeta,\bar{\zeta}\rangle(\bar{\zeta})=0,
\ \ x\in V.
$$
We note that the duality between $V$ and
$V^*$ allows to consider the element $x^*\otimes x$ as a linear map in $V^*$
as follows $(x^*\otimes x)(y^*)=\langle y^*,x\rangle x^*$, and to build the
corresponding boundary operators in $V^*$. Also, in the finite dimensional
case we always get $dim(V)-2=dim[Ker(x^*\otimes x)/Im(x^*\otimes x)]$, where
$\langle x^*,x\rangle=0$.

The above mentioned property that the image space of any such boundary
operator $x^*\otimes x$, where $\langle x^*, x\rangle =0$, is 1-dimensional,
implies that the natural extensions of these boundary operators to
derivations in the graded exterior algebras $\Lambda(V)$ and $\Lambda(V^*)$
define boundary operators of degree zero in these graded algebras (see
further).

\subsection{The null field electromagnetic energy tensor as boundary operator}
Let now the 2-forms $(F,*F)$ represent an electromagnetic field on Minkowski
space-time with the corresponding energy tensor
$$
Q_\mu^\nu=-\frac12\Big[F_{\mu\sigma}F^{\nu\sigma} +
(*F)_{\mu\sigma}(*F)^{\nu\sigma}\Big].
$$
We recall the property of the null field electromagnetic
stress-energy-momentum tensor $Q(F)_\mu^\nu$ established by Rainich
$$
Q_\mu^\sigma Q_\sigma^\nu=\frac14(I_1^2+I_2^2)\delta
_\mu^\nu,                                                      %53%
$$
where $I_1=\frac12 F_{\mu\nu}F^{\mu\nu}$ and
$I_2=\frac12 F_{\mu\nu}(*F)^{\mu\nu}$ are the two invariants.

In the frame of Extended Electrodynamics (EED), the
2-forms $(F,*F)$ satisfiy the nonlinear equations ($\mathbf{d}$
denotes the exterior derivative and $\delta=*\mathbf{d}*$ denotes the
coderivative):
\begin{eqnarray*}
&&F^{\mu\nu}(\mathbf{d}F)_{\mu\nu\sigma}=0, \ \
(*F)^{\mu\nu}(\mathbf{d}*F)_{\mu\nu\sigma}=0, \\
&&F^{\mu\nu}(\mathbf{d}*F)_{\mu\nu\sigma}+(*F)^{\mu\nu}(\mathbf{d}F)_{\mu\nu\sigma}=0,
\end{eqnarray*}
in terms of the coderivative $\delta$ these equations are given
correspondingly by
$$
(*F)_{\mu\nu}(\delta *F)^\nu=0, \ \ F_{\mu\nu}(\delta F)^\nu=0,
\ \ F_{\mu\nu}(\delta *F)^\nu+(*F)_{\mu\nu}(\delta F)^\nu=0.
$$
It was shown that all nonlinear solutions, i.e. those satisfying
$\mathbf{d}F\neq 0$ and $\mathbf{d}*F\neq 0$, have zero invariants:
$I_1=I_2=0$. Therefore, considering $Q_\mu^\nu$ as a linear map in the module
of vector fields, or 1-forms, over the Minkowski space-time, we see that if the
two invariants are equal to zero, we obtain a boundary operator at all points
of $M$ where the field is nonzero. Our purpose now is to consider how the
corresponding homology is connected with the structure of the nonlinear
solutions of the vacuum EED equations.

Recalling that in this isotropic case the energy tensor has only one isotropic
eigen direction defined by the vector field $\bar{\zeta}, \bar{\zeta}^2=0$, and
that the 2-forms $F$ and $*F$ are represented in an appropriate coordinate
system (called $\zeta$-adapted) by two 1-forms $A$ and $A^*$ as:
$F=A\wedge\zeta $,  $*F=A^*\wedge\zeta$, we have the following result: The
image space $Im(Q_F)$ coincides with the only isotropic eigen direction of
$Q_F$. In fact, in the $\zeta$-adapted coordinate system we have
$$
A=u\,dx+p\,dy, \ \ A^*=-\varepsilon p\,dx+\varepsilon u\,dy, \ \
\zeta=\varepsilon dz+d\xi,
 \ \ \varepsilon=\pm 1,
$$
so, the linear map $Q_F$ is given in this $\zeta$-adapted coordinate system by
$$
Q_F=-\phi^2 dz\otimes \frac{\partial}{\partial z}+ \varepsilon
\phi^2 dz\otimes\frac{\partial}{\partial\xi} - \varepsilon \phi^2
d\xi\otimes\frac{\partial}{\partial z} + \phi^2 d\xi\otimes
\frac{\partial}{\partial \xi},
$$
where $\phi^2=u^2+p^2$. Clearly,
$Q_F=\phi^2\zeta\otimes\bar{\zeta}$, where $\bar{\zeta}=-\varepsilon
\partial_z+\partial_\xi$ defines the only isotropic eigen direction of $Q_F$.
Let in this coordinate system the arbitrary vector field $X$ be presented by
its components $(X^\mu),\ \mu=1,\dots,4$. We obtain
\begin{eqnarray*}
Q_F(X)&=&-\phi^2 dz(X)\frac{\partial}{\partial z}+
\varepsilon \phi^2 dz(X)\frac{\partial}{\partial\xi}
- \varepsilon \phi^2 d\xi(X)\frac{\partial}{\partial z} +
\phi^2 d\xi(X)\frac{\partial}{\partial \xi} \\
&=&-\phi^2 X^3\frac{\partial}{\partial z}+
\varepsilon \phi^2 X^3\frac{\partial}{\partial\xi}
- \varepsilon \phi^2 X^4\frac{\partial}{\partial z} +
\phi^2 X^4\frac{\partial}{\partial \xi} \\
&=&\varepsilon\phi^2X^3\bar{\zeta}+\phi^2 X^4\bar{\zeta}
=\phi^2(\varepsilon X^3+X^4)\bar{\zeta}.
\end{eqnarray*}

If $\alpha=\alpha_{\mu}dx^\mu$ is a 1-form then in the same way we obtain
$$
Q_F(\alpha)=(Q_F)^\nu_\mu\alpha_\nu dx^\mu=
\phi^2(-\varepsilon\alpha_3+\alpha_4)\zeta.
$$
Hence the image space of $Q_F$ coincides with the only isotropic eigen space of
$Q_F$. Further we continue denoting the vectors and the $\eta$-corresponding
forms by the same letter.

Another important moment is that the kernel space $Ker(Q_F)$ coincides
with the 3-space spanned by the $A, A^*$ and $\zeta$.
In fact, we know that $Q_F(A)=Q_F(A^*)=Q_F(\zeta)=0$. Now, if $X$ is an
arbitrary vector, then from the above it follows that
$Q_F(X)=\phi^2(\varepsilon X^3+X^4)\zeta$, so we conclude that $Q_F(X)$ will be
equal to zero only if $X$ is a linear combination of $A, A^*$ and $\zeta$.
\noindent

Hence, we may write $Ker(Q_F)=\{A\}\oplus\{A^*\}\oplus\{\zeta\}$. The
corresponding factor space
$$
H(Q_F)=Ker(Q_F)/Im(Q_F)
$$
is isomorphic to $\{A\}\oplus\{A^*\}$.  The classes defined by $A$ and $A^*$
are given by $[A]=A+f\zeta$ and $[A^*]=A^*+f^*\zeta$, where $f$ and $f^*$ are
functions.

Recall now that $\delta F$ is an eigen vector of
$F$ and $\delta*F$ is an eigen vector of $*F$, so, $\delta F$ and $\delta*F$
are eigen vectors of $Q_F$ corresponding to the $0$-eigen values of $Q_F$.
Therefore $Q_F(\delta F)=Q_F(\delta*F)=0$, i.e. $\delta F$ and $\delta*F$
define corresponding homology classes: $[\delta F]=\delta F +h\zeta$ and
$[\delta *F]=\delta *F+h^*\zeta$, where $h$ and $h^*$ are functions.

According to the above mentioned property, every symmetry of a boundary
operator induces a linear map inside the homology space. Therefore,
the homology spaces are invariant with respect to
the linear isomorphisms which commute with the boundary operators. In our
case we have to find those linear maps $\Phi$ in the module of vector fields
over $M$, which commute with $Q_F$, i.e.  $\Phi\circ Q_F=Q_F\circ\Phi$. It is
readily obtained that in the $\zeta$-adapted coordinate system every such $\Phi$
is given by a matrix of the following kind:
$$
\Phi=\begin{Vmatrix}a & b & c & -\varepsilon c\\
		    m & n & q & -\varepsilon q\\
		    r & s & w &  0\\
	 \varepsilon r & \varepsilon s & 0 & w\end{Vmatrix},
$$
where all nine independent entries of this matrix are functions of the
coordinates. It follows that the $Q_F$- homology spaces are invariant with
respect to all diffeomorphisms $\varphi: M\rightarrow M$ which generate
isomorphisms $d\varphi: TM\rightarrow TM$ of the tangent bundle of $M$ given
in the $\zeta$-adapted coordinate system by a nondegenerate matrix of the above
kind.

An important property of the boundary operator $Q_F$ is that its image space
$Im(Q_F)$ is 1-dimensional. As it was mentioned earlier, this allows to
extend $Q_F$ as a boundary operator in the graded exterior algebra of
differential forms over $M$. In fact, recall that a linear map $\varphi$ in a
linear space $V$ induces derivation $\varphi^{\wedge}$ in
the exterior algebra $\Lambda(V)$ according to the rule
$$
\varphi^{\wedge}(x_1\wedge x_2\wedge \dots \wedge
x_p)
$$
$$
= \varphi(x_1)\wedge x_2\wedge\dots \wedge x_p +x_1\wedge
\varphi(x_2)\wedge\dots \wedge x_p+ \dots +x_1\wedge x_2\wedge\dots
\wedge\varphi(x_p).
$$
\vskip 0.3cm
\noindent
{\bf Remark}.\ If we try to extend $Q_F$ to antiderivation with respect to the
usual involution
$\omega(\alpha)=(-1)^p \alpha,\  \alpha\in\Lambda^p(M)$, we'll
find that this is not possible since the necessary condition for this, given
by $Q_F(\alpha)\wedge\alpha+\omega(\alpha)\wedge Q_F(\alpha)=0$, does not
hold for every $\alpha\in\Lambda^1(M)$.
\vskip 0.3cm
Hence, if $Im(\varphi)=\varphi(V)$ is 1-dimensional, then
every summond of $\varphi^{\wedge}\circ\varphi^{\wedge}(x_1\wedge x_2\wedge
\dots \wedge x_p)$
will contain two elements of the kind $\varphi(x_i)$ and
$\varphi(x_j)$, and if these two elements are collinear, their exterior
product is zero and the corresponding summond is zero. In our case
$\varphi=Q_F$ and $Im(Q_F)=\{\zeta\}$ is 1-dimensional, so we shall have
$Q_{F}^{\wedge}\circ Q_{F}^{\wedge}(\alpha)=0, \alpha\in \Lambda(M)$.
\vskip 0.3cm
{\bf Corollary}. The extension $Q_F^{\wedge}$ defines a boundary operator
of degree zero in $\Lambda(M)$.
\vskip 0.3cm
\noindent
{\bf Remark}. Further the extension $Q_F^{\wedge}$ will be denoted just by
$Q_F$.
\vskip 0.3cm
{\bf Corollary}. The extension of $Q_F$ to derivation
in $\Lambda (M)$ introduces in $\Lambda(M)$ some structure of graded
{\it differential} algebra with corresponding {\it graded homology algebra}
$H(Q_F)(\Lambda(M))$.
\vskip 0.3cm
The following relations are readily verified:
$$
Q_F(F)=0,\ \ Q_F(*F)=0,\ \
Q_F(\delta F\wedge F)=0.
$$
For example,
$Q_F(F)=Q_F(A\wedge\zeta)=Q_F(A)\wedge\zeta+A\wedge Q_F(\zeta)=0$.  So, $F,
*F$ and $\delta F\wedge F$ are $Q_F$-cycles.

We recall that every nonlinear solution satsfies the conditions:
$(\delta F)^2=(\delta *F)^2<0$, $A^2=(A^*)^2<0$, $(\delta F).(\delta*F)=0$,

Then recalling the scale factor $\mathcal{L}_o=|A|/|\delta F|$
we have the following result:

 {\bf The scale factor $\mathcal{L}_o=|A|/|\delta F|$
depends only on the classes of $A$ and} $\delta F$.
\vskip 0.2cm
In fact,
since $|[A]|=|A+f\zeta|=|A|$ and $|[\delta F]|=|\delta F+h\zeta|$ we obtain
$\mathcal{L}_o=|A|/|\delta F|=|[A]|/|[\delta F]|$.  \vskip 0.3cm Clearly,
$([A],[A^*])$ and $([\delta F],[\delta *F])$ represent two bases of
$H(Q_F)$ in $\Lambda^1(M)$.

Consider now the coframe $(A/\phi, A^*/\phi, dz, d\xi)$ and denote by $\chi^o$
its matrix with respect to the coordinate frame $(dx,dy,dz,d\xi)$. Let
$\psi=arccos(\frac12tr(\chi^o))$.

 \vskip 0.3cm {\bf Corollary.} The transformation
$([A],[A^*])\rightarrow ([\delta F],[\delta *F])$ is given by
$$
([A],[A^*])\begin{Vmatrix}0 & \varepsilon L_{\bar{\zeta}}\psi \\
-\varepsilon L_{\bar{\zeta}}\psi & 0\end{Vmatrix}=
(-\varepsilon L_{\bar{\zeta}} \psi[A^*], \varepsilon L_{\bar{\zeta}}\psi [A])=
([\delta F],[\delta *F]).
$$
The above formula shows that the transformation matrix, further denoted
by $\mathcal{M}$, between these two bases is
$(\pm\varepsilon\mathcal{L}_o)^{-1}J$,
where $J$ is the canonical complex structure in a real 2-dimensional space.
This fact may give another look on the duality,
because of the invariance of $J$ with respect to the transformation
$J\rightarrow S.J.S^{-1}$, where $S$ is given earlier  :
$$
S.J.S^{-1}=
\begin{Vmatrix}a & b \\-b & a\end{Vmatrix}
\begin{Vmatrix}0 & 1 \\-1 & 0\end{Vmatrix}
\begin{Vmatrix}a &-b \\ b & a\end{Vmatrix}\frac{1}{a^2+b^2}=
\begin{Vmatrix}0 & 1 \\-1 & 0\end{Vmatrix}.
$$
One could say that the duality symmetry of the nonlinear solutions is a
consequence of the null-field homology presented. In other words, every
initial null-field configuration given by $(A,A^*,\zeta)$, with $|\delta
F|\neq 0, |\delta *F|\neq 0, \zeta^2=0$, compulsory has
rotational-translational {\it dynamical} nature, so, it is {\it intrinsically
forced} to propagate with rotational component of propagation in space-time,
because, the nonzero $F_{\mu\nu}$, i.e. the
nonzero $(A,A^*)$, imply nonzero values of the derivatives of $F_{\mu\nu}$
including the nonzero value of $L_{\bar{\zeta}} \psi$ even if $\psi$ is
time-independent, and the basis $([A],[A^*)]$ is continuously forced to
rotate. In fact, the running-wave character of $\phi=|A|$ drags the solution
along the coordinate $z$ and the nonzero $L_{\bar{\zeta}} \psi$ implies
$\mathrm{cos}\,\psi\neq 0, \mathrm{sin}\,\psi\neq 0$. The evolution obtained
is strongly connected with the nonzero {\it finite} value of the scale factor
$\mathcal{L}_o=|L_{\bar{\zeta}} \psi|^{-1}$, which, in turn, determines
rotation in the homology space $H_F(Q)$. This rotation is determined entirely
by the Lie derivative of the phase $\psi$ with respect to $\bar{\zeta}$, and it
is intrinsically consistent with the running wave translational propagation of
the energy-density $\phi^2$. It is seen that the field configuration has a
rotational component of propagation, while the energy density has just
translational component of propagation.

It is interesting to see the action of $Q_F$ as {\it derivation}
in $\Lambda(M)$. We shall do this in a
$\zeta$-adapted coordinate system. Let $F$ define a nonlinear solution and
$Q_F: \Lambda(M)\rightarrow \Lambda(M)$ be the corresponding derivation
with $\phi^2=u^2+p^2$ the corresponding energy-density.  We give first the
action of $Q_F$ as {\it derivation} on the bases elements.

\begin{align*}
&Q_F(dx)=Q_F(dy)=0,\quad
Q_F(dz)=-\varepsilon\phi^2 \zeta,\quad
Q_F(d\xi)=\phi^2 \zeta
\end{align*}
\begin{alignat*}{2}
Q_F(dx\wedge dy)&=0, &\qquad Q_F(dx\wedge d\xi) &=\phi^2 dx\wedge\zeta\\
Q_F(dx\wedge dz) &=-\varepsilon\phi^2 dx\wedge\zeta, &\qquad
Q_F(dy\wedge d\xi) &=\phi^2 dy\wedge\zeta, \\
Q_F(dy\wedge dz) &=-\varepsilon\phi^2 dy\wedge\zeta, &\qquad
Q_F(dz\wedge d\xi) &=0.
\end{alignat*}
\begin{alignat*}{2}
Q_F(dx\wedge dy\wedge dz)&=-\varepsilon\phi^2 dx\wedge dy\wedge\zeta, &\qquad
Q_F(dx\wedge dy\wedge d\xi)&=\phi^2 dx\wedge dy\wedge\zeta, \\
Q_F(dx\wedge dz\wedge d\xi)&=0, & \qquad
Q_F(dy\wedge dz\wedge d\xi)&=0,
\end{alignat*}
$$
Q_F(dx\wedge dy\wedge dz\wedge d\xi)=0.
$$

Let now the arbitrary 2-form $G$ be represented in this coordinate system by
$G=G_{\mu\nu}dx^\mu\wedge dx^\nu,\ \ \mu<\nu $. Making use of the above given
explicit form for the action of $Q_F$ on the basis elements as {\it
derivation} we obtain
\begin{eqnarray*}
Q_F(G)&=&\varepsilon\phi^2(-\varepsilon G_{13}+G_{14})dx\wedge dz+
\varepsilon\phi^2(-\varepsilon G_{23}+G_{24})dy\wedge dz\\
&+& \phi^2(-\varepsilon G_{13}+G_{14})dx\wedge d\xi+          %56%
\phi^2(-\varepsilon G_{23}+G_{24})dy\wedge d\xi.
\end{eqnarray*}
Since  $Q_F(G)$ is obviously null: $[Q_F(G)]^2=0$,
this result makes possible the following conclusions concerning 2-forms:

\hspace{1.5cm}1. The space $Im(Q_F)$ consists of null fields,
i.e. every nonlinear solution $F$ determines a subspace
$Im(Q_F)\subset \Lambda^2(M)$ of null-fields.

\hspace{1.5cm}2. The space $Ker(Q_F)$ consists of 2-forms, which in this
coordinate system satisfy:
$\varepsilon G_{13}=G_{14},\ \varepsilon G_{23}=G_{24}$,
and $(G_{12},G_{34})$-arbitrary.

\hspace{1.5cm}3. The eigen spaces of $Q_{F}(G)$ coincide with the eigen
spaces of $F$ for every (nonzero) $G\in \Lambda^2(M)$.

If $G$ is a 3-form with components $G_{123}, G_{124}, G_{134}, G_{234}$ in
the same $\zeta$-adapted coordinate system, we obtain
$$
Q_F(G)=\varepsilon \phi^2(-\varepsilon G_{123}+G_{124})dx\wedge dy\wedge dz+
\phi^2(-\varepsilon G_{123}+G_{124})dx\wedge dy\wedge d\xi.
$$
So, $Q_F(G)$ is isotropic, and a 3-form $G$ is in $Ker(Q_F)$ only if
$\varepsilon G_{123}=G_{124}$ in this coordinate system. Moreover, since
$Q_F(G)$ does not depend on $G_{134}dx\wedge dz\wedge d\xi$ and
$G_{234}dy\wedge dz\wedge d\xi$ we conclude that the kernel of $Q_F$ in this
case consists of time-like 3-forms.

Finally, if $G$ is a 4-form, then $Q_F(G)=0$.
\vskip 0.3cm
{\bf Corollary.} If $G\in\Lambda(M)$ lives in $Im(Q_F)$ where $F$ is a
nonlinear solution, then $Q_F(G)$ is isotropic.
\vskip 0.3cm
This may be extended to the smooth functions $f\in C^{\infty}(M)$ if we assume
$Q_F(f)=0$.

Let's summarize. Every space-like (straight-line) direction may be chosen for
$z$-coordinate on $M$, and the 1-form $\zeta=\varepsilon dz+d\xi$ determines an
isotropic direction along which a class of null-fields $F$ are defined. The
corresponding linear map $Q_F$ satisfies $Q_F\circ Q_F=0$ and defines homology
in the spaces of 1-forms and of vector fields.

Since the image space $Q_F(\Lambda^1 M)$ is 1-dimensional, $Q_F$ extends to a
boundary operator in the whole exterior algebras over the 1-forms and vector
fields.  The image space of the extended $Q_F$ consists of isotropic (null)
objects.  If $G$ is a 2-form then $Q_F(G)$ has, in general, the same eigen
properties as $F$.  Hence, every 2-form $F$ with zero invariants lives in
just one such subclass and the whole set of these 2-forms divides to such
nonintersecting subclasses. Moreover, every 2-form $G$ has its (null-field)
image in every such subclass.

For every nonlinear solution $F(u,p)$, ($|\delta F|=|\delta *F|\neq 0$) the
corresponding $\phi^2=u^2+p^2$ propagates translationally, i.e., is a running
wave, along the space-like direction chosen (considered as the coordinate
$z$). The 4-dimensional versions of the corresponding electric and magnetic
fields are presented by the nonisotropic parts of the homology classes
defined by the mutually orthogonal space-like 1-forms $A$ and $(-A^*)$. The
nonlinear solutions with isotropic $\delta\,F\neq 0$ and $\delta\,*F\neq 0$:
$|\delta F|=|\delta *F|=0$, propagate only
translationally, i.e. without rotation.  Rotational components of
propagation, or spin-momentum, may have just those nonlinear solutions having
nonzero finite scale factor $\mathcal{L}_o=|A|/|\delta F|$, or equivalently,
satisfying one of the conditions given erlier.  The isotropic
3-form $\delta F\wedge F$ defines a $Q_F$-homology class since
$Q_F(\delta F\wedge F)=0$, and it appears as a natural candidate representing
locally the spin-momentum if we assume the additional equation
$\mathbf{d}(\delta F\wedge F)=0$, which should reduce to an equation for the
phase $\psi$.  The two mutually orthogonal space-like 1-forms $\delta F$ and
$\delta *F$ define the same homology classes as $A^*$ and $A$ respectively.
The transformation matrix $\mathcal{M}$ between the two bases $(A,A^*)$ and
$(\delta *F,\delta F)$ defines a complex structure in the 2-dimensional
homology space through the scale factor:
$\mathcal{M}[A]=\mathcal{M}(A+f\zeta)=\mathcal{M}(A)+f\zeta=\delta
*F+f\zeta=[\delta *F]$
and $\mathcal{M}=\pm\varepsilon\mathcal{L}_o^{-1}J$, where $J$ is the canonical
complex structure in a 2-dimensional space. The 2-parameter duality symmetry
coincides with the symmetries of $\mathcal{M}$ and transforms solutions to
solutions inside the subclass of solutions propagating along the spatial
direction chosen.

\section{Explicit nonlinear solutions}
As it is clear from the above with every nonlinear solution $F$ of our
nonlinear equations a class of $\zeta$-adapted coordinate systems is associated,
such that $F$ and $*F$ acquire the form :
\begin{eqnarray*}
&&F=\varepsilon
udx\wedge dz + udx\wedge d\xi + \varepsilon pdy\wedge dz + pdy\wedge d\xi \\
&&*F=-pdx\wedge dz - \varepsilon pdx\wedge d\xi + udy\wedge dz +
\varepsilon udy\wedge d\xi.
\end{eqnarray*}
Since we look for non-linear solutions, after substitution of these $F$
and $*F$ and doing some elementary calculations we obtain:
\vskip 0.3cm
\noindent
Every couple $(F,*F)$ of the above kind satisfies the equation
$$
\delta F\wedge F-\delta *F\wedge *F=0,
$$
which in terms of the FN-bracket (Sec.8.2) coincides with the equation
$$
[A\otimes\bar{\zeta},A\otimes\bar{\zeta}]=
-[A^*\otimes\bar{\zeta},A^*\otimes\bar{\zeta}] .
$$
Further we obtain
$$
\delta F=(u_\xi-\varepsilon u_z)dx +(p_\xi-\varepsilon p_z)dy +
\varepsilon(u_x + p_y)dz +(u_x + p_y)d\xi,
$$
$$
\delta *F=-\varepsilon(p_\xi -\varepsilon p_z)dx
+\varepsilon(u_\xi-\varepsilon p_z)dy - (p_x - u_y)dz -
\varepsilon(p_x - u_y)d\xi,
$$
$$
F_{\mu\nu}(\delta F)^\nu
dx^\nu= (*F)_{\mu\nu}(\delta *F)^\nu dx^\nu=
$$
$$
=\varepsilon\left[p(p_\xi-\varepsilon p_z)+u(u_\xi-\varepsilon u_z)\right]dz+
\left[p(p_\xi-\varepsilon p_z)+u(u_\xi-\varepsilon u_z)\right]d\xi,
$$
$$
(\delta F)^2=(\delta *F)^2=
-(u_\xi-\varepsilon u_z)^2-(p_\xi-\varepsilon p_z)^2
=-\phi^2(\psi_\xi-\varepsilon \psi_z)^2.
$$
We infer that our equations reduce to only one equation, namely
$$
p(p_\xi-\varepsilon p_z)+u(u_\xi-\varepsilon u_z)=
\frac 12\left[(u^2+p^2)_\xi-\varepsilon(u^2+p^2)_z\right]=0,
$$
which in terms of FN-bracket (Sec.8.2) is equivalent to
$$
[A\otimes\bar{\zeta},A^*\otimes\bar{\zeta}]=0.
$$
The obvious solution to this equation is
$$
u^2+p^2=\phi^2 (x,y,\xi+\varepsilon z), \ \ \
u=\phi\,\varphi, \ \ \ p =\pm\phi\,\sqrt{1-\varphi^2}, \ \ |\varphi|<1,
$$
where $\phi$ is an arbitrary differentiable function of its arguments.
The solution obtained shows that the equations impose some limitations only
on the amplitude function $\phi$ and that the phase function
$\varphi$ is arbitrary except that it is bounded: $|\varphi|<1$. The
amplitude $\phi$ is a running wave along the specially chosen coordinate $z$,
which is common for all $\zeta$-adapted coordinate systems. Considered as a
function of the spatial coordinates, the amplitude $\phi$ is {\it arbitrary},
so it can be chosen {\it spatially finite}. The time-evolution does not
affect the initial form of $\phi$, so it will stay the same in time, but the
whole solution may change its form due to $\varphi$ . Since $|\varphi|<1$
and the two independent field components are given by
$F_{14}=u=\phi\,\varphi$, $F_{24}=p =\pm\phi\,\sqrt{1-\varphi^2}$ this shows,
that {\it among the nonlinear solutions of our equations there are (3+1)
spatially finite solutions}. The spatial structure of $\phi$ can be determined
by initial condition, and the phase function $\varphi$ can be used to describe
additional structure features and {\it internal dynamics} of the solution.

We compute $\delta F\wedge F=\delta *F\wedge *F$ and obtain
\begin{eqnarray*}
\delta F\wedge F&=&\Big[\varepsilon p(u_\xi-\varepsilon u_z)-
\varepsilon u(p_\xi-\varepsilon p_z)\Big]dx\wedge dy\wedge dz\\
&+&\Big[p(u_\xi-\varepsilon u_z)-
u(p_\xi-\varepsilon p_z)\Big]dx\wedge dy\wedge d\xi.
\end{eqnarray*}
In terms of $\phi$ and $\psi=\mathrm{arccos}(\varphi)$ we obtain
$$
p(u_\xi-\varepsilon u_z)-u(p_\xi-\varepsilon p_z)=
-\phi^2(\psi_\xi-\varepsilon \psi_z),
$$
and so
$$
\delta F\wedge F=-\varepsilon\phi^2(\psi_\xi-
\varepsilon \psi_z)dx\wedge dy\wedge dz-
\phi^2(\psi_\xi-\varepsilon \psi_z)dx\wedge dy\wedge d\xi.
$$
Applying $*$ from the left we get
$$
*(\delta F \wedge F)=-\phi^2(\psi_\xi-\varepsilon \psi_z)dz-
\varepsilon\phi^2(\psi_\xi-\varepsilon \psi_z)d\xi=
-\varepsilon\phi^2(\psi_\xi-\varepsilon\psi_z)\zeta.
$$
\vskip 0.3cm
{\bf Corollary.} $\delta F\wedge F$ is {\it null}: $|\delta F\wedge F|=0$,
and it is equal to zero iff $\psi$ is a running wave along $z$.
\vskip 0.3cm
{\bf Corollary.} $F$ is a running wave along $z$ iff $\psi$ is a running wave
along $z$, i.e. iff $\delta F\wedge F=0$.
\vskip 0.3cm
Computing the 4-forms
$\mathbf{d}A\wedge A\wedge\zeta$ and
$\mathbf{d}A^*\wedge A^*\wedge\zeta$ we obtain
\begin{eqnarray*}
&&\mathbf{d}A\wedge A\wedge\zeta=
 \mathbf{d}A^*\wedge A^*\wedge\zeta \\
&&=
\varepsilon\big[u(p_\xi-\varepsilon p_z)-p(u_\xi-\varepsilon u_z)\big]
dx\wedge dy\wedge dz\wedge d\xi \\
&&=\varepsilon\phi^2(\psi_\xi-
\varepsilon \psi_z)dx\wedge dy\wedge dz\wedge d\xi.
\end{eqnarray*}

Hence, since $(\delta F)^2=-\phi^2(\psi_\xi-\varepsilon \psi_z)^2$,
and $\phi\neq0$ is a running wave, the nonlinear solutions, satisfying
$|\delta F|=|\delta *F|=0$, imply $\psi_\xi-\varepsilon \psi_z=0$, i.e.
{\it absence of rotational component of propagation}.

The following relations are equivalent:
\vskip 0.1cm
\hspace{2cm}1. $\delta F\wedge F=0$.

\hspace{2cm}2. $|\delta F|=|\delta *F|=0$.

\hspace{2cm}3. $\psi$ is a running wave along $z$: $L_{\bar{\zeta}}\psi=0$.

\hspace{2cm}4. $\mathbf{d}A\wedge A\wedge\zeta=
\mathbf{d}A^*\wedge A^*\wedge\zeta=0.$
\vskip 0.1cm
We give now two other relations that are equivalent to the above four.
First, consider a nonlinear solution $F$, and the corresponding $(1,1)$ tensor
$F_\mu^\nu=\eta^{\nu\sigma}F_{\sigma\mu}$. We want to compute the
corresponding Fr\"olicher-Nijenhuis tensor $S_F=\big[F,F\big]$, which is
a 2-form on $M$ with values in the vector fields on $M$. The components of
$S_F$ in a coordinate frame are given by
$$
(S_F)_{\mu \nu }^\sigma =2\left[ F_\mu ^\alpha \frac{\partial F_\nu
^\sigma} {\partial x^\alpha }-F_\nu ^\alpha \frac{\partial F_\mu ^\sigma
}{\partial x^\alpha }-F_\alpha ^\sigma \frac{\partial F_\nu ^\alpha
}{\partial x^\mu } +F_\alpha ^\sigma \frac{\partial F_\mu ^\alpha }{\partial
x^\nu }\right].
$$
We recall the two unit vector fields $\mathbf{A}$ and
$\varepsilon\mathbf{A^*}$, given by (in a $\zeta$-adapted coordinate
system)
$$
\mathbf{A}=-\varphi \frac{\partial} {\partial x}-\sqrt{1-\varphi ^2}
\frac{\partial}{\partial y},\quad
\varepsilon\mathbf{A^*} =
\sqrt{1-\varphi^2}\frac{\partial}{\partial x}-
\varphi \frac{\partial} {\partial y},
$$
and we compute $S_F(\mathbf{A},\varepsilon\mathbf{A^*})$.
$$
(S_F)_{\mu \nu}^\sigma
\mathbf{A}^\mu\varepsilon\mathbf{A^*}^\nu
=(S_F)_{12}^\sigma (\mathbf{A}^1\varepsilon\mathbf{A^*}^2-
\mathbf{A}^2\varepsilon\mathbf{A^*}^1).
$$
We obtaian
$$
(S_F)_{12}^1=(S_F)_{12}^2=0,\quad
(S_F)_{12}^3=-\varepsilon (S_F)_{12}^4=2\varepsilon \{p(u_\xi -\varepsilon
u_z)-u(p_\xi -\varepsilon p_z)\}.
$$
It is easily seen that
${\bf A}^1{\bf \varepsilon A^*}^2-{\bf A}^2{\bf \varepsilon A^*}^1=1$, so,
the relation $S_F(\mathbf{A},\varepsilon\mathbf{A^*})=0$ is
equivalent to the above four.

Second, recall that if $\big[\mathcal{A}, (+,.)\big]$ is an algebra (may
graded), then the (anti)derivations
$\mathcal{D}:\mathcal{A}\rightarrow\mathcal{A}$ satisfy:
$\mathcal{D}(a.b)=\mathcal{D}a.b+\varepsilon_a a.\mathcal{D}b$, where $a,b\in
\mathcal{A}$ and $\varepsilon_a$ is the parity of $a\in\mathcal{A}$.  So,
the derivations are not morphisms of $\mathcal{A}$, and satisfy the
generalized Leibniz rule. The difference
$$
\Delta_D(a,b)=\Big[Da.b+\varepsilon_a a.Db\Big]-D(a.b),\quad
a,b\in\mathcal{A}
$$
is called the Leibniz bracket of the operator $D: \mathcal{A}\rightarrow\mathcal{A}$, and
$D$ is a (anti)deri\-vation if its Leibniz bracket vanishes. If $\mathcal{A}$
is the exterior algebra of differential forms on a (pseudo)riemannian manifold
$(M,g)$ and $D$ is the coderivative $\delta$ with respect to $g$,
then the corresponding Leibniz bracket is denoted by $\{,\}$. So, if $F$ is a
$p$-form, and $G$ is any form, then $$ \{F,G\}=\delta F\wedge G+(-1)^p F\wedge
\delta G -\delta(F\wedge G). $$ Note that the brackets $\{F,F\}$ do not vanish
in general.

Now, if the 2-form $F$ on the Minkowski space defines a nonlinear solution, then
$F\wedge F=*F\wedge *F=0$ and
$$
\{F,F\}=\delta F\wedge F+(-1)^2F\wedge \delta F-\delta(F\wedge F)
=2\delta F\wedge F=2\delta*F\wedge *F.
$$
So, the above relations $1 - 4$ are equivalent to the requirement that the
Leibniz brackets $\{F,F\},\{*F,*F\}$ vanish.

{\bf Remark}. In the B.Coll's paper (arXiv: gr-qc/0302056) the equations
$\nabla_\sigma\,Q^\sigma_\mu=0$ and $\{F,F\}+\{*F,*F\}=0$ have been proposed as
new vacuum field equations, but no further study of the solutions are known
to us.

Finally, all these conditions are equivalent to $\mathcal{L}_o=\infty$.

We note the very different nature of these seven conditions.  The complete
integrability of any of the two Pfaff 2-dimensional systems
$(A,\zeta)$ and $(A^*,\zeta)$ is equivalent to zero value of
$|\delta F|$ on the one hand, and to the zero value of the quantity
$S_F(\mathbf{A},\varepsilon\mathbf{A^*})$ on the other hand,
and both are equivalent to the vanishing of this Leibniz bracket and
to the infinite value of $\mathcal{L}_o$. This could hardly be occasional, so, a
physical interpretation of these quantities in the nonzero case, i.e. when
$\psi$ is not a running wave, is strongly suggested. In view of the above
conclusion that the condition $|\delta F|=0$ implies absence of rotational
component of propagation, our interpretation is the following:
\begin{center}
\hfill\fbox{
    \begin{minipage}{0.97\textwidth}
\begin{center}
\vskip 0.3cm
{\bf A nonlinear solution will carry rotational component of propagation,
i.e. {\it intrinsic angular (spin) momentum}, only if $\psi$ is NOT a running
wave along the direction of translational propagation}.
\end{center}
\vskip 0.3cm
\end{minipage}} \hfill \end{center}

Natural measures of this spin momentum appear to be $\delta
F\wedge F\neq 0$, or $|\delta F|\neq 0$, or $S_F(\mathbf{A},
\varepsilon\mathbf{A^*})$. The most attractive seems to be $\delta
F\wedge F$, because it is a 3-form, and imposing the requirement
$\mathbf{d}(\delta F\wedge F)=0$ we obtain both: the equation for $\psi$ and
the corresponding conserved (through the Stokes' theorem) quantity
$\mathcal{H}=\int{i^*(\delta F\wedge F)}dx\wedge dy\wedge dz$, where
$i^*(\delta F\wedge F)$ is the restriction of $\delta F\wedge F$ to
$\mathbb{R}^3$.

Making use of the relations $u=\phi\,\mathrm{cos}\psi$ and
$p=\phi\,\mathrm{sin}\psi$, we get
\begin{eqnarray*}
&&A=\phi\,\mathrm{cos}\,\psi\,dx+\phi\,\mathrm{sin}\,\psi\,dy +f\zeta,\\
&&A^*=-\varepsilon\phi\,\mathrm{sin}\,\psi\,dx+
\varepsilon\phi\,\mathrm{cos}\,\psi\,dy + f^*\zeta, \\
&&\delta F=-\phi\,\mathrm{sin}\,\psi\,(\psi_\xi-\varepsilon \psi_z)\,dx+
\phi\,\mathrm{cos}\,\psi\,(\psi_\xi-\varepsilon \psi_z)\,dy+
(u_x+p_y)\zeta \\
&&=-\varepsilon(L_{\bar{\zeta}} \psi)A^*+(\varepsilon f^*L_{\bar{\zeta}}
\psi+u_x+p_y)\zeta \\ &&\delta *F=-\varepsilon\phi\,\mathrm{cos}\,\psi\,
(\psi_\xi-\varepsilon \psi_z)\,dx-
\varepsilon\phi\,\mathrm{sin}\,\psi\,(\psi_\xi-\varepsilon \psi_z)\,dy-
\varepsilon (p_x-u_y)\zeta \\
&&=\varepsilon(L_{\bar{\zeta}} \psi)A-
\varepsilon(fL_{\bar{\zeta}} \psi+p_x-u_y)\zeta,
\end{eqnarray*}
where $L_{\bar{\zeta}}$ is the Lie derivative with respect to $\bar{\zeta}$.
We obtain also
\begin{eqnarray*}
&&|\delta F|=|\delta *F|=
\frac{\phi|\varphi_\xi-\varepsilon\varphi_z|}{\sqrt{1-\varphi^2}}=
\phi|\psi_\xi-\varepsilon \psi_z|=\phi|L_{\bar{\zeta}} \psi|,\\       %51%
&&\mathcal{L}_o =\frac{|A|}{|\delta F|}=
\frac{\sqrt{1-\varphi^2}}{|\varphi_\xi-\varepsilon\varphi_z|}=
\frac{1}{|\psi_\xi-\varepsilon \psi_z|}=|L_{\bar{\zeta}} \psi|^{-1}.
\end{eqnarray*}
Now we can write
$$
\delta F=\pm\varepsilon\frac{A^*}{\mathcal{L}_o}+
\left(\mp \varepsilon\frac{f^*}{\mathcal{L}_o}+(u_x+p_y)\right)\zeta,\
$$
$$                                                                 %52%
\delta *F=\mp\varepsilon\frac{A}{\mathcal{L}_o}+ \left(\pm
\varepsilon\frac{f}{\mathcal{L}_o}-\varepsilon (p_x-u_y)\right)\zeta.
$$
\vskip 0.3cm
{\bf Corollary.} Obviously, the following relations hold:
$$
A_\mu(\delta F)^\mu=0,\ \
(A^*)_\mu(\delta *F)^\mu=0,\ \
A_\mu(\delta *F)^\mu=\pm \varepsilon\frac{\phi^2}{\mathcal{L}_o}
=-(A^*)_\mu(\delta F)^\mu.
$$

Finally we note that since the propagating along the given $\bar{\zeta}$
nonlinear solutions in canonical coordinates are parametrized by one function
$\phi$ of 3 independent variables and one {\it bounded} function $\varphi$ of 4
independent variables, the separation of various subclasses of nonlinear
solutions is made by imposing additional conditions on these two functions.

\section{Structure of the Nonlinear Solutions}

\subsection{Properties of the duality matrices}
We consider the set $\mathbb{G}$ of matrices $\alpha$ of the kind
$$
\alpha=\begin{Vmatrix} a & b \\-b & a
\end{Vmatrix}, \quad \text{where}\quad a,b\in\mathbb{R}.
$$
The nonzero matrices of this kind form a 2-dimensional Lie group $\mathbb{G}$
with respect to the usual matrix multiplication. Together with the zero
$2\times 2$ matrix $I_o$ they also form a 2-dimensional linear space
$\mathcal{G}=\mathbb{G}\bigcup\{I_o\}$ over $\mathbb{R}$, and this linear space
is naturally recognized as the Lie algebra of $\mathbb{G}$. The linear
structure is with respect to the usual addition of matrices, so every element
of $\mathbb{G}$ may be considered as corresponding element of $\mathcal{G}$. As
is well known $\mathcal{G}$, endowed with the matrix multiplication, gives the
real representation of the field of complex numbers.

A natural basis of the linear space $\mathcal{G}$ is given by the two
matrices
$$
I=\begin{Vmatrix}1 & 0 \\ 0 & 1\end{Vmatrix},\quad
J=\begin{Vmatrix}0 & 1 \\-1 & 0\end{Vmatrix}.
$$
The group $\mathbb{G}$ is commutative, in fact,
$$
\begin{Vmatrix}a & b \\-b & a\end{Vmatrix}.  \begin{Vmatrix}m & n \\-n &
m\end{Vmatrix}= \begin{Vmatrix}m & n \\-n & m\end{Vmatrix}.
\begin{Vmatrix}a & b \\-b & a\end{Vmatrix}=
\begin{Vmatrix}am-bn & (an+bm) \\-(an+bm) & am-bn\end{Vmatrix} .
$$
Every element $\alpha\in\mathcal{G}$ can be
represented as $\alpha=aI+bJ,\ a,b\in \mathbb{R}$. Recall the  natural
representation $\rho_o: \mathbb{G}\rightarrow L_{\mathcal{G}}$
of $\mathbb{G}$ in $\mathcal{G}$ given by
$$
\rho_o(\alpha)(aI+bJ)=a\,(\alpha^{-1})^*(I)+b\,(\alpha^{-1})^*(J)
=\frac{1}{a^2+b^2}\Big[a\,\alpha(I)+b\,\alpha(J)\Big].
$$
From now on we shall consider
$\alpha(I)$ and $\alpha(J)$  just as matrix product, so we have
$$
\alpha(I)=\alpha.I=\alpha=aI+bJ,\quad
\alpha(J)=\alpha.J=-bI+aJ.
$$
Since $(J^{-1})^*=J$, $J$ generates a complex structure in $\mathcal{G}$:
$J\circ J(x)=-x, \ x\in \mathcal{G}$.

The product of two matrices $\alpha=aI+bJ$ and $\beta=mI+nJ$ looks like
$\alpha.\beta=(am-bn)I+(an+bm)J$. The commutativity of $\mathbb{G}$ means
symmetry, in particular, every $\alpha\in \mathbb{G}$ is a symmetry of $J$:
$\alpha.J=J.\alpha$.

Finally we note, that
the inner product $g_e$ in $\mathcal{G}=T_e(\mathbb{G})$, where $e$ is the
identity of $\mathbb{G}$, given by
$$
g_e(\alpha,\beta)=\frac12 tr(\alpha\circ\beta^*),
\quad \alpha,\beta\in\mathcal{G},
$$
generates a (left invariant) riemannian metric on $\mathbb{G}$ by means
of the (left) group multiplication:
$$
g_{\sigma}=(L_{\sigma^{-1}})^*g_{e},\quad \sigma\in\mathbb{G}.
$$

\subsection{Action of $\mathbb{G}$ in the space of 2-forms on
$M$} We consider now the space $\Lambda^2(M)$ - the space of 2-forms on
thr Minkowski space-time $M$ with its natural basis:
$$
dx\wedge dy,\ \ dx\wedge dz,\ \ dy\wedge dz,\ \
dx\wedge d\xi,\ \ dy\wedge d\xi,\ \ dz\wedge d\xi.
$$
We recall from Sec.5.3 that the Hodge
$*$ acts in $\Lambda^2(M)$ as a complex structure $*=\mathcal{J}$ and on the
above basis its action is given by:
\begin{align*}
\mathcal{J}(dx\wedge dy)
&=-dz\wedge d\xi & \mathcal{J}(dx\wedge dz) &=dy\wedge d\xi  &
\mathcal{J}(dy\wedge dz) &=-dx\wedge d\xi \\
\mathcal{J}(dx\wedge d\xi) &=dy\wedge dz &
\mathcal{J}(dy\wedge d\xi) &=-dx\wedge dz &
\mathcal{J}(dz\wedge d\xi) &=dx\wedge dy.
\end{align*}

Hence, in this basis the $(6\times 6)$-matrix of $\mathcal{J}$ is off-diagonal
with entries
$(\mathcal{J}_{16}=-\mathcal{J}_{25}=\mathcal{J}_{34}=
-\mathcal{J}_{43}=\mathcal{J}_{52}=-\mathcal{J}_{61}=-1$, (i.e. left action).

Let now $\mathcal{I}$ be the identity map in $\Lambda^2(M)$. We
define a representation $\rho$ of $\mathbb{G}\subset\mathcal{G}$ in
$\Lambda^2(M)$ as follows:
$$
\rho(\alpha)= \rho(aI+bJ)=a\mathcal{I}+ b\mathcal{J}, \quad
\alpha\in\mathbb{G}.
$$
Every $\rho(\alpha)$ is
a linear isomorphism, in fact, its determinant $\mathrm{det}||\rho(\alpha)||$
is equal to $(a^2+b^2)^3$.  The unity $I$ of $\mathbb{G}$ is sent to the
identity transformation $\mathcal{I}$ of $\Lambda^2(M)$, and the complex
structure $J$ of the vector space $\mathcal{G}$, considered as element of
$\mathbb{G}$, is sent to the complex
structure $\mathcal{J}$ of $\Lambda^2(M)$. This map is surely a
representation, because
$\rho(\alpha.\beta)=\rho(\alpha).\rho(\beta),\ \ \alpha,\beta\in\mathcal{G}$.
In fact,
$$
\rho(\alpha.\beta)=
\rho\big[(aI+bJ).(mI+ nJ)\big]
$$
$$
=\rho\big[(am-bn)I+(an+bm)J\big]
=(am-bn)\mathcal{I}+
(an+bm)\mathcal{J}.
$$
On the other hand
$$
\rho(\alpha).\rho(\beta)
=\rho(aI+bJ).\rho(mI+nJ)
$$
$$
=(a\mathcal{I}+b\mathcal{J}).
(m\mathcal{I}+ n\mathcal{J})=
(am-bn)\mathcal{I}+
(an+bm)\mathcal{J}.
$$

We consider now the space $\Lambda^2(M,\mathcal{G})$ of $\mathcal{G}$-valued
2-forms on $M$.  Every such 2-form $\Omega$ can be represented as
$\Omega=F_1\otimes I+F_2\otimes J$, where $F_1$ and $F_2$ are 2-forms. We
have the joint action of $\mathbb{G}$ in $\Lambda^2(M,\mathcal{G})$ as
follows:
$$
[\rho(\alpha)\times \alpha](\Omega)=
\rho(\alpha).F_1\otimes (\alpha^{-1})^*(I)+
\rho(\alpha).F_2\otimes (\alpha^{-1})^*(J).
$$
We obtain
$$
det(\alpha).[\rho(\alpha)\times \alpha](\Omega)
$$
$$
=\big[(a^2\mathcal{I}+ab\mathcal{J})F_1-
(b^2\mathcal{J}+ab\mathcal{I})F_2\big]\otimes I+
\big[(b^2\mathcal{J}+ ab\mathcal{I})F_1+
(a^2\mathcal{I}+ab\mathcal{J})F_2\big]\otimes J.
$$
In the special case $\Omega=F\otimes I+\mathcal{J}.F\otimes J$ it readily
follows that
$$
\left[\rho(\alpha)\times\alpha\right](\Omega)=\Omega.       %59%
$$
In this sense the forms $\Omega=F\otimes I+\mathcal{J}.F\otimes J$ are
{\it equivariant} with respect to this joint action of $\mathbb{G}$.

Explicitly for a general 2-form $F$ we have
\begin{equation*}
\begin{split}
\rho(\alpha).F & =a\mathcal{I}.F+b\mathcal{J}.F \\
& = (aF_{12}+bF_{34})dx\wedge dy+
(aF_{13}- bF_{24})dx\wedge dz+
(aF_{23}+\ bF_{14})dy\wedge dz \\
& +(aF_{14}-bF_{23})dx\wedge d\xi+
(aF_{24}+ bF_{13})dy\wedge d\xi
+(aF_{34}-bF_{12})dz\wedge d\xi.\\
\end{split}
\end{equation*}
If $F_{\varepsilon}$ is a nonlinear solution we
modify correspondingly the representation as follows:
$\rho(\alpha)=a\mathcal{I}+\varepsilon b\mathcal{J}$, and obtain (in the
$\zeta$-adapted coordinate system)
 \begin{equation*}
\begin{split}
\rho(\alpha).F_{\varepsilon} & =\varepsilon(au-bp)dx\wedge dz+
\varepsilon(ap+bu)dy\wedge dz \\ &+(au-bp)dx\wedge d\xi+               %60%
(ap+bu)dy\wedge d\xi.
\end{split}
\end{equation*}
It follows that if
$F_{\varepsilon}$ is a nonlinear solution then $\rho(\alpha).F_{\varepsilon}$
will be a nonlinear solution if the quantity
$$
\big[(\rho(\alpha).F_\varepsilon)_{14}\big]^2+
\big[(\rho(\alpha).F_\varepsilon)_{24}\big]^2=
(au-bp)^2+(ap+bu)^2
$$
is a running wave along $z$. But this quantity is equal to
$(a^2+b^2)(u^2+p^2)$ and since $(a^2+b^2)=const$, we see that
$\rho(\alpha).F_{\varepsilon}$ is again a nonlinear solution for any
$\alpha\in\mathbb{G}$. In other words, the group $\mathbb{G}$ acts as group of
symmetries of our nonlinear equations.  Moreover, in view of the conclusions
at the end of the preceding section, $\mathbb{G}$ acts inside every subclass
of solutions defined by the chosen space-like direction (the coordinate $z$).
Hence, if $F$ is a nonlinear solution, we may write $\mathbb{G}.F\subset
Q_F(\Lambda^2(M))$, i.e. any orbit $\mathbb{G}.F$ lives entirely and always
inside the subclass $Q_F(\Lambda^2(M))$.

Since for the phase $\psi$ of a product $(\alpha.\beta)$ we have
$\psi(\alpha.\beta)=\psi(\alpha)+\psi(\beta)$,
for the phase $\psi$ of the solution $\rho(\alpha).F_{\varepsilon}$
 we obtain (in this coordinate system)
$$
\psi=\mathrm{arccos}\left[\frac{\varepsilon(au-bp)}
{\sqrt{(a^2+b^2)(u^2+p^2)}}\right]=
 \psi(F_{\varepsilon}(u,p))+\psi(\alpha(a,b)),
$$
where
$$
\psi(F_{\varepsilon}(u,p))=\mathrm{arccos}\frac{\varepsilon u}{\sqrt{u^2+p^2}},
\ \  \psi(\alpha(a,b))=\mathrm{arccos}\frac{a}{\sqrt{a^2+b^2}}
$$
are respectively the phases of $F_{\varepsilon}(u,p))$ and
of the complex number $\alpha=aI+bJ$. Now, since $\psi(\alpha(a,b))=const$ we
obtain the

{\bf Corollary}. The 1-form $\mathbf{d}\psi_{F}$
and the scale factor
$\mathcal{L}_o=|L_{\bar{\zeta}}\psi(\rho(\alpha).F_{\varepsilon})|^{-1}$  are
 $\mathbb{G}$-invariants:
$\mathbf{d}(\psi_{F})=\mathbf{d}(\psi(\rho(\alpha).F)), \quad
\mathcal{L}_o(F)=\mathcal{L}_o(\rho(\alpha).F)$.
\vskip 0.3cm
{\bf Corollary}. If the scale factor
$\mathcal{L}_o=|L_{\bar{\zeta}}\psi)|^{-1}$ is {\it constant}, then, the
defined by a nonlinear solution $(F,*F)$ 3-form $F\wedge\delta F$ is closed:
$\mathbf{d}(F\wedge\delta F)=0$. \ \vskip 0.3cm In fact, we recall that $$
\delta F\wedge F= -\varepsilon\phi^2(\psi_\xi-\varepsilon\psi_z)dx\wedge
dy\wedge dz- \phi^2(\psi_\xi-\varepsilon\psi_z)dx\wedge dy\wedge d\xi. $$ So,
since $\psi_\xi-\varepsilon\psi_z=const$ and $\phi^2$ is runing wave along the
direction of $z$, i.e. $L_{\bar{\zeta}}(\phi^2)=0$, we obtain $$
\mathbf{d}(\delta F\wedge F)
=\varepsilon L_{\bar{\zeta}}(\phi^2)(L_{\bar{\zeta}}\psi)dx\wedge dy\wedge
dz\wedge d\xi=0.
$$
Hence, when the scale factor $\mathcal{L}_o$ is constant we
obtain another conservative quantity, namely, the integral of the restriction
of $\delta F\wedge F$ on $\mathbb{R}^3$ over the whole 3-space will not
depend on time.

We also note that under the action of $\alpha(a,b)$ we have
$$
A\rightarrow A'=[au-bp,ap+bu,0,0],\quad
A^*\rightarrow (A^*)'=[-(ap+bu),au-bp,0,0],
$$
and this is equivalent to
$$
A'=aA+bA^*,\quad (A^*)'=-bA+aA^*.
$$
Hence, in the $\zeta$-adapted coordinate systems the dual transformation, as
given above, restricts to transformations in the $(x,y)$-plane, so we have
\vskip 0.3cm
{\bf Corollary}. The Frobenius integrability of the 2-dimensional Pfaff
system $(A,A^*)$ is a $\mathbb{G}$-invariant property.
\vskip 0.3cm
{\bf Corollary}. The Frobenius NONintegrability of the 2-dimensional Pfaff
systems $(A,\zeta)$ and $(A^*,\zeta)$ is a $\mathbb{G}$-invariant property.
\vskip 0.3cm
\noindent
{\bf Remark}. For a possible connection of $\delta F\wedge F$ to the
Godbillon-Vey closed 3-form $\Gamma=\mathbf{d}\theta\wedge\theta$ see further.
\vskip 0.3cm

\subsubsection{\bf Point dependent group parameters}

We are going now to see what happens if the group parameters become functions
of the coordinates: $\alpha=u(x,y,z,\xi)I+p(x,y,z,\xi)J$, and to try to generate
nonlinear solution by means of defining appropriate action of $\alpha\in
\mathbb{G}$ in the linear spaces of 2-forms and 2-vectors.

 Denote by the same letter $\mathcal{I}$ the identity maps in $\Lambda^2(M)$
and in $\mathfrak{X}^2(M)$. The complex structure map in $\Lambda^2(M)$ defined
by the Hodge star we shall denote here by $\mathcal{J}$, and its dual map in
$\mathfrak{X}^2(M)$ shall be denoted by $\mathcal{J^*}$.

Let now a (smooth) map $M\rightarrow\mathbb{G}$ is given by
$$
\alpha(u,p)=u(x,y,z,\xi)I+\varepsilon p(x,y,z,\xi)J.
$$
We define the following actions:
$$
(\alpha,F)\rightarrow\rho(\alpha)(F)=
(u\,(x,y,z,\xi)\mathcal{I}+\varepsilon p\,(x,y,z,\xi)\mathcal{J})(F),
\ F\in\Lambda^2(M),
$$
$$
(\alpha,\bar{F})\rightarrow\rho(\alpha)(\bar{F})=
(u\,(x,y,z,\xi)\mathcal{I}+\varepsilon p\,(x,y,z,\xi)\mathcal{J^*})(\bar{F}),
\ \bar{F}\in\mathfrak{X}^2(M).
$$

Consider now the following objects:
$$
F_o=dx\wedge\zeta=\varepsilon\,dx\wedge dz+dx\wedge d\xi, \ \
\bar{F_o}=\bar{\zeta}\wedge\frac{\partial}{\partial x}=
\varepsilon\frac{\partial}{\partial x}\wedge\frac{\partial}{\partial
z}- \frac{\partial}{\partial x}\wedge\frac{\partial}{\partial \xi}.
$$
We obtain
$$
\rho(\alpha)(F_o)=
F=\varepsilon u\,dx\wedge dz + \varepsilon p\,dy\wedge dz+
u\,dx\wedge d\xi + p\,dy\wedge d\xi,
$$
and
$$
\rho(\alpha)(\bar{F_o})=\bar{F}=
\varepsilon\,u\frac{\partial}{\partial
x}\wedge\frac{\partial}{\partial z}+
\varepsilon\,p\frac{\partial}{\partial
y}\wedge\frac{\partial}{\partial z}-
u\frac{\partial}{\partial x}\wedge\frac{\partial}{\partial \xi}-
p\frac{\partial}{\partial y}\wedge\frac{\partial}{\partial \xi}.
$$
Similarly, $*F_o$ and $\bar{*F_o}$ generate
$$
\rho(\alpha)(*F_o)=*F=-p\,dx\wedge dz+u\,dy\wedge dz-
\varepsilon p\,dx\wedge d\xi+\varepsilon u\,dy\wedge d\xi
$$
and
$$
\rho(\alpha)(\bar{*F_o})=\bar{*F}=
-p\frac{\partial}{\partial x}\wedge\frac{\partial}{\partial z}+
u\frac{\partial}{\partial y}\wedge\frac{\partial}{\partial z}+
\varepsilon p\frac{\partial}{\partial x}\wedge\frac{\partial}{\partial \xi}-
\varepsilon u\frac{\partial}{\partial y}\wedge\frac{\partial}{\partial \xi}.
$$
Introducing now the objects
$$\Omega=\rho(\alpha)(F_o)\otimes e_1+\rho(\alpha)(*F_o)\otimes e_2
$$ and
$$\bar{\Omega}=\rho(\alpha)(\bar{F_o})\otimes e_1+
\rho(\alpha)(\bar{*F_o})\otimes e_2,
$$
we can impose our condition $\mathcal{L}^{\vee}_{\bar{\Omega}}\Omega=0$, and
to obtain in this way our nonlinear equations
$$
i(\bar{F})\mathbf{d}F=0, \ \ i(\bar{*F})\mathbf{d}*F=0, \ \
i(\bar{F})\mathbf{d}*F=-i(\bar{*F})\mathbf{d}F.
$$

The above consideration suggests
to check if we have a solution $(F,*F)$
defined by the two functions $u$ and $p$, and we consider a map
$\alpha:M\rightarrow\mathbb{G}\subset{\mathcal{G}}$,
such that the components $a(x,y,z,\xi)$ and
$b(x,y,z,\xi)$ of $\alpha=a(x,y,z,\xi)I + b(x,y,z,\xi)J$ determine another
solution, then whether the 2-form \linebreak $\tilde F=\alpha(x,y,z,\xi).F=
\big[a(x,y,z,\xi)\mathcal{I} +\varepsilon b(x,y,z,\xi)\mathcal{J}\big].F$
will define a solution?

For $\tilde F$ we obtain
$$
\tilde F=\rho(\alpha).F=(a\mathcal{I}+\varepsilon b\mathcal{J}).F=
$$
$$
\varepsilon (au-\varepsilon bp)dx\wedge dz +
\varepsilon (ap+\varepsilon bu)dy\wedge dz +
(au-\varepsilon bp)dx\wedge d\xi +
(ap+\varepsilon bu)dy\wedge d\xi,
$$
where $a$ and $b$ are functions of the coordinates.

Now, $\tilde F$ will define a solution iff
$$
\big[(au-\varepsilon bp)^2 +
(ap+\varepsilon bu)^2\big]_\xi -                   %62%
\varepsilon\big[(au-\varepsilon bp)^2 +
(ap+\varepsilon bu)^2\big]_z=0.
$$
This relation is equivalent to
$$
\big[(a^2+b^2)_\xi-\varepsilon(a^2+b^2)_z\big](u^2+p^2)+
\big[(u^2+p^2)_\xi-\varepsilon(u^2+p^2)_z\big](a^2+b^2)=0.
$$
This shows that if $F(u,p)$ defines a solution, then
$\tilde F(u,p\,;a,b)=\rho(\alpha(a,b)).F(u,p)$ will define a solution
iff  $F(a,b)$ defines a solution, i.e. iff $\rho(\alpha(a,b)).F_o$ defines a
solution. So we have the
\vskip 0.3cm
{\bf Corollary.} Every nonlinear solution $(F,*F)(a,b)$ defines a map
$$
\Phi(a,b):(F,*F)(u,p)\rightarrow \Phi(F,*F)(u,p\,;a,b)
$$
such that if $(F,*F)(u,p)$ is a nonlinear solution then $\Phi (F,*F)(u,p\,;a,b)$
is also a nonlinear solution.
\vskip 0.3cm

The above corollary says that the set $\Sigma(F_o,*F_o)$ of nonlinear
solutions, defined by the chosen $(F_o,*F_o)$, has a {\it commutative group
structure} with group multiplication
$$
F(a,b).F(u,p)=\rho(\alpha(a,b))F_o.\rho(\beta(u,p))F_o
$$
$$
=\rho(\alpha.\beta)F_o=\rho(\beta.\alpha)F_o=F(u,p).F(a,b),
$$
and a similar relation for $*F$.

In terms of the $F$-component of a nonlinear solution $(F,*F)(u,p)$ with the
corresponding map $\Phi(u,p)=\rho(\alpha(u,p))=u\mathcal{I}+\varepsilon
p\mathcal{J}$ we can introduce in an obviously invariant way the concept of
amplitude of a nonlinear solution. In fact,  $F(u,p)$ determines
the first component $F(u,-p)$ of a congugate solution. Making use of the two
corresponding linear maps $\Phi(u,p)$ and $\Phi(u,-p)$ we
compute the quantity
$\frac 16 tr\big[\Phi(u,p)\circ\Phi(u,-p)\big]$.
\begin{eqnarray*}
\frac 16 tr\big[\Phi(u,p)\circ \Phi(u,-p)\big]&=&
\frac 16 tr\left[\big(u\mathcal{I}+ \varepsilon p\mathcal{J}\big)\circ
\big(u\mathcal{I}- \varepsilon p\mathcal{J}\big)\right]\\
&=&
\frac 16 tr\big[(u^2+p^2)\mathcal{I}\big]=u^2+p^2=\phi^2(u,p).
\end{eqnarray*}

In the general case
$\alpha=\alpha_p.\alpha_{p-1}\dots\alpha_1$ we readily obtain
$$
F(\alpha)=F(\alpha_p).F(\alpha_{p-1})\dots F(\alpha_1),
$$
 and for the corresponding amplitude and phase
$$
\phi(F_p.F_{p-1}\dots F_1), \ \  \text{and} \ \ \
\psi(F_p.F_{p-1}\dots F_1)
$$ we obtain
$$
\phi(F_p.F_{p-1}\dots F_1))=\phi(F_p).\phi(F_{p-1})\dots\phi(F_1),
$$
$$
\quad \psi(F_p.F_{p-1}\dots F_1)=                     %63%
\sum_{i=1}^p(\psi_i).
$$
The corresponding scale factor is
$$
\mathcal{L}_o(F_p.F_{p-1}\dots F_1)=\frac{1}{|L_\zeta \psi|}=
\frac{1}{|\sum_{i=1}^p (\psi_i)_\xi-
\varepsilon\sum_{i=1}^p(\psi_i)_z|}.
$$

Clearly $\psi(F_o)=0$, which corresponds to $\mathcal{L}_o(F_o)=\infty$.

It deserves noting the following. If $\alpha(a,b)\neq 0$ defines a solution
 then $F^{-1}=\rho(\alpha^{-1}).F_o$ and
$(*F)^{-1}=\rho(\alpha^{-1}).(*F_o)$ will define a solution
given by
$$
(F)^{-1}=\left(\frac{u}{u^2+p^2}\mathcal{I}
-\frac{\varepsilon p}{u^2+p^2}\mathcal{J}\right)(F_o) , \
(*F)^{-1}=\left(-\frac{\varepsilon p}{u^2+p^2}\mathcal{I}
-\frac{u}{u^2+p^2}\mathcal{J}\right)(F_o),
$$
carrying energy density of $(u^2+p^2)^{-1}$.

%\vskip 0.5cm

Noting that $F_o=dx\wedge\zeta$, $*F_o=dy\wedge\zeta$,
$\bar{F_o}=\tilde{\eta}(F_o)$ and $\bar{*F_o}=\tilde{\eta}(*F_o)$,
the considerations above allow
the following conclusions and interpretations.  The whole set of nonlinear
solutions divides to subclasses, every such subclass is determined by the
spatial direction along which the solution propagates translationally, it is
the coordinate $z$ in our consideration, or by the corresponding $\zeta$.
Every solution $(F,*F)(u,p)$ of a given $\zeta$-subclass is obtained by means
of the action of a corresponding matrix $\alpha(u,p)$ on the corresponding
$(F_o,*F_o)$ through the representation $\rho$. We also note that every such
subclass divides to subsubclasses $\mathbb{G}.F$, determined by the two
functions $(u,p)$, and the action of $\mathbb{G}$ with {\it constant}
coefficients on the corresponding $(F,*F)(u,p)$. Every such subsubclass may be
considered as one solution represented in different bases of $\mathcal{G}$.

Clearly, since any $\alpha\in\mathbb{G}$ can be represented in various
ways in terms of other elements of $\mathbb{G}$, we see that every
solution $(F,*F)$ of a given subclass may be represented in
various ways in terms of other solutions of this class in the same domain $D$,
i.e., we have an example of a {\it nonlinear} "superposition" inside a given
subclass (let's not forget also about the Moivre relations in $\mathbb{C}$). We
could also say that the whole set of nonlinear solutions consists of orbits of
the (multiplicative) group of those complex valued functions
$\alpha(x,y,z,\xi)$ (the product is point-wise), the module
$|\alpha|=\sqrt{a^2+b^2}$ of which is a running wave along some fixed null
direction $\bar{\zeta}$.

We see also that the amplitude and the
phase of a solution in a natural way acquire the interpretations of
{\it amplitude} and {\it phase}
$
\phi=|\alpha|=\sqrt{a^2+b^2}, \ \ \ \psi=\mathrm{arccos}(\varphi)
$
of the corresponding complex field
$\alpha(x,y,z,\xi)=(\phi\,\mathrm{cos}\,\psi, \phi\,\mathrm{sin}\,\psi)$.

Finally, a suggestion comes to mind to consider the couple
$$
\Omega_o=F_o\otimes e_1+*F_o\otimes e_2, \ \
\bar{\Omega_o}=\bar{F_o}\otimes e_1+\bar{*F_o}\otimes e_2,
$$
where $(e_1,e_2)$ is the canoncal basis of $\mathbb{R}^2$, and $(F_o,*F_o)$ and
$(\bar{F_o},\bar{*F_o})$ depend on the chosen null direction $\bar{\zeta}$, as
{\it vacuum state}, appropriate to be acted upon by the {\it creation}
operators $\rho(\alpha)$, and the corresponding {\it annihilation} operators
$\rho(\alpha^{-1})$, considered as sections of the principal bundle
$M\times\mathbb{G}$.

The considerations made so far were limited, more or less, inside a given
subclass of (nonlinear) solutions, which propagate translationally along the
same isotropic 4-direction in $M$, or along a given spatial direction which
we choose for $z$-coordinate. A natural question arises: is it possible to
write down equations which would simultaneously describe a set of $N$ such
non-interacting solutions, which propagate
translationally along {\it different} spatial directions. The answer to this
question is positive, and the equations look like:
$$
\sum_{k=1}^{N}\left(\delta F^{2k-1}\wedge *F^{2k-1}\right)\otimes e_{2k-1}\vee
e_{2k-1}-
\sum_{k=1}^{N}\left(\delta *F^{2k-1}\wedge F^{2k-1}\right)\otimes e_{2k}\vee
e_{2k}+
$$
$$
\sum_{k=1}^{N}\left(-\delta F^{2k-1}\wedge F^{2k-1}+
\delta *F^{2k-1}\wedge *F^{2k-1}\right)\otimes e_{2k-1}\vee e_{2k}=0,
$$
where the index $k$ enumerates the 2-space $(\mathbb{R}^2)^k$ for the
corresponding couple $(F^k,*F^k)$. So, for every $k=1,2,\dots,N$ we obtain
the corresponding system of equations, i.e. the corresponding couple of fields
$(e_{2k-1},e_{2k})$, which defines the direction of translational
propagation of the solution $(F^k,*F^k)$.

\section{Two other views}
\subsection{$\Lambda^1(M,\mathcal{G})$ - view}

We  recall that on a principal bundle the triviality of the vertical
distribution allows the horizontal distribution to be defined by a connection
form $\omega$ on the bundle space $\mathcal{P}$, which is $\mathcal{G}$-valued
1-form and satisfies the conditions: $\omega(Z_h)=h, h\in \mathcal{G}$;\
$\omega\circ H_*=0$ and $R_a^*\omega=Ad(a^{-1})\circ \omega,\ a\in G$.  Then
the curvature $\Omega$ of the connection $\omega$ is given by
$\Omega=\mathbf{d}\omega\circ H_*=\mathbf{d}\omega+\frac12[\omega,\omega]$. If
the group is abelian, as it will be in our case, then
$\Omega=\mathbf{d}\omega$.

If the bundle is trivial, i.e. $\mathcal{P}=M\times G$ then the projection
$\pi$ is the projection on the first member: $\pi(x,a)=x, x\in M, a\in G$.
In this case with every connection form $\omega$ can be associated a
$\mathcal{G}$-valued 1-form $\theta$ on the base space such, that
$\omega(x,e;X,Z_h)=h+\theta(x;X)$, where $x\in M, X\in T_xM, h\in\mathcal{G}$
and $e$ is the identity of $G$. For the curvature in the abelian
case we obtain $\Omega=\pi^*\mathbf{d}\theta$.

This observation suggests to make use of the "connection-curvature machinary"
provided a $\mathcal{G}$-valued 1-form $\theta$ is at hand.

We consider the Lie algebra $\mathcal{G}$ of the group $\mathbb{G}$, given by the
$(2\times 2)$-real matrices
$$
\alpha(u,p) =\begin{Vmatrix}u & p\\-p &
u\end{Vmatrix},\quad u^2+p^2\neq 0.
$$
This Lie algebra has the natural basis $(I,J)$
(Sec.8.6.1) and, as a set, it differs from $\mathbb{G}$ just by adding the zero
$(2\times 2)$-matrix.  Recalling now that a nonlinear solution $(F,*F)$
in the $\zeta$-adapted coordinate system is determined by two functions
$(u,p)$, we define $\theta$ as
$$
\theta=u\zeta\otimes I+p\,\zeta\otimes J.
$$
For the "curvature" we obtain
$$
\Omega=\Omega^1\otimes I+\Omega^2\otimes
J=\mathbf{d}\theta= \big[u_xdx\wedge\zeta+u_ydy\wedge\zeta-
\varepsilon(u_\xi-\varepsilon u_z)dz\wedge d\xi\big]\otimes I
$$
$$
+\big[p_xdx\wedge\zeta+p_ydy\wedge\zeta-
\varepsilon(p_\xi-\varepsilon p_z)dz\wedge d\xi\big]\otimes J.
$$
So, we can find $*\Omega$ with respect to the Minkowski metric in $M$:
$$
*\Omega=(*\Omega^1)\otimes I+(*\Omega^2)\otimes J=
\varepsilon\big[u_xdy\wedge\zeta-u_ydx\wedge\zeta-
\varepsilon(u_\xi-\varepsilon u_z)dx\wedge dy\big]\otimes I
$$
$$
+\varepsilon\big[p_xdy\wedge\zeta-p_ydx\wedge\zeta-
\varepsilon(p_\xi-\varepsilon p_z)dx\wedge dy\big]\otimes J.
$$

We recall that the canonical conjugation $\alpha\rightarrow\alpha^*$ in
$\mathcal{G}$, given by $(I,J)\rightarrow (I,-J)$, defines the inner
product in $\mathcal{G}$ by $\langle\alpha,\beta\rangle=\frac12
tr(\alpha\circ\beta^*)$.  We have $\langle I,I\rangle=1, \langle J,J\rangle=1,
\langle I,J\rangle=0$.  We compute now the expressions
$$
*\langle\Omega,*\Omega\rangle, \ \
\langle\theta,*\Omega\rangle, \ \ \wedge(\theta,*\Omega),
$$
and obtain
respectively:

$$
*\langle\Omega,*\Omega\rangle= *(\Omega^1\wedge *\Omega^1)\langle I,I\rangle+
*(\Omega^2\wedge *\Omega^2)\langle J,J\rangle
$$
$$ =(u_\xi-\varepsilon
u_z)^2+(p_\xi-\varepsilon u_z)^2=-(\delta F)^2.
$$
$$ \langle\theta,*\Omega\rangle=
-\big[u(u_\xi-\varepsilon u_z)+
p(p_\xi-\varepsilon p_z)\big]dx\wedge dy\wedge dz
$$
$$
-\varepsilon\big[u(u_\xi-\varepsilon u_z)
+p(p_\xi-\varepsilon p_z)\big]dx\wedge dy\wedge d\xi ,
$$
%\newpage
$$
\wedge(\theta,*\Omega)=
\Big\{\big[p(u_\xi-\varepsilon u_z)-
u(p_\xi-\varepsilon p_z)\big]dx\wedge dy\wedge dz
$$
$$
+\varepsilon\big[p(u_\xi-\varepsilon u_z)
-u(p_\xi-\varepsilon p_z)\big]dx\wedge dy\wedge d\xi\Big\}\otimes I\wedge J
$$
$$
=\delta F\wedge F\otimes I\wedge J=\delta *F\wedge *F\otimes I\wedge J.
$$
\vskip 0.4cm
{\bf Corollary}. Equations
$F_{\mu\nu}\delta F^\nu=0, \ (*F)_{\mu\nu}\delta (*F)^\nu=0$
 are equivalent to the
equation $\langle\theta,*\Omega\rangle=0$; a non-linear solution may have rotational
component of propagation only if $\langle\Omega,*\Omega\rangle\neq 0$; the equation
$\mathbf{d}(\delta F\wedge F)=0$ is equivalent to
$\mathbf{d}\big[\wedge(\theta, *\Omega)\big]=0$.
\vskip 0.3cm
The two elements $Z_I=uI+pJ;\ \ Z_J=-pI+uJ$ of $\mathcal{G}$
have the following modules with respect to the above mentioned inner product in
$\mathcal{G}$: $|Z_I|=|Z_J|=\sqrt{u^2+p^2}$. Therefore, for the scale
factor $\mathcal{L}_o$ we readily obtain
$$
\mathcal{L}_o=\frac{|Z_I|}{|\Omega|}=\frac{|Z_J|}{|\Omega|}.
$$
\vskip 0.5cm
	{\bf Corollary}: The scale factor $\mathcal{L}_o$
is invariant with respect to the group action
$(M\times\mathbb{G},\mathbb{G})\rightarrow (M\times\mathbb{G}.\mathbb{G})$.
\vskip 0.5cm

Finally we note that we could write the 1-form $\theta$ in the form
$$
\theta=u\zeta\otimes I+p\zeta\otimes J=
\zeta\otimes uI+\zeta\otimes pJ=\zeta\otimes Z_I.
$$
If we start with the new 1-form
$$
\theta'=\zeta\otimes Z_J=\zeta\otimes (-pI)+\zeta\otimes (uJ)=
-p\zeta\otimes I+u\zeta\otimes J
$$
then, denoting $\mathbf{d}\theta'=\Omega'$, we obtain
\begin{eqnarray*}
*<\Omega',*\Omega'>&=&*<\Omega,*\Omega>,\\
<\theta',*\Omega'>&=&-<\theta,*\Omega>,\\
\wedge(\theta',*\Omega')&=&-\wedge(\theta,*\Omega).
\end{eqnarray*}
Hence the last corollary holds with respect to $\theta'$ too.
\vskip 0.5cm
	{\bf Remark}. The group $\mathbb{G}$ acts on the right on the basis
$(I,J)$ of $\mathcal{G}$. So, the transformed basis with
$\beta(u,\varepsilon p)\in\mathbb{G}$ is
$R_{\beta}(I,J)=(uI-\varepsilon pJ,\,\varepsilon pI+uJ)$.
Now let $\omega$ be 1-form on $M$ such that $\omega^2<0$ and $\omega\wedge
*\zeta=0$. Then the corresponding generalized field is given by
$$
F_\omega\otimes I+*F_\omega\otimes J=
=(\omega\wedge\zeta)\otimes (uI-\varepsilon pJ)+
*(\omega\wedge\zeta)\otimes (\varepsilon pI+uJ).
$$
In particular, $\omega=dx$ defines the field in an $F$-adapted coordinate
system.

\subsection{$\Lambda^1(M;L_{\Lambda^2(M)})$ - view}
Recall that linear connections $\nabla$ are 1st-order differential operators in
vector bundles. If in a vector bundle $\Sigma=(M,\pi,V^r)$, such
a connection $\nabla$ is given and $\sigma$ is a section of the bundle, then
$\nabla \sigma$ is 1-form on the base space valued in the space of sections of
the vector bundle, so if $X$ is a vector field on the base space then
$i(X)\nabla \sigma=\nabla_X \sigma$ is a new section of the same bundle. If $f$
is a smooth function on the base space then
$\nabla(f\sigma)=df\otimes\sigma+f\nabla\sigma$, which justifies the
differential operator nature of $\nabla$: the components of $\sigma$ are
differentiated and the basis vectors are lineary transformed. So, $\nabla$ is
$\mathbb{R}$-linear map $Sec(\Sigma)\rightarrow\Lambda^1(M,Sec(\Sigma))$
respecting the above differential property.

Let $e_a$ and $\varepsilon^b, a,b=1,2,\dots,r$ be two dual local bases of the
corresponding spaces of sections of $\Sigma$ and its dual:
$\langle\varepsilon^b,e_a\rangle=\delta_a^b$, then we can write
$$
\sigma=\sigma^a e_a,\quad
\nabla(e_a)=\Gamma_{\mu a}^b dx^\mu\otimes e_b,\quad
\nabla(\sigma^m e_m)=
\left[\mathbf{d}\sigma^b+\sigma^a\Gamma_{\mu a}^b dx^\mu\right]\otimes e_b.
$$

The components $\Gamma_{\mu a}^b$ with respect to the coordinates $\{x^\mu\}$
on the base space and with respect to the bases $\{e_a\}$ and
$\{\varepsilon^b\}$ are, in general, arbitrary functions, they necessarily
satisfy corresponding NON-tensor transformation law under diffeomorphisms of
$M$, and by appropriate choice of the base space coordinates these components
$\Gamma_{\mu a}^b$ can be made equal to zero at an arbitrary point $p\in M$.
But the very construction of the curvature of $\nabla $ suggests to  use
$L_{V}$-valued 1-forms on $M$  in the following sense.

If $\Psi_1$ and $\Psi_2$ are two $\Lambda^1(M;L_V)$-valued 1-forms, then
a map $(\Psi_1,\Psi_2)\rightarrow (\wedge,\circledcirc)(\Psi_1,\Psi_2)$ is
defined by (we shall write just $\circledcirc$ for $(\wedge,\circledcirc)$ and
the usual $\circ$ will mean just composition)
\begin{eqnarray*}
\circledcirc(\Psi_1,\Psi_2)&=&
(\Psi_1)_{\mu a}^{b}(\Psi_2)_{\nu m}^{n}dx^\mu\wedge dx^\nu\otimes
\big[\circ(\varepsilon^a\otimes e_b,\varepsilon^m\otimes e_n)\big]\\
&=&(\Psi_1)_{\mu a}^{b}(\Psi_2)_{\nu m}^{n}dx^\mu\wedge dx^\nu\,
\otimes\big[\langle\varepsilon^a,e_n\rangle(\varepsilon^m\otimes e_b)\big]\\
&=&(\Psi_1)_{\mu a}^b(\Psi_2)_{\nu m}^a
dx^\mu\wedge dx^\nu\otimes(\varepsilon^m\otimes e_b),
\ \ \mu<\nu.
\end{eqnarray*}
Now, the "curvature" of such a $\Psi$, would read
$$
\left[\mathbf{d}(\Psi_{\mu a}^{b}dx^\mu)\right]\otimes(\varepsilon^a\otimes
e_a) +\circledcirc(\Psi,\Psi).
$$

We go back now to EED. The vector bundle under consideration is the
(trivial) bundle $\Lambda^2(M)$ of 2-forms on the Minkowski space-time $M$.
If $\alpha(u,p)\in\mathcal{G}$ then $\rho'(\alpha)$ is a linear map in
$\Lambda^2(M)$ (Sec.8.6.2). We recall the two linear
maps in $\Lambda^2(M)$ defined by the images of the matrices $I,J\in
\mathcal{G}$, which were denoted by $\mathcal{I}$ and $\mathcal{J}$
(Sec.8.6.2). Making use now of the 1-form $\zeta=\varepsilon dz+d\xi $ we can
define three $L_{\Lambda^2(M)}$-valued 1-forms on $M$ of the above kind:
$$
\Psi=\zeta\otimes\chi,\
\bar\Psi=\zeta\otimes\bar\chi,\ \Psi^*=\zeta\otimes\chi^*, $$ where $$
\chi=u\mathcal{I}+p\mathcal{J},\
\bar\chi=u\mathcal{I}-p\mathcal{J},\
\chi^*=-p\mathcal{I}+u\mathcal{J}.
$$
These $L_{\Lambda^2(M)}$-valued 1-forms satisfy
$$
\circledcirc(\Psi,\Psi)=
\circledcirc(\Psi,\bar\Psi)=
\circledcirc(\Psi,\Psi^*)=0.
$$
Now, since
\begin{eqnarray*}
&&\Psi=u\zeta\otimes\mathcal{I}+p\,\zeta\otimes\mathcal{J},\\
&&\bar{\Psi}=u\zeta\otimes\mathcal{I}-p\,\zeta\otimes\mathcal{J},\\
&&\Psi^*=-p\,\zeta\otimes\mathcal{I}+u\,\zeta\otimes\mathcal{J},
\end{eqnarray*}
for the corresponding "curvatures" we obtain
\begin{eqnarray*}
&&\mathcal{R}=
\mathbf{d}(u\,\zeta)\otimes\mathcal{I}+
\mathbf{d}(p\,\zeta)\otimes\mathcal{J},\\
&&\bar{\mathcal{R}}=
\mathbf{d}(u\,\zeta)\otimes\mathcal{I}-
\mathbf{d}(p\,\zeta)\otimes\mathcal{J},\\
&&\mathcal{R}^*=
\mathbf{d}(-p\,\zeta)\otimes\mathcal{I}+
\mathbf{d}(u\,\zeta)\otimes\mathcal{J}.
\end{eqnarray*}
\noindent
{\bf Remark}. We have omitted here $\varepsilon$ in front of $p\mathcal{J}$,
but this is not essential since, putting $p\rightarrow\varepsilon p$ in the
expressions obtained, we easily restore the desired generality.
\vskip 0.3cm

By direct calculation we obtain:
\begin{eqnarray*}
*\frac16 Tr\left[ \circledcirc(\bar\Psi,*\mathbf{d}\Psi)\right]
&=&-\varepsilon\big[u(u_\xi-\varepsilon u_z)+
p(p_\xi-\varepsilon p_z)\big]dz\\
&-&\big[u(u_\xi-\varepsilon u_z)+
p(p_\xi-\varepsilon p_z)\big]d\xi;\\
\frac16Tr\left[\circledcirc(\Psi^*,*\mathbf{d}\Psi)\right]
&=&\varepsilon\Big[p(u_\xi-\varepsilon u_z)-
 u(p_\xi-\varepsilon p_z)\Big]dx\wedge dy\wedge dz\\
&+&
\Big[p(u_\xi-\varepsilon u_z)-
u(p_\xi-\varepsilon p_z)\Big]dx\wedge dy\wedge d\xi
=\delta F\wedge F;
\end{eqnarray*}
Denoting by $|\mathcal{R}|^2$ the quantity
$\frac16|*Tr\left[\circledcirc(\mathcal{R}\wedge*\bar{\mathcal{R}})\right]|$
we obtain
$$
|\mathcal{R}|^2=
\frac16|*Tr\left[\circledcirc(\mathbf{d}\Psi,
*\mathbf{d}\bar{\Psi})\right]|=
(u_\xi-\varepsilon u_z)^2+(p_\xi-\varepsilon p_z)^2=|\delta F|^2.
$$
Finally, since in our coordinates
$$
\frac16 tr(\chi)=\frac16 tr(u\mathcal{I}+p\mathcal{J})=u,\quad
$$
$$
\frac16 tr\big[(\chi\circ\bar\chi)\big]=\frac16
tr\big[(u\mathcal{I}+p\mathcal{J})\circ(u\mathcal{I}-p\mathcal{J})\big]
=u^2+p^2,
$$
for the phase $\psi$ and for the scale factor
$\mathcal{L}_o$ we obtain respectively
$$
\psi=\mathrm{arccos}\frac{\frac16 tr\chi}
{\sqrt{\frac16 tr(\chi\circ\bar\chi)}},
\quad
\mathcal{L}_o=\frac{\sqrt{\frac16
tr(\chi\circ\bar\chi)}}{\sqrt{\frac16}|\mathcal{R}|}=
\frac{\sqrt{tr(\chi\circ\bar\chi)}}{|\mathcal{R}|}.
$$
These results allow to say that choosing such
$L_{\Lambda^2(M)}$-valued 1-forms then our nonlinear equations  are given by
$Tr\left[\circledcirc(\bar\Psi,*\mathbf{d}\Psi)\right]=0$, and that the
non-zero value of the squared "curvature" invariant $|\mathcal{R}|^2$
guarantees availability of rotational component of propagation.

\vskip 0.5cm
As a brief comment to these aspects of our
basic relations we would like to especially note the basic role of the
isotropic 1-form $\zeta$.
It also participates in defining
the 2-form $F_o=dx\otimes \zeta$, which gives the possibility to identify a
nonlinear solution $F(u,p)$ with an appropriately defined linear map
$\rho'(\alpha(u,p))=u\mathcal{I}+p\mathcal{J}$ in $\Lambda^2(M)$.

This special importance of $\zeta$  is based on the fact that it defines unique
direction of translational propagation of the solution, and its uniqueness
is determined by our equations: {\it all nonlinear solutions have zero
invariants}: $F_{\mu\nu}F^{\mu\nu}=F_{\mu\nu}(*F)^{\mu\nu}=0$.

For all nonlinear solutions we have $(\delta F)^2<0$, and all finite nonlinear
solutions have finite amplitude $\phi$: $0<\phi^2=\frac16 tr(F\circ\bar{F})=
\frac16 tr(\chi\circ\bar\chi)<\infty$. The scale factor $\mathcal{L}_o$
separates the finite nonlinear solutions to two subclasses: if
$\mathcal{L}_o=\infty$, i.e. $|\delta F|=|\Omega|=|\mathcal{R}|=0$, the
solution has no spin properties; if $0<\mathcal{L}_o<\infty$, i.e. $|\delta
F|=|\Omega|=|\mathcal{R}|\neq 0$, the solution carries spin momentum.

Hence, we can say that, the corresponding invariants $|\Omega|$ and
$|\mathcal{R}|$  are responsible for availability of rotational, or spin,
component of propagation.

\section{Nonlinear solutions with spin} \index{spin of nonlinear solutions}
Before to start with spin-carrying solutions we briefly comment the nonlinear
solutions with running wave character, these solutions require $|\delta
F|=|\delta *F|=0$, so, for the two spatially finite functions $u$ and $p$ we
get in the $\zeta$-adapted coordinate system $u=u(x,y,\xi+\varepsilon z)$ and
$p=p(x,y,\xi+\varepsilon z)$. Whatever the spatial shape and spatial
structure of these two finite functions could be the whole solution will
propagate {\it only translationally} along the coordinate $z$ with the
velocity of light $c$ without changing its shape and structure. In this sense
this class of nonlinear solutions show soliton-like behavior: finite 3d
spatial light-like formations propagate translationally in vacuum. If we forget
about the spin properties of electromagnetic radiation, we can consider such
solutions as mathematical models of classical finite electromagnetic
macro-formations of any shape and structure, radiated by ideal parabolic
antennas. Maxwell equations can NOT give such solutions.

Now we turn to spin-carrying solutions. The crucial moment here is to find
reasonable additional conditions for the phase function $\varphi$, or
for the phase $\psi=\mathrm{arccos}\,\varphi$.

\subsection{The Basic Example}
The reasoning here follows the idea that
these additional conditions {\it have to express some internal compatibility
among the various characteristics of the solution}. A suggestion what kind of
internal compatibility to use comes from the observation that {\it the
amplitude function $\phi$ is a first integral of the vector field}
$\bar{\zeta}$, i.e.
$$
 \bar{\zeta}(\phi)=\left(-\varepsilon \frac{\partial}{\partial z}+\frac
{\partial} {\partial \xi}\right)(\phi)=-\varepsilon \frac{\partial}{\partial
z}\phi(x,y,\xi+\varepsilon z) +\frac {\partial}{\partial
\xi}\phi(x,y,\xi+\varepsilon z)=0. $$

In order to extend
this compatibility between $\bar{\zeta}$ and $\phi$ we require the phase
function $\varphi$ to be first integral of some of the available $F$-generated
vector fields.  Explicitly, we require the following (recall ${\bf R,S}$ from
Sec.8.3.2, also, $\mathbf{S}=\frac{\partial}{\partial \xi}$ and
$\mathbf{R}=\frac{\partial}{\partial z}$ in $\zeta$-adapted frame):
 \vskip
0.5cm \noindent
{\it The phase function}\ $\varphi$\ {\it is a first integral
of the three unit vector fields} ${\bf A,A^*},{\bf S}$:
$$
{\bf A}(\varphi)= {\bf A^*}(\varphi)={\bf S}(\varphi)=0,
$$
{\it and the scale
factor $\mathcal{L}_o$ is a first integral of}
 $\mathbf{R}: \mathbf{R}(\mathcal{L}_o)=0$.
\vskip 0.5cm
The first two requirements ${\bf
A}(\varphi)={\bf A^*}(\varphi)=0$ define the following system of differential
equations for $\varphi$:
$$
-\varphi\frac{\partial \varphi}{\partial
x}-\sqrt{1-\varphi^2}\frac{\partial \varphi}{\partial y}=0,\
\sqrt{1-\varphi^2}\frac{\partial \varphi}{\partial x}- \varphi\frac{\partial
\varphi}{\partial y}=0.
$$
Noticing that the matrix
$$
\begin{Vmatrix} -\varphi
&-\sqrt{1-\varphi^2}\\ \sqrt{1-\varphi^2}    &-\varphi \end{Vmatrix}
$$
has non-zero determinant, we see that the only solution of the above system
is the zero-solution:
$$
\frac{\partial \varphi}{\partial x}=\frac{\partial
\varphi}{\partial y}=0.
$$
We conclude that in the coordinates used the phase
function $\varphi$ may depend only on $(z,\xi)$. The third equation
$\mathbf{S}(\varphi)=0$ requires $\varphi$ not to depend on $\xi$ in this
coordinate system, so, $\varphi=\varphi(z)$. For
$\mathcal{L}_o=|L_{\bar{\zeta}}\psi|^{-1}$ in terms of $\varphi$ we get $$
\mathcal{L}_o=\frac{\sqrt{1-\varphi^2}}{|\varphi_z|}.
$$
Now, the last requirement, which in these coordinates reads
$$
\mathbf{R}(\mathcal{L}_o)=\frac{\partial \mathcal{L}_o}{\partial z}=
\frac{\partial}{\partial z}\frac{\sqrt{1-\varphi^2}}{|\varphi_z|}=0,
$$
means that the scale factor $\mathcal{L}_o$ is a pure constant:
$\mathcal{L}_o=const$.  In this way the defining relation for $\mathcal{L}_o$
turns into a differential equation for $\varphi$:
$$
\mathcal{L}_o=\frac{\sqrt{1-\varphi^2}}{|\varphi_z|}\  \rightarrow
\frac{\partial \varphi}{\partial z}=                               %75%
\mp \frac{1}{\mathcal{L}_o}\sqrt{1-\varphi^2}.
$$
The obvious solution to this equation is
$$
\varphi(z)=\mathrm{cos}\left(\kappa\frac{z}{\mathcal{L}_o}+const\right),
$$                                    %76%
where $\kappa=\pm 1$. We note that the naturally arising in this case spatial
periodicity $2\pi\mathcal{L}_o$ and {\it characteristic frequency}
$\nu=c/2\pi\mathcal{L}_o$ have nothing to do with the corresponding concepts
in classical vacuum electrodynamics. In fact, {\it our scale factor
$\mathcal{L}_o$ can not be defined in Maxwell's theory}.

The above considerations may be slightly extended and put in terms of the
phase $\psi$, and in these terms they look simpler. In fact, we have the
equation
$$
\psi_\xi-\varepsilon \psi_z=\kappa\frac{1}{\mathcal{L}_o}, \ \ \kappa=\pm 1,
$$
where $\mathcal{L}_o=const$. So, we get the two basic solutions
$$
\psi_1=-\varepsilon\kappa\frac{z}{\mathcal{L}_o}+const, \quad
\psi_2=\kappa\frac{\xi}{\mathcal{L}_o}+const.
$$

We get two kinds of periodicity: spatial periodicity along the coordinate $z$
and time-periodicity along the time coordinate $\xi$. The two values of
$\kappa=\pm 1$ determine the two possible rotational structures: left-handed
(left polarized), and right-handed (right polarized).
Further we are going to
concentrate on the spatial periodicity because it is strongly connected with
the spatial shape of the solution. In particular, it suggests to localize the
amplitude function $\phi$ inside a helical cylinder of height
$2\pi\mathcal{L}_o$, so, the solution will propagate along the prolongation of
this finite initial helical cylinder in such a way that all points of the
spatial support shall follow their own helical trajectories without
crossings. For such solutions with $|\delta F|\neq 0$ we are going to
consider various ways for quantitative description of the available intrinsic
rotational momentum, or the {\it spin momentum}, of these solutions. We call
it {\it spin-momentum} by obvious reasons: it is of intrinsic nature and does
not depend on any {\it external} point or axis as it is the case of angular
momentum.

\subsection{The $\mathcal{G}$-Approach}
In this approach we make use of the corresponding
scale factor $\mathcal{L}_o=const$, of the isotropic 1-form $\zeta$ and
of the two objects $Z_I=uI+pJ$ and $Z_J=-pI+uJ$, considered as
$\mathcal{G}$-valued functions on $M$. By these quantities we build the
following $\mathcal{G}\wedge\mathcal{G}$-valued 1-form $H$:
$$
H=\kappa \frac{2\pi\mathcal{L}_o}{c} \zeta \otimes(Z_I\wedge Z_J).
$$
In components we have
$$
H_\mu^{ab}=\kappa\frac{2\pi\mathcal{L}_o}{c} \zeta_\mu(
Z_I^aZ_J^b-Z_I^bZ_J^a).
$$
In our system of coordinates we get
$$
H=\kappa\frac{2\pi\mathcal{L}_o}{c}\phi^2\
(\varepsilon dz+d\xi)\otimes I\wedge J,
$$
hence, the only non-zero components are
$$
H_3^{12}=\kappa\varepsilon\frac{2\pi\mathcal{L}_o}{c} \phi^2,
\ H_4^{12}=\kappa\frac{2\pi\mathcal{L}_o}{c} \phi^2.
$$
It is easily seen that the 3-form $*H$ is closed: $\mathbf{d}*H=0$. In fact,
$$
\mathbf{d}*H=\kappa\frac{2\pi\mathcal{L}_o}{c}\big[\big(\phi^2\big)_\xi-
\varepsilon\big(\phi^2\big)_z\big](dx\wedge dy\wedge d\wedge d\xi)\otimes
(I\wedge J)=0
$$
because $\phi^2$ is a running wave along the coordinate $z$. We reduce now
$*H$ to $\mathbb{R}^3$ and obtain
$$
(*H)_{\mathbb{R}^3}=
\kappa\frac{2\pi\mathcal{L}_o}{c}
\phi^2(dx\wedge dy\wedge dz)\otimes(I\wedge J).
$$
According to Stokes theorem, for finite solutions, we obtain the finite
(conserved) quantity
$$
{\bf H}=\int_{\mathbb{R}^3}(*H)_{\mathbb{R}^3}
=\kappa\frac{2\pi\mathcal{L}_o}{c} E=\kappa ET\,I\wedge J,
$$
which is a volume form in $\mathcal{G}$, $T=2\pi\mathcal{L}_o/c$, and $E$ is
the integral energy of the solution. The module $|{\bf H}|$ of {\bf H} is
$|{\bf H}|=ET$.

We see the basic role of the two features of the solutions: their spatially
finite/concentrated nature, giving finite value of all spatial integrals,
and their translational-rotational dynamical nature with
$|\delta F|=|\delta*F|\neq 0$,
allowing finite value of the scale factor $\mathcal{L}_o$.

\subsection{The FN-Bracket Approach}
We proceed to the next approach to introduce spin-momentum. We recall the
components of the Fr\"oliher-Nijenhuis bracket $S_F$ of the finite nonlinear
solution $(F,*F)$:
$$
(S_F)_{\mu \nu }^\sigma =2\left[ F_\mu ^\alpha \frac{\partial F_\nu
^\sigma} {\partial x^\alpha }-F_\nu ^\alpha \frac{\partial F_\mu ^\sigma
}{\partial x^\alpha }-F_\alpha ^\sigma \frac{\partial F_\nu ^\alpha
}{\partial x^\mu } +F_\alpha ^\sigma \frac{\partial F_\mu ^\alpha }{\partial
x^\nu }\right].
$$
When evaluated on the two unit vector fields $\mathbf{A}$ and
$\varepsilon\mathbf{A}^*$ we obtain
$$
(S_F)_{\mu \nu }^\sigma{\bf A}^\mu {\bf \varepsilon A^*}^\nu =
(S_F)_{12}^\sigma ({\bf A}^1{\bf \varepsilon A^*}^2-
{\bf A}^2{\bf \varepsilon A^*}^1).
$$
For $(S_F)_{12}^\sigma $ we get
$$
(S_F)_{12}^1=(S_F)_{12}^2=0,\quad
(S_F)_{12}^3=-\varepsilon (S_F)_{12}^4=2\varepsilon \{p(u_\xi -\varepsilon
u_z)-u(p_\xi -\varepsilon p_z)\}.
$$
It is easily seen that the following relation holds:
${\bf A}^1{\bf \varepsilon A^*}^2-{\bf A}^2{\bf \varepsilon A^*}^1=1.$
Now, for the above obtained solution for $\varphi$ we have
$$
u=\phi (x,y,\xi +\varepsilon z)
\cos\left(\kappa \frac{z}{\mathcal{L}_o} +const\right),\quad
p=\phi (x,y,\xi +\varepsilon z)
\sin \left(\kappa\frac{z}{\mathcal{L}_o} +const\right).
$$
We obtain
$$
(S_F)_{12}^3=-\varepsilon (S_F)_{12}^4=
-2\varepsilon \frac{\kappa}{\mathcal{L}_o} \phi ^2,
$$
$$
(S_F)_{\mu\nu }^\sigma {\bf A}^\mu {\bf \varepsilon A^*}^\nu
=\left[0,0,-2\varepsilon
\frac{\kappa}{\mathcal{L}_o}\phi ^2,
2\frac{\kappa}{\mathcal{L}_o}\phi ^2\right].
$$
Since $\phi^2$ is a running wave along the $z$-coordinate, the vector
field $S_F({\bf A,\varepsilon A^*})$ has zero divergence:
$\nabla_\nu \left[S_F({\bf A,\varepsilon A^*})\right]^\nu=0$.
Now, defining the {\it helicity 1-form} of the solution $F$ by
$$
\Sigma_F=\frac{\mathcal{L}_o}{2}\frac{2\pi\mathcal{L}_o}{c}
\tilde{\eta}(S_F({\bf A,\varepsilon A^*})),
$$
 then $*\Sigma_F$ is closed 3-form, and the
integral of the $\mathbb{R}^3$-reduced $*\Sigma_F$
$$
\int_{\mathbb{R}^3}(*\Sigma_F)_{\mathbb{R}^3}=
\int_{\mathbb{R}^3}{\left(\Sigma_F\right)_4}dx\wedge dy\wedge dz
$$
does not depend on time and is equal to $\kappa ET$.

A coordinate free version of this approach makes use of the bracket relation for
$A\otimes\bar{\zeta}$ with itself as given in Sec.1.4.3.
In fact, the computation gives
$$
[A\otimes\bar{\zeta},A\otimes\bar{\zeta}]
$$
$$
=\big[u(p_\xi-\varepsilon\,u_z)-
p(u_\xi-\varepsilon\,u_z)\big]dx\wedge dy\otimes\bar{\zeta}
=\mathbf{R}dx\wedge dy\otimes\bar{\zeta}.
$$
Computing now the quantity
$$
\frac{2\pi\mathcal{L}_o^2}{c}\frac12[A\otimes\bar{\zeta},A\otimes\bar{\zeta}]
(\mathbf{\bar{A}}\wedge \varepsilon\mathbf{\bar{A}^*})
$$
we come to the same $\Sigma_F$.

\subsection{The $\mathbf{d}(F\wedge\delta F)=0$ Approach}
Here we make use of the equation $\mathbf{d}(F\wedge\delta F)=0$ and see what
restrictions this equation imposes on $\psi$, and what conservation law this
closed 3-form will give. In our system of coordinates this equation reeds
$$
\mathbf{d}(F\wedge \delta F)=\varepsilon\Phi^2\left(\psi_{\xi\xi}+\psi_{zz}-
2\varepsilon\psi_{z\xi}\right)dx\wedge dy\wedge dz\wedge d\xi=0,   %77%
$$
i.e.
$$
\psi_{\xi\xi}+\psi_{zz}-2\varepsilon\psi_{z\xi}=
\left(\psi_\xi-\varepsilon\psi_z\right)_\xi -
\varepsilon\left(\psi_\xi-\varepsilon\psi_z\right)_z =0.           %78%
$$
This equation has the following solutions:

1$^o $. Running wave solutions $\psi=\psi(x,y,\xi+\varepsilon z)$,

2$^o $. $\psi = \xi.g(x,y,\xi+\varepsilon z)+b(x,y)$,

3$^o $. $\psi = z.g(x,y,\xi+\varepsilon z)+b(x.y)$,

4$^o $. Any linear combination of the above solutions with coefficients which
are allowed to depend on $(x,y)$.

The functions $g(x,y,\xi+\varepsilon z)$ and $b(x,y)$ are arbitrary
in the above expressions.

The running wave solutions $\psi_1$, defined by $1^o$, lead to $F\wedge \delta
F=0$ and to $|\delta F|=0$.

The solutions $\psi_2$ and $\psi_3$, defined respectively by 2$^o$ and 3$^o$,
give the scale factors $\mathcal{L}_o=1/|g|$, and since $\mathcal{L}_o$ is
invariant with respect to the rotation action of $\mathbb{G}$ on the plane
$(x,y)$, it should not depend on $(x,y)$ in this coordinate system. Hence, we
obtain $g=g(\xi+\varepsilon z)$, so, the most natural choice seems $g=const$,
which implies also $\mathcal{L}_o=const$. A possible dependence of $\psi$ on
$(x,y)$ may come only through $b(x,y)$.  Note that the physical dimension of
$\mathcal{L}_o$ is {\it length} and $b(x,y)$ is dimensionless.

We turn now to the integral spin-momentum computation.
In this approach its density is given by the correspondingly
normalized Leibniz bracket $\{F,F\}=\delta F\wedge F=
\delta*F\wedge *F=\{*F,*F\}$ (Sec.8.5). We normalize it as follows:
\begin{eqnarray*} \beta
&=&2\pi\frac{\mathcal{L}_o^2}{c}\,\delta F\wedge F\\
&=&2\pi\frac{\mathcal{L}_o^2}{c}[-\varepsilon                    %79%
\phi^2(\psi_\xi-\varepsilon\psi_z)dx\wedge dy\wedge dz
-\phi^2(\psi_\xi-\varepsilon\psi_z)dx\wedge dy\wedge d\xi].
\end{eqnarray*}
The physical dimension of $\beta$ is "energy-density $\times $ time".
Since $\beta$ is closed: $\mathbf{d}\beta=0$,
we may use the Stokes' theorem.  The restriction of
$\beta$ to $\mathbb{R}^3$ is:
$$
\beta_{\mathbb{R}^3}=2\pi\frac{\mathcal{L}_o^2}{c}\left[-\varepsilon
\phi^2(\psi_\xi-\varepsilon \psi_z)dx\wedge dy\wedge dz\right].
$$
We note that on the nonlinear solutions the 3-form
$\delta F\wedge F=\delta *F\wedge *F$ is dually
invariant:
$$
\delta F\wedge F=\delta (Fcos\,\alpha-*Fsin\,\alpha)\wedge
(Fsin\,\alpha+*Fcos\,\alpha), \ \alpha=const.
$$
 Let's consider first the solutions $3^o$ above with
$\mathcal{L}_o=const$ and $b(x,y)=const$.  The corresponding phase
$\psi=\kappa\frac{z}{\mathcal{L}_o}+const, \kappa=\pm1$, requires {\bf spatial}
periodicity along the coordinate $z$ with period $2\pi\mathcal{L}_o$. So, if we
restrict the spatial extension of the solution along $z$  to one such period
$2\pi\mathcal{L}_o$, our solution will occupy at every moment a smoothed out
one-step part of a helical tube. Its time evolution will be a
translational-rotational propagation along this helical tube. So, we have
an example of an object with helical spatial structure and with intrinsical
rotational component of propagation, and this rotational component of
propagation does NOT come from a rotation of the object as a whole around some
axis.

On the contrary, the solutions defined by $2^o$, are NOT obliged to have
spatial periodicity. Their evolution includes $z$-translation and rotation
around the $z$-axis {\it as a whole}.

For the case $3^o$ with $\mathcal{L}_o=const$ we can integrate
$$
\beta_{\mathbb{R}^3}=
2\pi\frac{\mathcal{L}_o}{c}\kappa\phi^2 dx\wedge dy\wedge dz
$$
over the 3-space and obtain
$$
\int_{\mathbb{R}^3}\beta_{\mathbb{R}^3}=                                       %71%
\kappa E\frac{2\pi\mathcal{L}_o}{c}=
\kappa ET=\pm ET,
$$
where $E$ is the integral energy of the solution, $T=2\pi\mathcal{L}_o/c$ is
the intrinsically defined time-period, and $\kappa=\pm 1$ accounts for the
two polarizations.  According to our interpretation this is the integral
spin-momentum of the solution for one period $T$.

\subsection{The Nonintegrability Approach}
Here we make use of the observation that the two Pfaff systems $(A,\zeta)$
and $(A^*,\zeta)$ are nonintegrable when $\mathcal{L}_o\neq 0$. We
have
\begin{eqnarray*}
\begin{split}
\mathbf{d}A\wedge A\wedge\zeta=
 \mathbf{d}A^*\wedge A^*\wedge\zeta
&=\varepsilon\big[u(p_\xi-\varepsilon p_z)-p(u_\xi-\varepsilon u_z)\big]
dx\wedge dy\wedge dz\wedge d\xi \\
&=\varepsilon\phi^2(\psi_\xi-
\varepsilon \psi_z)dx\wedge dy\wedge dz\wedge d\xi.\\
\end{split}
\end{eqnarray*}

Integrating the 4-form
$$
\frac{2\pi\varepsilon\mathcal{L}_o}{c}\, \mathbf{d}A\wedge A\wedge\zeta
$$
on the 4-volume $\mathbb{R}^3\times \mathcal{L}_o$ we obtain $(-\kappa ET)$.
\vskip 0.5cm

We recall also that for finite solutions the electromagnetic volume form
$\omega_\chi=-\frac1c
A\wedge\varepsilon A^*\wedge\mathbf{R}\wedge\mathbf{S}$
gives the same quantity $ET$ when integrated over the 4-volume
$\mathbb{R}^3\times \mathcal{L}_o,\  \mathcal{L}_o=const$.

\subsection{The Godbillon-Vey 3-form as a Conservative\\ Quantity}
According to the Frobenius integrability theorem having a completely
integrable $p-$dimensional differential system on a $n$-manifold $M$ is
equivalent to having a suitable completely integrable $(n-p)-$dimensional Pfaff
system on the same manifold. In case of 1-dimensional Pfaff system it is
determined by a suitable 1-form $\omega$, defined up to a nonvanishing
function: $f\omega, f(x)\neq 0, x\in M$, and $\omega$ satisfies the equation
$\mathbf{d}\omega\wedge\omega=0$ (so obviously, $f\omega$ also satisfies
$\mathbf{d}(f\omega)\wedge f\omega=0$). From this last equation it follows that
there is 1-form $\theta$ such, that $\mathbf{d}\omega=\theta\wedge\omega$.
Now, the Godbillon-Vey theorem says that the 3-form
$\beta=\mathbf{d}\theta\wedge\theta$ is closed:
$\mathbf{d}\beta=\mathbf{d}(\mathbf{d}\theta\wedge\theta)=0$, and, varying
$\theta$ and $\omega$ in an admissible way:
$$
\theta\rightarrow
(\theta+g\omega); \ \omega\rightarrow f\omega,
$$
where $g$ is a function, leads to adding an
exact 3-form to $\beta$, so we have a cohomological class $\Gamma$ defined
entirely by the integrable 1-dimensional Pfaff system. From physical point
of view the conclusion is that each completely integrable 1-dimensional Pfaff
system on Minkowski space may be tested as  generator of conservation law
through the restriction of $\beta$ on $\mathbb{R}^3$.

Recall now the following objects on our Minkowski space-time: $A$, $A^*$ and
$\zeta$. These are 1-forms. We form the corresponding vector fields through
the Lorentz-metric and denote them by $\vec{A}, \vec{A^*}, \vec{\zeta}$.
Let's consider the 1-form $\omega=f\zeta=\varepsilon fdz+fd\xi$, where $f$ is
a nonvanishing function on $M$. We have the relations:
$$
\omega(\vec{A})=0,\ \ \ \omega(\vec{A^*})=0,\ \ \ \omega(\vec{\zeta})=0.
$$
Moreover, since $\zeta$ is closed, $\omega=f\zeta$ satisfies the Frobenius
integrability condition:
$$
\mathbf{d}\omega\wedge\omega=f\mathbf{d}f\wedge\zeta\wedge\zeta=0.
$$
Therefore, the corresponding 1-dimensional Pfaff system, defined by
$\omega=f\zeta$, is completely integrable, and there exists a new 1-form
$\theta$, such that $\mathbf{d}\omega=\theta\wedge\omega$, and
$\mathbf{d}(\mathbf{d}\theta\wedge\theta)=0$.

From the point of view of generating a conservative quantity through
integrating the restriction $i^*\beta$ of
$\beta=\mathbf{d}\theta\wedge\theta$ to $\mathbb{R}^3$, through the
imbedding $i: (x,y,z)\rightarrow(x,y,z,0)$ it is not so important whether
$\Gamma$ is trivial or nontrivial.  The important point is the 3-form $\beta$
to have appropriate component $\beta_{123}$ in front of the basis element
$dx\wedge dy\wedge dz$, because only this component survives after the
restriction to $\mathbb{R}^3$ is performed, which formally means that we put
$d\xi=0$ in $\beta$.  The value of the corresponding conservative quantity
will be found provided the integration can be carried out successfully, i.e.
when $(i^*\beta)_{123}$ has no singularities and $(i^*\beta)_{123}$ has cimpact
finite 3d support in $\mathbb{R}^3$.

In order to find appropriate $\theta$ in our case we are going to take
advantage of the freedom we have when choosing $\theta$: the 1-form $\theta$
is defined up to adding to it an 1-form $\gamma=g\,\omega$, where $g$
is an arbitrary function on $M$, because $\theta$ is defined by the relation
$\mathbf{d}\omega=\theta\wedge\omega$, and
$(\theta+g\,\omega)\wedge\omega=\theta\wedge\omega$
always. The freedom in choosing $\omega$ consists in choosing the
function $f$, and we shall show that $f$ may be chosen in such a way:
$\omega=f\,\zeta$, that the corresponding integral of $i^*\beta$ to present a
finite conservative quantity.

Recalling that $\mathbf{d}\zeta=0$, we have
$$
\mathbf{d}\omega=\mathbf{d}(f\zeta)=\mathbf{d}f\wedge\zeta+f\mathbf{d}\zeta=
\mathbf{d}f\wedge\zeta.
$$
Since $\mathbf{d}\omega$ must be equal to $\theta\wedge\omega$ we obtain
$$
\mathbf{d}\omega=\mathbf{d}f\wedge\zeta=\theta\wedge\omega=
\theta\wedge(f\zeta)=f\theta\wedge\zeta.
$$
It follows
$$
\theta\wedge\zeta=
\frac{1}{f}\mathbf{d}f\wedge\zeta=\mathbf{d}(ln\,f)\wedge\zeta=
\Big[\mathbf{d}(ln\,f)+h\zeta\Big]\wedge\zeta,
$$
where $h$ is an arbitrary function. Hence, in general, we obtain
$\theta=\mathbf{d}(ln\,f)+h\zeta$. Therefore, since
$\mathbf{d}\theta=\mathbf{d}h\wedge\zeta$, for $\mathbf{d}\theta\wedge\theta$
we obtain
$$
\mathbf{d}\theta\wedge\theta=\mathbf{d}(ln\,f)\wedge\mathbf{d}h\wedge\zeta.
$$
Denoting for convenience $(ln\,f)=\varphi$ for the restriction $i^*\beta$ we
obtain
$$
i^*\beta=\varepsilon(\varphi_x h_y-\varphi_y h_x)dx\wedge dy\wedge dz.
$$

In order to find appropriate interpretation of $i^*\beta$ we recall that
$$
*(\delta F\wedge F)=-\varepsilon\phi^2(\psi_\xi-\varepsilon \psi_z)\zeta,
$$
so, $*(\delta F\wedge F)$ is of the kind $f\,\zeta$, and it
defines the same 1-dimensional Pfaff system as $f\,\zeta$ does. We recall also
that if the scale factor $\mathcal{L}_o=1/|\psi_\xi-\varepsilon \psi_z|$ is a
nonzero constant then $\phi^2(\psi_\xi-\varepsilon \psi_z)$ is a running
wave and the 3-form $\delta F\wedge F$ is closed. Hence, the
interpretation of $i^*\beta$ as
$i^*(\delta F\wedge F)=
-\frac{1}{\mathcal{L}_o}\varepsilon\kappa\phi^2 dx\wedge dy\wedge dz$
requires appropriate definition of the two functions $f$ and $h$. So we must
have
$$
\varphi_x h_y -h_x\varphi_y=-\frac{\kappa}{\mathcal{L}_o}\phi^2.
$$
If we choose
$$
f=exp(\varphi)=exp\left[\int{\phi^2}dx\right],\ \
h=-\frac{\kappa}{\mathcal{L}_o}y+const
$$
all requirements will be fulfilled, in particular,
$\varphi_{xy}=\varphi_{yx}=(\phi^2)_y$ and $h_{xy}=h_{yx}=0$.

Hence, the above choice of $f$ and $h$ allows
 the spatial restriction of the Godbillon-Vey 3-form $\beta$ to be
interpreted as the spatial restriction of $F\wedge \delta F$. So,
the curvature expressions found in the previous sections, as well as the
corresponding {\it spin}-properties of the nonlinear solutions being
available when $\delta F\wedge F\neq 0$, are being connected with the
integrability of the Pfaff system $\omega=f\,\zeta$.

Finally, consider the Maxwell-Minkowski "energy" tensor $P_\mu^\nu$ generated
by the 2-form $S=\sqrt{2\gamma}\mathbf{A}\wedge\mathbf{A^*}$, where $\gamma$ is
a constant with appropriate physical dimension. We get in our coordinates
$S=\sqrt{2\gamma}\,dx\wedge dy$. Also, $P_{\mu\sigma}P^{\sigma\nu}=id_\mu^\nu$,
so, $P$ is involution and it is easily verified that $P$ commutes (formally)
with the standard energy tensor $Q_\mu^\nu: P\circ Q=Q\circ P$. The eigen
values of $P$ are $\pm\gamma$. Hence, assuming $dim(\gamma)=action$, then the
invariance of the eigen values and the invariance of $P$ with respect to
duality transformations suggest to introduce the characteristic integral time
period $T=\frac{E}{\gamma}$, where $E$ is the full energy of the solution, i.e.
to consider $\gamma$ as {\it proper integral unit action} of the
solution considered. \index{photon-like solutions figures}
\vskip 0.3cm
On the two figures below are given two
theoretical examples with $\kappa=-1$ and $\kappa=1$ respectively, amplitude
function $\phi$ filling in a smoothed out tube around a circular helix of height
$2\pi\mathcal{L}_o$ and pitch $\mathcal{L}_o$, and phase
function $\varphi=\mathrm{cos}(\kappa z/\mathcal{L}_o)$. The solutions
propagate left-to-right along the coordinate $z$.

\begin{center}
\begin{figure}[ht!]
\centerline{
{\mbox{\psfig{figure=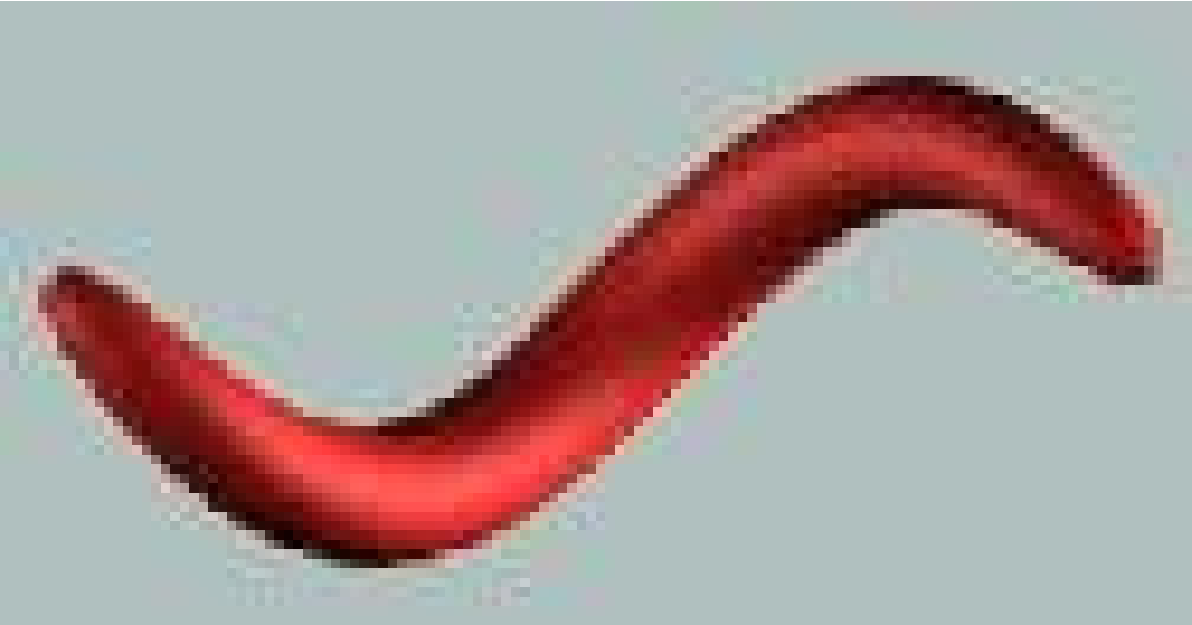,height=1.8cm,width=3.5cm}}
\mbox{\psfig{figure=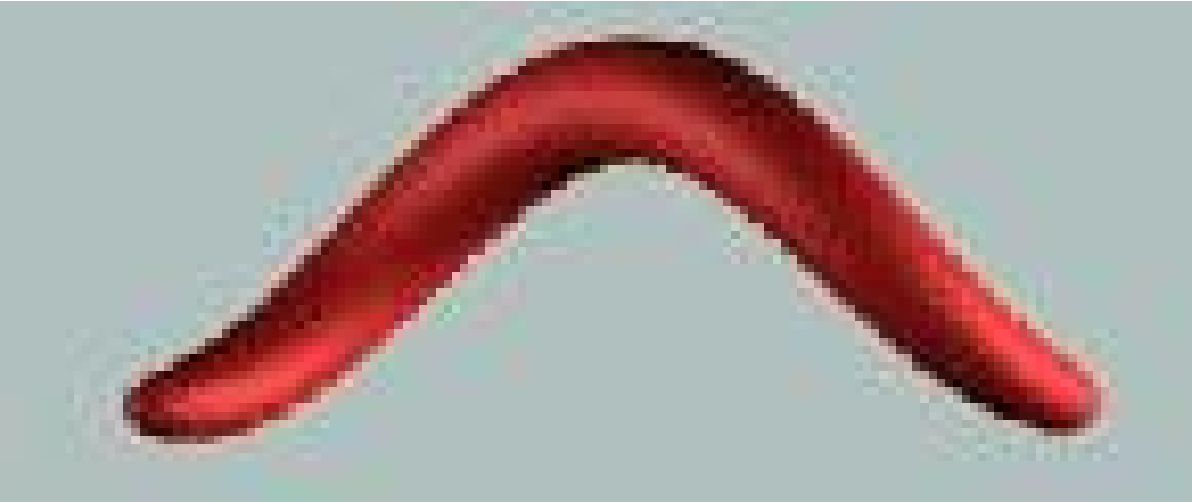,height=1.8cm,width=4.2cm}}
\mbox{\psfig{figure=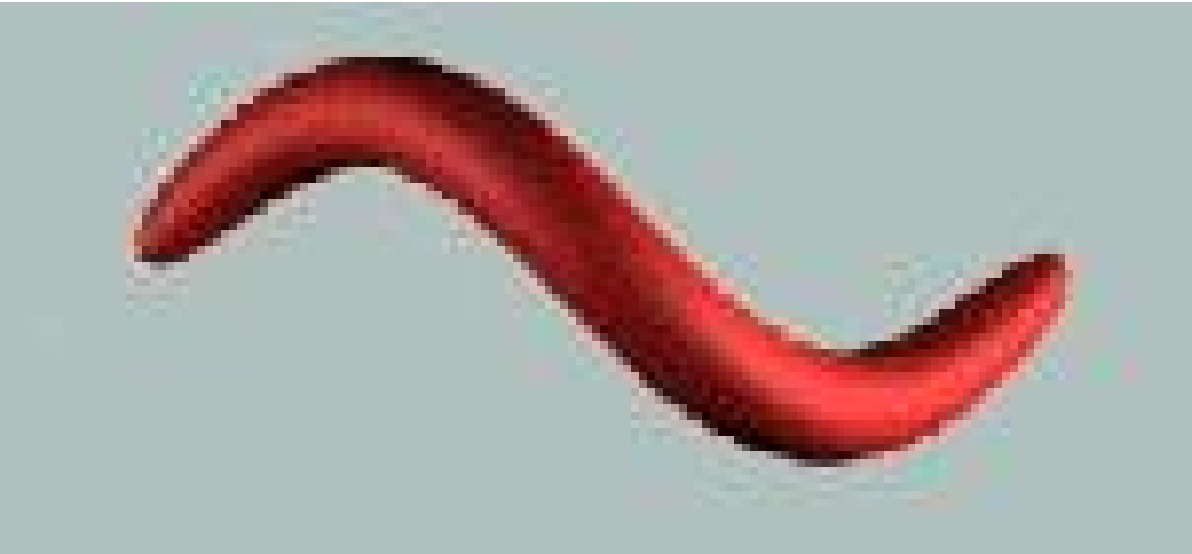,height=1.8cm,width=4.2cm}}}}
\caption{Theoretical example with $\kappa=-1$. The Poynting vector is
directed left-to-right.}
\end{figure}
\end{center}
\begin{center}
\begin{figure}[ht!]
\centerline{
{\mbox{\psfig{figure=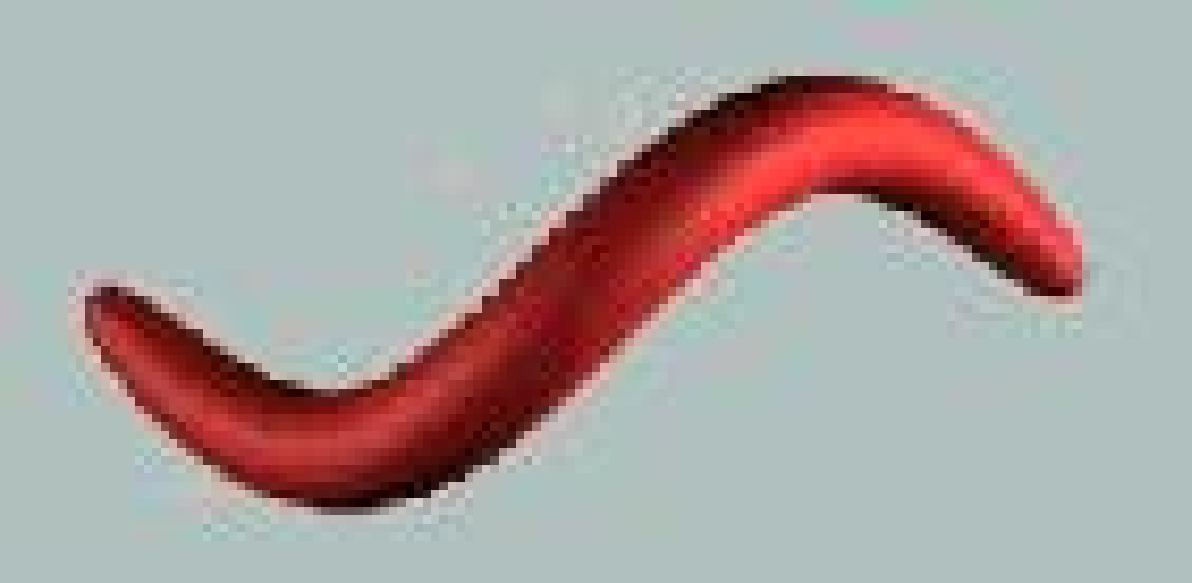,height=1.8cm,width=3.5cm}}
\mbox{\psfig{figure=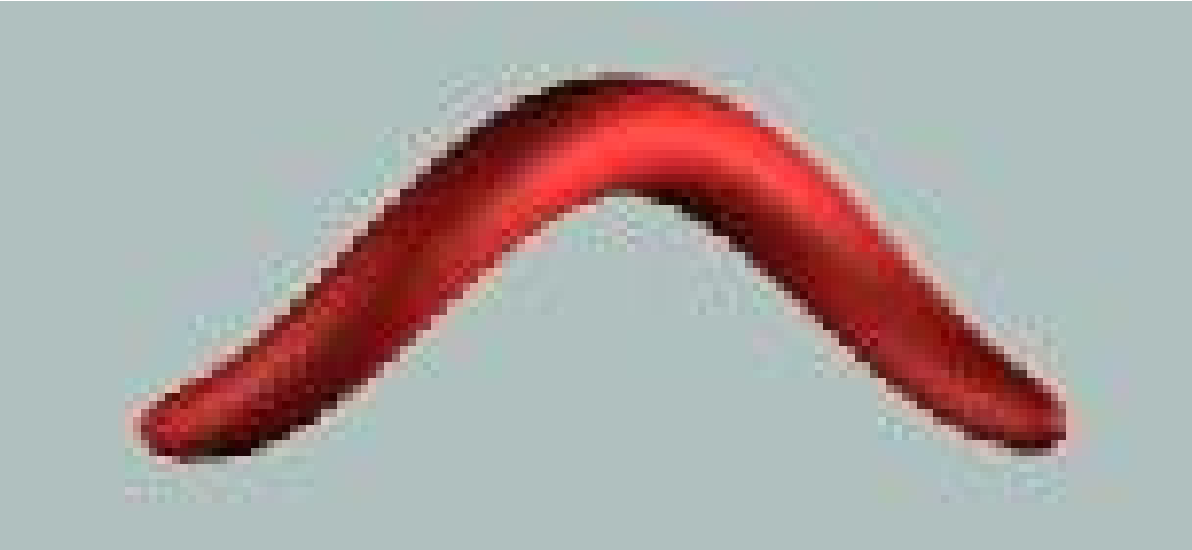,height=1.8cm,width=4.2cm}}
\mbox{\psfig{figure=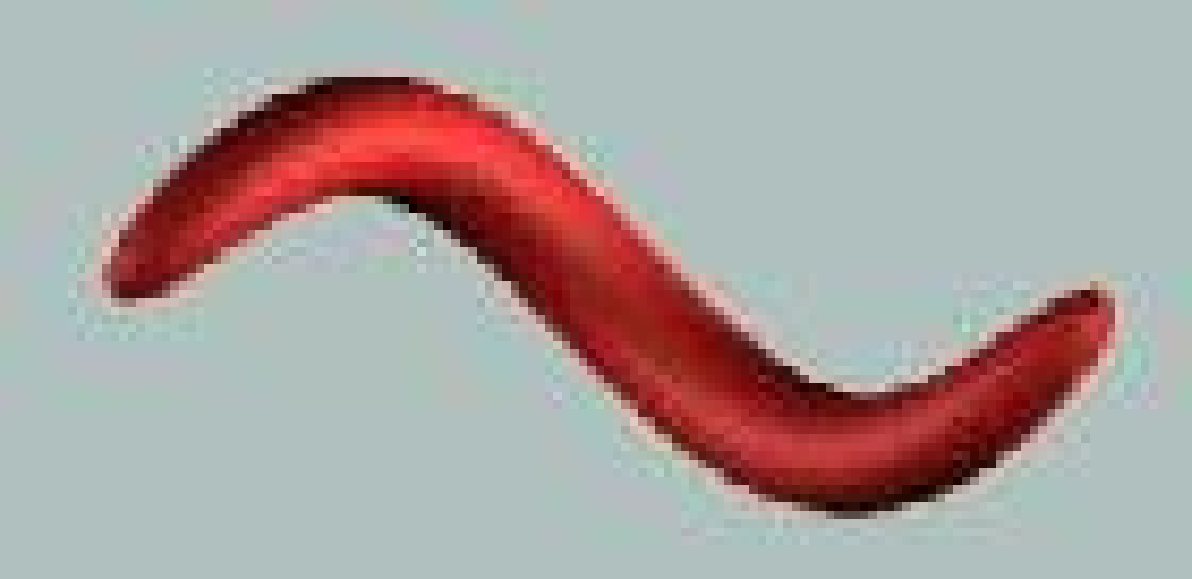,height=1.8cm,width=4.2cm}}}}
\caption{Theoretical example with $\kappa=1$. The Poynting vector is
directed left-to-right.}
\end{figure}
\end{center}
In the case $\kappa=-1$ the magnetic vector is always directed to the
rotation axis, i.e. it is normal to the rotation, and the electric vector
is always tangent to the rotation, so, looking from behind (i.e. along the
Poynting vector) we find clock-wise rotation. In the case $\kappa=1$ the
two vectors exchange their roles and, looking from behind again, we find
anti-clock-wise rotation. From structural point of view the case $\kappa=-1$
is obtained from the case $\kappa=1$ through rotating the couple
$(\mathbf{E},\mathbf{B})$ anti-clock-wise to the angle of $\pi/2$, hence we
get the (dual) transformation
$(\mathbf{E},\mathbf{B})\rightarrow(\mathbf{-B},\mathbf{E})$.
The dynamical roles of the two vectors are exchanged: now the electric vector
drags the points of the object towards rotation axis, the magnetic vector
generates rotation. In both cases the Poynting vector "pushes" the object
along the rotation axis.

The above pictures suggest the interpretation  that the rotation axis
directed vectors keep the object from falling apart.

If we project the object on the plane orthogonal to the Poynting vector we
shall obtain a sector between two circumferences with the same center, and this
sector has nontrivial topology. One of the two vectors is always directed to
the center of the circumferences and this stabalizes the solution, and the
other is tangent to the circumferences and correspondingly oriented.  The stability
of the construction is in accordance also with the fact that, when $z$ runs
from zero to $2\pi\mathcal{L}_o$ any of the two vectors performs {\it just one}
full rotation.  Any other rotaional evolution would make the center-directed
vector leave its directional behavior and this would bring to structural
changes and, most probably, to falling apart of the structure.  Choosing
orientation and computing the corresponding {\it rotation numbers} of
$\mathbf{E}$ and $\mathbf{B}$ we shall obtain, depending on the orientation
chosen, $(+1)$, or $(-1)$, and in definite sense these values guarantee from
mathematical viewpoint the dynamical stability of the solution-object.

\part{Photon-like Objects}
\chapter{Physical Notion}
\section{The Notion of Photon-like Object(s)}

\subsection{Introduction} At the very dawn of the 20th century Planck
proposed [1] and a little bit later Einstein appropriately
used [2] the well known and widely used through the whole last century simple
formula $E=h\nu$, $h=const>0$.  This formula marked the beginning of a new era
and became a real symbol of the physical science during the following years.
According to the Einstein's interpretation it gives the full energy $E$ of {\it
really existing} light quanta of frequency $\nu=const$, and in this way a new
understanding of the nature of the electromagnetic field made first steps:
{\it the field consists of individualized recognizable entities}, which
does not go along with the description given by Maxwell vacuum equations.

After De Broglie's suggestion [3] for the
particle-wave nature of the electron obeying the same energy-frequency
relation, one could read Planck's formula in the following way:
\begin{center}
\hfill\fbox{
    \begin{minipage}{0.97\textwidth}
\begin{center}
\vskip 0.3cm
{\bf There are
physical objects in Nature the very existence of which is strongly connected to
some periodic (with time period $T=1/\nu =const$) process of intrinsic for the
object nature and such that the Lorentz invariant product $ET$ is equal to
the Planck constant $h$}.
\end{center}
\vskip 0.3cm
\end{minipage}} \hfill \end{center}
%\vskip 0.3 cm
Such a reading should suggest that these objects {\bf do NOT admit
point-like approximation since the relativity principle for free particles
requires straight-line uniform motion}, hence, periodicity of any kind should
not be allowed.

Although the great (from pragmatic point of view) achievements of the developed
theoretical approach, known as {\it quantum theory}, the great challenge to
build an adequate description of individual representatives of these objects,
especially of light quanta called by Lewis {\it photons} [4], is still to be
appropriately met since  the efforts made in this direction, we have to admit,
still have not brought satisfactory results. Recall that Einstein in his late
years recognizes that
[5] "{\it All these fifty years of pondering have
not brought me any closer to answering the question What are light quanta}",
and now, more than half a century later, theoretical physics still
needs progress to present a satisfactory answer to the question "{\it what are
photons}". We consider the corresponding theoretically directed efforts as
necessary and even {\it urgent} in view of the growing amount of definite
experimental needs in manipulation with individual photons, for example, in
connection with the experimental advancement in the "quantum computer" project.

The dominating modern theoretical view on microobjects is based on the notions
and concepts of quantum field theory (QFT) where the structure of the photon
(as well as of any other microobject) is accounted for mainly through the so
called {\it structural function}, and highly expensive and delicate collision
experiments are planned and carried out namely in the frame of these concepts
and methods. Going not in details we just note a special feature of this QFT
approach: if the study of a microobject leads to conclusion that it has
structure, i.e., it is not point-like, then the corresponding constituents of
this structure are considered as point-like, so the point-likeness stays in the
theory just in a lower level.

According to our view on photon-like object(s) (PhLO) we follow here, an
approach based on the assumption that the description of the available (most
probably NOT arbitrary) spatial structure of photon-like objects can be made by
{\it continuous finite/localized} functions of the three space variables.  The
difficulties met in this approach consist mainly, in our view, in finding
adequate enough mathematical objects and solving appropriate PDE.  The lack of
sufficiently reliable corresponding information made us look into the problem
from as general as possible point of view on the basis of those properties of
photon-like objects which may be considered as most undoubtedly trustful, and
in some sense, {\it identifying}. The analysis made suggested that such a
property seems to be {\it the available and
intrinsically compatible translational-rotational dynamical structure}, so we
shall focus on this property in order to see what useful for our purpose
suggestions could be deduced and what appropriate structures could be
constructed. All these suggestions and structures should be the building
material for a step-by-step creation of a {\it self-consistent} system. From
physical point of view this should mean that {\it the corresponding properties
may combine to realize a dynamical harmony in the inter-existence of
appropriately defined time-recognizable subsystems of a finite and time stable
physical system}. \vskip 1cm

\subsection{The notion of photon-like object} \index{photon-like object}
 We begin with recalling our view
that any notion of a physical object must unify two kinds of properties of the
object considered: {\it identifying} and {\it kinematical}. The identifying
properties, being represented by quantities and relations, stay unchanged
throughout the existence, i.e., throughout the time-evolution, of the object,
they represent all the intrinsic structure and relations.  The kinematical
properties describe those changes, called {\it admissible}, which do NOT lead
to destruction of the object, i.e., to the destruction of any of the identifying
properties. Correspondingly, physics introduces two kinds of quantities and
relations: identifying and kinematical. From theoretical point of view the more
important quantities used turn out to be the {\it dynamical} quantities which,
as a rule, are functions of the identifying and kinematical ones, and the joint
relations they satisfy represent the necessary interelations between them in
order this object to survive under external influence. This view suggests to
introduce the following notion of Photon-like object(s)(PhLO):

\begin{center}
\hfill\fbox{
    \begin{minipage}{0.97\textwidth}
\begin{center}
\vskip 0.3cm
{\bf PhLO are real massless time-stable physical objects with an
intrinsically compatible and time-recognizable translational-rotational
dynamical structure.}
\end{center}
\vskip 0.3cm \end{minipage}} \hfill
\end{center}

We would like to emphasize that {\it this is a notion} and {\it not a
definition}.
\vskip 0.3cm
\noindent
We give now some explanatory comments concerning this notion.

\subsection{Reality}
We begin with the term {\bf real}.

{\bf
First} we emphasize that this term means that we consider PhLO as {\it really}
existing {\it physical} objects, not as appropriate and helpful but imaginary
(theoretical) entities.  Accordingly, PhLO {\bf necessarily carry
energy-momentum}, otherwise, they could hardly be detected by physical
means/devices, since every such physical detection requires energy-momentum
exchange.

{\bf Second}, PhLO can undoubtedly be {\it created} and {\it destroyed}, so, no
point-like and infinite models are reasonable: point-like objects are assumed
to have no structure, so they can not be destroyed since there is no available
structure to be destroyed; creation of spatially infinite physical objects
(e.g. plane waves) requires infinite time periods, and, most probably, infinite
quantity of energy to be transformed from one kind to another during finite
time-periods, which seems also unreasonable. Accordingly, PhLO are {\it
spatially finite} and have to be modeled like such ones, which is the only
possibility to be consistent with their "created-destroyed" nature. It seems
hardly reasonable to believe that PhLO can not be created and destroyed, and
that spatially infinite and indestructible physical objects may exist at all.

{\bf Third}, "spatially finite" implies that PhLO may carry only {\it
finite values} of physical (conservative or non-conservative) quantities.  In
particular, the most universal physical quantity seems to be the
energy-momentum, so the model must allow finite integral values of
energy-momentum to be carried by the corresponding solutions.

{\bf Fourth},
"spatially finite" means also that PhLO do not
"move" like classical particles along trajectories, PhLO {\it
propagate}, therefore, partial differential equations should be used to describe
their evolution in space-time.

\subsection{Masslessness}
The term "{\bf massless}" characterizes physically the way of propagation in
terms of appropriate dynamical quantities: the {\it integral} 4-momentum $P$ of
a PhLO should satisfy the relation $P_\mu P^\mu=0$, meaning that its integral
energy-momen\-tum vector {\it must be isotropic}, i.e., to have zero module with
respect to Minkowski (pseudo)metric $\eta$ in $\mathbb{R}^4$. The object
considered has appropriate spatial and time-stable structure, such that the
translational velocity of every point where the corresponding field functions
are different from zero is equal to $c$ and does not change its direction from
point to point, i.e., the 2-planes defined by the couple
$(\mathbf{E},\mathbf{B})$ do not itersect. Thus, we have in fact null geodesic
direction in the space-time {\it intrinsically determined} by a PhLO. Such a
direction is formally defined by a null vector field
$\bar{\zeta},\bar{\zeta}^2=0$. The integral trajectories of this vector field
are isotropic (or null) {\it straight lines} as is traditionally assumed in
physics, except in presence of special kind of interaction NOT leading to
destruction. It follows that with every PhLO a null straight line direction is
{\it necessarily} associated, so, canonical coordinates
$(x^1,x^2,x^3,x^4)=(x,y,z,\xi=ct)$ on $(\mathbb{R}^4,\eta)$ may be chosen such
that in the corresponding coordinate frame $\bar{\zeta}$ to have only two
non-zero components of magnitude $1$: $\bar{\zeta}^\mu=(0,0,-\varepsilon, 1)$,
where $\varepsilon=\pm 1$ accounts for the two directions along the coordinate
$z$ (we recall that such a coordinate system we call $\bar{\zeta}$-adapted,
or $\zeta$-adapted).

We'd like to emphasize that our PhLO propagates {\it as a whole} along the
$\bar{\zeta}$-direction, so the corresponding energy-momentum tensor field
$T_{\mu\nu}(x,y,z,\xi)$ of the model must satisfy the corresponding {\it local
isotropy (null) condition}, namely, $T_{\mu\nu}T^{\mu\nu}=0$ (summation over
the repeated indices is throughout used), and this null conditin must be
compatible with its time-recognizable dynamical structure.

\subsection{Translational-Rotational Compatability}
The term "{\bf translational-rotational}" means that besides translational
component along $\bar{\zeta}$, the PhLO propagation necessarily demonstrates
some rotational (in the general sense of this concept) component in such a way
that {\it both components are compatible and exist simultaneously}, and this is
an {\it intrinsic} property. It seems reasonable to expect that such kind of
dynamical behavior should require some distinguished spatial shape and
structure. Moreover, if the Planck relation $E=h\nu$ must be respected
throughout the evolution, the rotational component of propagation should have
{\it time-periodical} nature with time period $T=\nu^{-1}=h/E=const$, and one
of the two possible, {\it left} or {\it right}, orientations. It seems
reasonable also to expect appropriate spatial structure of PhLO, which somehow
to be related to the time periodicity.

\subsection{Dynamical Structure. Systems and Subsystems.}
The term "{\bf dynamical structure}" means that the supposed propagational kind
of existence of PhLO is necessarily accompanied by an {\it
internal energy-momentum redistribution}, which may be considered in the model
as energy-momentum exchange between (or among) some appropriately defined
time-recognizable subsystems.  It could also mean that PhLO live in a dynamical
harmony with the outside world, i.e. {\it any outside directed energy-momentum
flow should be accompanied by a parallel inside directed energy-momentum flow}.

\subsection{Spin structure}
Note that the time periodicity and the possible spatial periodicity
could be consistent with each other somehow, determining in this way
corresponding integral time-stability and constant spin structure through the
integral above mentioned Planck relation $ET=h$, only if PhLO has appropriate
local dynamical structure. From integral viewpoint the simplest integral
feature of such compatability would seem like this: the spatial size $\lambda$
along the translational component of propagation is equal to $cT$:
$\lambda=cT$, where $\lambda$ is some finite positive characteristic constant
of the corresponding solution. From local viewpoint, however, this would
require appropriate local time-stable dynamical structure, such that
the corresponding intrinsically determined translational-rotational time-satble
compatability to permanently guarantee available and adequate to our
experimental knowledge spin structure.

%\newpage
All this would mean that every individual nonperturbed PhLO
\vskip 0.2cm
	-carries the same elementary action equal to the Planck constant $h$,

	-determines its own length/time scale,

	-keeps its massless nature through propagating translationally with
quantitatively and directionally constant velocity,

	-keeps appropriate local time-stable dynamical structure,

	-exists in a permanent equilibrium with the enironment,

	-under interaction with classical mass objects no reflection of the
same PhLO should be expected in general, although another PhLO could be created
as a result of the interaction.

\vskip 0.4cm
It is important to note now the following. As far as we know, today's
theoretical physics has not come to a well motivated and sufficiently trustful
enough oppinion about which mathematical object is most appropriate for
describing individual PhLO. The next Section is devoted namely to find
mathematical structures that may be considered as adequate enough initial
steps to the above introduced notion for PhLO and carrying rich enough
flexability to meet all requirements for a field theory of spatially finite and
massless time-stable physical objects with time-recognizable dynamical
structure. We strongly hope that the ideas and concepts connected with the
Frobenius integrability theory seem to represent adequate enough part of
mathematics for this purpose.

\vskip 1cm
 {\bf References}
\vskip 0.3cm
[1] {\bf Planck, M.}, {\it Ann. d. Phys.}, {\bf 4}, 553 (1901)

[2] {\bf Einstein, A.}, {\it Ann. d. Phys.}, {\bf 17}, 132 (1905),

[3] {\bf De Broglie, L.},  {\it Ondes et quanta}, C. R. {\bf 177}, 507 (1923)

[4] {\bf Lewis, G. N.}, {\it Nature}, {\bf 118}, 874 (1926)

[5] {\bf Einstein, A.}, see Abraham Pais, {\it Subtle is the Lord. The Science
and Life of Albert Einstein}, Oxford University Press 1982 (Reissued 2005 with
a Forward by R. Penrose), p.382.

\chapter{Frobenius Curvature and Internal Dynamics}
\section{Curvature of Distributions and \\ Physical Interaction}
\subsection{The general idea for geometrization of local \\physical
interaction} We begin with a short motivation for this choice of mathematics
directed to the readers already acquainted with Frobenius integrability theory,
and right after this we shall carefully introduce the necessary mathematics.

Any physical system with a dynamical structure is characterized by some
internal energy-momentum redistributions among its subsystems, i.e., internal
energy-momentum fluxes, during evolution. Any time-stable compatible system of
recognizable energy-momentum fluxes (as well as fluxes of other interesting for
the case physical quantities subject to change during evolution, but we limit
ourselves just to energy-momentum fluxes here) can be considered mathematically
as a compatible system of vector fields, defining a linear space. Hence, a {\it
physically isolated} and {\it interelated time-stable} system of
energy-momentum fluxes can be considered to correspond directly or indirectly
to a linear space, defining a {\it completely integrable distribution} $\Delta$
of vector fields (or differential system) according to the principle: {\it some
local objects can generate integral object}. The corresponding distribution
must contain at least one completely integrable space-like subdidstribution,
determining corresponding spatial stress-strain structure, i.e., physical
appearance. Every nonintegrable distribution on a manifold defines, as we know,
its own curvature form, so, the nonintegrable subdistributions of $\Delta$ may
"communicate" through their curvature forms. These "communications" define the
{\it internal} dynamics of the physical system considered.

 Let $\Delta_1$ and $\Delta_2$ be two nonintegrable distributions on
the same manifold with corresponding curvature forms $\Omega_1$ and $\Omega_2$.
Each of them carries couples of vector fields inside their distributions
outside $\Delta_1$ and $\Delta_2$ correspondingly, i.e. $\Omega_1(Y_1,Y_2)\neq
0$ is out of $\Delta_1$ and $\Omega_2(Z_1,Z_2)\neq 0$ is out of $\Delta_2$,
where $(Y_1,Y_2)$ live in $\Delta_1$ and $(Z_1,Z_2)$ live in $\Delta_2$. Let
now $\Delta_1$ and $\Delta_2$ characterize two locally interacting physical
systems, or two locally interacting subsystems of a larger physical system. It
seems reasonable to assume as a working tool the following geometrization of
the concept of local physical interaction:
\vskip 0.3cm
{\it Two nonintegrable distributions
$\Delta_1$ and $\Delta_2$ on a manifold will be said to interact
infinitesimally (or locally) if some of the nonzero values of the corresponding
two curvature forms $\Omega_1$/$\Omega_2$ live respectively in
$\Delta_2$/$\Delta_1$}.
\vskip 0.3 cm
The above geometric concept of {\it infinitesimal interaction} is motivated by
the fact that, in general, an integrable distribution $\Delta$ may contain
various {\it nonintegrable} subdistributions $\Delta_1, \Delta_2, \dots$, which
subdistributions may be associated physically with interacting subsystems of a
larger time stable physical system. Any physical interaction between 2
subsystems is necessarily accompanied with available energy-momentum exchange
between them, this could be understood mathematically as nonintegrability of
each of the two subdistributions of $\Delta$ and could be naturally measured
directly or indirectly by the corresponding curvatures. For example, if
$\Delta$ is an integrable 3-dimensional distribution represented by the vector
fields $(X_1,X_2,X_3)$ then we may have, in general, three non-integrable, i.e.
geometrically interacting, 2-dimensional subdistributions $(X_1,X_2),
(X_1,X_3), (X_2,X_3)$. Finally, some interaction with the outside world can be
described by curvatures of distributions (and their subdistributions) in which
elements of $\Delta$ and vector fields outside $\Delta$ are involved (such
processes will not be considered in this book).

The above considerations launch the general idea to {\it consider the
concept of Frobenius curvature as a natural and universal mathematical
tool for describing local physical interaction between/among the relatively
stable subsystems of the physical world}. In other words, the {\bf Frobenius
curvature appears as appropriate mathematical tool describing formally the
possible ability two continuous systems to recognize each other as physically
interacting partners}.

Two formal aspects of the above idea will be considered. The first applies
directly the {\bf Frobenius integrability machinery}, while the second one
(been developed recently) is known as {\bf nonlinear connections}. \vskip 0.3cm
We proceed now with the first one.

\subsection{Frobeniuss integrability, curvature and \\local physical
interaction} We recall some facts from Sec.3.2 and Sec.3.3. A
$p$-dimensional distribution $\Delta_{p}$ on a $n$-dimensional manifold $M^n$
is defined by associating to each point $x\in M^n$ a $p$-dimensional subspace
of the tangent space at this point: $\Delta^{p}_{x}\subset T_xM^n, x\in M^n,
1\leq p<n.$ Let the system of vector fields $\left\{X_1,X_2,\dots, X_p\right\}$
represent this distribution, so $\left\{X_1(x),X_2(x),\dots, X_p(x)\right\}$,
$x\in M^n$, $1\leq p<n$, satisfy $X_1(x)\wedge X_2(x)\wedge \dots ,\wedge\,
X_p(x)\neq 0, \,x\in M^n$, and represent a basis of $\Delta^{p}_{x}$. According
to the Frobenius integrability theorem (further all manifolds are assumed
smooth and finite dimensional and all objects defined on $M^n$ are also assumed
smooth) $\Delta_p$ is completely integrable, i.e., through every point $x\in
M^n$ passes a $p$-dimensional submanifold $N^p$ such that all elements of
$\Delta_p$ are tangent to $N^p$, iff all Lie brackets $\left[X_i,X_j\right], \
i,j=1,2,\dots, p$, are representable lineary through the very $X_i,
i=1,2,\dots, p: \left[X_i,X_j\right]=C^k_{ij}X_k$, where $C^k_{ij}$ are
functions. Clearly, an easy way to find out if a distribution is completely
integrable is to check if the exterior products
$$
[X_i,X_j](x)\wedge X_1(x)\wedge
X_2(x)\wedge \dots ,\wedge\, X_p(x), \,x\in M^n;\ \ \ i,j=1,2,\dots,p
$$
are identically zero. If this is not the case (which means
that at least one such Lie bracket "sticks out" of the distribution $\Delta_p$)
then the corresponding coefficients, which are multilinear combinations of the
components of the vector fields and their derivatives, represent the
corresponding curvatures. We note finally that if two subdistributions contain
at least one common vector field it seems naturally to expect interaction.

In the dual formulation of Frobenius theorem in terms of differential 1-forms
(i.e. Pfaff forms), having the distribution $\Delta_p$ , we
look for $(n-p)$-Pfaff forms $(\alpha^{p+1}, \alpha^{p+2}, \dots,
\alpha^{n}$), i.e. a $(n-p)$-codistribution $\Delta^*_{n-p}$ , such that
$\langle\alpha^m,X_j\rangle=0,\ \ \text{and} \ \
\alpha^{p+1}(x)\wedge\alpha^{p+2}(x)\wedge\dots \wedge\alpha^{n}(x)\neq 0, $
$m=p+1,p+2,\dots,n, \ \ j=1,2,\dots,p ; x\in M^n. $ Then the integrability of
the distribution $\Delta_p$ is equivalent to the requirements
$$
\mathbf{d}\alpha^m\wedge\alpha^{p+1}\wedge\alpha^{p+2}
\wedge\dots\wedge\alpha^{n} =0,\ \ \ m=p+1,p+2,\dots, n,
$$
where $\mathbf{d}$ is the exterior derivative.

Since the idea of curvature associated with, for example, an arbitrary
2-dimensional
distribution $(X,Y)$ is to find out if the Lie bracket $[X,Y](x)$ has components
along vectors outside the 2-plane defined by $(X_x,Y_x)$, in our case
 we have to evaluate
the quantities $\langle\alpha^m,[X,Y]\rangle$, where all lineary independent
1-forms $\alpha^m$ annihilate
$(X,Y):\langle\alpha^m,X\rangle=\langle\alpha^m,Y\rangle=0$. In view of the
formula
$$ \mathbf{d}\alpha^m(X,Y)=X(\langle\alpha^m,Y\rangle)-
Y(\langle\alpha^m,X\rangle) -\langle\alpha^m,[X,Y]\rangle=
-\langle\alpha^m,[X,Y]\rangle
$$
we may introduce explicitly the curvature 2-form for the distribution
$\Delta(X)=(X_1,\dots,X_p)$. In
fact, if $\Delta(Y)=(Y_{p+1},\dots,Y_{n})$ define a distribution which is
complimentary (in the sense of direct sum) to $\Delta(X)$ and
$\langle\alpha^m,X_i\rangle=0$,
$\langle\alpha^m,Y_l\rangle=\delta^m_l, l=p+1,...,n$,
i.e., $(Y_{p+1},\dots,Y_{n})$ and
$(\alpha^{p+1}, \dots, \alpha^{n})$ are dual bases, then the
corresponding curvature 2-form $\Omega_{\Delta(X)}$ should be defined by
$$
\Omega_{\Delta(X)}=-\mathbf{d}\alpha^m\otimes Y_m,
$$
since
$$
\Omega_{\Delta(X)}(X_i,X_j)=-\mathbf{d}\alpha^m(X_i,X_j) Y_m=
\langle\alpha^m,[X_i,X_j]\rangle Y_m ,
$$
where it is meant here that
$\Omega_{\Delta(X)}$ is restricted to the distribution $(X_1,\dots,X_p)$.

Hence, if we call the distribution $(X_1,\dots,X_p)$ {\it horizontal} and the
complimentary distribution $(Y_{p+1},\dots,Y_{n})$ {\it vertical}, then the
corresponding curvature 2-form acquires the status of {\it vertical bundle
valued 2-form}.

We see that the curvature 2-form distinguishes those couples of
vector fields inside $\Delta(X)$ the Lie brackets of which define outside
$\Delta(X)$ directed flows, and so, not allowing to find integral manifold
of $\Delta(X)$.

Clearly, the supposition here for dimensional complementarity
of the two distributions $\Delta(X)$ and $\Delta(Y)$ is not essential for the
idea of geometrical interaction, i.e., the distribution $\Delta(Y)\neq\Delta(X)$
may be any other distribution on the same manifold with dimension less than
$(n-p)$, so that $m=1,2,\dots,q<(n-p)$ in general, the important moment is that
the two distributions (or subdistributions) can "communicate" {\it
differentially} through their curvature 2-forms.

Hence, from physical point of view, if the quantities
$\Omega_{\Delta(X)}(X_i,X_j)$ are meant to be used for building the components
of the energy-momentum locally transferred from the system $\Delta(X)$ to the
system $\Delta(Y)$, naturally, we have to make use of the quantities
$\Omega_{\Delta(Y)}(Y_m,Y_l)$ to build the components of the energy-momentum
transferred from $\Delta(Y)$ to $\Delta(X)$.

It deserves to note that this formalism allows a {\it dynamical
equilibrium} between the two systems $\Delta(Y)$ and
$\Delta(X)$ to be described: each system to gain from the other as much
energy-momentum as it loses, and this to take place at every space-time point.
Therefore, if $W_{(X,Y)}$ denotes the energy-momentum transferred locally from
$\Delta(X)$ to $\Delta(Y)$, $W_{(Y,X)}$ denotes the energy-momentum transferred
locally from $\Delta(Y)$ to $\Delta(X)$, and $\delta W_{(X)}$ and $\delta
W_{(Y)}$ denote respectively the local energy-momentum changes of the two
systems $\Delta(X)$ and $\Delta(Y)$, then according to the local
energy-momentum conservation law we can write
$$ \delta
W_{(X)}=W_{(Y,X)}+W_{(X,Y)}, \ \ \delta W_{(Y)}=-(W_{(X,Y)}+W_{(Y,X)})=-\delta
W_{(X)} ,
$$
i.e. $\Delta(X)$ and $\Delta(Y)$ are {\it physically compatible},
or {\it able to interact}, therefore, we may call them {\it interacting
partners}.

For the case of dynamical equilibrium we have
$W_{(X,Y)}=-W_{(Y,X)}$, so in such a case we obtain
$$
 \delta W_{(X)}=0,\ \ \ \delta W_{(Y)}=0,\ \ \ W_{(Y,X)}+W_{(X,Y)}=0.
$$
As for how to build explicitly the corresponding representatives of the
energy-momentum fluxes, probably, universal procedure can not be offered. The
simplest procedure seems to "project" the curvature values
$\Omega_{\Delta(X)}(X_i,X_j)$ and $\Omega_{\Delta(Y)}(Y_m,Y_l)$ on the
corresponding co-distribution volume forms, i.e. to consider the corresponding
interior products
$i(\Omega(X_i,X_j))(\alpha^{p+1}\wedge\alpha^{p+2}\wedge\dots\wedge\alpha^{n})$
and
$i(\Omega(Y_m,Y_l))(\alpha^{1}\wedge\alpha^{2}\wedge\dots\wedge\alpha^{p})$
(Sec.3.2.3), which we implemented in the $\varphi$-extended Lie derivative.

\section{PhLO Dynamical Structure in Terms of Frobenius Curvature}
We consider the Minkowski space-time $M=(\mathbb{R}^4,\eta)$ with
signature $sign(\eta)=(-,-,-,+)$ related to the standard global
coordinates $(x^1,x^2,x^3,x^4)=\linebreak(x,y,z,\xi=ct)$, the natural volume
form $\omega_o=\sqrt{|\eta|}dx^1\wedge dx^2\wedge dx^3\wedge dx^4=
dx\wedge dy\wedge dz\wedge d\xi$, and the
Hodge star $*$ defined by $\alpha\wedge *\beta=-\eta(\alpha,\beta)\omega_o$.

In view of our concept of PhLO which requires the couple
$(\mathbf{E},\mathbf{B})$ to define nonintersecting space-like 2-planes,
further identified as the $(x,y)$-planes, we introduce the null vector field
$\bar{\zeta},\ \bar{\zeta}^2=0$, which must define the translational space-time
propagation. In the $\bar{\zeta}$-adapted coordinates (throughout used further)
$\bar{\zeta}$ shall look as follows: $$
\bar{\zeta}=-\varepsilon\frac{\partial}{\partial z} + %1%
\frac{\partial}{\partial \xi}, \ \ \varepsilon=\pm 1. $$ Let's denote the
corresponding to $\bar{\zeta}$ completely integrable 3-dimensional Pfaff system
by $\Delta^*(\bar{\zeta})$. Thus, $\Delta^*(\bar{\zeta})$ can be generated by
any three lineary independent 1-forms $(\alpha_1,\alpha_2,\alpha_3)$ which
annihilate $\bar{\zeta}$, i.e. $$
\alpha_1(\bar{\zeta})=\alpha_2(\bar{\zeta})=\alpha_3(\bar{\zeta})=0; \ \
\alpha_1\wedge \alpha_2\wedge \alpha_3\neq 0.
$$
Instead of $(\alpha_1,\alpha_2,\alpha_3)$ we introduce the notation
$(A, A^*, \zeta)$ and define $\zeta$ to be the $\eta$-corresponding 1-form to
$\bar{\zeta}$:
$$
\zeta=\varepsilon dz+d\xi, \ \ \text{so},
\ \ \langle\zeta,\bar{\zeta}\rangle=0 ,                 %2%
$$
where $\langle\, , \rangle$ is the coupling between forms and vectors.

Now, since $\zeta$ is closed, it defines 1-dimensional completely
integrable Pfaff codistribution, so, we have the corresponding completely
integrable distribution $(\bar{A},\bar{A^{*}},\bar{\zeta}):
\langle\zeta,\bar{A}\rangle=\langle\zeta,\bar{A^{*}}\rangle=0$.
We shall restrict our further study to
PhLO of electromagnetic nature according to the following
\vskip 0.2cm
{\bf{Definition}}: We shall call a PhLO {\it electromagnetic} if the following
conditions hold:

1. the vector fields $(\bar{A},\bar{A^{*}})$ have no components along
$\bar{\zeta}$,

2. $(\bar{A},\bar{A^{*}})$ are $\eta$-corresponding to $(A, A^*)$
respectively .

3. $\langle A,\bar{A^{*}}\rangle=0,\ \
\langle A,\bar{A}\rangle =\langle A^{*},\bar{A^{*}}\rangle$ .
\vskip 0.2cm
\noindent
{\bf Remark}. These relations formalize knowledge from Classical
electrodynamics (CED). In fact, our vector fields $(\bar{A},\bar{A^{*}})$ are
meant to represent what we call in CED {\it electric and magnetic
components} of a free time-dependent electromagnetic field, where, as we have
mentioned several times, the translational propagation of
the field energy-momentum along a fixed null direction with the velocity "$c$"
is possible only if the two invariants $I_1=\mathbf{B}^2-\mathbf{E}^2$ and
$I_2=2\mathbf{E}.\mathbf{B}$ are zero, because only in such a case the
electromagnetic energy-momentum tensor $T_{\mu\nu}$ satisfies
$T_{\mu\nu}T^{\mu\nu}=0$ and has {\it unique} null eigen direction. So it seems
naturally to consider this property as {\it intrinsic} for the field and to
choose it as a starting point. Moreover, in such a case the relation
$T_{\mu\nu}T^{\mu\nu}=(I_1)^2+(I_2)^2=0$ is equivalent to
$\mathbf{E}^2+\mathbf{B}^2=2|\mathbf{E}\times\mathbf{B}|$ and this relation
shows that this is the only case when the field momentum can
not be made equal to zero by means of frame change. Together with the fact
that the spatial direction of translational energy-momentum propagation is
determined by $\mathbf{E}\times\mathbf{B}$, this motivates to introduce the
vector field $\bar{\zeta}$ in this form and to assume the properties 1-3 in the
above definition. \vskip 0.3cm
 From the above conditions it follows that in the
$\bar{\zeta}$-adapted coordinate system we have
$$
A=u\,dx + p\,dy, \ \
A^*=-\varepsilon\,p\,dx + \varepsilon\,u\,dy;
$$
$$
\bar{A}=-u\,\frac{\partial}{\partial x} -          %1%
p\,\frac{\partial}{\partial y}, \ \
\bar{A^*}=\varepsilon\,p\,\frac{\partial}{\partial x} -            %1%
\varepsilon\,u\,\frac{\partial}{\partial y},
$$
where $\varepsilon=\pm 1$, and $(u,p)$ are two smooth functions on $M$.

The completely integrable 3-dimensional Pfaff system $(A, A^*, \zeta)$
contains three 2-dimensional subsystems: $(A,A^{*}), (A,\zeta)$ and
$(A^*,\zeta)$. Now, the following relations can be immediately verified:
\begin{eqnarray*}
\mathbf{d}A\wedge A\wedge A^*&=&0;\\
\mathbf{d}A^*\wedge A^*\wedge A&=&0;\\                              %3%
 \mathbf{d}A\wedge A\wedge \zeta&=& \varepsilon\big[u(p_\xi-\varepsilon
p_z)- p\,(u_\xi-\varepsilon u_z)\big]\omega_o;\\
\mathbf{d}A^*\wedge A^*\wedge \zeta&=&                                  %4%
\varepsilon\big[u(p_\xi-\varepsilon p_z)- p\,(u_\xi-\varepsilon
u_z)\big]\omega_o.
\end{eqnarray*}
\vskip 0.3cm
\noindent
These relations say that the 2-dimensional Pfaff system
$(A,A^*)$ is completely integrable for any choice of the two functions
$(u,p)$, while the two 2-dimensional Pfaff systems $(A,\zeta)$ and
$(A^*,\zeta)$ are NOT completely integrable in general, and the same
curvature factor $$ \mathbf{R}=u(p_\xi-\varepsilon
p_z)-p\,(u_\xi-\varepsilon u_z) $$ determines their nonintegrability.

Correspondingly, the 3-dimensional completely
integrable distribution (or differential system)
$\Delta(\bar{A},\bar{A^*},\bar{\zeta})$
contains three 2-dimensional subdistributions:
$(\bar{A},\bar{A^*})$, $(\bar{A},\bar{\zeta})$ and $(\bar{A^*},\bar{\zeta})$.

The following relations can also be easily verified to hold:
(recall that $[X,Y]$ denotes the Lie bracket):
\begin{eqnarray*}
&&\left[\bar{A},\bar{A^*}\right]\wedge\bar{A}\wedge\bar{A^*}=0,\\
&&L_{\bar{\zeta}}\bar{A}=
\left[\bar{\zeta},\bar{A}\,\right]=-(u_\xi-\varepsilon u_z)
\frac{\partial}{\partial x}-(p_\xi-\varepsilon p_z)
\frac{\partial}{\partial y},\\
&&L_{\bar{\zeta}}\bar{A^*}=
\left[\bar{\zeta},\bar{A^*}\,\right]=
\varepsilon(p_\xi-\varepsilon p_z)\frac{\partial}{\partial x}-
\varepsilon(u_\xi-\varepsilon u_z)\frac{\partial}{\partial y}.
\end{eqnarray*}

\vskip 0.3cm

From these last relations it follows that the distribution $(\bar{A},\bar{A^*})$
is completely integrable, and it can be easily shown that the
two distributions $(\bar{A},\bar{\zeta})$
and
$(\bar{A^*},\bar{\zeta})$ would be completely integrable only if the same
curvature factor
$$
\mathbf{R}=u(p_\xi-\varepsilon p_z)-p\,(u_\xi-\varepsilon u_z)     %8%
$$
is zero (the elementary proof is omitted).

As it should be, the two projections
$$
\langle A,[\bar{A^*},\bar{\zeta}]\rangle=
-\langle A^*,[\bar{A},\bar{\zeta}]\rangle=
-\varepsilon u(p_\xi-\varepsilon p_z)+\varepsilon p(u_\xi-\varepsilon u_z)=
-\varepsilon\,\mathbf{R}
$$
are nonzero and give (up to a sign) the same factor $\mathbf{R}$. The same curvature factor
appears, of course, as coefficient in the exterior products
$
[\bar{A^*},\bar{\zeta}]\wedge \bar{A^*}\wedge\bar{\zeta}$ and
$[\bar{A},\bar{\zeta}]\wedge \bar{A}\wedge\bar{\zeta}$.
In fact, we obtain
$$
[\bar{A^*},\bar{\zeta}]\wedge \bar{A^*}\wedge\bar{\zeta}=
-[\bar{A},\bar{\zeta}]\wedge \bar{A}\wedge\bar{\zeta}=
-\varepsilon\mathbf{R}\,\frac{\partial}{\partial x}
\wedge\frac{\partial}{\partial y}\wedge
\frac{\partial}{\partial z}+
\mathbf{R}\,\frac{\partial}{\partial x}\wedge\frac{\partial}{\partial y}\wedge
\frac{\partial}{\partial \xi} .
$$
On the other hand, for the other two projections we obtain
$$
\langle
A,[\bar{A},\bar{\zeta}]\rangle=                          %9%
\langle
A^*,[\bar{A^*},\bar{\zeta}]\rangle= \frac12\big[(u^2+p^2)_\xi-\varepsilon
(u^2+p^2)_z\big].
$$
Clearly, the last relation may be put in
terms of the Lie derivative $L_{\bar{\zeta}}$ as
$$
\frac 12L_{\bar{\zeta}}(u^2+p^2)=
-\frac12L_{\bar{\zeta}}\langle A,\bar{A}\rangle=
-\langle A,L_{\bar{\zeta}}\bar{A}\rangle=
-\langle A^*,L_{\bar{\zeta}}\bar{A^*}\rangle.
$$

{\bf Remark}. Further we shall denote $\sqrt{u^2+p^2}\equiv
\phi$.

We notice now that there is a function $\psi(u,p)$ such, that

$$L_{\bar{\zeta}}\psi=
\frac{u(p_\xi-\varepsilon p_z)-p(u_\xi-\varepsilon u_z)}{\phi^2}=
\frac{\mathbf{R}}{\phi^2} .
$$
\vskip 0.4cm
It is immediately verified that $\psi=\arctan\frac pu$ is such one.
\vskip 0.4cm
We note that the function $\psi$ has a natural
interpretation of {\it phase} because of the easily verified now relations
$u=\phi\cos\psi$, $p=\phi \sin\psi $,
and $\phi$ acquires the status of {\it amplitude}, i.e. energy density. Since
the transformation $(u,p)\rightarrow (\phi,\psi)$ is non-degenerate this allows
to work with the two functions $(\phi,\psi)$ instead of $(u,p)$.

From the above we have
$$
\mathbf{R}=\phi^2L_{\bar{\zeta}}\psi= \
\phi^2(\psi_\xi-\varepsilon\psi_z) \ \ \
\rightarrow \ \ L_{\bar{\zeta}}\psi=
\frac{\mathbf{R}}{T(\partial_{\xi},\partial_{\xi})}=
\frac{*\varepsilon(\mathbf{d}A\wedge A\wedge\zeta)}{T(\partial_{\xi},\partial_{\xi})},
$$
where $T(\partial_{\xi},\partial_{\xi})$ is the coordinate-free definition of
the energy density $\phi^2$.

 This last formula shows something very important: at
any $\phi\neq 0$ the curvature $\mathbf{R}$ will NOT be zero only if
$L_{\bar{\zeta}}\psi\neq 0$, which admits in principle availability of
rotation. In fact, lack of rotation would mean that $\phi$ and $\psi$ are
running waves along $\bar{\zeta}$. The relation $L_{\bar{\zeta}}\psi\neq 0$
means, however, that rotational properties are possible in general, and some of
these properties are carried by the phase $\psi$. It follows that in such a
case the translational component of propagation along $\bar{\zeta}$ (which is
supposed to be available) must be determined essentially, and most probably,
entirely, by $\phi$.  In particular, we could expect the relation
$L_{\bar{\zeta}}\phi=0$ to hold, and if this happens, then the rotational
component of propagation will be represented entirely by the phase $\psi$, and,
more specially, by the curvature factor $\mathbf{R}\neq 0$, so, the objects we
are going to describe may have compatible translational-rotational dynamical
structure. Finally, this relation may be considered as a definition for the
phase function $\psi$.

We are going now to represent some relations, analogical to the energy-momentum
relations in classical electrodynamics, determined by some 2-form $F$, in
terms of the Frobenius curvatures given above.

The two nonintegrable Pfaff systems $(A,\zeta)$ and
$(A^*,\zeta)$ carry two volume 2-forms:
$$
G=A\wedge\zeta \ \
\text{and} \ \ G^*=A^*\wedge\zeta ,
$$
and by the $\eta$-correspondence we define the 2-vectors
$$
\bar{G}=\bar{A}\wedge\bar{\zeta}, \ \ \text{and} \ \ \
\bar{G^*}=\bar{A^*}\wedge\bar{\zeta} .
$$
Making use now of the Hodge $*$-operator,
 we can verify the relation: $G^*=*G$.
Now $G$ and $G^*$ define the (1,1)-tensor, called
stress-energy-momentum tensor $T_{\mu}^{\nu}$, according to the rule
$$
T_{\mu}^{\nu}=-\frac12\big[G_{\mu\sigma}\bar{G}^{\nu\sigma}+
(G^*)_{\mu\sigma}(\bar{G}^*)^{\nu\sigma}\big] ,
$$
and the divergence of this tensor field can be represented in the form
$$
\nabla_\nu T_{\mu}^{\nu}=\big[i(\bar{G})\mathbf{d}G\big]_{\mu}+
\big[i(\bar{G^*})\mathbf{d}G^*\big]_\mu,
$$
where $\bar{G}$ and $\bar{G^*}$ coincide with the metric-corresponding
contravariant tensor fields, and $i(\bar{G})=i(\bar{\zeta})\circ i(\bar{A})$,
$i(\bar{G^*})=i(\bar{\zeta})\circ i(\bar{A^*})$,
$i(X)$ is the standard insertion operator in the exterior
algebra of differential forms on $\mathbb{R}^4$ defined by the vector field
$X$.
So, we shall need the quantities
$$
i(\bar{G})\mathbf{d}G, \ \ i(\bar{G^*})\mathbf{d}G^*,
\ \
i(\bar{G^*})\mathbf{d}G, \ \ i(\bar{G})\mathbf{d}G^* .
$$
Having in view the
explicit expressions for $A,A^*,\zeta,\bar{A},\bar{A^*}$ and $\bar{\zeta}$ we
obtain
$$
i(\bar{G})\mathbf{d}G=i(\bar{G}^*)\mathbf{d}G^*=
\frac12 L_{\bar{\zeta}}\left(\phi^2\right).
\,\zeta \ ,
$$
also we obtain
$$
i(\bar{G^*})\mathbf{d}G=-i(\bar{G})\mathbf{d}G^* =
$$
$$
=\Big[u(p_\xi-\varepsilon p_z)-p(u_\xi-\varepsilon u_z)\Big]dz+
\varepsilon\Big[u(p_\xi-\varepsilon p_z)-p(u_\xi-\varepsilon u_z)\Big]d\xi
=\varepsilon\mathbf{R}\,\zeta .
$$
If $F$ and $H$ are correspondingly 2 and 3 forms on $M$ we recall the relation
$$
*(F\wedge *H)=i(\bar{F})H=F^{\mu\nu}H_{\mu\nu\sigma}dx^\sigma, \mu<\nu.
$$
Therefore, since $G^*=*G$ (Sec.6.3.1),
$$
i(\bar{G^*})\mathbf{d}G=-*(\delta *G\wedge *G), \ \
-i(\bar{G})\mathbf{d}G^*=-*(\delta G\wedge G),
$$
so,
$$
\delta *G\wedge *G=\delta G\wedge G=\varepsilon\mathbf{R}*\zeta .
$$

In the following formulae we must keep in mind the relations
$$
\mathbf{d}\zeta=0, \langle A,\bar{A^*}\rangle=\langle A^*,\bar{A}\rangle=
\langle\zeta,\bar{A^*}\rangle=\langle\zeta,\bar{A}\rangle=0,
$$ and
$$(\bar{A})^2=
(\bar{A}^*)^2=\langle A,\bar{A}\rangle=\langle A^*,\bar{A^*}\rangle=
$$
$$
-(u^2+p^2)=-\Phi^2=-|A|^2=-|A^*|^2=-|\bar{A}|^2=-|\bar{A^*}|^2.
$$
In view of these formulae and the required duality in the definition of the
curvature form, the two distributions $(\bar{A},\bar{\zeta})$ and
$(\bar{A^*},\bar{\zeta})$ determine the following two curvature forms $\Omega$
and $\Omega^*$:
$$
\Omega=-\mathbf{d}A^*\otimes\frac{\bar{A^*}}{|\bar{A^*}|^2}
,\ \ \
\Omega^*=-\mathbf{d}A\otimes\frac{\bar{A}}{|\bar{A}|^2}.
$$
Denoting $Z_{\Omega}\equiv\Omega(\bar{\zeta},\bar{A})$,
$Z^*_{\Omega}\equiv\Omega(\bar{\zeta},\bar{A^*})$,
$Z_{\Omega^*}\equiv\Omega^*(\bar{\zeta},\bar{A})$ and
$Z^*_{\Omega^*}\equiv\Omega^*(\bar{\zeta},\bar{A^*})$
we obtain
$$
Z_{\Omega}=-\frac{\varepsilon\mathbf{R}}{\phi^2}\bar{A^*},\ \
Z^*_{\Omega}=-\frac{\bar{A^*}}{2\phi^2}L_{\bar{\zeta}}(\phi^2), \ \
Z_{\Omega^*}=-\frac{\bar{A}}{2\phi^2}L_{\bar{\zeta}}(\phi^2), \ \
Z^*_{\Omega^*}=\frac{\varepsilon\mathbf{R}}{\phi^2}\bar{A}.
$$
The following relations express the connection between the curvatures and the
energy-momentum characteristics.
\begin{center}
\hfill\fbox
   {\begin{minipage}{0.98\textwidth}
\begin{center}
$$
i(Z_{\Omega})(A\wedge\zeta)=0, \ \
i(Z_{\Omega})(A^*\wedge\zeta)=\varepsilon\mathbf{R}.\zeta=
-i(\bar{G})\mathbf{d}G^*=i(\bar{G^*})\mathbf{d}G,
$$
$$
i(Z_{\Omega^*})(A^*\wedge\zeta)=0, \ \
i(Z^*_{\Omega^*})(A\wedge\zeta)=-\varepsilon\mathbf{R}.\zeta=
i(\bar{G})\mathbf{d}G^*=-i(\bar{G^*})\mathbf{d}G,
$$
$$
i(Z^*_{\Omega})(A\wedge\zeta)=0, \ \
i(Z^*_{\Omega})(A^*\wedge\zeta)=\frac12L_{\bar{\zeta}}(\phi^2).\zeta=
i(\bar{G})\mathbf{d}G=i(\bar{G^*})\mathbf{d}G^{*},
$$
$$
i(Z^*_{\Omega^*})(A^*\wedge\zeta)=0, \ \
i(Z_{\Omega^*})(A\wedge\zeta)=\frac12L_{\bar{\zeta}}(\phi^2).\zeta=
i(\bar{G})\mathbf{d}G=i(\bar{G^*})\mathbf{d}G^{*}.
$$
\end{center}
\hfill
   \end{minipage}}
\end{center}

It follows from these relations that in case of dynamical equilibrium we shall
have
$$
L_{\bar{\zeta}}(\phi^2)=0,\ \
i(\bar{G})\mathbf{d}G=0,\ \ i(\bar{G^*})\mathbf{d}G^*=0, \ \
i(\bar{G^*})\mathbf{d}G+i(\bar{G})\mathbf{d}G^*=0.
$$

Summerizing, we can say that Frobenius integrability viewpoint suggests to
make use of one completely integrable 3-dimensional distribution (resp.
Pfaff system) consisting of one isotropic and two space-like vector fields
(resp. 1-forms), such that the corresponding 2-dimensional spatial
subdistribution $(\bar{A},\bar{A^*})$ (resp. Pfaff system $(A,A^*)$)
defines a completely integrable system, and the rest two 2-dimensional
subdistributions $(\bar{A},\bar{\zeta})$ and $(\bar{A^*},\bar{\zeta})$
(resp. Pfaff systems $(A,\zeta )$ and $(A^*,\zeta )$) are NON-integrable
in general and give the same curvature. This curvature may be used to
build quantities, physically interpreted as energy-momentum
internal exchanges between the corresponding
two subsystems $(\bar{A},\bar{\zeta})$ and $(\bar{A^*},\bar{\zeta})$
(resp.$(A,\zeta)$ and $(A^*,\zeta))$. Moreover, rotational component of
propagation will be available only if the curvature $\mathbf{R}$ is nonzero,
i.e. only if an internal energy-momentum exchange takes place. We see that
all physically important characteristics and relations, describing the
translational and rotational components of propagation, can be expressed in
terms of the corresponding Frobenius curvature. We'll see that this
holds also for some integral characteristics of PhLO.

\section{Photon-like nonlinear connections}
We are going to make use of the concepts and relations from Sec 3.3 in this
subsection, and the usual notations: our manifold is the Minkowski space-time
$M=(\mathbb{R}^4,\eta)$, endowed with standard coordinates
$(x^1,x^2,x^3,x^4=x,y,z,\xi=ct)$. We give some preliminary considerations in
order to make the choice of two projections:
$V,\tilde{V}:TM\rightarrow TM$ consistent with the
introduced concept of PhLO.

The intrinsically defined straight-line
translational component of propagation of the PhLO will be assumed to be
parallel to the coordinate plane $(z,\xi)$. Also, $\frac{\partial}{\partial x}$
and $\frac{\partial}{\partial y}$ will be vertical coordinate fields, so every
vertical vector field $Y$ can be represented by  $Y=u\,\frac{\partial}{\partial
x}+p\,\frac{\partial}{\partial y}$, where $(u,p)$ are two functions on
$M$. It is easy to check that any two such lineary independent
vertical vector fields $Y_1$ and $Y_2$ define an integrable distribution,
hence, the corresponding curvature will be zero. It seems very natural to
choose $Y_1$ and $Y_2$ to coincide correspondingly with the vertical
projections $V(\frac{\partial}{\partial z})$ and
$\tilde{V}(\frac{\partial}{\partial z})$, or with $V(\frac{\partial}{\partial
\xi})$ and $\tilde{V}(\frac{\partial}{\partial\xi})$, since these images are
meant to represent the electric and magnetic components, which have to be
smoothly straight-line translated along the plane $(z,\xi)$ with the velocity
of light. Now we know from classical electrodynamics that the situation
described corresponds to zero invariants of the electromagnetic field,
therefore, we may assume that $Y_1$ and $Y_2$ are orthogonal to each other and
with the same norms with respect to the induced euclidean metric in the
2-dimensional space spent by $\frac{\partial}{\partial x}$ and
$\frac{\partial}{\partial y}$. It follows that the essential components of
$Y_1$ and $Y_2$ should be expressible only with two independent functions
$(u,p)$. The conclusion is that our projections must have the same image space
and should depend only on $(u,p)$. Finally, we note that these assumptions lead
to the horizontal nature of $dz$ and $d\xi$.

Note that if the translational component of propagation is along the vector
field $\bar{\zeta}$ then we can define two new distributions :
$(Y_1,\bar{\zeta})$ and $(Y_2,\bar{\zeta})$, which do not seem to be integrable
in general even if $\bar{\zeta}$ has constant components as it will be in our
case. Since these two distributions are nontrivially intersected (they have a
common member $\bar{\zeta}$), it seems natural to consider them as geometrical
images of two interacting physical subsystems of our PhLO. Hence,
our two projections will have the same image space,
and the components of both projections must depend only on the
two functions $(u,p)$.

Let now $(u,p)$ be two smooth functions on $M$ and
$\varepsilon=\pm 1$ . We introduce two projections, i.e., two nonlinear
connections, $V$ and $\tilde{V}$ in $TM$ as follows:
$$
V=dx\otimes\frac{\partial}{\partial
x}+dy\otimes\frac{\partial}{\partial y}
-\varepsilon\,u\,dz\otimes\frac{\partial}{\partial x}-
u\,d\xi\otimes\frac{\partial}{\partial x}-
\varepsilon\,p\,dz\otimes\frac{\partial}{\partial y}-
p\,d\xi\otimes\frac{\partial}{\partial y},
$$
$$
\tilde{V}=
dx\otimes\frac{\partial}{\partial
x}+dy\otimes\frac{\partial}{\partial y} +p\,dz\otimes\frac{\partial}{\partial
x} +\varepsilon p\,d\xi\otimes\frac{\partial}{\partial x} -
u\,dz\otimes\frac{\partial}{\partial y}- \varepsilon
u\,d\xi\otimes\frac{\partial}{\partial y}.
$$
So, in both cases we consider $(\frac{\partial}{\partial
x},\frac{\partial}{\partial y})$ as vertical vector fields, and $(dz,d\xi)$ as
horizontal 1-forms. By corresponding transpositions we can determine
projections $V^*$ and $\tilde{V}^*$ in $T^*M$.
$$
 V^*=dx\otimes\frac{\partial}{\partial x}+ dy\otimes\frac{\partial}{\partial
y} -\varepsilon\,u\,dx\otimes\frac{\partial}{\partial z}-
u\,dx\otimes\frac{\partial}{\partial \xi}-
\varepsilon\,p\,dy\otimes\frac{\partial}{\partial z}-
p\,dy\otimes\frac{\partial}{\partial \xi},
$$
$$
\tilde{V}^*=
dx\otimes\frac{\partial}{\partial x}+ dy\otimes\frac{\partial}{\partial y}+
p\,dx\otimes\frac{\partial}{\partial z} + \varepsilon
p\,dx\otimes\frac{\partial}{\partial \xi} -
u\,dy\otimes\frac{\partial}{\partial z}- \varepsilon
u\,dy\otimes\frac{\partial}{\partial \xi}.
$$
 The corresponding horizontal
projections, denoted by $(H,\tilde{H};H^*\tilde{H}^*)$ look as follows:
$$
H=dz\otimes\frac{\partial}{\partial z}+d\xi\otimes\frac{\partial}{\partial \xi}
+\varepsilon\,u\,dz\otimes\frac{\partial}{\partial x}+
u\,d\xi\otimes\frac{\partial}{\partial x}+
\varepsilon\,p\,dz\otimes\frac{\partial}{\partial y}+
p\,d\xi\otimes\frac{\partial}{\partial y},
$$
$$
\tilde{H}=dz\otimes\frac{\partial}{\partial
z}+d\xi\otimes\frac{\partial}{\partial \xi}-
p\,dz\otimes\frac{\partial}{\partial x} - \varepsilon
p\,d\xi\otimes\frac{\partial}{\partial x}+ u\,dz\otimes\frac{\partial}{\partial
y}+ \varepsilon u\,d\xi\otimes\frac{\partial}{\partial y},
$$
$$
H^*=dz\otimes\frac{\partial}{\partial z}+ d\xi\otimes\frac{\partial}{\partial
\xi} +\varepsilon\,u\,dx\otimes\frac{\partial}{\partial z}+
u\,dx\otimes\frac{\partial}{\partial \xi}+ \varepsilon
p\,dy\otimes\frac{\partial}{\partial z}+ p\,dy\otimes\frac{\partial}{\partial
\xi},
$$
$$
\tilde{H}^*=dz\otimes\frac{\partial}{\partial z}+
d\xi\otimes\frac{\partial}{\partial \xi}- p\,dx\otimes\frac{\partial}{\partial
z} - \varepsilon p\,dx\otimes\frac{\partial}{\partial \xi}+
u\,dy\otimes\frac{\partial}{\partial z}+ \varepsilon
u\,dy\otimes\frac{\partial}{\partial \xi}.
$$

The corresponding matrices look like:
$$ V=
\begin{Vmatrix}1 & 0 & -\varepsilon\,u & -u \\
0 & 1 & -\varepsilon\,p  & -p \\
0 & 0 & 0 & 0 \\
0 & 0 & 0 & 0 \end{Vmatrix} ,
\ \ H=
\begin{Vmatrix}0 & 0 & \varepsilon\,u & u \\
0 & 0 & \varepsilon\,p & p \\
0 & 0 & 1 & 0 \\
0 & 0 & 0 & 1 \end{Vmatrix} ,
$$

$$ V^*=
\begin{Vmatrix}1 & 0 & 0 & 0\\
0 & 1 & 0 & 0 \\
-\varepsilon\,u & -\varepsilon\,p & 0 & 0 \\
-u & -p & 0 & 0
\end{Vmatrix} ,
\ \ H^*=
\begin{Vmatrix}0 & 0 & 0 & 0\\
0 & 0 & 0 & 0 \\
\varepsilon\,u & \varepsilon\,p & 1 & 0 \\
u & p & 0 & 1 \end{Vmatrix} ,
$$

\vskip 0.4cm
$$
\tilde{V}= \begin{Vmatrix}1 & 0 & p & \varepsilon\,p \\
0 & 1 & -u & -\varepsilon\,u \\
0 & 0 & 0 & 0 \\
0 & 0 & 0 & 0 \end{Vmatrix} ,
\ \
\tilde{H}= \begin{Vmatrix}
0 & 0 & -p & -\varepsilon\,p \\
0 & 0 & u & \varepsilon\,u \\
0 & 0 & 1 & 0 \\
0 & 0 & 0 & 1 \end{Vmatrix} ,
$$
\vskip 0.4cm
$$
\tilde{V}^*= \begin{Vmatrix}1 & 0 & 0 & 0 \\
0 & 1 & 0 & 0\\
p & -u & 0 & 0 \\
\varepsilon\,p & -\varepsilon\,u & 0 & 0 \end{Vmatrix} ,
\ \
\tilde{H}^*= \begin{Vmatrix}
0 & 0 & 0 & 0 \\
0 & 0 & 0 & 0 \\
-p & u & 1 & 0 \\
-\varepsilon\,p & \varepsilon\,u & 0 & 1 \end{Vmatrix} .
$$
The
projections of the coordinate bases are:
$$
\left(\frac{\partial}{\partial x},
\frac{\partial}{\partial y}, \frac{\partial}{\partial z},
\frac{\partial}{\partial \xi}\right).V=
\left(\frac{\partial}{\partial x}, \frac{\partial}{\partial y},
-\varepsilon u\frac{\partial}{\partial x}
-\varepsilon p\frac{\partial}{\partial y},
-u\frac{\partial}{\partial x}
-p\frac{\partial}{\partial y}\right);
$$
\vskip 0.3cm
$$
\left(\frac{\partial}{\partial x},\frac{\partial}{\partial y},
\frac{\partial}{\partial z},\frac{\partial}{\partial \xi}\right).H=
\left(0,0,\varepsilon u\frac{\partial}{\partial x}
+\varepsilon p\frac{\partial}{\partial y}+\frac{\partial}{\partial z},
u\frac{\partial}{\partial x}
+p\frac{\partial}{\partial y}+
\frac{\partial}{\partial \xi}\right);
$$
%\newpage
%\vskip 0.3cm
$$
\left(dx,dy,dz,d\xi\right).V^*=
\left(dx-\varepsilon udz-ud\xi, dy-\varepsilon pdz-pd\xi,0,0\right)
$$
%\vskip 0.1cm
$$
\left(dx,dy,dz,d\xi\right).H^*=
\left(\varepsilon udz+ud\xi,\varepsilon pdz+pd\xi,dz,d\xi\right)
$$
$$
\left(\frac{\partial}{\partial x},
\frac{\partial}{\partial y}, \frac{\partial}{\partial z},
\frac{\partial}{\partial \xi}\right).\tilde{V}=
\left(\frac{\partial}{\partial x}, \frac{\partial}{\partial y},
p\frac{\partial}{\partial x}
-u\frac{\partial}{\partial y},
\varepsilon\,p\frac{\partial}{\partial x}
-\varepsilon\,u\frac{\partial}{\partial y}\right);
$$
$$
\left(\frac{\partial}{\partial x},\frac{\partial}{\partial y},
\frac{\partial}{\partial z},\frac{\partial}{\partial \xi}\right).\tilde{H}=
\left(0,0,-p\frac{\partial}{\partial x}
+u\frac{\partial}{\partial y}+\frac{\partial}{\partial z},
-\varepsilon\,p\frac{\partial}{\partial x}
+\varepsilon\,u\frac{\partial}{\partial y}+
\frac{\partial}{\partial \xi}\right);
$$
$$
\left(dx,dy,dz,d\xi\right).\tilde{V}^*=
\left(dx+p\,dz+\varepsilon\,pd\xi, dy-u\,dz-\varepsilon\,ud\xi,0,0\right)
$$
$$
\left(dx,dy,dz,d\xi\right).\tilde{H}^*=
\left(-p\,dz-\varepsilon\,p\,d\xi,u\,dz+\varepsilon\,u\,d\xi,dz,d\xi\right) .
$$

We compute now the two curvature 2-forms $\mathcal{R}$ and
$\tilde{\mathcal{R}}$. The components $\mathcal{R}^\sigma_{\mu\nu}$ of
$\mathcal{R}$ in coordinate basis are given by
$V^\sigma_\rho\Big(\big[H\frac{\partial}{\partial
x^{\mu}},H\frac{\partial}{\partial x^{\nu}}\big]^{\rho}\Big)$, and the only
nonzero components are just
$$
\mathcal{R}^{x}_{z\xi}=\mathcal{R}^{1}_{34}=
-\varepsilon(u_{\xi}-\varepsilon\,u_z),\ \ \
\mathcal{R}^{y}_{z\xi}=\mathcal{R}^{2}_{34}=
-\varepsilon(p_{\xi}-\varepsilon\,p_z)  .
$$
For the nonzero components of $\tilde{\mathcal{R}}$ we obtain
$$
\tilde{\mathcal{R}}^{x}_{z\xi}=\tilde{\mathcal{R}}^{1}_{34}=
(p_{\xi}-\varepsilon\,p_z),\ \ \
\tilde{\mathcal{R}}^{y}_{z\xi}=\tilde{\mathcal{R}}^{2}_{34}=
-(u_{\xi}-\varepsilon\,u_z)  .
$$

The corresponding two curvature forms are:
\begin{center}
\hfill\fbox
   {\begin{minipage}{0.98\textwidth}
\begin{center}
$$
\mathcal{R}=-\varepsilon(u_\xi-\varepsilon u_z)dz\wedge d\xi\otimes
\frac{\partial}{\partial x}-
\varepsilon(p_\xi-\varepsilon p_z)dz\wedge d\xi\otimes
\frac{\partial}{\partial y}
$$
$$
\mathcal{\tilde{R}}=(p_\xi-\varepsilon p_z)dz\wedge d\xi\otimes
\frac{\partial}{\partial x}-
(u_\xi-\varepsilon u_z)dz\wedge d\xi\otimes
\frac{\partial}{\partial y} .
$$
\end{center}
\hfill
   \end{minipage}}
\end{center}
We obtain (in our coordinate system):
$$
\langle V^*(dx)\wedge \tilde{V}^*(dy), V\left(\frac{\partial}{\partial
z}\right)\wedge \tilde{V}\left(\frac{\partial}{\partial
\xi}\right)\rangle
=u^2+p^2=\phi^2,
$$
and
$$
V\left(\left[H\left(\frac{\partial}{\partial z}\right),
H\left(\frac{\partial}{\partial \xi}\right)\right]\right)=
\left[H\left(\frac{\partial}{\partial z}\right),
H\left(\frac{\partial}{\partial \xi}\right)\right]
$$
$$=
-\varepsilon(u_{\xi}-\varepsilon u_z)\frac{\partial}{\partial x}
-\varepsilon(p_{\xi}-\varepsilon p_z)\frac{\partial}{\partial y}\equiv Z_1,
$$
$$
\tilde{V}\left(\left[\tilde{H}\left(\frac{\partial}{\partial z}\right),
\tilde{H}\left(\frac{\partial}{\partial \xi}\right)\right]\right)
=\left[\tilde{H}\left(\frac{\partial}{\partial z}\right),
\tilde{H}\left(\frac{\partial}{\partial \xi}\right)\right]
$$
$$
=(p_{\xi}-\varepsilon p_z)\frac{\partial}{\partial x}
-(u_{\xi}-\varepsilon u_z)\frac{\partial}{\partial y}\equiv Z_2,
$$
where $Z_1$ and $Z_2$ coincide with the values of the two curvature forms
$\mathcal{R}$ and $\tilde{\mathcal{R}}$ on the coordinate vector fields
$\frac{\partial}{\partial z}$ and $\frac{\partial}{\partial \xi}$ respectively:
$$
Z_1=\mathcal{R}\left(\frac{\partial}{\partial z},
\frac{\partial}{\partial \xi}\right),\ \ \
Z_2=
\tilde{\mathcal{R}}\left(\frac{\partial}{\partial z},
\frac{\partial}{\partial \xi}\right) .
$$

We evaluate now the vertical 2-form
$V^*(dx)\wedge V^*(dy)$ on the bivector $Z_1\wedge Z_2$ and obtain
$$
\langle V^*(dx)\wedge V^*(dy), Z_1\wedge Z_2\rangle
=(u_{\xi}-\varepsilon u_z)^2+(p_{\xi}-\varepsilon p_z)^2.
$$
An important parameter, having dimension
of length (the coordinates are assumed to have dimension of length)
and denoted by $\mathcal{L}_o$, \index{scale factor}  can be defined by
$$
\mathcal{L}_o^2=\frac {\langle V^*(dx)\wedge \tilde{V}^*(dy),
V\left(\frac{\partial}{\partial z}\right)\wedge
\tilde{V}\left(\frac{\partial}{\partial \xi}\right)\rangle}{\langle
V^*(dx)\wedge V^*(dy),Z_1\wedge Z_2\rangle}=
\frac{u^2+p^2}{(u_{\xi}-\varepsilon
u_z)^2+(p_{\xi}-\varepsilon p_z)^2}.
$$
Clearly, if $\mathcal{L}_o$ is finite constant it could be interpreted as some
parameter of extension of the PhLO described, so it could be used as
identification parameter in the dynamical equations and in lagrangians, but
only if $(u_{\xi}-\varepsilon u_z)\neq 0$ and $(p_{\xi}-\varepsilon p_z)\neq
0$. This goes along with our concept of PhLO which does not admit spatially
infinite extensions.

The parameter $\mathcal{L}_o$ has the following symmetry. Denote by
$$
V_o=dx\otimes\frac{\partial}{\partial x}+
dy\otimes\frac{\partial}{\partial y},
$$
 then $V=V_o+V_1$ and
$\tilde {V}=V_o+\tilde{V}_1$, where, in our coordinates, $V_1$ and $\tilde
{V}_1$ can be seen above how they look like. We form now
$$
W=aV_1-b\tilde{V}_1 \ \text{and} \ \ \tilde{W}=bV_1+a\tilde{V}_1,
$$
where $(a,b)$ are two arbitrary real
numbers. The components of the corresponding linear maps $P_W=V_o+W$ and
$P_{\tilde{W}}=V_o+\tilde{W}$ can be obtained through the substitutions:
$u\rightarrow(au+\varepsilon bp); \ p\rightarrow (\varepsilon bp-ap)$,
and, obviously, $P_W$ and $P_{\tilde{W}}$ are projections. Now, it is easily
checked that the above $\mathcal{L}_o$-defining relation stays invariant, so,
$$
\mathcal{L}_o(V,\tilde{V})=\mathcal{L}_o(W,\tilde{W}).
$$
This corresponds in some sense to the dual
symmetry of classical vacuum electrodynamics and of our nonlinear equations.
We shall see that this parameter is in fact our {\it scale factor} when it is
constant.

 We note also that the squared modules of the two curvature forms by
means of the Minkowski metric $\eta$ satisfy
$$
|\mathcal{R}|^2=
|\mathcal{\tilde{R}}|^2=(u_{\xi}-\varepsilon
u_z)^2+(p_{\xi}-\varepsilon p_z)^2
$$
in our coordinates, therefore, the nonzero
values of $|\mathcal{R}|^2$ and $|\mathcal{\tilde{R}}|^2$, as well as the
finite value of $\mathcal{L}_o$ guarantee that the two functions $u$ and $p$
are NOT plane waves.

Finally, the phase function $\psi$ may be defined by the
relations
\begin{eqnarray*}
L_{\bar{\zeta}}\psi&=&
\frac{\eta(V(\frac{\partial}{\partial\xi}),Z_2)}
{\phi^2}
=-\frac{\eta(\tilde{V}(\frac{\partial}{\partial\xi}),Z_1)}
{\phi^2}\\
&=&
\frac{u(p_\xi-\varepsilon p_z)-p(u_\xi-\varepsilon u_z)}{\phi^2}
=\frac{\mathbf{R}}{\phi^2} ,
\end{eqnarray*}
where $\phi^2$ is defined above .

 \vskip 0.5cm
\subsection{Electromagnetic PhLO in terms of nonlinear\\ connections}

Recall that the relativistic formulation of classical
electrodynamics in vacuum ($\rho=0$) is based on the following assumptions. The
configuration space is the Minkowski space-time $M=(\mathbb{R}^4,\eta)$ where
$\eta$ is the pseudometric with $sign(\eta)=(-,-,-,+)$ with the corresponding
volume 4-form $\omega_o=dx\wedge dy\wedge dz\wedge d\xi $ and
Hodge star $*$ defined by $\alpha\wedge *\beta=-\eta(\alpha,\beta)\omega_o$. The
electromagnetic filed is described by two closed 2-forms $(F,*F):\mathbf{d}F=0,
\ \mathbf{d}*F=0$. The physical characteristics of the field are deduced from
the following stress-energy-momentum tensor field
$$
T_{\mu}{^\nu}(F,*F)=-\frac12\big[F_{\mu\sigma}F^{\nu\sigma}+
(*F)_{\mu\sigma}(*F)^{\nu\sigma}\big].
$$
In the non-vacuum case the allowed energy-momentum exchange with other physical
systems is given in general by the divergence
$$
\nabla_\nu\,T_{\mu}^{\nu}
=\frac12 \Big[F^{\alpha\beta}(\mathbf{d}F)_{\alpha\beta\mu}
+(*F)^{\alpha\beta}(\mathbf{d}*F)_{\alpha\beta\mu}\Big] =
F_{\mu\nu}(\delta F)^\nu + (*F)_{\mu\nu}(\delta *F)^\nu,
$$
where $\delta=*\mathbf{d}*$ is the coderivative.
If the field is free: $\mathbf{d}F=0, \mathbf{d}*F=0$, this divergence is
obviously equal to zero on the vacuum solutions.
Therefore, energy-momentum exchange between the two partner-fields $F$ and
$*F$, which should be expressed by the terms
$(*F)^{\alpha\beta}(\mathbf{d}F)_{\alpha\beta\mu}$ and
$F^{\alpha\beta}(\mathbf{d}*F)_{\alpha\beta\mu}$ is NOT allowed on the
solutions of $\mathbf{d}F=0, \mathbf{d}*F=0$. This shows that the
widely used 4-potential approach (even if two 4-potentials $A,A^*$ are
introduced so that $\mathbf{d}A=F, \ \mathbf{d}A^*=*F$ locally)
to these equations excludes any possibility to individualize two
energy-momentum exchanging time-stable subsystems of the field that are
mathematically represented by $F$ and $*F$.

On the contrary, as we have mentioned several times, our concept of PhLO does
NOT exclude such two physically interacting subsystems of the field to really
exist, and therefore, to be mathematically individualized. The intrinsically
connected two projections $V$ and $\tilde{V}$ and the corresponding two
curvature forms give the mathematical realization of this idea: $V$ and
$\tilde{V}$ individualize the two subsystems, and the corresponding two
curvature 2-forms $\mathcal{R}$ and $\mathcal{\tilde{R}}$ represent the
instruments by means of which the available mutual local energy-momentum
exchange between these two subsystems could be described. We should not forget
that, as we have already emphasized several times, the energy-momentum tensor
for a PhLO must satisfy the additional local isotropy (null) condition
$T_{\mu\nu}(F,*F)T^{\mu\nu}(F,*F)=0$.

So, we have to construct appropriate quantities and relations
having direct physical sense in terms of the introduced and considered two
projections $V$ and $\tilde{V}$. The above well established in electrodynamics
relations say that we need two 2-forms to begin with.

Recall that our coordinate 1-forms $dx$ nd $dy$ have the following $V$-vertical
and $H$-horizontal projections:
$$
 V^*(dx)=dx-\varepsilon u\,dz-u\,d\xi, \ \
H^*(dx)=\varepsilon u\,dz+u\,d\xi\ ,
$$
$$
 V^*(dy)=dy-\varepsilon
p\,dz-p\,d\xi, \ \ H^*(dy)=\varepsilon p\,dz+p\,d\xi .
$$
We form now the 2-forms
$V^*(dx)\wedge H^*(dx)$ and $V^*(dy)\wedge H^*(dy)$:
$$
V^*(dx)\wedge
H^*(dx)=\varepsilon\,u\,dx\wedge dz+u\,dx\wedge d\xi,
$$
$$ V^*(dy)\wedge
H^*(dy)=\varepsilon\,p\,dy\wedge dz+p\,dx\wedge d\xi .
$$
Summing up these last
two relations and denoting the sum by $F$ we obtain
$$
F=\varepsilon\,u\,dx\wedge dz+u\,dx\wedge d\xi+
\varepsilon\,p\,dy\wedge dz+p\,dy\wedge d\xi .
$$
Doing the same steps with $\tilde{V}^*$ and $\tilde{H}^*$ we obtain
$$
\tilde{F}=-p\,dx\wedge dz-\varepsilon\,p\,dx\wedge d\xi+
u\,dy\wedge dz+\varepsilon u\,dy\wedge d\xi .
$$
Noting that our definition of the Hodge star requires
$(*F)_{\mu\nu}=-\frac12\,\varepsilon_{\mu\nu}\,^{\sigma\rho}F_{\sigma\rho}$,
it is now easy to verify that $\tilde{F}=*F$. Moreover, introducing the
notations
$$
A=u\,dx+p\,dy, \ \ A^*=-\varepsilon\,p\,dx+\varepsilon\,u\,dy ,\ \
\zeta=\varepsilon\,dz+d\xi ,
$$
we can represent $F$ and $\tilde{F}$ in the form
$$
F=A\wedge \zeta , \ \ \tilde{F}=*F=A^*\wedge \zeta .
$$
From these last relations
we see that $F$ and $*F$ are isotropic:
$$
F\wedge F=0,\quad F\wedge *F=0,
$$ i.e. the
field $(F,*F)$ has zero invariants:
$F_{\mu\nu}F^{\mu\nu}=F_{\mu\nu}(*F)^{\mu\nu}=0$. The following relations are
now easy to verify:
$$
V^*(F)=H^*(F)=V^*(*F)=H^*(*F)
$$
$$
=\tilde{V}^*(F)=\tilde{H}^*(F)=
\tilde{V}^*(*F)=\tilde{H}^*(*F)=0,
$$
i.e. $F$ and $*F$ have zero vertical and horizontal projections with respect
to $V$ and $\tilde{V}$. Since, obviously, $\zeta $ is horizontal with respect
to $V$ and $\tilde{V}$ it is interesting to note that $A$ is vertical with
respect to $\tilde{V}$ and $A^*$ is vertical with respect to $V$:
$\tilde{V}^*(A)=A$, $V(A^*)=A^*$. In fact, for example,
$$
\tilde{V}^*(A)=\tilde{V}^*(u\,dx+p\,dy)=u\tilde{V}^*(dx)+p\tilde{V}^*(dy)=
$$
$$
u[dx+p\,dz+\varepsilon p\,d\xi]+p[dy-u\,dz-\varepsilon u\,d\xi]=
u\,dx+p\,dy.
$$

We are going to establish now that there is real energy-momentum exchange
between the $F$-component and the $*F$-component of the field,
computing $i(Z_1)F,\ \ i(Z_2)*F,\ \ i(Z_1)*F,\ \
i(Z_2)F$, where $Z_1,Z_2$ are given in the preceding section (10.2, p.306).
We obtain:
\begin{center}
\hfill\fbox
   {\begin{minipage}{0.99\textwidth}
\begin{center}
$$
i(Z_1)F=i(Z_2)*F=\langle A,Z_1\rangle\zeta=\langle A^*,Z_2\rangle\zeta=
\frac12\big[(u^2+p^2)_{\xi}-\varepsilon\,(u^2+p^2)_{z}\big]\zeta
$$
$$
=\frac12F^{\sigma\rho}(\mathbf{d}F)_{\sigma\rho\mu}dx^{\mu}=
\frac12(*F)^{\sigma\rho}(\mathbf{d}*F)_{\sigma\rho\mu}dx^{\mu}
=\frac12\nabla_\nu\,T_{\mu}^{\nu}(F,*F)dx^\mu ,
$$
$$
i(Z_1)*F=-i(Z_2)F=\langle A^*,Z_1\rangle\zeta =-\langle A,Z_2\rangle\zeta=
\big[u(p_{\xi}-\varepsilon\,p_{z})-p(u_{\xi}-\varepsilon\,u_{z})\big]\zeta
$$
$$
=-\frac12F^{\sigma\rho}(\mathbf{d}*F)_{\sigma\rho\mu}dx^{\mu}=
\frac12(*F)^{\sigma\rho}(\mathbf{d}F)_{\sigma\rho\mu}dx^{\mu} .
$$
\end{center}
\hfill
   \end{minipage}}
\end{center}

If our field is free then $\nabla_\nu\,T_{\mu}^{\nu}(F,\tilde{F})dx^\mu=0$.
Moreover, in view of the divergence of the stress-energy-momentum tensor given
above, these last relations show that some real energy-momentum exchange between
$F$ and $*F$ takes place: the magnitude of the energy-momentum, transferred from
$F$ to $*F$ and given by
$$
i(Z_1)*F=\frac12(*F)^{\sigma\rho}(\mathbf{d}F)_{\sigma\rho\mu}dx^{\mu},
$$
is equal to that, transferred from $*F$ to $F$, which is
given by
$$
-i(Z_2)F=-\frac12F^{\sigma\rho}(\mathbf{d}*F)_{\sigma\rho\mu}dx^{\mu}.
$$
On the other hand, as it is well known, in case of zero invariants we have
$$
F_{\mu\sigma}F^{\nu\sigma}=(*F)_{\mu\sigma}(*F)^{\nu\sigma},
$$
so, $F$ and $*F$ dynamically keep the stress-energy-momentum they carry.

We interpret physically this as follows. The electromagnetic PhLO exist through
a special internal dynamical equilibrium between the two subsystems of the
field, represented by $V$ and $\tilde{V}$, namely, both subsystems carry the
same stress-energy-momentum and the mutual energy-momentum exchange between
them is simultanious and always in equal quantities. This individualization
does NOT mean that any of the two subsystems can exist separately,
independently on the other. Moreover, NO spatial "part" of PhLO should be
considered as potentially able to represent a real physical object.

\section{Strain and Photon-like Objects.}
The mathematical concept of (infinitesimal) strain was introduced in Sec.5.2.3
as Lie derivative of the metric tensor on a manifold with respect to a vector
field $X$ on the same manifold. The physics behind this definition is to
introduce a local measure of the change of distance  in a continuous material
when it is subject to external action by a physical field and this action does
not lead to some irreversible changes in the material, like wholes for example.
In other words, the external action leads to bearable perturbation, meaning
mathematically to smooth and reversible deformations of the metric tensor.
Such deformations of the material are called in physics {\it elastic}, they
transform one admissible configuration of the material to another admissible
configuration. Since these configuration transformations are reversible, they
can not transform a flat metric tensor $g$, i.e., $\mathcal{R}(g)=0$, to a new
metric tensor $g'$ giving non-zero Riemann curvature tensor
$\mathcal{R}(g')\neq 0$.

The concept of curvature, however, considered as a measure of nonintegrability
of a system of partial differential equations, has much more general sense, as
we have presented it and treated so far. Our aim now is to show that the Lie
derivatives of our flat Minkowski metric $\eta$ on $M$ with respect
to the naturally arized vector fields $\bar{A}$ and  $\bar{A}^*$ in our
approach to formal description of photon-like objects, generate non-zero
curvature as a measure of nonintergability of the distributions
$(\bar{A},\bar{\zeta})$ and $(\bar{A}^*,\bar{\zeta})$. Hence, the metric
changes along these vector fields, but does not "get curved" in riemannian
sense. The reason for this we see in the very nature of the Lie derivatve: it
accounts for the infinitesimal changes of both, the differentiated object-the
metric $\eta$ in our case, as well as the referent object(s)-the just mentioned
vector fields $(\bar{A},\bar{A}^*)$ in our case.

In our further study we shall call the infinitesimal strain
tensors $L_{\bar{A}}\eta$, $L_{\bar{A}^*}\eta$ and $L_{\bar{\zeta}}\,\eta$ just
{\it strain tensors} for breavity. We have to note, however, that the term
"material" is not quite appropriate for PhLO because {\it no static situations
are admissible, our objects of interest are of entirely dynamical nature}, so
the corresponding {\it relativistic strain} tensors must take care of this.

According to the preliminary considerations important vector fields in our
approach to describe electromagnetic PhLO are
$\bar{\zeta},\,\bar{A},\,\bar{A^*}$, so, we consider the corresponding three
 strain tensors: $ L_{\bar{\zeta}}\,\eta; \,L_{\bar{A}}\,\eta;
\,L_{\bar{A^*}}\,\eta$.
\vskip 0.3cm
The three strain tensors look in our coordinates as follows:
$$
L_{\bar{\zeta}}\,\eta=0,
$$
$$
(L_{\bar{A}}\,\eta)_{\mu\nu}\equiv D_{\mu\nu}= \begin{Vmatrix}2u_x & u_y+p_x &
u_z & u_{\xi} \\ u_y+p_x & 2p_y & p_z & p_{\xi} \\ u_z & p_z & 0 & 0 \\ u_{\xi}
& p_{\xi} & 0 & 0 \end{Vmatrix} ,
$$
$$
(L_{\bar{A^*}}\,\eta)_{\mu\nu}\equiv D^*_{\mu\nu} =
\begin{Vmatrix}-2\varepsilon p_x & -\varepsilon(p_y+u_x) & -\varepsilon p_z &
-\varepsilon p_{\xi} \\ -\varepsilon(p_y+u_x) & 2\varepsilon u_y & \varepsilon
u_z & \varepsilon u_{\xi} \\ -\varepsilon p_z & \varepsilon u_z & 0 & 0 \\
-\varepsilon p_{\xi} & \varepsilon u_{\xi} & 0 & 0 \end{Vmatrix} .
$$
The following two relations are immediately verified:
$$
det\|D_{\mu\nu}\|=det\|D^*_{\mu\nu}\|=(p_\xi u_z-u_\xi p_z)^2\geq 0.
$$
\vskip 0.3cm
We give now some important from our viewpoint relations.
$$
D(\bar{\zeta},\bar{\zeta})=D^*(\bar{\zeta},\bar{\zeta})=0,
$$
$$
D(\bar{\zeta})\equiv D(\bar{\zeta})_\mu dx^\mu\equiv D_{\mu\nu}\bar{\zeta}^\nu
dx^\mu =(u_\xi-\varepsilon u_z)dx + (p_\xi-\varepsilon p_z)dy,
$$
$$
D(\bar{\zeta})^\mu\frac{\partial}{\partial x^\mu}\equiv
D^{\mu}_{\nu}\bar{\zeta}^\nu\frac{\partial}{\partial x^\mu}=
-(u_\xi-\varepsilon u_z)\frac{\partial}{\partial x} -
(p_\xi-\varepsilon p_z)\frac{\partial}{\partial y}=-[\bar{A},\bar{\zeta}],\ \
$$
$$
D_{\mu\nu}\bar{A}^\mu\bar{\zeta}^\nu=
-\frac12\Big[(u^2+p^2)_\xi -\varepsilon(u^2+p^2)_z\Big]=
-\frac12L_{\bar{\zeta}}\phi^2 ,
$$
$$
D_{\mu\nu}\bar{A^*}^\mu\bar{\zeta}^\nu=
-\varepsilon\Big[u(p_\xi-\varepsilon p_z)-p(u_\xi-\varepsilon u_z)\Big]=
-\varepsilon\mathbf{R}=-\varepsilon \phi^2\,L_{\bar{\zeta}}\psi.
$$
We also have:
$$
D^*(\bar{\zeta})=\varepsilon\Big[-(p_\xi-\varepsilon p_z)dx+
(u_\xi-\varepsilon u_z)dy\Big] ,
$$
$$
D^*(\bar{\zeta})^\mu\frac{\partial}{\partial x^\mu}\equiv
(D^*)^{\mu}_{\nu}\bar{\zeta}^\nu\frac{\partial}{\partial x^\mu}=
-\varepsilon(p_\xi-\varepsilon p_z)\frac{\partial}{\partial x} +
(u_\xi-\varepsilon u_z)\frac{\partial}{\partial y}=[\bar{A^*},\bar{\zeta}],\ \
$$
$$
D^*_{\mu\nu}\bar{A^*}^\mu\bar{\zeta}^\nu=
-\frac12\Big[(u^2+p^2)_\xi -\varepsilon(u^2+p^2)_z\Big]=
-\frac12L_{\bar{\zeta}}\phi^2 ,
$$
$$
D^*_{\mu\nu}\bar{A}^\mu\bar{\zeta}^\nu=
\varepsilon\Big[u(p_\xi-\varepsilon p_z)-p(u_\xi-\varepsilon u_z)\Big]=
\varepsilon\mathbf{R}=\varepsilon \phi^2\,L_{\bar{\zeta}}\psi.
$$
Clearly, $D(\bar{\zeta})$ and $D^*(\bar{\zeta})$ are lineary independent in
general:
$$
D(\bar{\zeta})\wedge D^*(\bar{\zeta})=\varepsilon
\Big[(u_\xi-\varepsilon u_z)^2+(p_\xi-\varepsilon p_z)^2\Big]dx\wedge dy
=\varepsilon\phi^2(\psi_\xi-\varepsilon \psi_z)^2\,dx\wedge dy\neq 0.
$$
Recall now that every 2-form $F$ defines a linear map $\tilde{F}$ from
1-forms to 3-forms through the exterior product:
$\tilde{F}(\alpha):=\alpha\wedge F$, where $\alpha\in \Lambda^1(M)$.
Moreover, the Hodge $*$-operator, composed now with $\tilde{F}$, gets
$\tilde{F}(\alpha)$ back to $*\tilde{F}(\alpha)\in\Lambda^1(M)$. In the
previous section we introduced two 2-forms $G=A\wedge\zeta$ and
$G^*=A^*\wedge\zeta$ and noticed that $G^*=*G$. We readily obtain now
$$
D(\bar{\zeta})\wedge G=D^*(\bar{\zeta})\wedge G^*= D(\bar{\zeta})\wedge
A\wedge\zeta=D^*(\bar{\zeta})\wedge A^*\wedge\zeta
$$
$$
=-\varepsilon\Big[u(p_\xi-\varepsilon p_z)- p(u_\xi-\varepsilon
u_z)\Big]dx\wedge dy\wedge dz- \Big[u(p_\xi-\varepsilon p_z)-
p(u_\xi-\varepsilon u_z)\Big]dx\wedge dy\wedge d\xi
$$
$$
=-\phi^2\,L_{\bar{\zeta}}\psi\,(\varepsilon\,dx\wedge dy\wedge dz+
dx\wedge dy\wedge d\xi)=-\mathbf{R}\,(\varepsilon\,dx\wedge dy\wedge dz+
dx\wedge dy\wedge d\xi),
$$
$$
 D(\bar{\zeta})\wedge
G^*=-D^*(\bar{\zeta})\wedge G= D(\bar{\zeta})\wedge
A^*\wedge\zeta=-D^*(\bar{\zeta})\wedge A\wedge\zeta
$$
$$
=\frac12\Big[(u^2+p^2)_\xi-\varepsilon(u^2+p^2)_z\Big] (dx\wedge dy\wedge
dz+\varepsilon\,dx\wedge dy\wedge d\xi).
$$
Thus we get
\begin{center}
\hfill\fbox
   {\begin{minipage}{0.98\textwidth}
\begin{center}
$$
*\Big[D(\bar{\zeta})\wedge
A\wedge\zeta\Big]= *\Big[D^*(\bar{\zeta})\wedge A^*\wedge\zeta\Big]=
-\varepsilon\mathbf{R}\,\zeta =-i(\bar{G^*})\mathbf{d}G=i(\bar{G})\mathbf{d}G^*,
$$
$$
*\Big[D(\bar{\zeta})\wedge
A^*\wedge\zeta\Big]= -*\Big[D^*(\bar{\zeta})\wedge A\wedge\zeta\Big]=
\frac12L_{\bar{\zeta}}\phi^2\,\zeta=
i(\bar{G})\mathbf{d}G=i(\bar{G^*})\mathbf{d}G^*.
$$
\end{center}
\hfill
   \end{minipage}}
\end{center}
The above relations show various dynamical aspects of the energy-momen\-tum
redistribution during evolution of our PhLO. In particular, we see that it is
possible the translational and rotational components of the energy-momentum
redistribution during propagation to be represented in form depending on the
$\bar{\zeta}$-directed strains $D(\bar{\zeta})$ and $D^*(\bar{\zeta})$. So, the
admissible local translational changes of the energy-momentum carried by the
two field components $G$ and $G^*$ of our PhLO are given by the two 1-forms
$$
*\big[D(\bar{\zeta})\wedge A^*\wedge\zeta\big] \ \ \text{and} \ \
*\big[D^*(\bar{\zeta})\wedge A\wedge\zeta\big]),
$$
 and the admissible local rotational ones are given by the 1-forms
$$
*\big[D(\bar{\zeta})\wedge A\wedge\zeta\big] \ \ \text{and} \ \
*\big[D^*(\bar{\zeta})\wedge A^*\wedge\zeta\big].
$$
In fact, the 1-form
$*\big[D(\bar{\zeta})\wedge A\wedge\zeta\big]$ determines the strain that
"leaves" the 2-plane defined by $(\bar{A},\bar{\zeta})$ and the 1-form
$*\big[D^*(\bar{\zeta})\wedge A^*\wedge\zeta\big]$ determines the strain that
"leaves" the 2-plane defined by $(\bar{A}^*,\bar{\zeta})$. Since the PhLO is
free, i.e., it does not lose or gain energy-momentum, this means
that the two (null-field) components $G$ and $G^*$ exchange locally {\it equal}
energy-momentum quantities:
$$
 *\Big[D(\bar{\zeta})\wedge A\wedge\zeta\Big]=
*\Big[D^*(\bar{\zeta})\wedge A^*\wedge\zeta\Big].
$$
Now, the local energy-momentum conservation law
$$
\nabla_{\nu}\big[G_{\mu\sigma}\bar{G}^{\nu\sigma}+
(G^*)_{\mu\sigma}(\bar{G}^*)^{\nu\sigma}\big]=0
$$
requires
$L_{\bar{\zeta}}\phi^2=0$, and the corresponding strain-fluxes
$$
*\big[D^*(\bar{\zeta})\wedge A\wedge\zeta\big]=
D^*(\bar{\zeta})^{\mu}(A\wedge\zeta)_{\mu\nu}dx^{\nu},
$$
$$
*\big[D(\bar{\zeta})\wedge A^*\wedge\zeta\big]=
 D(\bar{\zeta})^{\mu}(A^*\wedge\zeta)_{\mu\nu}dx^{\nu}
$$
become zero.

It seems important to note that, only dynamical relation between  the local
energy-momentum change and strain fluxes exists, so NO analog of the assumed in
elasticity theory generalized Hooke law, (i.e., linear relation between the
stress tensor and the strain tensor) seems to exist. This clearly goes along
with the fully dynamical nature of PhLO, i.e., linear relations exist between
the divergence terms of our stress tensor $
\frac12\big[-G_{\mu\sigma}\bar{G}^{\nu\sigma}-
(G^*)_{\mu\sigma}(\bar{G}^*)^{\nu\sigma}\big] $ and the $\bar{\zeta}$-directed
strain fluxes as given above.

\chapter{PhLO as Solutions of linear equations}

{\it In this chapter instead of $*F$ we write $\tilde{F}$, and show that
appropriate solutions for PhLO can be obtained by solving {\bf linear}
equations}.

 \section{The approach based on the notion for PhLO} Every system of
equations describing the time-evolution of some physical system should be
consistent with the very system in the sense that all identification
characteristics of the system described must not change. In the case of
electromagnetic PhLO we assume the couple $(F,\tilde{F})$ to represent the
field, and in accordance with our notion for PhLO one of the identification
characteristics is straight-line translational propagation of the
energy-density with constant velocity "$c$", therefore, with every PhLO we may
associate appropriate direction, i.e. a geodesic null vector field
$\bar{\zeta}, \bar{\zeta}^2=0$ on the Minkowski space-time. On the other hand,
the complex of field functions $(F_{\mu\nu},\tilde{F}_{\mu\nu})$ admits both
translational and rotational components of propagation. We choose further
$\bar{\zeta}=-\varepsilon \frac{\partial}{\partial z}+\frac{\partial}{\partial \xi}$,
which means that we have chosen the
coordinate system in such a way that the translational propagation is parallel to
the plane $(z,\xi)$. For another such parameter we assume that
the finite longitudinal extension of any PhLO is fixed and is given by an
appropriate positive number $\lambda $. In accordance with the "compatible
translational-rotational dynamical structure" of PhLO we shall assume that {\it
no translation is possible without rotation, and no rotation is possible
without translation}, and in view of the constancy of the translational
component of propagation we shall assume that the rotational component of
propagation is periodic, i.e. it is characterized by a constant frequency. The
natural period $T$ suggested is obviously $T=\frac{\lambda}{c}$. An obvious
candidate for "rotational operator" is the linear map $\mathcal{J}$
transforming $F$ to $\tilde{F}$, which map coincides with the reduced to
2-forms Hodge-$*$. Geometrically, $*$ rotates the 2-frame $(A,A^*)$ to
$\frac\pi 2$, so if such a rotation is associated with a translational
advancement of $\mathcal{L}_o$, then a full rotation should correspond to
translational advancement of $2\pi\mathcal{L}_o=\lambda $. The simplest and
most natural translational change of the field $(F,\tilde{F})$ along
$\bar{\zeta}$ should be given by the Lie derivative of the field along
$\bar{\zeta}$. Hence, the simplest and most natural equations should read $$
\kappa \mathcal{L}_o\,L_{\bar{\zeta}}(F)=\varepsilon\tilde{F}, $$ where $F$ and
$\tilde{F}$ are given in Sec.10.3., $\kappa=\pm 1$ is responsible for
left/right orientation of the rotational component of propagation, and
$\mathcal{L}_o=const$. Vice versa, since $*_2\circ *_2=-id$ and $*_2^{-1}=-*_2$
the above equation is equivalent to $$
\kappa\mathcal{L}_o\,L_{{\bar{\zeta}}}(\tilde{F})=-\varepsilon F.
$$
It is easy to show that these equations are equivalent to $$ \kappa
\mathcal{L}_o\,L_{{\bar{\zeta}}}(V-V_o)=\varepsilon(\tilde{V}-V_o),
$$
where $V$ is given
in Sec.10.3.1, and $V_o=dx\otimes\frac{\partial}{\partial x}+
dy\otimes\frac{\partial}{\partial y}$
in our coordinates is the identity map in $Im(V)=Im(\tilde{V})$.
Another equivalent form is given by
$$
\kappa\mathcal{L}_oZ_1=\bar{A^*}, \ \ \ \ \text{or} \ \ \ \  \kappa
\mathcal{L}_oZ_2=-\bar{A}, $$ where $\bar{A^*}$ and $\bar{A}$ are $\eta$-corresponding
vector fields to the 1-forms $A^*$ and $A$, and $Z_1,Z_2$ are given in
Sec.10.3.

\section{The Lagrangian Approach}
Appropriate lagrangian for the above equations ($\mathcal{L}_o$=const.) is
$$
\mathbb{L}=\left(
\kappa\mathcal{L}_o \bar{\zeta}^\sigma\frac{\partial
F_{\alpha\beta}}{\partial x^\sigma}+\tilde{F}_{\alpha\beta}\right)
\tilde{F}^{\alpha\beta}
-\left(
\kappa\mathcal{L}_o\bar{\zeta}^\sigma                          %4%
\frac{\partial \tilde{F}_{\alpha\beta}}{\partial x^\sigma}-
F_{\alpha\beta}\right)F^{\alpha\beta},  \ \ \ \alpha<\beta ,
$$
$F$ and $\tilde{F}$ are considered as independent, $\kappa=\pm 1$.
The corresponding Lagrange equations read
$$
\kappa\mathcal{L}_o\bar{\zeta}^\sigma                          %4%
\frac{\partial \tilde{F}_{\alpha\beta}}{\partial x^\sigma}-
F_{\alpha\beta}=0,\ \
\kappa\mathcal{L}_o\bar{\zeta}^\sigma
\frac{\partial F_{\alpha\beta}}{\partial x^\sigma}+
\tilde{F}_{\alpha\beta}=0 , \ \ \alpha<\beta .
$$
Note that on the solutions the lagrangian becomes zero:
$\mathbb{L}(solutions)=0$.
The null character of the objects described require
$$
F\wedge F=F\wedge \tilde{F}=-\tilde{F}\wedge \tilde{F}=0,
$$
which lead to
$$
\bar{\zeta}^\sigma\frac{\partial
F_{\alpha\beta}}{\partial x^\sigma}\tilde{F}^{\alpha\beta}=
\bar{\zeta}^\sigma                          %4%
\frac{\partial \tilde{F}_{\alpha\beta}}{\partial x^\sigma}F^{\alpha\beta}=0.
$$
The stress-energy-momentum tensor, in view of the
null character of $F$ and $\tilde{F}$, is the same, where $*F$ has to be
replaced by $\tilde{F}$. It deserves noting that the above null conditions lead
to $F_{\mu\sigma}F^{\nu\sigma}=\tilde{F}_{\mu\sigma}\tilde{F}^{\nu\sigma}$ and
to $F_{\mu\sigma}\tilde{F}^{\nu\sigma}=0$. Hence, the two subsystems
represented by $F$ and $\tilde{F}$ carry the same stress-energy-momentum,
therefore, $F\rightleftarrows \tilde{F}$ energy-momentum exchange is possible
only in equal quantities.

In our coordinates the above equations reduce to
$$
\varepsilon\kappa\mathcal{L}_o(u_{\xi}-\varepsilon\,u_z)=p,\ \ \
\varepsilon\kappa\mathcal{L}_o(p_{\xi}-\varepsilon\,p_z)=-u,
$$
it is seen that the constant $\mathcal{L}_o$ satisfies the corresponding relation
in Sec.10.3. From these last equations we readily obtain the relations
$$
(u^2+p^2)_{\xi}-\varepsilon\,(u^2+p^2)_z=0, \ \
u\,(p_{\xi}-\varepsilon\,p_z)-p\,(u_{\xi}-\varepsilon\,u_z)=
\frac{\kappa}{\mathcal{L}_o}(u^2+p^2),
$$
which represent our equations in energy-momentum terms. Now,
the substitution $u=\Phi\cos\,\psi,\ \  p=\Phi\sin\,\psi $, leads to the
relations
$$
L_{{\bar{\zeta}}}\Phi=0, \ \
L_{{\bar{\zeta}}}\psi=\frac{\kappa}{\mathcal{L}_o}.
$$

\section{Translational-rotational compatability\\ approach}

In order to look at the translational-rotational compatability as a generating
tool for writing equations of motion we recall first the concept of
local symmetry of a distribution: a vector field $Y$ is a local (or
infinitesimal) symmetry of a p-dimensional distribution $\Delta$ defined by the
vector fields $(Y_1,\dots,Y_p)$ if every Lie bracket $[Z,Y]$ is in $\Delta$:
$[Z,Y]\in\Delta $, where $Z=f^iY_i\in\Delta$. Clearly, if $\Delta$ is completely
integrable, then every $Y_i$ is a symmetry of $\Delta$, and the flows of these
vector fields move the points of each integral manifold of $\Delta$ inside this
integral manifold, that's why they are called sometimes internal symmetries. If
$Y$ is outside $\Delta$ then it is called {\it shuffling} symmetry , and in
such a case the flow of $Y$ transforms a given integral manifold to another
one, i.e. the flow of $Y$ "shuffles" the lists of the corresponding foliation.
We are going to find the conditions for our vector field
$\bar{\zeta}=-\varepsilon \frac{\partial}{\partial z}+\frac{\partial}{\partial
\xi}$ to define a shuffling symmetry for the distribution $\Delta_o$ defined by
the vector fields $(\bar{A},\bar{A}^*)$. The suggestion comes from the
observation that $\Delta_o$ coincides with our vertical distribution generated
by $(\frac{\partial}{\partial x}, \frac{\partial}{\partial y})$. From physical
point of view this should be expected in view of the intrinsically required
stability of our PhLO under translational propagation along null straight
lines: this propagation just transforms the 2-plane $(x,y)$ passing through the
point $(z_1,\xi_1)$ to a parallel to it 2-plane passing through the point
$(z_2,\xi_2)$, and these two points lay on the same straight line trajectory of
our field $\bar{\zeta}$.

The corresponding Lie brackets are
$$
[{\bar{\zeta}},\bar{A}]=-(u_\xi-\varepsilon\,u_z)\frac{\partial}{\partial
x}- (p_\xi-\varepsilon\,p_z)\frac{\partial}{\partial y},
$$
$$
[\bar{\zeta},\bar{A}^*]=
\varepsilon\,(p_\xi-\varepsilon\,p_z)\frac{\partial}{\partial x}-
\varepsilon\,(u_\xi-\varepsilon\,u_z)\frac{\partial}{\partial y}.
$$
We see that $[\bar{\zeta},\bar{A}]\wedge[\bar{\zeta},\bar{A}^*]\neq 0$, and
that they are generated by $(\frac{\partial}{\partial x},
\frac{\partial}{\partial y})$.

We notice now that at each point we have two different time-changing frames:
$(\bar{A},\bar{A}^*,\partial_z,\partial_{\xi})$ and
$([\bar{A},\bar{\zeta}],[\bar{A^*},\bar{\zeta}],\partial_z,\partial_{\xi})$.
Since physically we have internal energy-momentum redistribution during
propagation, we could try to interpret formally this physical process in terms
of these two intrinsically connected time-changing frames. Taking into account
that only the first two vectors of these two frames change during propagation
we may write down the corresponding linear transformation as follows: $$
([\bar{\zeta},\bar{A}],[\bar{\zeta},\bar{A^*}])=(\bar{A},\bar{A}^*)
\begin{Vmatrix}\alpha & \beta \\ \gamma & \delta\end{Vmatrix} .
$$
Denoting $\Phi^2=u^2+p^2$ and solving this system with respect to
$(\alpha,\beta,\gamma,\delta)$ we obtain
\begin{eqnarray*}
\begin{Vmatrix}\alpha & \beta \\ \gamma & \delta \end{Vmatrix}&=&
\frac{1}{\Phi^2}
 \begin{Vmatrix}
\frac12 L_{\bar{\zeta}}\Phi^2 & \varepsilon \mathbf{R} \\ -\varepsilon
\mathbf{R} & \frac12 L_{\bar{\zeta}}\Phi^2 \end{Vmatrix}\\&=&
\frac12\frac{L_{\bar{\zeta}}\Phi^2}{\Phi^2}
\begin{Vmatrix} 1 & 0 \\ 0 & 1\end{Vmatrix}+
\varepsilon L_{\bar{\zeta}}\psi\begin{Vmatrix} 0 & 1 \\ -1 & 0
\end{Vmatrix}\\&=&
\frac12\frac{L_{\bar{\zeta}}\Phi^2}{\Phi^2}I+
\varepsilon L_{\bar{\zeta}}\psi J,
 \end{eqnarray*}
so,
$$
[\bar{\zeta},\bar{A}]=\frac{1}{2\Phi^2}L_{\bar{\zeta}}\Phi^2\,\bar{A}+
\varepsilon\frac{\mathbf{R}}{\Phi^2}\,\bar{A^*},\ \ \ \
[\bar{\zeta},\bar{A^*}]=-\varepsilon\frac{\mathbf{R}}{\Phi^2}\,\bar{A}+
\frac{1}{2\Phi^2}L_{\bar{\zeta}}\Phi^2\,\bar{A^*},
$$
where $\mathbf{R}=u\,(p_\xi-\varepsilon\,p_z)-p\,(u_\xi-\varepsilon\,u_z)$ is
the Frobenius curvature and $\psi$ is the phase. Hence, $\bar{\zeta}$ is
infinitesimal shuffling symmetry of the distribution $(\bar{A},\bar{A^*})$.
If the translational propagation respects by the conservation law
$L_{\bar{\zeta}}\Phi^2=0$, then we obtain that the rotational component of
propagation is governed by the matrix $\varepsilon L_{\bar{\zeta}}\psi\,J$,
where $J$ denotes the canonical complex structure in $\mathbb{R}^2$, and since
$\mathbf{R}=\Phi^2\,L_{\bar{\zeta}}\psi=u\,(p_\xi-\varepsilon\,p_z)-p\,(u_\xi-\varepsilon\,u_z)\neq
0$ we conclude that the rotational component of propagation would be available
in such a case if and only if $\mathbf{R}\neq 0$. We may also say that a
compatible translational-rotational dynamical structure is available if the
amplitude $\Phi^2=u^2+p^2$ is a running wave along $\bar{\zeta}$ and the phase
$\psi$ is NOT a running wave along
$\bar{\zeta} : L_{\bar{\zeta}}\psi\neq 0$.
Physically this means that the rotational component of propagation is
entirely determined by the available internal energy-momentum exchange:
$i(\eta^{-1}(\tilde{F}))\mathbf{d}F=-i(F)\mathbf{d}\eta^{-1}(\tilde{F})$.

Now, assuming $L_{\bar{\zeta}}\Phi^2=0$, if we have to
{\it guarantee the
intrinsic compatability between the translational and
rotational aspect} of the PhLO nature, we should assume
$L_{\bar{\zeta}}\psi=const=\kappa\mathcal{L}_o^{-1}, \kappa=\pm 1$.
Thus, the frame rotation
$(\bar{A},\bar{A^*},\partial_z, \partial_\xi)\rightarrow
([\bar{A},\bar{\zeta}],[\bar{A^*},\bar{\zeta}],\partial_z, \partial_\xi)$,
i.e. $[\bar{\zeta},\bar{A}]=\varepsilon\bar{A^*}\,L_{\bar{\zeta}}\psi$
and $[\bar{\zeta},\bar{A^*}]=-\varepsilon\bar{A}\,L_{\bar{\zeta}}\psi $,
gives the following equations for the two functions $(u,p)$:
$$
u_\xi-\varepsilon u_z=\frac{\varepsilon\kappa}{\mathcal{L}_o}\,p, \ \ \
p_\xi-\varepsilon p_z=-\frac{\varepsilon\kappa}{\mathcal{L}_o}\,u \ .
$$

The quantity
$\mathbf{R}=u\,(p_\xi-\varepsilon\,p_z)-p\,(u_\xi-\varepsilon\,u_z)=
\Phi^2L_{\bar{\zeta}}\psi=\kappa\mathcal{L}_o^{-1}\Phi^2$ suggests
to find an integral characteristic of the PhLO rotational nature. In fact, the
two co-distributions $(A,\zeta)$ and $(A^*,\zeta)$ define the two (equal in our
case) Frobenius 4-forms
$\mathbf{d}A\wedge A\wedge \zeta=\mathbf{d}A^*\wedge A^*\wedge \zeta$.
Each of these two 4-forms is equal to
$\varepsilon\mathbf{R}\omega_o=
\varepsilon\mathbf{R}dx\wedge dy\wedge dz\wedge d\xi$.
Now, multiplying by $\mathcal{L}_o/c$ each of them we obtain:
$$
\frac{\mathcal{L}_o}{c}\,\mathbf{d}A\wedge A\wedge \zeta=
\frac{\mathcal{L}_o}{c}\,\mathbf{d}A^*\wedge A^*\wedge \zeta=
\frac{\mathcal{L}_o}{c}\varepsilon\mathbf{R}\omega_o =
\varepsilon\kappa\frac{\Phi^2}{c}\omega_o\ .
$$
Integrating over the 4-volume $\mathbb{R}^3\times(\lambda=2\pi\mathcal{L}_o)$ (and
having in view the spatially finite nature of PhLO) we obtain the finite
quantity $\mathcal{H}=\varepsilon\kappa ET$, where $E$ is the integral energy
of the PhLO, $T=\frac{\lambda}{c}$, which clearly is the analog of the Planck
formula $E=h\nu$, i.e. $h=ET$. The combination $\varepsilon\kappa$ means that
the two orientations of the rotation, defined by $\kappa=\pm 1$, may be
observed in each of the two spatial directions of translational propagation
of the PhLO along the $z$-axis: from $-\infty$ to  $+\infty$, or from $+\infty$
to $-\infty$.

Finally, we can easily see that in case of
$L_{\bar{\zeta}}\psi=\kappa\mathcal{L}_o^{-1}$ and $L_{\bar{\zeta}}\Phi^2=0$ the 3-form
$\delta F\wedge F=\varepsilon \mathbf{R}*\zeta=
\frac{\varepsilon \kappa}{\mathcal{L}_o}\Phi^2*\zeta$
becomes closed: $\mathbf{d}(\delta F\wedge F)=0$, which also gives
an integral conservation law. In fact, the 3-integral of the reduced on
$\mathbb{R}^3$ 3-form $\frac{\mathcal{L}_o^2}{c}(\delta F\wedge F)$ gives
$\varepsilon\kappa ET$, where $E$ is the integral energy, so, the Planck
formula holds.

\vskip 0.5cm
\section{Photon-like Solutions} \index{photon-like solutions figures}
\subsection{Analytical form}
We consider the equations obtained in terms of the two functions
$\Phi=\sqrt{u^2+p^2}$ and $\psi=\mathrm{arctg}\frac{p}{u}$.
The equation for $\Phi$
in our coordinates is $\Phi_{\xi}-\varepsilon \Phi_{z}=0$, therefore,
$\Phi=\Phi(x,y,\xi+\varepsilon z)$, where $\Phi$ is allowed to be spatially
finite, as assumed further, or spatially localized function. The equation for
$\psi $ is $\psi_{\xi}-\varepsilon\psi_{z}=\frac{\kappa}{\mathcal{L}_o}$. Two families of
solutions for $\psi$, depending on an arbitrary function $\varphi$ can be given
by
$$
\psi_1=-\frac{\varepsilon\kappa}{\mathcal{L}_o}z+\varphi(x,y,\xi+\varepsilon z),\ \ \
\text{and}\ \ \ \psi_2=\frac{\kappa}{\mathcal{L}_o}\xi+\varphi(x,y,\xi+\varepsilon z) .
$$
Since $\Phi^2$ is a spatially finite function representing the energy density
we see that the translational propagation of our PhLO is represented by a
{\it spatially finite running wave} along the $z$-coordinate. Let's assume that
the phase is given by $\psi_1$ and, for simplicity, $\varphi=0$. The form of
this solution suggests to choose the initial condition $u_{t=0}(x,y,\varepsilon
z),p_{t=0}(x,y,\varepsilon z)$ in the following way. Let for $z=0$ the initial
condition be located on a disk $D=D(x,y;a,b;r_o)$ of small radius $r_o$, the
center of the disk to have coordinates $(a,b)$, and the value of
$\Phi_{t=0}(x,y,0)=\sqrt{u_{t=0}^2+p_{t=0}^2}$ to be
proportional to some appropriate for the case bump function $f>0$ on $D$ of the
distance $\sqrt{(x-a)^2+(y-b)^2}$ between the origin of the coordinate system
and the point $(x,y,0)$, such that it is centered at the point $(a,b)$, so,
$f(x,y)=f(\sqrt{(x-a)^2+(y-b)^2}\,)$, $D$ is defined by
$D=\{(x,y)|\sqrt{(x-a)^2+(y-b)^2}\leq r_o\}$, and $f(x,y)$ is zero outside $D$.
Let also the dependence of $\Phi_{t=0}$ on $z$ be given by be the corresponding
bump function $\theta(z;\lambda)>0$ of an interval $(z,z+\lambda)$ of length
$\lambda=2\pi\mathcal{L}_o$ on the $z$-axis. If $\gamma>0$ is the proportionality
 coefficient we obtain
\begin{align*}
u=\gamma\,f(x,y;a,b)\,
\theta(ct+\varepsilon z;\lambda)\,\cos(\psi_1), \\
p=\gamma\,f(x,y;a,b)\,
\theta(ct+\varepsilon z;\lambda)\,\sin(\psi_1).
\end{align*}
We see that because of the available $z$-dependent {\it sine} and {\it cosine}
factors in the solution, the initial condition for the solution will occupy a
$3d$-spatial region of shape that is close to a helical tube of height
$\lambda$, having internal radius of $r_o$ and wrapped up around the $z$-axis.
Also, its central helix will always be $\sqrt{a^2+b^2}$-distant from the
$z$-axis. Hence, the solution will propagate translationally along the
coordinate $z$ with the velocity $c$, and, rotationally, inside the
corresponding infinitely long helical tube because of the $z$-dependence of
the available periodical multiples.

\subsection{Figures}
We recall the  figures from Sec.8.8.6 giving a solution with
amplitude function $\Phi$ filling in at every moment a smoothed out finite tube
$\mathcal{H}$ around a circular helix, the pitch of the central helix is
$\mathcal{L}_o$, and phase $\psi=-\varepsilon\kappa \frac{z}{\mathcal{L}_o}$.
The solutions with $\varepsilon=-1$ will propagate left-to-right along the
coordinate $z$.

\begin{center}
\begin{figure}[ht!]
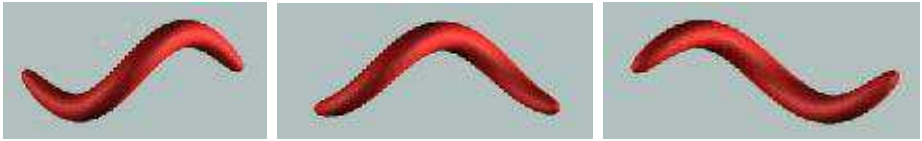

\centerline{
{\mbox{\psfig{figure=photon1.eps,height=1.8cm,width=3.5cm}}
\mbox{\psfig{figure=photon2.eps,height=1.8cm,width=4.2cm}}
\mbox{\psfig{figure=photon3.eps,height=1.8cm,width=4.2cm}}}}
\caption{Theoretical example with $\kappa=-1$. The Poynting vector is
directed left-to-right.}
\end{figure}
\end{center}
\begin{center}
\begin{figure}[ht!]
\centerline{
{\mbox{\psfig{figure=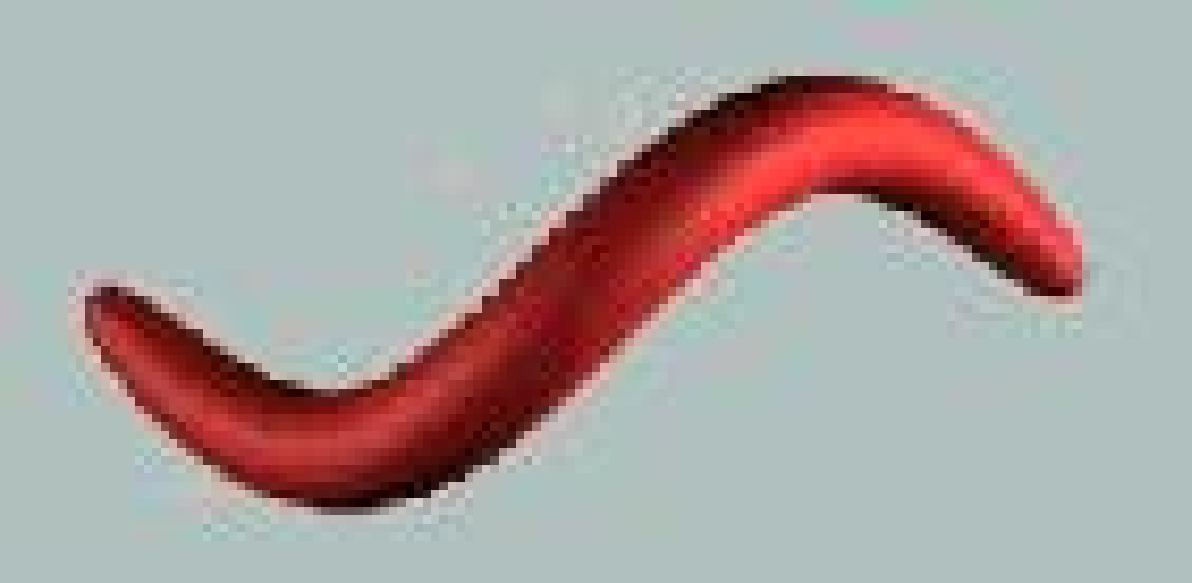,height=1.8cm,width=3.5cm}}
\mbox{\psfig{figure=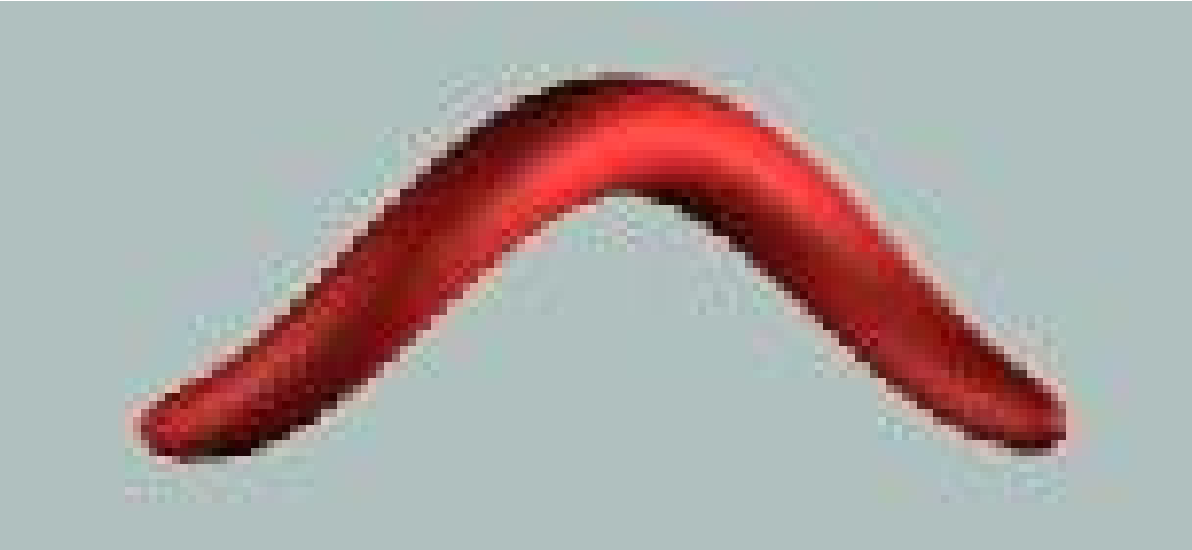,height=1.8cm,width=4.2cm}}
\mbox{\psfig{figure=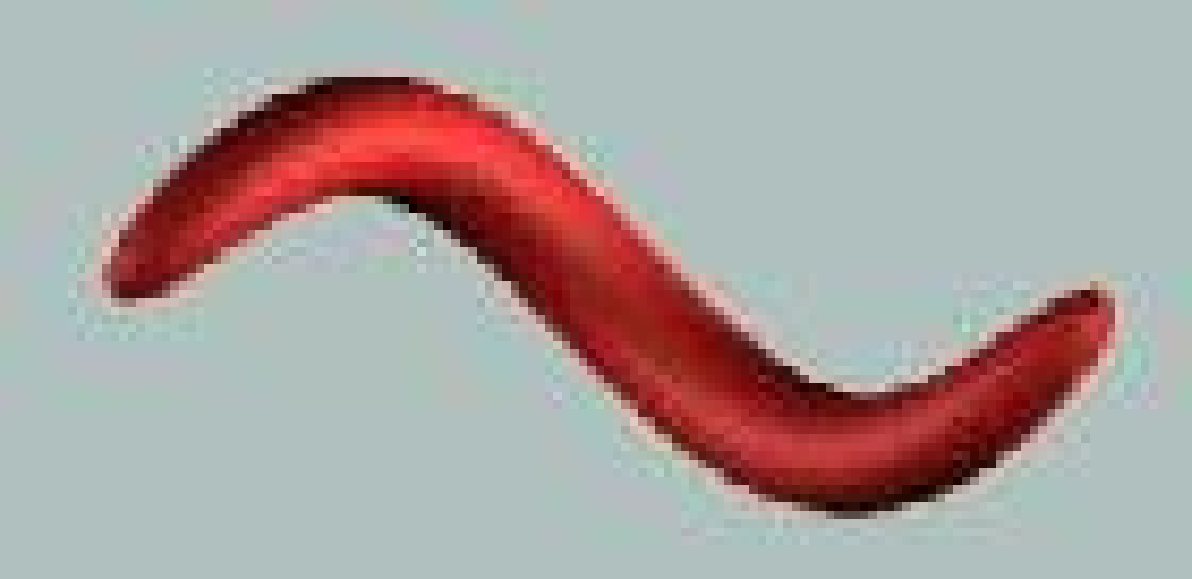,height=1.8cm,width=4.2cm}}}}
\caption{Theoretical example with $\kappa=1$. The Poynting vector is
directed left-to-right.}
\end{figure}
\end{center}

The curvature $K$ and the torsion $\tau$ of the helix line inside
$\mathcal{H}$ through the point $(x,y,0)\in D$ will be
$$
K=\frac{\gamma\,f\theta}{(\gamma\,f\theta)^2+b_o^2},\ \ \ \
\tau=\frac{\kappa\,b_o}{(\gamma\,f\theta)^2+b_o^2} \ ,
$$
where $b_o=\mathcal{L}_o$. The
rotational frequency $\nu$ will be $\nu=c/2\pi\mathcal{L}_o$, so we can introduce
period $T=1/\nu$ and elementary action $h=E.T$, where $E$ is the
(obviously finite) integral energy of the solution defined as 3d-integral
of the energy density $\Phi^2=(\gamma\,f\theta)^2$.

Finally we recall that for photon-like solutions
$$
u=\Phi(x,y,\xi+\varepsilon z)\cos\left(-\varepsilon
\kappa\frac{z}{\mathcal{L}_o}+const\right),
$$
$$
p=\Phi(x,y,\xi+\varepsilon z)\sin\left(-\varepsilon
\kappa\frac{z}{\mathcal{L}_o}+const\right)
$$
the matrices $D_{\mu\nu}$ and $D^*_{\mu\nu}$ from Sec.10.4 have
positive determinants,
$$
det\|D_{\mu\nu}\|=
det\|D^*_{\mu\nu}\|=\frac{1}{\mathcal{L}_o^2}(\Phi.\Phi_\xi)^2>0,
$$
so they define on the Minkowski space-time two definite bilinear forms .

\newpage
{\bf Retrospect}
\vskip 0.3cm
\addcontentsline{toc}{chapter}{{\bf Retrospect}}
\vskip 0.2cm
Modern differential geometry has become a very powerful branch of mathematics
ranging its applicability from simple derivative to differential topology, so,
we would not mistake to call it also {\it invariant analysis}, local, as well
as, global. Extracting trustful integral properties of an object through
studying its local properties is a great step in positive science, and the
serious role of modern differential geometry in this respect could be hardly
disputed and in no way neglected. Its 20th century development has brought a
real intelectual pleasure to all people being able to feel its wonderful
logical harmony, so theoretical physics should make all necessary efforts to
make use of this harmony in trying to understand and describe the harmony of
the physical world. Being part of this physical world we, the people, building
mathematical harmony, we create in fact a very little part of the real world,
called intelectual human knowledge, and in this way we, more or less, justify
ourselves as reality building creatures. Such kind of activity suggests, and
even requires, to pay due respect to every scientific truth and never to turn
scientific truths into dogmas.

One of the most interesting and fruitful initiatives in the development of 20th
century mathematics was to formulate, introduce and appropriately use the
concept of {\it mathemetical structure}. So, the mathematical sets now reveal
their nature through the relations among their elements and with elements of
other sets. Moreover, some elements may belong to various sets according to
their properties, such elements may suggest how to connect the corresponding
mathematical structures and to realize the important idea for {\it
compatibility of structures} carried by a given set. In this way, mathematics
sets a real claim to be used as appropriate logical world for creating adequate
models of the real world.

Let's peep into some of these structures that have demonstrated corresponding
adequacy.

The mostly studied and used such algebraic structure seems to be that of {\it
linear/vector space} $V$ over a field/ring $\mathbb{R}$. Inside such a space we
can represent each element/vector in terms of other vectors through the {\it
adding/subtracting and scalar multiplication} machinery, but we can not get out
of the space in this way. However, this structure allows to consider {\it
subspaces} of $V$, and to introduce additional algebraic structure, usually
called {\it multiplication}, such that the corresponding product of two vectors
of a given subspace $W\subset V$ to leave this subspace. Moreover, the products
of vector spaces have also been introduced and used. Paying due respect to the
idea for compatibility and trying to resolve all arising problems, mathematics
has built the so called tensor algebra over $V$, and its {\it
antisymmetric/exterior} and {\it symmetric} subalgebras.

A very important moment in the whole mathematics is to find procedures to
produce elements of $\mathbb{R}$ from such structures. In these
algebraic structures this is achieved mainly through the concept of {\it
duality} : to every linear space $(V,\mathbb{R})$ is associated another linear
space $(V^*,\mathbb{R})$, the elements of which are linear mappings from $V$ to
$\mathbb{R}$. If $V$ is finite dimensional, we can find a basis
$(e_1,e_2,...,e_n)$ of $V$, and unique basis
$(\varepsilon^1,\varepsilon^2,...,\varepsilon^n)$ of $V^*$ satisfying the
relation $\langle\varepsilon^i,e_j\rangle=(id_V)^i_j, i,j=1,2,...,n$. The
identity map of $V$ acquires the representation $id_V=\varepsilon^i\otimes
e_i$, and this representation of $id_V$ has the same form for every couple of
such dual bases.

Considering now the two nonintersecting subspaces $W_1\subset V$ and
$W_2\subset V$ we immediately see that two other subspeces $W_1^*$ and
$W_2^*$ of $V^*$ appear: these are the corresponding duals for $W_1$ and $W_2$,
with corresponding dual bases $(k_1,...,k_p),(\varepsilon^1,...,\varepsilon^p)$
for $(W_1,W_1^*)$ and $(r_1,...,r_q),(\theta^1,...,\theta^q)$ for $(W_2,W_2^*)$.

If we have a multiplication $"\times"$ in $V$, and for some $x,y\in W_1$ we get
$x\times y$ is out of $W_1$, then in order to check if $x\times y$ is
in $W_2$ we have to find if some of the terms of the kind
$\langle\theta^1,x\times y\rangle,...,\langle\theta^q,x\times y\rangle$
are not zero, or equivalently, to check if
$i_{x\times y}(\theta^1\wedge\theta^2\wedge...\wedge\theta^q)$ is
different from zero. If $"\times"$ is the exterior product we can check also if
$i_{x_1\wedge...\wedge x_r}(\theta^1\wedge\theta^2\wedge...\wedge\theta^q),
r<q$, is different from zero.

The same check can be made if the restriction of $"\times"$ to $W_2$ takes
values in $W_1$, and if so, then we may consider $"\times"$ as "communication
agent" between $W_1\subset V$ and $W_2\subset V$. Hence, with respect to the
linear structure in $V$ the subspaces $W_1$ and $W_2$ keep their identities and
do not recognize each other, while with respect to the new algebraic sructure
$"\times"$ these two subspaces intercommunicate, i.e. they recognize each
other: the $q$-form $\theta^1\wedge\theta^2\wedge ... \wedge \theta^q$ is
attractive for at least some of the products $(x\times y)$, where $x,y,\in
W_1$, and the $p$-form $\varepsilon^1\wedge\varepsilon^2\wedge ... \wedge
\varepsilon^p$ becomes attractive for at least some of the products $(u\times
v)$, where $u,v\in W_2$. In the 3-dimensional case the 4-linear map
$R$ (Sec. 6.2.3) represents (with respect to the usual "vector cross product")
a measure of such algebraic intercommunication between subspaces.

Recall now that each subspace in $V$, resp.$V^*$, can be represented by
corresponding multivector $\Psi$, resp.$\Phi$, and since the action of the
communication agent $"\times"$ does not lead to destruction of any subspace of
$V$, we can anumerate those couples of subspaces of $(V_i,V_j^*)$,
which are interesting for us from a definite point of view, by the
basis vectors $J_1,J_2,...$ of another appropriate vector space $H$. Let now
$\Psi_i=X_{i1}\wedge X_{i2}\wedge...\wedge X_{ip}$ represent $V_i$, and let
$\Phi^j=\alpha^{j1}\wedge\alpha^{j2}\wedge...\wedge\alpha^{jq}$ represent
$V^*_j$ and $p\leq q$. In this way we come to objects of the kind
$(\Psi_1\otimes J_1, \Psi_2\otimes J_2,...)$ and $(\Phi^1\otimes J_1,
\Phi^2\otimes J_2,...)$. This allows to separate some of the
$"\times"$-intercommunicating subspaces of $V$ by means of appropriate
multiplication in the space $H$, e.g., by means of an appropriate bilinear map
$\varphi:H\times H\rightarrow H$, according to
$$
i_{\Psi_i}\Phi^j\otimes\varphi(J_i,J_j), \ \
i,j=1,2,...dim(H).
$$
So, two subspaces $W_1,W_2$ of $V$ will intercommunicate
if the corresponding $(W_1,W_2)$-representing $H$-valued multivectors
$\Psi_a\otimes J_a, a=1,2$ (no summation) have nonzero flows across the
$(W_2^*,W_1^*)$-representing $H$-valued exterior forms $\Phi^b\otimes J_b,
b=2,1$, and the corresponding
value of $\varphi(J_a,J_b)$ is nonzero: if $p<q$ then we expect
 $i_{\Psi_1}\Phi^2\neq 0, \varphi(J_1,J_2)\neq 0$, and if $q<p$, then we expect
$i_{\Psi_2}\Phi^1\neq 0, \varphi(J_2,J_1)\neq 0$.
This suggests to talk about $\varphi$-{\it algebraic
equilibrium} between two intersecting subspaces of the same dimension if the
mutual flows differ just in sign: $$ i_{\Psi^a}\Phi_b=-i_{\Psi^b}\Phi_a , \ \
a\neq b. $$

Clearly, every element $x$ of a subspace $W\subset V$ can be considered as a
symmetry generator of $W$ with respect to the linear structure of $W$, since
the sum $(x+v), v\in W$, is always in $W$. If $"\times"$ is a product in $V$,
then $x\in W$ may, or may not, be a $"\times"$-symmetry generator of $W$, i.e.,
the set $x\times v, v\in W$, shortly denoted by $x\times W$,
 may, or may not, coinside with $W$. A similar question may be raised also
for some $x$ outside $W$.

Let now all elements of the series $x_1,x_2,...$ be outside $W\subset V$ and
the subsets
$$
W_o=W, \ W_1=x_1\times W_o, \ W_2=x_2\times W_1, \ W_3=x_3\times W_2, \ ...
$$
be isomorphic subspaces of $V$
with nonzero consecutive intersection between $W_i$ and $W_{i-1}$,
, i.e., the elements $x_1,x_2,x_3,...$
generate through the introduced product $"\times"$ a number of correspondingly
intersected isomorphic subspaces of $V$. In such a case we may say that the
series $x_1,x_2,...$ defines an {\it external} series of isomorphic to $W$
subspaces, or, that $W\subset V$ is {\it algebraically}, or {\it
discontinuously}, extensible along the series $x_1,x_2,...\in V$.

We turn now to the corresponding addaptation of this purely algebraic picture
of subspace intercommunication to the manifold case, i.e., to available smooth
structure on appropriate topological spaces, where tensor algebras in the
tangent/cotangent bundles over the manifold are considered. Making use of
the corresponding differential geometric structures and relations we can
construct mathematical objects, carrying appropriate infinitesimal symmetries.
Then, an adequate riemannian, or pseudo-riemannian, structure would allow such
mathematical objects to be interpreted as models of really existing,
time-stable and carrying dynamical structure physical objects. The basic idea
that we have followed throughout the book is to extend  available Frobenius
integrability of space-like distributions $(X_1,X_2,...,X_p)$ along appropriate
from physical viewpoint infinitesimal symmetry $\bar{\zeta}$ $$
(X_1,X_2,...,X_p)\rightarrow (X_1,X_2,...,X_p\,;\bar{\zeta}),
$$
where the new integrable
distribution $(X_1,X_2,...,X_p\,;\bar{\zeta})$ will represent the
entire space-time propagational nature of the physical object, and
the object's internal local dynamical structure, i.e. local energy-momentum
exchange among recognizable time-stable subsystems of the object, will be
represented by the curvature forms of available nonintegrable
subdistributions, including elementary ones of the kind $(X_i,\bar{\zeta})$.
This opens the door before the curvature object as a measure of
nonintegrability to enter theoretical physics as adequate and appropriate tool
for local measure of field interaction, i.e. of local energy-momentum exchange.

 We briefly present now the main steps we have made.

The starting point we have paid due respect was the {\it intrinsic dynamical
nature} of a vector field $X$ on a manifold: every vector field
determines a flow, i.e., a family of local (in some cases global)
diffeomorphisms $\varphi_t$ of the manifold, and, conversely, every such family
of diffeomorphisms is determined by a vector field.

The next point we have paid due respect was the interior product (sometimes
called substitution operator, or contraction operator, or insertion operator)
of a vector field $X$ and a differential form $\alpha$ on the manifold denoted
by $i(X)\alpha$, or $i_X\alpha$. This interior product we called {\it local
flow of $X$ across $\alpha$}, and we paid equal respect to both: the vector
field and the differential form. The (anti)derivation property of this product
with respect to the exterior product in the algebra of differential forms we'd
like to specially mention. The natural extention of this interior product to
$p$-vectors according to
$$
i(X_1\wedge X_2\wedge ... \wedge X_p)\alpha=
i(X_p)\circ i(X_{p-1})\circ ... \circ i(X_1)\alpha ,  \ \ \ p\neq 1,
$$
however, loses this (anti)derivation property.

In view of the supposed purpose to make use of this extension in defining local
physical interaction, or field interaction, we made additional
$\varphi$-extension of the interior product to vector valued $p$-vectors
$\Psi_i\otimes e_i$ and vector valued $q$-forms $\alpha^j\otimes k_j$ according
to
$$
i^{\varphi}(\Psi_i\otimes e_i)(\alpha^j\otimes k_j):=
i(\Psi_i)(\alpha^j)\otimes\varphi(e_i,k_j),
$$
where $\{e_i\}$ is a basis of a vector space $V_1$; $\{k_j\}$ is a basis of a
vector space $V_2$. The values $\varphi(e_i,k_j)$ of the bilinear map
$\varphi: V_1\times V_2\rightarrow W, i=1,2,...,dim\,V_1, j=1,2,...,dim\,V_2$,
distinguish those local flows $i(\Psi_i)(\alpha^j)$ which are nonzero, and in
this way define local interaction.

Recalling the existing extension of the Lie derivative of differential forms
with respect to $p$-vectors $\Psi$ (Sec.2.8.3), denoted by $L_{\Psi}\alpha$,
this $\varphi$-extension of the interior product allowed to define
$\varphi$-extended Lie derivative $\mathcal{L}^{\varphi}$ of a $V_2$-valued
differential form $\Phi=\alpha^j\otimes k_j$ with respect to a $V_1$-valued
$p$-vector $\Psi_i\otimes e_i$ according to
$$
\mathcal{L}^{\varphi}_{\Psi_i\otimes e_i}(\alpha^j\otimes k_j) :=
(L_{\Psi_i}\alpha^j)\otimes\varphi(e_i,k_j).
$$
In particular, such extension is, clearly, applicable to Lie algebra
$\mathfrak{g}$ valued $\Phi$ and $\Psi$, and $\varphi$ may, or may not,
coincide with the available Lie bracket in $\mathfrak{g}$. So, the Yang-Mills
theories are easily extensible to $\mathfrak{g}$-valued multivectors $\Psi$ and
differential forms $\Phi$ provided some connection form $\mathcal{A}$ is given.

This $\varphi$-extended Lie derivative was further extended
to $\varphi$-extended covariant Lie derivative of vector bundle valued
differential forms, where $\varphi$ is a vector bundle morphism $\varphi\,:
\eta_1\times\eta_2\rightarrow\eta_3$, these three vector bundles are on the
same base manifold and are endowed with linear connections (Sec.3.7.1).

The next step was to connect the Frobenius integrability of
distributions with the extended Lie derivative (Sec. 3.2.3) by means of
the introduced there explicit expressions for the values of
the associated to the distribution two curvature forms, whose values we called
 in Sec.6.3.3 {\it flow generators} $\mathfrak{D}_{(1,p)}^{(p+1,n)}$ and
$\mathfrak{D}_{(p+1,n)}^{(1,p)}$
of the intercommunication between the distribution
considered and its transversal, and finally, the $CI$-{\it operators}
$\mathbb{D}_{(1,p)}^{(n-p)}$ and $\mathbb{D}_{(1,n-p)}^{(p)}$ giving
explicitly the local values of the exchanged quantities between the two
distributions, i.e., characterizing quantitatively the intercommunication
between the two distributions. In particular, if $\varphi$ is the symmetrized
tensor product we may obtain a picture of "partnership": each distribution
to gain locally as much as it localy loses, in this way both distributions
protect their individuality and recognizability.

The next important moment was to make use of the concept of infinitesimal
shuffling symmetry $\bar{\zeta}$ of an integrable distribution $\Delta^p$ and
to interprit such a symmetry as a generator of allowed formal propagation of
our distribution along the flow of diffeomorphisms defined by $\bar{\zeta}$.

We extended the concept of infinitesimal symmetry of a distribution $\Delta$
with respect to a vector field to infinitesimal (or, local) symmetry of
$\Delta$ with respect to a $p-$vector field, and, finally, with respect to
another distribution $\Delta_1$ (Sec.3.2.2).

The above mentioned intercommunication among the nonintegrable 2-dimen\-sional
subdistributions of $\Delta^p$ of the kind $(F_i=X_i\wedge\bar{\zeta})$ through
the corresponding curvature forms and $CI$-operators, we considered as
appropriate mathematical picture of the interaction of time recognizable
subsystems of a time-stable and space propagating along $\bar{\zeta}$ physical
object, the subsystems of which are represented by the communicating partners
among these $F_i$ through the corresponding subcodistributions
$F_j^*=\eta^{-1}(X_j)\wedge\zeta$.

The zero value of each {\it balance}-operator: $i(F_i)\mathbf{d}F_i^*=0$, (no
summation along $i$), (Sec.3.2.3), we consider as appropriate formal
requirement for a permanent recognizability of the subsystem formaly described
by $F_i$. Under this assumption, in terms of the $\varphi$-extended Lie
derivative, the equation (summation on $i$ and $j$) $$
\mathcal{L}^{\varphi}_{F_i\otimes e_i}(F_j^*\otimes e_j)=0 $$ will reduce to
corresponding mutual local $\varphi$-balance among the interacting couples of
subsystems.

Let's turn now to physics.
%\newpage

The viewpoint we paid due respect in this book consists in the following.
Physical reality demonstrates itself through creating spatially finite entities
called by us {\it physical objects}. These entities show two aspect nature:
{\it physical appearance} and {\it time existence and recognizability}. The
physical appearance of a physical object is understood as corresponding {\it
stress-strain abilities} presenting the spatial structure of the object. The
{\it time existence and recognizability} require {\it survival abilities},
which demonstrate the {\it dynamical appearance} of the oblect through building
stress-energy-momentum potential, formally represented by the tensor
$Q^{\mu\nu}$ on one hand, and corresponding {\it acting instruments}, i.e.,
local flows, formally represented by the interior products of vector valued
$p$-vector fields across appropiate vector valued differential forms, on the
other hand. A distinguished part of these local flows appear as constituents of
the divergence $\nabla_{\mu}Q^{\mu\nu}$.

Following this view, in trying to understand our observational knowledge of the
real world we must be able to separate the {\it important} structural and
behavioral properties of the real objects, i.e., their {\it physical
appearance} and their {\it time existence and recognizability}, i.e., their {\it
dynamical appearance}.

One of the basic in
our view lessons that we more or less have been taught is that the physical
objects demonstrate physical appearance as {\it spatially finite entities}, and
that for their detection and further study we must get knowledge of their {\it
dynamical appearence} making use of
some sufficiantly universal physical quantities. Historically, the physical quantity called {\it
stress-energy-momentum}, has proved to satisfy the basic needed requirements,
since it is universal, i.e., every physical object necessarily carries
energy-momentum and every interaction between two physical objects has such an
energy-momentum exchange aspect. The second lesson concerning any interaction
is that, beyond its {\it universality}, energy-momentum is a {\it conserved}
quantity, so NO loss of it is allowed: it may only pass from one object to
another. This means that {\it annihilation} processes shall give birth to {\it
creation} processes, and the full energy-momentum that has been in posession of
the annihilated objects, to be carried away by the newly created ones.
Energy-momentum always needs carriers, as well as every physical object always
carries energy-momentum.  Hence, the energy-momentum exchange abilities of any
physical object realize its protection against external influence on one hand,
and reveal its intrinsic nature when appropriately viewed, on the other hand.
Therefore, our knowledge about the entire complex of properties of a physical
object relies on getting information about its abilities in this respect and on
finding corresponding quantities describing quantitatively these abilities.

The spatially finite nature of a physical object implies spatial structure and
finite quantity of energy-momentum needed  for its creation, so NO
structureless and infinite objects may exist. The approximations for "point
object" and "infinite field", although useful in some respects, seem
theoretically inadequate [1], and we follow the opinion that they should not be
considered as basic ones. More reliable appears to be the notion for {\it finite
continuous object being in continuous dynamical equilibrium with the physical
environment, that is guaranteed by the time stable compatibility of an
appropriate internal dynamical structure}, which we tried to follow throughout
this book. This view suggested that nonlinear partial differential
equations should not be ignored as basic tools for building mathematical models
of local nature of such objects. The natural physical sense of these equations
is, therefore, supposed to be local energy-momentum exchange.

Keeping in mind that physical objects are {\it many-aspect
entities}, we paid due respect to the complicated structure they may have, and
that their very existence should be connected with {\it internal}
energy-momentum exchange/redistri\-bution among the various structural and
time-recognizable components/subsys\-tems. So, the mathematical model objects
should be many-component ones, and with appropriate mathematical structure.  Of
basic help in finding appropriate mathematical objects is having knowledge of
the internal symmetry properties of the physical object under consideration.
This "step by step" process of getting and accumulating important information
about the physical properties of natural objects reflects in the "step by step"
process of further refining the corresponding mathematical models.

The supposed {\it many-aspect} nature of a physical object $\Phi$ sets the
question: which aspects are {\it identifying} for the object, and which aspects
are allowed to change without influencing the nature of the object? In other
words, which changes are {\it admissible}, i.e., leaving the object {\it
recognizable} as the same after been subject to external influence, and which
changes concern the object's nature, i.e., leading to object(s) of different
nature. In order to answer this question theoreticians must elaborate
theoretical rule(s). One of the ways that mathematics approaches this problem
is by means of building appropriate for the case {\it coupling} : from the
change object $\mathcal{D}\Phi$ and the initial object $\Phi$ (both having
tensor nature) is built another object $\mathfrak{P}(\mathcal{D}\Phi,\Phi)$,
and if $\mathfrak{P}(\mathcal{D}\Phi,\Phi)=0$ it is said that the change
$\mathcal{D}\Phi$ is {\it admissible}, otherewise, the change is {\it not
admissible}, so, $\mathfrak{P}(\mathcal{D}\Phi,\Phi)=\Phi'$, where
$\Phi'\neq\Phi$. For example, the Lie bracket $[X,X]=0$ always, and the Lie
bracket $[X,Y]$ may be not zero. Also, if the covariant derivative $\nabla \sigma$ of a
vector bundle section $\sigma$ is zero, it is said that the section is
parallel with respect to the corresponding linear connection, if $\nabla
\sigma\neq 0$ but $\nabla_{X}\sigma=0$, it is said that $\sigma$ is
$\nabla$-parallel with respect to $X$, finally, if $\nabla_{X}\sigma\neq 0$,
then a new object has been produced. The general idea here is that the object
and its change {\it must be refered somehow to each other} in order to find
corresponding compatibility, or noncompatibility. This view made us make use of
the concept of flow of a $p$-vector field $\Psi$ across a differential
form $\alpha: i_{\Psi}\alpha$, or $i_{\Psi}\mathbf{d}\alpha$, and its
$\varphi$-extension $i^\varphi_{\Psi}\alpha; i^\varphi_{\Psi}\mathbf{d}\alpha$, as
measures of local mathematical influence, and to physically interpret the result
as local physical interaction, i.e., when the physical influence between two
recognizable physical objects, formally represented by $\Psi$ and $\alpha$, may
be ignored, and when it may not be ignored. The happy moment here is, that
similar quantities are used in the Frobenius integrability theory, so, the
Frobenius curvature forms appear as natural formal quantities to be used for
describing available local physical interaction.  In this way we came to the
above mentioned $\varphi$-extended Lie derivative
$\mathcal{L}^{\varphi}_{\Psi_i\otimes e_i}(\alpha^j\otimes k_j)$ as a basic
mathematical tool able to represent local dynamical interaction between/among
appropriate subsystems of a time-stable and, possibly, space-time propagating
real physical field system.

The concept of field object entered theoretical physics through corresponding
interpretation of the Newton law of gravitation and through the Coulomb law of
interaction of two electric charges. According to the traditional view these
laws say that each mass/charge particle generates field around it, this field
has vector nature, it depends on the distance $r$ from the source object as
$r^{-2}$, it is spherically symmetric with respect to the source object, it is
static and is able to physically act on other  appropriate mass/charged
objects. As discussed in Sec.6.1.1, we can not accept the theoretical
assumption that a static field object can act upon other physical objects when
the whole system is isolated. With respect to the corresponding source-frame,
such a field can not carry momentum at all, so, in our view, the still met in
textbooks standard way of inroducing the concepts of static electric, magnetic
and gravitational fields as able to realize direct mechanical action, is not
adequate to the reality and, therefore, has to be reconsidered. Another open
question is: why in such an interpretation of Coulomb law:
$\mathbf{F}=q\mathbf{E}_Q$, only one of the two fields presents, so, isn't it a
theoretical absurd to ignore the really existing physical field
$\mathbf{E}_q$ in terms of which the charge $q$ is theoretically defined
according to the Gauss-Stokes theorem, recognizing at the same time the field
$\mathbf{E}_Q$ of the other particle? Isn't it clear that close enough to each
of the particles the field of the other one is much weaker than its proper, so
that no neglect is admissible?

From a slightly more general point of view when we consider an isolated
mechanical system consisting of time-recognizable subsystems like particles and
potential fields theoretical physics should answer mainly two questions:
{\it first}, what is the physical reason making this isolated system develop
from one configuration to another following the least action principle; {\it
second}, what is the nature of the physical factor determining the direction of
the configurational changes, so that corresponding time parametrization to be
adequatly introduced.

Our approach to electrostatic fields allowed to pay equal respect to both
fields when we consider regions away enough from the two charged particles. The
introduced integral character of the quantity $qQ/R$, considered as integral
interaction energy of the two fields, gives two things: equal treating of the
two particles and the two fields, and suggestion to associate the required
relation $\frac{d}{dR}\frac{qQ}{R}<0$, when the system is isolated, with the
idea that the system  as a whole aims at configurations with less integral
interaction energy. We must not forget, however, that such a consideration is
made under the supposition "{\it far enough from the charges}". We do not
consider as appropriate to approach in this same way charged microsystems like,
for example, electron-proton system, where the hypoteses for spherical symmetry
and time-independence of the two fields should be reconsidered. It is hardly
believable that the available spin structure of such objects as electrons is
not of dynamical nature.

Let's turn now to the concept of field as presented in classical
electrodynamics by means of Maxwell equations. The great discovories of
Faraday,  Ampere and others in 19th century  have been theoretically summarized
in a system of partial differential equations known as Maxwell equations. These
equations clearly say : {\it the electromagnetic field has two recognizable
constituents}, formally represented by : {\it electric} $\mathbf{E}(x,y,z;t)$
and {\it magnetic} $\mathbf{B}(x,y,z;t)$ vector fields on $\mathbb{R}^3$.
According to the equations

	-the propagation of the field is its intrinsic property and it does not
affect the standard euclidean volume in the 3-space: $\mathrm{div}\mathbf{E}=0,
\mathrm{div}\mathbf{B}=0$,

	-these two constituents represent corresponding stresses, they have
vector nature and their time evolution is strongly interdependent: the time
change of each presumes appropriate spatial nonhomogenity of the other:
$$ {\rm
rot}\,\mathbf{E}+ \frac 1c      \frac{\partial {\mathbf{B}}} {\partial t}=0, \
\ \ {\rm rot}\,\mathbf{B} -
	\frac 1c \frac{\partial {\mathbf{E}}} {\partial t}=0 .
$$
These equations have the following two features:

	-they can not be directly verified since they do not represent
verifiable relations between/among appropriate physical quantities, e.g.,
energy-momentum exchange sense,

	-every scalar component of each of the two vector fields
$(\mathbf{E},\mathbf{B})$ necessarily satisfies the D'Alembert wave equation,
so, no time stability of a spatially finite initial condition in free space
should be expectable, and no lightray-like propagation of a finite impuls could
be understood.

Moreover, the recognition of the vector
$\frac1c\mathbf{E}\times\mathbf{B}$ as momentum density of a propagating field
object does not allow any of the two space-time recognizable constituents
$(\mathbf{E},\mathbf{B})$ to carry momentum separately from the other, which,
on one hand, does not go along with the assumption for space-time
recognizability of $\mathbf{E}$ and $\mathbf{B}$, i.e., as real and propagating
subsystems of the field, on the other hand, suggests reconsideration of the
question: which are the real and keeping their identity during propagation
components/subsystems of the field.

The greatest discovery at the very beginning of the last century was that the
notion of electromagnetic field as suggested by the vacuum solutions of Maxwell
equations is inadequate: the time dependent electromagnetic field is not an
infinite smooth perturbation of the aether, on the contrary, it consists of
many individual time-stable objects, called later {\it photons}, which are
created/destroyed mainly during intra-atomic state-transition processes.
Photons are finite objects, they carry energy-momentum, and after they have
been radiated outside their atom-creator, they propagate as a whole
translationally by the speed of light.  Moreover, their propagation is not
just translational, it includes rotational component, been appropriatly
compatible with the translational one, and which is of {\it
intrinsic} and {\it periodical} nature. The corresponding intrinsic action
for one period $T$ is $h=ET$, where $E$ is the full energy of the photon, and
all photons carry the same intrinsic action $h$. During the entire 20th
century physicists have tried to understand the dynamical structure/nature of
photons from various points of view, and this process is still going on
today.

The developed by the authors Extended Electrodynamics and presented in this
book, is an attempt in this direction. The basic starting observation for
approaching the problem is that the energy-momentum local quantities and
relations of Maxwell-Minkowski mathematical approach do agree with the
experiment, but the free field equations $\mathbf{d}F=0,\ \ \mathbf{d}*F=0$
give non-realistic free time-dependent solutions: they are either {\it strongly
time-unstable}, or {\it infinite}. Hence, these solutions can not be used as
mathematical models of photons, since the latter are time-stable and
spatially finite objects. The basic idea of writing down new nonlinear equations
was to pass to local energy-momentum relations, describing how the internal
energy-momentum exchanges during propagation are carried out. The formal
structure of Maxwell-Minkowski equations and local conservation relations
allowed the extension procedure to be used, so the new nonlinear equations we
came to, contain all Maxwell solutions as exact solutions, a feature that we
consider important from the point of view of applications, but surely not from
theoretical point of view.

In order to come to the new equations we paid due and equal respect to the
$\mathbb{R}^2$-valued differential 2-form $\Omega=F\otimes e_1+*F\otimes e_2$
and its $\tilde{\eta}$ 2-vector image
$\Bar{\Omega}=\bar{F}\otimes e_1+\bar{*F}\otimes e_2$.
We had in view also the dual symmetry
$F\rightarrow *F$ of the
Maxwell-Minkowski stress-energy-momentum tensor
$Q_\mu^\nu$ which satisfies the isotropy (null field) condition $Q_{\mu\nu}Q^{\mu\nu}=0$,
and the mentioned symmetry is obvious from its apparent form
$$
Q_\mu^\nu=-\frac12\big[F_{\mu\sigma}F^{\nu\sigma}+
(*F)_{\mu\sigma}(*F)^{\nu\sigma}\big].
$$
This form of $Q_\mu^\nu$ clearly suggested also to consider the field as having
two dynamically interconnected vector components represented by the two
differential 2-forms $(F,*F)$ on Minkowski space-time. The divergence of this
tensor
$$
\nabla_\nu Q_\mu^\nu=\Big[F_{\mu\nu}(\delta F)^\nu+
(*F)_{\mu\nu}(\delta *F)^\nu\Big]=
\frac{1}{2}\Big[F^{\alpha\beta}\mathbf{d}F_{\alpha\beta\mu}+
(*F)^{\alpha\beta}\mathbf{d}(*F)_{\alpha\beta\mu}\Big]
$$
consists also of two recognizable summond-components:
$$
F_{\mu\nu}\delta F^\nu
=\frac12(*F)^{\alpha\beta}(\mathbf{d}*F)_{\alpha\beta\mu} \ \ \ \text{and}
\ \ \  (*F)_{\mu\nu}(\delta
*F)^\nu=\frac12F^{\alpha\beta}(\mathbf{d}F)_{\alpha\beta\mu}.
$$
In the free field case these two components are zero:
$$
F_{\mu\nu}\delta F^\nu=
(*F)_{\mu\nu}(\delta *F^\nu)=0,
$$
they determine and guarantee in energy-momentum exchange terms the
time-recognizability
for each of the two subsystems, formally represented by $F$ and $*F$, during
propagation. The available internal local process of losing and gaining
energy-momentum in equal quantities between $F$ and $*F$ is quantitatively
described by the third equation: $$ F_{\mu\nu}(\delta
*F)^\nu=-(*F)_{\mu\nu}\delta F^\nu, \ \ \ \text{or}, \ \ \
F^{\mu\nu}(\mathbf{d}*F)_{\mu\nu\sigma}=-(*F)^{\mu\nu}(\mathbf{d}F)_{\mu\nu\sigma}.
$$
So, the dynamics described by these equations is of {\it intrinsic} for the
field nature, it respects the recognizability of each of the two subsystems,
and establishes {\it local dynamical equilibrium} between them. These
relations represented the dynamical appearance of PhlO.

It was found that such an internal local recognizability and
local dynamical equilibrium, i.e., physical and dynamical appearance,
 was successfully represented formally as
constant value of the $"\vee"$-flow of $\bar{\Omega}$ across $\Omega$, and
zero value of the $"\vee"$-flow of the introduced by the authers
extended Lie derivative of $\Omega=F\otimes e_1+*F\otimes e_2$
along its $\bar{\eta}$ image
$\bar{\Omega}=\bar{F}\otimes e_1+\bar{*F}\otimes e_2$:
$$
i_{\bar{\Omega}}^{\vee}\Omega \ \ \text{is constant}, \ \ \
\mathcal{L}^{\vee}_{\bar{\Omega}}\Omega=0.
$$

In studying the nonlinear solutions to these equations with
$|\delta F|\neq 0$ and $|\delta *F|\neq 0$, we found their basic property:
every nonlinear solution has zero-invariants:
$F_{\alpha\beta}F^{\alpha\beta}=(*F)_{\alpha\beta}F^{\alpha\beta}=0$. The
formal identity (Sec.6.3.1)
$$
\frac12F_{\alpha\beta}F^{\alpha\beta}\delta_\mu^\nu=
F_{\mu\sigma}F^{\nu\sigma}-(*F)_{\mu\sigma}(*F)^{\nu\sigma}
$$
now requires the two components $F$ and $*F$ to carry always equal
energy-momentum, so energy-mementum exchange between $F$ and $*F$ during
propagation is allowed only {\it simultaneously} and in {\it equal} quantities.
Moreover, the eigen properties of such $F$, $*F$ and of the
corresponding energy-momentum tensor $Q_\mu^\nu$  determine {\it unique}
isotropic eigen direction $\bar{\zeta}$ along which the solution necessarily
propagates translationally as a whole, which fits well with the photons' way of
propagation.  The corresponding simple form of $F=A\wedge\zeta$ and
$*F=A^*\wedge\zeta$ allowed a complete analysis of the nonlinear solutions to
be made. The whole set of nonlinear solutions consists of nonintersecting
subsets, and each subset is characterized by the corresponding isotropic eigen
direction. Every solution of a given subclass is uniquely determined by two
functions: the amplitude function $\phi$ which is arbitrary with respect to the
spatial variables and is a running wave along $\bar{\zeta}$; the phase function
$\varphi$, where $\varphi$ may depend arbitrarily on all
space-time variables (we made use of the function $\psi=arccos(\varphi)$). The
field scalar components have the form $\phi\,cos\psi$ and $\phi\,sin\psi$, so,
finite solutions with photon-like behavior are allowed.

Two basic characteristics of these nonlinear solutions deserve noting. First,
nonlinear analogs $\vec{\mathcal{F}}$ and $\vec{\mathcal{M}}$ (Sec.7.6) of
classic electric and magnetic components $(\mathbf{E},\mathbf{B})$ were found
such, that our nonlinear equations were represented as $$
\vec{\mathcal{F}}\times\mathbf{E}=0, \ \  \vec{\mathcal{M}}\times\mathbf{B}=0,
\ \ \ \vec{\mathcal{F}}\times\mathbf{B}=-\vec{\mathcal{M}}\times\mathbf{E}.
$$
Second, the natural appearence of the {\it scale factor} $\mathcal{L}_o$,
which we showed that may be introduced in theory in various ways. When defined
by the relation $\mathcal{L}_o=|A|/|\delta F|$ its physical dimension of
[length] becomes obvious. Hence, every such nonlinear solution
 defines {\it its own scale}, and the relation $\mathcal{L}_o\neq 0$
guarantees available rotational component of propagation. As it was shown
further, the case $\mathcal{L}_o=const$ is allowed, which we consider as
intrinsic consistency with the constant translational velocity. So,
corresponding intrinsically defined time-period $T=2\pi\mathcal{L}_o/c$ and
frequency $\nu=T^{-1}$ can be introduced, and the corresponding finite
solutions with integral energy $E$ acquire the characteristic {\it intrinsic
action} given by $ET$.

Hence, the natural question "do there exist nonlinear spatially finite solutions
with compatible rotational and translational components of propagation" was
answered positively. Seven equivalent conditions of quite different nature,
determining when this is possible, were found. It is remarkable that
the condition $\mathcal{L}_o\neq 0$ is one of them. So, for spatially finite
solutions with $\mathcal{L}_o=const$, we have a natural and intrinsically
defined measure of this rotational component of propagation, namely, the
elementary action $ET$, which recalls the corresponding invariant
characteristic of photons, the Planck constant $h$. Anyway, these solutions
deserve to be called "photon-like".

It was very interesting to find that the energy-momentum tensor $Q$ of
a nonlinear solution defines an algebraic boundary operator in the tangent and
cotangent bundles of the Minkowski space-time. The corresponding
homology/cohomology spaces are 2-dimensional, the classes represented by the
electric and magnetic components of the field form a basis of the corresponding
homology/cohomology space. Moreover, since for each nonlinear solution the
corresponding image spaces of $Q$ are 1-dimensional, $Q$ is extensible to
boundary operator in the whole algebra of exterior forms and multivectors.  The
$Q$-image of any differential 2-form is collinear to the nonlinear solution
that generates $Q$, and the $Q$-image $Q(\alpha)$ of any $p$-form $\alpha$ is
isotropic:  $[Q(\alpha)]^2=0$.

The natural representation of the dual group $\mathbb{G}$ in the space of
2-forms leaves the scale factor $\mathcal{L}_o=const$ invariant. As a
consequence we obtained that the 3-form $\delta F\wedge F$ is closed :
$\mathbf{d}(\delta F\wedge F)=0$, which
generates a conserved quantity through Stokes theorem, and this conserved
quantity is proportional to the elementary action $ET$. If the group parameters
depend on the space-time points, then the commutative group structure of
$\mathbb{G}$ generates a group structure inside the subset of solutions with
the same $\bar{\zeta}$. In such a case a "vacuum state" $(F_o,*F_o)$ can be
defined such, that every solution of the subclass is defined by an action upon
$(F_o,*F_o)$ of a point-dependent group element with determinant $(a^2+b^2)$ having
running wave character along the intrinsically defined direction $\bar{\zeta}$.

It worths specially noting that among the seven ways to compute the integral
spin $ET$ of a nonlinear solution an appropriate representative of the
Godbillon-Vey class determined by the completely integrable 1-dimensional Pfaff
system $\zeta$, or $f\zeta$, can be made equal to $\delta F\wedge F$, and so it
can be used to compute the elementary action $ET$.

Finally, in presence of external fields (see Apendix A), we showed that our
general system of nonlinear equations (together with the additional Pfaff
equations) is compatible, and we found a large family of solutions. This family
is parametrized by one function of one-space (say $z$) and one-time independent
variables: $V(z,\xi)$, and two other functions $g(x)$ and $h(y)$. So,
choosing $V(z,\xi)$ to be any (one, or many)-soliton solution of any soliton
equation, and $g(x)$ and $h(y)$ to be finite, we obtain its (3+1) image as a
finite/concentrated solution of our nonlinear equations with well defined
energy-momentum quantities. We illustrated our approach with examples from
the 1-soliton solutions of the well known Sine-Gordon, KdV and NLS
equations.

We would like to note also that the nonlinear vacuum equations obtained follow
the idea that the admissible changes (in our case $\delta *F$ and $\delta F$,
or correspondingly, $\mathbf{d}F$ and $\mathbf{d}*F$) are metrically projected
upon the field components $(F,*F)$, and these projections are assumed to be
zero, or intrinsically connected. This leads directly to nonlinear equations
with corresponding physical sense of local energy-momentum balance relations.
From mathematical point of view this resulted in finding a natural
$\varphi$-extension of the Lie derivative, on one hand, and to generalization
of the geometrical concept of parallelism (Sec.3.7.3), on the other hand, as
worked out and illustrated with many examples from differential geometry and
theoretical physics in Appendix C. This, in turn, suggests natural ways to
nonlinearization of important physical linear and nonlinear equations.

The development of theoretical physics during the last century, and especially
during the last 60 years, shows growing interest to nonlinearization of the
widely used linear equations. At the beginning of the last century Einstein
declared that the fundamental equation of optics $\square \phi =0$, and linear
field equations at all, must be replaced by nonlinear equation(s)
[1], moreover, he launched the creed that a particle may appear only as a
bounded space region where the field intensity and the energy density are
particularly big. It deserves noting that he worked hard in trying to find
such appropriate equations, General Relativity made a decisive step along the
road of nonlinearization. The well known Mie's [2] and Born-Infeld's [3]
nonlinearizations of CED are also steps in this direction. One of the serious
later achievements was the Yang-Mills approach, which dominates nowadays the
various models of field theory. Nevertheless, spatially finite solutions of the
vacuum field equations with appropriate spatial structure and time behaviour are
still rarely met today.

The most frequently met way to nonlinearization usually follows the rule: "{\it
add nonlinear interaction term to the "free" lagrangian giving linear
equations, and see what happens}". We do not share the view that this is the
right way to pass to appropriate (3+1)-dimensional nonlinear equations. We
showed that our approach: "{\it pass to local energy-momentum exchange
relations}", works in the important case of the vacuum photon-like finite
solutions, as well as, in presence of media(Appendix A). Another, in some sense                                re general approach to find appropriate nonlinarizations of the available
linear field equations, was launched (Sec.3.7.3) and tested as working tool
in Appendix C.

If physisists acknowledge that {\bf all free and not-free time-stable objects
in Nature are spatially finite entities carrying dynamical structure}, they
should pay due respect to this philosophy in building theories and must be very
grateful to modern mathematics which gives everything necessary.
%to meet all experimentally motivated visions.

In short, our physical visions and their formal presentation read:
\vskip 1cm
\begin{center}
\hfill\fbox{
    \begin{minipage}{0.97\textwidth}
\begin{center}
\vskip 0.3cm
{\bf
{\it The Visions}:
\vskip 0.3cm

{\bf Physical reality demonstrates itself through creating spatially finite
entities called by us physical objects. Each of these entities
exists among the other ones, keeping its identity and showing two aspect
nature: physical appearance and time existence and recognizability.
The physical appearance of a physical object is understood as corresponding
stress-strain abilities, presenting the spatial structure and
corresponding abilities for internal dynamics. The  time
existence and recognizability require survival abilities, which
demonstrate the dynamical appearance of the oblect through building
corresponding  acting instruments, called local flows, presenting its
interaction abilities with the rest of the world}.
\vskip 0.5cm
{\it The Formal Presentation}:
\vskip 0.3cm
{\bf Space propagation of a physical system as a whole corresponds to
mathematical integrability of symmetry-extended spatial integrable geometric
distribution} $\Delta$, {\bf and any consistent with this spatial propagation
internal dynamical structure corresponds to curvature intercommunication among
the nonintegrable subdistributions of} $\Delta$}.
 \end{center} \vskip 0.3cm
\end{minipage}} \hfill \end{center}
%\newpage
%\vskip 0.5cm
{\bf References}
\addcontentsline{toc}{subsection}{{\bf References}}
\vskip 0.2cm
{\bf 1. A. Einstien}

{\it On the generalized theory of gravitation, Sci.
Amer}., {\bf 182}, 13-17 (1950);

{\it Physics and Reality}, Journ. Franklin
Inst., {\bf 221}, 349-382 (1936);

{\it Remarks concerning the essays brought
together in this co-operate volume}, in "Albert Einstein
philosopher-scientist", ed. by P.A.Schillp, The library of living
philosophers, v.{\bf 7}, Evanston, Illinois, 665-688 (1949);

{\it Zum
gegewartigen Stand des Strahlungsproblems}, Phys.Zs., {\bf 10}, 185-193 (1909)
 \vskip 0.2cm
{\bf 2. G. Mie}
{\it Ann. der Phys.} Bd.{\bf 37}, 511 (1912); Bd.{\bf 39}, 1 (1912);
Bd.{\bf 40}, 1 (1913)
 \vskip 0.2cm
{\bf 3. M. Born, L. Infeld}, {\it Nature}, {\bf 132}, 970 (1932);
 {\it Proc.Roy.Soc.}, {\bf A 144}, 425 (1934).

\newpage
{\bf Appendix A}
\vskip 0.2cm
\addcontentsline{toc}{section}{{\bf Appendix A.}\\
{Extended Electrodynamics in presence of external
fields(media)}}

\vskip 0.2cm
{\bf Extended Electrodynamics in presence of media}
\vskip 0.3cm
{\bf 1. Introduction.}

Let's recall how this situation is treated in standard classical
electrodynamics (CED). In this case of propagation of EM-field inside {\it
macroscopic bodies/media} the approximation {\it physically small volume} is
introduced. Let $l$ denote the average distance among the particles,
creating a given medium, and let $\Delta V$ denote the so called
{\it physically small volume},  now if $L$ denotes some typical linear scale of
the macroscopic object/medium, the following relations are required to hold:
$$
l^3\ll\Delta V\ll L^3 .
$$
Further we assume these conditions satisfied.  It is important, however, to
understand, to realize and to be conscious of the concequences of this
assumption because it strongly changes the region of valididty of concepts like
{\it continuity, differentiability and smoothnes}, compare to the use of these
concepts in the pure field case. Therefore, under these conditions, the very
concept of {\it EM-field} is seriously changed and some well known
and widely used invariant characteristics of pure EM-fields may be fully lost.

From practical point of view important class of media are those, which can be
{\it electrified} and {\it magnetized} when placed in external EM-fields. Such
media are called {\it dielectrics}. According to classical electrodynamics this
additional electrifying is due to the presence of {\it bound charges} in these
media. Subject to the action of the external field these charges perform
limited in small regions displacements, which leads to
appearance of additional charges, of currents and dipole moments. After an
averaging over the volume $\Delta V$, they are denoted respectively by
$\rho_b$-{\it bound charge density}, ${\bf j}_b$-{\it bound current density},
and $\mathbf{P}$-{\it polarization vector}. The additional magnetization is due
to the circle-like displacements of the charges, generating in this way new
magnetic moments. The corresponding averaging of these new magnetic moments over
the volume $\Delta V$ defines the {\it magnetization vector} $\mathbf{M}$.
Further we assume that $\mathbf{E},\mathbf{B},\mathbf{P},\mathbf{M}$ have the
same dimension.

In analogy with the case {\it free charges in vacuum} the following relations
among these new quantities are assumed:
$$
\rho_b=-\mathrm{div}\mathbf{P},\quad {\bf j_b}=\mathrm{rot}\mathbf{M}+
       \frac1c \frac{\partial\mathbf{P}}{\partial t}.
$$
After replacing in Maxwell equations {\bf j} and $\rho$ by
$({\bf j}+{\bf j}_b)$ and $(\rho +\rho_b)$ respectively,
the Maxwell's equations for continuous media are obtained:
$$
\frac{1}{c}\frac{\partial\mathbf{D}}{\partial t}=\mathrm{rot}\mathbf{H} -
 {\bf j},\quad \mathrm{div}\mathbf{B}=0,
$$
$$ \frac1c\frac{\partial\mathbf{B}}{\partial t}
=-\mathrm{rot}\mathbf{E},\quad \ \
\mathrm{div}\mathbf{D}=\rho,
$$
where
$$
\mathbf{H}=\mathbf{B}-\mathbf{M},\quad
\mathbf{D}=\mathbf{E}+\mathbf{P}.
$$

When passing from one medium to another, the dielectric properties of which
strongly differ from each other, it is naturally to expect a
violation of the continuous properties of $\mathbf{H}$ and $\mathbf{D}$.
Therefore it is necessary to define the behaviour of these quantities on the
corresponding boundary surfaces. To this end, two new quantities are
introduced: {\it surface density of the electric charge}-$\sigma$ and {\it
surface density of the current}-$i$. Then the analysis of the above equations
brings us to the following relations:
$$
(\mathbf{D}_n)_2-(\mathbf{D}_n)_1=\sigma,
\quad (\mathbf{E}_n)_2-(\mathbf{E}_n)_1=0,
$$
$$
(\mathbf{H}_n)_2-(\mathbf{H}_n)_1= i,
\quad (\mathbf{B}_n)_2-(\mathbf{B}_n)_1=0,
$$
where the index "n" denotes the normal to the boundary surface component of
the corresponding vector at some point.

Assuming that the quantity of electromagnetic energy, transformed to mechanical
work or heat during 1 second in the volume $V$ is equal to
$\int_V({\bf j}.\mathbf{E})dV$, and making use of the above Maxwell's equations
for medium, we get
$$
({\bf j}.\mathbf{E})=
-\frac1c\biggl[\biggl(\mathbf{E}.\frac{\partial\mathbf{D}}{\partial
t}\biggr)+ \biggl(\mathbf{H}.\frac{\partial\mathbf{B}}{\partial
t}\biggr)\biggr]-\mathrm{div}\biggl[\mathbf{E}\times\mathbf{H}\biggr].
$$
Replacing now
$\mathbf{D}=\mathbf{E}+\mathbf{P}$ and
$\mathbf{H}=\mathbf{B}-\mathbf{M}$ in this
relation we obtain
$$ ({\bf j}.\mathbf{E})=-\frac1c\frac{\partial}{\partial t}
\frac{\mathbf{E}^2+\mathbf{B}^2}{2}-
\mathrm{div}\biggl[\mathbf{E}\times\mathbf{B}\biggr]-
$$
$$
-\frac1c\biggl[\mathbf{E}.\frac{\partial\mathbf{P}}{\partial t}-
{\mathbf{M}}.\frac{\partial\mathbf{B}}{\partial t}\biggr]+
\mathrm{div}\biggl[\mathbf{E}\times{\mathbf{M}} \biggr].
$$
These relations describe the local energy-momentum balance.

The above 2-vector and 2-scalar equations have to determine 15 functions
$E_i,B_i,H_i$, $D_i,j_i$. Clearly, more relations among these functions
are needed, in order to determine them. The usual additional relations
assumed are of the kind
$$
P^i=P^i\biggl(E^j,\frac {\partial E^j}{\partial x^k},...;B^j,\frac {\partial B^j}{\partial x^k},...\biggr),\quad
M^i=M^i\biggl(E^j,\frac {\partial E^j}{\partial x^k},...;B^j,\frac {\partial B^j}{\partial x^k},...\biggr).
$$

The most frequently met additional assumption is
$\mathbf{P}=\mathbf{P}(\mathbf{E}),\ \mathbf{M}=\mathbf{M}(\mathbf{B})$
together with the requirement $\mathbf{P}(0)=0,\ \ \mathbf{M}(0)=0$.
 A series development gives

$$ P^i=\kappa^i_j E^j+\frac12 \kappa^i_{jk}E^j E^k+. .
.;\quad M^i=\alpha^i_j B^j+\frac12 \alpha^i_{jk}B^j B^k+. . .
$$ The
tensors $\kappa^i_j,\ \kappa^i_{jk}, . . . $ are called {\it polarization
tensors} (of corresponding rank), and $\alpha^i_j,\alpha^i_{jk},. . . $
are called {\it magnetization tensors } (of corresponding rank). For $D^i$
and $H^i$ we obtain respectively
$$
D^i=E^i+(\kappa^i_j E^j+\frac12\ \kappa^i_{jk}E^j E^k+. . .)=(\delta^i_j +\kappa^i_j)E^j+. . .
$$
$$
H^i=B^i-(\alpha^i_j B^j+\frac12 \alpha^i_{jk}B^j B^k+. . .)=(\delta^i_j-\alpha^i_j)B^j-. . .
$$
If the medium is homogeneous and isotropic and the EM-field is weak, the
nonlinearities in these developments are neglected, so, for such menium,
$$
D^i=(1+\kappa)\delta^i_jE^j =\varepsilon^i_j E^j=\varepsilon\delta^i_jE^j
$$
and
$$
H^i=(1-\alpha)\delta^i_jB^j =\alpha^i_j B^J=\alpha\delta^i_jB^j.
$$

The constants $\varepsilon$ and $\mu=\alpha^{-1}$
are called
{\it dielectric} and
{\it magnetic } permeabilities respectively. In case of nonisotropic media
the two tensors $\varepsilon^i_j$ and $\mu^i_j=(\alpha^i_j)^{-1}$ are used.

In the relativistic formulation  on Minkowski space-time
$M=(\mathbb{R}^4,\eta; *_{\eta})$ besides the 2-form $F$, a new 2-form $S$ is
introduced, namely
$$
S=M_3dx\wedge dy-M_2 dx\wedge dz+M_1dy\wedge dz- P_1
dx\wedge d\xi-P_2 dy\wedge d\xi-P_3 dz\wedge d\xi
$$
as well as a new 4-current
$$
J^\mu_b=({\bf j}_b,\rho_b).
$$
With these notations the "medium
part" of the equations acquires the following compact form
$$
\frac {\partial
S^{\sigma\nu}}{\partial x^\sigma}=-J^\nu_b.
$$
If we introduce now the 2-form
$G=F-S$, then the equations look as follows ($J$ denotes the vacuum
4-current)
$$
\delta *F=0,\quad \delta G=J.
$$
The two relations
$D^i=\varepsilon^i_jE^j$ and $B^i=\mu^i_jH^j$  may be unified in one relation
of the kind
$$
G_{\mu\nu}=R_{\mu\nu}^{..\alpha\beta}F_{\alpha\beta},\ \mu<\nu,\
\alpha<\beta .
$$
Obviously,
$R_{\mu\nu}^{..\alpha\beta}=-R_{\nu\mu}^{..\alpha\beta},\
R_{\mu\nu}^{..\alpha\beta}=-R_{\mu\nu}^{..\beta\alpha}.$
Explicitly,
$$
R_{i4}^{..kl}=0, \ R_{kl}^{..j4}=0,\ R_{i4}^{..j4}=\varepsilon_i^j,
$$

$$
R_{kl}^{..mn}=\tilde{\varepsilon}_{klr}\chi^r_s \tilde{\varepsilon}^{smn},\ \chi^r_s=(\mu^r_s)^{-1},\ k<l,\ m<n.
$$
The equations  $\varepsilon^i_j=\varepsilon^j_i,\ \mu^i_j=\mu^j_i$ lead to
$R_{\mu\nu}^{..\alpha\beta}=R_{\alpha\beta}^{..\mu\nu}$. It is immediately
verified that
$$
R_{\mu\nu}^{..\alpha\beta}+R_{\mu\alpha}^{..\beta\nu}+R_{\mu\beta}^{..\nu\alpha}=0.
$$
The  $(6\times 6)$ matrix $R_{\mu\nu}^{..\alpha\beta}$ looks as follows:
$$
R_{\mu\nu}^{..\alpha\beta}=
\begin{Vmatrix}
\chi_3^3   &-\chi_2^3  &\chi_1^3      &0               &0               &0\\
-\chi_3^2  &\chi^2_2   &-\chi_1^2     &0               &0               &0\\
\chi_3^1   &\chi_2^1   &\chi^1_1      &0               &0               &0\\
0          &0          &0             &\varepsilon_1^1 &\varepsilon_1^2
&\varepsilon_1^3 \\ 0          &0          &0             &\varepsilon_2^1
&\varepsilon_2^2 &\varepsilon_2^3 \\ 0          &0          &0
&\varepsilon_3^1 &\varepsilon_3^2 &\varepsilon_3^3
\end{Vmatrix}
$$
For the invariant $R=R_{\mu\nu}^{..\mu\nu}$ we obtain
$$
R=2(\varepsilon_1^1+\varepsilon_2^2+\varepsilon_3^3+\chi_1^1+\chi_2^2+\chi_3^3).
$$
These algebraic properties of the tensor  $R_{\mu\nu}^{..\alpha\beta}$
are the same as those of the Riemann curvature tensor. Since for vacuum we
have  $\varepsilon_i^j=\chi_i^j=\delta_i^j$
for the vacuum $R_{\mu\nu}^{..\alpha\beta}$ we get
$$
R_{\mu\nu}^{..\alpha\beta}=\delta_\mu^\alpha \delta_\nu^\beta - \delta_\mu^\beta\delta_\nu^\alpha,
$$
or
$$
R_{\mu\nu,\alpha\beta}=\eta_{\mu\alpha} \eta_{\nu\beta} - \eta_{\mu\beta}\eta_{\nu\alpha},
$$
which is exactly the induced by $\eta$ metric in the bundle of 2-forms over
the Minkowski space-time.

Now we are going to consider the energy-momentum distribution
of the field in presence of an active medium. Recall that in case of vacuum,
these quantities are described by the energy-momentum tensor
$$
Q_\mu^\nu=\biggl[\frac 14 F_{\alpha\beta}F^{\alpha\beta}\delta_\mu^\nu-F_{\mu\sigma}F^{\nu\sigma}\biggr]=\\
\frac{1}{2}\biggl[-F_{\mu\sigma}F^{\nu\sigma}-(*F)_{\mu\sigma}(*F)^{\nu\sigma}\biggr].
$$
The natural generalization of this tensor in presence of a new 2-form $S$, or $G$,
looks as follows
$$
W_\mu^\nu=\frac {1}{2}\biggl[\frac 12
F_{\alpha\beta}G^{\alpha\beta}\delta_\mu^\nu-F_{\mu\sigma}G^{\nu\sigma}-G_{\mu\sigma}F^{\nu\sigma}\biggr].
$$
Using the identity, which holds for any two 2-forms $(F,G)$ in the Minkowski
space
$$ \frac 12
F_{\alpha\beta}G^{\alpha\beta}\delta_\mu^\nu=F_{\mu\sigma}G^{\nu\sigma}-(*G)_{\mu\sigma}(*F)^{\nu\sigma},
$$
for $W_\mu^\nu$ is obtained
$$
W_\mu^\nu=\frac
{1}{2}\biggl[-F_{\mu\sigma}G^{\nu\sigma}-(*F)_{\mu\sigma}(*G)^{\nu\sigma}\biggr]=
\frac
{1}{2}\biggl[-G_{\mu\sigma}F^{\nu\sigma}-(*G)_{\mu\sigma}(*F)^{\nu\sigma}\biggr].
$$
Obviously, $W_{\mu\nu}=W_{\nu\mu}$, and if $S_{\mu\nu}$ $\rightarrow 0$,
or equivalently, $G=F$, we get $W_\mu^\nu$$\rightarrow Q_\mu^\nu$. Here are
the explicit expressions of $W_\mu^\nu$ by means of the components of the
 3-vectors $\mathbf{E,B,D,H}$:
$$
W_i^j=\frac{1}{2}\biggl[E_i D_j +E_j D_i+B_i H_j+B_jH_i+
\delta_i^j(\mathbf{B}.\mathbf{H}-\mathbf{E}.\mathbf{D})\biggr],
$$
$$
W_i^4=\frac{1}{2}\biggl[(\mathbf{E}\times \mathbf{H})_i +
(\mathbf{B}\times \mathbf{D})_i\biggr],
\quad W_4^4=\frac{1}{2}(\mathbf{E}.\mathbf{D}+\mathbf{B}.\mathbf{H}).
$$
It is easily verified the
following relation
$$
\nabla_\nu W_\mu^\nu=\frac
{1}{2}\biggl[F_{\mu\nu}(\delta G)^\nu+G_{\mu\nu}(\delta F)^\nu+
(*F)_{\mu\nu}(\delta *G)^\nu+(*G)_{\mu\nu}(\delta *F)^\nu\biggr].
$$
Note that, if we require at $J=0$, i.e.,  $\delta G=0$, the following local
conservation law to hold:
$$
\nabla_\nu W_\mu^\nu =0,
$$
then, making use of the above introduced definitions, we come to the equation
$$
 S_{\mu\nu}(\delta S)^\nu=
F_{\mu\nu}(\delta S)^\nu-(*F)_{\mu\nu}(\delta *S)^\nu,
$$
or in coordinate free form ($\bar{F},\bar{*F},\bar{*S}$
denote the corresponding bivectors)
$$
i(\bar{*S})\mathbf{d}*S=i(\bar{*F})\mathbf{d}*S-
i(\bar{F})\mathbf{d}S.
 $$

One of the informations that we get from
this last relation is, that some of the energy-momentum exchange
between the field and the medium is performed through the flows of the vector
fields $\bar{\delta S}$ and $\bar{\delta *S}$ across the 2-forms $F$ and $*F$.
In view of the following general relation on Minkowski space-time
$M=(\mathbb{R}^4,\eta)$
$$
F_{\mu\nu}X^\nu=(*F)^{\alpha\beta}(*\tilde{\eta}(X))_{\alpha\beta\mu},
 \ \ \text{so,} \ \ i(\bar{\delta S})F=-i(\bar{*F})*\delta S,
$$
where $i(.)$ denotes the interior product by a (multi)vector,
this relation suggests also an alternative
view: the field subsystems, represented by the 2-vectors $(\bar{F},\bar{*F})$
flow across appropriately generated by the medium 3-forms $\beta^1, \beta^2,
...$, and since the medium survives during this interaction, these 3-forms
should generate, in turn, appropriate 1-forms $\alpha^1, \alpha^2, ...$ , in
terms of which this survival to be formally expressed by corresponding
Frobenius integrability relations.

Further we are going to consider how this view to be explicitly realized.

\vskip 0.3cm

 {\bf 2. The new equations.}

We assume here that our field $(F,*F)$, although seriously modified, propagates
suxessfully inside a medium, and a permanent local energy-momentum exchange
between the field $(F,*F)$ and the medium takes place. In order to describe
formally this exchange we need the mathematical image(s) of the medium's ability
instruments in this respect. The corresponding local quantities describing the
local balance between flow-out and flow-in of the energy-momentum concerning
both partners, in accordance with the local energy-momentum conservation law,
might be expressible in terms of $(F,*F)$, in terms of the mathematical images
of the medium's instruments, or in terms of both $(F,*F)$ and the medium's
instruments.

As we mentioned above, Classical electrodynamics (CED) approaches this
situation following the assumption that most of the interesting in this respect
media react to the "invasion" of the external $(F,*F)$-field through creating
appropriate {\it proper} field $(S,*S)$ in terms of two space-like vector
fields $(\mathbf{P},\mathbf{M})$, and $(S,*S)$ are expressed in terms of
$(\mathbf{P},\mathbf{M})$ almost in the same way as $(F,*F)$ are expressed in
terms of $(\mathbf{E},\mathbf{B})$:
$$
S=M_3dx\wedge dy-M_2 dx\wedge
dz+M_1dy\wedge dz- P_1 dx\wedge d\xi-P_2 dy\wedge d\xi-P_3 dz\wedge
d\xi,
$$
$$ *S=-P_3dx\wedge dy+P_2 dx\wedge
dz-P_1dy\wedge dz- M_1 dx\wedge d\xi-M_2 dy\wedge d\xi-M_3 dz\wedge
d\xi.
$$
It is seen that $(-\mathbf{P})$ plays the role of electric constituent, and
$\mathbf{M}$ plays the role of magnetic constituent inside $(S,*S)$.
Also, it is assumed that $(S,*S)$ appear and change only in presense
and change of $(F,*F)$, and $(S,*S)\rightarrow 0$ when the external field
$(F,*F)$ is absent. From this viewpoint, it seems more natural the flows of
$\bar{F}$ and $\bar{*F}$ across the differentials $\mathbf{d}S$ and
$\mathbf{d}*S$ to be considered as local energy-momentum characteristics of
the local physical interaction between the field and the medium.

It is important to specially stress that CED allows energy-momentum exchanges
between the external field and the medium to be performed {\it only}
with one of the two field constituents $(F,*F)$. The traditional
justification/motivation for this is the absence of magnetic charges. In our
view this motivation is insufficient and has to be reconsidered.

In our approach, based on Extended electrodynamics (EED), we
shall keep in mind the following:

$\mathbf{1^o}$. Compare to the free field case,
the nature of the field $(F,*F)$ may now {\it significantly} change.

$\mathbf{2^o}$. The interaction, i.e. the energy-momentum exchange
{\it field $\leftrightarrow$ medium}, must NOT destroy the medium.

So, according to $\mathbf{1^o}$, admitting {\it significant change of the
nature of the field $(F,*F)$}, the two vacuum ivariants
$\frac12F_{\mu\nu}F^{\mu\nu}$ and $\frac12(*F)_{\mu\nu}F^{\mu\nu}$ may be NOT
zero. In view of the great diversity of electromagnetically active media, we
are going to consider for now only those media, the structure of which allows
to establish a local time-stable energy-momentum exchange with the field just
by means of the nonzero flows of the two 2-vectors $\bar{F}$ and $\bar{*F}$
accross corresponding 3-forms of the medium considered. Fomally this means that
the available nonzero differential flows of $\bar{F}$ and $\bar{*F}$ accross
$\mathbf{d}F$ and $\mathbf{d}*F$ must be accompanied now by nonzero flows of
$\bar{F}$ and $\bar{*F}$ accross two 3-forms $\beta^i$:
$F^{\mu\nu}\beta^1_{\mu\nu\sigma}dx^\sigma,
(*F)^{\mu\nu}\beta^2_{\mu\nu\sigma}dx^\sigma, \mu<\nu,$ describing the
$(F,*F)$-attractive abilities of the medium, or, the corresponding sensitive
abilities of the field $(F,*F)$. As we mentioned above, such 3-forms in
the frame of CED may be represented by the differentials $\mathbf{d}S$ and
$\mathbf{d}*S$. In the frame of EED, we admit also new
criteria for choosing such 3-forms, in particular, these 3-forms
$(\beta^i,\beta^2)$ will not be required to be exact differentials in general.

Also, according to $\mathbf{2^o}$, {\it definite integrability properties of
the medium MUST be available during interaction and propagation of the field
inside the medium}. These intergability properties, (futher under integrability
properties we understand {\it complete} integrability in the sense of
Frobenius) include "preinteraction" part, "interaction extending" part and
"interaction" part. The "preinteraction" part of these integrability properties
should necessarily establish some kind of initially existing dynamical
stress-equilibrium among the internal stress components of the media, the
"interaction extending" part should suggest these stress components to be
naturally extensible to incorporate in an integrable way the interaction
components, and the "interaction" part should guarantee the internal stability
of the newly created components. In CED, where the charged particles represent
through the sum $J=j_{free}+j_{bound}$ any medium, this integrability property
implicitly presents through the implied stability of the charged particles, and
it is mathematically represented by the local conservation: $\delta J=0$, on
one hand, and by integrability of the electric current vector field $J$: the
vector field $J$ always generates (local) 1-parameter family of
diffeomorphisms, on the other hand.

Another difference with CED which we'd like to stress is that, in our approach,
a medium is allowed to exchange energy-momentum with the field through $F$, as
well as, through $*F$. Moreover, it is not forbidden, in general,
some media to influence the intra-field energy-momentum exchange between $F$
and $*F$, but further we assume NO such influence to be present, so,
possible flows of $\bar{F}$ across $\beta^2$, and of $\bar{*F}$ across
$\beta^1$, as well as, possible entropy and temperature effects, will be
neglected.

We briefly sketch now our approach.

The field instruments, describing its abilities for admissible energy-momen\-tum
change are assumed to look formally the same as in the vacuum case :
$$
F_{\mu\nu}\delta F^\nu\equiv
(*F)^{\alpha\beta}(\mathbf{d}*F)_{\alpha\beta\mu}, \ \ \
(*F)_{\mu\nu}(\delta *F)^\nu\equiv
F^{\alpha\beta}(\mathbf{d}F)_{\alpha\beta\mu},
$$
$$
(*F)_{\mu\nu}\delta F^\nu\equiv -F^{\alpha\beta}(\mathbf{d}*F)_{\alpha\beta\mu},
\ \ \ F_{\mu\nu}(\delta
*F)^\nu\equiv -(*F)^{\alpha\beta}(\mathbf{d}F)_{\alpha\beta\mu}, \ \
\alpha<\beta.
$$

In view of the above assumptions the medium generates TWO 3-forms $\beta^1,
\beta^2$, which will regulate the corresponding
energy-momentum exchange with field. So, our equations, describing this local
energy-momentum exchange in terms of the flows of $(\bar{F},\bar{*F})$ across
$(\mathbf{d}F,\mathbf{d}*F)$ and $(\beta^1,\beta^2)$,
acquire the following general form:
$$
i(\bar{F})(\mathbf{d}F-\beta^1)=0,\ \
i(\bar{*F})(\mathbf{d}*F-\beta^2)=0,\ \
$$
$$
i(\bar{F})(\mathbf{d}*F)+i(\bar{*F})(\mathbf{d}F)=0 .
$$
The sense of the equations is obvious: what the field loses goes
to the medium. The equations say also that the resulted field still
keeps 2-component structure: the interaction with the media changes in general
the nature of the field, but does not destroy the recognizability of the two
field subsystems $(F,*F)$.

Denoting the corresponding $*$-duals of these 3-forms $\beta^1,\beta^2$ as
$(-\alpha,\beta)$: $(*\beta^1=-\alpha,*\beta^2=\beta)$, these equations are
respectively equivalent to:
$$
\delta *F\wedge F=\alpha\wedge F, \ \ \delta
F\wedge *F=\beta\wedge *F, $$ $$ \delta *F\wedge *F-\delta F\wedge F=0.
$$

The 1-forms $\alpha, \beta$  represent now the abilities of the
corresponding medium, on one hand, to "protect itself against the external
invasion" through building a "self-guarding" local system, on the other hand,
to "friendly" communicate with the external field by means of appropriate local
energy-momentum exchange. Moreover, these differential 1-forms represent the
interaction part of the corresponding to the medium
integrability/nonintegrability properties.

Here is the 3-dimensional form of the above equations (the bold
$\mathbf{a,b}$ denote the spatial parts of $\tilde{\eta}^{-1}(\alpha)$ and
$\tilde{\eta}^{-1}(\beta)$, and $(a^4,b^4)$ denote their time components):
$$
\left({\rm rot}{\bf E}+\frac{\partial {\bf B}}{\partial \xi}\right)\times
{\bf E}+{\bf B}{\rm div}{\bf B}=
{\bf a}\times {\bf E}-{\bf B}a^4,
$$
$$
{\bf B}.\left({\rm rot}{\bf E}+\frac{\partial {\bf B}}{\partial \xi}\right)=
{\bf B}.{\bf a},
$$
$$
\left({\rm rot}{\bf B}-\frac{\partial {\bf E}}{\partial \xi}\right)\times
{\bf B}+{\bf E}{\rm div}{\bf E}=
{\bf b}\times {\bf B}+{\bf E}.b^4,
$$
$$
{\bf E}.\left({\rm rot}{\bf B}-\frac{\partial {\bf E}}{\partial
\xi}\right)={\bf E}.{\bf b},
$$
$$
\left({\rm rot}{\bf E}+\frac{\partial {\bf B}}{\partial \xi}\right)\times
{\bf B}+ \left({\rm rot}{\bf B}-\frac{\partial {\bf E}}{\partial
\xi}\right)\times {\bf E}- {\bf B}{\rm div}{\bf E}-{\bf E}{\rm div}{\bf B}=0.
$$

If the physical system "electromagnetic field in medium" is energy-momen\-tum
isolated, and the field and the medium considered survive in definite
sense during interaction, in order to pay due respect to the medium surviving,
we shall assume the following rule/principle,
determining the "interaction" integrability/noninteg\-rability properties of
the medium:

\begin{center}
\hfill\fbox{
    \begin{minipage}{0.97\textwidth}
\begin{center}
\vskip 0.3cm
{\bf
The couple of 1-forms $(\alpha, \beta)$
defines a completely integrable 2-dimensional Pfaff system}.
 \end{center} \vskip 0.3cm
\end{minipage}} \hfill \end{center}

This assumption means that the following
equations holds:
$$
{\bf d}\alpha\wedge\alpha\wedge\beta=0, \ \ \
{\bf d}\beta\wedge\beta\wedge\alpha=0
$$

This integrability system and the above system connecting $(F,*F)$ with
$\alpha, \beta$, constitute the basic system of equations
in case of "field $\oplus$ medium". Of course, the various special cases can be
characterized by adding some new consistent with these equations new relations.

\vskip 0.3 cm
{\bf 3. Solutions}

{\bf 3.1. Remarks.}

Turning to searching solutions with nonzero $\alpha, \beta$, we
must keep in mind that the field $(F,*F)$ now will certainly be of {\it quite
different nature compare to the vacuum case}, and its interpretation as
electromagnetic field is {\it much conditional}. Any solution is meant
to represent a field interacting {\it continuously} with other
continuous physical system, so the situation is quite different and,
correspondingly, the properties of the solution may differ drastically from the
vacuum solutions' properties.  For example, contrary to the nonlinear vacuum
case where the solutions propagate translationally with the velocity of light,
here it is not excluded to find solutions which do not propagate at all with
respect to an appropriate Lorentz frame.  In {\it
integrability/nonintegrability} terms this would mean that at least some of the
proper integrability properties of the vacuum solutions have been lost. On the
other hand, the interaction integrability properties of the
medium are, at least partially, guaranteed to hold through the requirement
for the Frobenius integrability of the 2-dimensional Pfaff
systems $(\alpha, \beta)$.

From purely formal point of view finding a solution, whatever it is,
legitimizes the equations considered as a compatible system. Our purpose in
looking for solutions in the nonvacuum case,
however, is not purely formal, we'd like to consider the corresponding
solutions as physically meaningful, in other words, we are interested in
solutions, which can be, more or less, {\it physically interpretable}, i.e.
presenting more or less reasonable properties of real objects and processes.
That's why we'll try to meet the following.

First, the solutions must be somehow {\it physically clear}, which means that
the anzatz assumed should be comparatively simple and its choice should be made
on the base of a preliminary analysis of the physical situation in view of the
mathematical model used.

Second, {\it it is absolutely obligatory} the solutions to have
well defined local and integral energy and momentum.

Third, {\it existence of solutions of soliton-like nature} is,
of course, highly desirable, especially if (3+1)-extension of "popular" and well
known soliton solutions of "well liked" equations could be established.

\vskip 0.3 cm

{\bf 3.2. On the preinteraction an interaction integrability}
\vskip 0.2cm

In order to illustrate what is meant under preinteraction and interaction
integrability, we give the following consideration. We work on Minkowski
space-time $M=(\mathbb{R}^4,\eta)$ and shall use the previously used notations.

Let the stress in the medium considered be represented by the following two
lineary independent space-like vector fields (the field $(F,*F)$ is still abcent)
$$
\bar{P}=-\frac{f}{f^2+g^2}\,\frac{\partial}{\partial
x}-\frac{g}{f^2+g^2}\,\frac{\partial}{\partial y}, \ \
\bar{Q}=\frac{g}{f^2+g^2}\,\frac{\partial}{\partial x}
-\frac{f}{f^2+g^2}\,\frac{\partial}{\partial y},
$$
where $(f,g)$ are two nonvanishing at least on an open set $U\subset M$
functions. The corresponding 1-forms $(P,Q)$ that form dual to
$(\bar{P},\bar{Q})$ basis are
$$
P=-f\,dx-g\,dy, \ \ Q=g\,dx-f\,dy,
$$
$$
\langle P,\bar{P}\rangle=1,\ \ \langle P,\bar{Q}\rangle=0, \ \
\langle Q,\bar{P}\rangle=0, \ \ \langle Q,\bar{Q}\rangle=1.
$$
Since $\bar{P}\wedge\bar{Q}\neq 0$, these two vector fields define a
2-dimensional distribution on $M$. Moreover, a direct check shows
$[\bar{P},\bar{Q}]\wedge \bar{P}\wedge \bar{Q}=0$, so this distribution is
integrable, the dual codistribution $\{P,Q\}$ is also integrable:
$\mathbf{d}P\wedge P\wedge Q=0, \ \ \mathbf{d}Q\wedge Q\wedge P=0.$

Assuming that our medium is homogeneous with respect to the stress, generated
by the above integrable distribution, we assume that the course of time is the
same throughout the 3d-volume occupied by the medium, so, let
$\frac{\partial}{\partial \xi}$ be the time-like vector field along the time
coordinate. We consider now the integrability properties of the two distributions
$\{\bar{P},\frac{\partial}{\partial \xi}\}$ and
$\{\bar{Q},\frac{\partial}{\partial \xi}\}$. It turns out that these two
distributions are {\it nonintegrable} in general, and the corresponding
curvature forms are
$$
\mathcal{R}_{\left(\bar{P},\frac{\partial}{\partial \xi}\right)}=
-\mathbf{d}Q\otimes{\bar{Q}} \ \ ,\ \
\mathcal{R}_{\left(\bar{Q},\frac{\partial}{\partial \xi}\right)}=
-\mathbf{d}P\otimes{\bar{P}}\cdot
$$
We evaluate now these curvature forms on the representing vector fields and
obtain
$$
Z_1\equiv\mathcal{R}_{\left(\bar{P},\frac{\partial}{\partial \xi}\right)}
\left(\bar{P},\frac{\partial}{\partial \xi}\right)=
\frac{gf_{\xi}-fg_{\xi}}{f^2+g^2}\,\bar{Q} ,
$$
$$
Z_2\equiv\mathcal{R}_{\left(\bar{Q},\frac{\partial}{\partial \xi}\right)}
\left(\bar{Q},\frac{\partial}{\partial \xi}\right)
=\frac{fg_{\xi}-gf_{\xi}}{f^2+g^2}\,\bar{P} \cdot
$$
Now, the two Pfaff forms $(Q,dz)$ annihilate
$\left(\bar{P},\frac{\partial}{\partial \xi}\right)$,
and the two Pfaff forms $(P,dz)$ annihilate
$\left(\bar{Q},\frac{\partial}{\partial \xi}\right)$. For the flows of $Z_1$
and $Z_2$ across $Q\wedge dz$ and $P\wedge dz$ we obtain:
$$
i_{Z_1}(P\wedge dz)=i_{Z_2}(Q\wedge dz)=0, \ \
i_{Z_1}(Q\wedge dz)=-i_{Z_2}(P\wedge dz) .
$$
So, these two distributions are in dynamical equilibrium .

Noticing that
$\frac{\partial}{\partial \xi}$ is local symmetry of the distribution
$\{\bar{P},\bar{Q}\}: L_{\frac{\partial}{\partial \xi}}\{\bar{P},\bar{Q}\}$ is
inside $\{\bar{P},\bar{Q}\}$, we come to the conclusion that the
3-dimensional distribution $(\bar{P},\bar{Q},\frac{\partial}{\partial \xi}\}$
is integrable, and {\it this is the preinteraction integrability: the medium
is time-stable}.

If under the attack of $(F,*F)$ our medium creates another distribution
$(\bar{X},\bar{Y})$, such that the corresponding 3-dimensional distributions
$(\bar{P},\bar{Q},\bar{X})$ and $(\bar{P},\bar{Q},\bar{Y})$ are integrable,
in this sense we talk about {\it interaction extensible integrability}.

Finally, the above required Frobenius integrability of the codistribution
$(\alpha,\beta)$ we call {\it interaction integrability}, since it
describes the surviving abilities of the medium just during interaction.

 \vskip 0.3cm {\bf 3.3. A class of solutions}
\vskip 0.2cm
Let our attacked by $(F,*F)$ medium creates the "self-guarding" distribution
$(\bar{X},\bar{Y})$, and let's choose
$\alpha=\tilde{\eta}(\bar{Y}),\beta=\tilde{\eta}(\bar{X})$.

Let now the medium's self-guarding and
additional stress generating system be formally represented by the 2-dimensional
distribution $\bar{X}, \bar{Y}$, where
$$
\bar{X}=-b\,\frac{\partial}{\partial y}+B\,\frac{\partial}{\partial z} \ \ , \
\bar{Y}=A\frac{\partial}{\partial \xi} ,
$$
correspondingly,
$$
\beta=\tilde{\eta}(\bar{X})=b\,dy-B\,dz , \ \ \
\alpha=\tilde{\eta}(\bar{Y})=A\,d\xi,
$$
and $(A,b,B)$ are three functions. Clearly, the nontrivial function
$A(x,y,z,\xi)$ is meant to take care of the different impact of the attacking
field $(F,*F)$ on the local time course through making the referent time
measuring process to be point-dependent.

It is easily verified that the two 3-dimensional distributions
$(\bar{P},\bar{Q},\bar{X})$ and $(\bar{P},\bar{Q},\bar{Y})$, where
$\bar{X},\bar{Y}$ are given above, are integrable, so,
we have the case of {\it interaction extensible integrability}.

Fanally we recall that according to our assumption
the two 1-forms $(\alpha,\beta)$ must define integrable
2-dimensinal codistribution.

We turn now to the difficult problem to find how the surviving field will look
like when it propagates inside this medium. We shall need the field to keep
the following two properties:
\vskip 0.2cm
1. Since propagation inside the medium is allowed,
we shall be interested in time-dependent solutions.

2. The "electric" and the "magnetic" constituents of the field must be present.
\vskip 0.2cm
The simplest $(F,*F)$, meeting these requirements, look as follows (we use the
above assumed notations):
$$
F=-udy\wedge dz -vdy\wedge d\xi,\ \ \ *F=vdx\wedge dz + udx\wedge d\xi,   %72%
$$
where $u(x,y,z,\xi)$ and $v(x,y,z,\xi)$ are two functions on Minkowski
space-time $M=(\mathbb{R}^4,\eta)$ satisfying $u^2\neq v^2$, so
$F\wedge *F\neq 0$, and $F\wedge F=0$.

We begin now studying the compatibility of the assumptions made.

At these conditions our equations
$$
\delta *F\wedge F=\alpha\wedge F,\quad
\delta F\wedge* F=\beta\wedge * F,\quad
\delta *F\wedge *F-\delta F\wedge F=0,
$$
$$
{\bf d}\beta\wedge\beta\wedge\alpha=0,\quad
{\bf d}\alpha\wedge \alpha\wedge\beta=0
$$
take the form: $\delta *F\wedge *F-\delta F\wedge F=0$ reduces to
$$
-vu_y+uv_y=0,\quad  -uv_x +vu_x=0,
$$
so, $\delta *F\wedge *F=0, \ \delta F\wedge F=0$.

The Frobenius equations  ${\bf d}\alpha\wedge \alpha\wedge\beta=0,\
{\bf d}\beta\wedge \alpha\wedge\beta=0$ reduce to
$$
\left(-b_xB+B_xb\right).A=0,
$$
$\delta *F\wedge F=\alpha\wedge F$ reduces to
$$
u\left(u_\xi-v_z\right)=0,\quad
v\left(u_\xi-v_z\right)=0,\quad
uu_x-vv_x=Au,
$$
finally, $\delta F\wedge* F=\beta\wedge *F$ reduces to
$$
v\left(v_\xi-u_z\right)=-bv,\quad
u\left(v_\xi-u_z\right)=-bu,\quad
uu_y-vv_y=Bu.
$$
In this way we obtain 7 equations for 5 unknown functions $u,v,A,B,b$.

The two equations
$$
-vu_y+uv_y=0,\  -uv_x +vu_x=0
$$
 have the following solution:
$$
u(x,y,z,\xi)=f(x,y)U(z,\xi),\quad v(x,y,z,\xi)=f(x,y)V(z,\xi).
$$
That's why
$$
AU=f_x\left(U^2-V^2\right), \ \ BU=f_y\left(U^2-V^2\right), \ \
f\left(V_\xi-U_z\right)=-b,\ \ U_\xi-V_z=0.
$$
It follows that $b(x,y,z,\xi)$ should be of the kind $f(x,y)b^o(z,\xi)$,
so the equation $B_xb-Bb_x=0$ takes the form
$$
ff_{xy}=f_xf_y.
$$
The general solution of this last equation is $f(x,y)=g(x)h(y)$.
The equation $gh\left(V_\xi-U_z\right)=-b$ reduces to
$$
V_\xi-U_z=-b^o.
$$

The relations obtained show how to build an appropriate for us solution of this
class. Namely, first, we choose the function $V(z,\xi)$ to be $z$-finite or
$z$-localized, then we determine the function $U(z,\xi)$ by $$
U(z,\xi)=\int{V_z d\xi} +l(z),
$$
where $l(z)$ is an arbitrary function, which may be assumed equal to $0$.
After that we define $b^{o}=U_z-V_\xi$. The functions $g(x)$ and $h(y)$ are
chosen also finite or localized, and for $A$ and $B$ we find
$$ A(x,y,z,\xi)=g'(x)h(y)\frac{U^2-V^2}{U},\quad
B(x,y,z,\xi)=g(x)h'(y)\frac{U^2-V^2}{U}\cdot
$$
In this way we obtain a family of solutions, which is parametrized by one
function $V$ of the two variables $(z,\xi)$ and two functions $g(x),\ h(y)$,
each depending on one variable. Clearly, the spatial dependence of these
functions is arbitrary, so they are allowed to be finite/localized.

In order to find corresponding conserved quantities we sum up the nonzero
right-hand sides of the equations and obtain
$(\alpha\wedge F+\beta\wedge *F)$. The $*$-image of this expression
is representable in divergence form as follows:
$$
*(\alpha\wedge F+\beta\wedge *F)=Audx-Budy-budz-bvd\xi
$$
$$
=\frac12 (U^2-V^2)\left[(gh)^2\right]_x dx-
\frac12 (U^2-V^2)\left[(gh)^2\right]_y dy
$$
$$
-(gh)^2\left(\int{Ub^{o}dz}\right)_{z}dz-(gh)^2\left(\int{Vb^{o}d\xi}\right)_{\xi}d\xi
=-\left\{\frac{\partial}{\partial x^\nu}H_\mu^\nu\right\}dx^\mu,
$$
where the interaction energy-momentum tensor is defined by the matrix
$$
H_\mu^\nu=\begin{Vmatrix}
-\frac12(gh)^2 Z  &0                &0                     &0                      \cr
0                 &\frac12(gh)^2 Z  &0                     &0                      \cr
0                 &0                &(gh)^2\int{Ub^{o}dz}  &0                      \cr
0                 &0                &0                     &(gh)^2\int{Vb^{o}d\xi}
\end{Vmatrix},
$$
and the notation $Z\equiv U^2-V^2$ is used. From the equations it follows that
the divergence of the tensor $T_\mu^\nu=Q_\mu^\nu+H_\mu^\nu$ must be zero,
where
$$
 T_\mu^\nu=
-\frac12\Big[F_{\mu\sigma}F^{\nu\sigma}+(*F)_{\mu\sigma}(*F)^{\nu\sigma})\Big]
+H_\mu^\nu .
$$
For the components we obtain
$$
T_3^3=(gh)^2\left[\int{Ub^{o}dz} - \frac12(U^2+V^2)\right],
$$
$$
T_3^4=-T_4^3=(gh)^2 UV,
$$
$$
T_4^4=(gh)^2\left[\int{Vb^{o}d\xi}+\frac12(U^2+V^2)\right],
$$
and all other components are zero.

\vskip 0.5cm
{\bf 3.4. Examples}

Here we consider some of the well known and well studied
(1+1)-dimensional soliton equations as generating tools for choosing explicit
forms of the function $V(z,\xi)$, and only some 1-soliton solutions will be
explicitly elaborated. Of course, there is nothing standing in our way to
consider other (e.g. breather, multisoliton) solutions.

We turn to the soliton equations mainly because of three reasons. {\bf First},
many of the solutions have clear physical sense in definite parts of physics
and, according to our opinion, they are attractive for building models of
real physical objects with internal structure. {\bf Second}, all soliton
solutions are intrinsically connected to the concept of integrability.
{\bf Third}, soliton solutions may describe
interacting field objects with {\it no dissipation} of energy and momentum.

\vskip 0.5cm
\noindent{\it 1. Nonlinear (1+1) Klein-Gordon Equation}.
In this example we define our functions $U$ and $V$ through the derivatives
of the function $k(z,\xi)=k(z,ct)$ in the following way: $U=k_z,\ V=k_\xi$.
Then the equation $U_\xi-V_z=k_{z\xi}-k_{\xi z}=0$ is satisfied automatically,
and the equation $U_z-V_\xi=b^{o}$ takes the form $k_{zz}-k_{\xi\xi}=b^{o}$.
Since $b^o$ is unknown, we may assume $b^o=b^o(k)$, which reduces the whole
problem to solving the general nonlinear (1,1)-Klein-Gordon equation when $b^o$
depends nonlineary on $k$. Since in this case $V=k_\xi$ we have
$$
\int{Vb^o(k)d\xi}=\int{k_\xi b^o(k)d\xi}=
\int{\left[\frac{\partial}{\partial \xi}\int{b^o(k)dk}\right]d\xi}=
\int{b^o(k)dk}.
$$
For the full energy density we get
$$
T_4^4=\frac12(gh)^2\left\{k_z^2+k_\xi^2+2\int{b^o(k)dk}\right\}.
$$
Choosing $b^o(k)=m^2 sin(k)$, $m=const$, we get the well known and widely
used in physics Sine-Gordon equation [2], and accordingly, we can use {\it
all} solutions of this (1+1)-dimensional nonlinear equation to generate
(3+1)-dimensional solutions of our equations following the above described
procedure.  When we consider the (3+1) extension of the soliton solutions of
this equation, the functions $g(x)$ and $h(y)$ should be localized too. The
determination of all five functions in our approach is straightforward, so
we obtain a (3+1)-dimensional version of the soliton solution chosen. As it
is seen from the above given formulas, the energy density of the solution
differs from the energy density of the corresponding (1+1)-dimensional solution
just by the $(x,y)$-localizing factor $[g(x)h(y)]^2$.

For the 1-soliton solution (kink) we have ($c$ is the velocity of light):
$$
k(z,\xi)=4arctg\left\{exp\left[\pm \frac{m}{\gamma}(z-\frac wc
\xi)\right]\right\},\quad \gamma=\sqrt{1-\frac{w^2}{c^2}}
$$
$$
U(z,\xi)=k_z=\frac1\gamma\frac{\pm 2m}{
ch\left[\pm\frac{m}{\gamma}\left(z-\frac{w}{c} \xi \right)\right]},\quad
V(z,\xi)=k_\xi=\frac{1}{c\gamma}\frac{\pm 2mw}{
ch\left[\pm\frac{m}{\gamma}\left(z-\frac{w}{c} \xi \right)\right]},
$$
$$
A=g'(x)h(y)\frac{\pm 2m\gamma}{ch\left
[\pm\frac{m}{\gamma}\left(z-\frac{w}{c} \xi \right)\right]},\quad
B=g(x)h'(y)\frac{\pm 2m\gamma}{ch\left[\pm\frac{m}{\gamma}\left(z-\frac{w}{c} \xi \right)\right]},\
$$
$$
b^o=U_z-V_\xi=
\frac{-2m^2 sh\left[\pm\frac{m}{\gamma}
\left(z-\frac{w}{c} \xi \right)\right]}
{ch\left[\pm\frac{m}{\gamma}\left(z-\frac{w}{c} \xi \right)\right]},\quad
T_4^4=\frac{1}{\gamma^2}\frac{(gh)^2 4m^2}{
ch^2\left[\pm\frac{m}{\gamma} \left(z-\frac{w}{c} \xi \right)\right]}
$$
and for the 2-form $F$ we get
$$
F=-\frac1\gamma\frac{\pm2mg(x)h(y)}{ch
\left[\pm\frac{m}{\gamma}(z-\frac wc \xi)\right]}\,dy\wedge dz+
\frac{w}{c\gamma} \frac{\pm2mg(x)h(y)}{ch\left[\pm\frac{m}{\gamma}
(z-\frac wc \xi)\right]}\,dy\wedge d\xi.
$$
From symmetry considerations, i.e. at
homogeneous and isotropic media, we come to the most natural (but not
necessary) choice of the functions $g(x)$ and $h(y)$:
$$
g(x)=\frac{1}{ch(mx)},\quad h(y)=\frac{1}{ch(my)}\cdot
$$

\vskip 0.5cm
2.{\it Korteweg-de Vries equation.}  This nonlinear equation has the
following general form: $$ f_\xi+a_1ff_z+a_2f_{zzz}=0, $$ where $a_1$ and $a_2$
are two constants. The well known 1-soliton solution is $$
f(z,\xi)=\frac{a_o}{ch^2\left[\frac zL -\frac{w}{cL}\xi\right]},\quad
L=2\sqrt{\frac{3a_2}{a_o a_1}},\quad w=\frac{ca_oa_1}{3},
$$
where $a_o$ is a constant. We choose $V(z,\xi)=f(z,\xi)$ and get
$$
U=-\frac{a_oc}{w}\frac{1}{ch^2\left[\frac zL
-\frac{w}{cL}\xi\right]},\quad
b^o=U_z-V_\xi=\left(\frac{c}{Lw}-
\frac wc\right)\frac{2a_o}{ch^3\left[\frac zL -\frac{w}{cL}\xi\right]},
$$
$$
T_4^4=(gh)^2\frac{a_o^2c^2(1+L)}{2w^2Lch^4\left[\frac zL -
\frac{w}{cL}\xi\right]}.
$$
\vskip 0.5cm

3. {\it Nonlinear Schr\"odinger equation} [1]. In this case we have an equation
for a complex-valued function, i.e. for two real valued functions. The
equation reads
$$
if_\xi+f_{zz}+2|f|^2 f=0,
$$
and its 1-soliton solution, having oscillatory character, is
$$
f(z,\xi)=2\beta\frac{exp\left[-2i\alpha z-
4i(\alpha^2-\beta^2)\xi -i\theta\right]}
{ch\left(2\beta z+8\alpha\beta \xi -\delta\right)},
$$
where $\alpha, \beta, \delta$ and $\theta$ are constants. Further computations
with
$$
f(z,\xi)=\sqrt{\rho(z,\xi)}.exp\,(i\varphi(z,\xi))
$$
we leave to the reader.

%\vskip 0.5cm
Following this procedure we can generate a spatially finite solution to our
system of equations making use of {\it every} known
soliton solution to {\it any} (1+1)-soliton equation, as well as to compute
the corresponding conserved quantities. We are not going to do this here, payng
due respect to all interested in the subject and creatively inclined readers.

\vskip 0.3cm
 {\bf References}
\vskip 0.2cm
\addcontentsline{toc}{section}{{\bf References}}
1. {\bf G. L. Lamb, Jr.}, {\it Elements of Soliton Theory}, John Wiley and Sons,
New York, 1980

2. {\bf F. Calogero, A. Degasperis}, {\it Spectral transform and Solitons I},
North Holland Publishing, 1982

\newpage
{\bf Appendix B}
\vskip 0.2cm
\addcontentsline{toc}{section}{{\bf Appendix B.} \\
 Do PhLO solutions interfere?}
\vskip 0.2cm
{\bf Do PhLO solutions interfere?}
\vskip 0.2cm
{\bf 1. Remarks}

Having at hand the photon-like solutions a natural next step is to try to
describe the situation when two such solutions occupy the same (or partially the
same) 3-region in some period of time. It is clear, that if such two
photon-like solutions meet somewhere, i.e., their cilinder-like world-tubes
intersect appropriately, the interesting case is when they {\it move along the
same spatial straight line} and in {\it the same direction}. Since they move by
the same velocities they will continue to overlap each other until some
outer agent causes a change. What kind of an object is obtained in this way,
is it photon-like or not, what kind of interaction takes place, what is its
integral energy , its momentum and its angular momentum? Many
challenging and still not answered questions may be set in this direction
before the theoretical physics. And this section is devoted to consideration
of some of these problems in the frame of our approach to electrodynamics.

Almost all experiments set to find some immediate mutual
interaction of two (or more) electromagnetic fields in vacuum, causing some
observable effects (e.g.  frequency or amplitude changes), as far as we know,
have faild, exept when the two fields satisfy the so called {\it coherence
conditions}. In the frame of classical electrodynamics (CED) and working with
plane waves this simply means, that their {\it phase difference must be a
constant quantity}. The usual way of consideration is limited to {\it
cosine}-like running waves with the {\it same} frequency. The physical
explanation is based on the linearity of Maxwell's equations, which require any
linear combination of solutions to be again a solution, so the "building
points" of the medium, subject to the field pressure of the two independent
fields, go out of their equilibrium state obeying simoultaneously the two
forces applied in the overlaping 3-region.  After getting out of this
overlaping 3-region the fields stay what they have been before the interaction.
In order to describe the interaction, i.e. the observed redistribution of the
energy-momentum density inside the overlaping 3-region, CED uses the
corresponding mathematical expressions in Maxwell's theory and gets
comparetively good results. Most frequently the Poynting vector $S\sim
\left[(E_1+E_2)\times (B_1+B_2)\right]$ is used and the cross-terms $(E_1\times
B_2)+(E_2\times B_1)$ are held responsible for the interaction, in fact, the
very {\it interference} is defined by the condition that these cross-terms,
usually called "interference terms", are different from zero.

Our nonlinear equations make us approach this physical situation in a new way.
First, let's specify the situation more in detail and in terms of the notion
for $EM$-field in our approach. Roughly speaking, this notion is based on the
idea for discreteness, i.e., the real electromagnetic fields consist of many
noninteracting, or very weakly interacting, photon-like objects (PhLO),
propagating as a whole in {\it various} directions. Because of the great
velocity of their straight line motion it is hardly possible to observe and say
what happens when two photons meet somewhere.  The experiment suggests that in
most cases they pass through each other and forget about the meeting.  As we
mentioned above, the interesting case is when they move along the same
direction and the regions, they occupy, overlap nontrivially.

The nonlinear solutions we have obtained can not
describe such set of PhLO, moving in {\it various} directions. Even if we
choose the amplitude function $\phi$ of a solution to consist of many "3-bubles"
these "bubles" have to move in the same direction, which is a special, but not
the general, case of the situation we consider here.  So, in order to
incorporate for description more general situations, some perfection is needed.
As before, this perfection shall consist of two steps:  first, elaboration of
the algebraic character of the mathematical field, second, elaboration of the
equations.  The second step, besides its dynamical task, must define also the
necessary conditions for interference of photon-like solutions, which should
coincide with the above mentioned, experimentally established and repeatedly
confirmed coherence conditions.

\vskip 0.5cm
{\bf 2. Elaborating the mathematical object}

Recall that our mathematical object that represents the field is a
2-form $\Omega$ with values in $\mathbb{R}^2$. We want to elaborate it in order
to reflect more fully the physical situation. The new moment is that inside the
3-region under consideration we have {\it many} photon-like objects. Each of
these photon-like objects, considered as independent object, is described by a
pfoton-like solution as given in the preceding sections, i.e., each of them has
its own spatial structure, its own scale factor (or frequency) and its own
direction of motion as a whole. Of course, the velocity of motion is the same
for all of them. To this physical situation we have to juxtapose {\it one}
mathematical object, which have to generalize in a natural way our old object
$\Omega$. The idea for this generalization is very simple and consists in the
following. With every single PhLO we associate its own
$\mathbb{R}^2$-space, so if the number of the presenting PhLO is $N$, we'll have $N$
such spaces. Denoting this vector space
by ${\cal N}$, our object becomes a 2-form $\Omega$ with values in the vector
space $\mathbb{R}^2\otimes {\cal N}$:
$\ \Omega\in \Lambda^2\left(M,\mathbb{R}^2\otimes {\cal N}\right)$.
We recall now how this vector space ${\cal N}$ is explicitly built [1].

If ${\cal K}$ is an arbitrary set, finite or infinite, we consider those
mappings of this set into a given field, e.g. $\mathbb{R}$, which are different
from zero only for finite number of elements of ${\cal K}$. Such kind of
mappings will be the elements of the space ${\cal N}$. A basis of this space is
built in the following way. We consider the elements $f\in {\cal N}$, having
the property: if $a\in {\cal K}$ then $f(a)=1$ and $f$ has zero values for all
othe elements of ${\cal K}$. So, with every element $a\in {\cal K}$ we
associate the corresponding element $f_a \in {\cal N}$, therefore, an
arbitrary element $f\in {\cal N}$ is represented as follows:
$$
f=\sum_{i=1}^N \left(\lambda^i f_{a_i}\right),
$$
where $\lambda^i$, $i=1,2,...,N$, are the values, aquired by $f$, when $i$
runs from 1 to $N$ (of course, some of the $\lambda $'s may be equal to
zero). The linear structure in ${\cal N}$ is naturally introduced, making use
of the linear structure in $\mathbb{R}$ in the well known way. The linear
{\it independence} of $f_{a_i} $ is easily shown. In fact, assuming the
opposite, i.e. that there exist such $\lambda^i$, among which at least one is
not zero and the following relation holds
$$
\sum_{i=1}^{N}\lambda^{i}f_{a_i}=0,
$$
then for any $j=1,2,...,N$ we'll have
$$
\sum_{i=1}^{N}\lambda^{i}f_{a_i}(a_j)=\lambda^j=0,
$$
which contradicts the assumption. Hence, $f_{a_i}$ define really a basis of
${\cal N}$. Now we form the injective mapping $i_N :N\rightarrow {\cal N}$,
defined by
$$
i_N (a)=f_a,\ a\in N,
$$
so the set $N$ turns into a basis of ${\cal N}$. If such a construction is
made, then ${\cal N}$ is called a {\it free vector space over the set $N$}.
Further on the corresponding basis of our set of PhLO will be
denoted by $E_a$. So, our mathematical object will look as follows (summing up
over the repeating index $a$)
$$
\Omega=\Omega^a \otimes E_a =
\left[F^a \otimes e_1^a + (*F)^a \otimes e_2^a\right]\otimes E_a,
$$
where $(e^a_1,e^a_2)$ is the associated with the field $(F^a,*F^a)$ basis. If we
work in an arbitrary basis of $\mathbb{R}^2$, the full writing reads ($i$=1,2)
$$
\Omega=\Omega^a \otimes E_a =\Omega_i^a\otimes k^i_a\otimes E_a.
$$
We define now the $"\vee"$ product of two
2-forms of this kind. For erxample, if
$\Phi=\Phi^a_i\otimes k_a^i\otimes E_a,
\Psi=\Psi^b_j\otimes l_b^j\otimes E_b$, we'll have
$$
(\vee,\vee)(\Phi,\Psi)=(\vee,\vee)(\Phi^a_i\otimes k_a^i\otimes E_a,
 \Psi^b_j\otimes l_b^j\otimes E_b)=
$$
$$
 =\sum_{a=1}^{N}\Big[\Phi^a_1\wedge \Psi^a_1\otimes k^1_a\vee
 l^1_a+\Phi^a_2\wedge \Psi^a_2\otimes k^2_a\vee l^2_a+
$$
$$
 +(\Phi^a_1\wedge \Psi^a_2+\Phi^a_2\wedge \Psi^a_1)
 \otimes k^1_a\vee l^2_a\Big]\otimes E_a\vee E_a +
$$
$$
 +\sum_{a<b=1}^{N}\Big[(\Phi^a_1\wedge \Psi^b_1+\Phi^b_1\wedge \Psi^a_1)
 \otimes k_a^1\vee l_b^1+ (\Phi^a_2\wedge \Psi^b_2+\Phi^b_2\wedge \Psi^a_2)
 \otimes k_a^2\vee l_b^2+
$$
$$
+(\Phi^a_1\wedge \Psi^b_2+\Phi^a_2\wedge \Psi^b_1+
\Phi^b_1\wedge \Psi^a_2+\Phi^b_2\wedge \Psi^a_1)\otimes k_a^1\vee l_b^2\Big]
\otimes E_a\vee E_b.
$$

Let now $\Omega$ be of the kind $\Omega=(F^a\otimes e_a^1+*F^a \otimes e_a^2)
\otimes E_a$. Then, forming $*\Omega$ and $\delta \Omega$, for
$(\vee,\vee)(\delta\Omega,*\Omega)$ we obtain
$$
(\vee,\vee)(\delta\Omega,*\Omega)=
 \sum_{a=1}^{N}\Big[\delta F^a\wedge *F^a \otimes e^1_a\vee e^1_a+
\delta *F^a\wedge **F^a\otimes e^2_a\vee e^2_a+
$$
$$
 +(\delta F^a\wedge **F^a+\delta *F^a\wedge *F^a)\otimes e^1_a\vee e^2_a\Big]
\otimes E_a\vee E_a+
$$
$$
 +\sum_{a<b=1}^{N}\Big[(\delta F^a\wedge *F^b+\delta F^b\wedge *F^a)
\otimes e^1_a\vee e^1_b+
(\delta *F^a\wedge **F^b+\delta *F^b\wedge **F^a)\otimes e^2_a\vee e^2_b
$$
$$
 +(\delta F^a\wedge **F^b+\delta *F^a\wedge *F^b+
 \delta F^b\wedge **F^a+\delta *F^b\wedge *F^a)\otimes e^1_a\vee e^2_b\Big]
\otimes E_a\vee E_b.
$$

\vskip 0.5cm
{\bf 3. Elaborating the field equations}

If we want to consider a set of independent solutions, then in the above
expression we take the trace $tr$ over the indeces of $E_a\vee E_b$. The
compact writing of this condition reads
$$
tr(\vee,\vee)(\delta\Omega,*\Omega)=0,
$$
which is equivalent to the equations
$$
\delta F^a\wedge *F^a=0,\ \delta *F^a\wedge **F^a=0,\
\delta F^a\wedge **F^a +\delta *F^a\wedge *F^a=0.
$$
Clearly, in this case the full energy-momentum tensor $Q_\mu^\nu$ will be a
sum of all energy tensors $(Q^a)_\mu^\nu$ of the single solutions.

The general equations are written down as follows:
$$
(\vee,\vee)(\delta\Omega,*\Omega)=0.
$$
The equivalent (component-wise) form reads
$$
{\delta F^a\wedge *F^a=0,\ \delta *F^a\wedge **F^a=0,\
\delta F^a\wedge **F^a +\delta *F^a\wedge *F^a=0},
$$
$$
{\delta F^a\wedge *F^b +\delta F^b\wedge *F^a=0,\
\delta *F^a\wedge **F^b +\delta *F^b\wedge **F^a=0},
$$
$$
{\delta F^a\wedge **F^b +\delta *F^b\wedge **F^a+
\delta *F^a\wedge *F^b +\delta *F^b\wedge *F^a=0}.
$$

Let now $F^a$, $a=1,2,...,N$ define a solution of the above system of
equations. We are going to show that the linear combination with
constant coefficients $\lambda_a$
$$
F=\sum_{a=1}^{N}\lambda_a F^a
$$
satisfies the equations:
$$\delta F\wedge *F=0,\ \delta *F\wedge **F=0,\ \ \delta
F\wedge **F+\delta*F\wedge *F=0.
$$
In fact
$$
\delta F\wedge
*F=\sum_{a=1}^{N}(\lambda_a)^2(\delta F^a\wedge *F^a)+ \sum_{a<b=1}^N
\lambda_a \lambda_b (\delta F^a\wedge *F^b + \delta F^b\wedge *F^a),
$$
$$
\delta *F\wedge **F=\sum_{a=1}^{N}(\lambda_a)^2(\delta *F^a\wedge **F^a)+
\sum_{a<b=1}^N \lambda_a \lambda_b (\delta *F^a\wedge **F^b +
\delta *F^b\wedge **F^a),
$$
$$
\delta F\wedge **F +\delta *F\wedge *F=
\sum_{a=1}^N (\lambda_a)^2 (\delta F^a\wedge **F^a)+
$$
$$
+\sum_{a<b=1}^N \lambda_a \lambda_b(\delta F^a\wedge **F^b+
\delta F^b\wedge **F^a)+
$$
$$
+\sum_{a=1}^N (\lambda_a)^2 (\delta *F^a\wedge *F^a)+
\sum_{a<b=1}^N \lambda_a \lambda_b(\delta *F^a\wedge *F^b+
\delta *F^b\wedge *F^a)=
$$
$$
=\sum_{a=1}^N(\lambda_a)^2(\delta F^a\wedge **F^a +\delta *F^a\wedge *F^a)+
$$
$$
+\sum_{a<b=1}^N \lambda_a\lambda_b(\delta F^a\wedge **F^b +
\delta F^b\wedge **F^a+\delta *F^a\wedge *F^b +\delta *F^b\wedge *F^a).
$$
Obviously, the component-wise writing down of the equations shows that
every addend is equal to zero. This result can be interpreted as some
particular "superposition principle", i.e. if we have finite number of
solutions $F^a$ of the system
$$
\delta F\wedge *F=0,\ \delta *F\wedge **F=0,\
\delta F\wedge **F+\delta*F\wedge *F=0,
$$
which solutions satisfy additionally the equations
$$
{\delta F^a\wedge *F^b +\delta F^b\wedge *F^a=0,\ \delta *F^a\wedge **F^b +\delta *F^b\wedge **F^a=0},
$$
$$
{\delta F^a\wedge **F^b +\delta *F^b\wedge **F^a+\delta *F^a\wedge *F^b +\delta *F^b\wedge *F^a=0},
$$
then the 2-form $F=\sum_{a=1}^{N}\lambda_a F^a$ is again a solution.
 Then, clearly, if $F$ and $G$ are 2 solutions of all the
equations, the new solution $(F+G)$ is naturally endowed
with the following energy-momentum tensor
$$
Q_{\mu\nu}=\frac{1}{4\pi}\Big[-(F+G)_{\mu\sigma} (F+G)_\nu^\sigma \Big].
$$
In the general case we'll have
$$
Q_{\mu\nu}=\frac{1}{4\pi}\left[-\left(\sum_{a=1}^{N}\lambda_a F^a \right)_{\mu\sigma}
\left(\sum_{a=1}^{N}\lambda_a F^a \right)_\nu^\sigma \right]
$$
In this way we can compute the corresponding "interference terms". In
\linebreak
particular, the "interference" energy density is obtained proportional
\linebreak to $-2F_{4\sigma}G^{4\sigma}$.

\vskip 0.5cm
{\bf 4. Spatial coherence and interference}

We consider now two photon-like solutions determined by $F_1$ and $F_2$,
propagating along the same direction. We choose this direction for
the $z$-axis of our coordinate system. We are going to find what additional
conditions on these solutions come from the additional equations.
We assume also, that the 3-regions, where the two amplitudes $\phi_1$
and $\phi_2$ are different from zero have non-empty intersection, because
otherwise, the interference term is equal to zero. Explicitly,
$$
F_1=\varepsilon_1 u_1 dx\wedge dz + u_1 dx\wedge d\xi + \varepsilon_1 p_1
dy\wedge dz +p_1dy\wedge d\xi
$$
$$
F_2=\varepsilon_2 u_2 dx\wedge dz + u_2 dx\wedge d\xi + \varepsilon_2 p_2
dy\wedge dz +p_2 dy\wedge d\xi,
$$
where
$$
u_1=\phi_1cos\left(-\kappa_1\frac{z}{\mathcal{L}^1_o}+const_1\right),\
p_1=\phi_1sin\left(-\kappa_1\frac{z}{\mathcal{L}^1_o}+const_1\right),\
$$
$$
u_2=\phi_2cos\left(-\kappa_2\frac{z}{\mathcal{L}^2_o}+const_2\right),\
p_2=\phi_2sin\left(-\kappa_2\frac{z}{\mathcal{L}^2_o}+const_2\right),
$$
and $\kappa_1=\pm 1, \kappa_2=\pm 1$. Assuming further $const_1=0=const_2$ for
the first additional equation we obtain
$$
\delta F_1\wedge *F_2+\delta
F_2\wedge *F_1=\left(-\frac{\kappa_1}{\mathcal{L}_o^1}+
\frac{\kappa_2}{\mathcal{L}_o^2}\right)(u_1p_2-u_2p_1)dx\wedge dy\wedge
dz+
$$
$$
+\left(-\varepsilon_1\frac{\kappa_1}{\mathcal{L}_o^1}+
\varepsilon_2\frac{\kappa_2}{\mathcal{L}_o^2}
\right)(u_1p_2-u_2p_1)dx\wedge dy\wedge d\xi+
$$
$$
+\left[p_1\left(u_{1x}+p_{1y}\right)+p_2\left(u_{2x}+p_{2y}\right)\right]
(\varepsilon_1\varepsilon_2 -1)dx\wedge dz\wedge d\xi+
$$
$$
+\left[u_1\left(u_{1x}+p_{1y}\right)+u_2\left(u_{2x}+p_{2y}\right)\right]
(1-\varepsilon_1\varepsilon_2)dy\wedge dz\wedge d\xi=0.
$$
Since
$$
u_1p_2-u_2p_1=\phi_1\phi_2sin\left[\left(\frac{\kappa_2}{\mathcal{L}_o^2}-
\frac{\kappa_1}{\mathcal{L}_o^1}\right).z\right]\neq 0,
$$
the coefficient before $dx\wedge dy\wedge dz$ will be equal to zero only if
$$
\frac{\kappa_1}{\mathcal{L}_o^1}=\frac{\kappa_2}{\mathcal{L}_o^2}+k\pi, \
k=0,1,2,....
$$
Under this condition the coefficient in front of $dx\wedge
dy\wedge d\xi$ will become zero if $\varepsilon_1=\varepsilon_2$. From this
last relation it follows that the other two coefficients, obviously, are also
zero. A corresponding computation shows that for $k=0$ the so obtained conditions
$$
\mathcal{L}_o^1=\mathcal{L}_o^2,\ \varepsilon_1=\varepsilon_2,\
\kappa_1=\kappa_2
$$
are sufficient for $F_1$ and $F_2$ to satisfy the
additional equations. Hence, if the 2-form
$$
\Omega=(F_1\otimes e_1
+*F_1\otimes e_2)\otimes E_1+ (F_2\otimes k_1 +*F_2\otimes k_2)\otimes E_2
$$
satisfies the full system of equations, then the 2-form
$F=F_1+F_2$ could be further studied as a possible solution of our initial
equations
$$
\delta F\wedge *F=0,\ \delta *F\wedge **F=0,\ \delta F\wedge
**F+\delta *F\wedge *F=0.
$$  Some spatial analog of
coherence conditions appears and the "interference" of the two fields $F_1$ and
$F_2$ has some chance, provided the other two additional equations will also be
satisfied. As for the "interference" energy density $W_{12}$, if $const_1\neq
const_2$, the conditins obtained lead to
$$
 W_{12}=\phi_1^2+\phi_2^2+2\phi_1\phi_2cos(const_2-const_1),
$$
which suggests that some spatial analog of the classical interference picture
could be expected.

\vskip 0.2cm
{\bf Reference}
\vskip 0.2cm
\addcontentsline{toc}{subsection}{{\bf Reference}}

1. {\bf W.H. Greub}, {\it Linear Algebra}, third edition, Springer, 1967

\newpage
{\bf Appendix C}
\vskip 0.2cm
\addcontentsline{toc}{section}{{\bf Appendix C.}\\
 Generalized parallelism - examples}
\vskip 0.2cm
{\bf Generalized parallelism - examples}
\vskip 0.2cm
We begin studying the potential strength of the {\bf Generalized parallelism}
({\bf GP}) as formulated in Sec.3.7.3. \vskip 0.4cm {\bf 1. Integral invariance
relations} \vskip 0.2cm These relations have been introduced and studied from
the point of view of applications in mechanics by Lichnerowicz.

We specify the bundles over the real finite dimensional manifold
$M$ introduced in Sec.3.7.3.:

$\xi_1=TM;\  \xi_2=T^*(M);\  \
\eta_1=\eta_2=\xi_3=\eta_3=M\times\mathbb{R},\  \text{denote}\
Sec(M\times \mathbb{R})\equiv C^{\infty}(M)$

$\Phi$=substitution operator, denoted by\  $i(X), X\in Sec(TM)$;

$\varphi$=point-wise product of functions.

We denote by $1$ the function $f(x)=1, x\in M$. Consider the sections
\newline $X\otimes 1\in Sec(TM\otimes(M\times \mathbb{R}));\ \ \alpha\otimes
1\in Sec(T^*M\otimes(M\times\mathbb{R}))$.  Then the {\bf GP} leads to
$$
(\Phi,\varphi)(X\otimes 1, \alpha\otimes 1)=i(X)\alpha\otimes 1      %3%
=i(X)\alpha=0.
$$
We introduce now the differential operator $\mathbf{d}$: if $\alpha$ is an
exact 1-form, $\alpha=\mathbf{d}f$, so that $\tilde{\xi}=M\times\mathbb{R}$,
and obtain
$$
i(X)\alpha=i(X)\mathbf{d}f=X(f)=0,
$$
i.e. the derivative of $f$ along the vector field $X$ is equal to zero. So,
we obtain the well known relation, defining the first integrals  $f$ of the
dynamical system determined by the vector field $X$.  In this sense $f$ may
be called $(\Phi,\varphi,\mathbf{d})$-{\it parallel} with respect to $X$,
where $\Phi$ and $\varphi$ are defined above. If $\alpha$ is a $p$-form,
$\alpha\in Sec(\Lambda^p(T^*M))$, but this does not change the validity of
the above relation.
\vskip 0.3cm
%\noindent
{\bf 2. Absolute and relative integral invariants}
\vskip 0.2cm
These quantities have been introduced and studied in mechanics by Cartan.
By definition, a $p$-form $\alpha$ is called an {\it absolute integral
invariant} of the vector field $X$ if $i(X)\alpha=0$ and
$i(X)\mathbf{d}\alpha=0$. And $\alpha$ is called a {\it relative integral
invariant} of the field $X$ if $i(X)\mathbf{d}\alpha=0$. So, in our
terminology (the same bundle picture as above), we can call the relative
integral invariants of $X$ $(\Phi,\varphi;\mathbf{d})$-{\it parallel} with
respect to $X$, and the absolute integral invarians of $X$ have additionally
$(\Phi,\varphi)$-{\it parallelism} with respect to $X$, with $(\Phi,\varphi)$
as defined above.  A special case is when $p=n$, and $\omega\in\Lambda^n(M)$
is a volume form on $M$.
\newpage
%\vskip 0.5cm
%\noindent
{\bf 3. Symplectic mechanics}
\vskip 0.2cm
Symplectic manifolds are even dimensional and have a
distinguished nondegenerate closed $2$-form $\omega$, $\mathbf{d}\omega=0$.
This structure may be defined in terms of the {\bf GP} in the following way.
Choose $\xi_1=\eta_1=\eta_2=M\times\mathbb{R}$, $\xi_2=\Lambda^2(T^*M)$, and
$\mathbf{d}$ as a differential operator. Consider now the section $1\in
Sec(M\times\mathbb{R})$ and the section $\omega\otimes 1\in
Sec(\Lambda^2(T^*M))\otimes Sec(M\times\mathbb{R})$, with $\omega$ -
nondegenerate.  The map $\Phi$ is the product $f.\omega$ and the map
$\varphi$ is the product of functions. So, we have
$$
(\Phi,\varphi;\mathbf{d})(1\otimes1, \omega\otimes 1)=
1.\mathbf{d}\omega\otimes 1=\mathbf{d}\omega=0.
$$
Hence, the relation $\mathbf{d}\omega=0$ is equivalent to the requirement
$\omega$ to be $(\Phi,\varphi;\mathbf{d})$-{\it parallel} with respect to the
section $1\in Sec(M\times\mathbb{R})$.

The hamiltonian vector fields $X$ are defined by the condition
$L_X\omega=\mathbf{d}i(X)\omega=0$.
If $\Phi=\varphi$ is the point-wise product of functions
we have
$$
(\Phi,\varphi;\mathbf{d})(1\otimes 1,i(X)\omega\otimes 1)=
(\Phi,\varphi)(1\otimes 1,\mathbf{d}i(X)(\omega)\otimes 1)=
L_X\omega\otimes 1=L_X\omega=0.
$$
In terms of the {\bf GP} we can say that $X$ is hamiltonian if $i(X)\omega$ is
$(\Phi,\varphi;\mathbf{d})$-{\it parallel}.

The induced Poisson structure $\{f,g\}$, is
given in terms of the {\bf GP} by setting $\Phi=\omega^{-1}$,
where $\omega^{-1}.\omega=id_{TM}$,
$\varphi$=point-wise product of functions,
and $1\in Sec(M\times\mathbb{R})$. We get
$$
(\Phi,\varphi)(\mathbf{d}f\otimes 1, \mathbf{d}g\otimes 1)=
\omega^{-1}(\mathbf{d}f,\mathbf{d}g)\otimes 1.
$$
A closed 1-form $\alpha,\ \mathbf{d}\alpha=0$, is a first integral of the
hamiltonian system $Z$, $\mathbf{d}i(Z)\omega=0$, if $i(Z)\alpha=0$. In terms
of the {\bf GP} we can say that the first integrals $\alpha$ are
$(i,\varphi)$-parallel with respect to $Z$:
$(i,\varphi)(Z\otimes 1,\alpha\otimes 1)=i(Z)\alpha\otimes 1=0$.
From $L_Z\omega=0$ it follows
$L_Z\omega^{-1}=0$. The Poisson bracket $(\alpha,\beta)$ of two first
integrals $\alpha$ and $\beta$ is equal to
$(-\mathbf{d}\omega^{-1}(\alpha,\beta))$ [5]. The well known property that the
Poison bracket of two first integrals of $Z$ is again a first integral of $Z$
may be formulated as: the function $\omega^{-1}(\alpha,\beta)$ is
$(i,\varphi;\mathbf{d})$-parallel with respect to $Z$,

$$
(i,\varphi;\mathbf{d})(Z\otimes 1, \omega^{-1}(\alpha,\beta)\otimes 1)=
i(Z)\mathbf{d}\omega^{-1}(\alpha,\beta)\otimes 1=0.
$$

\vskip 0.4cm
%\noindent
{\bf 4. Frobenius integrability theorems}
\vskip 0.2cm
Let $\Delta=(X_1,\dots,X_r)$ be a differential system on $M$, i.e. the vector
fields $X_i, i=1,\dots,r$ define a locally stable submodule of $Sec(TM)$ and
at every point $p\in M$ the subspace $\Delta_p^r\subset T_p(M)$ has dimension
$r$. Then $\Delta^r$ is called integrable if $[X_i,X_j]\in \Delta^r,
i,j=1,\dots,r$. Denote by $\Delta^{n-r}_p\subset T_p(M)$ the complimentary
subspace: $\Delta_p^r\oplus\Delta^{n-r}_p=T_p(M)$, and let $\pi:
T_p(M)\rightarrow \Delta^{n-r}_p$ be the corresponding projection. So,
the corresponding Frobenius integrability condition means $\pi([X_i,X_j])=0,
i,j=1,\dots,r$.

In terms of the {\bf GP} we set $D(X_i)=\pi\circ L_{X_i}$, $\Phi$="product of
functions and vector fields",  and
$\varphi$ again the pruduct of functions. The integrability
condition now is
\[
\begin{split}
&(\Phi,\varphi;D(X_i))
(1\otimes 1, X_j\otimes 1)\\&=(\Phi,\varphi)
(1\otimes 1,\pi([X_i,X_j]\otimes 1))=1.\pi([X_i,X_j])\otimes 1.1
=0,
\quad i,j=1,\dots,r.
\end{split}
\]

In the dual formulation we have the Pfaff system
$\Delta^*_{n-r}$, generated by the lineary independent 1-forms
$(\alpha_1,\dots,\alpha_{n-r})$, such that
$\alpha_m(X_i)=0, i=1,\dots r; m=1,\dots n-r$.
Then $\Delta^*_{n-r}$ is
integrable if $\mathbf{d}\alpha\wedge \alpha_1\wedge\dots \wedge
\alpha_{n-r}=0, \alpha\in \Delta^*_{n-r}$. In terms of {\bf GR} we set
$\varphi$ the same as above, $\Phi=\wedge$ and $\mathbf{d}$ as differential
operator.
$$
 (\Phi,\varphi;\mathbf{d})(\alpha_1\wedge\dots
\wedge\alpha_{n-r}\otimes 1, \alpha\otimes 1)= \mathbf{d}\alpha\wedge
\alpha_1\wedge\dots \wedge\alpha_{n-r}\otimes 1=0.
$$

\vskip 0.4cm
%\noindent
 {\bf 5. Linear connections}
\vskip 0.2cm
The concept of a linear connection in a vector bundle has proved to be of
great importance in geometry and physics. In fact, it allows to differentiate
sections of vector bundles along vector fields, which is a basic operation in
differential geometry, and in theoretical physics the physical fields are
represented mainly by sections of vector bundles. We recall now how one comes
to it.

Let $f:\mathbb{R}^n\rightarrow\mathbb{R}$ be a differentiable function. Then
we can find its differential $\mathbf{d}f$. The map $f\rightarrow\mathbf{d}f$
is $\mathbb{R}$-linear: $\mathbf{d}(\kappa.f)=\kappa.\mathbf{d}f$,
$\kappa \in \mathbb{R}$, and it has the
derivative property $\mathbf{d}(f.g)=f\mathbf{d}g+g\mathbf{d}f$. These two
properties are characteristic ones, and they are carried to the bundle
situation as follows.

Let $\xi$ be a vector bundle over $M$. We always have the trivial bundle
$\xi_o=M\times\mathbb{R}$. Consider now $f\in C^{\infty}(M)$ as a section of
$\xi_o$. We note that $Sec(\xi_o)=C^{\infty}(M)$ is a module over itself, so
we can form $\mathbf{d}f$ with the above two characteristic  properties. The
new object $\mathbf{d}f$ lives in the space $\Lambda^1(M)$ of 1-forms on $M$,
so it defines a linear map $\mathbf{d}f: Sec(TM)\rightarrow Sec(\xi_o),
\mathbf{d}f(X)=X(f)$.  Hence, we have a map $\nabla$ from $Sec(\xi_o)$ to
the 1-forms with values in $Sec(\xi_o)$, and this map has the above two
characteristic properties.  We say that $\nabla$ defines a linear connection
in the vector bundle $\xi_o$.

In the general case the sections $Sec(\xi)$ of the vector bundle $\xi$ form a
module over $C^{\infty}(M)$. So, a linear connection $\nabla$ in $\xi$ is a
$\mathbb{R}$-linear map $\nabla: Sec(\xi)\rightarrow \Lambda^1(M,\xi)$. In
other words, $\nabla$ sends a section $\sigma\in Sec(\xi)$ to a 1-form
$\nabla \sigma$ valued in $Sec(\xi)$ in such a way, that
$$
\nabla(k\,\sigma)=k\,\nabla(\sigma), \quad                     %4%
\nabla(f\,\sigma)=df\otimes\sigma+f\,\nabla(\sigma),
$$
where $k\in \mathbb{R}$ and $f\in C^{\infty}(M)$. If $X\in Sec(TM)$ then we
have the composition $i(X)\circ\nabla$, so that
$$
i(X)\circ\nabla(f\,\sigma)=X(f)\,\sigma+f\,\nabla_X(\sigma),
$$
where $\nabla_X(\sigma)\in Sec(\xi)$.

In terms of the {\bf GP} we put $\xi_1=TM=\tilde\xi$ and
$\xi_2=\Lambda^1(M)\otimes\xi$,
and $\eta_1=\eta_2=\xi_o$. Also, $\Phi(X,\nabla\sigma)=\nabla_X \sigma$ and
$\varphi(f,g)=f.g$. Hence, we obtain
$$
(\Phi,\varphi;\nabla)(X\otimes 1, \sigma\otimes 1)           %5%
(\Phi,\varphi)(X\otimes 1,(\nabla \sigma)\otimes 1)
=\nabla_X \sigma\otimes 1=\nabla_X \sigma,
$$
and the section $\sigma$ is called $\nabla$-{\it parallel} with respect to
$X$ if $\nabla_X\sigma=0$.

\vskip 0.4cm
%\noindent
{\bf 6. Covariant exterior derivative}
\vskip 0.2cm
The space of $\xi$-valued $p$-forms $\Lambda^p(M,\xi)$ on $M$ is isomorphic
to $\Lambda^p(M)\otimes Sec(\xi)$. So, if $(\sigma_1,\dots,\sigma_r)$ is a
local basis of $Sec(\xi)$, every $\Psi\in \Lambda^p(M,\xi)$ is represented by
$\psi^i\otimes \sigma_i, i=1,\dots,r$, where $\psi^i\in \Lambda^p(M)$.
Clearly the space $\Lambda(M,\xi)=\Sigma^n_{p=0}\Lambda^p(M,\xi)$, where
$\Lambda^o(M,\xi)=Sec(\xi)$, is a
$\Lambda(M)=\Sigma^n_{p=0}\Lambda^p(M)$-module:
$\alpha.\Psi=\alpha\wedge\Psi=(\alpha\wedge\psi^i)\otimes \sigma_i$.

A linear connection $\nabla$ in $\xi$ generates covariant exterior
derivative $\mathbf{D}: \Lambda^p(M,\xi)\rightarrow\Lambda^{p+1}(M,\xi)$ in
$\Lambda(M,\xi)$ according to the rule
\[
\begin{split}
\mathbf{D}\Psi&=\mathbf{D}(\psi^i\otimes \sigma_i)=
\mathbf{d}\psi^i\otimes \sigma_i+(-1)^p \psi^i\wedge\nabla(\sigma_i)\\
&=(\mathbf{d}\psi^i+(-1)^p \psi^j\wedge\Gamma_{\mu j}^i dx^\mu)\otimes\sigma_i
=(\mathbf{D}\Psi)^i\otimes\sigma_i.
\end{split}
\]
We may call now a $\xi$-valued $p$-form $\Psi$ $\nabla$-{\it parallel} if
$\mathbf{D}\Psi=0$, and $(X,\nabla)$-{\it parallel} if
$i(X)\mathbf{D}\Psi=0$. This definition extends in a natural way to
$q$-vectors with $q\le p$. Actually, the substitution operator $i(X)$ extends
to (decomposable) $q$-vectors $X_1\wedge X_2\wedge\dots\wedge X_q$ as
follows:
$$
i(X_1\wedge X_2\wedge\dots\wedge X_q)\Psi =i(X_q)\circ
i(X)_{q-1}\circ\dots\circ i(X_1)\Psi,
$$
and extends to nondecomposable
$q$-vectors by linearity. Hence, if $\Theta$ is a section of
$\Lambda^q(TM)$ we may call $\Psi$ $(\Theta,\nabla)$-{\it parallel} if
$i(\Theta)\mathbf{D}\Psi=0$.

Denote now by $L_\xi$ the vector bundle of (linear) homomorphisms $(\Pi,id):
\xi\rightarrow \xi$, and let $\Pi\in Sec(L_\xi)$. Let
$\chi \in Sec(\Lambda^q(TM)\otimes L_\xi)$ be represented as
$\Theta\otimes\Pi$. The map $\Phi$ will act as: $\Phi(\Theta,\Psi)=
i(\Theta)\Psi$, and the map $\varphi$ will act as: $\varphi(\Pi,\sigma_i)=
\Pi(\sigma_i)$.
So, if $\nabla(\sigma_k)=\Gamma^j_{\mu k}dx^\mu\otimes
\sigma_j$, we may call $\Psi$ $(\nabla)$-{\it parallel} with respect to
$\chi$ if
$$
(\Phi,\varphi;\mathbf{D})(\Theta\otimes\Pi,\Psi=\psi^i\otimes\sigma_i)
$$
$$
=(\Phi,\varphi)(\Theta\otimes\Pi,(\mathbf{D}\Psi)^i\otimes\sigma_i)=     %6%
i(\Theta)(\mathbf{D}\Psi)^i\otimes\Pi(\sigma_i)=0.
$$
If we have isomorphisms $\otimes^p TM\backsim
\otimes^p T^*M, p=1,2,\dots$, defined in some natural way (e.g. through a
metric tensor field), then to any $p$-form $\alpha$ corresponds unique
$p$-vector $\tilde\alpha$. In this case we may talk about "$\backsim$"-
{\it autopaparallel} objects with respect a (point-wise) bilinear map
$\varphi:  (\xi\times\xi)\rightarrow \eta$, where $\eta$ is also a vector
bundle over $M$. So, $\Psi=\alpha^k\otimes\sigma_k\in \Lambda^p(M,\xi)$ may
be called $(i,\varphi;\nabla)$-{\it autoparallel} with respect to the
isomorphism "$\backsim$" if
\[
\begin{split}
&(i,\varphi;\nabla)
(\tilde\alpha^k\otimes \sigma_k,\alpha^m\otimes\sigma_m)\\
&=i(\tilde\alpha^k)\mathbf{d}\alpha^m\otimes\varphi(\sigma_k,\sigma_m)+
(-1)^p i(\tilde\alpha^k)(\alpha^j\wedge \Gamma^m_{\mu j}dx^\mu)
\otimes\varphi(\sigma_k,\sigma_m)\\                                   %7%
&=\big[i(\tilde\alpha^k)\mathbf{d}\alpha^m+
(-1)^p i(\tilde\alpha^k)(\alpha^j\wedge \Gamma^m_{\mu j}dx^\mu)\big]
\otimes\varphi(\sigma_k,\sigma_m)=0.
\end{split}
\]
\noindent
Although the above examples do not, of course, give a complete list of the
possible applications of the {\bf GP} (2), they will serve as a good basis
for the physical applications we are going to consider further.
\vskip 0.3cm
%\noindent
{\bf 7. Autoparallel vector fields and 1-forms}
\vskip 0.2cm
In nonrelativistic and relativistic mechanics the vector fields $X$ on a
manifold $M$ are the local representatives (velocity vectors) of the
evolution trajectories for point-like objects.  The condition that a particle
is {\it free} is mathematically represented by the requirement that the
corresponding vector field $X$ is autoparallel with respect to a given
connection $\nabla$ (covariant derivative) in $TM$:
$$
i(X)\nabla X=\nabla_XX=0,\quad
\text{or in components},\quad
X^\sigma \nabla_\sigma X^\mu +\Gamma^\mu_{\sigma\nu}X^\sigma X^\nu=0.   %8%
$$
In view of the physical interpretation of $X$ as velocity vector field the
usual latter used instead of $X$ is $u$. The above equation presents a
system of nonlinear partial differential equations for the components
$X^\mu$, or $u^\mu$. When reduced to 1-dimensional submanifold which is
parametrised locally by the appropriately chosen parameter $s$, we get a
system of ordinary differential equations:
$$
\frac{d^2 x^\mu}{ds^2}+\Gamma^\mu_{\sigma\nu}\frac{dx^\nu}{ds}       %9%
\frac{dx^\nu}{ds}=0,
$$
which are known as ODE defining the geodesic (with respect to $\Gamma$)
lines in $M$. When $M$ is reimannian with metric tensor $g$ and $\Gamma$ the
corresponding Levi-Civita connection, i.e. $\nabla g=0$ and
$\Gamma^\mu_{\nu\sigma}=\Gamma^\mu_{\sigma\nu}$, then the solutions
give the extreme (shortest or longest) distance $\int^b_a ds$
between the two points $a,b\in M$:
$$
\delta\left(\int^b_a ds\right)=
\delta\left(\int^b_a
\sqrt{g_{\mu\nu}\frac{dx^\mu}{ds}\frac{dx^\nu}{ds}}\right)=0.
$$
A system of particles that move along such solutions  with $g$-the
Minkowski metric and $g_{\mu\nu}\frac{dx^\mu}{ds}\frac{dx^\nu}{ds}>0$,
is said to form an {\it inertial frame of reference}.

It is interesting to note that the above system $\nabla_XX=0$
has (3+1)-soliton-like
(even spatially finite) solutions on Minkowski space-time
$(\mathbb{R}^4,\eta)$ with respect to the corresponding to $\eta$ Levi-Civita
connection $\Gamma$. In fact, in canonical coordinates
$(x^1,x^2,x^3,x^4)=(x,y,z,\xi=ct)$ we have $\Gamma_{\mu\nu}^\sigma=0$, and let
$u^\mu=(0,0,\pm \frac vc f,f)$ be the components of $u$, where $0<v=const<c$,
and $c$ is the velocity of light, so $\frac vc < 1$ and $u^\sigma u_\sigma =
\left(1-\frac{v^2}{c^2}\right)f^2>0$. Then every function $f$ of the kind $$
f(x,y,z,\xi)=f\left(x,y,\alpha.(z\mp\frac vc \xi)\right),\ \alpha=const,
\quad\text{for example}\quad \alpha=\frac{1}{\sqrt{1-\frac{v^2}{c^2}}}, $$
defines a slution.  If $u_\sigma u^\sigma =0$ then the equations
 are equivalent to $u^\mu(\mathbf{d}u)_{\mu\nu}=0$, where
$\mathbf{d}$ is the exterior derivative. In fact, since the connection used
is riemannian, we have $0=\nabla_\mu\frac12(u^\nu u_\nu)=u^\nu\nabla_\mu
u_\nu$, so the relation $u^\nu\nabla_\nu u_\mu -u^\nu\nabla_\mu u_\nu=0$
holds and is obviously equal to $u^\mu(\mathbf{d}u)_{\mu\nu}=0$. The
soliton-like solution is defined by $u=(0,0,\pm f,f)$ where the function $f$
is of the form
$$
f(x,y,z,\xi)=f(x,y,z\mp \xi).
$$
Clearly, for every autoparallel vector field $u$ (or one-form $u$) there
exists a canonical coordinate system on the Minkowski space-time, in
which $u$ takes such a simple form: $u^\mu=(0,0,\alpha f,f), \alpha=const$.
The dependence of $f$ on the three spatial coordinates $(x,y,z)$ is arbitrary
, so it is allowd to be chosen {\it soliton-like} and, even, {\it finite}.

The properties described give a connection between free point-like objects
and (3+1) soliton-like autoparallel vector fields on Minkowski space-time.
Moreover, they suggest that extended free objects with more complicated
space-time dynamical structure may be described by some appropriately
generalized concept of autoparallel mathematical objects.
\vskip 0.4cm
%\noindent
{\bf 8. Electrodynamics}
\vskip 0.2cm
{\bf 8.1 Maxwell equations}

The Maxwell equations $\mathbf{d}F=0, \mathbf{d}*F=0$ in their 4-dimensional
formulation on Minkowski space-time $(M,\eta), sign(\eta)=(-,-,-,+)$ and the
Hodge $*$ is defined by $\eta$, make use of the exterior derivative as a
differential operator.  The field has, in general, 2 components
$(F,*F)$, so the interesting bundle is $\Lambda^2(M)\otimes V$, where $V$ is
a real 2-dimensional vector space. Hence the adequate mathematical field
will look like $\Omega=F\otimes e_1+*F\otimes e_2$, where $(e_1,e_2)$ is a
basis of $V$. The exterior derivative acts on $\Omega$ as:
$\mathbf{d}\Omega=\mathbf{d}F\otimes e_1+\mathbf{d}*F\otimes e_2$, and the
equation $\mathbf{d}\Omega=0$ gives the vacuum Maxwell equations.

In order to interpret in terms of the above given general view ({\bf GP})
on parallel objects with respect to given sections of vector bundles and
differential operators we consider the sections (see the above introdused
notation) $(1\times1, \Omega\times 1)$ and the differential operator
$\mathbf{d}$. Hence, the {\bf GP} acts as follows:
$$
(\Phi,\varphi;\mathbf{d})(1\times1, \Omega\times 1)=
(\Phi,\varphi)(1\times1,\mathbf{d}\Omega\times 1)=
(1.\mathbf{d}\Omega\otimes 1.1)
$$
The corresponding $(\Phi,\varphi;\mathbf{d})$-parallelism leads to
$\mathbf{d}\Omega=0$. In presence of electric $\mathbf{j}$ and magnetic
$\mathbf{m}$
currents, considered as 3-forms, the parallelism condition does not hold and
on the right-hand side we'll have non-zero term, so the full condition is
$$
(\Phi,\varphi)(1\times1, (\mathbf{d}F\otimes e_1+
\mathbf{d}*F\otimes e_2)\times 1)=
(\Phi,\varphi;\mathbf{d})(1\times1, (\mathbf{m}\otimes e_1+        %10%
\mathbf{j}\otimes e_2)\times 1)
$$
The case $\mathbf{m}=0, F=\mathbf{d}A$ is, obviously a special case.

\vskip 0.2cm
{\bf 8.2 Extended Maxwell equations}

The extended Maxwell equations (on Minkowski space-time) in vacuum read:
$$
F\wedge *\mathbf{d}F=0,\quad (*F)\wedge(*\mathbf{d}*F)=0,\quad      %11%
F\wedge(*\mathbf{d}*F)+(*F)\wedge(*\mathbf{d}F)=0 .
$$
They may be expressed through the {\bf GP} in the following way.
On $(M,\eta)$ we have the bijection between $\Lambda^2(TM)$ and
$\Lambda^2(T^*M)$ defined by $\eta$, which we denote by
$\tilde F\leftrightarrow F$. So, the equations are equivalent to
$$
i(\tilde F)\mathbf{d}F=0,\quad i(\widetilde{*F})\mathbf{d}*F=0,\quad
i(\tilde F)\mathbf{d}*F+i(\widetilde{*F})\mathbf{d}F=0.
$$
We consider the
sections $\tilde\Omega=\tilde F\otimes e_1+\widetilde{*F}\otimes e_2$ and
$\Omega=F\otimes e_1+*F\otimes e_2$ with the differential operator
$\mathbf{d}$. The maps $\Phi$ and $\varphi$ are defined as:
$\Phi$ is the substitution operator $i$, and $\varphi=\vee$ is
the symmetrized tensor product in $V$. So we obtain
\[
\begin{split} &(\Phi,\varphi;\mathbf{d})(\tilde F\otimes
e_1+\widetilde{*F}\otimes e_2, F\otimes e_1+*F\otimes e_2)\\ &=i(\tilde
F)\mathbf{d}F\otimes e_1\vee e_1+ i(\widetilde{*F})\mathbf{d}*F\otimes  %12%
e_2\vee e_2\\ &+(i(\tilde F)\mathbf{d}*F+i(\widetilde{*F})\mathbf{d}F)
\otimes e_1\vee e_2=0.
\end{split}
\]

\vskip 0.4cm
%\noindent
{\bf 9. Yang-Mills theory}
\vskip 0.2cm
{\bf 9.1 Yang-Mills equations}

In this case the field is a connection, represented locally by its connection
form $\omega\in \Lambda^1(M)\otimes\mathfrak{g}$, where $\mathfrak{g}$ is the
Lie algebra of the corresponding Lie group $G$.
If $\mathbf{D}$ is the corresponding
covariant derivative, and $\Omega=\mathbf{D}\omega$ is the curvature,
then Yang-Mills equations read
$\mathbf{D}*\Omega=0$. The formal difference with the Maxwell case
is that $G$ may NOT be commutative, and may have, in general, arbitrary
finite dimension. So, the two sections are $1\otimes 1$ and
$*\Omega\otimes 1$, the maps $\Phi$ and $\varphi$ are product of functions
and the differential operator is $\mathbf{D}$. So, we may write
$$
(\Phi,\varphi;\mathbf{D})(1\otimes 1, *\Omega\otimes 1)=      %15%
\mathbf{D}*\Omega\otimes 1=0.
$$

\vskip 0.3cm
{\bf 9.2 Extended Yang-Mills equations}
\vskip 0.2cm
The extended Ynag-Mills equations are written down in analogy
with the extended Maxwell equations.
The field of interest is an arbitrary 2-form  $\Psi$ on $(M,\eta)$ with
values in a Lie algebra $\mathfrak{g}$, $\dim(\mathfrak{g})=r$. If
$\{E_i\}, i=1,2,\dots,r$ is a basis of $\mathfrak{g}$ we have
$\Psi=\psi^i\otimes E_i$ and $\tilde\Psi=\tilde\psi^i\otimes E_i$. The map
$\Phi$ is the substitution operator, the map $\varphi$ is the corresponding
Lie product $[,]$, and the differential operator is the exterior covariant
derivative with respect to a given connection $\omega$:
$\mathbf{D}\Psi=\mathbf{d}\Psi+[\omega,\Psi]$. We obtain
$$
(\Phi,\varphi;\mathbf{D})(\tilde\psi^i\otimes E_i,\psi^j\otimes E_j)=
i(\tilde\psi^i)(\mathbf{d}\psi^m+\omega^j\wedge\psi^k\,C_{jk}^m)      %16%
\otimes[E_m,E_i]=0,
$$
where $C_{jk}^m$ are the corresponding structure constants.
If the connection is the trivial one, then $\omega=0$ and
$\mathbf{D}\rightarrow \mathbf{d}$, so, this equation reduces to
$$
i(\tilde\psi^i)\mathbf{d}\psi^j\,C_{ij}^k\otimes E_k=0 .        %17%
$$
If, in addition, instead of $[,]$ we assume for $\varphi$ some bilinear map
$f:\mathfrak{g}\times\mathfrak{g}\rightarrow\mathfrak{g}$, such that in this
basis  $f$ is given by $f(E_i,E_i)=E_i$, and $f(E_i,E_j)=0$ for
$i\neq j$ the last relation reads
$$
i(\tilde\psi^i)\mathbf{d}\psi^i\otimes E_i=0,\quad i=1,2,\dots,r.  %18%
$$
The last equations define the components $\psi^i$ as independent 2-forms
(of course $\psi^i$ may be arbitrary $p$-forms).
If the bilinear map
$\varphi$ is chosen to be the symmetrized tensor product
$\vee:\mathfrak{g}\otimes\mathfrak{g}\rightarrow \mathfrak{g}\vee
\mathfrak{g}$, we obtain
$$
i(\tilde{\psi^i})\mathbf{d}\psi^j\otimes E_i\vee E_j=0,\quad i\leqq
j=1,\dots,r.
$$
These equations may be used to model bilinear
interaction among the components of $\Psi$. If the terms
$i(\psi^i)\mathbf{d}\psi^j\otimes E_i\vee E_j$ have the physical sense of
energy-momentum exchange we may say that every component $\psi^i$ gets locally
as much energy-momentum from $\psi^j$ as it gives to it. Since
$C^k_{ij}=-C^k_{ji}$, the former equations consider only the case $i<j$, while
the latter equations consider $i\leq j$, in fact, for every $i,j=1,2,\dots,r$ we
obtain
 $$ i(\tilde\psi^i)\mathbf{d}\psi^i=0,\ \ \text{and}\ \
i(\tilde\psi^i)\mathbf{d}\psi^j+i(\tilde\psi^j)\mathbf{d}\psi^i=0.
$$
Clearly, these last equations may be considered as a natural generalization of
the extended electrodynamics equations, so spatial soliton-like solutions
are expectable.

\vskip 0.4cm
%\noindent
{\bf 10. General Relativity}
\vskip 0.2cm
In General Relativity the field function of interest is in a definite
sense identified with a pseudometric $g$ on a 4-dimensional manifold,
and only those $g$ are considered as appropriate to describe the real
gravitaional fields which satisfy the equations
$R_{\mu\nu}=0$, where $R_{\mu\nu}$ are the components of the Ricci tensor.
The main mathematical object which detects possible gravity is the Rieman
curvature tensor $R_{\alpha\mu,\beta\nu}$, which is a second order nonlinear
differential operator $R:g\rightarrow R(g)$. The map $\Phi$ is just a
contraction:
$$
\Phi:(g_{\alpha\beta},R_{\alpha\mu,\beta\nu})
=g^{\alpha\beta}R_{\alpha\mu,\beta\nu}=R_{\mu\nu}
$$
and is obviouly bilinear. The map $\varphi$ is a product of functions, so
the {\bf GP} gives
$$
(\Phi,\varphi;R)(g\otimes 1,g\otimes 1)=\Phi(g,R(g))\otimes 1    %20%
=Ric(R(g))\otimes 1=0.
$$
In presence of matter fields $\Psi^a, a=1,2,\dots,r$, the system of equations
is
$$
R_{\mu\nu}-\kappa\left(T_{\mu\nu}-\frac12 g_{\mu\nu}T\right)=0.
$$
It is easily obtained through the {\bf GP} if we modify the differential
operator $R_{\alpha\mu,\beta\nu}$ to
$$
R_{\alpha\mu,\beta\nu}-\frac{\kappa}{2}\left(T_{\alpha\beta}g_{\mu\nu}+
T_{\mu\nu}g_{\alpha\beta}-T_{\alpha\nu}g_{\mu\beta}-
T_{\mu\beta}g_{\alpha\nu}\right)+
\frac{\kappa}{3}\left(g_{\alpha\beta}g_{\mu\nu}-
g_{\alpha\nu}g_{\mu\beta}\right)T,
$$
where $\kappa$ is the gravitational constant, $T_{\mu\nu}(\Psi^a)=
T_{\nu\mu}(\Psi^a)$ is the corresponding stress energy momentum tensor, and
$T=g^{\mu\nu}T_{\mu\nu}$.
\vskip 0.4cm

\vskip 0.4cm
%\noindent
{\bf 11. Schr\"odinger equation}
\vskip 0.2cm
The object of interest in this case is a map $\Psi: \mathbb{R}^4\rightarrow
\mathbb{C}$, and $\mathbb{R}^4=\mathbb{R}^3\times\mathbb{R}$ is
parametrized by the canonical coordinates $(x,y,z;t)$, where $t$ is the
(absolute) time "coordinate". The operator $D$ used here is
$$
D=i\hbar\frac{\partial }{\partial t}-\mathbf{H},
$$
where $\mathbf{H}$ is the corresponding {\it hamiltonian}. The maps $\Phi$
and $\varphi$ are products of functions, so the {\bf GP} gives
$$
(\Phi,\varphi;\mathbf{D})(1\otimes 1,\Psi\otimes 1)=
\left(1\otimes\left(i\hbar\frac{\partial \Psi}                  %21%
{\partial t}-\mathbf{H}\Psi\right)\right)\otimes 1=0.
$$

\vskip 0.4cm
%\noindent
{\bf 12. Dirac equation}
\vskip 0.2cm
The original free Dirac
equation on the Minkowski space-time $(M,\eta)$ makes use of the following
objects:
$\mathbb{C}^4$ - the canonical 4-dimensional complex vector space,
$L_{\mathbb{C}^4}$-the space of $\mathbb{C}$-linear maps
$\mathbb{C}^4\rightarrow\mathbb{C}^4$, $\Psi\in Sec(M\times \mathbb{C}^4)$,
$\gamma\in Sec(T^*M\otimes L_{\mathbb{C}^4})$, and the usual differential
$\mathbf{d}: \psi^i\otimes e_i\rightarrow \mathbf{d}\psi^i\otimes e_i$, where
$\{e_i\}, i=1,2,3,4$, is a basis of $\mathbb{C}^4$. We identify further
$L_{\mathbb{C}^4}$ with $(\mathbb{C}^4)^*\otimes \mathbb{C}^4$ and if
$\{\varepsilon^i\}$ is a basis of $(\mathbb{C}^4)^*$, dual to $\{e_i\}$, we
have the basis $\varepsilon^i\otimes e_j$ of $L_{\mathbb{C}^4}$. Hence, we
may write
$$
\gamma=\gamma_{\mu i}^j dx^\mu\otimes(\varepsilon^i\otimes e_j),
$$
and
\[
\begin{split}
\gamma(\Psi)&=
\gamma_{\mu i}^j dx^\mu\otimes(\varepsilon^i\otimes e_j)(\psi^k\otimes e_k)\\
&=\gamma_{\mu i}^j dx^\mu\otimes \psi^k<\varepsilon^i,e_k>e_j=
\gamma_{\mu i}^j dx^\mu\otimes \psi^k\delta^i_k e_j=
\gamma_{\mu i}^j \psi^i dx^\mu\otimes e_j.
\end{split}
\]
The 4 matrices $\gamma_\mu$ satisfy
$\gamma_\mu\gamma_\nu+\gamma_\nu\gamma_\mu=\eta_{\mu\nu}id_{\mathbb{C}^4}$,
so they are nondegenerate: $det(\gamma_\mu)\neq 0, \mu=1,2,3,4$,
and we can find $(\gamma_\mu)^{-1}$ and introduce $\gamma^{-1}$ by
$$
\gamma^{-1}=((\gamma_\mu)^{-1})_i^jdx^\mu\otimes(\varepsilon^i\otimes e_j)
$$
We introduce now the differential operators
$\mathcal{D}^{\pm}:Sec(M\times\mathbb{C}^4)\rightarrow Sec(T^*M\otimes
\mathbb{C}^4)$ through the formula:
$\mathcal{D}^{\pm}=i\mathbf{d}\pm\frac 12 m\gamma^{-1}, i=\sqrt{-1},
m\in \mathbb{R}$.
The corresponding maps are: $\Phi=\eta$, $\varphi:
L_{\mathbb{C}^4}\times\mathbb{C}^4\rightarrow\mathbb{C}^4$ given by
$\varphi(\alpha^*\otimes\beta,\rho)=<\alpha^*,\rho>\beta$.  We obtain

\begin{eqnarray*}
&&(\Phi,\varphi;\mathcal{D}^{\pm})(\gamma,\Psi)\\
&&=
(\Phi,\varphi)(\gamma^j_{\mu i}dx^\mu\otimes(\varepsilon^i\otimes e_j),
i\frac{\partial \psi^k}{\partial x^\nu}dx^\nu\otimes e_k\\
&&\pm\frac 12 m
(\gamma_\nu^{-1})^s_r dx^\nu\otimes(\varepsilon^r\otimes e_s)\psi^m e_m)\\
&&=i\gamma^j_{\mu i}\frac{\partial \psi^k}{\partial x^\nu}\eta(dx^\mu,dx^\nu)
<\varepsilon^i,e_k>e_j\\
&&\pm\frac 12 m\gamma^j_{\mu i}(\gamma_\nu^{-1})^s_r
\psi^r\eta(dx^\mu,dx^\nu)<\varepsilon^i,e_s>e_j\\
&&=i\eta^{\mu\nu}\gamma^j_{\mu i}\frac{\partial \psi^k}
{\partial x^\nu}\delta^i_k e_j\pm                                        %22%
\frac 12 m\eta^{\mu\nu}\gamma^j_{\mu i}(\gamma_\nu^{-1})^s_r\psi^r
\delta^i_s e_j\\
&&=i\gamma^{\mu j}_i\frac{\partial \psi^i}{\partial x^\mu} e_j\pm
\frac 12 m(-2\delta^j_r\psi^r)e_j
=\left(i\gamma^{\mu j}_i \frac{\partial \psi^i}{\partial
x^\mu}\mp m\psi^j\right)e_j=0.
\end{eqnarray*}

In terms of parallelism we can say that the Dirac equation is equavalent to
the requirement the section $\Psi\in Sec(M\times \mathbb{C}^4)$ to be
($\eta,\varphi;\mathcal{D}^{\pm}$)-parallel with respect to the given
$\gamma\in Sec(M\times L_{\mathbb{C}^4})$. Finally, in presence of external
gauge field $\mathbf{A}=A_\mu dx^\mu$ the differential operators
$\mathcal{D}^{\pm}$ modify to
$$
\mathfrak{D}^{\pm}=i\mathbf{d}-
e\mathbf{A}\otimes id_{\mathbb{C}^4}\pm \frac 12 m \gamma^{-1},
$$
where $e$ is the corresponding charge.

In conclusion, it was shown that the {\bf GP} defined, naturally generalizes
the geometrical concept of parallel transport,  and that it may be successfully
used as a unified tool to represent formally important equations in theoretical
physics.

\newpage
{\bf Studies of the authors related to the subject}
\vskip 0.2cm
\addcontentsline{toc}{section}{{\bf Studies of the authors related to the
subject}}
\vskip 0.3cm

1. "On the Equivariance of Some Conserved Quantities in Classical
Yang-Mills Theory", /Donev, S./, Compt. Rend. Bulg. Acad. Sci.,vol.33,
No.10, 1980.
\vskip 0.25cm
2. "Equivariance of conserved Quantities in Yang-Mills Theory",
/Donev, S./ IX Intern. Conf. on General Relativity and Gravitation, July
14-19,1980, Jena, Germany.
\vskip 0.25cm
3. "Symmetries of the Hodge
$*$-operator and Conserved Quantities in Some Classical Field Theories",
/Donev, S./, Physica Scripta, vol.25, 601 (1982).
\vskip 0.25cm
4. "A Particular Non-linear Generalization  of   Maxwell Equations Admitting
Spatially Localized Wave  Solutions",/Donev, S./, Compt.Rend.Bulg.Acad.Sci.,
vol.34, No.4 (1986).
\vskip 0.25cm
5. "A  Covariant  Generalization  of  Sine-Gordon Equation on Minkowski
Space-Time",/Donev, S./, Bulg.Journ.Phys.,vol.13, 295 (1986).
\vskip 0.25cm
6. "Geodesic  Vector Fields on  Minkowski Space-Time and  (3+1)-Solitary
Waves",/Donev, S./,Commun.JINR - Dubna, E2-88-107.
\vskip 0.25cm
7. "Autoclosed Differential Forms and (3+1)-Solitary Waves"/Donev, S./,
Bulg.Journ.Phys.,vol.15, 419 (1988).
\vskip 0.25cm
8. "On the Description of
Single Massless Quantum Objects" /Donev, S./,  Helvetica Physica Acta, vol.65,
910 (1992).
\vskip 0.25cm
9. "Energy-Momentum Directed Nonlinearization
of  Maxwell's Pure   Field Equations",/Donev, S., Tashkova, M./,
Proc.R.Soc.Lond.A 443, 301,  (1993).
\vskip 0.25cm
10. "Energy-Momentum
Directed Nonlinearization of Maxwell's  Equations in the Case of a
Continuous Medium" /Donev, S., Tashkova, M./,   Proc.R.Soc. Lond.A 450, 281
(1995).
\vskip 0.25cm
11. "Extended Electrodynamics: I. Basic Notions,
Principles and Equations", /Donev, S., Tashkova, M./,
 Annales de la Fondation Louis de Broglie, vol.23, No.2, 1998
\vskip 0.25cm
12. "Extended Electrodynamics: II. Properties and invariant characteristics
of the non-linear vacuum solutions", /Donev, S., Tashkova, M./,
Annales de la Fondation  Louis de Broglie, vol.23, No.3, 1998
\vskip 0.25cm
13. "Extended Electrodynamics: III. Free Photons and
$(3+1)$-Soliton-like Vacuum Solutions", /Donev, S., Tashkova, M./,
 Annales de la Fondation Louis de Broglie, vol.23, No.4, 1998
\vskip 0.25cm
\newpage
14. "How to Describe Photons as (3+1)-Solitons", /Donev,S., Trifonov,D./,
in "Complex Analysis, Differential Geometry, Mathematical Physics and
Applications", ed. by Sekigawa, K., Dimiev, St., World Scientific, 1999,
pp.246-261; arXiv: physics/9812009
\vskip 0.25cm
15. "EXTENDED ELECTRODYNAMICS: Basic Equations and Photon-Like (3+1)-Soliton
Solutions", /Donev, S./, in "Photon: Old Problems in Light of New Ideas", ed.
V. Dvoeglazov, Nova Science Publishers, 1999, pp.32-56.
\vskip 0.25cm
16. "A New Look on Electromagnetic Duality. Suggestions and Developments.",
/Donev, S./, Annales de la Fondation Louis de Broglie, vol.27, No.4 (2002), pp.
621-640, arXiv: hep-th/0006208
\vskip 0.25cm
17. "From Electromagnetic Duality to Extended Electrodynamics",
/Donev, S./, Annales de la Fondation Louis de Broglie, vol.29, No.3 (2004),
pp.375-392; arXiv: hep-th/0101137
\vskip 0.25cm
18. "Screw Photon-Like (3+1) Solitons in Extended Electrodynamics", /Donev, S./,
The EPJ "B", vol. 29, No.2 (2002),pp.233-236 (a larger version: arXiv:
hep-th/0104088)
\vskip 0.25cm
19. "On the Structure of the Nonlinear Vacuum Solutions in EED",  /Donev, S./,
arXiv: hep-th/0204217
\vskip 0.25cm
20. "Parallel Objects and Field Equations", /Donev, S./, arXiv:/math-ph/0205046
\vskip 0.25cm
21. "Structure and Spin of Photon-like Objects in Extended Electrodynamics",
/Donev,S./, Hadronic Journal, vol.26, No.3-4, pp.523-536 (2003)
\vskip 0.25cm
22. "Generlized Parallelism and Field Equations", /Donev,S., Tashkova,M./,
Contemporary Aspects of Complex Analysis, Differential Geometry and
mathematical Physics, pp.49-63 (2005), Proc. of 7th Intern.Workshop on Complex
structures and vector Fields, August-September 2004, Plovdiv, Bulgaria;
\vskip 0.25cm
23. "Extended Electrodynamics: A Brief Review", /Donev.S., Tashkova,M./,
arXiv: hep-th/0403244
\vskip 0.25cm
24. "Extended Objects in Minkowski Space-time", /Donev,S., Tashkova,M./, Prof.
G.Manev's Legacy in Contemporary Astronomy, Theoretical and Gravitational
Physics, Ed.by V,Gerdjikov and M.Tsvetkov, pp.318-330, Heron press, Sofia, 2005
\vskip 0.25cm
25. "Integrability-Nonintegrability Structures and Individual Photons'
Description as Finite Field Objects", /Donev, S., Tashkova, M./, arXiv:
hep-th/058091 (2005)
\vskip 0.25cm
26. "Complex Structures in Electrodynamics", /Donev,S., Tashkova,M./,
Journal of Geometry and Symmetry in Physics, {\bf 7} (2006),pp.13-36,
arXiv: math-ph/0106008
\vskip 0.25cm
27. "From Maxwell Stresses to Nonlinear
Field Equations",/Donev,S., Tashkova,M./, arXiv :physics/0604021
\vskip 0.25cm
28. "Integrability, Curvature and Description of Photon-like Objects",
/Donev,S., Tashkova,M./,
Contemporary Aspects of Complex Analysis, Differential Geometry and
mathematical Physics, pp.57-65, (2007),
Proc. of 8th Intern.Workshop on Complex structures and vector Fields,
August 2006, World Scientific, Sofia,Bulgaria,
\vskip 0.25cm
29. "Frobenius Curvature, Electromagnetic Strain and Description of Photon-like
Objects", /Donev,S., Tashkova,M./, arXiv: hep-th/0705.4170
\vskip 0.25cm
30. "Nonlinear Connections and Description of Photon-like Objects"
/Donev, S., Tashkova, M./, in Geometry, Integrability and Quantization, Ed.by
I.Mladenov, Softex 2008, Proc. of the 9th Conference, June 2007, Varna,
Bulgaria; arXiv: math-ph/0806.4058
\vskip 0.25cm
31. "Relativistic Strain
and Electromagnetic Photon-like Objects", /Donev, S., Tashkova, M./, in 'Trends
in Differential Geometry, Complex Analysiss and mathematical Physics', Ed. by
K.Sekigawa, V. Gerdjikov, S. Dimiev, Proc.9-th Int.Workshop, August 25-29,
2008, Sofia, Bulgaria, World Scientific 2009.
\vskip 0.25cm
32. "From Maxwell
Stresses to Photon-like Objects through Frobenius Curvature Geometrisation of
Local Physical Interaction", /Donev, S., Tashkova, M./,
arXiv :math-ph/0902.3924
\vskip 0.25cm
33. "From Maxwell
Stresses to Photon-like Objects" (Toward dynamical interpretation of Frobenius
Nonintegrability), /Donev, S., Tashkova, M./, Scientific Monograph, VDM Verlag
Dr.Muller (Scientific Publishing House Ltd.) 2010, Germany, ISBN:
978-3-639-23542-5,
\vskip 0.25cm
34. "On the Homology Defined by the Electromagnetic Energy
Tensor", /Donev, S., Tashkova, M./, International Workshop on Complex
Structures, Integrability and Vector Fields, September 13-17, 2010, Sofia,
Bulgaria, ed. E.Sekigawa,... AIP/Conf. Proc.-1340, Melville,
New York, 2011, pp.23-31.
\vskip 0.25cm
35. "Curvature forms and interaction of fields", /Donev, S., Tashkova, M./,
Journal of Geometry and Symmetry in Physics, vol.21, pp.41-59, 2011
\vskip 0.25cm
36. "A nonlinear prerelativistic approach to mathematical representation
of vacuum electromagnetism", /Donev, S., Tashkova, M./, arXiv: hep-ph/1303.2808.
 \vskip 0.25cm
37. "A nonlinear relativistic approach to mathematical representation of
vacuum electromagnetism based on extended Lie derivative", /Donev, S.,
Tashkova, M./, arXiv: hep-ph/1303.3451

\newpage
\begin{center}
 STUDIES ON NONLINEARIZATION OF \\MAXWELL VACUUM EQUATIONS
\end{center}
\vskip 0.2cm
\addcontentsline{toc}{section}{{\bf Important papers on nonlinearization of
Maxwell equations}}
\vskip 0.3cm

[1]. {\bf M. Born, L. Infeld}, {\it Nature}, {\bf 132}, 970 (1932)

[2]. {\bf M. Born, L. Infeld}, {\it Proc.Roy.Soc.}, {\bf A 144}, 425 (1934)

[3]. {\bf E. Schrodinger}, {\it Contribution to Born's new theory of
electromagnetic feld}, Proc. Roy. Soc. Lond. {\bf A 150}, 465 (1935).

[4]. {\bf W. Heisenberg, H. Euler}, {\it Zeit.Phys.}, {\bf 98}, 714 (1936)

[5]. {\bf M. Born}, {\it Ann. Inst. Henri Poincare}, {\bf 7}, 155-265 (1937).

[6]. {\bf J. Schwinger}, {\it Phys.Rev}. ,{\bf 82}, 664 (1951).

[7]. {\bf H. Schiff}, {\it Proc.Roy.Soc.} {\bf A 269}, 277 (1962).

[8]. {\bf J. Plebanski}, {\it Lectures on Nonlinear Electrodynamics}, NORDITA,
Copenhagen, 1970.

[9]. {\bf G. Boillat}, {\it Nonlinear Electrodynamics: Lagrangians and
Equations of Motion}, J.Math.Phys. {\bf 11}, 941 (1970).

[10]. {\bf B. Lehnert, S. Roy}, {\it Extended Electromagnetic Theory}, World
Scientific, 1998.

[11]. {\bf D.A. Delphenich}, {\it Nonlinear Electrodynamics and QED},
arXiv:hep-th/0309108, (good review article).

[12]. {\bf B. Lehnert}, {\it A Revised Electromagnetic Theory with Fundamental
Applications}, Swedish Physic Arhive, 2008.

[13]. {\bf D. Funaro}, {\it Electromagnetism and the Structure of Matter},
Worldscientific, 2008; also: {\it From photons to atoms}, arXiv: gen-ph/1206.3110
(2012).

[14]. {\bf G. Gibbons, D. Rasheed}, {\it Electric-magnetic duality rotations in
non-linear electrodynamics}, Nucl. Phys. {\bf B 454} 185 (1995) hep-th/9506035.

[15] {\bf R. Kerner, A.L. Barbosa, D.V. Gal'tsov}, {\it Topics in Born-Infeld
Electrodynamics}, arXiv: hep-th/0108026 v2

[16]. {\bf A. Sowa}, arXiv: physics/0103061

\newpage
\addcontentsline{toc}{section}{{\bf Index}}
\printindex
\vskip 0.3cm

%\printindex

\end{document}